\documentstyle[11pt,leqno]{book} 
\newcommand{\nc}{\newcommand} 
\nc{\dps}{\displaystyle} 

\newtheorem{theorem}{Theorem}

\newcommand{\loota}{\hbox{\enspace{\vrule height 7pt depth 0pt width 
      7pt}}} 
\nc{\RR}{\mbox{\rm I$\!$R}}

 \newcommand{\beqn}{\begin{eqnarray}} 
 \newcommand{\eeqn}{\end{eqnarray}} 
 \newcommand{\be}{\begin{equation}} 
 \newcommand{\ee}{\end{equation}} 
 \newcommand{\ba}{\begin{array}} 
 \newcommand{\ea}{\end{array}} 
  
 \newcommand{\pa}{\partial} 
 \newcommand{\re}{\ref} 
 \newcommand{\ci}{\cite} 
 \newcommand{\ds}{\displaystyle} 
 \newcommand{\la}{\label} 
 \newcommand{\bfr}{\begin{flushright}} 
 \newcommand{\efr}{\end{flushright}} 
  
 \newcommand{\rIm}{{\rm Im\5}} 
 \newcommand{\rRe}{{\rm Re\5}} 
\newcommand{\bfl}{\begin{flushleft}} 
\newcommand{\efl}{\end{flushleft}} 
\newcommand{\fr}{\frac} 
\newcommand{\spec}{{\rm spec\5}} 
 
\newcommand{\ov}{\overline} 
 
\newcommand{\lra}{\leftrightarrow} 
\newcommand{\ti}{\tilde} 
\newcommand{\rot}{{\rm rot~}} 
\newcommand{\dv}{{\rm div~}} 
\newcommand{\co}{{\rm const}} 
\newcommand{\bo}{{\hfill\loota}} 
\newcommand{\supp}{{\rm supp~}} 
\renewcommand{\Pr}{{\bf Proof~}} 
\newcommand{\bh}{{\bf h}} 
\newcommand{\n}{{\bf n}} 
\newcommand{\e}{{\bf e}} 
\newcommand{\x}{{\bf x}} 
\newcommand{\y}{{\bf y}} 
\newcommand{\p}{{\bf p}} 
\newcommand{\s}{{\bf s}}\newcommand{\bbl}{{\bf l}} 
\newcommand{\w}{{\bf w}}\newcommand{\bu}{{\bf u}} 
\newcommand{\bk}{{\bf k}} 
\newcommand{\m}{{\bf m}}\newcommand{\g}{{\bf g}} 
\newcommand{\bv}{{\bf v}} 
\newcommand{\q}{{\bf q}} 
\newcommand{\X}{{\bf X}} 
\newcommand{\f}{{\bf f}} 
\newcommand{\da}{{\dagger}} 
\newcommand{\A}{{\cal A}} 
\newcommand{\bA}{{\bf A}} 
\newcommand{\bB}{{\bf B}} 
\newcommand{\bC}{{\bf C}} 
\newcommand{\bD}{{\bf D}} 
\newcommand{\bH}{{\bf H}} 
 
\newcommand{\bM}{{\bf M}}\newcommand{\bG}{{\bf G}} 
\newcommand{\bP}{{\bf P}} 
\newcommand{\cH}{{\cal H}} 
\newcommand{\E}{{\cal E}} 
\newcommand{\bE}{{\bf E}}\newcommand{\bF}{{\bf F}} 
\newcommand{\bS}{{\bf S}} 
\newcommand{\I}{{\cal I}} 
\newcommand{\J}{{\cal J}} 
\newcommand{\cL}{{\cal L}} 
\newcommand{\bL}{{\bf L}} 
\newcommand{\M}{{\cal M}} 
\newcommand{\cO}{{\cal O}}\newcommand{\cP}{{\cal P}} 
 
\newcommand{\F}{{\cal F}} 
\newcommand{\ccT}{{\cal T}} 
\newcommand{\cS}{{\cal S}} 
\newcommand{\V}{{\cal V}} 
\newcommand{\bj}{{\bf j}} 
\newcommand{\bJ}{{\bf J}} 
\newcommand{\eps}{\epsilon}\newcommand{\ka}{\kappa} 
\newcommand{\ve}{\varepsilon} 
\newcommand{\vp}{\varphi} 
\newcommand{\we}{\wedge} 
\newcommand{\De}{\Delta} 
\newcommand{\de}{\delta} 
 
\newcommand{\al}{\alpha} 
\newcommand{\ga}{\gamma} 
\newcommand{\Ga}{\Gamma} 
\newcommand{\si}{\sigma} 
\newcommand{\om}{\omega} 
\newcommand{\Om}{\Omega} 
\newcommand{\na}{\nabla} 
\newcommand{\Si}{\Sigma} 
\newcommand{\lam}{\lambda} 
\newcommand{\Lam}{\Lambda} 
 \newcommand{\h}{{h^{\hspace{-2.5mm}-}}} 
 \newcommand{\tr}{{\rm tr}} 
\newcommand{\5}{{\hspace{0.5mm}}} 

\newcommand{\bre}{|\kern-.15em|\kern-.15em|} 
 \def\R{{\rm I\kern-.1567em R}}                              % Doppel R 
 \def\C{{\rm C\kern-4.7pt                                    % Doppel C 
 \vrule height 7.pt width 0.4pt depth -0.5pt \phantom {.}}\5} 
 \def\Z{{\sf Z\kern-4.5pt Z}}                                % Doppel Z 

\newtheorem{qtheorem}{QTheorem}[section]

\newtheorem{defin}[theorem]{Definition} 
 
\newtheorem{lemma}[theorem]{Lemma} 
\newtheorem{example}[theorem]{Example} 
\newtheorem{exercice}[theorem]{Exercise} 
\newtheorem{remark}[theorem]{Remark} 
\newtheorem{remarks}[theorem]{Remarks} 
\newtheorem{cor}[theorem]{Corollary} 
\newtheorem{pro}[theorem]{Proposition} 
\newtheorem{com}[theorem]{Comment} 
\newtheorem{coms}[theorem]{Comments}

\newcommand{\bd}{\begin{defin}} 
 \newcommand{\ed}{\end{defin}} 
\newcommand{\bt}{\begin{theorem}} 
 \newcommand{\et}{\end{theorem}} 
\newcommand{\bqt}{\begin{qtheorem}} 
 \newcommand{\eqt}{\end{qtheorem}}

\newcommand{\bp}{\begin{pro}} 
 \newcommand{\ep}{\end{pro}} 
 
\newcommand{\bl}{\begin{lemma}} 
 \newcommand{\el}{\end{lemma}} 
\newcommand{\bc}{\begin{cor}} 
 \newcommand{\ec}{\end{cor}} 
 
\newcommand{\bex}{\begin{example}} 
 \newcommand{\eex}{\end{example}} 
\newcommand{\bexs}{\begin{examples}} 
 \newcommand{\eexs}{\end{examples}}

\newcommand{\bexe}{\begin{exercice}} 
 \newcommand{\eexe}{\end{exercice}}

\newcommand{\br}{\begin{remark} } 
 \newcommand{\er}{\end{remark}} 
\newcommand{\brs}{\begin{remarks}} 
 \newcommand{\ers}{\end{remarks}} 
 
\newcommand{\bcom}{\begin{com}} 
\newcommand{\ecom}{\end{com}} 
\newcommand{\bcoms}{\begin{coms}} 
\newcommand{\ecoms}{\end{coms}}

\newcommand{\ctg}{\mathop{\rm ctg}\nolimits} 
 
\newcommand{\const}{\mathop{\rm const}\nolimits}

\begin{document}

%%\cleardoublepage 

\begin{center}  
{\huge\bf Lectures on Quantum Mechanics}  
\vspace{4mm}\\
{\Large\bf (nonlinear PDE point of view)} 
\vspace{1cm}\\ 
 
\bigskip
{\Large\bf A.I.Komech}
\\ 
\vspace{5mm} 
 {\large\bf Wien 2002/2004} 
 
\end{center} 
\bigskip

\begin{center}
{\Large\bf Abstract} 
\end{center} 

We expose the Schr\"odinger  quantum mechanics
with traditional applications to Hydrogen atom: 
the calculation of the Hydrogen atom spectrum
via  Schr\"odinger, Pauli and Dirac equations,
the Heisenberg representation,
the  selection rules,  
the calculation of
quantum and classical scattering of light (Thomson cross section), 
photoeffect (Sommerfeld cross section), 
quantum and classical scattering of electrons (Rutherford cross section), 
normal and anomalous Zeemann effect (Land\'e factor),
polarization and dispersion (Kramers-Kronig formula), 
diamagnetic susceptibility (Langevin formula).

We discuss carefully the experimental and theoretical 
background for the introduction of the Schr\"odinger,
Pauli and Dirac equations, as well as for the 
Maxwell equations.
We explain in detail
all basic theoretical concepts: 
the introduction of the 
quantum stationary states,
charge density 
and electric current density,
quantum magnetic moment, electron spin
and spin-orbital coupling in ``vector model'' and
in the Russel-Saunders approximation,
differential cross section of scattering,
the Lorentz theory 
of polarization and magnetization,
the Einstein special relativity
and  covariance of the Maxwell Electrodynamics.

We explain all details of the calculations
and mathematical tools: 
Lagrangian and Hamiltonian formalism 
for the systems with finite degree of freedom and 
for fields, 
Geometric Optics, the Hamilton-Jacobi equation and WKB approximation,
Noether theory of invariants including the theorem on currents,
four conservation laws  (energy, momentum, angular momentum and charge),
Lie algebra of angular momentum and spherical functions, 
scattering theory (limiting amplitude principle and limiting absorption
principle), 
the Lienard-Wiechert formulas,
Lorentz group and Lorentz formulas, Pauli theorem
and relativistic covariance of the Dirac equation, etc.

We give a detailed oveview of the 
conceptual development of the 
quantum mechanics, and expose main achievements of the 
``old quantum mechanics'' in the form of exercises.

One of our basic aim in writing this book,
is an open and concrete discussion of the problem 
of a  mathematical description of 
the following two fundamental 
quantum phenomena:
i) Bohr's quantum transitions and ii) de Broglie's 
wave-particle duality. 
Both phenomena cannot be described by 
autonomous linear dynamical equations, and 
we give them a new mathematical treatment 
related with  recent progress
in the theory of global attractors of nonlinear
hyperbolic PDEs. 
Namely, we suggest that i) the quantum stationary states form a global 
attractor of the coupled Maxwell-Schr\"odinger or 
Maxwell-Dirac equations, in the presence of
 an external confining potential, 
and ii) 
the wave-particle duality corresponds to the soliton-like asymptotics
for the solutions of the translation-invariant 
coupled equations without an external potential.

We emphasize, in the whole of our exposition, that
the coupled equations are nonlinear, and just this 
nonlinearity lies behind all traditional perturbative
calculations that is known as the {\it Born approximation}.
We suggest that 
both fundamental quantum phenomena could be described by this 
nonlinear coupling. The suggestion is confirmed by 
recent results on  the global attractors and soliton asymptotics 
for model nonlinear
hyperbolic PDEs.

%%\authorhead{Alexander Komech} 
 
%%\titlehead{Classical Fields and Quantum Mechanics} 
%%\RRnbpage{87} 
%%\RRtheme{4} 
 
%%\RRprojet{Ondes} 
%%\RRdate{Fevrier 2003} 
 
%%\URRocq 
%%\makeRR 
 
\newpage 
\tableofcontents 
 
%   \textwidth =165mm 
%  \textheight =236mm 
%  \topmargin =-15mm 
%  \oddsidemargin =-1cm 
%  \evensidemargin =-1cm 

%%\input{comm} 

%%%%%%%%%%%%%%%%%%%%%%%%%%%%%%%%%%%%%%%%%%%%%%%% 
%%%%%%%%%%%%%%%%%%%%%%%%%%%%%%%%%%%%%%%%%%%%%% 
 
 \newpage 
\setcounter{equation}{0} 
\setcounter{section}{-1} 
\section 
{Preface} 
\setcounter{section}{0} 
These lectures 
correspond to a three-semester 
course 
given by the author
 at the Faculty of Mathematics of the 
 Vienna University in the academic years 2002/2003 and 2003/2004. 
We expose well known Schr\"odinger  quantum mechanics
with traditional applications to Hydrogen atom,
but the form 
of the exposition is intended for a mathematically 
oriented reader of a graduate level. 
 
The reader might ask why this new textbook could be useful 
as there are many other well established introductions to quantum mechanics. 
Let us explain our motivations 
in writing this book. 

Our principal aim is to give a reasonable introduction which
provides a unified mathematical strategy in applications 
to different problems. 
Many modern textbooks mainly focus on 
calculations of `matrix elements' of a more formal nature, 
which are referred to as a {\it first approximation} 
of a perturbative procedure.
We go one small step further in 
discussing the full nonperturbative problems which 
form the background of such calculations. 
This makes the strategy of the applications 
of the theory to specific problems more transparent. 

 Almost all existing texts avoid dealing with such questoins,
 merely mentioning them.
 This makes the subject less accessible for understanding,
 and also impoverishes both its mathematical
 and physical aspects.
 On the other hand,
the recognition of the status of the 
original nonperturbative problems leads to many 
questions which are open mathematical problems at the moment. 
Some of them are suggested by the Heisenberg Program \ci{Heis}.
The open discussion of these problems is also one of the principal aims 
of this book. 
 \medskip\\ 
{\bf On Open Questions} 
Quantum Mechanics 
exists as an axiomatic theory operating with Quantum Stationary States, 
Elementary Particles, Probabilities, etc. 
It is well established as 
a set of rules which give an excellent description 
of many experimental facts: atomic and molecular spectra, 
light and particle scattering, 
periodic system of elements, chemical reactions, etc. 
However, a rigorous foundation, i.e. a mathematical model 
of the axiomatics, is unknown at the moment 
because there are many open mathematical questions. 
The cornerstone of the theory is 
Schr\"odinger's dynamical equation
$$ 
(i\h\pa_t-e\phi(\x))\psi(t,\x) 
=\fr 1{2\mu} (-i\h\na_\x-\ds\fr ec \bA(\x))^2\psi(t,\x) \eqno{(S)} 
$$ 
(or, analogously, the Klein-Gordon, Dirac Eqn, etc) for the wave function
$\psi(t,\x)$, 
where $\phi(\x)$ and $\bA(\x)$ are external 
electrostatic and magnetic vector potentials, respectively.

 In a number of important particular cases,
 this equation could be solved exactly.
 Besides, the perturbation theory of the first, second,
 or fourth orders sometimes allows
 to find an approximate solution with an amazing accuracy.
 This gives an impression that, in principle,
 if one did not restrict oneself with the first
 order approximation,
 then the computations could be carried out
 with an arbitrarily high precision.
 Yet, this impression leads to a very deep
 confusion, since many of the fundamental quantum effects
 could not be described by this linear equation.
Let us mention  some of them.
\\ 
{\bf I.} Bohr's 
transitions between Quantum Stationary States, 
\be\la{Bji} 
 |E_-\rangle\,\,\mapsto\,\,|E_+\rangle, 
\ee 
{\bf II.} De Broglie's 
Wave-Particle Duality: diffraction of the electrons, etc.
\\ 
{\bf III.} 
Born's probabilistic interpretation of the wave function.
\smallskip

The transitions (\re{Bji}) 
 are responsible for the spectra of 
atom radiation which coincide with the eigenvalues of the 
stationary equation corresponding to $(S)$. 
However the transitions are not an inherent property of 
the solutions of the linear equation $(S)$.

Similarly, 
the equation  $(S)$ explains 
the diffraction pattern 
 in the Davisson-Germer "double-slit" experiment
by the Bragg rules.
However, the {\it discrete registration of electrons} 
(known as "reduction of wave packets"),
 with the {\it counting rate} corresponding to this pattern, 
is not related to a property of the solutions. 
Note that the stability of elementary particles 
is not explained yet 
as was pointed out by Heisenberg \ci{Heis}
 (see \ci{BL,FGJS, FTY} for a recent progress 
in this direction for nonlinear equations).

Finally, Born
identified $|\psi(t,\x)|^2$
with the density of probability just to explain 
the  counting rate in the  Davisson-Germer experiment.
The identification  plays the key role 
in quantum mechanical scattering problems. 
However, it has never been 
justified in the sense of Probability Theory. 
\smallskip
 
Among other open questions: 
the explanation of statistic equilibrium in
atom radiation (see Comment \re{rstat}), in the photoeffect
(see Comment \re{pstat}), 
etc.
\medskip\\ 
{\bf On Dynamical Treatment} 
The common strategy in applications of Quantum Mechanics is a skilful 
combination 
of the postulates {\bf I}-{\bf III} 
with the dynamical equation $(S)$ in the description of 
various quantum phenomena. The strategy 
is not mathematically selfconsistent 
since the postulates formally are not compatible with the 
linear autonomous equation $(S)$. 
Hence, the equation 
 requires a suitable (nonlinear) 
modification which would imply 
the postulates as inherent 
properties of the modified dynamics.

Fortunately, an obvious choice for the nonlinear version is well
established: it is the system of coupled 
Maxwell-Schr\"odinger equations. 
The coupled equations are known 
since the Schr\"odinger paper \ci[IV]{Sch}, 
where the charge and current densities have 
been expressed in terms of the wave function. 
 Namely, the static potentials $\phi(\x)$ and $\bA(\x)$ in $(S)$ 
should be replaced by the time-dependent potentials 
$\phi(t,\x)$ and $\bA(t,\x)$ which obey the 
Maxwell equations with the charge and current densities corresponding 
to the Schr\"odinger equation. 
The coupled  equations 
 constitute a {\bf nonlinear system} for the electron wave function 
and the Maxwell field, though the Schr\"odinger equation is linear 
with respect to the wave function. 
The coupling is inevitable in a description of the atom spectra 
since the spectral lines correspond to the wavelengths of 
the atom electromagnetic 
radiation which is a solution to the Maxwell equations. 
Hence, the coupled equations give an authentic  
framework within which 
Quantum Mechanics, or at least some of its aspects, may be 
mathematically selfconsistent
formulated.

We suggest a new treatment of basic quantum phenolena
based on the coupled  
nonlinear dynamical equations. Namely,
the phenomena  {\bf I} and {\bf II} 
inspire the following {\it dynamical treatment} respectively: 
\\ 
{\bf A.} The  transitions (\re{Bji}) can be treated mathematically 
as the long-time asymptotics of the solutions to the coupled
equations,
\be\la{Bjii} 
(\psi(t,\x), A(t,\x))\sim 
(\psi_\pm(\x)e^{-i\om_\pm t}, A_\pm(\x)), ~~~~~~~t\to\pm\infty. 
\ee 
Here $A(t,\x)=(\phi(t,\x), \bA(t,\x))$, and
the 
limit functions $(\psi_\pm(\x)e^{-i\om_\pm t}, A_\pm(\x))$ 
correspond to the 
stationary states $|E_\pm\rangle$. 
The asymptotics would mean that the set of 
all Quantum Stationary States 
is the {\bf global point attractor} of the coupled dynamical equations
(see \ci{BV,Hale, Hen, Te}).
\\ 
{\bf B.} 
The  elementary particles seem to correspond to 
traveling waves (or "solitons") which are the solutions of type
$(\psi(\x-\bv t)e^{i\Phi(\bv,\x,t)},A(\x-\bv t))$, to the coupled equations. 
Respectively, the 
de Broglie's wave-particle duality 
can be treated mathematically as the 
soliton-type asymptotics 
\be\la{scatai} 
(\psi(t,\x), A(t,\x))\sim 
\ds\sum\limits_{k} (\psi^k_\pm(\x-\bv_\pm^k t) 
e^{i\Phi(\bv_\pm^k,\x,t)},A^k_\pm(\x-\bv_\pm^k t)), 
~~~~t\to\pm\infty.
\ee 
 
Note that the asymptotics {\bf A} correspond to the {\it bound system},
with an external confining potential, like atom with a Coulomb nuclear
potential, while {\bf A} correspond to the translation invariant systems,
witout an external potential. More detailed asymptotics would include also
a dispersive wave which is a solution to the corresponding 
{\it free} linear
system (see \ci{Kanwe}). The dispersive wave in (\re{Bjii}) 
should describe the electromagnetic radiation
(the ``photon emission'') following the quantum transitions for an atom:
the radiation of the wave function is impossible by the charge 
conservation and the neutrality of the atom.

The long time asymptotics of type {\bf A} and {\bf B} are 
proved at present time 
for some model nonlinear hyperbolic partial differential equations 
(see below).
For the coupled Maxwell-Schr\"odinger and Maxwell-Dirac equations,
the  
proving of the asymptotics {\bf A} and {\bf B} are open problems. 
Only 
the existence of the solutions is proved 
for the coupled equations,
\ci{Bour, GNS}, and 
the existence of solitons is proved for the Maxwell-Dirac Equations, 
\ci{EGS}. 
Note that 
the coupling of the fields is the main subject of 
Quantum Field Theory: Quantum Electrodynamics, etc. 
We consider only the {\it semiclassical} coupled equations and 
do not touch the {\it second quantized version} which is 
the subject of 
{\it Quantum Electrodynamics}. 
The coupled second quantized Maxwell-Dirac system of 
Quantum Electrodynamics is also nonlinear 
(see \ci{BD,IZ,Sak, Scharf}). 
The corresponding analogs of the conjectures 
are also open questions, and are very likely more difficult than 
in the semiclassical context. 

A mathematical treatment of the Born 
statistical interpretation  {\bf III} is still unknown.
Note that the Born identification of the 
counting rate with a probability is equivalent to the 
ergodicity of the corresponding random process.
The 
statistical description is also necessary in the analysis 
of atom radiation (see Comments \re{rstat} and \re{pstat}). 
\medskip\\
%%%%%%%%%%%%%%%%%%%%%%%%%%%%%%%%%%%%%%%%%%%%%%%
{\bf On Perturbative Approach}
The coupled equations are commonly used by perturbation 
techniques. 
The formal, perturbative approach is very 
successful as far as the `classical' quantum mechanical results 
on the electron-nucleon interactions through the Maxwell field 
are concerned.
Our above discussion mainly serves to 
complement this approach with a slightly more mathematical viewpoint. 
Our suggestion is equivalent to the explicit 
recognition of the status of the coupled dynamics which is 
responsible for the Quantum Transitions and Wave-Particle Duality. 
The recognition is inevitable since the both phenomena 
seem to be  genuine nonperturbative properties of the 
coupled nonlinear equations.

Classical textbooks (for example, \ci{Born,Schiff,Som}) 
also use the coupled equations implicitly. 
That is, they treat both equations 
separately which corresponds to a perturbation approach 
for the coupled system. 
We follow the same strategy {\it explicitly}
adding certain {\it comments} 
on possible relations 
to the 
coupled system and the suggested 
long-time asymptotics {\bf A} and {\bf B}. 
For example, 
\\ 
$\bullet$ The asymptotics {\bf A} 
clarify Schr\"odinger's identification of the 
Quantum Stationary States with eigenfunctions. 
\\ 
$\bullet$ 
The asymptotics {\bf B} 
claim an inherent mechanism of the "reduction of wave packets". 
It clarifies  the de Broglie's wave-particle duality 
in the Davisson-Germer diffraction experiment 
and in the 
description of the electron beam 
by plane waves.

The great success of the perturbative approach to the 
electron-nucleon interactions, however, crucially depends on 
the following two main facts: 
\\ 
i) the linear part of the coupled 
Maxwell-Schr\"odinger equations is 
completely known from the Rutherford experiment, 
which both detects the universal fact of the positive charge concentration and 
uniquely identifies the Coloumbic potential of the nuclei, 
\\ 
ii) the nonlinear terms are `small' due to the  smallness of the 
Sommerfeld coupling constant $\al\approx 1/137$, thereby  providing the 
numerical 
convergence of the perturbation series (corresponding to 
the {\it Feynman diagrams} 
in the {\it second-quantized version}). 
 
Both of these facts no longer hold in the case of the strong nuclear 
interaction. Therefore, a genuine nonlinear approach might be 
necessary even from a more phenomenological point of view in this 
latter case. 
\medskip\\ 
%%%%%%%%%%%%%%%%%%%%%%%%%%%%%%%%%%%%%%%%%%%%%%%%%%%%%%%%
{\bf On a Mathematical Justification of the Asymptotics} 
It is natural to think that the long-time asymptotics 
of the type {\bf A} and {\bf B} are common features of a very 
general class of nonlinear Hamiltonian equations. 
Otherwise, the dynamical equations of Quantum Mechanics 
would be very exceptional that does not correspond to 
the universal character of the physical theory. 
 
Note that the soliton-type asymptotics of type {\bf B} 
have been discovered initially for {\it integrable equations}: 
KdV, sine-Gordon, cubic Schr\"odinger, etc. 
(see \ci{NMPZ} for a survey of these results). 
Let us list the results  on the asymptotics of type {\bf A} and   {\bf B} 
obtained in the last decade 
for nonintegrable nonlinear Hamiltonian equations. 

The asymptotics have been proved 
for 
all finite energy solutions of 
i) 1D nonlinear wave  equations 
(see \ci{K1}-\ci{K3}) 
for a 1D singular relativistic 
Klein-Gordon equation, \ci{K10}), 
ii) nonlinear systems 
of 3D wave, Klein-Gordon and Maxwell equations  coupled to a classical 
particle (see  \ci{K11}-\ci{K12} and \ci{K5}-\ci{sp4}), 
iii) the 
Maxwell-Landau-Lifschitz-Gilbert Equations \cite{JKV} 
(see \ci{Kpla} for a survey of these results).

 For $U(1)$-invariant  3D nonlinear 
 Schr\"odinger equations, 
the first result on the attraction of type {\bf A} 
has been established in \ci{SW1,SW2}  (see also \ci{PW}). 
The first results on the asymptotics  of type {\bf B} 
have been established 
in \ci{BP, BS} for translation-invariant 
$U(1)$-invariant 1D nonlinear Schr\"odinger equations. 
The results \ci{BP, BS} are extended in \ci{Cu} 
to  the  dimension $n=3$. 
An extension to the relativistic-invariant 
nonlinear Klein-Gordon equation 
is still an open problem (see \ci{Cukin, PSW} for a progress in  this
direction). 
All the results  \ci{BP,BS,Cu,PW, SW1, SW2} concern 
  initial states which are {\it sufficiently close 
to the solitary manifold}.
 
The {\it global attraction}  of type {\bf A} 
of {\it all finite energy states} 
is established for the first time in 
\ci{Kis3} 
for the $U(1)$-invariant 1D nonlinear Klein-Gordon equation 
with 
a nonlinear interaction concentrated at one point.

In \ci{K5} 
an {\it adiabatic effective dynamics} is established, 
for solitons of a 3D wave equation 
coupled to a classical particle, 
in a {\it slowly varying external potential}. 
The effective dynamics explains the increment of the 
mass of the particle caused by its interaction 
with the field. 
The effective dynamics is extended in \ci{FGJS,FTY} to the 
solitons of the  nonlinear Schr\"odinger  and Hartree equations. 
An extension to relativistic-invariant 
equations 
is still an open problem. On the other hand, 
the existence of  solitons and 
the Einstein mass-energy identity for them are proved, 
respectively, 
in \ci{BL} and \ci{edks}, 
for general  relativistic nonlinear Klein-Gordon equations. 
Numerical experiments \ci{KVin} 
demonstrate that the soliton-type asymptotics 
{\bf B} 
hold  for ``all'' 1D relativistic-invariant equations, however the 
proof is still an open problem. 
 
There are also known results concerning stability of 
solitary waves for general nonlinear Hamiltonian equations, 
\ci{GSS}. A large variety of the stability results can be found in 
the survey \ci{SuSu} concerning 
the nonlinear  Schr\"odinger equations. 
 
Note that the mathematical theory of the linear 
Schr\"odinger and Dirac equations is well established now: 
see for example, \ci{BSh,GS,HS,JvN,RS,Thal}. 
However, the asymptotics {\bf A} and {\bf B} 
generally 
do not hold for the linear autonomous  Schr\"odinger and 
Dirac equations. 
%%%%%%%%%%%%%%%%%%%%%%%%%%%%%%%%%%%%%%%%%%%%%5
\medskip\\ 
{\bf Main Goals of the Exposition} 
We pursue the following two principal goals: 
\smallskip\\ 
{\bf I.} To explain why the theory has admitted its present form 
of a dynamical system described by the Schr\"odinger, Pauli or Dirac 
equations coupled to the Maxwell equations. 
\smallskip\\ 
We follow all details of the development of 
the coupled dynamical equations, 
from experimental facts to the related mathematical context. 
We introduce the Schr\"odinger equation as a wave equation 
for which rays coincide with trajectories 
of the Lorenz equation for the classical electron. 
This introduction is based on the {\it geometric optics} 
and WKB short wave asymptotics, and is close to the original 
Schr\"odinger's idea on the Hamilton {\it optical-mechanical analogy}, 
\ci[II]{Sch}. 
We explain in detail the geometric optics and 
the WKB asymptotics. For the introduction of the Pauli equation 
we analyze the double splitting in the Stern-Gerlach experiment, 
the Einstein-de Haas effect and the anomalous Zeemann effect. 
The Dirac equation corresponds to  a relativistic 
energy-momentum relation similarly to the Schr\"odinger equation 
which corresponds to a non-relativistic one. 
\smallskip\\ 
{\bf II.} To demonstrate that the coupled dynamical equations 
allow us to describe the basic quantum phenomena 
of interaction of matter and electromagnetic radiation 
as inherent properties of the dynamics 
if the asymptotics {\bf A} and {\bf B} would hold. 
\smallskip\\ 
Our analysis shows that the asymptotics would play the key role 
in a mathematical foundation of Quantum Mechanics. Our exposition 
cannot be rigorous when we solve the nonlinear 
coupled Maxwell-Schr\"odinger equations in the first order approximation. 
We try to be careful in the solution 
of the corresponding linear problems 
but we did not strain to be everywhere mathematically 
rigorous in order not to overburden the exposition.
\medskip\\ 
{\bf Among Other Novelties of the Exposition} are the following: 
\\
$\bullet$ 
An 
application of the Lagrangian formalism 
for the identification i) of the energy, momentum and angular momentum 
for the  Schr\"odinger equation (Lecture 6), ii) 
of the coupling of the  Maxwell and Schr\"odinger 
equations (Lecture 7, cf. \ci{Born, Schiff, Weyl}). 
\\ 
$\bullet$ 
The straightforward derivation of the {\it Rydberg-Ritz 
combination principle} and 
{\it intensities for the dipole radiation} 
from the coupled Maxwell-Schr\"odinger equations (Lecture 10). 
The derivation is a formalized version of the  approach \ci{Som}. 
\\ 
$\bullet$ 
An update version of the 
Lorentz theory of molecular polarization and magnetization (Lecture 42), 
necessary for the quantum theory of dispersion and diamagnetism. 
We follow mainly \ci{Bek} applying the theory of distributions. 
The theory gives a framework for quantum versions of 
the Kramers-Kronig dispersion theory  (Lecture 14) 
and the Langevin theory of diamagnetism (Lecture 18). 
\\ 
$\bullet$ 
The application of the {\it nonstationary scattering theory} 
to the explanation of the 
Einstein formula for the 
{\it photoeffect}, and the calculation 
of the {\it limiting amplitude} 
via the {\it limiting absorption principle} 
(Lecture 15, cf. \ci{Som}). 
We demonstrate the fundamental role 
of {\it retarded potentials} in this calculation.
\\ 
$\bullet$ A concise presentation of the Russell-Saunders 
theory of the 
coupling between the orbital and spin angular momentum 
(Lecture 21). We follow mainly \ci{Be,CS,LVM,Schiff} and comment on 
probable relations with the coupled Maxwell-Pauli equations. 
\\ 
$\bullet$ 
An update form of the 
Noether Theorem on Currents with the complete proof
(Lecture 40, cf. \ci{GF,Z}). 
\smallskip\\ 
We demonstrate a parallelism of quantum and classical description 
to clarify their relations and to motivate an introduction of the 
corresponding quantum phenomenology: for example, 
the classical and quantum description of diamagnetism, 
Zeemann effect, 
scattering of light and particles, the introduction 
of {\it differential cross section}, {\it magnetic moment}, etc. 
\smallskip\\ 
We explain carefully all details of the calculations 
and all necessary methods of modern Mathematical Physics: 
the Lagrange and Hamilton theory for the fields, 
the Maxwell Electrodynamics
and the Einstein Special Relativity, 
scattering theory and 
representation theory of the rotation group. 
Let us note that we consider only one-electron problems 
(Hydrogen atom, alkali atoms, etc). We 
do not touch multi-particle problems, 
Hartree-Fock methods, etc (see 
\ci{CS, DB, Lieb, SSA}). 
\medskip\\ 
{\bf Further Reading} Our main goal is a concise explanation 
of basic theoretical concepts. More technical details and 
a systematic comparison with the experimental data can be found 
in \ci{BeS,CS,Schiff,Som}. 
\medskip\\ 
{\bf Plan of Exposition} 
In the Introduction we describe the 
chronology 
of the conceptual development of Quantum Mechanics. 
Then  in 
Lectures 2-4 we provide the mathematical background 
for a concise introduction of  the Schr\"odinger equation 
in Lectures 5-7. 
Lectures 8-21 concern various applications of the 
Schr\"odinger and Pauli equations. 
The relativistic Dirac theory is exposed in Lectures 22-35, 
where we benefit a lot from the book of Hannabus \ci{Han}. 
In  Lectures 36-43 we collect 
mathematical appendices. 
In Sections 44-45 we solve numerous exercises
containing 
main achievements of the 
``old quantum mechanics''. 
Let us explain the plan of the lectures in some detail. 
\medskip \\ 
{\bf Quantum Chronology} 
The  genesis of the Schr\"odinger equation 
has been inspired by the lack of a matter equation in classical 
electrodynamics. 
The Schr\"odinger theory is the result of a synthesis 
of a theoretical development with 
 various experimental observations. 
The ``quantum chronology'' starts from 
Kirchhoff's {\it spectral law}  (1859) and the invention of the 
{\it vacuum tube}  by 
Croockes  (1870). 
The next main steps are the identification of the {\it cathode 
rays} 
in the vacuum tube 
with the electrons by Thomson (1897), and 
the fundamental Planck relation $E=\h\om$ (1901) 
inspired by the 
comparizon of the   
experimental Wien formula for the spectral density 
in the Kirchhoff law, with the Boltzmann-Gibbs canonical distribution. 
The Planck relation has been applied by 
Einstein to the photoeffect (1905). 
The  Einstein theory has been developed further by 
Bohr (1913) to explain the 
Rydberg-Ritz combination principle and the 
stability of 
Rutherford's planetar model for the 
atom. 
Finally, Bohr's theory of {\it atom stationary states} 
has lead to 
the {\it quantum rules} of 
Debye and Sommerfeld-Wilson (1913 and 1916) 
for the {\it action function}, 
which is a solution to the Hamilton-Jacobi equation. The quantum rules 
provide the atom stationary states and 
allow to reproduce some basic experimentally 
observed spectral lines.

In 1926 Schr\"odinger introduced an equation for the wave function 
 \ci{Sch}. The equation extends de Broglie's wave-particle theory 
(1922) from free particles to bound particles. 
De Broglie's theory is based on an algebraic argument 
relying on Einstein's special relativity and 
the Planck relation. 
Schr\"odinger's extension combines the algebraic argument 
with Hamilton's optical-mechanical analogy. 
The analogy is justified by the WKB short wave asymptotics 
for the solutions to the Schr\"odinger equation. 
That is, the 
 corresponding ``eikonal equation'' 
coincides with the Hamilton-Jacobi equation of the 
Maxwell-Lorentz 
theory for the classical electron in the Maxwell field. 
This means that the  Schr\"odinger equation describes the cathode 
rays 
as short-wave solutions. 
An important role of the 
Hamilton-Jacobi equation has been recognized previously related to the 
Debye-Sommerfeld-Wilson quantization rules for the action.

Formally, a wave equation of the  Schr\"odinger type 
can be 
introduced directly by an identification of the trajectories of the 
classical electron in the Maxwell field with the rays of 
short-wave solutions. However, 
the magnitude $\h$ of the corresponding small parameter, and 
the central role of 
Quantum Stationary States, can be recognized  only from 
the whole development, starting with the Kirchhoff spectral law and 
leading to 
de Broglie's wave-particle theory. 
\medskip\\ 
{\bf Mathematical Background} 
In Lectures 2-4 we expose 
 the 
Lagrangian field theory, 
and the Lagrangian form of Maxwell's electrodynamics. 
Lecture 5 concerns the {\it geometric optics} 
and short-wave WKB asymptotics. 
In Lecture 6 we introduce the Schr\"odinger equation, 
the  Heisenberg representation, and prove 
conservation laws. 
In 
Lecture 7 we derive the coupling of 
the Schr\"odinger wave function 
 to the Maxwell field through the Lagrangian formalism. 
%%%%%%%%%%%%%%%%%%%%%%%%%%%%%%%%%%%%%%%%%%%
\medskip\\ 
{\bf Applications} 
In Lectures 8-18 we apply 
the  Schr\"odinger-Maxwell equations to 
the 
derivation 
of the spectrum of the hydrogen atom,  the {\it dipole 
radiation} and  {\it selection rules}, 
the {\it differential cross sections} of the 
scattering of light and particles by an atom, 
the {\it refraction coefficient}, the {\it diamagnetism}, 
and  the {\it normal} Zeemann effect. 
Lecture 9 contains a complete theory of  {\it quantum angular momentum},
including the {\it representations} of the corresponding {\it Lie algebra}. 
In Lecture 15 we explain the {\it photoeffect} by the 
{\it limiting amplitude principle} for the scattering 
in the {\it continuous spectrum}. 
In  Lectures 19-21 we introduce 
the {\it Pauli equation} with 
{\it electron spin} 
and apply it to  the 
calculation of the 
{\it gyromagnetic ratio}. 
In Lectures 22-35 we expose 
the {\it special relativity} (Lorentz Transformations and 
Covariant Electrodynamics), introduce the relativistic {\it Dirac 
equation}, prove its covariance and the {\it Pauli theorem}, and 
calculate the spectrum of the hydrogen atom via the Dirac equation. 
\medskip\\ 
{\bf Mathematical Appendices} In Lectures 36-43 
we give an introduction to the 
Lagrange theory for finite-dimensional systems 
and for fields: 
variational principle, conservation of energy, momentum, 
angular momentum, 
Noether invariants, 
Hamilton equations, Hamilton-Jacobi theorem. 
In Lecture 40 we give a new simple proof of the Noether 
theorem on currents. 
In Lecture 42 we expose the 
Lorentz theory of molecular polarization and magnetization. 
In Lecture 43 we explain 
 the 
limiting amplitude principle,
limiting absorption principle
and the  role 
of retarded potentials 
for the calculation of limiting amplitudes. 
\bigskip\\ 
{\bf Exercises} In Part VIII, written by C.Adam,
 we collect the solutions 
to the related classical problems: the Kepler problem, 
Bohr-Sommerfeld quantization, energy and momentum in the 
Maxwell field, 
electromagnetic plane waves and Fresnel's formulae, 
Hertzian dipole radiation, 
vector model for the spin-orbital coupling, 
etc. 
\bigskip\\ 
{\bf Acknowledgments} 
I thank H.Brezis, I.M.Gelfand,  P.Joly,
J.Le\-bo\-witz, E.Lieb, I.M.Sigal, 
A.Shni\-rel\-man, M.Shubin, H.Spohn, M.I.Vishik and E.Zeidler 
for fruitful discussions. I thank very much Dr. C.Adam who wrote the 
Part VIII, for the collaboration in running the seminar 
following my lectures, and also
for his help in preparation of this book.
Finally, I am indebted to 
the Faculty of Mathematics of Vienna University, 
the Max-Planck Institute for Mathematics in the Sciences (Leipzig), 
the Project ONDES (INRIA, Rocqencourt), 
and Wolfgang Pauli Institute
for their hospitality. 
\bigskip\bigskip\bigskip\\ 
A.Komech
\bigskip\bigskip\\ 
Leipzig,~~~~~~~~~~~~10.04.2005

%%%%%%%%%%%%%%%%%%%%%%%%%%%%%%%%%%%%%%%%%%%%%% 
%%%%%%%%%%%%%%%%%%%%%%%%%%%%%%%%%%%%%%%%%%%%%%% 
%%%%%%%%%%%%%%%%%%%%%%%%%%%%%%%%%%%%%%%%%%%%%%% 
 
\newpage 
 
\section{Introduction: Quantum Chronology 1859-1927} 
\subsection{Missing  ``Matter Equation''} 
We mark important points of the development of Quantum Theory. 
Every point is significant in either obtaining new experimental 
results or inventing a new treatment, or developing new 
mathematical methods. 
 
The Maxwell theory (1865) perfectly describes the motion of charged 
particles in a {\it given} electromagnetic field and also the propagation 
of electromagnetic waves generated by {\it known} charge and 
current densities $\rho,j$. However, generally it cannot describe the 
{\bf simultaneous} evolution of the unknown 
densities and fields since the {\it microscopic} 
evolution  equation for the densities is 
missing. 
The representation of the densities as an aggregation of moving charged 
particles does not help, since the corresponding {\it mass density} $\mu$ 
is unknown. That is, the ratio $\rho/\mu$ is not known and even does not 
have a reasonable meaning, since the ratio $e/m$ takes different 
values for different elementary particles. 
 
The situation is better 
on the {\it macroscopic} level 
in  {\it simple} media 
with known electric and magnetic permeability and conductivity, 
since then the 
the 
{\it macroscopic} charge and current densities 
$\rho_{\rm mac},j_{\rm mac}$ are 
{\it functions} of the fields 
({\it polarization}, {\it magnetization}, 
{\it Ohm's} law, etc). 
However, this is not the case for the vacuum. 
Hence, classical electrodynamics is not sufficient to 
explain the structure of matter at the {\it microscopic} level. 
\bigskip\\ 
Quantum mechanics  just provides various 
matter equations: 
Schr\"odinger, Klein-Gordon, Dirac  equations, etc. 
The equations arise {\it inside} classical electrodynamics, 
 thermodynamics, optics and atomic physics from 
experimental observations of various aspects of the 
field-matter interaction and their theoretical 
treatment. 
Let us briefly sketch the chronology of the development. 
%%%%%%%%%%%%%%%%%%%%%%%%%%%%%%%%%%%%%%%%%%%%%%%%%%%%%%%%%%%% 
%%%%%%%%%%%%%%%%%%%%%%%%%%%%%%%%%%%%%%%%%%%%%%%%%%%%%%%%%%%% 
\subsection{Thermodynamics, Optics and Electrodynamics} 
%%%%%%%%%%%%%%%%%%%%%%%%%%%%%%%%%%%%%%%%%%%%%%%%%%%%%%%%%%%% 
\subsubsection{Kirchhoff-Planck: The Black-Body Radiation} 
In 1859 Kirchhoff stated 
the radiation law 
which predicts a specific spectral density 
of light waves radiated by 
a black body at a fixed temperature. 
Light was identified with the electromagnetic field 
by Maxwell in 1865. Hence, the Kirchhoff law 
concerns the spectrum of the equilibrium 
distribution of the electromagnetic field at a fixed temperature. 
Therefore, it provides a deep indirect information on the 
interaction of matter with the Maxwell field.

The experimental measurements have been performed by Tyndall in 
1865, Crova in 1880, Langley in 1886, Weber in 1887, and 
Paschen in 1895-1899. Very precise measurements 
were made in 1899 by 
 Lummer and Pringscheim, 
and  Kurlbaum and Rubens. They confirmed the Wien formula (1896)
\be\la{W} 
I(\om)\sim \om^3 
\ds \exp({-\fr {\beta\om} T})\,. 
\ee 
Note that the traditional reference to the {\it black} body 
just means that its equilibrium radiation coincides with the 
equilibrium 
Maxwell field since the absorption of the black body is zero 
by definition. 
The 
comparison of (\re{W}) 
with the general equilibrium Boltzmann-Gibbs 
distribution $\exp (-\ds\fr E {kT})$ 
(where $k$ is the Boltzmann constant) 
suggests the famous Planck relation (1901) 
$$ 
E= \h \om \, , \eqno{(P)} 
$$ 
where $E$ is the energy of the ``emitted photon'' and 
$\h=k\beta\approx 1.05\cdot 10^{-27}erg\cdot sec$ is the Planck constant. 
Using this relation, Planck 
adjusted the formula (\re{W}) as 
$$ 
I(\om)\sim \om^2 
\ds \fr{\exp({-\ds\fr {a\om} T})} 
{1-\exp({-\ds\fr {a\om} T})}.  \eqno{(KP)} 
$$ 
%%%%%%%%%%%%%%%%%%%%%%%%%%%%%%%%%%%%%%%%%%%%%%%%%%%%%%%%%%%% 
\subsubsection{Rydberg-Ritz: Atom Spectra} 
Atom spectra provide extremely important 
information on the structure of the atom. 
In 1885 Balmer discovered the representation 
$\om_{2n}=R\Big(\ds\fr 1{2^2}-\fr 1{n^2}\Big)\,\,(n\ge 3)$ 
for a spectral series in the spectrum of the 
hydrogen atom. 
 Later, similar representations 
were found for other series by 
Paschen (1908) $\om_{3n}=R\Big(\ds\fr 1{3^2}-\fr 1{n^2}\Big)\,\, 
(n\ge 4)$, 
Lyman (1909) 
$\om_{1n}=R\Big(\ds 1-\fr 1{n^2}\Big)\,\,(n\ge 2)$, and 
Brackett (1914) 
$\om_{4n}=R\Big(\ds \fr1{4^2}-\fr 1{n^2}\Big)\,\,(n\ge 5)$. 
Similar structure 
$$ 
\om_{mn}=\om_m-\om_n,\,\,\,\,\eqno{(R)} 
$$ 
has been discovered experimentally by Rydberg (1900) for all 
the lines in  several series of other elements. 
The importance of these observations was also stressed by Ritz (1908), 
so it is now commonly known as the 
{\it Rydberg-Ritz combination principle}, 
and the numbers $\om_m$ are called {\it terms}. 
%%%%%%%%%%%%%%%%%%%%%%%%%%%%%%%%%%%%%%%%%%%%%%%%%%%%%%%%%%%% 
\subsubsection{Crookes-Herz-Perrin-Thomson: The Cathode Rays and the Electron} 
The  cathode rays were discovered first 
in {\it vacuum tube} by Crookes in 1870 
(he made {\it a discharge tube} with a vacuum 
level higher than that of {\it the Geissler tube} 
used by Faraday in 1836-1838). 
 The rays demonstrated 
the continuous motion of charge in the vacuum in  the 
presence of a Maxwell field. 
This is just one of the situations which is not covered by 
classical electrodynamics. 
 
The deflection of cathode rays in a magnetic field 
has been observed 
in 1880-1890 
by Hertz, Lenard, Perrin and many others. 
 Some physicists thought, like Goldstein, Hertz, and Lenard, 
that this phenomenon is like light, due to vibrations of the ether 
 or even that it is light of short wavelength. 
It is easily understood that such rays may have a 
rectilinear path, excite phosphorescence, 
and effect photographic plates. 
 Others thought, like Crookes, J.J. Thomson, Perrin and others, 
 that these rays are formed by matter which is negatively charged and 
moving with great velocity, and on this hypothesis their 
mechanical properties, as well 
as the manner in which they become curved in a magnetic field, 
are readily explicable. 
In 1895, Perrin collected the cathode rays, 
obtaining a negative charge. 
 
In 1897 J.J.Thomson 
 showed that the rays are also deflected by an electrostatic field. 
He systematized all previous observations 
 and demonstrated the particle-like behavior of the cathode rays 
which is described by the 
Lorentz equation, 
$$ 
a=\fr e\mu  (E+v\times B), \eqno{(L)} 
$$ 
where $\ds\fr e\mu  <0$. 
Concretely, he identified the cathode rays with a beam of 
particles with negative charge and introduced the 
name {\it electron} for these particles. 
This study led to the first measurement of the ratio 
$\ds\fr e\mu $ close to its 
present value. 
Kauffmann \ci{Ka} also observed the magnetic 
deflection of cathode rays and obtained a ratio  $\ds\fr e\mu $ 
which is close to the value of J.J.Thomson. 
 
J.J.Thomson's identification of the cathode rays led to many 
fundamental problems concerning the size and the 
structure of the electron: 
\\ 
i) Abraham 
(1906) noted  that 
the energy and the mass of the electron are 
infinite if its radius is zero. 
He introduced the model 
of the {\it extended} electron and calculated its radius. 
\\ 
ii) The extended electron cannot be stable because of the electrostatic 
repulsion (Poincar\'e 1908). 
 
So, classical electrodynamics had to be complemented with a 
{\it matter equation} which could describe i) the cathode rays and 
their {\it particle-like} behavior, and ii) the stability of the 
electron. 
%%%%%%%%%%%%%%%%%%%%%%%%%%%%%%%%%%%%%%%%%%%%%%%%%%%%%%%%%%%%  
\subsubsection{Herz-Einstein: Photoeffect} 
In 1887 Herz discovered the photoeffect 
(the "light electricity") 
via the generation of electric charge by the sun radiation. 
Later, the photoeffect has been observed with different 
types of electromagnetic radiation by 
Stoletov, Elster, Geitel, Righi, Townsend, Rutherford, 
Compton and many others. 
The experimental observations led to the relation 
$$ 
\h\om=E_{\rm el}-A. \eqno{(E)} 
$$ 
Here, $E_{\rm el}$ is the (maximal) energy of the photoelectrons 
detached from the metal by 
light of the frequency $\om$. 
The constant $A$ depends on the metal. 
The photoeffect occurs only for large frequencies 
of light $\om>\om_{\rm red}=A/\h$, where $\om_{\rm red}$ 
is called  the ``red bound'' of the photoeffect. 
 
In 1905 
Einstein proposed the theory of the photoeffect \ci{Ei}: 
he identified the relation $(E)$ 
with energy conservation. 
That is, Einstein 
\medskip\\ 
{\bf I.} Identified the quantity $\h\om$ with the energy of the 
{\it absorbed} photon with frequency $\om $ in accordance with the 
Planck relation $(P)$ (which concerns the {\it emitted} photon !), and 
\\ 
{\bf II.} Identified $A$ with the {\it escape energy} 
of the metal. 
\medskip\\ 
This explanation treats light as a collection of particle-like 
``photons'' that 
cannot be 
explained by using a wave picture of light and the classical 
representation 
of electrons as particles. 
%%%%%%%%%%%%%%%%%%%%%%%%%%%%%%%%%%%%%%%%%%%%%%%%%%%%%%%%%%%%  
\subsection{Atomic Physics} 
\subsubsection{Rutherford: The Nucleus of the Atom and Atom Stability} 
In 1913 Rutherford discovered the nucleus of the atom 
in an experiment on the scattering of 
$\al$-particles. This discovery suggested to him 
the classical 
model of the atom, where a finite number of  electrons 
moves around a 
point-like nucleus with positive charge. 
The electrons are governed by the classical Lorentz eq. $(L)$. 
However, 
the model 
is unstable due to the radiation of the rotating electrons 
in accordance with Maxwell electrodynamics. 
Therefore, the Maxwell theory is insufficient to explain 
the stability of the atoms. 
%%%%%%%%%%%%%%%%%%%%%%%%%%%%%%%%%%%%%%%%%%%%%%%%%%%%%%%%%%%% 
\subsubsection{Bohr:
 Quantum Stationary States and Transitions} 
In 1913 Niels Bohr has proposed a new phenomenology 
for description of the atom stability. 
Namely, he represented 
the Rydberg-Ritz combination principle $(R)$ in the form 
\be\la{BRR} 
\h\om_{mn}=E_m-E_n, 
\ee 
which was suggested by the comparison of $(R)$ with the Planck relation 
 $(P)$ and  the Einstein 
treatment of the 
relation for the photoeffect, $(E)$. 
Moreover,   Bohr 
interpreted $(B)$ generalizing the Planck and Einstein 
ideas: 
\medskip\\ 
{\bf I.} For an atom, there exist  {\bf Stationary States} $|E_n\rangle$ 
with the energies $E_n$. The atom is ``always'' in a stationary state, and, 
sometimes, make {\bf transitions} (or ``jumps'') 
from one Stationary State to another, 
\medskip 
\be\la{BT} 
|E_m\rangle\mapsto |E_n\rangle. 
\ee 
\medskip 
{\bf II.} The transition is accompanied by the radiation or absorption 
of light 
with frequency $\om_{mn}$. 
\\ 
{\bf III.} The identity (\re{BRR}) is energy balance in the 
transition, 
in accordance 
with  the identification of Planck and Einstein of the quantum 
$\h\om_{mn}$ with the energy of an emitted or absorbed photon. 
\medskip\\ 
Both, the role of the Planck constant $\h$ and the discreteness of 
the energies $E_n$ of the stationary states, 
cannot be explained by the Maxwell theory. 
The discreteness of the energies is related to a restriction to certain 
{\it stable} orbits of the 
electron in the atom. 
%%%%%%%%%%%%%%%%%%%%%%%%%%%%%%%%%%%%%%%%%%%%%%%%%%%%%%%%%%%% 
\subsubsection{Debye-Sommerfeld-Wilson: ``Old Quantum Theory''} 
In 1913 Debye stated the {\it quantum rule} for the determination 
of stable 
{\it periodic} orbits of the electrons in the atom, 
$$ 
\De S=2\pi n\h,\,\,\,\,n=1,2,3,... \eqno{(D)} 
$$ 
where $\De S$ is the action integral corresponding to the 
time-periodic orbit of the electron. 
The rule was motivated by the Ehrenfest idea of adiabatic 
invariance. 
The  quantum rules allowed to find the hydrogen spectral terms 
$\om_n=\ds\fr R{n^2}$, $n=1,2,...$ which exactly agree with the 
series of Lyman, Balmer etc. 
In 1916 Sommerfeld and Wilson extended the rule to more general 
{\it quasiperiodic} orbits. 
 
In 1923 Bohr has developed the {\it correspondence principle} which 
allowed him to discover the {\it selection rules} for the {\it magnetic 
and azimuthal quantum numbers}. 
The selection rules play the key role in the explanation 
of atom spectra and agree with experimental observations, 
see \ci{Born} and \ci[Vol.I]{Som}.

%%%%%%%%%%%%%%%%%%%%%%%%%%%%%%%%%%%%%%%%%%%%%%%%%%%%%%%%%%%%  
\subsubsection{Zeemann-Stern-Gerlach: Atoms in  Magnetic Fields} 
In 1896 Zeemann discovered the splitting of the spectral lines 
of atoms in a magnetic field. 
Lorentz explained the splitting by the Maxwell theory 
in the simplest case 
of the {\it normal} Zeemann effect 
when the line $\om$ 
splits into three lines: $\om$ and $\om_\pm =\om\pm \De\om$, 
where $\De\om$ is proportional to the magnetic field. 
However, the explanation of the general 
{\it anomalous} Zeemann effect 
cannot be deduced from the Maxwell theory. 
One example is the double splitting 
of the spectra of alkali 
atoms.

The Maxwell theory predicts a unique value $r$ for the gyromagnetic ratio 
$|\m|/|\bJ|= 
\mu_B$, where 
 $\mu_B=\ds\fr{|e|\h}{2mc}$ is the {\it Bohr magneton} 
and 
$|\m|$ resp. $|\bJ|$ are the magnetic and mechanical momenta of the electron 
in an atom. 
In 1915 Einstein and de Haas first measured the gyromagnetic ratio 
by an observation of a magnetization of an iron in an external 
weak magnetic field. 
However, the observed ratio was $2r$, i.e., 
two times larger than the 
theoretical value. 
 
In 1921 Stern and Gerlach observed the 
double splitting of a beam of  silver atoms in a 
strong non-uniform magnetic field. This implies 
that the stationary state of the atom is split into 
two states with different 
gyromagnetic ratios $|\m|/|\bJ|$, which again contradicts the 
Maxwell theory. 
\subsubsection{Compton: Scattering of Light by Electrons} 
In 1923,  Compton 
discovered that the scattered light has a wavelength $\lam'$ 
different 
from the wavelength $\lam$ 
of the incident light: 
$$ 
\lam'-\lam\sim \fr {2\h}{\mu c}\sin^2\fr\theta 2 \, , 
$$ 
where 
$\theta$ is the angle between the incident and scattered waves, 
and $\mu$ is the electron mass. 
Similarly to the photoeffect, the scattering also cannot be 
explained by using a wave picture of light, 
where the wavelength does not change. 
%%%%%%%%%%%%%%%%%%%%%%%%%%%%%%%%%%%%%%%%%%%%%%%%%%%%%%%%%%%%  
\subsubsection{De Broglie: Wave-Particle Duality for Free Particles} 
In 1924 de Broglie, in his PhD thesis, 
introduced a wave function for 
a possible description of matter by waves, \ci{Jam}, 
in analogy with the particle-wave duality of light 
which is demonstrated by the Maxwell theory and the photoeffect. 
Namely, he  has applied Einstein's Special Relativity Theory 
to a beam of free particles with the  energy-momentum vector $(E,\p)$. 
\bigskip\\ 
{\bf I.} The beam is identified 
with a plane wave by the following ``wave-particle'' relation: 
\be\la{dB1} 
 \!\!\!\!\!\!\!\!\psi(t,\x)=Ce^{i(\bk\x-\om t)}~~\lra~~ 
\mbox{\bf beam of free particles.}
\ee 
{\bf II.} The Einstein relativity principle 
and the Planck relation $(P)$ imply the identity 
\be\la{dB} 
(E,\p)=\h(\om, \bk). 
\ee 
The identity plays a crucial role everywhere in quantum theory. 
In particular, it implies the famous de Broglie 
relation for 
the ``particle wave length'' $\lam=2\pi/|\bk|$, 
$$ 
\lam=\fr{2\pi \h}{|\p|}\,\,. 
$$ 
It also implies the relativistic dispersion relation 
\be\la{drr}
\fr{\h^2\om^2}{c^2}=\h^2\bk^2+\mu^2c^2, 
\ee
where $\mu$ is the particle mass and $c$ is the speed of light. 
It follows from the expression for the Hamiltonian of the 
relativistic particle (see (\re{relps})) 
\be\la{EPr} 
\fr{E^2}{c^2}=\p^2+\mu ^2c^2. 
\ee 
iv) For small values of $|\p|\ll \mu c$ the 
non-relativistic approximation holds, 
\be\la{EPnr} 
E=\sqrt{\p^2c^2+\mu ^2c^4}\approx \mu c^2+\fr{\p^2}{2\mu }\,\,. 
\ee 
Dropping  here the 
``unessential'' additive constant $\mu c^2$, 
we get the non-relativistic dispersion relation 
\be\la{drnr} 
\h\om=\fr {\h^2\bk^2}{2\mu}\,\,. 
\ee 
The dispersion relations 
(\re{drr}) resp. (\re{drnr}) 
implies the {\it free} 
Klein-Gordon resp. Schr\"odinger 
equation for the corresponding  wave function 
$\psi(t,\x)=e^{i(\bk\x-\om t)}$: 
$$ 
\fr 1{c^2}[i\h\pa_t]^2\psi(t,\x)=[(-i\h\na_\x)^2+ 
\mu^2c^2]\psi(t,\x),\eqno{(KG_0)} 
$$ 
$$ 
i\h\pa_t\psi(t,\x)=\fr 1{2\mu} [-i\h\na_\x]^2\psi(t,\x).\eqno{(S_0)} 
$$ 
%%%%%%%%%%%%%%%%%%%%%%%%%%%%%%%%%%%%%%%%%%%%%%%%%%%%%%%%%%%% 
\subsubsection{Klein-Gordon-Schr\"odinger: Wave Equation for Bound Particles} 
In 1925-1926 Klein, Gordon and Schr\"odinger extended de Broglie's 
wave equation to the bound electron in an external 
Maxwell field. 
The {\it free} equations 
$(KG_0)$ resp. $(S_0)$ formally follow from the energy-momentum 
relations (\re{EPr}) resp. (\re{EPnr}) 
by the substitutions 
\be\la{corp} 
E\mapsto i\h\pa_t,~~~~~~~~~~~~~~~~ \p\mapsto -i\h\na_\x. 
\ee 
For an electron in the external scalar potential $\phi(t,\x)$ and 
magnetic 
vector potential $\bA(t,\x)$, 
the (conserved) energy $E$ is given by 
$[E-e\phi(t,\x)]^2/c^2=[\p- \ds\fr ec \bA(t,\x)]^2+\mu^2c^2$, where 
$e$ is  the 
charge of the electron (see (\re{Hamfrts})). 
Then  Klein, Gordon and Schr\"odinger generalized $(KG_0)$ to 
$$ 
\fr 1{c^2}[i\h\pa_t-e\phi(t,\x)]^2\psi(t,\x)=[-i\h\na_\x- \ds\fr ec 
 \bA(t,\x)]^2\psi(t,\x)+ 
\mu^2c^2\psi(t,\x). 
\eqno{(KG)} 
$$ 
Schr\"odinger also generalized 
the nonrelativistic approximation (\re{drnr}) to 
$$
E-e\phi(t,\x)=(\p- \ds\fr ec \bA(t,\x))^2/(2\mu),
$$
 which transforms into
the wave equation 
$$ 
[i\h\pa_t-e\phi(t,\x)]\psi(t,\x) 
=\fr 1{2\mu} [-i\h\na_\x-\ds\fr ec \bA(t,\x)]^2\psi(t,\x).\eqno{(S)} 
$$ 
The next crucial step of Schr\"odinger's theory is the identification 
of stationary states with 
solutions of the type 
$\exp  (-i\om t)\psi(\x)$ for the static external Maxwell fields 
$\phi(t,\x)\equiv\phi(\x)$ and $A(t,\x)\equiv \bA(\x)$. 
This identification is suggested by the 
de Broglie plain wave $\exp(-i\om t)\exp(i\bk\x)$, where only 
the spatial factor has to be modified since 
the external field 
``twists'' space but not time. The energy is again $E=\h\om$. 
This identification leads to the corresponding 
{\it stationary equations} which are the 
{\it eigenvalue problems}, 
$$ 
\fr 1{c^2}[\om-e\phi(\x)]^2\psi(\x)=[-i\h\na_\x- \ds\fr ec \bA(\x)]^2\psi(\x)+ 
\mu^2c^2\psi(\x), 
$$ 
$$ 
[\om-e\phi(\x)]\psi(x)=\fr 1{2\mu}[-i\h\na_\x-\ds\fr ec \bA(\x)]^2\psi(\x) 
$$ 
for the determination of the energies $E=\h\om$ and the amplitudes 
$\psi(\x)$ of the stationary states. 
 
Schr\"odinger calculated all solutions to the last equation 
for the 
hydrogen atom: 
$\phi(\x)=-e/|\x|$ is the Coulomb potential of the nucleus, and 
 $\bA(\x)=0$. The agreement with the experimentally observed spectrum 
was perfect. 
The calculation uses the standard 
separation of variables in  spherical 
coordinates, which involves some integer numbers as in 
the Debye quantum condition $(D)$. 
It was just this analogy which suggested to 
Schr\"odinger an eigenvalue 
problem for the determination of the stationary states of the atom. 
%%%%%%%%%%%%%%%%%%%%%%%%%%%%%%%%%%%%%%%%%%%%%%%%%%%%%%%%%%%% 
\subsubsection{Heisenberg: Matrix Mechanics} 
In 1925 Heisenberg have extended significally the 
Bohr correpondence principle. He assigned the {\it infinite matrix} 
i.e. the operator in the Hilbert space, 
 to each classical observable: energy, mometum, coordinate, 
angular momentum, etc. 
This assignment is called now {\it the quantization} of the 
corresponding classical dynamical system. 
In particular, the substitutions 
(\re{corp}) give an example of the assignment at a fixed time. In this 
{\it matrix mechanics} the operators depend on time and obeys the 
dynamical equations which formally coincide with the 
equations for the classical observables. 
The Heisenberg approach allowed him to discover the key 
{\bf Uncertainty Principle} (see Lecture 7) which plays a fundamental role 
in the theoretical analysis and applications 
of the quatum theory. 
 
The Heisenberg ideas were developed by Born, Jordan and others. 
Later on it was proved that the Heisenberg approach is completely 
equivalent to the Schr\"odinger theory. 
The equivalence has played a crucial role in further development 
of the quantum theory leading to the 
{\it second quantization} and 
{\it quantum field theory}. 
 %%%%%%%%%%%%%%%%%%%%%%%%%%%%%%%%%%%%%%%%%%%%%%%%%%%%%%%%%%%% 
\subsubsection{Uhlenbeck-Goudsmith-Pauli: Nonrelativistic Theory of Spin} 
In 1925 Uhlenbeck and Goudsmith introduced the hypothesis 
of the existence of the {\it spin} of the electron, 
i.e., of its {\it own angular momentum} $\s$        
(i.e. not related to its rotation)
and a
magnetic moment $\m$ 
(not related to the  corresponding convection current)
with 
the gyromagnetic ratio $g:=|\m|/|\s|=2\mu_B$. 
The hypothesis 
had been inspired by  the double splitting in the 
Stern-Gerlach experiment, the {\it anomalous} Zeemann effect 
and the Einstein-de Haas experiment. 
In 1927 Pauli obtained the wave 
equation 
which takes the spin of the electron into account, 
\be\la{PE} 
~~~~~~[i\h\pa_t-e\phi(t,\x)]\Psi(t,\x) 
=\fr 1{2\mu} [-i\h\na_\x-\ds\fr ec \bA(t,\x)]^2\Psi(t,\x)+ 
g\mu_B\sum_1^3 \hat\s_k \bB_k\Psi(t,\x), 
\ee 
where 
the wave function $\Psi(t,\x)=(\psi_1(t,\x),\psi_2(t,\x))$ 
with two complex-valued functions $\psi_j (t,\x)$. 
Further, 
$\bB_k$ are the components of the {\it uniform} magnetic field and 
$\hat\s_k=\si_k/2$, where $\si_k$ are the 
complex $2\times 2$ {\it Pauli matrices}. 
Pauli obtained the equation by just postulating the double splitting, 
the {\it gyromagnetic ratio} $g=2$, 
and covariance with respect to space rotations. 
 
This equation 
leads to the correct gyromagnetic ratio observed in 
the Einstein-de Haas experiment. It also 
explains the anomalous Zeemann effect and the 
Stern-Gerlach double splitting. 
The agreement with many experimental observations 
was a great triumph of the  quantum theory. 
 %%%%%%%%%%%%%%%%%%%%%%%%%%%%%%%%%%%%%%%%%%%%%%%%%%%%%%%%%%%% 
\subsubsection{Dirac: Relativistic Theory of Spin} 
In 1927 Dirac discovered the relativistic invariant 
equation which 
generalizes  the Pauli and Klein-Gordon equations, 
$$ 
 \sum_{\al=0}^3 \ga_\al[i\h\na_\al-\ds\fr ec\A_\al(x)]=
\mu c\psi(x), ~~~~~~~~x\in\R^4, 
$$ 
where $\na_0=\pa_t/c$, $(\na_1,\na_2,\na_3)=\na_\x$, 
$\A_0(x)=\phi(x)$ and 
$\A_k=-\bA_k(x)$, $k=1,2,3$, $\ga_\al$ are the 
$4\times 4$ Dirac matrices, 
and $\psi(x)\in\C^4$ for $x\in\R^4$. 
 
The Dirac equation automatically provides 
the correct gyromagnetic ratio $g=2$ 
for the electron. It gives a much more precise description 
of the hydrogen atom spectrum than the Schr\"odinger and Pauli equations. 
%%%%%%%%%%%%%%%%%%%%%%%%%%%%%%%%%%%%%%%%%%%%%%%%%%%%%%%%%%%% 
\subsubsection{Davisson-Germer: Interference of Electrons} 
In 1927 
Davisson and Germer 
observed the interference 
of electron beams. 
Later the experiments were repeated and confirmed by many 
authors: 
Thomson, 
Rupp, 
Kikouchi 
and others. 
In 1949 Biberman, Sushkin and Fabrikant observed 
the interference pattern with a weak beam 
with a very low rate of registration of the electrons. 
%%%%%%%%%%%%%%%%%%%%%%%%%%%%%%%%%%%%%%%%%%%%%%%%%%%%%%%%%%%% 
\subsubsection{Born: The Probabilistic Interpretation of the Wave Function} 
In 1927 Born proposed the following 
interpretation of the wave function to explain the Davisson-Germer experiment: 
{\it 
$|\psi(t,\x)|^2$ is a 
density of probability.}

\newpage 
 
\part{Lagrangian Field Theory} 
%%%%%%%%%%%%%%%%%%%%%%%%%%%%%%%%%%%%%%%%%%%%%%%%%% 

%%%%%%%%%%%%%%%%%%%%%%%%%%%%%%%%%%%%%%%%%%%%%%%%%% 

%%%%%%%%%%%%%%%%%%%%%%%%%%%%%%%%%%%%%%%%%%%%%%%%%% 

%%\setcounter{section}{-1} 
\setcounter{subsection}{0} 
\setcounter{theorem}{0} 
\setcounter{equation}{0} 
\section 
{Euler-Lagrange Field Equations} 
We introduce two equations important for Quantum Mechanics, 
namely, the Klein-Gordon and Schr\"odinger equations. 
We prove that they are of Lagrangian form. 
Then we introduce an action functional and prove 
the Hamilton least action principle.

The quantum mechanical evolution equations for charged matter in an 
electromagnetic field are provided by the Schr\"odinger equation 
and Klein-Gordon equation in the non-relativistic and relativistic 
cases, respectively. Both equations may be derived from an action 
functional via the Hamiltonian least action principle, and, therefore, 
a Lagrangian description exists for both cases. 
 
%%\section{Euler-Lagrange Field Equations} 
\subsection{Klein-Gordon and Schr\"odinger Equations} 
Let us define $x_0=ct$, $\x=(x_1,x_2,x_3)$, $x=(x_0,\x)$ and 
consider the Klein-Gordon 
equation 
%%%($\h=c=1$, $V(t,\x)=e\phi(t,\x)$ and the magnetic field is zero), 
\beqn 
&&[i\h\na_0-\ds\fr ec\phi(x)]^2\psi(x)\nonumber\\ 
&&=[-i\h\na_\x- \ds\fr ec \bA(x)]^2\psi(x)+ 
\mu^2c^2\psi(x),\,\,\,~~~~x\in\R^4, 
\la{NKG} 
\eeqn 
where the function 
$\psi(x )$ takes complex values and $\mu>0$ is the electron mass. 
Let us also consider the Schr\"odinger equation 
\beqn 
&&[i\h\na_0-e\phi(x)]\psi(x)\nonumber\\ 
&&=\fr 1{2\mu} [-i\h\na_\x-\ds\fr ec \bA(x)]^2\psi(x), 
~~~~~~~~~~~\,\,\,\,\,\,\,\,\,x\in\R^4,  \la{NS} 
\eeqn 
where $x_0:=t$. 
\subsection{Lagrangian Density} 
\bd\la{cri} 
We will identify the complex vectors $\psi\in\C^M$ with 
the real vectors $\R\psi:=(\rRe\psi,\rIm\psi)\in\R^{2M}$ and 
the multiplication by a complex number with 
an application of the corresponding matrix. 
We will denote by 
$\cdot$ the real scalar product in $\R^{2M}$. 
\ed 
This definition implies the formulas 
\beqn 
&& 
\R u\cdot\R v=\rRe (u\ov v),\la{real}\\ 
\nonumber\\ 
&&\na_{\R u}(u\cdot iv)=iv,~~\na_{\R v}(u\cdot iv)=-iu,~~~~~~\na_{\R u}(iu\cdot v)=-iv, 
~~\na_{\R v}(iu\cdot v)=iu, 
~~~~~~~~~\la{cdot} 
\eeqn 
for $u,v\in\C$ 
since $u\cdot iv=-iu\cdot v$ and $iu\cdot v=-u\cdot iv$.

Let us introduce the 
{\it Lagrangian densities} $\cL$ for Eqs. (\re{NKG}) and (\re{NS}) 
as the {\it real} functions 
defined by 
(cf. (\re{E1}), (\re{EN})), 
\beqn 
\cL_{KG}(x,\psi,\na\psi) 
&=& 
\fr{|[i\h\na_0-\ds\ds\fr ec\phi(x)]\psi|^2}2- 
\fr{|[-i\h\na- \ds\ds\fr ec \bA(x)]\psi|^2}2 
\nonumber\\ 
&&-\mu^2c^2\fr{|\psi|^2}2, 
\la{LdNKG}\\ 
~\nonumber\\ 
\cL_{S}(x,\psi,\na\psi) 
&=& 
[i\h\na_0-e\phi(x)]\psi\cdot \psi- 
\fr1{2\mu}|[-i\h\na- \ds\fr ec \bA(x)]\psi|^2. 
\la{LdNS} 
\eeqn 
We will demonstrate below that 
 the field equations 
(\re{NKG}), (\re{NS}) can be represented in the 
 {\bf Euler-Lagrange} form, 
\be\la{ELf} 
\cL_\psi(x,\psi(x),\na\psi(x)) 
-\sum\limits_{\al=0}^3 \na_\al\cL_{\na_\al\psi}(x,\psi(x),\na\psi(x))=0, 
\,\,\,\,x\in\R^4, 
\ee 
where $\cL$ is the corresponding Lagrangian density. 
 
\br 
In  (\re{LdNKG}) -- (\re{ELf}) the $x$ and $\psi,\na_\al\psi$ are considered 
as independent variables with the values in $\R^{4}$ and  $\R^2$, 
respectively. 
\er

\bd\la{dLf} 
The Lagrangian field  $\psi(x)$ with values in $\R^N$ 
is the dynamical system described by the 
 $N$ 
real scalar equations 
(\re{ELf}) 
with a given Lagrangian density 
 $\cL(x,\psi,\na\psi)\in C^2(\R^{4}\times\R^N\times\R^{4N})$. 
\ed

\bd 
The {\bf canonically conjugate fields} 
$\pi_\al(x)$ 
are defined by 
\be\la{cLdf} 
\pi_\al(x)=\cL_{\na_\al\psi}(x,\psi(x),\na\psi(x)),\,\,\,\,\al=0,...,3. 
\ee 
\ed 
With these notations the Euler-Lagrange equations (\re{ELf}) 
read 
\be\la{ELfp} 
\na_\al\pi_\al(x)=\cL_\psi(x,\psi(x),\na\psi(x)),\,\,\,\,x\in\R^{4}. 
\ee 
{\bf Here and below we use the Einstein convention } 
$\na_\al\pi_\al(x):=\sum_\al \na_\al\pi_\al(x)$ etc. 
Also $\al=0,1,2,3$ and  $k=1,2,3$. 
\subsection{Free Equations} 
First consider the {\it free} equations without Maxwell field 
and without nonlinear self-interaction, and with $\h=1$: 
\beqn 
&\na_0^2\psi(x)&=\na_\x^2\psi(x)- 
\mu^2c^2\psi(x),\,\,\,~~~~x\in\R^{4}, 
\la{NKGf} 
\\ 
~&&\nonumber\\ 
&-i\na_0\psi(x)&=\fr 1{2\mu} \na_\x^2\psi(x), 
~~~~~~~~~~~~\,\,\,\,\,\,\,\,\,x\in\R^{4},  \la{NSf} 
%%\ddot \psi(t,\x )&=&\De\psi(t,\x )-\mu^2\psi(t,\x ) 
%%-V(t,\x)\psi(t,\x )+F(\psi(t,\x )),\,\,\,(t,\x)\in\R^{d+1}, 
%%\la{NKG}\\ 
%%-i\dot \psi(t,\x )&=&\fr1{2\mu}\De\psi(t,\x ) 
%%-V(t,\x)\psi(t,\x )+F(\psi(t,\x )),\,\,\,\,\,\,\,\,\, 
%%(t,\x)\in\R^{d+1},\la{NS} 
\eeqn 
Then the Lagrangian densities (\re{LdNKG}), (\re{LdNS}) become 
\beqn 
\cL_{KG}^0(x,\psi,\na\psi) 
&=& 
\fr{|\na_0\psi|^2}2- 
\fr{|\na\psi|^2}2-\mu^2c^2\fr{|\psi|^2}2, 
\la{LdNKGf}\\ 
~\nonumber\\ 
\cL_{S}^0(x,\psi,\na\psi) 
&=& 
i\na_0 \psi\cdot \psi- 
\fr1{2\mu}|\na\psi|^2 
\la{LdNSf} 
\eeqn

\bexe 
Check the   Euler-Lagrange form (\re{ELfp}) 
for the Klein-Gordon and Schr\"odinger equations 
 (\re{NKGf}),  (\re{NSf}). 
\eexe 
{\bf Solution}\\ 
{\bf I} For the Klein-Gordon equation (\re{NKGf}): 
by Formulas (\re{cdot}) we get 
\be\la{piKGf} 
\pi_0(x)=\na_0\psi(x), 
\,\,\,\,\pi_k(x)=-\na_k\!\psi(x),~~~~~~~~ 
\cL_\psi=-\mu^2c^2\psi. 
\ee 
Hence (\re{ELfp}) is equivalent to (\re{NKGf}). 
\\ 
{\bf II} For the Schr\"odinger  equation (\re{NSf}): 
by Formulas (\re{cdot}) we get 
\be\la{piSf} 
\pi_0(x)=-i\psi(x),\,\,\,\,\pi_k(x)= 
-\fr 1{\mu} \na_k\psi(x),~~~ 
\cL_\psi=i\na_0\psi(x). 
\ee 
Hence (\re{ELfp}) is equivalent to (\re{NSf}).

\subsection{The Equations with Maxwell Field}

\bexe 
Check the   Euler-Lagrange form (\re{ELfp}) 
for the Klein-Gordon and Schr\"odinger equations 
 (\re{NKG}),  (\re{NS}). 
\eexe 
{\bf Solution}\\ 
{\bf I} For the Klein-Gordon equation (\re{NKG}): 
by Formulas (\re{cdot}) we get 
\be\la{piKG} 
\left\{\ba{ll} 
\pi_0(x)&\!\!\!\!=-i\h[i\h\na_0-\ds\fr ec\phi(x)]\psi(x), 
\,\,\,\,\pi_k(x)\!=\!-i\h[\!-\!i\h\na_k\!-\!\ds\fr ec \bA_k(x)]\psi(x), 
\\~\\ 
\cL_\psi&\!\!\!\!=\!-\ds\fr ec\phi(x)[i\h\na_0-\ds\fr ec\phi(x)]\psi(x) 
+\ds\fr ec \bA_k(x)[\!-\!i\h\na_k\!-\! \ds\ds\fr ec \bA_k(x)]\psi(x)\!- 
\!\mu^2c^2\psi. 
\ea\right. 
\ee 
Hence (\re{ELfp}) is equivalent to (\re{NKG}). 
\\ 
{\bf II} For the Schr\"odinger  equation (\re{NS}): 
by Formulas (\re{cdot}) we get 
\be\la{piS} 
~~~~~\left\{\ba{ll} 
\pi_0(x)&=-i\h\psi(x),\,\,\,\,\pi_k(x)= 
-\ds\fr 1{\mu} i\h[-i\h\na_k-\ds\fr ec \bA_k(x)]\psi(x), 
\\~\\ 
\cL_\psi&\!=i\h\na_0\psi(x)-2e\phi(x)\psi(x)+\ds\fr e{\mu c} \bA_k(x) 
[-i\h\na_k- \ds\ds\fr ec \bA_k(x)]\psi(x). 
\ea\right. 
\ee 
Hence (\re{ELfp}) is equivalent to (\re{NS}).

\subsection{Action Functional} 
\bd\la{Cf} 
For $k=1,2,...$ and $\si>0$ the space 
$C^k_\si$ is the set of the functions 
$\psi(t,\x )\in C^k([0,\infty)\times \R^3, \R^N)$ 
with the space-decay 
\be\la{sde} 
\sum_{|\al|\le k}|\na^\al\psi(t,\x )|\le C_T 
(1+|\x |)^{-\si}, 
\,\,\,\,(t,\x )\in C^k([0,T]\times\R^3). 
\ee 
\ed 
We will consider the real-valued functionals ${\cal F}$ on $C^1_\si$. 
By definition, ${\cal F}$ is a map $C^1_\si\to\R$.

\bd \la{dGdf} 
For $\psi\in C^1_\si$ 
the Gateau derivative $D{\cal F}(\psi)$ is the {\bf linear} functional 
\be\la{Gdf} 
\langle D{\cal F}(\psi), h \rangle=\fr{d}{d\ve}\Bigg|_{\ve=0} 
{\cal F}(\psi+\ve h) 
\ee 
for $h(\cdot)\in C^1_\si$ 
if the derivative exists. 
\ed 
 
We will assume 
the following bounds for the Lagrangian density, 
\be\la{bLd} 
\left.\ba{rcl} 
|\cL(x,\psi,\na\psi)|&\le& C 
(|\psi|+|\na\psi|)^2\\ 
|\cL_\psi(x,\psi,\na\psi)|+ 
|\cL_{\na\psi}(x,\psi,\na\psi)|&\le& C 
(|\psi|+|\na\psi|) 
\ea\right|, 
\,\,\,\,|\psi|\!+\!|\na\psi|\!\le\! \co . 
\ee 
Let us fix a $T>0$. 
\bd The action for the field is the functional on $C^1_\si$, 
 $\si>3/2$ defined by 
\be\la{Sf} 
S_T(\psi)=\int_0^T [\int_{\R^3} 
\cL(x,\psi(x),\na\psi(x))d\x ]dx_0,\,\,\,\,\psi(\cdot)\in C^1_\si. 
\ee 
\ed 
Note that for $\si>3/2$ the action is defined on the whole of $C^1_\si$ 
by (\re{sde}) and the first inequality of (\re{bLd}). 
Moreover, the functional is 
differentiable: 
\bl 
The Gateau derivative 
$\langle DS_T(\psi),h\rangle$ exists for $\psi,h\in C^1_\si$ if $\si>3/2$. 
\el 
\Pr From Definition \re{dGdf} we get by the theorem 
of the differentiation of integrals, 
\beqn\la{DSf} 
&&\langle DS_T(\psi), h \rangle:=\fr{d}{d\ve}\Bigg|_{\ve=0} 
\int_0^T [\int_{\R^3} 
\cL(x,\psi(x)+\ve h(x),\na\psi(x)+\ve \na h(x))d\x ]dx_0 
\nonumber\\ 
&&=\int_0^T [\int_{\R^3} 
\Big(\cL_\psi(x,\psi(x),\na\psi(x))h(x) 
+ 
\sum_{\al=0}^3 
\cL_{\na_\al\psi}(x,\psi(x),\na\psi(x))\na_\al h(x)\Big)d\x ]dx_0 
\eeqn 
since the integrals converge uniformly by (\re{sde}) and 
(\re{bLd}).\bo

\subsection{Hamilton Least Action Principle} 
Let us introduce the space of {\it variations}. 
\bd\la{Cf0} 
$C^1(T)$ is the space of functions 
$h(\cdot)\in C^1([0,T]\times\R^3,\R^N)$ such that 
\beqn 
&&h(0,\x )=h(T,\x )=0,\,\, \x \in\R^3,\\ 
\la{h0T} 
&&h(t,\x )=0,\,\,\,\,\,|\x |\ge\co ,\,\, 
t\in[0,T]. 
\la{hx} 
\eeqn 
\ed 
 
\bd The function $\psi\in C^1_\si$ satisfies the 
Hamilton Least Action Principle (LAP) if for any $T>0$, 
\be\la{LAPf} 
\langle DS_T(\psi), h \rangle=0,\,\,\,\,\forall h(\cdot) 
\in C^1(T). 
\ee 
\ed 
\bt\la{tLAPf} 
For $\psi\in 
C^2_\si$ with $\si>3/2$ the Hamilton LAP 
is equivalent to the 
Euler-Lagrange equations (\re{ELf}). 
\et 
\Pr (\re{DSf}) implies by 
partial integration in $x_\al$, $\al=0,...,3$, 
\beqn\la{paLAPf} 
&&\langle DS_T(\psi), h \rangle\nonumber\\ 
&=& 
\int_0^T [\int_{\R^3} 
\Big(\cL_\psi(x,\psi(x),\na\psi(x))h(x) 
- 
\sum_{\al=0}^3\na_\al \cL_{\na_\al\psi}(x,\psi(x),\na\psi(x))\Big) 
h(x)d\x ]dx_0. 
\eeqn 
Therefore, (\re{LAPf}) is equivalent to (\re{ELf}) 
by the main lemma of the calculus of variations. 
\bo

%%%%%%%%%%%%%%%%%%%%%%%%%%%%%%%%%%%%%%%%%%%%%%%%%%%%%%%%%%%%%% 
%%%%%%%%%%%%%%%%%%%%%%%%%%%%%%%%%%%%%%%%%%%%%%%%%%%%%%%%%%%%%% 
%%%%%%%%%%%%%%%%%%%%%%%%%%%%%%%%%%%%%%%%%%%%%%%%%%%%%%%%%%%%%% 
%%%%%%%%%%%%%%%%%%%%%%%%%%%%%%%%%%%%%%%%%%%%%%%%%%%%%%%%%%%%%% 
\newpage 

\setcounter{subsection}{0} 
\setcounter{theorem}{0} 
\setcounter{equation}{0} 
\section{Four Conservation Laws for Lagrangian Fields} 
 
By the general Noether theorem \re{tN2}, 
an invariance of the Lagrangian density of a field 
with respect to a {\it symmetry group} provides a conservation law. 
In particular, the theorem implies four classical conservation laws: 
conservation of total energy, momentum, angular momentum and charge. 
Here, the energy conservation follows from an invariance 
under time translations, the momentum conservation follows from 
translation invariance in space, the angular momentum 
conservation follows from  rotation invariance, and 
the charge conservation follows from a phase invariance. 
 
Here we state the four conservation laws. All these conservation laws 
are deduced from  the Noether theorem of Section 27.2. 
Let us fix a $\si>3/2$. 
\subsection{Time Invariance: Energy Conservation} 
\bd 
\la{LmEf} 
The energy of the Lagrangian field $\psi(x)\in C^1_\si$ 
is defined by 
(cf.  (\re{eLd}), (\re{eLdN})) 
\be\la{eLdf} 
E(t)= \int_{\R^3}\Big[\pi_0(x)\na_0\psi(x)- 
\cL(x,\psi(x),\na\psi(x))\Big]d\x , 
\,\,\,\,t\in\R. 
\ee 
\ed 
\br 
By definition, $t=x_0$ and $\pi_0(x)\,\na_k\psi(x)= 
\pi_0^j(x)\,\na_k\psi^j(x)$. 
\er 
Note that by (\re{sde}) and (\re{bLd}) 
the energy and momentum are well 
 defined for the solutions $\psi(t)\in C^1_\si$ since $\si>3/2$. 
 
\bex 
The linear {\it wave equation} 
\be\la{dE} 
\ddot \psi(t,\x )=\De\psi(t,\x ),\,\,\,\,\x \in\R^3 
\ee 
is a particular case of the Klein-Gordon equation 
(\re{NKG}) with $m=0$. 
The Lagrangian density $\cL$ 
is given by (\re{LdNKG}), i.e. 
\be\la{dELd} 
\cL(t,\x ,\psi,\dot \psi, \psi')=\fr {|\dot \psi|^2} 2 
-\fr {|\na\psi|^2} 2. 
\ee 
Then the canonically conjugate field is $\pi_0=\dot\psi$. 
Hence the 
energy is given by 
\be\la{eLdfe} 
E(t)= \int_{\R^3}\Big[\fr {|\dot \psi(t,\x )|^2} 2 
+\fr {|\na\psi(t,\x )|^2} 2 \Big]d\x , 
\,\,\,\,t\in\R. 
\ee 
\eex

\bt\la{ECLf} Let 
the Lagrangian density $\cL$  not depend on 
$x_0:=t$, 
\be\la{Lt} 
\cL(x, \psi,\na\psi)\equiv 
\cL_1(\x ,\psi,\na\psi). 
\ee 
Then for any trajectory $\psi(x)\in C^2_\si$ 
of the equations (\re{ELfp}) 
the energy is 
 conserved, 
$E(t)=\co$. 
 
\et 
{\bf Proof for a particular case} 
Let us prove the theorem for an equation of type (\re{dE}) 
in  dimension one, 
\be\la{d1} 
\ddot \psi(t,x )=\psi''(t,x ),~~~~~x\in\R. 
\ee 
First, 
$E(t)= \lim_{R\to\infty}E_R(t)$ where 
\be\la{eLdfRd} 
E_R(t)= 
\int_{-R}^R 
\Big[\fr {|\dot \psi(t,x )|^2} 2 
+\fr {|\psi'(t,x )|^2} 2 \Big]dx . 
\ee 
Differentiating, we get 
\be\la{eLdfRdd} 
\dot E_R(t)= 
\int_{-R}^R 
\Big[\dot \psi(t,x )\ddot \psi(t,x ) 
+\psi'(t,x )\dot\psi'(t,x ) \Big]dx . 
\ee 
Substituting here $\ddot \psi(t,x )=\psi''(t,x )$, 
we get by partial integration 
\be\la{derd} 
\dot E_R(t)= 
\Big[\dot \psi(t,x )\psi'(t,x )\Big]\Big|_{-R}^R 
=\dot \psi(t,R )\psi'(t,R )-\dot \psi(t,-R )\psi'(t,-R ). 
\ee 
Now, Eq.  (\re{sde}) implies that for every fixed $t$ we have 
$\dot E_R(t)\to 0$ as 
$R\to\infty$, hence 
$E(t)=$const. Indeed, for any $T>0$, 
\be\la{ERt} 
E_R(T)-E_R(0)=\int_0^T \dot E_(t)dt, 
\ee 
hence in the limit $R\to\infty$, 
\be\la{ERtl} 
~~~~~~~~~~~~~~~~~~E(T)-E(0)=\int_0^T \lim_{R\to\infty}\dot E_R(t)dt=0. 
~~~~~~~~~~~~~~~~~~~~~~~~~\loota 
\ee 
{\bf Proof for the general case} By Definition  (\re{eLdf}), 
$E(t)= \lim_{R\to\infty}E_R(t)$ where 
\be\la{eLdfR} 
E_R(t)\!= \! 
\int_{|\x |<R}\Big[\pi_0(x)\na_0\psi(x)- 
\cL(x, \psi(x),\na\psi(x))\Big]\Big| 
_{x_0=ct}d\x . 
\ee 
Differentiating, we get 
\beqn\la{eLdfd} 
\dot E_R(t)\!\!&\!\!=\!\!&\!\! 
\int_{|\x |<R}\Big[\na_0\pi_0(x)\na_0\psi(x) 
+\pi_0(x)\na_0^2\psi(x) 
\nonumber\\ 
&&\!-\! 
\cL_\psi (x,\psi(x),\na\psi(x))\na_0\psi(x) 
\!-\! 
\sum\limits_{\al=0}^{3} 
\pi_\al(x)\na_0{\na_\al\psi} 
\Big] d\x . 
\eeqn 
By (\re{ELfp}), we have 
$\na_0\pi_0(x) 
=\cL_\psi (x,\psi(x),\na\psi(x)) 
-\sum_1^3 
\na_k\pi_k(x)$. 
Substituting into (\re{eLdfd}), we get 
by the Stokes theorem, 
\beqn\la{der} 
\nonumber\\ 
\!\!\!\!\dot E_R(t)\!\!&\!\!=\!\!&\!\! 
-\!\! \int\limits_{|\x |<R}\!\!\!\!\Big[\sum_1^3 
\na_k\pi_k(x)\na_0\psi\!\!+\!\! 
\sum\limits_{\al=1}^{3} 
\pi_\al(x)\na_0\na_\al\psi 
\Big] d\x 
\nonumber\\ 
\!\!&\!\!=\!\!&\!\! 
-\!\! \int\limits_{|\x |<R}\sum_1^3 
\na_k\Big[\pi_k(x)\na_0\psi(x) 
\Big] d\x 
\nonumber\\ 
\!\!&\!\!=\!\!&\!\! 
-\!\! \int\limits_{|\x |=R}\sum_1^3 
n_k(\x )\Big[ 
\pi_k(x)\na_0\psi(x) 
\Big] dS, 
\eeqn 
where $n_k(\x ):=x_k/|\x |$, and $dS$ is the 
Lebesgue measure on the sphere $|\x |=R$. 
Now, Eq. (\re{sde}) implies that for every fixed $t$ we have 
$\dot E_R(t)\to 0$ as 
$R\to\infty$, hence 
$E(t)=$const.\bo 
\br 
The identity (\re{der}) means that the vector-function 
\be\la{enc} 
S_k(x):=\pi_k(x)\na_0\psi(x):=\pi_k^j(x)\na_0\psi^j(x),\,\,\,\,\,\,\,\, 
k=1,...,3,\,\,\,\,x\in\R^{4} 
\ee 
is the {\bf energy current density}. 
\er 
 
%%%%%%%%%%%%%%%%%%%%%%%%%%%%%%%%%%%%%%%%%%%%%%%%%55 
\subsection{Translation Invariance: Momentum Conservation} 
\bd 
\la{LmEfp} 
The momentum of the Lagrangian field at time $t$ 
is the vector 
\be\la{mLdf} 
\p_n(t)= -\int_{\R^3} 
\Big[\pi_0(x)\,\na_k\psi(x)\Big] d\x , 
\,\,\,\,\,\,n=1,2,3,\,\,\,\,t\in\R. 
\ee 
\ed 
 
\bex 
For the linear wave equation (\re{dE}), the 
 momentum is given by 
\be\la{mLdfe} 
\p (t)= -\int_{\R^3} 
\dot\psi(t,\x )\,\na\psi(t,\x )\,d\x , 
\,\,\,\,\,\,\,t\in\R. 
\ee 
\eex

\bt\la{mCLf} Assume that 
the Lagrangian density $\cL(x, \psi,\na\psi)$ 
does not depend on the coordinate 
$x_n$. 
Then for any trajectory $\psi(x)\in C^2_\si$ 
of Eq. (\re{ELfp}) 
the corresponding component of momentum is 
 conserved, $\p _n(t)=\co$ . 
\et 
{\bf Proof for a particular case} 
Let us prove the theorem for the d'Alembert equation (\re{d1}). 
 By definition, 
$P (t)= \lim_{R\to\infty}P _R(t)$ where 
\be\la{eLdfRdp} 
P _R(t)= - 
\int_{-R}^R 
\dot \psi(t,x )\psi'(t,x )dx. 
\ee 
Differentiating, we get 
\be\la{eLdfRddp} 
\dot P_R(t)= - 
\int_{-R}^R 
\Big[\ddot \psi(t,x ) \psi'(t,x ) 
+\dot\psi(t,x )\dot\psi'(t,x ) \Big]dx . 
\ee 
Substituting here $\ddot \psi(t,x )=\psi''(t,x )$, we get, 
\be\la{derdp} 
\dot P_R(t)=- 
\Big[\fr{|\psi'(t,x )|^2}2+ 
\fr{|\dot\psi(t,x )|^2}2 
\Big]\Bigg|_{-R}^R. 
\ee 
Now, Eq. (\re{sde}) implies that, for every fixed $t$, we have 
$\dot P_R(t)\to 0$ as 
$R\to\infty$, hence 
$P(t)=$const.\bo

\bexe 
Prove Theorem \re{mCLf} for the general case. 
{\bf Hint:} Differentiate in time and apply partial integration 
in $\x$. 
\eexe

\subsection 
{Rotation Invariance: Angular Momentum Conservation} 
 
Let us denote by  $O_n(s)$ 
the rotations of the space $\R^3$ 
around a unit vector $\e_n\in\R^3$ ($\e_1:=(1,0,0)$, etc), 
with an angle of $s$ {\it radian}. 
Let us consider 
the Lagrangian densities 
which are invariant with respect to 
rotations: 
\be\la{LrN} 
~~~~~~~~\cL(x_0,\x, \psi,\na_0\psi,\na_\x\psi)\equiv 
\cL(x_0,O_n(-s)\x, \psi,\na_0\psi,O_n^t(s)\na_\x\psi ). 
\ee 
\bex 
Let us consider the case of $\e_3:=(0,0,1)$. Then 
the invariance (\re{LrN}) holds for Lagrangian densities of the 
following structure: 
\be\la{Lr} 
~~~~~~~~\cL(x, \psi(x),\na\psi(x))\equiv 
\cL_*(x_0,|(x_1,x_2)|,x_3,\psi(x),\na_0\psi(x),|(\na_1\psi(x),\na_2\psi(x))|, 
\na_3\psi(x)). 
\ee 
\eex 
\bd 
\la{amEf} 
The angular momentum of the Lagrangian field at time $t$ 
is vector 
\be\la{MLdf} 
\bJ_n(t)= \int_{\R^d}\Big[\pi_0(x)\,\, 
(\x\times\na_\x)_n\psi(x)\Big] d\x , 
\,\,\,\,\,\,n=1,2,3,\,\,\,\,t\in\R, 
\ee 
where ${\na_\x}:=(\na_1,\na_2,\na_3)$: for example, 
$(\x\times\na_\x)_3:=x_1\na_2-x_2\na_1$ etc. 
\ed 
Note that the integral converges 
for solutions $\psi(x)\in C^1_\si$ with $\si>2$, 
by (\re{sde}) and (\re{bLd}). 
\br 
By definition, $\pi_0(x)\,\, 
(\x\times\na_\x)_n\psi(x)=\pi_0^j(x)\,\, 
(\x\times\na_\x)_n\psi^j(x)$. 
\er 
 
\bt\la{MCLf} Let the Lagrangian density satisfy 
(\re{Lr})  and 
$\si>5/2$. Then, for 
any trajectory $\psi(x)\in C^2_\si$ 
of Eq. (\re{ELfp}), 
the corresponding component 
 of angular momentum is 
 conserved, $\bJ_n(t)=\co$. 
\et 
 
\bexe 
Prove Theorem \re{MCLf}. 
\eexe

\bexe 
Calculate the 
angular momentum for the Klein-Gordon and Schr\"odinger equations 
 (\re{NKG}),  (\re{NS}). 
\eexe 
%%%%%%%%%%%%%%%%%%%%%%%%%%%%%%%%%%%%%%%%%%%%%%%%%%%%%%%%%% 
\subsection{Phase Invariance: Charge Conservation} 
Consider complex-valued fields $\psi$ with Lagrangian densities 
which are invariant with respect to 
rotations in $\psi$: 
\be\la{Lri} 
\cL(x, e^{i\theta}\psi(x),e^{i\theta}\na\psi(x))\equiv 
\cL(x,\psi(x),\na\psi(x)),\,\,\,\,\,\theta\in\R. 
\ee 
\bexe 
Check  that (\re{Lri}) holds for Eqns  (\re{NKG}) and   (\re{NS}). 
\eexe 
 
\bd 
\la{dQ} 
The charge of the Lagrangian field at time $t$ 
is defined by 
\be\la{LQ} 
Q(t)= \int_{\R^d}\Big[\pi_0(x)\cdot 
i\psi(x)\Big] d\x , 
\,\,\,\,\,\,\,\,\,\,t\in\R, 
\ee 
where $\pi_0(x)$ and $i\psi(x)$ are identified 
with real vectors in $\R^2$, and $\cdot$ 
is the real scalar product in $\R^2$ (cf. (\re{cdot})).

\ed 
Note that the integral converges 
for the solutions $\psi(t)\in C^1_\si$ 
by (\re{sde}) and (\re{bLd}) since $\si>3/2$.

\bt\la{Qc} Let the Lagrangian density satisfy 
(\re{Lri}) and $\si>3/2$. Then for 
any trajectory $\psi(x)\in C^2_\si$ 
of Eq. (\re{ELfp}) 
the charge is  conserved, $Q(t)=\co$ . 
\et 
{\bf Proof for a particular case} 
Let us prove the theorem for the {\it free linear} 1D Klein-Gordon equation 
(\re{NKG}), 
\be\la{dQd} 
\ddot \psi(t,x )=\psi''(t,x )-\mu^2\psi(t,x ), 
\,\,\,\,\x \in\R. 
\ee 
Then $\pi=\dot\psi$ and 
$Q(t)= \lim_{R\to\infty}Q_R(t)$, where 
\be\la{QR} 
Q_R(t)= 
\int_{-R}^R 
\dot\psi(t,x )\cdot i\psi(t,x )dx . 
\ee 
Differentiating, we get 
\be\la{QRd} 
\dot Q_R(t)= 
\int_{-R}^R 
\ddot \psi(t,x )\cdot i \psi(t,x )dx 
+\int_{-R}^R\dot\psi(t,x )\cdot i\dot\psi(t,x ) dx . 
\ee 
The second integral on the RHS is zero since $z_1\cdot iz_2$ 
is an antisymmetric bilinear form in $\C$ by 
(\re{cdot}). 
Hence, 
substituting $\ddot \psi(t,x )=\psi''(t,x )-\mu^2\psi(t,x )$ 
from (\re{dQd}), we get by partial integration, 
\beqn\la{QRdg} 
\dot Q_R(t)&=& 
\psi'(t,x )\cdot i \psi(t,x )\Big|_{-R}^{R} 
-\int_{-R}^R 
\psi'(t,x )\cdot i \psi'(t,x )dx 
-\mu^2 
\int_{-R}^R 
\psi(t,x )\cdot i \psi(t,x )dx \nonumber\\ 
&=& 
\psi'(t,x )\cdot i \psi(t,x )\Big|_{-R}^{R} 
=j(t,-R)-j(t,R),\,\,\,\,\,j(t,x):=-\psi'(t,x )\cdot i \psi(t,x ). 
\eeqn 
since both integrals are zero by antisymmetry. 
Therefore, Eq. (\re{sde}) implies that for every fixed $t$ we have 
$\dot Q_R(t)\to 0$ as 
$R\to\infty$, hence 
$Q(t)=$const.\bo

\bexe 
Prove Theorem \re{Qc} for the {\it nonlinear} 1D 
Schr\"odinger equation 
(\re{NS}), 
\be\la{dQdS} 
i\dot \psi(t,\x )=\fr 1{2\mu}\psi''(t,\x )+F(\psi), 
\,\,\,\,\x \in\R, 
\ee 
where $F(\psi)=-\na U(|\psi|)$, $\psi\in\C\equiv\R^2$. 
\eexe 
 
\bexe 
Prove Theorem \re{Qc} for  the general case and check the 
{\bf continuity equation}, 
\be\la{j} 
\dot \rho(t,x)+\dv \bj(t,x)=0, 
\ee 
where $\rho(t,x):=\pi_0(x)\cdot i\psi(x)$ is the 
{\bf charge density} and 
$\bj_k(t,x):=\pi_k(t,\x )\cdot i \psi(t,\x )$ is the {\bf current 
density}. 
\eexe 
 
\bexe 
Calculate the charge- and current 
densities for the Klein-Gordon and Schr\"odinger equations 
 (\re{NKG}),  (\re{NS}). 
\eexe

\newpage

%%%%%%%%%%%%%%%%%%%%%%%%%%%%%%%%%%%%%%%%%%%%%%%%%%%%%% 
 
%%%%%%%%%%%%%%%%%%%%%%%%%%%%%%%%%%%%%%%%%%%%%%%%%%%%%% 
 
%%%%%%%%%%%%%%%%%%%%%%%%%%%%%%%%%%%%%%%%%%%%%%%%%%%%%% 

%%\setcounter{section}{10} 
\setcounter{subsection}{0} 
\setcounter{theorem}{0} 
\setcounter{equation}{0} 
\section 
{Lagrangian Theory for the Maxwell Field} 

The evolution of the electromagnetic field is governed by the Maxwell 
equations, which describe, among other features, the propagation of 
light. The Maxwell equations already obey the symmetries of 
special relativity. The formulation of the Maxwell equations in terms of 
{\em potentials} rather than fields is especially useful, because 
these potentials show up in the matter equations (e.g., in the 
Schr\"odinger and Klein-Gordon equations). 
 
The Maxwell equations, as well, can be derived from a least action principle, 
and, therefore, a Lagrangian formulation of the Maxwell theory may be 
given.  A classical, point-like charged particle in an electromagnetic 
field is described by the non-relativistic or relativistic Lorentz equations, 
respectively. Again, a Lagrangian and Hamiltonian formulation for the 
Lorentz equations can be found. 
 
%%%%%%%%%%%%%%%%%%%%%%%%%%%%%%%%%%%%%%%%%%%%%%%%%%%%%%%%%% 
 
\subsection{Maxwell  Equations and Potentials. Lagrangian Density} 
In 1862, Maxwell completed the Coulomb, Faraday and Biot-Savart-Laplace 
equations by the {\it displacement current } and wrote the complete 
system of {\it classical electrodynamics}. 
 In 
{\it Gaussian units}  it reads 
\beqn\la{meq} 
~~~~~~\left\{ 
\ba{ll} 
\dv \bE(t,\x)= 4\pi\rho (t,\x),&\rot \bE(t,\x)= - \ds\fr 1c\dot  \bB(t,\x),\\ 
~\\ 
\dv \bB(t,\x)= 0,&\rot \bB(t,\x)= 
\ds\fr{4\pi}c 
\,\bj(t,\x)+ 
\ds\fr 1c \,\dot \bE(t ,\x), 
\ea 
\right|~~~~~~(t,\x)\in\R^4. 
\eeqn 
where $\rho (t,\x)$ and $\bj(t,\x)$ stand for the charge 
and current density, respectively. 
\br 
The Maxwell equations imply the {\it charge continuity equation}, 
\be\la{cce} 
\dot\rho(t,\x)+\dv \bj(t,\x)=0,\,\,\,\,(t,\x)\in\R^4. 
\ee 
\er 
\subsubsection{Potentials and gauge invariance} 
Let us introduce scalar and vector potentials to rewrite 
(\re{meq}) in a relativistic covariant four-dimensional form. 
Namely, $\dv \bB(t,\x)= 0$ implies that $\bB(t,\x)= \rot \bA(t,\x)$. 
Then 
$\rot\,\bE(t,\x)= - \ds\fr 1c\dot \bB\b(t,\x)$ implies 
$\rot\,[\bE(t,\x)+\ds\fr 1c\,\dot  \bA(t,\x)]=0$ hence 
$\bE(t,\x)+\ds\fr 1c\,\dot  \bA(t,\x)=-\na_\x\phi(t,\x)$. Finally, 
\be\la{pot} 
\bB(t,\x)= \rot \bA(t,\x),\,\,\,\,\,\,\, 
\bE(t,\x)=-\na_\x\phi(t,\x)-\ds\fr 1c\,\dot 
\bA(t,\x),~~~~~~~\,\,\,(t,\x)\in\R^4. 
\ee 
The justification of all these relations follows from 
the Fourier transform. 
The choice of the potentials is not unique since the {\it gauge transformation} 
\be\la{gt} 
\phi(t,\x)\mapsto \phi(t,\x)+\fr 1c\,\dot\chi(t,\x),\,\,\,\, 
\bA(t,\x)\mapsto \bA(t,\x)-\na_\x\chi(t,\x) 
\ee 
does not change the fields $\bE(t,\x)$,  $\bB(t,\x)$ for any function 
$\chi(t,\x)\in C^1(\R^4)$. 
Therefore, 
it is possible to 
satisfy an additional {\it gauge} condition. Let us choose for example 
the {\it Lorentz gauge} 
\be\la{Lg} 
\fr 1c\,\dot\phi(t,\x)+\dv \bA(t,\x)=0,\,\,\,(t,\x)\in\R^4. 
\ee 
\bexe 
Prove the existence of the potentials satisfying (\re{Lg}). 
{\bf Hint:} Derive an equation for the function $\chi(t,\x)$. 
\eexe 
Let us express the Maxwell equations (\re{meq}) in terms of 
the potentials. 
Substituting Eq. (\re{pot}) into the first Maxwell equation leads to 
$4\pi\rho (t,\x)=\dv \bE(t,\x)= 
-\De\phi(t,\x) 
-\ds\fr 1c\, 
\dv\dot  \bA(t,\x)$. 
Eliminating $\dv\dot  \bA(t,\x)$ 
by 
the differentiation of (\re{Lg}) 
in time, 
$\ds\fr 1c\,\ddot\phi(t,\x)+\dv \dot \bA(t,\x)=0$, we get 
\be\la{dp} 
\Box\phi(t,\x):=[\fr 1{c^2}\pa_t^2-\De]\phi(t,\x)= 
4\pi\rho (t,\x),\,\,\,(t,\x)\in\R^4. 
\ee 
Similarly,  substituting Eq. (\re{pot}) into 
the last Maxwell equation, we get 
\be\la{rrA} 
\rot\rot \bA(t,\x)=\ds\fr{4\pi}c\,\bj(t,\x)+\ds\fr 1c \,\dot \bE(t ,\x)= 
\ds\fr{4\pi}c\,\bj(t,\x) 
-\ds\fr 1c \,\na_\x\dot\phi(t,\x) 
-\ds\fr 1{c^2}\,\ddot  \bA(t,\x). 
\ee 
\bexe Prove the identity 
\be\la{rr} 
\rot\rot =-\De+\na_\x\dv \, . 
\ee 
\eexe 
Substituting  Eq. (\re{rr}) into (\re{rrA}) and eliminating 
$\ds\fr 1c \,\na_\x\dot\phi(t,\x)$ 
by application of $\na_\x$ to (\re{Lg}), we get 
\be\la{dA} 
\Box \bA(t,\x)= 
\ds\fr{4\pi}c\,\bj(t,\x),\,\,\,(t,\x)\in\R^4. 
\ee 
\br 
The arguments above show that the 
Maxwell equations (\re{meq}) 
are equivalent to the 
system of two wave equations (\re{dp}), (\re{dA}) for the potentials 
with the Lorentz gauge condition (\re{Lg}). 
\er 
\subsubsection{4D vector potential} 
Let us introduce the four-dimensional notations 
\beqn\la{4dn1} 
\left\{\ba{lll} 
\,\,x_0=ct,\,\,\,\,\,\,\,x_\mu=(x_0,...,x_3),\,\,\,&\pa_\mu=\na_\mu\,\,\,=\,\,\,(\,\,\,\pa_0,\,\,\,\pa_1, 
\,\,\,\pa_2,\,\,\,\pa_3),&\\ 
x^\mu:=g^{\mu\nu}x_\nu=(x_0,-x_1,-x_2,-x_3) 
,&\pa^\mu:=g^{\mu\nu}\pa_\nu=(\pa_0,-\pa_1, 
-\pa_2,-\pa_3),&\\ 
\ea\right. 
\eeqn 
where $g_{\mu\nu}=g^{\mu\nu}={\rm diag}(1,-1,-1,-,1)$. 
Let us also introduce the four-dimensional fields and currents 
\beqn\la{4vf} 
~~~~~\left\{\ba{ll} 
\A^\mu(x)=(\phi(t,\x), \bA(x)),& 
\A_\mu(x):=g_{\mu\nu}\A^\nu(x)=(\phi(x), -\bA(x)),\\ 
\J^\mu(x)=(\rho(x), \ds\fr 1c\,\bj(x)),& 
\,\J_\mu(x):=g_{\mu\nu}\J^\nu(x)=(\rho(x), -\ds\fr 1c\,\bj(x)). 
\ea\right|\,\,\,\,\,\,\,\,\,x\in\R^4 
\eeqn 
Then the Maxwell equations 
(\re{dp}), (\re{dA}) become 
\be\la{4m} 
\Box \A^\mu(x)=4\pi \J^\mu(x),\,\,\,\,\,\,x\in\R^4. 
\ee 
Similarly, the charge continuity equation (\re{cce}), 
gauge transformation (\re{gt}) and 
Lorentz gauge 
(\re{Lg}) become 
\be\la{cLg} 
\pa_\mu\J^\mu(x)=0, 
\,\,\,\,\,\,\,\, 
\A_\mu(x)\mapsto \A_\mu(x)+\pa^\mu\chi(x), 
\,\,\,\,\,\,\,\, 
\pa_\mu\A^\mu(x)=0,\,\,\, 
\,\,\,\,\,\,x\in\R^4. 
\ee 
\subsubsection{Tensor field} 
\bd 
The Maxwell tensor is defined by 
\be\la{tenf} 
\F^{\mu\nu}(x)=\pa^\mu\A^\nu(x)-\pa^\nu\A^\mu(x), 
~~~~ 
x\in\R^4. 
\ee 
\ed 
\bexe 
Check the formula 
\be\la{Maxten} 
\left(\F^{\mu\nu}(x)\right)_{\mu,\nu=0}^{3}= 
\left( 
\ba{llll} 
0&-\bE_1(x)&-\bE_2(x)&-\bE_3(x)\\ 
\bE_1(x)&0&-\bB_3(x)&\bB_2(x)\\ 
\bE_2(x)&\bB_3(x)&0&-\bB_1(x)\\ 
\bE_3(x)&-\bB_2(x)&\bB_1(x)&0 
\ea\right) 
\ee 
{\bf Hint:} 
Use the formulas (\re{pot}) and (\re{4dn1}), (\re{4vf}). 
\eexe

\bp 
The Maxwell equations (\re{4m}) are equivalent to 
\be\la{mt} 
\pa_\mu\F^{\mu\nu}(x)=4\pi\J^\nu(x),\,\,\,\,\,\,x\in\R^4. 
\ee 
\ep 
\Pr The Maxwell tensor does not depend on the choice of gauge 
since $\pa^\mu\pa^\nu\chi(x)-\pa^\nu\pa^\mu\chi(x)=0$. 
Therefore, we can assume the Lorentz gauge (\re{Lg}) 
without loss of generality. Then (\re{4m}) implies 
\be\la{mtp} 
~~~~~~~~~~ 
\pa_\mu\F^{\mu\nu}(x)\!=\!\pa_\mu(\pa^\mu\A^\nu(x)\!-\!\pa^\nu\A^\mu(x)) 
\!=\!\pa_\mu\pa^\mu\A^\nu(x)\!=\!\Box \A^\nu(x)\!=\! 
4\pi\J^\nu(x),\,\,\,\,x\in\R^4.~~~~~~\loota 
\ee 
\subsubsection{Lagrangian density} 
\bd 
The Lagrangian density for the Maxwell equations (\re{mt}) with 
{\bf given} external 
charge-current densities $\J^\nu(x)$ is defined by 
\be\la{mtL} 
\cL(x,\A_\mu,\na\A_\mu) 
=-\fr 1{16\pi}\,\F^{\mu\nu}\F_{\mu\nu}-\J^\nu(x)\A_\nu, 
\,\,\,\,\,\,(x,\A_\mu,\na\A_\mu)\in\R^4\times\R^4\times\R^{16}. 
\ee 
where $\F^{\mu\nu}:=\pa^\mu\A^\nu-\pa^\nu\A^\mu$ and 
$\F_{\mu\nu}:=\pa_\mu\A_\nu-\pa_\nu\A_\mu$. 
\ed 
 
\bp 
The Maxwell equations (\re{mt}) with given 
charge-current densities $\J^\nu(x)$ 
are equivalent to 
the Euler-Lagrange equations (\re{ELfp}) 
with the Lagrangian density (\re{mtL}) 
for the fields $\A_\mu(x)$. 
\ep 
\Pr $\F_{\mu\nu}\F^{\mu\nu}$ is a quadratic form 
in $\na\A$. Therefore, 
the canonically conjugate fields $\pi_\al$ with 
components $\pi_{\al\beta}$ are given by 
\be\la{pab} 
\pi_{\al\beta}:=\na_{\pa_\al \A_\beta}{\cL}=-\fr 1{8\pi}\,\F^{\mu\nu} 
\na_{\pa_\al \A_\beta}\F_{\mu\nu}=-\fr 1{8\pi}\,(\F^{\al\beta}-\F^{\beta\al}) 
=-\fr 1{4\pi}\,\F^{\al\beta}. 
\ee 
Therefore, 
\be\la{napab} 
\na_\al\pi_{\al\beta}(x)= 
-\fr 1{4\pi}\,\na_\al\F^{\al\beta}(x),\,\,\,\,\,\,x\in\R^4. 
\ee 
On the other hand, $\pa_{\A_\beta}\cL=-\J_\beta(x)$, 
hence (\re{mt}) is equivalent to 
(\re{ELfp}). 
\bo 
 
%%%%%%%%%%%%%%%%%%%   %%%%%%%%%%%%%%%%%%%%%% 
\subsection{Lagrangian for Charged Particle in Maxwell Field} 
In a {\bf continuous} Maxwell field 
the motion of a charged particle with small velocity 
$|\dot \x(t)|\ll c$ 
is governed by the 
Lorentz equation $(L)$ from the Introduction, 
\be\la{Le} 
\mu \ddot \x(t)=e [\bE(t,\x(t))+\fr 1c\,\dot \x(t)\times \bB(t,\x(t))], 
\,\,\,\,t\in\R, 
\ee 
where $\mu $ is the mass of the particle and $e$ its charge. 
For large velocities $|\dot \x(t)|\sim c$ 
the equation must be replaced by 
\be\la{Ler} 
\dot \p(t)=e [\bE(t,\x(t))+\fr 1c\,\dot \x(t)\times \bB(t,\x(t))], 
\,\,\,\,t\in\R, 
\ee 
where $\p:=\mu \dot \x/\sqrt{1-(\dot \x/c)^2}$ is the relativistic momentum 
of the particle (see (\re{relps})). 
Let us assume that the fields $\bE,\bB$ are  $C^1$ vector functions 
in $\R^4$. Then the dynamical equations (\re{Le}), (\re{Ler}) 
define the corresponding dynamics uniquely. 
Let us define the charge-current 
densities $\J^\nu(t,\x)$ corresponding  to the trajectory $\x(\cdot)$: 
by (\re{4vf}), 
\be\la{ccd} 
\J^0(t,\x)=e\de(x-\x(t)),\,\,\,\,\,\,\,\,\,\, 
~~~~~~~~\J^k(t,\x)=\fr 1c\, 
\dot \x_k(t)e\de(x-\x(t)),\,\,\,k=1,2,3. 
\ee 
Let us show  that the dynamical equations (\re{Le}), (\re{Ler}) 
{\it automatically follow} 
from the Hamilton LAP applied to the Lagrangian density 
(\re{mtL}) with  fixed fields $E(t,\x),B(t,\x)$. 
We consider  (\re{Le}) for  concreteness. 
Let $\A(t,\x)$ be the 4-potential corresponding to the fields 
$\bE(t,\x),\bB(t,\x)$. 
The Lagrangian density (\re{mtL}) consists of two parts: 
the field part 
$\cL_f$ and the {\it field-matter interaction} part 
$\cL_{fm}$. 
Substituting (\re{ccd}) into $\cL_{fm}$, 
we get the field-matter action in the form 
\beqn\la{Laga} 
S^T_{fm}:&=&\int_0^T\int_{\R^3}\J^\nu(t,\x)\A_\nu(t,\x)d\x\, dt 
\nonumber\\ 
&&\nonumber\\ 
&=&e\int_0^T\Big[\phi(t,\x(t))-\fr 1c\,\dot \x(t)\cdot 
\bA(t,\x(t))\Big]dt. 
\eeqn 
The interaction term 
corresponds to the following Lagrangian function for the 
nonrelativistic particle, 
\be\la{Lagf} 
L(\x,\bv,t)=\fr{\mu \bv^2}2-e\phi(t,\x)+\ds\fr ec\bv\cdot \bA(t,\x). 
\ee 
\bt 
Let  the potential  $\A(t,\x)$ be fixed. 
Then the Lorentz equation (\re{Le}) for the trajectory 
$\x(\cdot)$ is 
equivalent to the Euler-Lagrange equations corresponding 
to the 
Lagrangian  (\re{Lagf}) 
\et 
\Pr 
First let us evaluate the momentum. By definition, 
$\p\!:=L_\bv=\mu \bv+\ds\ds\fr ec\,\bA(t,\x)$, hence 
\be\la{mo} 
\p(t):=L_\bv(\x(t),\dot \x(t),t)=\mu \dot \x(t)+\ds\fr ec\,\bA(t,\x(t)). 
\ee 
Now (\re{ELN}) becomes, 
\be\la{ELM} 
\dot \p_k(t)=L_{\x_k}(\x(t),\dot \x(t),t) 
=-e\na_k\phi(t,\x)+\ds\fr ec\dot \x\cdot \na_k \bA(t,\x)], 
~~~~~~k=1,2,3. 
\ee 
Let us calculate the derivative on the LHS: 
\be\la{pkd} 
\dot \p_k(t)=\fr d{dt}(\mu \dot \x_k(t)+\ds\fr ec\,\bA_k(t,\x(t)))= 
\mu \ddot \x_k+\ds\fr ec\,[\dot \bA_k(t,\x)+\na_j \bA_k(t,\x)\dot \x_j]. 
\ee 
Substituting this expression into the LHS of (\re{ELM}), we get 
\be\la{ELMsl} 
\mu \ddot \x_k+\ds\fr ec\,[\dot \bA_k(t,\x)+\na_j \bA_k(t,\x)\dot \x_j] 
=-e\na_k\phi(t,\x)+\ds\fr ec\dot \x_j\na_k \bA_j(t,\x). 
%%%\mu \ddot \x_1 
%%%=e[-\fr 1c\,\dot A_1(t,\x)-\na_1\phi(t,\x)] 
%%%+\ds\fr ec\,[\dot \x\cdot \na_1 A(t,\x)- \na_\x A_1(t,\x)\cdot\dot \x]. 
\ee 
Let us rewrite this as follows: 
\be\la{ELMslr} 
\mu \ddot \x_k=e[-\na_k\phi(t,\x)-\fr 1c\,\dot \bA_k(t,\x)] 
+\ds\fr ec\,\dot \x_j[\na_k \bA_j-\na_j \bA_k]. 
\ee 
The first square bracket on the RHS is $\bE(t,\x)$ by 
(\re{pot}). Hence, it remains to check that 
$\dot\x_j[\na_k \bA_j-\na_j \bA_k]=\dot \x\times \rot \bA(t,\x)$. 
Let us note that $\na_k \bA_j-\na_j \bA_k=(\rot \bA)_l\ve_{kjl}$ 
where $\ve_{kjl}$ is the antisymmetric tensor. Therefore, 
$\dot\x_j[\na_k \bA_j-\na_j \bA_k]=\dot\x_j(\rot \bA)_l\ve_{kjl}= 
[\dot \x\times \rot \bA(t,\x)]_k$ by definition of the vector 
product.\bo

\br 
The derivation of the expression for the Lorentz force from 
the Maxwell equations 
is not very surprising since the expression also follows from the 
Coulomb and Biot-Savart-Laplace equations. 
\er 
\subsection{Hamiltonian for Charged Particle in Maxwell Field} 
 
\subsubsection{Nonrelativistic particle} 
We evaluate the Hamilton function as the Legendre transform 
of the nonrelativistic Lagrangian (\re{Lagf}): first, 
\be\la{Hamfnr} 
~~~~~~~~~~H:=\p\bv-L=\p\bv-\fr{\mu \bv^2}2+e\phi(t,\x)-\ds\fr ec\bv\cdot A(t,\x)= 
e\phi(t,\x)+\bv(\p-\ds\fr ec \bA(t,\x))-\fr{\mu \bv^2}2. 
\ee 
Next, we eliminate $\bv$ by the relation $\p-\ds\fr ec \bA(t,\x)=\mu \bv$. 
Then we get finally, 
\be\la{Hamfnrt} 
H= 
e\phi(t,\x)+\fr{\mu \bv^2}2=e\phi(t,\x)+\fr1{2\mu }(\p-\ds\fr ec \bA(t,\x))^2. 
\ee 
\subsubsection{Relativistic particle} 
Let us consider the relativistic  Lagrangian (cf. (\re{relps})) 
\be\la{Lagfr} 
L(\x,\bv,t)=-\mu c^2\sqrt{1-\beta^2}-e\phi(t,\x)+\ds\fr ec\bv\cdot A\b(t,\x), 
\ee 
where $\beta:=|\bv|/c$. Let us note that the first term on the RHS 
is asymptotically $-\mu c^2+\mu \bv^2/2$ for $\beta\ll 1$. 
First we evaluate the momentum: by definition, 
$\p\!:=L_\bv=\mu \bv/\sqrt{1-\beta^2}+\ds\ds\fr ec\,\bA(t,\x)$, hence 
\beqn\la{Hamfr} 
H:&=&\p\bv-L=\p\bv+\mu c^2\sqrt{1-\beta^2} 
+e\phi(t,\x)-\ds\fr ec\bv\cdot \bA(t,\x)\nonumber\\ 
&=& 
e\phi(t,\x)+\bv(\p-\ds\fr ec \bA(t,\x))+\mu c^2\sqrt{1-\beta^2}. 
\eeqn 
Next, we eliminate $\p$ by the relation 
$\p-\ds\fr ec \bA(t,\x)=\mu \bv/\sqrt{1-\beta^2}$. 
Then we get, 
\beqn\la{Hamfrt} 
H&=& 
e\phi(t,\x)+\fr{\mu \bv^2}{\sqrt{1-\beta^2}}+\mu c^2\sqrt{1-\beta^2}= 
e\phi(t,\x)+\fr{\mu c^2}{\sqrt{1-\beta^2}}\nonumber\\ 
&=&e\phi(t,\x)+\mu c^2\sqrt{1+(\p-\ds\fr ec \bA(t,\x))^2/(\mu c)^2} 
\,. 
\eeqn 
We can rewrite this relation in the following standard form 
\be\la{Hamfrts} 
(H/c-\ds\fr ec\phi(t,\x))^2 
= 
\mu ^2c^2+(\p-\ds\fr ec A(t,\x))^2 
\,. 
\ee 
\br 
This expression coincides with (\re{EPr}) for $\phi=0$, $A=0$. 
\er

\part{Schr\"odinger Equation}

\newpage 

\setcounter{subsection}{0} 
\setcounter{theorem}{0} 
\setcounter{equation}{0} 
\section 
{Geometric Optics and Schr\"odinger Equation} 

In this lecture we show the wave propagation along straight lines 
for the simple cases of free Klein-Gordon and Schr\"odinger 
equations. For the Schr\"odinger equation coupled to the 
Maxwell field we analyze the propagation along rays. 
The proof is based on constructing a formal Debye expansion.

Wave equations of type (\re{dp}) describe the 
wave propagation in electrodynamics, 
acoustics and many other fields. 
They describe well the diffraction 
and the interference of waves. On the other hand, 
the wave processes also demonstrate the straight-line 
propagation of waves, thereby justifying 
geometric optics. The mathematical description 
of this feature by the wave equation 
has been discovered by Hamilton around 1830 and developed 
by Liouville in 1837, Debye in 1911, 
Rayleigh in 1912, Jeffreys in 1923, 
and 
Schr\"odinger, Wentzel, Kramers and Brillouin in 1926.

\subsection{Straight Line Propagation for the Free Equations} 
Let us analyze the straight line propagation 
in the concrete example of 
the {\bf free} Klein-Gordon equation 
\be\la{KGf} 
\left\{ 
\ba{l} 
\ddot\psi(t,\x)=\De\psi(t,\x)\,-\, \mu^2\psi(t,\x)\\ 
\psi(0,\x)\!=\!\psi_0(\x),~\5\dot\psi(0,\x)\!=\!\pi_0(\x) 
\ea\right| 
~~~~~~~~\,\,\,\, 
(t,\x)\in\R\times\R^3. 
\ee 
Let us choose the initial data $\psi_0(\x),\pi_0(\x)$ from the 
{\it Schwartz space} $\cS(\R^3)$ of test functions. 
\bd 
 $\cS(\R^3)$ is the space of functions 
$\psi(\x)\in C^\infty (\R^3)$ such that 
\be\la{Sc} 
\sup\limits_{\x\in\R^3} (1+|\x|)^N|\na_\x^\al\psi(\x)|<\infty 
\ee 
for any $N=1,2,...$ and multiindices $\al=(\al_1,\al_2,\al_3)$. 
\ed 
\bd 
For $\psi\in\cS(\R^3)$ the Fourier transform 
is defined by 
\be\la{Ft} 
F\psi(\bk):=\hat\psi(\bk):=(2\pi)^{-3}\int_{\R^3} 
e^{i\bk\x}\psi(\x)d\x,\,\,\,\,\,\,\,\bk\in\R^3. 
\ee 
\ed 
\bp 
Let $\psi_0,\pi_0\in \cS(\R^3)$. Then the Cauchy problem 
(\re{KGf}) admits a unique solution $\psi(t,\x)$ 
satisfying the bounds 
\be\la{Scs} 
\sup\limits_{\x\in\R^3} (1+|\x|)^N|\pa_t^{\al_0} 
\na_\x^\al\psi(t,\x)|<C(\al_0,\al, N)(1+|t|)^N,\,\,\,\,t\in\R 
\ee 
for any $N=1,2,...$, $\al_0=0,1,2,...$ 
 and multiindices $\al=(\al_1,\al_2,\al_3)$. 
\ep\Pr 
Let us calculate the solution to (\re{KGf}) 
with $\psi_0,\pi_0\in \cS(\R^3)$ 
by using the Fourier 
transform. 
Let us apply the transform to Equations (\re{KGf}) 
using the well-known formulas 
\be\la{Ftf} 
(F\pa_1\psi)(t,\bk)\!=\!-ik_1\hat\psi(t,\bk),~~~ 
\,\,\,\,(F\De\psi)(t,\bk)\!=\!-k^2\hat\psi(t,\bk),\,~~~~~ 
~~~\,\,\,\,(t,\bk) 
\in \R\times\R^3. 
\ee 
The bounds (\re{Scs}) imply also 
that 
$(F\dot\psi)(t,\bk)=\dot{\hat\psi}(t,\bk)$ 
and $(F\ddot\psi)(t,\bk)=\ddot{\hat\psi}(t,\bk)$ hence 
(\re{KGf}) becomes 
\be\la{KGfF} 
\left\{ 
\ba{l} 
\ddot{\hat\psi}(t,\bk)=-k^2\hat\psi(t,\bk)\,-\, \mu^2\hat\psi(t,\bk)\\ 
\hat\psi(0,\bk)\!=\!\hat\psi_0(\bk), 
~\5\dot{\hat\psi}(0,\bk)\!=\!\hat\pi_0(\bk) 
\ea\right| 
~~~~~~~~\,\,\,\, 
(t,\bk)\in\R\times\R^3. 
\ee 
This is the Cauchy problem 
for an ordinary differential equation 
which depends on the parameter 
$\bk\in\R^3$. The solution is well-known, 
\be\la{KGsol} 
\hat\psi(t,\bk)=\hat\psi_0(\bk)\cos \om t+ 
\hat\pi_0(\bk)\fr{\sin \om t}\om,\,\,\,\,~~~~~~ 
\om=\om(k):=\sqrt{k^2+\mu^2}. 
\ee 
Therefore, the solution $\psi(t,\x)$ is given by the inverse 
Fourier transform, 
\beqn\la{inF} 
\psi(t,\x)&=&\int_{\R^3}e^{-i\x\bk}\Big[\hat\psi_0(\bk)\cos \om t+ 
\hat\pi_0(\bk)\fr{\sin \om t}\om\Big]d\x\nonumber\\ 
\nonumber\\ 
&=&\fr12\,\int_{\R^3}e^{-i\x\bk} 
\Big[ 
e^{i\om t} 
\Big(\hat\psi_0(\bk)+\fr{\hat\pi_0(\bk)}{i\om}\Big)+ 
e^{-i\om t} 
\Big(\hat\psi_0(\bk)-\fr{\hat\pi_0(\bk)}{i\om}\Big) 
\Big]d\bk\nonumber\\ 
\nonumber\\ 
&=&\psi_+(t,\x)+\psi_-(t,\x). 
\eeqn 
The representation obviously implies the bounds (\re{Scs}) 
since $\hat\psi_0, \hat\pi_0\in \cS(\R^3)$. 
\bo

Now let us choose the initial functions $\psi_0,\pi_0$ with a 
localized spectrum, i.e. 
\be\la{ls} 
\supp\hat\psi_0,\,  \supp\hat\pi_0\subset B_r(\bk_*), 
\ee 
where $B_r(\bk_*)$ is an {\it open} 
ball with center $\bk_*\in\R^3$ and 
a small radius $r\ll |\bk_*|$. Then the same is true for the 
spectra of the functions $\psi_\pm(t,\x)$ by (\re{inF}): 
\be\la{lspm} 
\supp\hat\psi_\pm(t,\cdot)\subset B_r(\bk_*),~~~~~~~~t\in\R. 
\ee 
The solutions of type $\psi_\pm(t,\x)$ are called 
{\it wave packets}. 
\bt 
Let $\psi_0,\pi_0\in \cS(\R^3)$, and (\re{ls}) holds. 
Then the corresponding  wave packets $\psi_\pm(t,\x)$ 
are localized solutions moving with the {\bf group velocities} 
$\bv_\pm=\pm\na\om(\bk_*)$ in the following sense: 
\\ 
i) For a constant $a>0$ and any $N>0$, 
\be\la{ppm} 
|\psi_\pm(t,\x)|\le C_N (1+|t|+|\x|)^{-N},\,\,\,\, 
|\x-\bv_\pm t|>\fr{ar}{|\bk_*|}|t|. 
\ee 
ii) For any constant $A>0$, 
\be\la{ppma} 
~|\psi_\pm(t,\x)|\le C (1+|t|)^{-3/2},\,\,\,\,~~~~~~~ 
|\x-\bv_\pm t|\le \fr{Ar}{|\bk_*|}|t|. 
\ee 
\et 
\Pr 
Let us prove the theorem for $\psi_+$ since for $\psi_-$ 
just the same arguments hold. 
Let us consider the function $\psi_+(t,\x)$ along a ray 
$\x=\bv t$ with an arbitrary $\bv\in\R^3$: by (\re{inF}), 
\be\la{pmv} 
\psi_+(t,\bv t)= 
\int e^{-i \phi_+(\bk)t}  \Psi_+ (\bk) d\bk. 
\ee 
Here the {\it phase function}  is given by 
$\phi_+(\bk):=\bv\bk- \om(\bk)$ and the amplitude 
$\Psi_+ (\bk):=\hat\psi_0(\bk)/2+{\hat\pi_0(\bk)}/({2i\om(\bk)})$. 
Let us apply the {\it method of stationary phase} 
\ci{F} to the 
 integral (\re{pmv}). 
Then we get that the asymptotics for $t\to\infty$ 
depends on the existence of the critical points $\bk\in\supp \Psi_+ $ 
of the phase function $\phi_+(\bk)$, 
\be\la{cr} 
\na\phi_+(\bk)=\bv\mp\na\om(\bk)=0,\,\,\,\,\,\bk\in\supp \Psi_+. 
\ee 
In other words, $\bv=\na\om(\bk)$ with a $\bk\in\supp \Psi_+ $. 
Now let us take into account 
 that $\supp\Psi_+\subset B_r(\bk_*)$ by  (\re{ls}). 
Then the system (\re{cr}) admits a solution iff 
$\bv\in {\bf V}_r(\bk_*):=\{\na\om(\bk):\bk\in B_r(\bk_*)\}$. 
Now let us analyze two distinct situations separately. 
\medskip\\ 
i) First let us consider $\bv\not\in {\bf V}_r(\bk_*)$. Then 
the asymptotics of the integral (\re{pmv}) is $|t|^{-N}$ 
for any $N>0$ 
which corresponds to (\re{ppm}). Let us deduce the asymptotics 
by a partial integration in (\re{pmv}) with the help of 
an 
obvious identity 
\be\la{iden} 
e^{-i \phi_+(\bk)t} =\fr D{-it} e^{-i \phi_+(\bk)t},\,\,\,\,~~~ 
\bk\in\supp\Psi_+, 
\ee 
where $D$ is the following differential operator, 
$ 
D=(\na\ov\phi(\bk)/|\na\phi(\bk)|^2)\cdot\na. 
$ 
It is important that $\na\phi(\bk)\ne 0$ for $\bk\in\supp\Psi_+$ 
since $\bv\not\in {\bf V}_r(\bk_*)$. 
Applying the identity (\re{iden}) $N$ times  in (\re{pmv}), 
we get by partial integration, 
\be\la{pmvp} 
\psi_+(t,\bv t)= 
 (-it)^{-N} 
\int D^N 
e^{-i \phi_+(\bk)t}  \Psi_+ (\bk) d\bk= 
(-it)^{-N} 
\int e^{-i \phi_+(\bk)t}  (D^*)^N 
\Psi_+ (\bk) d\bk, 
\ee 
where $D^*\Psi(\bk)=\na \cdot[ 
{\na\ov\phi(\bk)}\Psi(\bk)/{|\na\phi(\bk)|^2}]$ 
is the adjoint operator to $D$. Therefore, 
\be\la{pmvpt} 
\psi_+(t,\bv t)\le C_N(\bv)(1+|t|)^{-N},\,\,\,\,t\in\R 
\ee 
if $\bv\not\in {\bf V}_r(\bk_*)$. Hence, for any $B>0$, 
\be\la{pmvptx} 
\psi_+(t,\bv t)\le C_N(B)(1+|t|+|\bv t|)^{-N},\,\,\,\,t\in\R 
\ee 
if $\bv\not\in {\bf V}_r(\bk_*)$ and $|\bv|\le B$. 
\bexe 
Prove that the bounds (\re{pmvptx}) hold 
for all $\bv\not\in {\bf V}_r(\bk_*)$ 
with a uniform constant 
$C_N$ instead of $C_N(B)$. \\{\bf Hint:} Consider the function 
$\psi_+(t,\x)$ along the ray $t=w\tau,x=v\tau$ with $|w|<1$, $|v|=1$ 
and apply the partial integration with the phase function 
$\bk v-w\om(\bk)$ and large parameter $\tau\to\infty$. 
\eexe 
Finally, the bounds (\re{pmvptx}) with $C_N$ instead of $C_N(B)$ 
imply 
(\re{ppm}) since 
the diameter of the set ${\bf V}_r(\bk_*)$ does not exceed 
$\ds ar/{|\bk_*|}$. The last fact follows from the bound 
$|\na\na\om(\bk)|\le a/|\bk|$ which is obvious since 
\be\la{nom} 
\na\om(\bk)=\fr \bk{\om(\bk)}. 
\ee 
\medskip 
ii) It remains to  consider $\bv\in {\bf V}_r(\bk_*)$. 
This means that $\bv=\na\om(\bk)$ with a point $\bk\in B_r(\bk_*)$ 
which 
is a solution to the system (\re{cr}). Then 
the integral  (\re{pmv}) is called the {\it Fresnel} 
integral and its 
asymptotics is $\sim |t|^{-3/2}$ (see \ci{F}).\bo 
\br 
The asymptotics (\re{ppm}), (\re{ppma}) means that the energy 
of the field $\psi_\pm(t,\x)$ 
{\bf outside} 
a ball $|\x-\bv_\pm t|\le ar|t|/|\bk_*|$ decays rapidly, 
while {\bf inside} 
it is about constant since the energy is a {\bf quadratic form}. 
Therefore, 
the wave packet 
$\psi_\pm(t,\x)$ of the {\bf free} Klein-Gordon equation 
(\re{KGf}) 
moves like a free particle 
of the size $\cO( r|t|/|\bk_*|)$ 
and with the {\bf group velocity} 
$\bv_\pm=\pm\na\om(\bk_*)$. 
Formula (\re{nom}) means that the velocity $\bv_\pm$ corresponds 
to the relativistic 
momentum $\pm\bk_*$ of the particle. 
\er 
\bexe 
Analyze the wave packet propagation for  the {\bf free} 
 Schr\"odinger equation 
\be\la{Sfn} 
\left\{ 
\ba{rcl} 
-i\dot\psi(t,\x)&=&\ds\fr 1{2\mu}\,\De\psi(t,\x)\\ 
~\\ 
\psi(0,\x)&=&\psi_0(\x), 
\ea\right| 
~~~~~~~~\,\,\,\, 
(t,\x)\in\R\times\R^3. 
\ee 
Prove that the packets move like {\bf free} non-relativistic particles 
of the size $\cO( r|t|/|\bk_*|)$ 
with mass $m$ and momentum $\bk_*$.\\ 
{\bf Hint:} $\om=\bk^2/2m$, hence the {\bf group velocity} 
$\bv$ equals $\na\om=\bk$. 
\eexe 
 
\subsection{WKB Asymptotics for 
Schr\"odinger Equation with a Maxwell Field} 
Let us write the Lorentz equation (\re{Le}) in Hamilton form 
with the Hamiltonian (\re{Hamfnrt}): 
\be\la{LH} 
\left\{ 
\ba{rl} 
\dot \x&=~H_\p(\x,\p,t)=\ds\fr 1\mu(p-\ds\fr ec A(t,\x)),\\ 
\dot \p&=\!\!-H_\x(\x,\p,t)=\ds-e\phi_\x(t,\x)+ 
\fr e{\mu c} A_\x(t,\x), 
\ea\right. 
\ee 
where $e$ is the charge of the particle and $\mu$ its mass. 
E.Schr\"odinger associated with the Hamilton system the 
wave equation 
(\re{NS}): 
\be\la{SMm} 
(i\h\pa_t-e\phi(t,\x))\psi(t,\x) 
=\fr 1{2\mu} (-i\h\na_\x-\ds\fr ec A(t,\x))^2\psi(t,\x),~~~~~~(t,\x)\in\R^4. 
\ee 
Let us demonstrate that the short-wave solutions to (\re{SMm}) 
are governed by the Hamilton equations (\re{LH}). 
More precisely, let us consider the Cauchy problem for 
(\re{SMm}) 
with the initial condition 
\be\la{SMi} 
\psi|_{t=0}=a_0(\x) e^{\ds{iS_0(\x)}/\h}\,\,,~~~~~~\x\in\R^3, 
\ee 
where $S_0(\x)$ is a real function. 
Let us denote by $(\x(t,\x_0),\p(t,\x_0))$ the solution to the Hamilton 
equations (\re{LH}) with the initial conditions (\re{CHE}): 
\be\la{LHi} 
\x|_{t=0}=\x_0,~~~~~\p|_{t=0}=\na S_0(\x_0). 
\ee 
The solution exists for $|t|<T(\x_0)$ where $T(\x_0)>0$. 
\bd 
The curve $x=\x(t,\x_0)$ is the {\bf ray} of the 
Cauchy problem (\re{SMm}), (\re{SMi}) starting at the point 
$x_0$. 
\ed 
\bd 
The {\bf ray tube} or {\bf ray beam emanating from 
the initial function} (\re{SMi}) 
is the set 
$\ccT=\{(t,\x(t,\x_0))\in \R^4: |t|<T(\x_0),~~ 
\x_0\in\supp a_0 \}$. 
\ed 
The following theorem means roughly speaking that 
for $\h\ll 1$ 
the set $\ccT$ is 
the support of the wave function $\psi(t,\x)$, the solution 
to the Cauchy problem (\re{SMm}), (\re{SMi}). 
Let us 
assume that the potentials 
$\phi(t,\x), A(t,\x)\in C^\infty(\R^4)$ 
and $a_0,S_0\in C^\infty(\R^3)$. 
Then 
the map $\x_0\mapsto \x(t,\x_0)$ is a local 
$C^\infty$-diffeomorphism 
of $\R^3$ for small $|t|$. 
We will construct the formal {\it Debye expansion} 
\be\la{Deb} 
\psi(t,\x)\sim 
\Big(\sum\limits_{k=0}^\infty\h^k a_k(t,\x)\Big) 
e^{\ds{iS(t,\x)}/\h}\,\,,~~~~~~\h\to 0. 
\ee 
 \bt\la{tWKB} 
Let the map $\x_0\mapsto \x(t,\x_0)$ be a 
diffeomorphism 
of $\R^3$ for $|t|<T$. 
Then the formal expansion (\re{Deb}) exists 
for $|t|<T$ and is identically zero 
outside $\ccT$, i.e. 
\be\la{Deba} 
a_k(t,\x)=0,~~~~~~(t,\x)\not\in\ccT, ~~|t|<T,~~k=0,1,2,... 
\ee 
\et 
\Pr 
First, let us define the phase function $S(t,\x)$ as the 
solution to the Cauchy problem (\re{Cp}) with $N=3$: 
\be\la{Cp3} 
\left\{ 
\ba{rcl} 
-\dot S(t,\x)&=&H(\x,\na S(t,\x),t),\,\,\,\,\x\in 
\R^3,~~~|t|<T, \\ 
~&&\\ 
S|_{t=0}   &=&S_0(\x),\,\,\,\, \x\in\R^3. 
\ea 
\right. 
\ee 
The solution exists by Theorem \re{tHJ}. 
Further, let us substitute 
$ 
\psi(t,\x)=a(t,\x) 
e^{\ds{iS(t,\x)}/\h} 
$ into the Schr\"odinger equation (\re{SMm}). Then Equation 
(\re{Cp3}) 
implies the following {\it transport equation} for the amplitude $a(t,\x)$: 
\be\la{eam} 
\left\{ 
\ba{rcl}\dot a(t,\x)&=& 
-\ds\fr 1{m}[\na S(t,\x)-\ds\fr ec A(t,\x)] 
\cdot\na a(t,\x) 
\\ 
&& 
+\ds\fr 1{2\mu}[\De S(t,\x)+\na\cdot A(t,\x)]a(t,\x)+ 
\ds\fr {i\h}{2\mu}\De a(t,\x)\\ 
&=:& 
-L a(t,\x)+B(t,\x)a(t,\x)+\ds\fr {i\h}{2\mu} d(t,\x), 
~~~|t|<T,\\ 
~&&\\ 
a|_{t=0}   &=&a_0(\x),\,\,\,\, \x\in\R^3, 
\ea 
\right. 
\ee 
where $L$  is the first order differential operator 
$L a(t,\x):=\ds\fr 1{m}[\na S(t,\x)-\ds\fr ec A(t,\x)] 
\cdot\na a(t,\x)$, 
$B(t,\x)$ 
is the function $\ds\fr 1{2\mu}[\De S(t,\x)+\na\cdot A(t,\x)]$ 
and 
$d(t,\x):=\De a(t,\x)$. 
Let us note that (\re{LH}) implies 
\be\la{den} 
\dot a(t,\x)+L a(t,\x)=\ds\fr d{dt}a(t,\x(t,\x_0)). 
\ee 
Let us express all functions in {\it ray coordinates} $(t, \x_0)$: 
   $\ti a(t,\x_0):= a(t,\x(t,\x_0))$, 
$\ti B(t,\x_0):=B(t,\x(t,\x_0))$ etc. 
Then  (\re{eam}) can be rewritten as 
\be\la{eamr} 
\left\{ 
\ba{rcl} 
\ds\fr d{dt}\ti a(t,\x_0) &= & 
\ti B(t,\x_0)\ti a(t,\x_0)+\ds\fr {i\h}{2\mu}\ti d(t,\x_0),~~~|t|<T,\\ 
~&&\\ 
\ti a(0,\x_0)&=&a_0(\x_0),\,\,\,\, \x_0\in\R^3. 
\ea 
\right. 
\ee 
Let us  substitute in the first equation 
the formal expansion 
$\ti a(t,\x_0)\sim 
\sum\limits_{k=0}^\infty\h^k \ti a_k(t,\x_0) 
$. Equating formally 
the terms with identical powers of $\h$, 
we get the recursive {\it transport equations} 
\be\la{req} 
\left. 
\ba{l} 
\ds\fr d{dt}\ti a_0(t,\x_0)= 
\ti B(t,\x_0)\ti a_0(t,\x_0),\\ 
\\ 
\ds\fr d{dt}\ti a_1(t,\x_0)= 
\ti B(t,\x_0)\ti a_1(t,\x_0)+\ds\fr {i}{2\mu}\ti d_0(t,\x_0),\\ 
\\ 
...\\ 
\\ 
\ds\fr d{dt}\ti a_k(t,\x_0)=\ti B(t,\x_0)\ti a_k(t,\x_0)+ 
\ds\fr {i}{2\mu}\ti d_{k-1}(t,\x_0),\\ 
\\ 
... 
\ea 
\right|~~~|t|<T, 
\ee 
where $\ti d_0(t,\x_0)$ 
is the function  $d_0(t,x) 
:=\De a_0(t,x)$ 
expressed in ray coordinates, etc. 
It remains to substitute the same expansion into the initial conditions 
(\re{eamr}) which gives 
\be\la{reqi} 
\ti a_0(0,\x_0)=a_0(\x_0),~~~ 
\ti a_1(0,\x_0)=0,~~~...,~~~\ti a_k(0,\x_0)=0,... 
\ee 
Now (\re{req}) and (\re{reqi}) imply that $\ti a_0(t,\x_0)=0$, 
$|t|<T$, if $\x_0\not\in \supp a_0$. Hence, also 
$\ti d_0(t,\x_0)=0$, 
$|t|<T$, if $\x_0\not\in \supp a_0$. 
Therefore, (\re{req}) and (\re{reqi}) imply that 
$\ti a_1(t,\x_0)=0$, 
$|t|<T$, if $\x_0\not\in \supp a_0$, etc.\bo

\bexe 
Prove the transport equation (\re{eam}). 
{\bf Hint:} Substitute 
$\psi=a e^{\ds{iS}/\h}$ 
into the Schr\"odinger equation (\re{SMm}) and divide by $e^{\ds{iS}/\h}$. 
Then Equation (\re{Cp3}) formally follows 
by setting $\h=0$, and afterwords, (\re{eam}) also follows. 
\eexe 
%%%%%%%%%%%%%%%%%%%%%%%%%%%%%%%%%%%%%%%%%%%%%%%%%%%%%%%%% 
%%%%%%%%%%%%%%%%%%%%%%%%%%%%%%%%%%%%%%%%%%%%%%%%%%%%%%%%% 

\newpage 
%%%%%%%%%%%%%%%%%%%%%%%%%%%%%%%%%%%%%%%%%%%%%%%%%%%%%%%%% 
%%%%%%%%%%%%%%%%%%%%%%%%%%%%%%%%%%%%%%%%%%%%%%%%%%%%%%%%% 

%%\setcounter{section}{16} 
\setcounter{subsection}{0} 
\setcounter{theorem}{0} 
\setcounter{equation}{0} 
\section 
{Schr\"odinger Equation and Heisenberg Representation} 
According to Schr\"odinger, the quantum mechanical electron 
in an external 
electromagnetic field is described by the corresponding 
Schr\"odinger 
equation for the electron field. The results of the last 
section then imply 
that, in the classical limit, a ray of electrons propagates according 
to the Lorentz dynamics. The symmetries of the Schr\"odinger equation 
lead to conserved physical observables of the electron field, 
like energy, momentum, angular momentum, and electric charge. 
All observables are quadratic forms of linear operators in the Hilbert 
space of electron wave functions, therefore the quantum observables 
are 
identified with linear operators in this Hilbert space. 
 
\subsection{Electrons and Cathode Rays} 
The {\it electron} was discovered by J.J.Thomson 
around 1897 in  {\it cathode rays}. He systematized 
observations of the 
deflection of rays in 
Maxwell fields. The observations demonstrated that 
the deflections may be described by the Lorentz dynamics 
(\re{Le}) or (\re{LH}), with a fixed ratio $e/\mu$ close to its present value. 
Later 
Kauffmann  \ci{Ka} confirmed these observations with high accuracy, 
$e/\mu\approx -1.76\cdot 10^8$C/g. 
The charge of the electron can be evaluated from Faraday's electrolytic 
law  to be $e\approx -96500$C$/6.06\cdot 10^{23} 
\approx 1.6\cdot 10^{-19}$C 
$\approx-4.77\cdot 10^{-10}$ esu. 
In 1913 Millikan confirmed this result, 
$e\approx 1.60 \cdot 10^{-19}$C$\approx-4.77\cdot 10^{-10}$ 
esu, leading to an electron mass of $\mu\approx 9.1\cdot 10^{-28}$g. 
Therefore, by Theorem \re{tWKB}, 
the {\it short-wave} 
cathode rays can also be described by the  Schr\"odinger 
 equation (\re{SMm}) 
with the electron 
mass $\mu$ and the 
{\it negative} electron charge $e<0$: 
\be\la{SM} 
[i\h\pa_t-e\phi(t,\x)]\psi(t,\x) 
=\fr 1{2\mu} [-i\h\na_\x-\ds\fr ec A(t,\x)]^2\psi(t,\x),
~~~~~~(t,\x)\in\R^4. 
\ee 
\br 
{\rm 
 Let us note that the charge of the cathode rays is not fixed, 
hence 
the equation (\re{SM}) describes the {\it short-wave} 
cathode rays 
of an arbitrary charge. 
It means that the Schr\"odinger 
 equation  describes rather the dynamics of the 
{\it electron field}, with a fixed ratio $e/\mu$, 
than the dynamics of a particle 
with the charge $e$.} 
\er 
\br 
 {\rm Theorem \re{tWKB} means that the Schr\"odinger equation 
(\re{SM}), 
with any small $\h\ll 1$, 
agrees with  the classical 
Lorentz equations (\re{Le}) or (\re{LH}). 
The actual value of the constant is fixed by 
the Planck relation $\h= a/k\approx 1.05\cdot 10^{-27}erg\cdot 
sec$. 
Here $a$ is the parameter 
in the 
Wien experimental formula (\re{W}) and 
$k$ is the Boltzmann constant. 
This identification of a small parameter $\h$ in the Schr\"odinger 
equation 
follows by  the experimental and theoretical development from 
the Kirchhoff spectral law 
to de Broglie's wave-particle duality. 
} 
\er

\subsection{Quantum Stationary States} 
Let us consider the 
case of a {\it static} Maxwell field with the potentials 
$\phi(t,\x)\equiv\phi(\x)$ and $\bA(t,\x)\equiv \bA(\x)$. 
The corresponding  Schr\"odinger 
equation (\re{SM}) becomes, 
\be\la{SMs} 
~~~~~~~~~~~~i\h\pa_t\psi(t,\x) 
=\cH\psi(t,\x):=\fr 1{2\mu} [-i\h\na_\x-\ds\fr ec \bA(\x)]^2\psi(t,\x) 
+e\phi(\x)\psi(t,\x) ,~~~~~~(t,\x)\in\R^4. 
\ee 
Then the energy is conserved (see (\re{eLdf})): 
\be\la{SMse} 
\int_{\R^3}\Big(\ds\fr 1{2\mu} 
{|[\!-\!i\h\na_\x\!-\!\ds\fr ec \bA(\x)]\psi(t,\x)|^2}\!+\!{e\phi(\x)|\psi(t,\x)|^2} 
\Big)d\x\!=\!E, 
~~~~ 
t\in\R. 
\ee 
\bd\la{dQSS} 
{\bf Quantum Stationary States} for Equation (\re{SMs}) are 
nonzero finite energy 
solutions of the type 
\be\la{QSS} 
\psi(t,\x)=\psi_\om(\x)e^{-i\om t} 
\ee 
with an $\om\in\R$. 
\ed 
Substituting (\re{QSS}) into (\re{SMs}), we get the {\it stationary 
 Schr\"odinger 
 equation} 
\be\la{SMss} 
\h\om\psi_\om(\x) 
=\cH \psi_\om(\x), 
~~~~~~\x\in\R^3, 
\ee 
which constitutes an {\it eigenvalue problem}. 
Substituting (\re{QSS}) into (\re{SMse}) 
and using  (\re{SMss}), we get for the energy 
\be\la{SMses} 
E=E_\om= \h\om\int_{\R^3}|\psi_\om(\x)|^2d\x<\infty. 
\ee 
Hence, the quantum stationary states correspond to the 
eigenfunctions $\psi_\om$ from the Hilbert space $\E:=L^2(\R^3)$. 
Let us assume the standard {\bf normalization condition} (see Remark \re{rnc}) 
\be\la{Nor} 
\int_{\R^3}|\psi_\om(\x)|^2d\x=1. 
\ee 
Then (\re{SMses}) becomes (cf. $(P)$ from the Introduction) 
\be\la{SMsesb} 
E_\om= \h\om, 
\ee 
and (\re{SMss}) takes the form 
\be\la{SMsst} 
E_\om\psi_\om(\x) 
=\cH \psi_\om(\x), 
~~~~~~\x\in\R^3 . 
\ee 
 %%%%%%%%%%%%%%%%%%%%%%%%%%%%%%%%%%%%%%%%%%%% 
%%%%%%%%%%%%%%%%%%%%%%%%%%%%%%%%%%%%%%%%%%%% 
\subsection{Four Conservation Laws for Schr\"odinger Equation} 
Let us state the four classical conservation laws 
for the Schr\"odinger equation (\re{SMs}). 
We adjust the conserved quantities by some factors. 
\medskip\\ 
{\bf I. Energy Conservation:} The total energy is conserved 
if the Maxwell potentials 
do not depend on time $t$, i.e. 
 $\phi(t,\x)=\phi(\x), \bA(t,\x)=\bA(\x)$: 
\be\la{1ecs} 
E(t):=\int \Big( 
\ds\fr 1{2\mu} 
|[-i\h\na-\ds\fr ec \bA(\x)]\psi(t,\x)|^2+e\phi(\x)|\psi(t,\x)|^2\Big)d\x 
=\co,~~~~~t\in\R. 
\ee 
This follows from Theorem 
\re{ECLf} by (\re{eLdf}) and (\re{Nc1ecS}) . 
\medskip\\ 
{\bf II. Momentum Conservation} The 
component $\p_n$, $n=1,2,3$, of the total momentum, 
is conserved  if the Maxwell potentials 
do not depend on $\x_n$: 
\be\la{2mcs} 
\p_n(t):=-i\h\int \Big[\psi(t,\x)\cdot \na_n\psi(t,\x)\Big]d\x 
=\co,~~~~~t\in\R. 
\ee 
This follows from Theorem 
\re{mCLf} by (\re{mLdf}) and  (\re{Nc2ecS}). 
\medskip\\ 
{\bf III. Angular Momentum Conservation} 
Let us assume that the Maxwell fields 
are axially symmetric with respect to 
rotations around  the axis $\x_n$, i.e. 
\be\la{3axi} 
\phi(t,R_n(s)\x)\equiv\phi(t,\x)~~\mbox{and}~~ 
\bA(t,R_n(s)\x)\equiv R_n^t(s)\bA(t,\x),~~~~~s\in\R. 
\ee 
Then corresponding  component of the total angular momentum 
is conserved: 
\be\la{3mcs} 
\bL_n(t):=-i\h\int \Big[\psi(t,\x)\cdot (\x\times\na)_n\psi(t,\x)\Big]d\x 
=\co,~~~~~t\in\R. 
\ee 
This follows from Theorem 
\re{MCLf}  by (\re{MLdf}) and  (\re{Nc3ecS}) 
since the Lagrangian  density (\re{LdNS}) satisfies 
the rotation-invariance condition (\re{LrN}) by (\re{3axi}). 
\medskip\\ 
{\bf IV. Charge Conservation} The total charge 
is conserved: 
\be\la{4mcs} 
Q(t):=e\int |\psi(t,\x)|^2d\x 
=\co,~~~~~t\in\R. 
\ee 
This follows from Theorem 
\re{Qc} by 
(\re{LQ}) and (\re{Nc3ecS}). 
\medskip\\ 
\bexe 
Check the  rotation-invariance condition (\re{LrN}) for the 
Maxwell potentials satisfying  (\re{3axi}). 
\eexe 
\bexe 
Check that the axial symmetry  condition (\re{3axi}) is equivalent to the 
corresponding axial symmetry of the Maxwell field: for every $s\in\R$, 
\be\la{3axim} 
\bE(t,R_n(s)\x)\equiv R_n(s)\bE(t,\x),~~~~~~ 
\bB(t,R_n(s)\x)\equiv R_n(s)\bB(t,\x). 
\ee 
\eexe

\subsection{Quantum Observables and Heisenberg Representation} 
The charge conservation  (\re{4mcs}) implies that 
the {\it complex} Hilbert space 
$\E:=L^2(\R^3)$ is invariant with respect to the 
dynamics defined by the 
Schr\"odinger equation. 
For $\psi_{1,2}\in\E$, 
 let us denote by  $\langle\psi_1, \psi_2\rangle$ the Hermitian 
scalar product: 
\be\la{Hsp} 
\langle\psi_1, \psi_2\rangle:=\int \psi_1(\x) \ov\psi_2(\x) d\x. 
\ee 
All conserved quantities $E,\p_n,\bL_n,Q$ are quadratic forms in 
the phase space $\E$: 
\be\la{qvf} 
E=\langle\psi, \hat E\psi\rangle,~~~~~ 
\p_n=\langle\psi,\hat\p_n  \psi\rangle,~~~~~ 
\bL_n=\langle\psi, \hat\bL_n \psi\rangle,~~~~~ 
Q=\langle\psi, \hat Q \psi\rangle. 
\ee 
Here 
 $\hat E, \hat\p_n,\hat\bL_n,\hat Q $ stand for the corresponding 
operators of the quadratic forms: 
\be\la{qvfo} 
\hat E=\cH ,~~~~\hat\p_n=-i\h\na_n,~~~~~ 
\hat\bL_n=-i\h(\x\times\na)_n,~~~~~\hat Q =I. 
\ee 
This motivates the following 
\bd 
i) A {\bf quantum observable} is a linear  operator $\hat\bM$ 
in the Hilbert space $\E$. 
\\ 
ii) The quadratic form $\bM(\psi):=\langle\psi, \hat\bM\psi\rangle$ is 
a {\bf mean value} of the observable at the state $\psi\in\E$. 
\ed 
\bex 
The multiplication operator $\hat \x_k\psi=\x_k\psi(\x)$ 
is the quantum observable of the $k$-th {\bf coordinate}. 
\eex 
\bex 
 The (squared) {\it absolute value of the total angular momentum} 
is defined by 
$\hat\bL^2:=\hat\bL_1^2+\hat\bL_2^2+\hat\bL_3^2$. 
\eex

\bexe 
Prove that all mean values  $\langle\psi, \hat\bM\psi\rangle$ 
of a quadratic form are real iff the operator $\hat\bM$ is 
symmetric. 
\eexe 
\bexe 
Check that all the operators $\hat E,\hat\p_n,\hat\bL_n$ 
are symmetric. {\bf Hint:} All the mean values are real 
since they coincide with the total energy, momentum, etc, 
defined by the real Lagrangian density (\re{LdNS}). 
\eexe 
\bexe 
Check the formula 
\be\la{amd} 
\hat\bL_n=-i\h\fr\pa{\pa \vp_n}, 
\ee 
where $\vp_n$ is the angle of rotation around the vector $e_n$ 
in a positive direction. {\bf Hint:} Consider $n=3$ and choose 
the polar coordinate in the plane $\x_1,\x_2$. 
\eexe

Now let us assume that 
 $\phi(t,\x)=\phi(\x), \bA(t,\x)=\bA(\x)$ and denote by $U(t)$, $t\in\R$, 
the dynamical group of the Schr\"odinger equation in the 
Hilbert space $\E$: by definition, $U(t)\psi(0,\cdot)=\psi(t,\cdot)$ 
for any solution $\psi(t,\x)$ to the equation. 
The Schrodinger equation implies that 
\be\la{dUt} 
\dot U(t)=-\ds\fr i\h \cH U(t)=-\ds\fr i\h U(t)\cH . 
\ee 
\bexe 
Prove (\re{dUt}). 
\eexe 
The charge conservation  (\re{4mcs}) means that $U(t)$ is a 
unitary operator for all $t\in\R$.

Let us note that 
\be\la{HLn} 
\hat E=\cH = 
\fr 1{2\mu} [\hat\p-\ds\fr ec \bA(\hat\x)]^2 
+e\phi(\hat\x), 
~~~~~~~~~~~ 
\hat\bL=\hat\x\times\hat\p,~~~~~~~~~~[\hat\x_k,\hat\p_n]=i\h\de_{kn}, 
\ee 
where $[\cdot, \cdot]$ stands for the {\it commutator} 
$[A, B]:=AB-BA$. 
\bexe 
Check the identities  (\re{HLn}). 
{\bf Hint:} 
For 
polynomial potentials $\bA(\x),\phi(\x)$, 
the operator of multiplication by $\bA(\x))$ resp. $\phi(\x)$ 
coincides with $\bA(\hat\x))$ resp. $\phi(\hat\x)$. 
\eexe

\bd 
The {\bf Heisenberg representation} of a quantum observable 
$\hat\bM$ is an operator function 
$\hat\bM(t):=U(-t)\hat\bM U(t)$, $t\in\R$.\\

\ed 
The definition implies that 
\be\la{Hro} 
\bM(\psi(t))=\langle\psi(0), \hat\bM(t) 
\psi(0)\rangle,~~~~~~~~~~t\in\R 
\ee 
for any solution $\psi(t):=\psi(t,\cdot)$ 
to the the Schr\"odinger equation. 
\bexe 
Check (\re{Hro}). 
\eexe 
\bexe 
Check the identities 
\be\la{HLnH} 
~~~~~~~ 
\hat E(t)\!=\!\cH\!=\! 
\fr 1{2\mu} [\hat\p(t)\!-\!\ds\fr ec \bA(\hat\x(t))]^2 
\!+\!e\phi(\hat\x(t)), 
~~~ 
\hat\bL(t)\!=\!\hat\x(t)\times\hat\p(t), 
~~~[\hat\x_k(t),\hat\p_n(t)]\!=\!i\h\de_{kn}. 
\ee 
{\bf Hint:} First consider polynomial potentials 
$\bA(\x),\phi(\x)$ and use (\re{HLn}) together with the 
commutation 
$\cH U(t)=U(t)\cH $ from (\re{dUt}). 
\eexe

\bexe 
Check the identities 
\be\la{HLnHi} 
~~~~~~~[\hat\x_k(t),\hat\p_n^N(t)]\!\!=\!\!i\h\de_{kn}N\hat\p_n^{N-1}, 
[\bA(\hat\x(t)),\hat\p_n(t)]\!\!=\!\!i\h\de_{kn}\bA_{\x_n}(\hat\x(t)), 
[\phi(\hat\x(t)),\hat\p_n(t)]\!\!=\!\!i\h\de_{kn}\phi_{\x_n}(\hat\x(t)), 
\ee 
for any $N=1,2,...$. 
{\bf Hint:} First consider polynomial potentials 
$\bA(\x),\phi(\x)$ and apply the last formula of (\re{HLnH}) 
by induction. 
\eexe 
 
Let us derive a dynamical equation for $\hat\bM(t)$: 
(\re{dUt}) implies 
the {\bf Heisenberg equations} 
\be\la{He} 
\dot{ \hat\bM}(t)=\ds\fr i\h [\cH , \hat\bM(t)]=\ds\fr i\h 
[\hat E(t), \hat\bM(t)],~~~~t\in\R. 
\ee 
\bexe 
Check  (\re{He}). 
\eexe 
\bl\la{lHc} 
The mean value, $\bM(\psi(t))$, 
is conserved if $[\cH ,\hat \bM]=0$. 
\el 
\Pr By  (\re{SMs}), 
\beqn\la{Heb} 
\fr d{dt}\bM(\psi(t))\!\!&\!\!=\!\!&\!\!\langle\dot\psi(t), 
\hat\bM\psi(t)\rangle+\langle\psi(t), 
\hat\bM\dot\psi(t)\rangle= 
-\langle\ds\fr i\h\cH\psi(t), 
\hat\bM\psi(t)\rangle-\langle\psi(t), 
\hat\bM\fr i\h\cH\psi(t)\rangle\nonumber\\ 
\!\!&\!\!=\!\!&\!\! 
-\ds\fr i\h\Big[\langle\cH\psi(t), 
\hat\bM\psi(t)\rangle-\langle\psi(t), 
\hat\bM\cH\psi(t)\rangle\Big] 
=-\ds\fr i\h 
\langle\psi(t),[\cH , \hat\bM] 
\psi(t)\rangle,~~~~t\in\R. 
~~~\bo 
\eeqn

\bexe 
Prove the angular momentum conservation (\re{3mcs}), by using 
Lemma \re{lHc}. {\bf Hint:} 
Check the commutation 
$[\cH ,\hat\bL_n]=0$ under condition (\re{3axi}). 
\eexe 
Let us write the Heisenberg equations for the observables 
$\hat\x(t)$ and $\hat\p(t)$. Then we obtain 
the system 
\be\la{LHH} 
\left\{ 
\ba{rl} 
\dot {\hat\x}(t)& 
=\ds\fr 1\mu(\hat\p(t)-\ds\fr ec A(t,\hat\x(t))),\\ 
\dot {\hat\p}(t)&=\ds- 
e\phi_\x(t,\hat\x(t))+ 
\fr e{\mu c} A_\x(t,\hat\x(t)). 
\ea\right. 
\ee 
\bexe 
Check (\re{LHH}). {\bf Hint:} Use (\re{He}) and (\re{HLnHi}) 
with $N=1,2$. 
\eexe 
Let us note that Equations (\re{LHH}) formally coincide 
with the Hamilton system (\re{LH}). 
 
\newpage 
%%%%%%%%%%%%%%%%%%%%%%%%%%%%%%%%%%%%%%%%%%%%%%%%%%%%%%%%%%% 
%%%%%%%%%%%%%%%%%%%%%%%%%%%%%%%%%%%%%%%%%%%%%%%%%%%%%%%%%%% 
 
%%\setcounter{section}{18} 
\setcounter{subsection}{0} 
\setcounter{theorem}{0} 
\setcounter{equation}{0} 
\section{Coupling to the  Maxwell Equations} 
%%%%%%%%%%%%%%%%%%%%%%%%%%%%%%%%%%%%%%%%%%%%%%%%%%%%%%%%%%% 
The simultaneous 
evolution of the full system of electron wave function and Maxwell 
field is determined by the coupled Schr\"odinger and Maxwell 
equations. 
Stationary states of this coupled system would provide natural 
candidates 
for physical states and transitions, but the existence of these 
stationary states as well as the transition of general solutions 
to these 
stationary states for large times is an open problem. 
Approximate, iterative solutions to the coupled system are 
provided by the 
Born approximation. 
 
\subsection{Charge and Current Densities and Gauge Invariance} 
Let us determine the dynamics of the Maxwell field in presence 
of the wave field $\psi$ 
governed by the Schr\"odinger equation. 
The Lagrangian density of the free Maxwell field is 
known from (\re{mtL}), so we have to modify only the interaction 
term in (\re{mtL}). 
The interaction term gives the dynamics of the Maxwell field in 
the presence 
of given charge-current densities (\re{mtL}), 
so it remains to express the densities in terms of the wave field $\psi$. 
 
On the other hand, 
we have shown that  the interaction 
term in (\re{mtL}) uniquely determines the Lorentz dynamics 
(\re{Le}). 
The Schr\"odinger equation (\re{SM}) substitutes 
the Lorentz dynamics (\re{Le}). This suggests to 
identify the Maxwell-Schr\"odinger interaction 
with the interaction term from 
the Lagrangian density $\cL_S$ of the  Schr\"odinger 
equation (\re{SM}). 
That is, we add 
the Lagrangian density $\cL_f$ 
of the free Maxwell field to 
the Lagrangian density $\cL_S$ of the  Schr\"odinger equation 
(\re{SM}) 
and get the 
Lagrangian density $\cL_{MS}$ of the 
coupled Maxwell-Schr\"odinger equations, 
\be\la{cLMS} 
\cL_{MS} 
= 
{[i\h\pa_t\!-\!e\phi]\psi\cdot \psi}- 
\fr1{2\mu}\sum\limits_{k=1}^3{|[-i\h\na_k\!-\! 
\ds\fr ec \bA_k]\psi|^2} 
-\fr 1{16\pi}\,\F^{\al\beta}\F_{\al\beta}, 
\ee 
where $\F^{\mu\nu}:=\pa^\mu\A^\nu-\pa^\nu\A^\mu$ and 
$\F_{\mu\nu}:=\pa_\mu\A_\nu-\pa_\nu\A_\mu$, $\A^\nu=(\phi,\bA_1,\bA_2,\bA_3)$. 
The corresponding Euler-Lagrange equations read 
\be\la{SMe} 
~~~~~~~\left\{ 
\ba{l} 
\ds[i\h\pa_t-e\phi(t,\x)]\psi(t,\x) 
=\fr 1{2\mu} 
[-i\h\na_\x-\ds\fr ec \bA(t,\x)]^2\psi(t,\x),\\ 
\\ 
\ds\fr 1{4\pi}\,\na_\al\F^{\al\beta}(t,\x) 
= 
\left( 
\ba{ll} 
\rho:=e|\psi(t,\x)|^2,&\beta=0\\ 
\\ 
\ds\fr{\bj_\beta}c 
:=\ds\fr e{\mu c}[-i\h\na_\beta-\ds\fr ec \bA_\beta(t,\x)]\psi(t,\x ) 
\cdot\psi(t,\x ),&\beta=1,2,3 
\ea\right). 
\ea\right. 
\ee 
The system (\re{SMe}) describes the dynamics of the wave field 
$\psi$ in its ``own'', induced Maxwell field $\phi(t,\x), \bA(t,\x)$ 
generated 
by the charges and currents of the wave field. 
\br\la{rnc} 
i) The charge-current densities from (\re{SMe}) coincide, 
up to a factor, 
with the Noether currents (\re{Nc4ecS}) of the 
group of internal rotations (\re{gsir}). 
\\ 
ii) The factor is choosen in accordance to the normalization condition 
(\re{Nor}). Namely the condition means that the total charge 
of the corresponding quantum stationary state is equal to $e$. 
\er

Now let us introduce  external potentials 
$\phi^{\rm ext}(t,\x)$, $\bA^{\rm ext}(t,\x)$ 
of the Maxwell field 
generated by some external sources. We formalize the introduction through 
the Lagrangian density 
\be\la{cLMSe} 
~~~~~~\cL_{MS} 
\!=\! 
{[i\h\pa_t\!-\!e(\phi+\phi^{\rm ext})]\psi\cdot \psi} 
\!-\! 
\fr 1{2\mu}\sum\limits_{k=1}^3{|[\!-\!i\h\na_k\!-\! 
\ds\fr ec (\bA_k\!+\!\bA^{\rm ext}_k)]\psi|^2} 
\!-\!\fr 1{16\pi}\,\F^{\al\beta}\F_{\al\beta}. 
\ee 
The corresponding equations read 
\be\la{SMee} 
~~~~~\left\{ 
\ba{l} 
\ds[i\h\pa_t\!-\!e(\phi(t,\x)\!+\!\phi^{\rm ext}(t,\x))]\psi(t,\x) 
\!=\!\fr 1{2\mu} 
[\!-\!i\h\na_\x\!-\!\ds\fr ec (\bA(t,\x)\!+\!\bA^{\rm ext}(t,\x))]^2\psi(t,\x),\\ 
\\ 
\ds\fr 1{4\pi}\,\na_\al\F^{\al\beta}(t,\x) 
\!=\! 
\left( 
\ba{l} 
\rho\!:=e|\psi(t,\x)|^2\\ 
\\ 
\ds\fr{\bj_\beta}c 
\!:=\!\ds\fr e{\mu c}[\!-\!i\h\na_\beta\!-\!\ds\fr ec 
(\bA_\beta(t,\x)\!+\!\bA^{\rm ext}_\beta(t,\x))]\psi(t,\x ) 
\!\cdot\!\psi(t,\x ) 
\ea\right) 
\ea\right. . 
\ee 
\br 
The tensor  $\F^{\al\beta}$ is defined by the potentials 
$\phi(t,\x),\bA(t,\x)$, 
hence the system admits the solution $\psi(t,\x )=0,\phi(t,\x)=0,\bA(t,\x)=0$ 
corresponding to the absence of matter. 
\er 
 
The {\it gauge transformation} (\re{gt}) 
does not change the Maxwell fields $E(t,\x)$,  $B(t,\x)$ for any function 
$\chi(t,\x)\in C^1(\R^4)$. Therefore, it would be natural to expect that 
the solutions to the coupled equations (\re{SMe}), (\re{SMee}) also do 
not change too 
much under this transformation. Indeed, 
we can complete the transformation of the potentials (\re{gt}) with a 
corresponding transformation of the wave function: 
\be\la{gtw} 
\left|\ba{c} 
\ds\phi(t,\x)\mapsto \phi(t,\x)+\fr 1c\,\dot\chi(t,\x),\,\,\,\, 
\bA(t,\x)\mapsto \bA(t,\x)-\na_\x\chi(t,\x),\\ 
\psi(t,\x)\mapsto e^{-i\ds\fr e{c\h} \chi(t,\x)}\psi(t,\x). 
\ea 
\right. 
\ee 
It is easy to check that the new functions also provide a solution 
to Equations (\re{SMe}) resp. (\re{SMee}). 
Moreover, 
the transformations (\re{gtw}) do not change 
the electric charge and current densities 
$\rho(t,\x)$ and 
$\bj(t,\x)$ in (\re{SMe}) and (\re{SMee}).

\subsection{Electron Beams and 
Heisenberg's Uncertainty Principle} 
\subsubsection{Plane wave as an electron beam} 
According to the de Broglie's wave-particle duality 
(\re{dB1}), 
a free electron beam is described by a plane wave 
\be\la{plwH} 
\psi(t,\x)=Ce^{i(\bk\x-\om t)},\,\,\,\,\,\bk\ne 0 
\ee 
which 
is a solution to the {\it free} Schr\"odinger equation, i.e., without 
an external Maxwell field. Hence, 
\be\la{plhH} 
\h\om=\fr{\h^2}{2\mu}\bk^2>0. 
\ee 
We can define corresponding  density of the electrons 
from the charge density (\re{SMe}), 
\be\la{dennH} 
\rho(t,\x):= e|\psi(t,\x)|^2=e|C|^2. 
\ee 
The velocities of the electrons may be defined 
from the electric current  density (\re{SMe}), 
\be\la{velnH} 
\bj(t,\x):= \fr e\mu[-i\h\na\psi(t,\x)]\cdot\psi_{\rm in} 
(t,\x)=\fr{e\h \bk}\mu |C|^2. 
\ee 
Respectively, the electron density, velocity and momentum are 
defined by 
\be\la{pdvH} 
\left\{ 
\ba{l} 
d(t,\x)\!:=\!\rho(t,\x)/e\!=\! 
|C|^2, 
\\ 
\\ 
\bv(t,\x)\!:=\!\bj(t,\x)/\rho(t,\x) 
\!=\!\ds\fr{\h \bk}\mu, 
\\ 
\\ 
\p(t,\x)\!:=\!\mu\bv(t,\x)\!=\!\h \bk. 
\ea 
\right. 
\ee 
Then the energy density (see (\re{1ecs})) 
admits the following representation: 
\be\la{endH} 
~~~~~~~~~~~~e(t,\x)\!:=\! \fr 1{2\mu} |[-i\h\na] 
\psi(t,\x)|^2 
\!=\! 
\fr{\h^2}{2\mu} |\bk|^2|C|^2\!=\!\fr{\mu \bv^2}2 d_{\rm in}(t,\x). 
\ee 
\subsubsection{Heisenberg's Uncertainty Principle} 
The interpretation of the plane wave as a beam of free electrons 
cannot be completed by an assignment 
of coordinates to the electrons. 
This assignment is possible only asymptotically, for small $\h$. 
 Corresponding bounds were discovered by 
Heisenberg \ci{Han,New}: 
\be\la{HUP} 
\De \x\De \p\sim \h, 
\ee 
where 
$\De x:=\langle\psi,(\hat \x-\x)^2\psi\rangle^{1/2}$ resp. 
$\De \p:=\langle\psi,(\hat \p-\p)^2\psi\rangle^{1/2}$ is an 
{\bf uncertainty} of an electron coordinate resp. momentum 
(see (\re{qvf}) and (\re{qvfo}) for the notations). 
For the plane wave the momentum  is known exactly, 
hence 
$\De \p=0$. Now (\re{HUP}) implies that 
and $\De \x=\infty$, in other words, 
the coordinate is not well-defined.

\subsection{Quantum Stationary States and Attractors}

Let us consider the system (\re{SMee}) 
corresponding to a {\it static} external Maxwell field 
$\phi^{\rm ext}(\x)$, $\bA^{\rm ext}(\x)$: 
\be\la{SMees} 
~~~~~\left\{ 
\ba{l} 
\ds[i\h\pa_t\!-\!e(\phi(t,\x)\!+\!\phi^{\rm ext}(\x))]\psi(t,\x) 
\!=\!\fr 1{2\mu} 
[\!-\!i\h\na_\x\!-\!\ds\fr ec (\bA(t,\x)\!+\!\bA^{\rm ext}(\x))]^2\psi(t,\x),\\ 
\\ 
\ds\fr 1{4\pi}\,\na_\al\F^{\al\beta}(t,\x) 
\!=\! 
\left( 
\ba{l} 
\rho\!:=e|\psi(t,\x)|^2\\ 
\\ 
\ds\fr{\bj_\beta}c 
\!:=\!\ds\fr e{\mu c}[\!-\!i\h\na_\beta\!-\!\ds\fr ec 
(\bA_\beta(t,\x)\!+\!\bA^{\rm ext}_\beta(\x))]\psi(t,\x ) 
\!\cdot\!\psi(t,\x ) 
\ea\right) 
\ea\right. . 
\ee 
Let us generalize the definition of 
quantum stationary states for Eq. (\re{SMees}). 
\bd\la{dQSSM}(\ci{GSS})\, 
i) {\bf solitary waves} for Eq. (\re{SMees}) are 
nonzero finite energy 
solutions of the type 
\be\la{QSSM} 
\psi(t,\x)=\psi_\om(\x)e^{-i\om t},\,\,\,\phi(t,\x)=\phi_\om(\x) 
,\,\,\, 
\bA(t,\x)=\bA_\om(\x) 
\ee 
with an $\om\in\R$. 
\\ 
ii) The set $\Om:=\{\om\in\R:\mbox{there exist a solitary wave with ~} 
(\psi_\om(\x), \phi_\om(\x), \bA_\om(\x))\ne 0\}$ is the set 
of the {\bf solitary eigenfrequencies} 
and 
$\A:=\{(\psi_\om(\x), \phi_\om(\x), \bA_\om(\x)):\om\in\Om\}$ 
is the set of the {\bf solitary amplitudes}. 
 
\ed 
Substituting (\re{QSSM}) into  (\re{SMees}), we get the corresponding 
stationary {\it nonlinear eigenfunction problem}: 
in the Lorentz gauge we get, as in  (\re{mtp}), 
\be\la{SMst} 
~~~~~\left\{ 
\ba{l} 
\ds[\om\h\!-\!e(\phi_\om(\x)\!+\!\phi^{\rm ext}(\x))]\psi_\om(\x) 
\!=\!\fr 1{2\mu} 
[\!-\!i\h\na_\x\!-\!\ds\fr ec (\bA_\om(\x)\!+\!\bA^{\rm ext}(\x))]^2 
\psi_\om(\x),\\ 
\\ 
-\ds\fr 1{4\pi}\,\De \phi_\om(\x)=\rho\!:=e|\psi_\om(\x)|^2,\\ 
\\ 
-\ds\fr 1{4\pi}\,\De \bA_\om(\x)= 
\ds\fr{\bj_\om}c 
\!:=\!\ds\fr e {\mu c}[\!-\!i\h\na_\x\!-\!\ds\fr ec 
(\bA_\om(\x)\!+\!\bA^{\rm ext}(\x))]\psi_\om(\x ) 
\!\cdot\!\psi_\om(\x ). 
\ea\right. 
\ee 
The problem is nonlinear in contrast to (\re{SMss}). 
 For the coupled 
nonlinear Dirac-Maxwell equations with zero external potentials, 
the existence of solitary waves 
is proved in \ci{EGS}. 
 
Bohr's quantum transitions (\re{BT}) could be interpreted 
mathematically as the following convergence (or attraction) 
for every finite energy solution 
$\psi(x,t)$ to the Schr\"odinger equation: 
\be\la{BTS} 
\psi(t,\x)\sim \psi_\pm(\x)e^{-i\om_\pm t},\quad t\to\pm\infty, 
\ee 
where $\h\om_-=E_m$ and $\h\om_+=E_n$. 
This would mean that the set of all quantum stationary states 
$\{\psi(x)e^{-i\om t}\}$ is an {\it attractor} of the 
dynamical system defined by the Schr\"odinger equation (\re{SMs}). 
However, this is generally impossible for the linear 
autonomous equation by the principle of superposition. 
Namely, a linear combination of two linearly independent stationary states, 
$\{\psi_\pm(x)e^{-i\om_\pm t}\}$, 
\be\la{BTSl} 
\psi(t,\x)=C_- 
 \psi_-(\x)e^{-i\om_- t}+ 
C_+ 
 \psi_+(\x)e^{-i\om_+ t}, 
\ee 
is a solution to the Schr\"odinger equation, but the asymptotics 
(\re{BTS}) does not hold if $C_1\ne 0$ and $C_2\ne 0$. 
 
On the other hand, the asymptotics could hold for solutions 
to the nonautonomous Schr\"odinger equation (\re{SM}) with the 
potentials $\phi(t,\x),\bA(t,\x)\sim \sin(\om_+-\om_-) t$. 
Let us note that just this type of Maxwell field results from transition 
(\re{BTS}) according to the second Bohr postulate (see Introduction). 
\bcom 
{\rm 
%%{\bf Attraction to Solitary Waves} 
Summing up, we suggest that the transitions (\re{BTS}) hold for the 
{\it autonomous coupled 
nonlinear Maxwell-Schr\"odinger equations} (\re{SMees}): 
for every finite energy solution 
$(\psi(t,\x), \phi(t,\x),\bA(t,\x))$, the following long-time asymptotics hold 
(cf. (\re{Bjii})) 
\be\la{BTSa} 
(\psi(t,\x), \phi(t,\x),\bA(t,\x)) 
\sim 
(\psi_\pm(\x)e^{-i\om_\pm t}, \phi_\pm(\x),\bA_\pm(\x)), 
\quad t\to\pm\infty, 
\ee 
where $\om_\pm\in\R$ and $\psi_\pm, \phi_\pm(\x),\bA_\pm(\x)$ 
are the solutions to the stationary system (\re{SMst}) 
with $\om=\om_\pm$. This means that the set of all solitary amplitudes 
$\A$ is the {\it point attractor} of the  coupled 
equations. 
}

\ecom 
\medskip 
 
Let us note that an {\it analog} of these transitions 
was proved recently 
in \ci{K1} --\ci{K6} 
for some model nonlinear autonomous wave equations 
(see  the survey \ci{Kpla}): then the asymptotics 
of type (\re{BTSa}) hold in appropriate 
{\it local energy seminorms}. 
 However, the proof of the transitions 
(\re{BTSa}) for the coupled equations (\re{SMees}) is still an open problem.

\bcom 
{\rm 
The nonlinear eigenvalue problem (\re{SMst}) could be handled by 
the iterative procedure. Namely, first 
let us neglect the fields 
$\phi_\om(\x),\bA_\om(\x))$ 
in the Schr\"odinger equation of (\re{SMst}) and get the 
first {\it Born approximation} 
\be\la{SMssg} 
\om\h\psi_\om(\x) 
=\fr 1{2\mu} 
[-i\h\na_\x-\ds\fr ec \bA^{\rm ext}(\x)]^2 
\psi_\om(\x)+e\phi^{\rm ext}(\x)\psi_\om(\x) 
\ee 
which is the linear eigenvalue problem of type (\re{SMss}). 
Further we could express $\phi_\om(\x),\bA_\om(\x)$ from the Poisson 
equations in (\re{SMst}) (neglecting $\bA_\om(\x)$ in the RHS for the 
first approximation), 
substitute them into the first equation, 
and continue by induction. The "corrections"  $\phi_\om(\x),\bA_\om(\x))$ 
are small in a sense since the RHS of the Poisson 
equations contain the small factors $e$ and $e/c$. 
Hence we could expect that the iterations converge, and even the solution 
of the linear problem (\re{SMssg}) give a good (first) approximation 
to a solution of the nonlinear problem (\re{SMst}). 
If the iterations converge, the limit 
functions obviously give one of the solutions to  (\re{SMst}). 
} 
 
\ecom

\bcom 
{\rm 
i) The (orthodox)  {\it Copenhagen interpretation} of Quantum Mechanics 
(see \ci{JvN}) postulates that each observation of a quantum observable 
$\hat\bM$ gives one of its eigenvalue 
{\it with the probability one}. 
\\ 
ii) This postulate 
agrees with  our previous comment in the case when $\hat\bM$ commutes with 
the Schr\"odinger operator $\cH$ (with the external potentials 
$\phi^{\rm ext}(\x)$ and $\bA^{\rm ext}(\x)$): 
in particular, this is true for the energy $\hat\E=\cH$. 
Namely, between two successive macroscopic observations, 
the quantum system could be considered as unperturbed 
even if the observation provide  a disturbance. 
Therefore, the 
attraction 
(\re{BTSa}) implies that before the observation 
the wave function $\psi(t):=\psi(\cdot,t)$ is "almost sure" 
close to a solitary wave  $\psi_\om e^{-i\om t}$. 
Furthermore, $\psi_\om\approx\psi_k$, where 
$\psi_k$ 
is an 
eigenfunction 
of the linear operator $\cH$ (see previous comment). On the other hand, 
$\psi_k$ is also 
an eigenfunction of the  observable 
$\hat\bM$. Hence the (mean) value is equal to 
$\bM(t):=\langle\psi(t),\hat\bM\psi(t)\rangle 
\approx \langle\psi_k(t),\hat\bM\psi_k(t)\rangle$ 
that coincides with the 
corresponding eigenvalue. 
\\ 
iii) 
In the case when 
$\hat\bM$ does not commute with 
the Schr\"odinger operator $\cH$, this interpretation fails. 
One could expect then that the measuring process 
should modify the Schr\"odinger operator of the combined system 
(observable system + measuring instrument) 
to insure the commutation. 
} 
\ecom

\subsection{Charge Continuity Equations} 
The Lagrangian density  (\re{cLMSe}) 
is invariant with respect to the internal rotations 
(\re{gsir}). Furthermore, the  densities 
$\rho, \bj$ 
from  (\re{SMee}) 
coincide with the corresponding 
Noether currents 
of the form (\re{Nc4ecS}) up to a factor. 
 Therefore, the Noether theorem II implies 
 
\bl 
For any solution to the Schr\"odinger 
equation 
\be\la{SMese} 
\ds[i\h\pa_t-e\phi(t,\x)]\psi(t,\x) 
=\fr 1{2\mu} 
[-i\h\na_\x-\ds\fr ec \bA(t,\x)]^2\psi(t,\x), 
\ee 
the corresponding charge-current densities 
$\rho, \bj$ 
 satisfy the continuity equation (\re{cce}). 
\el

Let us prove a more general relation: 
\bl\la{lcc2} 
For any two solutions $\psi_1(t,\x),\psi_2(t,\x)$ 
to the Schr\"odinger equation in (\re{SMese}) the following 
identity holds: 
\be\la{cce2} 
\dot\rho_{12}(t,\x)+\dv\bj_{12}(t,\x)=0,~~~~~~ 
(t,\x)\in\R^4, 
\ee 
where 
\be\la{ccd2} 
\left\{\ba{rl} 
\rho_{12}(t,\x):=&e\psi_1(t,\x)\ov\psi_2(t,\x),\\ 
\\ 
\bj_{12}(t,\x):=&\ds\fr e{2\mu}\Big([-i\h\na_\x-\ds\fr ec \bA(t,\x)] 
\psi_1(t,\x)\Big)\ov\psi_2(t,\x)\\ 
\\ 
&\!+ 
\ds\fr e{2\mu}\Big([~i\h\na_\x-\ds\fr ec \bA(t,\x)] 
\ov\psi_2(t,\x)\Big)\psi_1(t,\x). 
\ea\right. 
\ee 
\el 
\br 
For $\psi_1=\psi_2$ 
the expressions (\re{ccd2}) 
become (\re{SMe}) by (\re{real}), 
and 
the identity (\re{cce2}) 
becomes (\re{cce}). 
\er 
{\bf Proof of Lemma \re{lcc2}} 
Let us write the equation (\re{SMese}) for $\psi_1$ and $\ov\psi_2$: 
\be\la{SMs2} 
~~~~~~~~\left\{ 
\ba{l} 
[i\h\pa_t-e\phi(\x)]\psi_1(t,\x) 
=\ds\fr 1{2\mu} [-i\h\na_\x-\ds\fr ec \bA(t,\x)]^2\psi_1(t,\x), 
\\~\\ 
\!\!~[ 
-i 
\h\pa_t-e\phi(\x)] 
\ov\psi_2(t,\x) 
=\ds\fr 1{2\mu} [i\h\na_\x-\ds\fr ec \bA(t,\x)]^2\ov\psi_2(t,\x) 
\ea\right| 
~(t,\x)\in\R^4. 
\ee 
Let us multiply the first equation by $i\ov\psi_2(t,\x)$ 
and add the second equation multiplied by $-i\psi_1(t,\x)$. 
Then we get, 
\beqn\la{SMs2g} 
&&i\h\pa_t\psi_1(t,\x) i\ov\psi_2(t,\x) 
+i\h\pa_t\ov\psi_2(t,\x) i\psi_1(t,\x)\nonumber\\ 
\nonumber\\ 
&&=\fr 1{2\mu} \Big([-i\h\na_\x-\ds\fr ec \bA(t,\x)]^2 
\psi_1(t,\x)\Big) 
 i\ov\psi_2(t,\x)\nonumber\\ 
\nonumber\\ 
&&\,-\fr 1{2\mu} \Big([i\h\na_\x-\ds\fr ec \bA(t,\x)]^2 
\ov\psi_2(t,\x)\Big) 
 i\psi_1(t,\x). 
\eeqn 
This may be rewritten as 
\beqn\la{SMs2gr} 
&&-\pa_t[\psi_1(t,\x)\ov \psi_2(t,\x)]\nonumber\\ 
\nonumber\\ 
&&=\fr 1{2\mu} \na\cdot 
\Big\{\Big([-i\h\na_\x-\ds\fr ec \bA(t,\x)] 
\psi_1(t,\x)\Big) 
\ov \psi_2(t,\x)\nonumber\\ 
\nonumber\\ 
&&~~~~~~~~~~~~+\Big([i\h\na_\x-\ds\fr ec \bA(t,\x)] 
\ov\psi_2(t,\x)\Big) 
\psi_1(t,\x)\Big\}. 
~~~~~~~~\loota 
\eeqn 
\bexe 
Check the identity (\re{SMs2gr}).

\eexe

\subsection{Born Approximation} 
Let us assume that the {\it induced Maxwell field} 
$\phi(t,\x), \bA(t,\x)$ in (\re{SMee}) 
is small compared to 
the external field. Then we can neglect the 
induced field in the first equation 
and consider the approximate equation (cf. (\re{SMssg})) 
\be\la{SMeea} 
\ds[i\h\pa_t-e\phi^{\rm ext}(t,\x)]\psi(t,\x) 
=\fr 1{2\mu} 
[-i\h\na_\x-\ds\fr ec \bA^{\rm ext}(t,\x)]^2\psi(t,\x). 
\ee 
Its solution is the {\it Born approximation} to the 
wave field $\psi$ in (\re{SMee}). 
Let us solve the equation and 
substitute the solution  $\psi$ 
into the RHS of the second equation 
of (\re{SMee}) also neglecting the induced Maxwell field: 
\be\la{SMeeaa} 
\ds\fr 1{4\pi}\,\na_\al\F^{\al\beta}(t,\x) 
= 
\left( 
\ba{l} 
\rho:=e|\psi(t,\x)|^2 \\ 
\\ 
\ds\fr{\bj_\beta}c 
:=\ds\fr e {\mu c}[-i\h\na_\beta-\ds\fr ec \bA^{\rm ext}_\beta(t,\x)]\psi(t,\x ) 
\cdot\psi(t,\x ) 
\ea\right) . 
\ee 
Then the solution $\F^{\al\beta}(t,\x)$ 
is the {\it Born approximation} to 
the Maxwell field radiated by the atom. 
This process of approximations 
can be iterated.

%%%%%%%%%%%%%%%%%%%%%%%%%%%%%%%%%%%%%%%%%%%%%%%%%%%%%%%%%%% 
%%%%%%%%%%%%%%%%%%%%%%%%%%%%%%%%%%%%%%%%%%%%%%%%%%%%%%%%%%% 
 
\part{Application of Schr\"odinger Theory} 
%%%%%%%%%%%%%%%%%%%%%%%%%%%%%%%%%%%%%%%%%%%%%%%%%%%%%%%%%%% 
%%%%%%%%%%%%%%%%%%%%%%%%%%%%%%%%%%%%%%%%%%%%%%%%%%%%%%%%%%% 
 
%%\setcounter{section}{18} 
\setcounter{subsection}{0} 
\setcounter{theorem}{0} 
\setcounter{equation}{0} 
\section{Spectrum of Hydrogen Atom} 
For a hydrogen atom we study the corresponding Schr\"odinger 
equation and find the spectrum as the solution to the eigenvalue 
problem. The proof is based on splitting the space $L^2(S^2)$ 
in a sum of orthogonal eigenspaces of the spherical Laplacian.

Let us consider the hydrogen atom, which has precisely one 
electron of negative charge $e$. 
The Rutherford experiment shows that the positive charge $-e$ 
is concentrated 
in a very small region called ``nucleus'', so its Maxwell field 
is Coulombic $\phi(t,\x)=-e/|\x|$. We assume the magnetic field 
of the 
nucleus to be zero: $ \bA(t,\x)= 0$. 
Then the Schr\"odinger equation 
(\re{SMs}) becomes 
\be\la{SMsb} 
i\h\pa_t\psi(t,\x) 
=\cH \psi(t,\x):=-\fr {\h^2}{2\mu} \De\psi(t,\x) 
- 
\ds\fr {e^2}{|\x|}\psi(t,\x),~~~~~~(t,\x)\in\R^4 \, , 
\ee 
and the 
stationary Schr\"odinger equation 
(\re{SMsst}) becomes, 
\be\la{SMssb} 
E_\om\psi_\om(\x)=\cH \psi(\x)_\om. 
\ee 
\bt\la{HS} The quantum stationary states 
$\psi_\om\in\E:=L^2(\R^3)$ 
of the hydrogen atom 
exist for energies  $E_\om=E_n:=-2\pi\h R/n^2$, 
where $R:=\mu e^4/(4\pi\h^3)$ is the 
{\bf Rydberg 
constant} and 
$n=1,2,3,...$. 
For other energies, stationary states do not exist.

\et 
We will prove the theorem in this lecture. 
%%%%%%%%%%%%%%%%%%%%%%%%%%%%%%%%%%%%%%%%%%%%%%%%%%%%%%%%%%%%%%%%%%%%% 
 
\subsection{Spherical Symmetry and Separation of Variables} 
 
\subsubsection{Rotation invariance} 
A basic issue in solving the eigenvalue problem 
(\re{SMssb}) 
 is its spherical symmetry which implies the angular momentum 
conservation. 
Namely, 
the Schr\"odinger  operator $\cH$ 
is invariant with respect to all rotations  of the space $\R^3$ 
since 
the Laplacian $\De$ and the Coulomb 
potential are invariant under the rotations. 
The Schr\"odinger  operator $\cH$ 
is invariant with respect to all rotations  of the space $\R^3$. 
This means that for any $k=1,2,3$, 
\be\la{cHni} 
\cH \hat R_k(\vp)=\hat R_k(\vp)\cH,~~~~\vp\in\R, 
\ee 
where $(\hat R_k(\vp)\psi)(\x):=\psi(R_k(\vp)\x)$ 
and $R_k(\vp)$ is a space rotation around the 
unit vector $\e_k$ with an angle of $\varphi$ radian 
in a positive direction, $\e_1=(1,0,0)$, etc. 
The commutation holds by (\re{SMssb}) 
since the Laplacian $\De$ and the Coulomb 
potential are invariant under all rotations of the space $\R^3$. 
Differentiating (\re{cHni}) 
in $\vp$ at $\vp=0$, we get 
\be\la{cHn1} 
[\cH,\na_{\vp_k}]=0,~~~~~~~k=1,2,3, 
\ee 
where 
$$ 
\na_{\vp_k}:=\left.\ds\fr d{d\vp}\right|_{\vp=0}\hat R_k(\vp). 
$$ 
Then $\cH$ also  commutes with  $\bH_k:=-i\na_{\vp_k}$ and the angular 
momentum operators $\hat\bL_k$ 
since $\hat\bL_k=\h\bH_k$ by (\re{amd}): 
\be\la{amco} 
[\cH,\bH_k]=0,~~~~~~~~ 
[\cH,\hat\bL_k]=0,~~~~~~~~~~k=1,2,3,. 
\ee 
Hence the 
angular momenta 
$\bL_k(t):=\langle\psi(t),\hat\bL_k \psi(t)\rangle$ are conserved. 
\br 
Let us note that the angular momentum conservation 
played a crucial role in the 
determination of the hydrogen spectrum 
in the Bohr-Sommerfeld "old quantum theory'' 
(see Exercise 5). 
\er

\subsubsection{Spherical harmonics} 
Now we can explain our general strategy in proving Theorem 
\re{HS}. Namely, 
the commutation (\re{cHn1}) provides a solution of the spectral problem 
(\re{SMssb})  by a separation of variables. More precisely, 
the strategy relies on the following three general arguments: 
\medskip\\ 
{\bf I.} The commutation (\re{amco}) obviously implies that the 
 operator $\bH^2:=\bH_1^2+\bH_2^2+\bH_3^2$ 
commutes with 
$\cH$ by (\re{amco}): 
\be\la{amco2} 
[\cH,\bH^2]=0. 
\ee 
Hence, 
each eigenspace of the 
Schr\"odinger operator $\cH$ is invariant with respect to 
each operators $\bH_k$ and $\bH^2$. 
Moreover, 
the operator 
$\bH^2$ commutes also 
with 
each operator $\bH_k$, for example, 
\be\la{amco3} 
[\bH_k,\bH^2]=0,~~~~~~~~~k=1,2,3. 
\ee 
\bexe 
Check (\re{amco3}). {\bf Hint:} 
First, prove the commutation relations 
$[\bH_k,\bH_j]=-i\h\epsilon_{kjl}\bH_l$ 
where $\epsilon_{kjl}$ is a totally antisymmetric tensor. 
\eexe 
Since the operators 
$\bH_k$, $k=1,2,3$,  do not commute, they cannot be diagonalized 
simultaneously. On the other hand, the operators 
$\cH$, $\bH^2$ and, for example, 
$\bH_3$,  commute with each other. 
Hence, we could expect that there is a basis 
of common eigenfunctions for them. 
\medskip\\ 
{\bf II.} 
First, let us diagonalize 
simultaneously $\bH_3$ and 
$\bH^2$. 
We consider the eigenvalue problem (\re{SMssb}) 
in the Hilbert space $\E:=L^2(\R^3)$. 
On the other hand, 
both operators, $\bH_3$ and  $\bH^2$, 
act only on the angular variables 
 in spherical coordinates. 
Hence, the operators act also in the Hilbert space 
$\E_1:=L^2(S,dS)$, 
where $S$ stands for the two-dimensional sphere 
$|\x|=1$. 
 
\bt\la{tDl} 
i) In $\E_1$ there exist an orthonormal basis 
of {\it Spherical Harmonics} $Y_l^m(\theta,\vp)$ 
which are 
common eigenfunctions 
of $\bH_3$ and $\bH^2$: 
\be\la{sev} 
\bH_3Y_l^m=m Y_l^m,~~~~~~~ 
\bH^2Y_l^m= l(l+1) Y_l^m,~~~~~~~~~m=-l,-l+1,...,l, 
\ee 
where $l=0,1,2,...$. 
\\ 
ii) $Y_l^m(\theta,\vp)=F_l^m(\theta)e^{im\vp}$, 
where $F_l^m(\theta)$ are real functions. 
\et 
 
We will prove this theorem in Section 10. 
It is important that each space of the common eigenfunctions 
is one-dimensional 
since the eigenvalues depend on $l$ and $m$. 
This suggests that we could construct the eigenfunctions of the 
Schr\"odinger operator $\cH$, by separation of 
variables, in the form 
\be\la{sevf} 
\psi(\x)=R(r)Y_l^m(\theta,\vp). 
\ee 
The following theorem justify this choice of a particular form 
for the eigenfunctions. 
\bl (On Separation of Variables) 
Each solution to the spectral problem (\re{SMssb}) is a sum 
(or a series) of the solutions of the particular form (\re{sevf}). 
\el 
{\bf Formal Proof} 
Let $\pi_l^m$ denote the orthogonal projection in $\E_1$ 
onto the linear span $\E_l^m$ of $Y_l^m$. 
Let us define its action also in  $\E$,  by the formula 
\be\la{ape} 
(\Pi_l^m\psi)(r,\cdot,\cdot):=\pi_l^m[\psi(r,\cdot,\cdot)],~~~~~~~~r>0, 
\ee 
in the spherical coordinates $r,\theta,\vp$. 
Then $\Pi_l^m$ 
commutes with 
the Schr\"odinger operator $\cH$ since the latter commutes with 
 $\bH_3$ and $\bH^2$: 
\be\la{prc} 
[\cH,\Pi_l^m]=0. 
\ee 
Hence, applying $\Pi_l^m$ to (\re{SMssb}), we get {\it formally}
\be\la{SMssbg} 
E_\om\Pi_l^m\psi_\om=\cH \Pi_l^m\psi_\om. 
\ee 
It remains to note that 
\\ 
i) The function $\Pi_l^m\psi_\om$ 
has the form (\re{sevf}) 
since 
$\pi_l^m[\psi_\om(r,\cdot,\cdot)] 
\in \E_l^m$, and 
the space $\E_l^m$ is one-dimensional; 
\\ 
ii) 
$\psi_\om=\sum_{l,m}\Pi_l^m\psi_\om$. 
\bo

\brs 
i) A complete solution to the spectral problem (\re{sev}) relies on 
an investigation of all commutation relations of the operators 
$\bH_k$, $k=1,2,3$, i.e.  the Lie algebra generated by them. 
\\ 
ii) It remains still to determine the radial function 
in (\re{sevf}). We will substitute (\re{sevf}) into the equation 
(\re{sev}). This  gives a radial eigenvalue problem 
which will be solved explicitly. 
\ers

%%%%%%%%%%%%%%%%%%%%%%%%%%%%%%%%%%%%%%%%%%%%%%%%%%%%%%%%%%%%%%%%%%%%% 

\subsection{Spherical Laplacian} 
To determine the radial functions 
in (\re{sevf}), 
let us express the Laplacian operator $\De$ in spherical coordinates 
$r,\theta,\vp$: by definition, 
\be\la{sco} 
x_3=r\cos\theta,\,\,\, 
x_1=r\sin\theta\cos\vp,\,\,\, 
x_2=r\sin\theta\sin\vp. 
\ee 
The operator $\De$ is symmetric in the real Hilbert 
space $L^2(\R^3)$. 
Hence, it is defined uniquely by the quadratic form 
$(\De\psi,\psi)$, where 
$\psi\in D:=\{\psi\in L^2(\R^3): 
\psi^{(\al)}\in L^2(\R^3)\cap C(\R^3),~~|\al|\le 2\}$. 
In spherical coordinates 
\be\la{Dsc} 
(\De\psi,\psi)=-(\na\psi,\na\psi)=-\int_0^\infty dr 
\int_0^\pi d\theta\int_0^{2\pi}d\vp |\na\psi(r,\theta,\vp)|^2 
r^2\sin\theta. 
\ee 
Geometrically it is evident that 
\be\la{nas} 
\na\psi(r,\theta,\vp)=\e_r\na_r\psi+\e_\theta\fr{\na_\theta\psi}r 
+\e_\vp\fr{\na_\vp\psi}{r\sin\theta}, 
\ee 
where $\e_r$, $\e_\theta$, $\e_\vp$ are the 
orthogonal unit vectors proportional to 
$\na_r$, $\na_\theta$, $\na_\vp$, respectively. 
Therefore, (\re{Dsc}) becomes 
\be\la{Dscb} 
(\De\psi,\psi)=-\int_0^\infty dr 
\int_0^\pi d\theta\int_0^{2\pi}d\vp \Big(\Big|\na_r\psi\Big|^2+ 
\Big|\fr{\na_\theta\psi}r\Big|^2+\Big|\fr{\na_\vp\psi}{r\sin\theta}\Big|^2 
\Big) 
r^2\sin\theta. 
\ee 
Integrating by parts, we get 
\beqn\la{Dscbp} 
(\De\psi,\psi)&=&\int_0^\infty dr 
\int_0^\pi d\theta\int_0^{2\pi}d\vp (r^{-2}\na_rr^2\na_r\psi+ 
\fr{\na_\theta\sin\theta\na_\theta\psi}{r^2\sin\theta}+ 
\fr{\na_\vp^2\psi}{r^2\sin^2\theta})\psi 
r^2\sin\theta\nonumber\\ 
&=&(r^{-2}\na_rr^2\na_r\psi+ 
\fr{\na_\theta\sin\theta\na_\theta\psi}{r^2\sin\theta}+ 
\fr{\na_\vp^2\psi}{r^2\sin^2\theta},\psi). 
\eeqn 
Therefore, we get the Laplacian operator in spherical coordinates, 
\be\la{Dscbpt} 
\De\psi=r^{-2}\na_rr^2\na_r\psi+ 
\fr{\na_\theta\sin\theta\na_\theta\psi}{r^2\sin\theta}+ 
\fr{\na_\vp^2\psi}{r^2\sin^2\theta}= 
r^{-2}\na_rr^2\na_r\psi+r^{-2}{\Lam\psi}~, 
\ee 
where $\Lam$ is the differential operator on the sphere $S$ with 
coordinates $\theta,\vp$: 
\be\la{Lam} 
\Lam= 
\fr{\na_\theta\sin\theta\na_\theta}{\sin\theta}+ 
\fr{\na_\vp^2}{\sin^2\theta}~, 
\ee 
\bexe 
Check the integration by parts (\re{Dscbp}). 
\eexe

\bd 
The operator 
$\Lam$ is called {\bf Spherical Laplacian Operator}. 
\ed 
 
\bexe 
Check the identity 
\be\la{laH} 
\Lam=-\bH^2. 
\ee 
{\bf Hint:} Both operators are second order spherically 
symmetric elliptic operators. 
 
\eexe

%%%%%%%%%%%%%%%%%%%%%%%%%%%%%%%%%%% 
\subsection{Radial Equation} 
Here we deduce Theorem \re{HS} from Theorem \re{tDl}. 
Let us consider a nonzero finite energy 
solution $\psi_\om(\x)\in\E:= L^2(\R^3)$ 
to the problem  (\re{SMssb}) of the form (\re{sevf}). 
Substituting into (\re{sev}), we get by 
(\re{Dscbpt}) and (\re{laH}) that 
\be\la{exho} 
-\fr{2\mu E_\om}{\h^2} R(r)= 
r^{-2}\na_rr^2\na_rR(r)-\fr{l(l+1)}{r^2}R(r)+\fr{2\mu e^2}{\h^2 r} 
 R(r), 
~~~r>0. 
\ee 
For  $r\to\infty$, the equation becomes 
\be\la{exhob} 
-\fr{2\mu E_\om}{\h^2}R(r)\sim 
R_l''(r). 
\ee 
This suggests that $E_\om<0$ and the asymptotics 
$R(r)\sim e^{-\ga r}$, $r\to\infty$ holds, 
where 
\be\la{gam} 
\ga=\sqrt{-2\mu E  }/\h>0. 
\ee 
Next, let us write $R(r)= e^{-\ga r}F(r)$. 
Substituting into (\re{exho}), we get 
\be\la{eF} 
F''+\Big[\fr 2r-2\ga\Big]F'+\Big[\fr{d}{r}-\fr{l(l+1)}{r^2}\Big]F=0,~~~~r>0, 
\ee 
where $d=b-2\ga$ with $b=2\mu e^2/\h^2$. Finally, let us introduce 
the new variable $\rho=2\ga r$, then (\re{eF}) becomes 
\be\la{eFb} 
f''+\Big[\fr 2\rho-1\Big]f'+\Big[\fr{\lam-1}{\rho}-\fr{l(l+1)}{\rho^2}\Big]f=0, 
~~~~\rho>0, 
\ee 
where $f(\rho)=F(r)$ and $\lam=b/(2\ga)$. 
Now let us seek for a solution $f$ of the form 
\be\la{fr} 
f(\rho)=\rho^s(a_0+a_1\rho+a_2\rho^2+...) 
\equiv L(\rho)\rho^s 
\ee 
with $a_0\ne 0$. We will find two 
linearly independent solutions: one with $s\ge 0$ and another 
with $s\le -1$. Only the solution with $s\ge 0$ is appropriate. 
For $s\le -1$ the corresponding eigenfunction $\psi_\om(x)$ 
is not a function of finite energy since $\na\psi_\om(x)\not\in 
L^2(\R^3)$ and $\ds\int \phi(\x)|\psi(\x)|^2d\x=\infty$. 
Substituting (\re{fr}) into (\re{eFb}), we get 
\be\la{eFbg} 
~~~~~~~~~~~~\rho^2 L'' 
+\Bigg[2s\rho+ 
\Big[\fr 2\rho-1\Big]\rho^2\Bigg]L'+ 
\Bigg[s(s-1)+\Big[\fr 2\rho-1\Big]s\rho+ 
\Big[\fr{\lam-1}{r}-\fr{l(l+1)}{\rho^2}\Big]\rho^2\Bigg]L=0 
\ee 
for $\rho>0$. 
After some evaluation, we have 
\be\la{eFbgh} 
~~~~~~\rho^2 L'' 
+\rho[2(s+1)-\rho]L'+[\rho(\lam-1-s)+s(s+1)-l(l+1)]L=0, 
~~\rho>0. 
\ee 
Setting $\rho=0$, we get {\it formally} that $s(s+1)-l(l+1)=0$. Hence, 
$s=l$ since $-l-1\le -1$. With  $s=l$, this equation 
becomes 
\be\la{eFbgp} 
\rho L'' 
+[2(l+1)-\rho]L'+[\lam-1-l]L=0, 
~~\rho>0. 
\ee 
Let us substitute here $L(\rho)=a_0+a_1\rho+a_2\rho^2+...$ and equate 
the coefficients with identical powers of $\rho$: 
\be\la{eFbgpe} 
\left\{\ba{ll} 
\rho^0:&2(l+1)+(\lam-1-l)a_0=0,\\ 
\rho^1:&2a_2+2(l+1)2a_2-a_1+(\lam-1-l)a_1=0,\\ 
\rho^2:&3\cdot 2a_3+2(l+1)3a_3-2a_2+(\lam-1-l)a_2=0,\\ 
...&\\ 
\rho^k:&(k+1)ka_{k+1}+2(l+1)(k+1)a_{k+1}-ka_k+(\lam-1-l)a_k=0,\\ 
...&\\ 
\ea\right. 
\ee 
Therefore, we get the recursive relation 
\be\la{rre} 
a_{k+1}=\fr{k-(\lam-1-l)}{(k+1)(k+2l+2)}a_k. 
\ee 
This implies $a_{k+1}/a_k\sim 1/k$ if all $a_k\ne 0$. 
Then we have 
$|L(\rho)|\ge C e^\rho=Ce^{2\ga r}$ 
with a $C>0$. 
Hence, $R(r)=F(r)e^{-\ga r} 
\to\infty$ as $r\to\infty$, which contradicts the definition 
of a quantum stationary state. 
Therefore, $a_{k+1}=0$ and $a_k\ne 0$ for some $k=0,1,2,...$. 
Then $k-(\lam-1-l)=0$, so 
\be\la{ak1} 
\lam=\fr b{2\ga}=k+l+1=n=1,2,... 
\ee 
Substituting here $b=2\mu e^2/\h^2 $ and $\ga= \sqrt{-2\mu E  }/\h  $, we get 
finally, 
\be\la{gf} 
E=E_n:=-\fr {2\pi\h R}{n^2}, ~~~~~~n=1,2,..., 
\ee 
where $R:=\mu e^4/(4\pi\h^3)$.\bo

\bc 
The eigenfunctions  $\psi_\om$ 
for $\om=-2\pi R/ n^2=-\mu e^4/(2\h^3n^2)$ 
have the following 
form in spherical coordinates: 
\be\la{eig} 
\psi_\om=\psi_{lmn}:=Ce^{-r/a_n}P_{ln}(r)F_l^m(\theta)e^{im\vp}. 
\ee 
Here 
\be\la{an} 
a_n:=1/\ga=\h/\sqrt{-2\mu E_n}=\h^2n/(\mu e^2), 
\ee 
and $P_{ln}(r)$ is a polynomial function of degree 
$n-1\ge l$, $ F_l^m(\theta)e^{im\vp}\in D(l)$ and  $m=-l,...,l$. 
\ec 
 
\bexe 
Calculate the multiplicity of the eigenvalue $E_n$.\\ 
{\bf Hint:} It is equal to $\sum_{0\le l \le n-1}(2l+1)$. 
\eexe 
\br{\rm For the stationary state (\re{eig}) 
\\ 
i) The function 
$L(\rho)=\rho^l(a_0+...+a_k\rho^k)$ is a polynomial 
of degree $k+l=n-1$. Hence, $n-1$ 
equals the number of zeros of the function 
$L(\rho)$, $\rho\ge 0$, and $n$ is called the 
{\it principal quantum number} of the stationary state. 
\\ 
ii) 
The component $\bL_3$ of angular momentum is equal to $m\h$. 
This follows from 
(\re{3mcs}),   (\re{qvf}) and (\re{amd}). 
The number $m$ is called the 
{\it magnetic quantum number} of the stationary state 
since it is responsible for the change of the energy in the magnetic field 
(see (\re{LL})). 
\\ 
iii) 
The  (squared) 
{\it absolute value of angular momentum} $\bL^2:=\bL_1^2+\bL_2^2+\bL_3^2$ 
is equal to $l(l+1)\h^2$. 
This follows from 
(\re{sev}) since $\bL^2=\h^2\bH^2$. 
The number $l$ is called the 
{\it azimuthal quantum number} of the stationary state. 
} 
\er 
 
%%%%%%%%%%%%%%%%%%%%%%%%%%%%%%%%%%%%%%%%%%%%%%%%%%%%%%%%%%%%% 
%%%%%%%%%%%%%%%%%%%%%%%%%%%%%%%%%%%%%%%%%%%%%%%%%%%%%%%%%%%%% 
 
\newpage 
%%%%%%%%%%%%%%%%%%%%%%%%%%%%%%%%%%%%%%%%%%%%%%%%%%%%%%%%%%%%% 
%%%%%%%%%%%%%%%%%%%%%%%%%%%%%%%%%%%%%%%%%%%%%%%%%%%%%%%%%%%%% 
 
%%\setcounter{section}{17} 
\setcounter{subsection}{0} 
\setcounter{theorem}{0} 
\setcounter{equation}{0} 
\section 
{Spherical Spectral Problem} 
%%\setcounter{equation}{0} 
%%%%%%%%%%%%%%%%%%%%%%%%%%%%%%%%%%%%%%%%%%%%%%%%%%%%%%%%%%%%% 
%%%%%%%%%%%%%%%%%%%%%%%%%%%%%%%%%%%%%%%%%%%%%%%%%%%%%%%%%%%%% 
We split the space $L^2(S^2)$ in a sum of orthogonal eigenspaces 
of the spherical Laplacian. The proof 
uses rotation invariance of the Schr\"odinger  operator 
and classification of 
all irreducible representations of commutation relations for a Lie 
algebra of angular momenta.

\subsection{Hilbert-Schmidt Argument} 
We start the proof of Theorem \re{tDl} 
with a diagonalization of the operator $\bH^2$. 
This lemma and the {\em elliptic theory} \ci{Sh} 
imply 
\bl 
The operator $\bH^2:=\bH_1^2+\bH_2^2+\bH_3^2$ 
in the space $\E_1:=L^2(S,dS)$ is selfadjoint 
and admits 
the {\bf spectral resolution} 
\be\la{sre} 
\E_1=\oplus_{n=0}^\infty L(n), 
\ee 
where $L(n)\subset C^\infty(S)$ 
are {\bf finite-dimensional} orthogonal subspaces of $\E_1$, 
and $\cH^2|_{L(n)}=\lam_n$ with $\lam_n\to\infty$ as $n\to\infty$. 
\el 
\Pr 
{\it Step i)} 
Each $\bH_k $ is a {\it symmetric} operator 
in $\E_1$. 
Indeed, 
the rotations  $\hat R_k(\vp)$ form a unitary group in  $\E_1=L^2(S,dS)$, 
hence its generators $\na_{\vp_k}$ are skew-adjoint. 
Therefore, the operator $\cH^2$ is a 
symmetric nonnegative operator in $\E_1$. 
\\ 
{\it Step ii)} 
The  operator $\cH^2$ is a nonnegative  elliptic second order 
operator on the sphere $S$. 
This follows from the formula (\re{Lam}) by the identity (\re{laH}). 
Hence, the operator $\bH^2+1: H^2(S)\to H^0(S)$ is invertible, 
where $H^k(S)$ stand for the Sobolev spaces on the sphere $S$. 
Respectively, the operator $(\bH^2+1)^{-1}: H^0(S)\to H^2(S)$ 
is selfadjoint and compact in $H^0(S)=L^2(S,dS)$ by 
the Sobolev embedding 
theorem. Hence, 
the resolution (\re{sre}) exists by the Hilbert-Schmidt 
theorem for the operator $(\bH^2+1)^{-1}$.\bo 
\medskip\\ 
 
We will prove that we can choose the resolution (\re{sre}) 
with $\dim L(n)=2l+1$ and 
$\lam_n=l(l+1)$, where $l=l(n)=0,1,2,...$. 
\bl 
All spaces $L(n)$ are invariant with respect to rotations 
of the sphere. 
\el 
\Pr This follows from the rotation invariance of $\bH^2$. 
\bc\la{Coi} 
All spaces $L(n)$ are invariant with respect to the 
operators  $\bH_k$, $k=1,2,3$. 
\ec 
 
\subsection{Lie Algebra of Angular Momenta}

The linear span $(\bH_1,\bH_2,\bH_3)$ 
is a Lie algebra since the following commutation relations hold: 
\be\la{com} 
[\bH_1,\bH_2]=i\bH_3,~~~~~~~[\bH_2,\bH_3]=i\bH_1,~~~~~~~[\bH_3,\bH_1]=i\bH_2. 
\ee 
\bexe 
Check the commutation relations. 
{\bf Hint:} $\bH_k=-i(\x\times\na_\x)_k$. 
\eexe 
\bl \la{lHk} 
For each $k=1,2,3$,\\ 
i) All spaces $L(n)$ are invariant with respect to $\bH_k$. 
\\ 
ii) $[\bH_k,\bH^2]=0$. 
\el 
\Pr 
{\it ad i)} The invariance holds by Corollary \re{Coi}. 
\\ 
{\it ad ii)} The commutation holds by (\re{amco3}).\bo 
\bd 
$\bH_\pm=\bH_1\pm i\bH_2$. 
\ed 
\bl  \la{lHpm} 
All spaces $L(n)$ are invariant with respect to $\bH_\pm$, and 
\beqn 
[\bH_3,\bH_\pm]&=&\pm \bH_\pm,\la{EH1} 
\\ 
\bH^2&=&\bH_+\bH_-+\bH_3(\bH_3-1)=\bH_-\bH_++\bH_3(\bH_3+1),\la{EH2} 
\\ 
\bH^2&=&\ds\fr 12(\bH_+\bH_-+\bH_-\bH_+)+\bH_3^2.\la{EH3} 
\eeqn 
\el 
\Pr 
The invariance 
 follows from Lemma \re{lHk} $i)$, (\re{EH1}) and (\re{EH2}) 
 follow from (\re{com}), (\re{EH3}) follows from (\re{EH2}).\bo\\ 
 
%%%%%%%%%%%%%%%%%%%%%%%%%%%%%%%%%%%%%%%%%%%%%%%%%%%%%%%%% 
\subsection{Irreducible Representations of Commutation Relations} 
Here we give a complete classification of all possible 
triples of Hermitian operators satisfying 
the commutation relations (\re{com}). 
It implies  a part of 
Theorem \re{tDl} for 
integer or half odd-integer values of $l$.

\bp\la{pNew}\ci{New} 
Let $E$ be a nonzero finite-dimensional complex linear space 
with a Hermitian scalar product, and 
\\ 
i) linear Hermitian operators 
 $\bH^k$, $k=1,2,3$, in $E$, 
 satisfying the commutation relations (\re{com}). 
\\ 
ii) $\bH^2$ is a scalar: $\bH^2=\al\in\C$, 
\\ 
iii) The space $E$ is irreducible, i.e. it does not contain any 
nontrivial subspace which is invariant under all $\bH_k$. 
\\ 
Then there exists a {\bf spin number} $J=0,\fr 12, 1,\fr 32,2,...$ and an 
orthonormal 
 basis in $E$, $\{e_m: m=-J,-J+1,...,J-1,J\}$,  such that 
$\al=J(J+1)$, and 
 the action of the operators 
is given by 
\be\la{act} 
~~~~~~\bH_1 e_m=\fr {s^+_{Jm}} 2e_{m+1}+\fr {s^-_{Jm}} 2e_{m-1}, 
~~~~\,\,\,\,\,\, 
\bH_2 e_m=\fr {s^+_{Jm}} {2i}e_{m+1}-\fr {s^-_{Jm}} {2i}e_{m-1}, 
~~~~\,\,\,\,\,\, 
\bH_3e_m=me_m, 
\ee 
where 
\be\la{actpm'} 
s^\pm_{Jm}=\sqrt{(J\mp m)(J\pm m+1)}. 
\ee 
 
\ep 
\Pr Set 
$\bH_\pm:=\bH_1\pm i\bH_2$, then 
all relations (\re{EH1})-(\re{EH3}) hold by (\re{com}), 
as in Lemma \re{lHk}. 
Since $\bH_3$ is a  Hermitian operator, there exists a basis in $E$, 
provided by its eigenvectors: for each vector $e_m$ of this basis 
we have 
$\bH_3e_m=me_m$ with a real eigenvalue $m\in\R$. 
 
\bl\la{lpm} 
 Either $e_\pm:=\bH_\pm e_m=0$, or 
$e_\pm$ is an eigenvector of $\bH_3$  with the eigenvalue 
$m \pm 1$. 
\el 
\Pr 
(\re{EH1}) implies that 
 $\bH_3e_\pm=\bH_3\bH_\pm e_m=\bH_\pm \bH_3e_m\pm 
\bH_\pm e_m 
=m e_\pm \pm e_\pm$.\bo 
\medskip\\ 
Further,  (\re{EH3}) implies that 
$\al$ is real and $\al\ge m^2$ since $\bH_\pm^*=\bH_\mp$. 
Therefore, 
$\al=J(J+1)$ for a unique $J\ge 0$. 
Finally, 
the identities (\re{EH2}) imply that 
$\bH_-^*\bH_-=\al-\bH_3(\bH_3-1)$ and $\bH_+^*\bH_+=\al-\bH_3(\bH_3+1)$. 
Hence, 
\beqn 
0\le\Vert \bH_-e_m\Vert^2&=&\langle e_m,\bH_-^*\bH_-e_m\rangle= 
\langle e_m,(\al-\bH_3(\bH_3-1))e_m \rangle\nonumber\\ 
&=&[J(J+1)-m(m-1)]\Vert e_m\Vert^2=[(J+m)(J-m+1)]\Vert e_m\Vert^2, 
\la{nor1} 
\\ 
\nonumber\\ 
0\le\Vert \bH_+e_m\Vert^2&=&\langle e_m,\bH_+^*\bH_+e_m\rangle= 
\langle e_m,(\al-\bH_3(\bH_3-1))e_m \rangle\nonumber\\ 
&=&[J(J+1)-m(m+1)]\Vert e_m\Vert^2=[(J-m)(J+m+1)]\Vert e_m\Vert^2.\la{nor2} 
\eeqn 
Therefore, $\bH_- e_m=0$ implies that either $m=-J$ 
or $m=J+1$. But the last value is impossible 
since   $J\ge 0$ and $J(J+1)=\al\ge m^2$. 
Consequently, $m=-J\le 0$. Similarly, 
$\bH_+e_m=0$ implies that $m=J\ge 0$. 
Therefore, $m$ has to run in integral steps from $-J$ to  $J$. 
Hence, $2J+1=1,2,...$. Further, 
$\{e_m: m=-J,-J+1,...,J-1,J\}$ is a basis since $E$ is irreducible. 
 
Finally, for the normalized vectors $e_m$, 
(\re{nor1}) and (\re{nor2}) imply that 
\be\la{actpm} 
~~~~~~\bH_+e_m=s^+_{Jm}\,e_{m+1},~~~~~~\,\,\,\, 
\bH_-e_m=s^-_{Jm}\,e_{m-1}. 
\ee 
Hence, the formulas (\re{act}) follow. 
\bo 
\bc\la{cZ} 
Let us set $Z^+_{Jm}=s^+_{J,-J}...s^+_{J,m-1}$. 
Then the vectors $Y_J^m=\bH_+^{J+m}e_{-J}/Z^+_{Jm}$, $m=-J,...J$, 
constitute an  orthonormal  basis in the space $E$. 
\ec 
\bd 
For $J=0,\fr 12, 1,\fr 32,2,...$, denote by 
$D(J)$ the space $E$ endowed with the 
operators  $\bH_k$, $k=1,2,3$, 
described by Proposition \re{pNew}. 
\ed 
\bex For $J=\fr 12$, the operators $\bH_k$ 
in the orthonormal basis $(Y_{1/2}^{-1/2},Y_{1/2}^{1/2})$ are represented 
by the matrices $\hat\s_k:=\fr 12 \si_k$, 
where $\si_k$ are the Pauli matrices: 
\be\la{mpa} 
\si_1=\left(\ba{rr}0&1\\1&0\ea\right),~~~~~ 
\si_2=\left(\ba{rr}0&-i\\i&0\ea\right),~~~~~ 
\si_3=\left(\ba{rr}1&0\\0&-1\ea\right). 
\ee 
\eex 
\bexe 
Deduce (\re{mpa}) from (\re{act}). 
\eexe

\subsection{Spherical Harmonics} 
Proposition \re{pNew} implies that 
each eigenspace $L(n)$ is isomorphic to a direct sum of 
a number $M(n,J)$ of 
 spaces $D(J)$ 
with a unique value of $J=J(n)=0,\fr 12, 1,\fr 32,2,...$. 
We can assume that $\lam_n$ are distinct for distinct $n$, then 
distinct values of $J$  for a fixed $n$ are impossible since the 
eigenvalue of $\bH^2$ in  $D(J)$, equal to $J(J+1)$, is a strictly 
 increasing 
function of  $J\ge 0$. 
Hence,  Theorem \re{tDl} i) will follow from the next lemma. 
\bl 
i) For every half-integer $J$  we have $M(n,J)=0$ for each $n$. 
\\ 
ii) For every integer $l=1,2,3,...$, 
there exists a unique $n=n(l)$ such that 
$M(n,l)=1$. 
\el 
\Pr 
{\it i}) 
Each eigenspace $L(n)$ is a direct sum of the irreducible 
subspaces $D(J)$. Let us consider one of the subspaces. 
For each eigenvector $e_m\in D(J)$, 
of the operator $\bH_3$, 
we have the following differential equations, 
\be\la{H3e} 
\bH_3e_m(\theta,\vp)=-i\na_\vp e_m(\theta,\vp)= 
m e_m(\theta,\vp). 
\ee 
Therefore, $e_m(\theta,\vp)=F_l^m(\theta)e^{im\vp}$, so 
$m$ is an integer number since the function $e_m(\theta,\vp)$ 
has to be single-valued. 
Hence all possible $J$ also are only integer: $J=l$. 
\\ 
{\it ii)} Let us consider the lowest eigenvalue $m=-l$ in the space 
$D(l)$. Then 
$\bH_-e_{-l}(\theta,\vp)=0$ by 
Lemma \re{lpm}. 
We will prove below that 
in spherical coordinates, 
\be\la{polc} 
\bH_-=-e^{-i\vp}[\na_\theta-i\cot\theta~\na_\vp], 
~~~~~ 
\bH_+=e^{i\vp}[\na_\theta+i\cot\theta~\na_\vp]. 
\ee 
Taking into account 
that $e_{-l}(\theta,\vp)=F_l^{-l}(\theta)e^{-il\vp}$, we get 
from $\bH_-e_{-l}(\theta,\vp)=0$ the differential equation 
$(\na_\theta-l\cot\theta )F_l^{-l}=0$. 
Hence, $F_l^{-l}=C\sin^{l}\theta$ which means that $-l$ 
is a simple eigenvalue. Hence, $M(n,l)\le 1$ for all $n$. 
 
It remains to check that $M(n,l)\ne 0$ 
for some $n$. This 
is 
equivalent to the existence of an eigenvector 
of the operator $\bH^2$ 
with eigenvalue  $l(l+1)$. 
However, this is obvious for the function $e_{-l}(\theta,\vp)$. 
Namely, 
Eqn. (\re{EH2}) implies that 
$$ 
\bH^2e_{-l}= 
\bH_+\bH_-e_{-l}+\bH_3(\bH_3-1)e_{-l}=l(l+1)e_{-l} 
$$ 
since $\bH_-e_{-l}=0$ and $\bH_3e_{-l}=-le_{-l}$.\bo 
 
Now Theorem \re{tDl} i) is proved. 
Further, Corollary \re{cZ} implies that 
the functions $\bH_+^{l+m} e_{-l}/Z^+_{lm}$, $m=-l,...,l$, provide an 
orthonormal basis 
in the space $D(l)$. This implies 
Theorem \re{tDl} ii) 
since 
\be\la{spf} 
\bH_+^{l+m} e_{-l}= 
\Big(e^{i\vp}[\na_\theta+i\cot\theta~\na_\vp]\Big)^{l+m} 
(\sin^{l}\theta e^{-il\vp})=F_l^m(\theta)e^{im\vp}, 
\ee 
where $F_l^m(\theta)$ is a real function. 
\bo 
\bd 
The normalized functions 
$Y_l^m(\theta,\vp)=F_l^m(\theta)e^{im\vp}/Z^+_{lm}$ 
are called  {\bf spherical harmonics}. 
\ed

\subsection{Angular Momenta in Spherical  Coordinates} 
We prove (\re{polc}). 
First, let us rewrite (\re{nas}) as 
\be\la{nasf} 
\na\psi(r,\theta,\vp)=\e_r\na_r\psi+\e_\theta\fr{\na_\theta\psi}r 
+\e_\vp\fr{\na_\vp\psi}{r\sin\theta}=\e_1\na_1\psi+\e_3\na_3\psi+\e_3\na_3\psi, 
\ee 
where $\e_1:=(1,0,0)$, etc. 
Then it is evident geometrically that 
\be\la{nasfg} 
\left\{\ba{l} 
\e_r=(\e_1\cos\vp+\e_2\sin\vp)\sin\theta+\e_3\cos\theta,\\ 
\e_\theta= 
(\e_1\cos\vp+\e_2\sin\vp)\cos\theta-\e_3\sin\theta,\\ 
\e_\vp=~\e_2\cos\vp-\e_1\sin\vp. 
\ea\right. 
\ee 
%%Solving this system, we get 
%%\be\la{sfgs} 
%%\left.\ba{l} 
%%e_1=(e_r\sin\theta+e_\theta\cos\theta)\cos\vp-e_\vp\sin\vp,\\ 
%%e_2=(e_r\sin\theta+e_\theta\cos\theta)\sin\vp+e_\vp\cos\vp,\\ 
%%e_3=~e_r\cos\theta-e_\theta\sin\vp. 
%%\ea\right| 
%%\ee 
Substituting this into (\re{nasf}), we get 
\be\la{fgsr} 
\left\{\ba{l} 
\ds\na_1=\sin\theta\cos\vp\na_r+\cos\theta\cos\vp\fr{\na_\theta}r-\sin\vp 
\fr{\na_\vp}{r\sin\theta},\\ 
\\ 
\ds\na_2=\sin\theta\sin\vp\na_r+\cos\theta\sin\vp\fr{\na_\theta}r+\cos\vp 
\fr{\na_\vp}{r\sin\theta},\\ 
\\ 
\ds\na_3=~\cos\theta\na_r-\sin\vp\fr{\na_\theta}r. 
\ea\right. 
\ee 
Substituting  (\re{fgsr}) and  (\re{sco}) into $\bH_k=-i(\x\times\na_\x)_k$, 
we get 
\be\la{Hpc} 
\left\{\ba{l} 
\bH_1=i(\sin\vp\na_\theta+\cot\theta\cos\vp\na_\vp)\\ 
\\ 
\bH_2=i(-\cos\vp\na_\theta+\cot\theta\sin\vp\na_\vp)\\ 
\\ 
\bH_3=-i\na_\vp. 
\ea\right. 
\ee 
At last, the first two formulas imply (\re{polc}).\bo

\newpage 
%%%%%%%%%%%%%%%%%%%%%%%%%%%%%%%%%%%%%%%%%%%%%%%%% 
 
%%%%%%%%%%%%%%%%%%%%%%%%%%%%%%%%%%%%%%%%%%%%%%%%% 
 
%%%%%%%%%%%%%%%%%%%%%%%%%%%%%%%%%%%%%%%%%%%%%%%%% 

\setcounter{subsection}{0} 
\setcounter{theorem}{0} 
\setcounter{equation}{0} 
\section{Atom Dipole Radiation and Selection Rules} \label{Sec-11} 
 
We  evaluate a radiation of an 
atom with an electrostatic potential 
$\phi^{\rm ext}(t,\x)=\phi^{\rm ext}(\x)$ in a static external 
magnetic field with vector potential 
$\bA^{\rm ext}(\x)$.Then the Schrodinger equation 
(\re{SMeea}) becomes 
\be\la{SMeeab} 
\ds[i\h\pa_t-e\phi^{\rm ext}(\x)]\psi(t,\x) 
=\fr 1{2\mu} 
[-i\h\na_\x-\ds\fr ec \bA^{\rm ext}(\x)]^2\psi(t,\x). 
\ee 
We apply the Bohr approximation and derive the Rydberg-Ritz 
combination principle and selection rules in the case of 
cylindric symmetry.

\subsection{Rydberg-Ritz 
Combination Principle } 
Consider the atom in a statistical equilibrium with an 
appropriate 
environment. 
Then the wave function $\psi(t,\x)$ 
admits an eigenfunction expansion 
\be\la{eex} 
\psi(t,\x)=\sum_k c_k\psi_k(\x)e^{-i\om_k t} 
+\int_0^\infty c_\om\psi_\om(\x)e^{-i\om t}d\om \, , 
\ee 
where $\psi_k$ and $\psi_\omega$ are the eigenmodes of the 
time-independent 
Schr\"odinger equation belonging to the discrete and continuous 
spectrum, 
respectively. Further, the $c_k$ and 
$c_\omega$ describe the 
degree of excitation of the individual eigenmodes. 
 
\bcom\la{rstat} 
{\rm
The amplitudes $c_k $ and $c_\omega$ are random variables 
corresponding to an equilibrium distribution. 
The equilibrium is maintained by an external 
source, like the electric discharge or heat bath, $X$-rays, 
 etc. 
The equilibrium distribution depends on the accelerating voltage 
(or the heat bath temperature, $X$-rays wave-lenght, etc): 
the number 
of the excited modes generally increases with the voltage (or with
the heat bath temperature, etc). 
A mathematical justification of the equilibrium distribution 
for the Maxwell-Schr\"odinger equations is an open problem. 
The justification of the convergence to equilibrium distribution 
in a thermal bath is done 
in 
\ci{DKKS}-\ci{DKSc} 
for a list of linear hyperbolic equations, 
and in \ci{JP} for a nonlinear one. 
The role of the thermal bath plays the initial wave field. 
The convergence to the  equilibrium distribution is caused by the 
fluctuations of the initial field. The fluctuations are provided by 
the mixing condition of Rosenblatt or Ibragimov-Linnik type. 
 
}
\ecom 
 
Let us analyze the Maxwell field  produced by the corresponding 
charge and current densities from (\re{SMeeaa}). 
The potentials of the 
field satisfy the Maxwell equations (\re{dp}), (\re{dA}) 
and their 
large time asymptotics for bounded $|x|$ 
is given by the retarded 
potentials (\re{A22b}): 
\be\la{A22bb} 
\left\{\ba{l} 
\ds\phi(t,\x)\sim~\ds\phi_{ret}(t,\x):= 
~~\ds\int \fr{\rho(t-|\x-\y|/c,\y)} { |\x-\y|}d\y\\~ \\ 
\ds \bA(t,\x)\sim\ds \bA_{ret}(t,\x):=\ds\fr 1c\,\int 
\fr {\bj(t-|\x-\y|/c,\y)} { |\x-\y|} 
d\y 
\ea\right|~~~~t\to\infty,\,~~|\x|\le R. 
\ee 
Let us assume that the atom is located at the point $\y=0$. 
Then the densities $\rho(t,\y)$, $\bj(t,\y)$ are localized 
in a ball $|\y|\le a\ll 1$, where 
$a>0$ is an {\it atom radius}, $a\sim 10^{-8}$cm:
\be\la{roji}
\rho(t,\y)=0, ~~~\bj(t,\y)=0, ~~~~~~~|\y| > a.
\ee
Therefore, any macroscopic observation 
at a distance 
 $|\x|\sim 1$ coincides with high precision with the 
{\it dipole approximation} (cf.  (\re{cpob})) 
\be\la{A22bd} 
\left\{\ba{l} 
\ds\phi(t,\x)\sim~ 
\ds\fr 1{|\x|}\int_{|\y|\le a} {\rho(t-|\x-\y|/c,\y)} d\y\\~ \\ 
\ds \bA(t,\x)\sim\ds\fr 1{c|\x|}\int_{|\y|\le a} 
{\bj(t-|\x-\y|/c,\y)} 
d\y 
\ea\right|~~~~t\to\infty,\,~~|\x|\le R. 
\ee 
\br 
The approximation $\sim |\x|^{-1}$ 
corresponds to the Hertzian dipole radiation 
(\re{ue-di-A}). 
\er 
Next, we calculate the charge-current densities 
corresponding to the wave function 
(\re{eex}) in the Born approximation (\re{SMeeaa}). 
We can write it using (\re{real}), in the form 
\be\la{ccdc} 
\left\{\ba{l} 
\rho(t,\y)= e\psi(t,\y)\ov{\psi(t,\y)}\\ 
\\ 
\bj(t,\y)= \rRe\Bigg[\Big(\ds\fr e{\mu}[-i\h\na_\y-\ds\fr ec \bA^{\rm ext}(\y)] 
\psi(t,\y)\Big)\ov{\psi(t,\y)}\Bigg] 
\ea\right. 
\ee 
The 
integral over the continuous 
spectrum in the RHS of (\re{eex}), 
is responcible only for the continuous spectrum of the 
atom radiation. 
Hence, the discrete spectrum is completely determined by the 
first sum. 
Substituting the sum into 
(\re{ccdc}), we get the `discrete components' 
of the charge and current densities in the form 
\be\la{roj} 
\left\{\ba{l} 
\rho_d(t-|\x-\y|/c,\y)\sim 
e\rRe\sum_{kk'} 
 c_{k}\ov c_{k'}e^{-i(\om_k-\om_{k'}) (t-|\x-\y|/c)} 
\psi_k(\y)\ov\psi_{k'}(\y)\\ 
\\ 
\bj_d(t-|\x-\y|/c,\y)\,\sim 
\\\\ 
\ds\fr e\mu\rRe\sum_{kk'} 
c_{k}\ov c_{k'}e^{-i(\om_k-\om_{k'})t} 
\Big([-i\h\na_\y-\ds\fr ec \bA^{\rm ext}(\y)] 
\psi_k(\y)e^{i\om_k |\x-\y|/c}\Big) 
\ov\psi_{k'}(\y)e^{-i\om_{k'}|\x-\y|/c}. 
\ea\right. 
\ee 
If we substitute this into (\re{A22bd}), we get the following approximation 
for the discrete component 
of the radiation: 
\be\la{A22bdg} 
\left\{\ba{l} 
\ds\phi(t,\x)\sim~~ 
\ds\fr 1{|\x|} 
\rRe\sum_{kk'} \phi_{kk'}(\x)e^{-i(\om_k-\om_{k'}) t} 
\\ 
\\ 
\ds \bA(t,\x)\sim\ds\fr 1{c|\x|} 
\rRe\sum_{kk'} 
\bA_{kk'}(\x)e^{-i(\om_k-\om_{k'}) t} 
\ea\right|~~~~t\to\infty,\,~~|\x|\le R. 
\ee 
\bc 
The Rydberg-Ritz combination principle holds: 
the {\bf discrete spectrum} of the atom radiation is contained in the set 
$\{\om_{k{k'}}:=\om_k-\om_{k'}\}$. 
\ec 
Further, applying the formula $\rRe \sum_{kk'} S_{kk'}= 
\ds\fr 12\sum_{kk'} (S_{kk'}+\ov S_{k'k})$, we get  the following expressions 
for 
the limiting amplitudes in (\re{A22bdg}): 
\be\la{A22bda} 
~~~~~\left\{\ba{ll} 
 \phi_{kk'}(\x)\sim  
c_{k} \ov c_{k'} 
~e\ds\int_{|\y|\le a} 
e^{i\om_{kk'} |\x-\y|/c} 
&\!\!\!\!\!\!\!\!\psi_k(\y)\ov\psi_{k'}(\y)d\y\\ 
\\ 
\!\bA_{kk'}(\x)\approx c_{k} \ov c_{k'} 
\ds\fr e{2\mu} 
\ds\int_{|\y|\le a} 
e^{i\om_{kk'}|\x-\y|/c} 
\Big\{&\!\!\!\!\!\!\Big( 
[-i\h\na_\y-\ds\fr ec \bA^{\rm ext}(\y)] 
\psi_k(\y)\Big)\ov 
\psi_{k'}(\y) 
\\&\!\!\!\!\!\!\!+ 
\Big( 
[i\h\na_\y-\ds\fr ec \bA^{\rm ext}(\y)] 
\ov\psi_{k'}(\y)\Big) 
\psi_k(\y) 
\Big\}d\y 
\ea\right. 
\ee 
with an error of order $\h\om_k/c$ in the last formula. 
Further let us assume that 
\be\la{dr} 
|\om_{kk'}|a/c\ll 2\pi 
\ee 
which means that the wave 
length $\lam_{kk'}:=2\pi c/|\om_{kk'}|\gg a$. 

Under condition (\re{dr}) 
the exponent in the integrands of (\re{A22bda}) 
is close to a constant $\exp(-i\om_{kk'}|\x|/c)$ since 
$|\y|\le a$. In addition, the ball of integration $|\y|\le a$ 
can be substituted by 
 the whole space since the eigenfunctions $\psi_k$, 
$\psi_{k'}$ are well localized. 
Therefore, (\re{A22bda}) becomes 
\be\la{A22bdb} 
\!\!\!\!\!\!\left\{\ba{ll} 
 \phi_{kk'}(\x)\approx ec_{k} \ov c_{k'} e^{i\om_{kk'}|\x|/c}
\ds\int 
&\!\!\!\!\!\!\!\!\psi_k(\y)\ov\psi_{k'}(\y)d\y\\ 
\\ 
\bA_{kk'}(\x)\approx \ds\fr e{2\mu}c_{k} \ov c_{k'}e^{i\om_{kk'}|\x|/c} 
\int 
\Big\{&\!\!\!\!\!\!\Big( 
[-i\h\na_\y-\ds\fr ec \bA^{\rm ext}(\y)] 
\psi_k(\y)\Big)\ov 
\psi_{k'}(\y) 
\\ 
\\&\!\!\!\!\!\!\!+ 
\Big( 
[i\h\na_\y-\ds\fr ec \bA^{\rm ext}(\y)] 
\ov\psi_{k'}(\y)\Big) 
\psi_k(\y) 
\Big\}d\y 
\ea\right. 
\ee 
First formula of  (\re{A22bdb}) implies that
$\phi_{kk'}(\x)\approx 0$ for $\om_{kk'}\ne 0$
 by the orthogonality of different 
eigenfunctions. 
Hence, the radiation is represented by the magentic potential
\be\la{A22bdgm} 
\ds \bA(t,\x)\sim\ds\fr 1{c|\x|}\ds\fr e{2\mu} 
\rRe\sum_{kk'} 
c_{k} \ov c_{k'}e^{-i\om_{kk'}(t-|\x|/c)}\bJ_{kk'}, 
~~~~~~t\to\infty,
\ee 
where $\bJ_{kk'}$ stands for the last integral in  (\re{A22bdb}).

Note that the radiation field (\re{A22bdgm})
 is identical with the sum of the 
{\it  Hertzian dipole  radiation fields} of type
(\re{ue-di-A}) corresponding to the {\it dipole moments}
$\sim\bJ_{kk'}$.

\bcom
The radiation  field (\re{A22bdgm}) brings the energy to infinity
like the  Hertzian dipole  radiation field. Hence, it cannot be 
stationary without an external source of the energy (cf. Comment
\re{rstat}).
\ecom

The formulas (\re{A22bdgm}) define the {\bf intensity 
 of the spectral line} $\om_{kk'}$ corresponding to 
simple eigenvalues $\om_{k}$ and $\om_{k'}$: 
\be\la{inten} 
\langle  \bA_{kk'}\rangle\sim 
\langle c_{k} \ov c_{k'} \rangle \bJ_{kk'}, 
\ee 
where $\langle \cdot\rangle$ stands for the {\it mathematical expectation}. 
For the multiple eigenvalues the intensity is also proportional to the 
{\bf multiplicity of the spectral line} 
which is defined by its splitting in a weak magnetic field (see \ci{Som}). 

Traditional identification of the 
relative intensities (see \ci{Som}) means that 
\be\la{relint} 
\fr{\langle  \bA_{kk'}\rangle} 
{\langle  \bA_{nn'}\rangle}= 
\fr{ \langle c_{k} \ov c_{k'} \rangle \bJ_{kk'}} 
{\langle c_{n} \ov c_{n'} \rangle \bJ_{nn'}} 
\approx 
\fr{ \bJ_{kk'}} 
{ \bJ_{nn'}}. 
\ee 
It holds if $\langle c_{k} \ov c_{k'} \rangle$ 
varies slowly in $k,k'\in (N_1,N_2)$ with $N_2-N_1$ 
sufficiently small.

%%%%%%%%%%%%%%%%%%%%%%%%%%%%%%%%%%%%%%%%%%%%%%%%%%%%% 

\subsection{Selection Rules for Cylindrical Symmetry} 
We consider the case of a {\bf radial} electrostatic potential 
$\phi^{\rm ext}(\x)=\phi^{\rm ext}(|\x|)$ 
and a static {\bf uniform} magnetic field $B$. 
Then  the vector potential $\bA^{\rm ext}(\x)=B\times \x/2$. 
Let us consider a fixed spectral line $\om_{kk'}$ corresponding to 
the frequencies 
$\om_k$ and $\om_{k'}$ 
and eigenfunctions (see (\re{eigc})) 
\be\la{eigf} 
\psi_\om=R_{nl}(r)F_l^m(\theta)e^{im\vp}, 
\qquad 
\psi_{\om'}=R_{n'l'}(r)F_{l'm'}(\theta)e^{im'\vp}. 
\ee 
The next theorem demonstrates that the spectrum of the atom radiation 
is a very small subset of the set of all differences 
$\{\om_{kk'}:=\om_k-\om_{k'}\}$ from the Rydberg-Ritz combination 
principle. 
\bt\la{tsr} 
Let the condition 
(\re{dr}) hold, $\phi^{\rm ext}(\x)=\phi^{\rm ext}(|\x|)$ 
and  $\bA^{\rm ext}(\x)=B\times \x/2$. 
Then for the dipole approximations of the 
limiting amplitudes  we have
$\bJ_{kk'}= 0$ if either $l'\ne l\pm 1$ or $m'\ne m,m\pm 1$. 
\et 
\Pr 
Let us use the identity 
(\re{cce2}) for the solutions 
$\psi_k(\x)e^{-i\om_k t}$, 
$\psi_{k'}(\x)e^{-i\om_{k'} t}$ 
to Equation (\re{SMeea}). 
Eqs. (\re{ccd2}) give the densities 
\be\la{ccd2n} 
\left\{\ba{rl} 
\rho_{kk'}(t,\x):=& 
e^{-i\om_{kk'} t} 
e\psi_k(\x)\ov{\psi_{k'}(\x)},\\ 
\\ 
\bj_{kk'}(t,\x):=& 
e^{-i\om_{kk'} t} 
\Big\{\Big( 
[-i\h\na_\x-\ds\fr ec \bA^{\rm ext}(\x)] 
\psi_k(\x)\Big)\ov 
\psi_{k'}(\x) 
\\ 
\\&~~~~~~~~~~+ 
\Big( 
[i\h\na_\x-\ds\fr ec \bA^{\rm ext}(\x)] 
\ov\psi_{k'}(\x)\Big) 
\psi_k(\x) 
\Big\} 
\ea\right. 
\ee 
Then (\re{cce2}) becomes 
\be\la{cce2n} 
-i\om_{kk'}\rho_{kk'}(t,\x)+\dv\bj_{kk'}(t,\x)=0,~~~~~~ 
(t,\x)\in\R^4. 
\ee 
We see that 
the integral in the second formula of (\re{A22bdb}) 
is proportional to the integral of $\bj_{kk'}(t,\y)$: 
\be\la{A22bdj} 
\bJ_{kk'}\approx e^{i\om_{kk'}|\x|/c}
e^{i\om_{kk'}t}\int 
\bj_{kk'}(t,\y) 
d\y. 
\ee 
It remains to prove 
\bl 
\be\la{ibj} 
\int\bj_{kk'}(t,\x)d\x=0,~~~~~~t\in\R, 
\ee 
if either $l'\ne l\pm 1$ or $m'\ne m,m\pm 1$. 
\el 
\Pr Let us check the case $m'\ne m,m\pm 1$. The case 
$l'\ne l\pm 1$ will be considered in next section.
Let us
multiply (\re{cce2n}) by the coordinate 
$\x^p$, $p=1,2,3$, and integrate 
over $\R^3$. Then we get by partial integration, 
\be\la{cce2p} 
-i\om_{kk'}\int \x^p\rho_{kk'}(t,\x)d\x 
-\int\bj_{kk'}^p(t,\x)d\x=0, 
\ee 
where $\bj_{kk'}^p$ is the $p$-th component of $\bj_{kk'}$. 
Hence, 
\be\la{cce2s} 
\int\bj_{kk'}^p(t,\x)d\x\sim c(t) 
\int \x^p\psi_{k}(\x)\cdot\psi_{k'}(\x)d\x. 
\ee 
Let us rewrite the last integral in spherical coordinates. 
Then we get by (\re{eigf}), 
\beqn\la{lsc} 
\!\!\!\!\!\!\!\!\!&&\!\!\!\!\!\! 
\int \x^p\psi_{k}(\x)\cdot\psi_{k'}(\x)d\x\nonumber\\ 
\!\!\!\!\!\!\!\!\!&&\nonumber\\ 
\!\!\!\!\!\!\!\!\!&&\!\!\!\!\!\!= 
\int_0^\infty R_{nl}(r)\ov{R_{n'l'}(r)}r^3dr 
\int_S\ds\fr {\x^p}r 
F_l^m(\theta)e^{im\vp} 
\ov{F_{l'm'}(\theta)e^{im'\vp}} 
dS. 
\eeqn 
Obviously, the last integral 
is equal to zero if $m'\ne m,m\pm 1$. \bo
\br 
The RHS of (\re{cce2s}) equals to the 
matrix element of the {\bf dipole moment 
operator} $\hat\x^p$. 
Hence, the intensities $\bJ_{kk'}$ 
are proportional to the  {\it dipole moment} like 
classical Hertzian dipole radiation (\re{ue-di-A}). 
This is an example of the Bohr 
 {\bf correspondence principle}.

\er 
\bcom 
The second Bohr postulate states that the radiation 
of the spectral line $\om_{kk'}$ is provided by the 
transition $\psi_k\mapsto\psi_{k'}$ between the stationary states. 
With this identification, the  formulas (\re{A22bdj}) 
and  (\re{cce2s}) mean that the RHS of (\re{cce2s}) 
is proportional to the probability 
of the transition.

\ecom

\subsection{Selection Rules for Orbital Momentum} 
%% 
%% %%CA 
%% 
%%For $l'\ne l\pm 1$ this property is proved in 
%%\ci[Appendix XXI]{Born}. 
%% 
%% Dear Sasha, here I inserted the proof for the $l' =l\pm 1$ selection rule. 
%% 
%% 
The proof of the selection rule $l' =l\pm 1$ is more involved. 
We follow the method of \ci[Appendix XXI]{Born}. 
First, let us re-express the $2l+1$ eigenvectors 
$F_l^m(\theta)e^{im\vp}$ of the spherical Laplacian with eigenvalue 
$-l(l+1)$ 
in terms of real eigenvectors, e.g., like 
\beqn 
Z_{l,m}=F_l^m(\theta)\cos m\vp & , & m=0,\ldots ,l \nonumber \\ 
Z_{l,m}=F_l^m(\theta)\sin (m-l)\vp & , & m=l+1,\ldots ,2l , 
\eeqn 
then let us prove the following 
\bl \la{Zml-hom} 
The $Z_{l,m}$ may be expressed as $r^{-l}U_{l,m}(\x)$, where each $U_{l,m}$ is 
a homogeneous polynomial of degree $l$ in the variables $x,y,z$ such that 
\be 
\Delta U_{l,m}=0 
\ee 
holds. 
\el 
\Pr {\bf of Lemma \ref{Zml-hom}} 
\begin{itemize} 
\item[1)] 
$r^{-l}U_{l,m}(\x)$ only depends on the angular coordinates $\theta ,\vp$. 
\item[2)] 
From 
$$ 
\Delta = \frac{1}{r^2}\partial^2_r + \frac{2}{r}\partial_r + 
\frac{1}{r^2}\Lambda 
$$ 
it follows that 
\be 
0=\Delta U_{l,m} =r^{l-2}\left( l(l+1) Z_{l,m} + \Lambda Z_{l,m}\right) 
\ee 
therefore $Z_{l,m}\equiv r^{-l}U_{l,m}$ are eigenvectors of $\Lambda$ 
with eigenvalues $-l(l+1)$. 
\item[3)] 
There are exactly $2l+1$ linearly independent $U_{l,m}$ 
with the property $\Delta U_{l,m}=0$. 
This may be shown as follows: There are $\frac{1}{2}(l+2)(l+1)$ linearly 
independent homogeneous polynomials of degree $l$, e.g., the monomials 
$$ 
x^l,(x^{l-1}y,x^{l-1}z),(x^{l-2}y^2,x^{l-2}yz,x^{l-2}z^2),\ldots , 
(xy^{l-1},\ldots ,xz^{l-1}),(y^l,\ldots ,z^l) 
$$ 
(here, all monomials with the same power of $x$ are grouped together). 
Further, the Laplacian $\Delta$ acting on a general homogeneous polynomial 
$V_l$ of degree $l$ produces a general homogeneous polynomial $\bar V_{l-2}$ 
of degree $l-2$, $\Delta V_l =\bar V_{l-2}$. The condition that 
$\bar V_{l-2}$ vanishes introduces therefore $\frac{1}{2}l(l-1)$ 
conditions on the $\frac{1}{2}(l+2)(l+1)$ coefficients of $V_l$, therefore 
there exist 
\be 
\frac{1}{2}(l+2)(l+1) - \frac{1}{2}l(l-1) =2l+1 
\ee 
linearly independent $V_l$ such that $\Delta V_l =0$.  Together with Theorem 
\ref{tDl} this proves the above lemma. \bo 
\end{itemize} 
From Theorem \ref{tDl} (orthogonality of the $D(l)$ subspaces) it follows that 
$$ 
\int \sin \theta d\theta d\vp Z_{l,m}Z_{l',m'}=0 
$$ 
unless $l=l'$. Therefore, all that remains to prove is that 
\be \label{l-pm-1Z} 
\frac{\x^p}{r} Z_{l,m}=Y_{l+1} +Y_{l-1} 
\ee 
where $Y_{l+1}$ and $Y_{l-1}$ are some general eigenvectors of $D(l+1)$ 
and $D(l-1)$, respectively, and $\x^p$ is one component of the coordinate 
vector $\x = (x,y,z)$. Eq. (\ref{l-pm-1Z}) may be re-expressed, e.g., for 
$\x^p = x$,  like 
\be \label{l-pm-1U} 
x U_{l} = U_{l+1} + r^2 U_{l-1} 
\ee 
where the $U_k$ are homogeneous polynomials of degree $k$ which obey 
$\Delta U_k =0$. Eq. (\ref{l-pm-1U}) may be proved with the help of the 
following 
\bl \label{hom-pol} 
Every homogeneous polynomial $F_n$ of degree $n$ may be uniquely 
expressed like 
\be \label{th-hom2} 
F_n = U_n + r^2 U_{n-2} + r^4 U_{n-4} + \ldots 
\ee 
where the $U_k$ are homogeneous polynomials of degree $k$ 
which obey $\Delta U_k=0$. 
\el 
The proof of this lemma will be given below. 
Certainly, $xU_l$ of Eq. (\ref{l-pm-1U}) is a homogeneous polynomial of 
degree $l+1$, therefore it may be re-expressed like 
\be \label{exp-U} 
xU_l = U_{l+1} +r^2 U_{l-1} +r^4 U_{l-3} +\ldots 
\ee 
Next, we act with the Laplace operator on both sides of Eq. (\ref{exp-U}). 
For the l.h.s. we get 
\be 
\Delta xU_l = 2\partial_x U_l 
\ee 
and for the r.h.s. we use 
\beqn \label{app-Lap} 
\Delta \left( r^{2h}U_{l+1-2h}\right) &=& r^{2h}\Delta U_{l+1-2h} + 
2\left( \nabla r^{2h} \right) \cdot \left( \nabla U_{l+1-2h} \right) 
+U_{l+1-2h}\Delta r^{2h} \nonumber \\ 
&=& 0+4h r^{2h-2}\x \cdot \nabla U_{l+1-2h} + 2h(2h+1)r^{2h-2}U_{l+1-2h} 
\nonumber \\ 
&=& 2h(2l+3-2h)r^{2h-2}U_{l+1-2h} 
\eeqn 
where we used Euler's theorem for homogeneous polynomials, 
$$ 
\x \cdot \nabla U_n = nU_n. 
$$ 
Therefore, we get altogether 
\be 
2\partial_x U_l = 2(2l+1)U_{l-1} + 4(2l-1)r^2 U_{l-3} +\ldots . 
\ee 
Now the important observation is that the l.h.s. vanishes upon a further 
application of the Laplace operator, $\Delta \partial_x U_l = 
\partial_x \Delta U_l =0$, therefore the r.h.s. must vanish, as well, when 
the Laplace operator is applied. But, according to (\ref{app-Lap}), 
this is possible only if all the $U_{l+1-2h}$ except for $U_{l-1}$ 
vanish (i.e., $U_{l-3}=U_{l-5}=\ldots =0$). This proves Eq. (\ref{l-pm-1U}) 
and, therefore, the original statement that the integral (\ref{lsc}) 
vanishes unless $l' = l\pm 1$. 
\hfill \bo 
\\ \\ 
We still have to prove Lemma (\ref{hom-pol}). \\ 
\Pr {\bf of Lemma \ref{hom-pol}} 
We prove the lemma by induction. It is obviously true for $n=0,1$. We assume 
that it is true for all $k<n$. Then it is also true for $\Delta F_n$, which 
is a homogeneous polynomial of degree $n-2$, i.e., 
\be 
\Delta F_n =\bar U_{n-2} +r^2 \bar U_{n-4} +\ldots +r^{2h}\bar U_{n-2h} 
+\ldots 
\ee 
The general solution to this equation, which is subject to the 
additional condition 
that it is a homogeneous polynomial 
of degree $n$, is 
\be 
F_n =G_n + \frac{1}{2(2n-1)}r^2 \bar U_{n-2} + \ldots + \frac{1}{2h(2n-2h+1} 
r^{2h}\bar U_{n-2h} 
\ee 
as may be shown easily with the help of Eq. (\ref{app-Lap}). Here $G_n$ 
is an arbitrary homogeneous polynomial of degree $n$ 
subject to the condition $\Delta 
G_n=0$. By choosing 
$$ 
U_n =G_n \, ,\quad U_{n-2h}=2h(2n-2h+1)\bar U_{n-2h} 
$$ 
this can be brought into the form stated in Lemma  (\ref{hom-pol}). \bo

%%%%%%%%%%%%%%%%%%%%%%%%%%%%%%%%%%%%%%%%%%%%%%%%%%%%%%%%%%%% 
%%%%%%%%%%%%%%%%%%%%%%%%%%%%%%%%%%%%%%%%%%%%%%%%%%%%%%% 
\newpage 
%%%%%%%%%%%%%%%%%%%%%%%%%%%%%%%%%%%%%%%%%%%%%%%%%%%%%%%%%%%%%%%%%%%%% 
%%%%%%%%%%%%%%%%%%%%%%%%%%%%%%%%%%%%%%%%%%%%%%%%%%%%%%%%%%%%%% 
%%\setcounter{section}{+9} 
\setcounter{subsection}{0} 
\setcounter{theorem}{0} 
\setcounter{equation}{0} 
\section{Classical Scattering of Light: Thomson formula} 
Electromagnetic waves may be scattered by a charged matter. 
Here, we 
describe a light as a plane wave satisfying the free Maxwell 
equation and a matter as composed of classical, point-like 
particles obeying the 
Lorentz equation. The inconsistencies inherent to the concept of 
point-like charged particles are circumvented by applying a 
certain approximation (neglecting self-interactions). We apply 
the Hertzian dipole radiation formula 
 and derive the 
Thomson differential cross-section.

\subsection{Incident Plane Wave} 
J.C.Maxwell identified light 
with electromagnetic waves which are solutions 
to the Maxwell equations in {\it free space} with 
$\rho(t,\x)=0$ and $\bj(t,\x)=0$. 
A general solution to the Maxwell equations is expressed through 
the potentials by (\re{pot}): 
\be\la{pots} 
~~~~ 
\bE(t,\x)=-\na_\x\phi(t,\x)-\ds\fr 1c\,\dot  \bA(t,\x),\,\,\,\,\,\,\, 
\bB(t,\x)= \rot \bA(t,\x),~~\,\,\,(t,\x)\in\R^4. 
\ee 
For the case $\rho(t,\x)=0$ and $\bj(t,\x)=0$ 
the potentials $\bA_0(t,\x)$ and $\phi_0(t,\x)$ are solutions to the 
homogeneous wave equations (\re{dA}) and (\re{dp}): 
\be\la{dAs} 
~~~~\Box \bA_0(t,\x)= 
0,\,\,\,(t,\x)\in\R^4. 
\ee 
\be\la{dps} 
~~~~\Box\phi_0(t,\x)=0,\,\,\,(t,\x)\in\R^4. 
\ee 
Let us choose $\phi(t,\x)=0$ for concreteness, and 
\be\la{A1s} 
~~~~\bA_0(t,\x)=A\sin k(x_3-ct)(1,0,0) 
\ee 
with a wave number $k>0$ and the corresponding frequency $\om=kc>0$. 
Then (\re{pots}) gives 
\be\la{BE} 
~~~~ 
\bE_0(t,\x)=\ds kA\cos k(x_3-ct)(1,0,0),\,\,\,\bB_0(t,\x)=kA\cos k(x_3-ct) 
(0,1,0). 
\ee 
The corresponding energy flux (i.e., Pointing vector) is 
\be\la{SPs} 
~~~~\bS_0(t,\x)=\fr c{4\pi}\bE_0(t,\x)\times \bB_0(t,\x)= 
\fr{cE_0^2}{4\pi}\cos^2k(x_3-ct)(0,0,1), 
~~~\,(t,\x)\in\R^4, 
\ee 
where $E_0:=kA$. 
Let us note that the energy flux is directed along $\e_3=(0,0,1)$ 
and its {\it intensity} is 
\be\la{Is} 
I_0:=\lim_{T\to\infty} \fr 1T|\int_0^T \bS_0(t,\x)dt|=\fr{cE_0^2}{8\pi} . 
\ee 
%%%%%%%%%%%%%%%%%%%%%%%%%%%%%%%%%%%%%%%%%%%%%%%%%%%%%%%%%%%%% 
\subsection{Scattering Problem} 
We consider the scattering of the plane wave (\re{BE}) 
by a classical electron. 
The scattering is described by the coupled Maxwell-Lorentz 
equations 
\be\la{meqs} 
~~~~~~~~~~~~\left\{ 
\ba{l} 
\dv \bE(t,\x)= 4\pi e\de(\x-\x(t)),\,\,\rot \bE(t,\x)= 
- \ds\fr 1c\dot  \bB(t,\x),\\ 
~\\ 
\dv \bB(t,\x)= 0,\,\,\rot \bB(t,\x)= \ds\fr 1c \,\dot \bE(t ,\x)+\ds\fr{4\pi}c 
\,e\dot \x\,\de(\x-\x(t)), 
\ea 
\right|~(t,\x)\in\R^4 , 
\ee 
\be\la{Les} 
~~~~\mu\ddot \x(t)=e [\bE(t,\x(t))+\fr 1c\,\dot \x(t)\times \bB(t,\x(t))], 
\,\,\,\,t\in\R . 
\ee 
%%%%%%%%%%%%%%%%%%%%%%%%%%%%%%%%%%%%%%%%%%%%%%%%%%%%%%%%%%%%%%%%%% 
%%%%\subsection{Scattering Initial Data} 
The {\it free electron} is governed by the Lorentz equation (\re{Les}) 
with the initial conditions 
\be\la{ices} 
\x(t)=0,\dot\x(t)=0,~~~~t<0. 
\ee 
The incident plane wave appears in the initial conditions for the fields, 
\be\la{ics} 
\bE(t,\x)=\Theta(-x_3+ct)\bE_0(t,\x)-e\fr {\x}{|\x|^2},\,\,\,\,\, 
\bB(t,\x)=\Theta(-x_3+ct)\bB_0(t,\x),~~~~t<0, 
\ee 
where $\Theta$ is the Heaviside function and 
$-e\x/|\x|$ is the static Coulomb field generated by 
the electron at the position (\re{ices}). 
Let us note that the incident wave 
is a solution to the homogeneous Maxwell equations.

\subsection{Neglecting the Self-Interaction} 
Let us split the solution to (\re{meqs}) like 
\be\la{sps} 
\left\{ 
\ba{l} 
\bE(t,\x)=\Theta(-x_3+ct)\bE_0(t,\x)+\bE_r(t,\x),\\ 
\bB(t,\x)=\Theta(-x_3+ct)\bB_0(t,\x)+\bB_r(t,\x), 
\ea 
\right|~~t\in\R, 
\ee 
where $\bE_r(t,\x),\bB_r(t,\x)$ stand for the {\it radiated fields}. 
The fields are defined by the splitting. 
Then the Maxwell equations (\re{meqs}) read 
\be\la{meqsb} 
~~~~~~~~~~~~~~~~~~~\left\{ 
\ba{l} 
\dv \bE_r(t,\x)= 4\pi e\de(\x-\x(t)),\,\,\rot \bE_r(t,\x)= - \ds\fr 1c\dot 
\bB_r(t,\x),\\ 
~\\ 
\dv \bB_r(t,\x)= 0,\,\,\rot \bB_r(t,\x)= \ds\fr 1c 
\,\dot \bE_r(t ,\x)+\ds\fr{4\pi}c 
\,e\dot \x\,\de(\x-x(t)), 
\ea 
\right|~(t,\x)\in\R^4 , 
\ee 
since the incident wave in (\re{ics}) 
is a solution to the homogeneous Maxwell equations. 
The initial conditions  (\re{ics}) become 
\be\la{icsb} 
\bE_r(t,\x)=-e\fr {\x}{|\x|^2},\,\,\,\,\, 
\bB_r(t,\x)=0,~~~~t<0. 
\ee 
The Lorentz equation (\re{Les}) now reads 
\beqn\la{Lesb} 
~~~~ \mu\ddot \x(t)&=&e [\Theta(-\x_3(t)+ct)\bE_0(t,\x(t))+ 
\bE_r(t,\x(t)) 
\nonumber\\ 
&+&\fr 1c\,\dot \x(t)\times 
(\Theta(-\x_3(t)+ct)\bB_0(t,\x(t))+\bB_r(t,\x(t)))], 
\,\,\,\,t\in\R . 
\eeqn 
Unfortunately, the problem  (\re{meqsb}), (\re{Lesb}) 
 is not well posed. Namely, 
the solutions $\bE_r(t,\x),\bB_r(t,\x)$ 
to (\re{meqsb}) are infinite at the points $(t,\x(t))$, 
which is obvious from (\re{icsb}). 
Therefore, 
the RHS of equation (\re{Lesb}) does not make sense 
for the fields $\bE_r(t,\x),\bB_r(t,\x)$, which are solutions to (\re{meqsb}). 
\br 
To make the problem well posed it is necessary to replace the point-like 
electron by the {\bf extended electron} suggested by M.Abraham \ci{Ab}. 
For this model, the well-posedness is proved in \ci{sp4}. 
\er 
Here we use another traditional approach 
to make the problem well posed, which is similar 
to the Born approximation. Concretely, 
we omit the radiation fields at 
the RHS of (\re{Lesb}): 
\be\la{Lesbn} 
~~~~\mu\ddot \x(t)=e [\Theta(-\x_3+ct)\bE_0(t,\x) 
+\fr 1c\,\dot \x(t)\times 
\Theta(-\x_3+ct)\bB_0(t,\x)], 
\,\,\,\,t\in\R. 
\ee 
Then we substitute the solution $\x(t)$ into the 
RHS of the Maxwell equations (\re{meqsb}) to calculate 
the radiated fields $\bE_r(t,\x),\bB_r(t,\x)$. 
 
The approximation (\re{Lesbn}) 
means that we neglect the self-interaction of the electron, 
which can be justified for the ``non-relativistic electron''. 
This means that the electron velocities are small compared to 
the speed of light: 
\be\la{sms} 
\beta:=\max_{t\in\R}|\dot \x(t)|/c\ll 1. 
\ee 
Then we can 
neglect the contribution of the magnetic field to the RHS 
of (\re{Lesbn}). 
Finally, we consider the equation 
\be\la{Lesbna} 
~~~~\mu\ddot \x(t)=e\bE_0(t,\x), 
\,\,\,\,t>0. 
\ee 
This equation and  the initial conditions (\re{ices}) 
define the trajectory $\x(t)$ uniquely: 
\be\la{smsu} 
\x(t)=\fr {eA}{\mu kc^2}(1-\cos kct)(1,0,0) . 
\ee 
The condition (\re{sms}) for the solution is equivalent to 
\be\la{smsa} 
\beta=\ds\fr {|e|A}{\mu c^2}\ll 1. 
\ee 
\br 
This relation means  that the amplitude of the oscillations 
$\ds\fr {|e|A}{\mu kc^2}$ 
is small compared to the 
wave length $2\pi/k$ of the incident wave. 
\er 
 
\subsection{Scattering in Dipole Approximation} 
Now we have to solve the Maxwell equations (\re{meqsb}) to define the radiation 
fields $\bE_r(t,\x),\bB_r(t,\x)$. 
Our goal is an analysis of the energy flux at infinity, i.e., of the 
Poynting vector 
$ 
\bS_r(t,\x):= ( c/{4\pi})\bE_r(t,\x)\times \bB_r(t,\x) 
$ 
as $|x|\to\infty$. We use the traditional 
{\it dipole  approximation} which leads to the well-known 
 {\it Thomson formula}.

For this purpose, let us 
expand the charge density 
in the Maxwell equations (\re{meqsb}) 
in a 
Taylor series of the type (\re{mdapn}): 
\be\la{exp} 
e\de(\x-\x(t))=e\de(\x)+e\x(t)\cdot\na\de(\x)+ 
\fr 12e(\x(t)\cdot\na)^2\de(\x)+...,\,\,\,t>0. 
\ee 
Here, the first term is static, and the corresponding Maxwell field 
is static with zero energy flux. 
The second term corresponds to the Hertzian dipole with dipole moment 
$\p(t):=e\x(t)$. 
The next terms of (\re{exp}) give small 
contributions to the total energy radiation to infinity by (\re{sms}). 
\bexe 
{\bf i)}  Prove the convergence of the Taylor series (\re{exp}) in the sense 
of (\re{mdas}) with $\psi\in\cH_a(\R^3)$ for $a>\max|\x(t)|= 
2\ds\fr {|e|A}{\mu kc^2}$.\\ 
{\bf ii)} Prove that the energy flux $S_r(t,\x)$ for large $|x|$ 
is determined by the second term of the expansion (\re{exp}) up to an error 
$\cO(\beta)/|\x|^2$. 
{\bf Hint:} Use the methods of Exercise \ref{ue-Ex-9}. 
\eexe 
So we can use the Hertzian formula (\ref{ue-hertz}) for the dipole radiation 
(Section \ref{ue-Ex-9}); 
\be\la{SHe} 
\bS_r(t,\x)\approx 
\fr{\sin^2\chi}{4\pi c^3|\x|^2}\ddot \p^2(t-|\x|/c)\n,\,\,\,|\x|\to\infty, 
\ee 
where $\chi$ is the angle between $\ddot \p(t-|\x|/c)$ 
and $\n:=\x/|\x|$. 
By (\re{Lesbna}) and (\re{BE}), we have 
\be\la{ddp} 
\ddot \p(t)=e\ddot \x(t)= \fr{e^2}\mu \bE_0\cos k(-ct). 
\ee 
Hence, 
\be\la{ddph} 
\ddot \p(t)^2= \Big(\fr{e^2}{\mu}\Big)^2 \bE_0^2\cos^2 kct. 
\ee 
 
\bd 
Denote by $\theta$ the angle between $\n$ and $\e_3$, and by 
$\vp$ the (azimuthal) angle between $\e_1$ and the plane $(\n,\e_3)$. 
\ed 
\bexe 
Check that $\cos\chi=\cos\vp\sin\theta$. 
{\bf Solution:} 
$\cos\chi$ is the projection of $\n$ onto the vector $\ddot \p(t-|\x|/c)$, 
which is parallel to $\e_1$. 
The projection of  $\n$ onto the plane  $\Pi=(\e_1,\e_2)$ 
has length $\sin\theta$. 
Finally, the angle between this  projection and $\e_1$ 
equals $\vp$. 
\eexe 
\bc 
$\sin^2\chi=1-\cos^2\vp\sin^2\theta$. 
\ec 
Therefore, 
\be\la{Sr} 
\bS_r(t,\x)\approx 
\fr{1-\cos^2\vp\sin^2\theta}{4\pi c^3|\x|^2} 
\Big(\fr{e^2}{\mu}\Big)^2 \bE_0^2\cos^2 kc(t-|\x|/c)\n,\,\,\,|\x|\to\infty. 
\ee 
Hence, for large $|\x|$,  the energy flux is directed along $\n$. 
The corresponding intensity is obtained by the standard 
replacement of $\cos^2 kc(t-|\x|/c)$ by $1/2$: 
\beqn\la{Ir} 
I_r(\x):=\lim_{T\to\infty} \fr 1T|\int_0^T\bS_r(t,\x)dt|&\approx& 
\fr{1-\cos^2\vp\sin^2\theta}{8\pi |\x|^2} 
\Big(\fr{e^2}{\mu c^2}\Big)^2 \bE_0^2\\ 
&=& 
\Big(\fr{e^2}{\mu c^2}\Big)^2\,\, 
\fr{1-\cos^2\vp\sin^2\theta}{|\x|^2} 
I_0.\nonumber 
\eeqn 
Therefore, the {\it intensity per unit angle} $\I_r:=I_r(\x)|\x|^2$ is 
\beqn\la{Ira} 
\I_r(\vp,\theta)\approx 
\Big(\fr{e^2}{\mu c^2}\Big)^2\,\, 
(1-\cos^2\vp\sin^2\theta) 
I_0. 
\eeqn 
Hence, the {\it differential cross-section} is given by 
the {\it Thomson formula} 
\be\la{dcs} 
D(\vp,\theta):=\fr{\I_r(\vp,\theta)}{I_0}\approx 
\Big(\fr{e^2}{\mu c^2}\Big)^2\,\, 
(1-\cos^2\vp\sin^2\theta). 
\ee 
Finally, the {\it total cross-section} is given by 
\beqn\la{tcs} 
T:=\int\I_r(\vp,\theta)d\Om/I_0 
&=&\int D(\vp,\theta)d\Om\approx 
\Big(\fr{e^2}{\mu c^2}\Big)^2\,\, 
\int 
 (1-\cos^2\vp\sin^2\theta)d\Om\nonumber\\ 
&=&\Big(\fr{e^2}{\mu c^2}\Big)^2 
\Big[4\pi-\int _0^{2\pi} \cos^2\vp d\vp\int_0^\pi 
\sin^3\theta d\theta\Big]=\Big(\fr{e^2}{\mu c^2}\Big)^2\,\,\fr{8\pi}{3} . 
\eeqn 
\br 
The differential cross-section (\re{dcs}) depends on $\vp$, hence it 
is not invariant with respect to rotations around $\e_3$. 
This reflects the fact that the incident wave is linearly polarized. 
If we consider light with random polarization, then the 
differential cross-section 
is given by (\re{dcs}) with $1/2$ instead of $\cos^2\vp$. 
\er

%%%%%%%%%%%%%%%%%%%%%%%%%%%%%%%%%%%%%%%%%%%%%%%%%%%%%%%%%%%%%%%%%%%%%%%%% 
%%%%%%%%%%%%%%%%%%%%%%%%%%%%%%%%%%%%%%%%%%%%%%%%%%%%%%%%%%%% 
%%%%%%%%%%%%%%%%%%%%%%%%%%%%%%%%%%%%%%%%%%%%%%%%%%%%%%% 

\newpage 
%%%%%%%%%%%%%%%%%%%%%%%%%%%%%%%%%%%%%%%%%%%%%%%%%%%%%%%%%%%%%%%%%%%%% 
%%%%%%%%%%%%%%%%%%%%%%%%%%%%%%%%%%%%%%%%%%%%%%%%%%%%%%%%%%%%%% 
%%\setcounter{section}{+9} 
\setcounter{subsection}{0} 
\setcounter{theorem}{0} 
\setcounter{equation}{0} 
\section{Quantum Scattering of Light. 
Zero Order Approximation} 
We study scattering of light described as a plane wave by a 
quantum hydrogen atom in its ground state, i.e. in the state 
of the lowest possible energy. We derive an energy flux and 
a differential cross-section. 
 
The ground state energy is 
$E_1=-2\pi\h R 
=-\mu e^4/(2\h^2)$, and the corresponding  wave function 
(\re{eig}) is 
$\psi_1(\x)=C_1e^{-|\x|/a_1}$ 
(we assume that the atom is situated at the origin). 
Then corresponding solution 
to the Schr\"odinger equation is 
\be\la{ceig} 
\psi_1(t,\x)=C_1e^{-|\x|/a_1}e^{-i\om_1t},\,\,\,\,\om_1=E_1/\h 
=-\mu e^4/(2\h^3). 
\ee 
We want to describe the scattering of the plane wave (\re{BE}) 
by the atom in the ground state. 
The scattering is described by the coupled Maxwell-Schr\"odinger 
equations in the Born approximation (\re{SMeea}), (\re{SMeeaa}). 
In the Lorentz gauge we have 
\be\la{SMes} 
\ds[i\h\pa_t-e\phi^{\rm ext}(\x)]\psi(t,\x) 
=\fr 1{2\mu} 
[-i\h\na_\x-\ds\fr ec \bA^{\rm ext}(t,\x)]^2\psi(t,\x), 
\ee 
\be\la{SMeas} 
\left\{\ba{l} 
\ds\fr 1{4\pi}\,\Box \phi(t,\x)=\rho(t,\x)=e|\psi(t,\x)|^2,\\ 
\\ 
\ds\fr 1{4\pi}\,\Box \bA(t,\x)=\ds\fr{\bj(t,\x)}c 
=\ds\fr e {\mu c}[-i\h\na-\ds\fr ec  \bA^{\rm ext}(t,\x)]\psi(t,\x ) 
\cdot\psi(t,\x ), 
\ea 
\right. 
\ee 
where $\phi^{\rm ext}=-e/|\x|$ is the Coulomb field of the nucleus and 
 $\bA^{\rm ext}$ stands for the incident wave 
(\re{A1s}) (cf. (\re{ics})), 
\be\la{A1t} 
~~~~\bA^{\rm ext}(t,\x)=A\sin k(x_3-ct)\Theta(-\x_3+ct)(1,0,0). 
\ee

%%%%%%%%%%%%%%%%%%%%%%%%%%%%%%%%%%%%%%%%%%%%%%%%%%%%%%%%%%% 
\subsection{Atom Form Factor} 
At zero order approximation (in $A$), the wave function 
is unperturbed, $\psi(t,\x)=\psi_1(t,\x)$. The corresponding 
approximation to the radiation fields $\phi, \bA$ is given by 
the solutions to the Maxwell equations (\re{SMeas}) with 
$\psi(t,\x)=\psi_1(t,\x)$. Then the charge density $\rho(t,\x)$ 
is static and the corresponding potential $\phi$ is static with zero 
energy flux. Therefore, it suffices to solve the equation for 
$\bA$ with the current 
\be\la{cur} 
\ds\fr{\bj(t,\x)}c 
=\ds\fr e{\mu c}[-i\h\na\psi_1(t,\x ) 
\cdot\psi_1(t,\x )-\ds\fr ec  \bA^{\rm ext}(t,\x)\psi_1(t,\x ) 
\cdot\psi_1(t,\x )]. 
\ee 
Here, the first term on the RHS is zero since the corresponding 
eigenfunction $\ds e^{-|x|/a_1}$ is real: 
\be\la{cur0} 
~~~~~~~~~i\h\na\psi_1(t,\x ) 
\cdot\psi_1(t,\x ):=\rRe\Big( i\h\na\psi_1(t,\x ) 
\ov\psi_1(t,\x )\Big)=|C|^2\rRe\Big( i\h(\na e^{- |x|/a_1}) 
e^{-|x|/a_1}\Big)=0. 
\ee 
Therefore, the current is reduced to 
\be\la{curr} 
\ds\fr{\bj(t,\x)}c 
=-\ds\fr{e^2} 
{\mu c^2} \bA^{\rm ext}(t,\x)|\psi_1(\x )|^2. 
\ee 
Let us split the solution 
\be\la{str} 
\bA(t,\x)=\bA^{\rm ext}(t,\x)+\bA_r(t,\x), 
\ee 
where $\bA_r(t,\x)$ is the {\it radiated field}. 
Then 
(\re{SMeas}) becomes 
\beqn\la{SMeasb} 
\Box \bA_r(t,\x)&=&-4\pi\ds\fr{e^2} 
{ \mu c^2} A\sin k(x_3-ct)|\psi_1(\x )|^2\Theta(-\x_3+ct)(1,0,0) 
\nonumber\\ 
&=& 
-4\pi\ds\fr{e^2} 
 {\mu c^2}\rIm Ae^{ ik(x_3-ct)}|\psi_1(\x )|^2\Theta(-\x_3+ct) 
(1,0,0)=:f(t,\x) 
\eeqn 
since $\bA^{\rm ext}$ is a solution to the homogeneous equation. 
 
The radiation field could be characterized uniquely by the initial condition
of type (\re{icsb}) if the atom radius would be zero as in the 
classical case of previous lecture. In our case  
the radiation field is of finite energy, i.e.
\be\la{fine}
E_r(t):=\int_{\R^3}\Big(|\dot\bA_r(t,\x)|^2+(|\na\bA_r(t,\x)|^2\Big)dx<\infty,
~~~~~~~~t\in\R.
\ee
An exact characterization of the radiation field would be given by the limit
$E_r(t)\to 0$, $t\to-\infty$, however (\re{fine})
 is sufficient for our purposes.
Namely, let us demonstrate that 
Theorem \re{trp} and (\re{fine}) with $t=0$ imply 
the 
{\it limiting amplitude principle} 
 holds, i.e. 
\be\la{wlap} 
\bA_r(t,\x)\sim \rIm A_r(\x)e^{-ikct},\,\,\,\,\,t\to\infty. 
\ee 
In particular, the zero order approximation to the 
radiation field  has the same frequency as the incident wave. 

Indeed, the asymptotics  (\re{wlap}) follow similarly to (\re{jex}). 
That is, Theorem \re{trp} implies 
that 
the long-time asymptotics of the finite-energy 
solution $\bA_r(t,\x)$ is given by 
the retarded potential 
\beqn\la{rpo} 
\bA_r(t,\x)&\sim& \int \fr{f(t-|\x-\y|/c,\y)d\y}{4\pi|\x-\y|} 
\,\nonumber\\ 
\nonumber\\ 
&=& 
-\ds\fr{e^2} 
{ \mu c^2} 
 A\, \rIm  \int\ds \fr{e^{ ik(\y_3-c(t-|\x-\y|/c))}|\psi_1(\y )|^2d\y} 
{|\x-\y|}(1,0,0) 
\,\nonumber\\ 
\nonumber\\ 
&=& 
-\ds\fr{e^2} 
 {\mu c^2}A\, \rIm 
e^{-ikct} 
\int\ds \fr{e^{ ik(\y_3+|\x-\y|)}|\psi_1(\y )|^2d\y} 
{|\x-\y|}(1,0,0). 
\eeqn 
Let us find the asymptotics of the integral as $|\x|\to\infty$. 
For any fixed $\y\in\R^3$, 
\be\la{mod} 
|\x-\y|=|\x|-\y\cdot \n+o(1),\,\,\,\,\,|\x|\to\infty, 
\ee 
where $\n=\n(x)=\x/|\x|$. Let us also write $\y_3=\y\cdot \e_3$. 
Then we get 
\be\la{rpog} 
\bA_r(t,\x)\sim 
-\ds\fr{e^2} 
 {\mu c^2} A\, \rIm 
e^{-ikct}\fr{e^{ik|\x|}}{|\x|} 
\int\ds e^{ ik\y\cdot(\e_3-\n)}|\psi_1(\y )|^2d\y 
(1,0,0),\,\,\,\,\,|\x|\to\infty. 
\ee 
Next, we evaluate the last integral. Set $K:=k|\e_3-\n|$ and 
denote by $\theta$ the angle between $\n$ and $\e_3$. Then 
\be\la{Kkn} 
~~~~~K=K(k,\theta)=k\sqrt{n_1^2+n_2^2+(1-n_3)^2}= 
k\sqrt{2(1-n_3)}=k\sqrt{2(1-\cos\theta)} 
=2k\sin\fr\theta2. 
 \ee 
Denote by $\al$ the angle between $\y$ and $\e_3-\n$, and by $\vp$ 
the azimuthal angle around $\e_3-\n$. 
Finally, let us take into account that the ground state $\psi_1(\y )=
\psi_1^r(|\y| )$ 
is spherically symmetric. Then 
the integral becomes 
\beqn\la{inb} 
&& 
\int_0^\infty |\y|^2d|\y| 
\int_0^\pi \sin\al d\al\int_0^{2\pi}d\vp 
\ds e^{ iK\cos\al|\y|}|\psi_1(\y )|^2 
\nonumber\\ 
&=& 
4\pi\int_0^\infty \fr{\sin K|\y|}{K|\y|}|\psi_1^r(|\y| )|^2|\y|^2d|\y| 
=:F_a(k,\theta) 
\eeqn 
which is called the {\it atom form factor}. 
\bexe 
Calculate the last integral. 
\eexe 
Since $F_a(\theta)$ is real,  the asymptotics 
(\re{rpog}) becomes 
\be\la{rpogb} 
\bA_r(t,\x)\sim 
-\ds\fr{e^2} 
 {\mu c^2} A 
\fr{\sin k(|\x|-ct)}{|\x|} 
F_a(k,\theta) 
(1,0,0),\,\,\,\,\,|\x|\to\infty. 
\ee 
\subsection{Energy flux} 
We still have to calculate the Maxwell field and the Poynting vector 
corresponding to this vector potential. 
It suffices to compare the expressions (\re{rpogb}) with the vector potential 
of the Hertzian dipole (formula (\ref{ue-di-A}) of Exercise \ref{ue-Ex-9}): 
\be\la{Hd} 
\bA(t,\x) 
=\fr 1c \fr{\dot \p(t-r/c)}r. 
\ee 
This is identical to (\re{rpogb}) if $F_a(\theta)=1$ and 
\be\la{ws} 
\p(t):=\ds\fr{e^2} 
 {\mu c^2 k} A 
\cos kct \, {\bf e}_3 . 
\ee 
Therefore, the energy flux $\bS(t,\x)$ 
corresponding to (\re{rpogb}) 
is given, up to an error $\cO(|x|^{-3})$, by the Hertzian 
formula (\re{SHe}), with a factor $|F_a(\theta)|^2$. 
This follows from the fact that any differentiation of the form factor 
$F_a(\theta)$ in $\x_k$ gives an additional factor with the 
decay $\cO(|x|^{-1})$ 
since the form factor is a homogeneous function of $\x$. 
Finally,  for the function (\re{ws}), 
$\ddot \p(t)$ coincides with (\re{ddp}). Therefore, 
(\re{Sr}) gives in our case 
\be\la{Src} 
 \bS(t,\x) \sim |F_a(\theta)|^2 
\fr{1-\cos^2\vp\sin^2\theta}{4\pi c^3|\x|^2} 
\Big(\fr{e^2}{\mu}\Big)^2 E_0^2\cos^2 kc(t-|\x|/c)\n,\,\,\,|\x|\to\infty. 
\ee 
Then the intensity 
and 
differential cross-section 
in our case 
also contain the additional factor  $|F_a(\theta)|^2$. 
Hence, 
the differential cross-section 
coincides with the Thomson formula (\re{dcs}) up to the atom form factor: 
\be\la{dcsb} 
D(k,\vp,\theta)= 
|F_a(k,\theta)|^2\Big(\fr{e^2}{\mu c^2}\Big)^2\,\, 
(1-\cos^2\vp\sin^2\theta). 
\ee

%%%%%%%%%%%%%%%%%%%%%%%%%%%%%%%%%%%%%%%%%%%%%%%%%%%%%%%%%%%%%%%%%%%%%%%%% 
%%%%%%%%%%%%%%%%%%%%%%%%%%%%%%%%%%%%%%%%%%%%%%%%%%%%%%%%%%%% 
%%%%%%%%%%%%%%%%%%%%%%%%%%%%%%%%%%%%%%%%%%%%%%%%%%%%%%% 

\newpage 
%%%%%%%%%%%%%%%%%%%%%%%%%%%%%%%%%%%%%%%%%%%%%%%%%%%%%%%%%%%%%%%%%%%%% 
%% 
%%%%%%%%%%%%%%%%%%%%%%%%%%%%%%%%%%%%%%%%%%%%%%%%%%%%%%%%%%%% 
%%\setcounter{section}{+9} 
\setcounter{subsection}{0} 
\setcounter{theorem}{0} 
\setcounter{equation}{0} 
\section{Light Scattering at Small Frequencies: Short-Range Scattering} 
Now we take into account the change of the ground state 
induced by the incident light (\re{A1s}), in the  first order 
with respect to $A$, for small frequencies of light. 
This change describes the {\it polarization} of the hydrogen atom 
(see previous lecture). 
It defines the corresponding {\it dispersion} and 
the {\it combinational scattering}. 
\subsection{First Order Approximation to the Ground State} 
To calculate the first approximation to the ground state, 
let us use the Schr\"odinger equation 
(\re{SMes}): 
for small amplitudes $|A|$, 
\be\la{fa} 
\psi(t,\x)=\psi_1(t,\x)+Aw(t,\x)+\cO(A^2). 
\ee 
Substituting into (\re{SMes}), we obtain in first order in $A$, 
\be\la{SMes1} 
A\ds[i\h\pa_t-e\phi^{\rm ext}(\x)]w(t,\x) 
=A\fr 1{2\mu} 
[-i\h\na_\x]^2 w(t,\x)+\fr{i\h e}{\mu c} \bA^{\rm ext}(t,\x)\cdot\na_\x 
\psi_1(t,\x), 
\ee 
since $\psi_1(t,\x)$ is a solution to Equation (\re{SMes}) with 
$\bA^{\rm ext}=0$ and the Coulomb potential $\phi^{\rm ext}=-e/|\x|$ 
of the nucleus. By 
(\re{A1s})  and  (\re{ceig}), we have the following long-time asymptotics for 
 the RHS of (\re{SMes1}), 
\beqn\la{SMes2} 
\fr{i\h e}{\mu c} \bA^{\rm ext}(t,\x)\cdot\na_\x 
\psi_1(t,\x)&\sim& \fr{i\h e}{\mu c} A\sin k(x_3-ct)(1,0,0)\cdot\na_\x 
\psi_1(\x) e^{-i\om_1 t}\nonumber\\ 
&=& 
\fr{A\h e}{2\mu c}  [e^{i k(x_3-ct)}-e^{-i k(x_3-ct)}]e^{-i\om_1 t} 
\na_1\psi_1(\x)\nonumber\\ 
&=&\psi_+(\x)e^{-i(\om_1+\om)t}-\psi_-(\x)e^{-i(\om_1-\om)t}, 
\eeqn 
where $\om:=kc$. 
Now let us apply the {\it limiting amplitude principle} 
(\re{lapr}): 
\be\la{lapra} 
~~~~~~~w(t,\x)= w_+(\x)e^{-i(\om_1+\om)t}-w_-(\x)e^{-i(\om_1-\om)t} 
+\sum_l C_l\psi_l(\x)e^{-i\om_l t}+r(t,\x), 
\ee 
where $w_\pm(\x)$ are the {\it limiting amplitudes}, 
$\psi_l(\x)$ stand for the eigenfunctions of the discrete 
spectrum of the homogeneous Schr\"odinger equation 
$ 
\ds[i\h\pa_t-e\phi^{\rm ext}(\x)]w(t,\x) 
=\fr 1{2\mu} 
[-i\h\na_\x]^2 w(t,\x) 
$, which corresponds to (\re{SMes1}), and 
$r(t,\x)\to 0$, $t\to\infty$, in an appropriate norm. 
Below, we will omit the sum over the discrete 
spectrum in the RHS of (\re{lapra}). 
According to Remark \re{rNH}, 
\\ 
i) the {\it limiting amplitudes} 
 $w_\pm(\x)$ 
characterize the response of the atom 
 to the incident wave and depend on the frequency $\om$ and 
Amplitude $A$ of the incident wave, and 
do not depend on  initial data. 
\\ 
ii) The sum over the discrete 
spectrum on the RHS of (\re{lapra}) depends only on initial data 
and does not depend on the incident wave. 
\\ 
Hence, the contribution of the incident wave to the 
electric current and magnetization (see below), in principle, 
can be singled out in an experimental observation 
of the considered scattering problem. 
Finally, we identify 
\be\la{was} 
w(t,\x)\sim w_+(\x)e^{-i(\om_1+\om)t}-w_-(\x)e^{-i(\om_1-\om)t}, 
\,\,\,\,t\to+\infty . 
\ee 
For $w_\pm$ we get the equations (cf. (\re{Sls})) 
\be\la{wequ} 
\ds[\h(\om_1\pm \om)-e\phi^{\rm ext}(\x)]w_\pm(\x) 
-\fr 1{2\mu} 
[-i\h\na_\x]^2 w_\pm(\x)=\psi_\pm(\x):=\fr{\h e}{2\mu c}e^{\pm i kx_3}\na_1\psi_1(\x). 
\ee 
Now we introduce the following condition: 
\be\la{fb} 
\mbox{\bf spectral bound:}~~~~~ 
\,\,\,\, |\om|<|\om_1| . ~~~~~~~~~~~~~~~~~ 
\ee 
\br 
Let us note that for the case of the 
hydrogen atom we have $|\om_1|=\ds\fr{\mu e^4}{2\h^3} 
\approx 20,5\cdot 10^{15} {\rm rad}/{\rm sec}$. Hence, the bound 
(\re{fb}) holds for the frequencies 
$|\om|< 3,27\cdot 10^{15} 
{\rm Hz}$ or wave numbers $k<|\om_1|/c\approx 68\cdot 10^{7}$ m$^{-1}$ 
and wave lengths 
$\lam > 0.91176\cdot 10^{-5}$ cm $=911.76 \stackrel\circ A$. 
\er 
The frequency bound implies that $\om_1\pm \om<0$, hence 
the values $\h(\om_1\pm \om)$ do not belong to the 
continuous spectrum of the stationary Schr\"odinger equation (\re{wequ}). 
We will also assume that $\h(\om_1\pm \om)$ 
do not belong to the discrete spectrum. The last assumption 
is unessential since the coincidence 
$\h(\om_1\pm \om)$ with an eigenvalue $\h\om_n$ 
is an event of codimension one, i.e. probability zero. 
Therefore, the solutions decay at infinity in the following sense 
\be\la{wpmL} 
w_\pm\in L^2(\R^3) 
\ee 
(see Remark \re{rlia} ii)). 
For a moment let us neglect 
the external potential $\phi^{\rm ext}(\x)$ 
in the equation (\re{wequ}). 
Then it can be rewritten in the form 
\be\la{weqt} 
[\ds z +\na_\x^2] w(\x)=f(\x),\,\,\,\,\x\in\R^3, 
\ee 
where $z<0$ and 
$|f(\x)|\le Ce^{-\ve|\x|}$ with an $\ve>0$. 
\bexe 
Prove that $w_\pm(\x)$ decay 
exponentially at infinity, like $\psi_\pm(\x)$: 
\be\la{wpmd} 
|w_\pm(\x)|\le Ce^{-\ve_1|\x|},~~~~~~~~\x\in\R, 
\ee 
where $\ve_1>0$. 
{\bf Hint:} 
 Apply the Fourier transform to prove that 
i) the solution is unique in the class of tempered distributions 
and ii) the solution is a convolution, $w=f*E$, where $E(\x)$ is 
the fundamental solution $E(\x)=-e^{-\kappa|\x|}/(4\pi|\x|)$ with 
$\kappa:=\sqrt{-z}>0$. Then $\ve_1=\min(\ve, \kappa)>0$. 
 
\eexe 
%% ????????????? The proof of the decay 
%%(\re{wpmd}) for the nonzero Coulombic potential $\phi^{\rm ext}(\x)$ 
\br\la{rsrs} 
{\rm We have supposed that the atom is in its groundstate, $\psi_1$, 
in the remote past, $t\to-\infty$. 
The representation (\re{fa}), the asymptotics (\re{lapra}) and  (\re{wpmL}) 
imply that, roughly speaking, the light scattering modify 
the groundstate in the first order in $A$, 
in the long-time limit  $t\to\infty$. 
The condition (\re{wpmL}) means that the scattering process 
is essentially  concentrated in a bounded region of space. 
Hence, the process corresponds to a {\it short-range scattering}. 
}

\er

Further, let us 
calculate the modified groundstate using the spectral resolution 
of the Schr\"odinger operator in the equation 
(\re{wequ}). First,  let us expand the RHS, 
\be\la{ese} 
\fr{\h e}{2\mu c}e^{\pm i kx_3}\na_1\psi_1(\x)={\sum}_l a_l^\pm\psi_l(\x), 
\ee 
where ${\sum}_l$ stands for the sum over the discrete spectrum and the 
integral over the continuous spectrum. 
Then the solutions $w_\pm$ have the form 
\be\la{wpm} 
w_\pm(\x)={\sum}_l \fr{a_l^\pm\psi_l(\x)}{\h(\om_1\pm \om-\om_l)}. 
\ee 
Therefore, (\re{was}) becomes 
\be\la{wasb} 
w(t,\x)\sim 
{\sum}_l \fr{a_l^+\psi_l(\x)}{\h(\om_1+ \om-\om_l)} 
e^{-i(\om_1+\om)t}- 
{\sum}_l \fr{a_l^-\psi_l(\x)}{\h(\om_1- \om-\om_l)} 
e^{-i(\om_1-\om)t}. 
\ee 
Let us calculate the coefficients $a_l^\pm$. 
Formally, 
\be\la{esef} 
a_l^\pm=\fr{\h e}{2\mu c}\int\ov\psi_l(\x)e^{\pm i kx_3}\na_1\psi_1(\x) 
 d\x. 
\ee 
Let us assume that $k\ll 1$ 
and set $e^{\pm i kx_3}=1$. 
Then $a_l^\pm$ are approximately identical, 
and by partial integration, 
\beqn\la{esefp} 
a_l^\pm&\approx& 
\fr{\h e}{4\mu c}\int 
[\ov\psi_l(\x)\na_1\psi_1(\x)-\na_1\ov\psi_l(\x)\psi_1(\x)] 
 d\x\nonumber\\ 
&=& 
\fr{\h e}{4\mu c}e^{i(\om_1-\om_l)t}\int 
[\ov\psi_l(t,\x)\na_1\psi_1(t,\x)-\na_1\ov\psi_l(t,\x)\psi_1(t,\x)] 
 d\x\nonumber\\ 
&=& 
i\fr{1}{2c}e^{i(\om_1-\om_l)t}\int \bj_{1l}^1(t,\x)d\x. 
\eeqn 
Here $\psi_l(t,\x):=e^{-i\om_l t}\psi_l(\x)$ and 
$\bj_{1l}^1(t,\x)$ is the first component of the current 
(\re{ccd2}) corresponding to $\psi_1(t,\x)$ and $\psi_l(t,\x)$, 
since the functions are solutions to the Schr\"odinger 
equation in (\re{SMee}) with $A(t,x)+A^{\rm ext}(t,x)=0$. Then the identity 
(\re{cce2p}) implies, 
\be\la{ccc} 
a_l^\pm\approx \fr{\om_{1l}}{2c} \int \x^1e\psi_1(\x)\ov\psi_l(\x)d\x 
=\fr{e\om_{1l}}{2c} \x^1_{1l}=:a_l, 
\ee 
where $\om_{1l}:=\om_1-\om_l$ and $\x^1_{1l}:= 
\ds\int \x^1\psi_1(\x)\ov\psi_l(\x)d\x$. 
Finally, (\re{wasb}) becomes, 
\be\la{wasbf} 
w(t,\x)\sim 
{\sum}_l 
a_l 
\psi_l(\x) 
\Big( 
\fr{e^{-i\om t}}{\h(\om_{1l}+ \om)} 
- 
\fr{e^{i\om t}}{\h(\om_{1l}- \om)} 
\Big)e^{-i\om_1t}, 
\ee 
and the wave function (\re{fa}) reads 
\beqn\la{far} 
\psi(t,\x)&=& 
\Big[\psi_1(\x)+A{\sum}_l 
a_l 
\psi_l(\x) 
\Big( 
\fr{e^{-i\om t}}{\h(\om_{1l}+ \om)} 
- 
\fr{e^{i\om t}}{\h(\om_{1l}- \om)} 
\Big)\Big]e^{-i\om_1t}+\cO(A^2)\nonumber\\ 
\nonumber\\ 
&=& 
\Big[\psi_1(\x)+A\Si(t,\x)\Big]e^{-i\om_1t}+\cO(A^2). 
\eeqn 
 
\subsection{Polarization and Dispersion: Kramers-Kronig Formula} 
Let us calculate the corresponding electric dipole moment. 
First, the charge density is given by 
\beqn\la{fad} 
\rho(t,\x)&=&e\psi(t,\x)\ov\psi(t,\x)= 
e(\psi_1(\x)+A\Si(t,\x)) 
(\ov\psi_1(\x)+A\ov \Si(t,\x))+\cO(A^2) 
\nonumber\\ 
&=& 
e|\psi_1(t,\x)|^2+eA\Big[\Si^+ e^{i\om t}+\Si^- e^{-i\om t}  \Big] 
+\cO(A^2), 
\eeqn 
where 
\be\la{Si} 
\Si^+={\sum}_l \Big(\fr{\psi_1\ov a_l\ov\psi_l}{\h(\om_{1l}+ \om)}- 
\fr{a_l\psi_l\ov\psi_1}{\h(\om_{1l}- \om)} 
   \Big) 
,\,\,\,\, 
\Si^-=\ov\Si^+. 
\ee 
Therefore, the electric dipole moment equals, mod$\cO(A^2)$, 
\be\la{dm} 
\p(t):=\int \x\rho(t,\x)d\x=\p_{11}+\bP(t), 
\ee 
where $\p_{11}:=\ds\int \x e|\psi_1(t,\x)|^2d\x=0$ by spherical 
symmetry, and 
\be\la{dmP} 
~~~~~~~~~\bP(t)=eA\Big[{\sum}_l \Big(\fr{\ov a_l \x_{1l}}{\h(\om_{1l}- \om)} 
-\fr{a_l\ov\x_{1l}}{\h(\om_{1l}+ \om)}   \Big)e^{i\om t}+ 
{\sum}_l \Big(\fr{ a_l \ov\x_{1l}}{\h(\om_{1l}- \om)} 
-\fr{ \ov a_l\x_{1l}}{\h(\om_{1l}+ \om)}   \Big)e^{-i\om t}\Big], 
\ee 
where $\x_{1l}:=\ds\int \x\psi_1 \ov\psi_ld\x$. 
By symmetry arguments, the vector $\bP(t)$ is directed along $\e_1$. 
Indeed, the invariance of $\p(t)$ with respect to the reflection 
$\x_2\mapsto -\x_2$ is obvious. The invariance with respect to the reflection 
$\x_3\mapsto -\x_3$ follows from 
 (\re{ese})--(\re{wasb}) since we set $k=0$. 
Therefore, 
substituting 
$a_l=e\om_{1l}\x^1_{1l}/(2c)$ and projecting $\x_{1l}$ 
onto $\e_1$, we get 
\beqn\la{dmg} 
\p(t)=A\e_1 
\fr {4 k e^2}{\h  }{\sum}_l 
\fr{\om_{1l}|\x_{1l}^1|^2}{\om_{1l}^2- \om^2}\cos \om t 
\eeqn 
At last, let us average this expression with respect to all 
orientations of the atom. Then we obtain 
\be\la{dmo} 
\ov\p(t):=A\e_1 
\fr {4ke^2}{\h }{\sum}_l 
\fr{\om_{1l}\ov{|\x_{1l}^1|^2}}{\om_{1l}^2- \om^2}\cos \om t, 
\ee 
since the average of $\p_{11}$ is obviously zero. 
Let us calculate the permittivity (\re{coee})  of the atomic 
hydrogen in the ground state $\psi_1$. We denote by $\bE(t)$ 
the electric field at the position $\x=0$ of the atom: 
by (\re{BE}), we have 
$\bE(t)=\ds kA\cos \om t \, \e_1$. 
\bc 
The permittivity of the atomic 
hydrogen in the ground state is given by the Kramers-Kronig formula 
\be\la{pg} 
\chi_e=n|\ov\p(t)|/|\bE(t)|= 
n\fr {4e^2}{\h }{\sum}_l 
\fr{\om_{1l}\ov{|\x_{1l}^1|^2}}{\om_{1l}^2- \om^2}, 
\ee 
where $n$ is the number of atoms per unit volume. 
\ec 
This formula 
has the same analytic structure as its classical analog (\ref{ue-perm}) 
of Exercise \ref{ue-Ex-10}. 
This allows us to express the hydrogen 
electric susceptibility 
$\ve=1+4\pi\chi$ (see (\re{coeeo})) 
and hence the refraction coefficient 
$n\sim \sqrt{\ve}$ 
%%%%??????????????????
in the case of the magnetic  susceptibility $\mu\sim 1$
(see (\re{nemu})). 
 
\subsection{Combinational Scattering} 
Formula (\re{far}) gives the first order correction to the 
unperturbed ground state 
$\ds\psi_1(\x)e^{-i\om_1 t}$ 
in the presence of the incident wave (\re{BE}). Similarly, for any 
unperturbed stationary state $\ds\psi_j(\x)e^{-i\om_j t}$, the 
corresponding 
 first order correction is obtained by changing the index from $1$ to $j$. 
Therefore, the first order correction to a 
general unperturbed solution of the type (\re{eex}) 
reads 
\be\la{farg} 
\psi(t,\x)={\sum}_j c_j(T) 
\Big[\psi_j(\x)+A{\sum}_l 
a_l 
\psi_l(\x) 
\Big( 
\fr{e^{-i\om t}}{\h(\om_{jl}+ \om)} 
- 
\fr{e^{i\om t}}{\h(\om_{jl}- \om)} 
\Big)\Big]e^{-i\om_jt}. 
\ee 
Hence, we can apply the methods of Section \ref{Sec-11} to 
calculate the spectrum of the dipole radiation of the hydrogen 
atom in the presence of light. 
Now we get 
a new set of frequencies: in the first order approximation 
in $A$, 
the spectrum is contained in the set 
$\{\om_{jj'}, \om_{jj'}\pm \om\}$. The corresponding selection rules 
also can be obtained by the methods of  Section \ref{Sec-11}. 
 
%%%%%%%%%%%%%%%%%%%%%%%%%%%%%%%%%%%%%%%%%%%%%%%%%%%%%%%%%%%%%%%%%%%%%%%% 
%%%%%%%%%%%%%%%%%%%%%%%%%%%%%%%%%%%%%%%%%%%%%%%%%%%%%%%%%%%%%%%%%%%%%%%%% 
%%%%%%%%%%%%%%%%%%%%%%%%%%%%%%%%%%%%%%%%%%%%%%%%%%%%%%%%%%%% 
%%%%%%%%%%%%%%%%%%%%%%%%%%%%%%%%%%%%%%%%%%%%%%%%%%%%%%% 

\newpage 
%%%%%%%%%%%%%%%%%%%%%%%%%%%%%%%%%%%%%%%%%%%%%%%%%%%%%%%%%%%%%%%%%%%%% 
%%%%%%%%%%%%%%%%%%%%%%%%%%%%%%%%%%%%%%%%%%%%%%%%%%%%%%%%%%%%%% 
%%\setcounter{section}{+9} 
\setcounter{subsection}{0} 
\setcounter{theorem}{0} 
\setcounter{equation}{0} 
\section{Light Scattering in Continuous Spectrum: Photoeffect} 
For light with a frequency which is larger than the binding energy of 
the stationary electron state, scattering of the electron into the 
continuous spectrum may occur. Within the coupled Maxwell-Schr\"odinger 
system this scattering is described by a weakly decaying contribution to the 
electron wave function, which corresponds to a non-zero electron 
current radiating to infinity. 
 
\subsection{Radiation in Continuous Spectrum} 
Let us consider the scattering of light with frequency 
\be\la{fbn} 
\om>|\om_1|. 
\ee 
Then $\om_1-\om<0$ and $\om_1+\om=\om-|\om_1|>0$. 
This means that $\h(\om_1+\om)$ belongs to the continuous spectrum 
of the  stationary Schr\"odinger equation (\re{wequ}) 
with a slow decay of the wave function $|w_+(\x)|\sim 1/|\x|$ 
at $|\x|\to\infty$. We will calculate the long-range asymptotics 
of  $w_+(\x)$ and obtain  the 
main term of the radiation in the form 
\be\la{mte} 
Aw_+(\x)e^{-i(\om_1+\om)t}\sim 
A 
\fr{a(\vp,\theta)}{|\x|}e^{i[k_r|\x|-(\om_1+\om)t]},\,\,\,\,|\x|\to\infty. 
\ee 
On the other hand, $\h(\om_1-\om)$ does not belong to the 
continuous spectrum. Let us also 
 assume that $\h(\om_1- \om)$ 
does not belong to the discrete spectrum, hence 
\be\la{wmL} 
w_-\in L^2 
\ee 
similarly to (\re{wpmL}). 
We will deduce from the asymptotics (\re{mte}) 
the following stationary 
electric current at infinity, 
\be\la{elc} 
\bj(t,\x)\sim A^2\fr{e\h k_r}{\mu} 
\fr{a^2(\vp,\theta)}{|\x|^2}\n(\x),\,\,\,\,|\x|\to\infty, 
\ee 
where $\n(\x):=\x/|\x|$. 
This describes the radiation of a non-zero electric current from the 
atom to infinity. 
\br 
i) The Schr\"odinger theory explains the ``red  bound'' 
of the photoeffect, $\om>\om_{\rm red}$, and provides the value 
$\om_{\rm red}=|\om_1|$. 
\\ 
ii) Similarly to (\re{pdvH}), 
the radiated wave function (\re{mte}) can be identified with a 
beam of electrons with the density 
\be\la{dens} 
d(\x)=\rho(t,\x)/e=|Aw_+(\x)|^2= 
A^2\fr{a^2(\vp,\theta)}{|\x|^2} 
\ee 
and the velocities 
\be\la{vel} 
\bv(\x)=\bj(t,\x)/(ed(\x))=\fr{\h k_r}{\mu}\n(\x). 
\ee 
\er 
\subsection{Limiting Amplitude} 
 
Let us calculate the limiting amplitude $w_+(\x)$. 
First, we rewrite Equation (\re{wequ}) 
in the form 
\be\la{weqf} 
[\na_\x^2 +k_r^2(\om)]w_+(\x)= 
\fr{e}{\h c}e^{i kx_3}\na_1\psi_1(\x)-\fr{2e^2\mu}{\h^2|\x|}w_+(\x), 
\ee 
where $k_r(\om)=\sqrt{2\mu(\om_1+\om)/\h}>0$. 
For a moment, 
let us  neglect the last term on the  RHS. 
So, we get the Helmholtz equation 
\be\la{weqg} 
[\na_\x^2 +k_r^2(\om)]w_+(\x)=f_+(\x):= 
\fr{ e}{\h c}e^{ i kx_3}\na_1\psi_1(\x). 
\ee 
The limiting amplitude is given by the convolution (cf. (\re{fus})) 
\be\la{cfs} 
w_+(\x)=-\int \fr{e^{ik_r(\om)|\x-\y|}}{4\pi|\x-\y|}f_+(\y)d\y. 
\ee 
This follows from the 
limiting absorption principle (\re{LAbP}) 
since 
the fundamental solution $\ds\fr{e^{ik_r(\om+i\ve)|\x-\y|}}{4\pi|\x-\y|}$ 
is a tempered distribution for small $\ve>0$. Indeed, 
  $\rIm k_r(\om+i\ve)>0$ for the fixed branch 
$k_r(\om)>0$.

Now we can calculate the asymptotics (\re{mte}). 
Let us substitute the expression 
(\re{weqg}) for $f_+$ into (\re{cfs}). Then, after partial integration, 
we get 
\beqn\la{cfp} 
w_+(\x)&=&\fr{e}{\h c}\int 
\na_1\fr{e^{ik_r|\x-\y|}}{4\pi|\x-\y|}e^{ik\y_3}\psi_1(\y)d\y 
\nonumber\\ 
\nonumber\\ 
&=&-\fr{ik_r e}{\h c}\int 
\fr{e^{ik_r|\x-\y|}(\x_1-\y_1)}{4\pi|\x-\y|^2}e^{ik\y_3}\psi_1(\y)d\y 
+\cO(|\x-\y|^{-2}). 
\eeqn 
Substituting here $\x_1-\y_1=\sin\theta\cos\vp|\x-\y|$, we get 
(\re{mte}) with the amplitude 
\be\la{amp} 
a(\vp, \theta)=C\sin\theta\cos\vp , 
\ee 
since the ground state $\psi_1(\y)$ decays rapidly at infinity.

\subsection{Long-Range Scattering: Wentzel Formula} 
Let us deduce (\re{elc}) from (\re{mte}).
In the Born approximation the current is given by (\re{SMeeaa}), 
\be\la{SMec} 
\bj 
:=-\ds\fr e \mu[i\h\na\psi(t,\x )] 
\cdot\psi(t,\x )-\ds\fr {e^2} {\mu c} \bA^{\rm ext}(t,\x)|\psi(t,\x )|^2 . 
\ee 
The function $w_-$ 
decays at infinity by (\re{wmL}). 
Therefore, (\re{fa}) becomes 
\be\la{fab} 
\psi(t,\x)\sim\psi_1(t,\x)+Aw_+(\x)e^{-i(\om_1+\om)t}. 
\ee 
The contribution of the function $\psi_1$ to the current 
at infinity 
also is negligible due to the exponential decay at infinity. 
Therefore, (\re{elc}) follows by substituting the asymptotics 
(\re{mte}) into the first term on the RHS. 

Finally, (\re{amp}) implies 
the following 
{\it stationary 
electric current} at infinity, 
\be\la{elcn} 
\bj(t,\x)\sim 
\fr{\sin^2\theta\cos^2\vp}{|\x|^2}\n(\x),\,\,\,\,|\x|\to\infty 
\ee 
which has been discovered first by G.Wentzel in 1927 (see \ci{We}).

\br 
{\rm The slow decay  (\re{mte}), (\re{elcn}) implies that the {\it total 
electric current to infinity} does not vanish, i.e. 
\be\la{teci} 
\lim_{R\to\infty}\int_{|\x|=R}\bj(t,\x)dS(\x)=J\ne 0. 
\ee 
This corresponds to the {\it long-range scattering} 
under the condition (\re{fbn}). If the condition fails, the 
`fast decay' (\re{wpmd}) implies that in this case 
the total 
electric current to infinity vanishies that 
corresponds to the {\it short-range scattering} 
under the condition (\re{fb}) (see Remark \re{rsrs}). } 
\er 
 
\bcom\la{pstat}
{\rm
The stationary nonvanishing current  (\re{teci})
formally contardicts the charge conservation for the atom.
The contradiction is provided by the perturbation strategy
 which leaves  the current (\re{cur}) unchanged.
To maintain the stationary photocurrent, one needs an external 
source (galvanic element, etc) to reimburse the charge
decay. 
}
\ecom

\subsection{Coulomb Potential} 
Now let us discuss the general equation (\re{weqf}) instead of (\re{weqg}). 
Then (\re{cfs}) changes to 
\be\la{cfst} 
w_+(\x)=\int G_{k_r}(\x,\y)f_+(\y)d\y, 
\ee 
where $G_{k_r}$ is the corresponding Green function. 
For each fixed $\y\in\R^3$ the following asymptotics holds (cf. 
\ci[Vol. II, formula (II.7.33)]{Som}), 
\be\la{cfsa} 
G_{k_r}(\x,\y)\sim C_1e^{i\ga\log |\x-\y|}\fr{e^{ik_r|\x-\y|}}{4\pi|\x-\y|}, 
\ee 
where $\ga=\ga(k_r)\in\R$. Therefore, the asymptotics (\re{mte}), (\re{elc}) 
follow by the same arguments as above, with the amplitude of the form 
(\re{amp}). 
\bexe 
Deduce (\re{mte}), (\re{elc}), (\re{amp}) from (\re{cfst}) and (\re{cfsa}). 
\eexe 

\subsection{Shift in Angular Distribution: Sommerfeld Formula}
Next correction to Wentzel formula has been obtained by Sommerfeld
and Shur \ci{SS}. Namely, they have taken into account the pressure of the 
incident light upon the ougoing photocurrent. The corresponding 
corrected formula reads (see \ci[Vol. II]{Som})
%%%%????????????????????
\be\la{elcnSS} 
\bj(t,\x)\sim 
\fr{\sin^2\theta\cos^2\vp}{|\x|^2}
\n(\x),\,\,\,\,|\x|\to\infty. 
\ee 
\bcom
The Sommerfeld-Shur calculation
continues the perturbation procedure for the coupled 
Maxwell-Schr\"odinger
equations.
It would be interesting to prove that the 
photoeffect is an inherent feature of the 
coupled Maxwell-Schr\"odinger equations and holds under 
the same spectral condition  (\re{fbn}). 
\ecom

\subsection{Photoeffect for Excited States} 
The angular dependence (\re{amp}) 
of the limiting amplitude is characteristic 
for the light scattering by spherically-symmetric ground state (\re{ceig}). 
The photoeffect is observed also in the light scattering 
by  excited stationary states of atoms which are not 
spherically-symmetric. 
In this case, the formula of type (\re{cfst}) 
for the limiting amplitude 
also holds and admits a  long-range asymptotics 
of type (\re{mte}). 
However, its angular distribution is different from  (\re{amp}): 
see \ci{Som}, where the distribution is obtained for 
the excited states from $K$-, $L$-, and $M$-shells of many-electron atoms.

%%%%%%%%%%%%%%%%%%%%%%%%%%%%%%%%%%%%%%%%%%%%%%%%%%%%%%%%%%%%%%%%%%%%%%%% 
%%%%%%%%%%%%%%%%%%%%%%%%%%%%%%%%%%%%%%%%%%%%%%%%%%%%%%%%%%%%%%%%%%%%%%%%% 
%%%%%%%%%%%%%%%%%%%%%%%%%%%%%%%%%%%%%%%%%%%%%%%%%%%%%%%%%%%% 
%%%%%%%%%%%%%%%%%%%%%%%%%%%%%%%%%%%%%%%%%%%%%%%%%%%%%%% 

\newpage 
%%%%%%%%%%%%%%%%%%%%%%%%%%%%%%%%%%%%%%%%%%%%%%%%%%%%%%%%%%%%%%%%%%%%% 
%%%%%%%%%%%%%%%%%%%%%%%%%%%%%%%%%%%%%%%%%%%%%%%%%%%%%%%%%%%%%% 
%%\setcounter{section}{+9} 
\setcounter{subsection}{0} 
\setcounter{theorem}{0} 
\setcounter{equation}{0} 
\section{Scattering of Particles: Rutherford Formula} 
Here, the scattering of charged particles at a heavy, charged object 
(``nucleus'') is considered. First, the classical case is considered, 
when the scattering particles are described by the trajectories of 
classical mechanics. Then, the quantum mechanical 
scattering of an electron with large 
momentum by a hydrogen atom is studied. 
 
\subsection{Classical Scattering by a Nucleus} 
 
Let us consider scattering of charged particles with charge 
$Q$ and mass $M$ by a nucleus 
with charge $|e|Z>0$. 
We will derive the Rutherford formula in both the repulsive case, 
when $Q>0$, and the attractive case, when $Q<0$. 
The repulsive case corresponds, for example, to the scattering of 
$\al$-particles with $Q=2|e|$. The attractive 
case corresponds, for example, to the scattering of 
electrons with $Q=e<0$. 
Let the heavy nucleus be situated at the origin. The 
incident particle moves along the 
trajectory $\x=\x(t)$, $-\infty <t<\infty$. 
We assume that the particles come from infinity, i.e. 
\be\la{inf} 
\lim_{t\to-\infty}|\x(t)|=\infty. 
\ee 
Then the trajectory is a hyperbola as described, e.g., in Exercise 
\ref{ue-Ex-4}. 
Let us choose the coordinates $\x_k$ in space in such a way 
that $\x_3(t)\equiv 0$ and 
\be\la{isc} 
\lim_{t\to-\infty}\dot \x(t)=(v,0,0),\,\,\, 
\lim_{t\to-\infty}\x_1(t)=-\infty,\,\,\,\,\,\lim_{t\to-\infty}\x_2(t)=b, 
\ee 
where $b$ is the {\it impact} parameter. 
Let us further choose standard polar coordinates in the plane $\x_3=0$, 
\be 
\x_1=r\cos\theta,\,\,\,\,\,\x_2=r\sin\theta 
\ee 
and denote by $r(t), \theta(t)$ the trajectory of the particle 
in these coordinates. 
Then the initial scattering  conditions (\re{isc}) 
imply 
\be\la{iscg} 
\lim_{t\to-\infty}\theta(t)=\pi,\,\,\,\,\,\, 
\lim_{t\to-\infty}r(t)\sin\theta(t)=b. 
\ee 
\subsubsection{Angle of scattering} 
Let us calculate the final scattering angle 
\be\la{asc} 
\ov\theta:=\lim_{t\to-\infty}\theta(t). 
\ee 
\bl 
The final scattering angle satisfies the equation 
\be\la{ien} 
\cot\fr{\ov\theta}2=\fr{Mbv^2}{Q|e|Z}. 
\ee 
\el 
\Pr 
First, let us write the angular momentum and energy conservation: 
\be\la{enec} 
r^2(t)\dot\theta(t)=bv,\,\,\,\,\,\,\,\,\,\,\,\,\,\,\,\,~~~~~~~~ 
\fr M2(\dot r^2(t)+ r^2(t)\dot\theta^2(t))+\fr{Q|e|Z}{r(t)}=\fr M2 v^2. 
\ee 
Let us substitute here 
\be\la{suh} 
\dot r(t):=\fr{dr}{dt}=\fr{dr}{d\theta}\fr{d\theta}{dt}=r'\dot\theta. 
\ee 
Then the energy conservation becomes, 
\be\la{ecb} 
\fr M2\dot\theta^2(t)(|r'(t)|^2+ r^2(t))+\fr{Q|e|Z}{r(t)}=\fr M2 v^2. 
\ee 
Now let us make the {\bf Clerot substitution} 
$r=1/\rho$. Then $r'=-\rho'/\rho^2$ and the momentum conservation 
gives 
$\dot\theta(t)=vb\rho^2$. Therefore, (\re{ecb}) reads, 
\be\la{enr} 
\fr M2b^2v^2 
(|\rho'|^2+ \rho^2)+{Q|e|Z}\rho=\fr M2 v^2. 
\ee 
Let us differentiate this expression in $\theta$. 
Then after division by $\rho'$, we get 
the {\bf Clerot equation} 
\be\la{enrg} 
\rho''+ \rho=C:=-\fr{Q|e|Z}{ Mb v^2}. 
\ee 
The general solution to this equation is 
\be\la{gse} 
\rho(\theta)=A\cos\theta+B\sin\theta+C. 
\ee 
The initial scattering conditions (\re{isc}) give, 
\be\la{isg} 
\lim_{\theta\to\pi}\rho(\theta)=0, 
\,\,\,\,\,\lim_{\theta\to\pi}\fr{\rho(\theta)}{\sin(\theta)}=\fr 1b. 
\ee 
Substituting here (\re{gse}), we get $-A+C=0$ and $B=1/b$, hence 
\be\la{gseh} 
\rho(\theta)=C(1+\cos\theta)+\fr 1b\sin\theta. 
\ee 
At last, for the final scattering angle we get from (\re{asc}) 
that $\rho(\ov\theta)=0$. Hence, 
\be\la{asch} 
C(1+\cos\ov\theta)+\fr 1b\sin\ov\theta=0. 
\ee 
This implies (\re{ien}).\hfill$\bo$ 
\br 
The solution $\ov\theta\in (0,\pi)\cup(\pi,2\pi)$ to Equation (\re{ien}) 
exists and is unique.  The repulsive and attractive 
cases correspond to $Q>0$, 
$\ov\theta\in (0,\pi)$ and  $Q<0$, 
$\ov\theta\in (\pi,2\pi)$, respectively. 
\er 
 
\subsubsection{Differential cross section} 
Now let us assume that the incident particles constitute 
 a beam with a 
flux density of $n$ particles per cm$^2$sec. Let us denote by $N=N(b,b+db)$ 
the number 
of incident  particles per sec with an impact parameter within the 
interval  $[b,b+db]$. 
By axial symmetry, we get for infinitesimal $db$, 
\be 
N(b,b+db)=n2\pi  bdb 
\ee 
The particles are scattered into the spatial angle 
$d\Om=2\pi \sin\ov\theta d\ov\theta $. 
\bd\la{dcsp} 
The differential cross section of the scattering is defined by 
\be\la{dD} 
D(\ov\theta):=\fr{N/d\Om}{n}=\fr{  bdb }{\sin\ov\theta d\ov\theta}. 
\ee 
\ed 
Let us calculate the  cross section. Rewriting (\re{ien}) in the form 
\be\la{ienr} 
b^2= 
\Big(\fr{Q|e|Z}{Mv^2}\Big)^2 
\cot^2\fr{\ov\theta}2 
\ee 
and differentiating, we get 
\be\la{iend} 
2bdb= 
\Big(\fr{Q|e|Z}{Mv^2}\Big)^2 
2\cot\fr{\ov\theta}2  \,\,\ds\fr 1{\sin^2\fr{\ds\ov\theta}2}\fr{d\ov\theta}2. 
\ee 
Substituting this into  (\re{dD}), we get the {\it Rutherford formula} 
\be\la{Rf} 
D(\ov\theta)=\fr{\ds\Big(\fr{Q|e|Z}{Mv^2}\Big)^2}{4\sin^4\fr{\ds\ov\theta}2} 
\, . 
\ee

\subsection{Quantum Scattering of Electrons by the Hydrogen Atom} 
We 
consider the scattering of the electron beam by a hydrogen atom 
in its ground state $\psi_1(t,\x)$, (\re{ceig}). 
The incident electron beam 
is described by the plane wave (\re{plwH}): 
\be\la{plw} 
\psi_{\rm in}(t,\x)=Ce^{i(\bk\x-\om t)},\,\,\,\,\,\bk\ne 0, 
\ee 
which 
is a solution to the {\it free} Schr\"odinger equation, i.e., without 
an external Maxwell field. Hence, (see (\re{plhH})) 
\be\la{plh} 
\h\om=\fr{\h^2}{2\mu}\bk^2>0. 
\ee 
Corresponding electric current  density is given by (\re{velnH}): 
\be\la{veln} 
\bj_{\rm in}(t,\x):= \fr e\mu[-i\h\na\psi_{\rm in}(t,\x)]\cdot\psi_{\rm in} 
(t,\x)=\fr{e\h \bk}\mu |C|^2. 
\ee 
We will assume that $|C|\ll C_1$ (see (\re{ceig})) and 
consider  the incident plane wave as a small perturbation. 
Hence, the total wave field, approximately, is a solution 
to the Schr\"odinger equation of the type (\re{SMe}), 
\be\la{SMet} 
\ds[i\h\pa_t-e\phi(\x)]\psi(t,\x) 
=\fr 1{2\mu} 
[-i\h\na_\x-\ds\fr ec  \bA(\x)]^2\psi(t,\x) , 
\ee 
where $\phi(\x)$, $A(\x)$ 
are the potentials of the total static Maxwell field corresponding to 
the ground state $\psi_1$. In particular (cf. (\re{SMeas})), 
\be\la{tmf} 
-\ds\fr 1{4\pi}\,\De \phi(\x)=\rho(\x)=e|\psi_1(t,\x)|^2 
+|e|\de(\x), 
\ee 
where $|e|\de(\x)$ is the charge density of the nucleus. 
The potential decays at infinity like $|x|^{-2}$, since 
$\ds\int\rho(\x)d\x=0$. For simplicity of calculations, 
we assume first that 
\be\la{tmfa} 
\phi(\x)=e\fr {e^{-\ve |\x}}{|\x|}, 
\ee 
where $\ve>0$ is small. At the end we will perform the limit 
$\ve\to 0$. 
 
\subsubsection{Radiated wave} 
Let us split the total wave field into three terms, 
\be\la{wf3} 
\psi(t,\x)=\psi_{\rm in}(t,\x)+\psi_1(t,\x)+\psi_{\rm r}(t,\x), 
\ee 
where $\psi_{\rm r}(t,\x)$ is a small radiated wave. 
Substituting (\re{wf3}) into the 
Schr\"odinger equation (\re{SMe}), we get 
\be\la{SMeg} 
\ds[i\h\pa_t-e\phi(\x)][\psi_{\rm in}(t,\x)+\psi_{\rm r}(t,\x)] 
=\fr 1{2\mu} 
[-i\h\na_\x-\ds\fr ec  \bA(\x)]^2[\psi_{\rm in}(t,\x)+\psi_{\rm r}(t,\x)] , 
\ee 
since $\psi_1(t,\x)$ is an exact solution. 
Neglecting the "relativistic'' corrections due to the small 
term $\ds\fr ec  \bA(\x)$, we rewrite the equation as 
\be\la{SMr} 
\ds\Big(i\h\pa_t-e\phi(\x) 
-\fr 1{2\mu} 
[-i\h\na_\x]^2\Big)\psi_{\rm r}(t,\x) 
= 
e\phi(\x)\psi_{\rm in}(t,\x)=e\phi(\x)Ce^{i(\bk\x-\om t)}. 
\ee 
Since $\om>0$ by (\re{plh}), 
the frequency $\om$ belongs to the continuous spectrum of the 
Schr\"odinger operator. 
Now let us apply the limiting amplitude principle (\re{lapr}) 
to derive the long-time asymptotics of the solution 
$\psi_{\rm r}(t,\x)$. 
Up to a contribution of the discrete spectrum 
(see Remark \re{rNH}), 
the solution admits the asymptotics 
\be\la{asy} 
\psi_{\rm r}(t,\x)\sim \psi_{\rm r}(\x)e^{-i\om t},\,\,\,\,\,\, 
t\to\infty. 
\ee 
Substituting 
this asymptotics into Equation (\re{SMr}), we get 
a stationary equation for the limiting amplitude 
\be\la{SMra} 
\ds\Big(\h\om-e\phi(\x) 
-\fr 1{2\mu} 
[-i\h\na_\x]^2\Big)\psi_{\rm r}(\x) 
= 
eC\phi(\x)e^{i\bk\x}. 
\ee 
Neglecting the term with the spatial decay at the LHS, we finally get 
the Helmholtz equation 
\be\la{SMrf} 
\ds\Big(\bk^2 
+ 
\na_\x^2\Big)\psi_{\rm r}(\x) 
= 
\fr{2\mu eC}{\h^2}\phi(\x)e^{i\bk\x} , 
\ee 
since $\bk^2=\ds\fr{2\mu\om}\h>0$ by (\re{plh}). 
This is an equation of the type (\re{weqg}). 
Therefore, the solution is given by a convolution similar to 
(\re{cfs}), 
\be\la{cfss} 
\psi_{\rm r}(\x)=-\fr{2\mu eC}{\h^2} 
\int \fr{e^{ik|\x-\y|}}{4\pi|\x-\y|}\phi(\y)e^{i\bk\y} 
d\y,\,\,\,\,\,\,~~~~~k:=|\bk|. 
\ee 
The convolution (\re{cfss}) is almost identical to the last integral (\re{rpo}) 
if we identify $\bk=(0,0,k)$. 
Evaluating this by the method  (\re{mod})-(\re{inb}), we get 
\be\la{cfsg} 
\psi_{\rm r}(\x) 
\sim 
-C 
\fr{e^{ik|\x|}}{|\x|} 
f(k,\theta),\,\,\,\,\,|\x|\to\infty, 
\ee 
where 
\be\la{inbw} 
f(k,\theta)=\fr{2\mu e}{\h^2} 
\int_0^\infty \fr{\sin K|\y|}{K|\y|}\phi(\y )|\y|^2d|\y|, 
\,\,\,\,\,\,K:=2k\sin\fr \theta 2. 
\ee 
Now 
(\re{asy}) becomes for large $|\x|$ 
\be\la{asyb} 
\psi_{\rm r}(t,\x) 
\sim 
-C 
\fr{e^{ik|\x|}}{|\x|} 
f(k,\theta)e^{-i\om t},\,\,\,\,\,t\to\infty. 
\ee 
The limiting amplitude $\psi_{\rm r}$ has a 
slow decay at infinity 
and infinite energy. 
This corresponds to the fact that the frequency $\om>0$ 
belongs to the continuous spectrum (see Remark \re{rlia} ii)). 
Physically this 
describes the radiation of electrons to infinity as we will see below.

\subsubsection{Differential cross section} 
Let us calculate the electric current corresponding to the radiated wave 
$\psi_{\rm r}(t,\x)$. As in (\re{veln}), we have 
for large $|\x|$, 
\be\la{velh} 
\bj_{\rm r}(t,\x):= \fr e\mu[-i\h\na\psi_{\rm r}(t,\x)] 
\cdot\psi_{\rm r}(t,\x)\sim\fr{e\h k\n(\x)}\mu \fr{ 
|f(k,\theta)|^2} 
{|\x|^2}|C|^2,\,\,\,\,\,\, 
\n(\x):=\fr \x{|\x|}. 
\ee 
Let us note that the current at infinity is radial. 
\bd\la{dcss} 
The differential cross section of the scattering is defined by 
\be\la{dDb} 
D(\n):=\lim\limits_{\x/|\x|=\n, |\x|\to\infty}\fr{|\bj_{\rm r}(t,\x)|} 
{|\bj_{\rm in}(t,\x)|}|\x|^2,\,\,\,\,\,\,\n\in\R^3. 
\ee 
\ed 
\br 
The definition obviously agrees with Definition \re{dcsp}. 
\er 
Now (\re{velh}) and (\re{veln}) imply that 
\be\la{dDi} 
D(\n)=|f(k,\theta)|^2. 
\ee 
Finally, substituting (\re{tmfa}) into (\re{inbw}), we get 
\be\la{inbt} 
f(k,\theta)=\fr{2\mu e^2}{K\h^2} 
\int_0^\infty \sin K|\y| e^{-\ve |\y|}d|\y|= 
\fr{2\mu e^2}{\h^2(K^2+\ve^2)} . 
\ee 
Here $K=\ds 2k\sin\fr \theta 2\gg \ve^2$ for $\theta\ne 0$ and 
large values of $k> k(\theta)$. Then we can drop $\ve^2$ 
in the limit $\ve\to 0$ and obtain 
\be\la{inbo} 
f(k,\theta)= 
\fr{2\mu e^2}{\h^2K^2}= 
\fr{\mu e^2}{2\h^2k^2\sin^2\ds\fr \theta 2}. 
\ee 
We rewrite this expression using (\re{pdvH}): 
\be\la{inbu} 
f(k,\theta)= 
\fr{e^2}{2\mu \bv^2\sin^2\ds\fr \theta 2} . 
\ee 
Now (\re{dDi}) implies the formula 
\be\la{dDii} 
D(\n)=\fr{\Big(\ds\fr{e^2}{\mu \bv^2}\Big)^2} 
{4\sin^4\ds\fr \theta 2}, 
\ee 
which coincides with the classical Rutherford formula (\re{Rf}) 
with $Q=e$ and $Z=1$. 
\br 
Born considered the agreement of (\re{dDii}) with the 
classical formula (\re{Rf}) as the crucial confirmation of the 
{\it probabilistic interpretation} 
of the wave function: $|\psi(t,\x)|$ is the {\bf density of the probability 
of the particle registration}, and the expression (\re{velh}) 
 is $e$ times the {\bf flux of the probability}, \ci{Born}. 
 
\er

%%%%%%%%%%%%%%%%%%%%%%%%%%%%%%%%%%%%%%%%%%%%%%%%%%%%%%%%%%%%%%%%%%%%% 
%%%%%%%%%%%%%%%%%%%%%%%%%%%%%%%%%%%%%%%%%%%%%%%%%%%%%%%%%%%%%% 
\newpage 
%%%%%%%%%%%%%%%%%%%%%%%%%%%%%%%%%%%%%%%%%%%%%%%%%%%%%%%%%%%%%%%%%%%%% 
%%%%%%%%%%%%%%%%%%%%%%%%%%%%%%%%%%%%%%%%%%%%%%%%%%%%%%%%%%%%%% 
%%\setcounter{section}{+9} 
\setcounter{subsection}{0} 
\setcounter{theorem}{0} 
\setcounter{equation}{0} 
\section{Hydrogen Atom in a Magnetic Field. Normal Zeemann Effect} 
We derive quantum stationary states and the corresponding 
energies for a hydrogen atom in a uniform magnetic 
field. 
We also analyze the normal Zeemann effect.

Let us consider a hydrogen atom 
in an external  static magnetic field $\bB(\x)$ with vector 
potential 
$A(\x)$. Then the 
Lorentz gauge condition (\re{Lg}) is equivalent to the Coulomb 
gauge condition 
\be\la{Cg} 
\na_\x  \bA(\x)=0. 
\ee 
The Schr\"odinger equation 
(\re{SMs}) becomes 
\be\la{BSb} 
i\h\pa_t\psi(t,\x) 
=\fr 1{2\mu} (-i\h\na_\x-\ds\fr ec  \bA(\x))^2\psi(t,\x) 
- 
\ds\fr {e^2}{|\x|}\psi(t,\x),~~~~~~(t,\x)\in\R^4. 
\ee 
Evaluating, we get by (\re{Cg}) 
\beqn\la{BSbe} 
i\h\pa_t\psi(t,\x)\!\!&=& 
-\fr 1{2\mu}\h^2\De\psi(t,\x) 
 + 
i\fr {\h e}{\mu c} 
\bA(\x)\cdot\na_\x \psi(t,\x) 
\nonumber\\ 
\nonumber\\ 
&&-\fr 1{2\mu}\h^2\bA^2(\x)\psi(t,\x) 
- 
\ds\fr {e^2}{|\x|}\psi(t,\x), 
~~~~~~\x\in\R^3. 
\eeqn 
Let us assume that the potential $\bA(\x)$ is small: 
\be\la{Bs} 
|\bA(\x)|\ll 1. 
\ee 
Then we can neglect the term with $\bA^2(\x)$ in 
(\re{BSMssbb}) and get the  equation 
\beqn\la{BSbeg} 
i\h\pa_t\psi(t,\x)\!\!&=& 
-\fr 1{2\mu}\h^2\De\psi(t,\x) 
 + 
i\fr {\h e}{\mu c} 
 \bA(\x)\cdot\na_\x \psi(t,\x) 
- 
\ds\fr {e^2}{|\x|}\psi(t,\x), 
~~~~~~\x\in\R^3. 
\eeqn 
 
\subsection{Uniform Magnetic Field} 
Now let us consider a hydrogen atom 
in an external  uniform static magnetic field $\bB$ with vector 
potential 
$ \bA(\x)=\bB\times \x/2$. 
Let us choose the coordinates in such a way that $\bB=(0,0,B)$ 
with $B\ge 0$. 
Then the vector 
potential is given by $ \bA(\x)=\ds\fr 12B(-\x_2,\x_1,0)= 
\ds\fr 12B|\x|\sin\theta 
\e_\vp$. 
Therefore, we have the 
{\it static} Maxwell field with the potentials 
\be\la{tM} 
\left. 
\ba{l} 
\phi(\x):=-\ds\fr e{|\x|},\\ 
\\ 
 \bA(\x):=\ds\fr 12B(-\x_2,\x_1,0) 
\ea 
\right| 
~~~~\x\in\R^3. 
\ee 
\br 
The angular momentum conservation 
 (\re{3mcs}) holds for $n=3$: 
\be\la{BM3bc} 
\bL_3(t):=\langle\psi(t,\x),\hat\bL_3\psi(t,\x)\rangle=\co, 
\ee 
where $\hat\bL_3$ is the operator in (\re{qvfo}) with $n=3$. 
The conservation reflects the axial symmetry of Equation 
(\re{BSbeg}) with the potentials (\re{tM}). 
\er 
Let us note that 
$i\h \bA(\x)\cdot\na_\x=Bi\h \na_\vp/2=-B\hat\bL_3/2$ 
by (\re{qvfo}). 
Here, $\vp$ is the angular 
coordinate of the rotation around the vector $\e_3:=(0,0,1)$, 
i.e. around $\bB$. 
Then  (\re{BSbeg}) becomes 
\beqn\la{BSbeb} 
i\h\pa_t\psi(t,\x)\!\!&=& 
-\fr 1{2\mu}\h^2\De\psi(t,\x) 
 - 
\fr {e}{2\mu c} 
 B\hat\bL_3\psi(t,\x) 
- 
\ds\fr {e^2}{|\x|}\psi(t,\x), 
~~~~~~\x\in\R^3. 
\eeqn 
Hence, the corresponding stationary equation reads 
\beqn\la{BSMssbb} 
E_\om\psi_\om(\x)= 
-\fr 1{2\mu}\h^2\De\psi_\om(\x) - 
\om_\cL 
\hat\bL_3 \psi_\om(\x) 
- 
\ds\fr {e^2}{|\x|}\psi_\om(\x), 
~~~~~~\x\in\R^3, 
\eeqn 
where $\om_\cL:= {e B}/({2\mu c})<0$ is the {\bf Larmor frequency}. 
Therefore, Theorem \re{HS}, Theorem \re{tDl} ii) and 
(\re{ak1}) 
imply the following theorem. 
\bt\la{BHS} 
i) The quantum stationary states $\psi_\om$  of the hydrogen atom 
in a uniform magnetic field coincide with  the functions 
$\psi_{lmn}$ from 
(\re{eig}).\\ 
ii) The corresponding energies are equal to 
\be\la{LL} 
E_{mn}:=-2\pi\h R/n^2+m\h\om_\cL, 
~~~~~~ 
n=1,2,3...,~~~~m=-l,...,l,~~~l\le n-1. 
\ee 
\et 
 
{\bf Radial scalar potential} 
Let us consider more general equations of the type 
(\re{tM}) with a static radial external potential $\phi(|x|)$ 
and the static uniform magnetic field $B$. 
Then the corresponding eigenvalue problem of the type 
(\re{BSbeb}) 
becomes 
\beqn\la{BSMc} 
E_\om\psi_\om(\x)= 
-\fr 1{2\mu}\h^2\De\psi_\om(\x) - 
i\h\om_\cL 
\na_\vp \psi_\om(\x) 
+ 
{e}{\phi(|\x|)}\psi_\om(\x),~~~~~~~~\x\in\R^3. 
\eeqn 
The following theorem can be proved 
by the same arguments as in 
Theorems \re{HS}, \re{BHS}. 
\bt\la{BHSc} 
i) The quantum stationary states of the electron 
in a static central electric and 
a uniform magnetic field 
have the  following 
form in spherical coordinates 
(cf. (\re{eig})): 
\be\la{eigc} 
\psi_{lmn}^*=R_{ln}(r)F_l^m(\theta)e^{im\vp}. 
\ee 
ii) 
The corresponding 
energies 
are equal to 
\be\la{elmn} 
E_{lmn}^*=E_{ln}^*+m\h\om_\cL,~~~~~~ 
n=1,2,3...,~~~~m=-l,...,l,~~~l\le n-1, 
\ee 
where $E_{ln}^*$ are the energies corresponding to $B=0$. 
 
\et

\subsection{Normal Zeemann Effect: Triplet Spectrum} 
In 1895  Zeemann observed the influence of a magnetic field 
on the spectral lines of atoms and molecules. 
Consider the spectrum of radiation of the 
hydrogen atom in a uniform magnetic field. 
Let us calculate the 
level splitting and the polarization of the radiation emitted by the 
atom: we will see that  both are as in the classical description 
(see Exercise \ref{ue-Ex-12}). 
Namely, the {\it unperturble spectral lines}, corresponding to the 
zero magnetic field $B=0$, split into the {\it normal triplet} 
in the case $B\ne 0$. 
 
%%\begin{itemize} 
First, by Theorem \re{BHS}, 
the bound state eigenfunctions are (\re{eig}): 
\be 
\psi_k =R_{nl}(r)F_l^m(\theta) e^{im\varphi}\quad ,\quad k=(nlm) , 
\ee 
and the eigenvalues are 
\be\la{nZe} 
\omega_k = -\frac{2\pi R}{n^2}+m\om_\cL \quad ,\quad \om_\cL = 
-\frac{eB}{2\mu c}. 
\ee 
In the case $B=0$, we  have also $\om_\cL=0$. Hence, 
the 
{\it unperturbed spectral lines}, 
$\om_{kk'}^0=\ds 2\pi R[\fr 1{n'^2}-\fr 1{n^2}]$, where 
 $k=(nlm)$ and $k'=(n'l'm')$. 
 
Next consider the case $B\ne 0$. 
The calculations (\re{A22bdj})-(\re{lsc}) imply the formula 
for the intensity of the spectral line $\om_{kk'}$ 
in the dipole approximation 
\be \label{CA-j-kk'} 
\bJ_{kk'}\sim 
\int_0^\infty 
R_{nl}(r)\overline{R_{n' l'}}(r) r^3 dr \int_S \x F_l^m(\theta) 
\overline{F_{l'}^{ m'}}(\theta) e^{i(m-m')\varphi}dS. 
\ee 
We know this integral is nonzero only for 
$m'=m,m\pm 1$ and $l'=l\pm 1$. Hence, for $B\ne 0$, the 
unperturbed spectral line 
$\omega_{kk'}^0$  generates the {\it triplet} of 
spectral lines with frequencies 
\be\la{trs} 
\left\{ 
\ba{lll} 
m' =m &:& \om_{kk'}=\omega_{kk'}^0\nonumber \\ 
m' =m-1 &:& \om_{kk'}=\omega_{kk'}^0 +\om_\cL \nonumber \\ 
m' = m+1 &:& \om_{kk'}=\omega_{kk'}^0 -\om_\cL, 
\ea 
\right. 
\ee 
precisely like in the classical case. 
To find the polarizations we re-write the vector $\x$ like 
\be 
\x =\sin \theta (e^{-i\varphi } \e_+ +e^{i\varphi }\e_- ) 
+\cos\theta \e_z,~~~~~\e_\pm =\frac{1}{2}(\hat e_x \pm i \hat e_y). 
\ee 
Inserting back into (\ref{CA-j-kk'}) we find that for $m' =m$ only the 
$\e_z$ component contributes, and for $m' =m\pm 1$ the $\e_\mp$ 
component. Therefore, we find from (\ref{A22bdgm}) the radiation in the form 
\be\left\{\ba{lll} 
m' =m &:& \bA_{kk'}(t,\x) \sim 
\ds\fr 1{|\x|} \rRe c_k \ov c_{k'}
 \e_z e^{-i\om_0 (t-|\x|/c)} \\\\ 
m' =m-1 &:& \bA_{kk'}(t,\x) \sim 
\ds\fr 1{|\x|} \rRe c_k \ov c_{k'}
 \e_+ e^{-i(\omega_0 +\om_\cL )(t-|\x|/c)} 
\\\\ 
m' =m+1 &:& \bA_{kk'}(t,\x) \sim 
\ds\fr 1{|\x|} \rRe c_k \ov c_{k'}
 \e_- e^{-i(\omega_0 -\om_\cL )(t-|\x|/c)}. 
\ea\right. 
\ee 
Note that
the resulting polarizations of the radiation fields are 
 exactly like in the classical case (see Exercise \re{ue-Ex-12}).

%%\end{itemize} 

\br\la{rZe} 
{\rm 
The spectral lines  (\re{trs}) do not depend on the {\it azimuthal 
quantum number} $l$ as well as the eigenvalues  (\re{nZe}). 
Therefore, the radiations, induced by the pairs 
 $k=(mln), k'=(m'l'n')$ with $m'=m,m\pm 1$ and all possible azimuthal 
quantum number $l,l'$, 
contribute to the same frequencies 
(\re{trs}). 
On the other hand, 
in many cases the atom spectra in a magnetic field 
demonstrate 
the {\it multiplet structure} drustically different from 
the triplet structure (\re{trs}). 
This  {\bf anomalous Zeemann effect} 
cannot be explained 
by the Schr\"odinger equation (\re{BSb}). 
The explanation 
is provided by the {\bf Pauli} equation 
which takes into account the  {\bf electron spin} 
(see Section 22.3).

} 
\er

%%%%%%%%%%%%%%%%%%%%%%%%%%%%%%%%%%%%%%%%%%%%%%%%%%%%%%%%%%%%%%%%%%%%%%%%% 
%%%%%%%%%%%%%%%%%%%%%%%%%%%%%%%%%%%%%%%%%%%%%%%%%%%%%%%%%%%% 
%%%%%%%%%%%%%%%%%%%%%%%%%%%%%%%%%%%%%%%%%%%%%%%%%%%%%%% 

\newpage 
%%%%%%%%%%%%%%%%%%%%%%%%%%%%%%%%%%%%%%%%%%%%%%%%%%%%%%%%%%%%%%%%%%%%% 
%%%%%%%%%%%%%%%%%%%%%%%%%%%%%%%%%%%%%%%%%%%%%%%%%%%%%%%%%%%%%% 
%%\setcounter{section}{+9} 
\setcounter{subsection}{0} 
\setcounter{theorem}{0} 
\setcounter{equation}{0} 
\section{Diamagnetism and Paramagnetism} 
We compute a magnetic moment of a hydrogen atom and split it in 
two parts, one of which describes a diamagnetism of the atom and 
the other describes a paramagnetism. 
This allows us to 
calculate the {\it magnetic susceptibility} of the hydrogen atom 
(see Lecture 14).

Namely, we will 
consider the stationary states of the atom in the uniform 
external magnetic 
field $\bB$ and calculate their magnetic moment (\re{mome}) 
\be\la{mg} 
\m=\fr 1{2c} \int \y\times \bj(\y)d\y. 
\ee 
Let us choose the coordinates in such a way that $\bB=(0,0,B)$ with $B\ge 0$. 
Then the corresponding vector potential is given by 
$ \bA(\x)=\ds\fr 12B(-x_2,x_1,0)=\ds\fr 12B|\x|\sin\theta\e_\vp$.

\subsection{Electric Current at Stationary States} 
First we  calculate the electric current $\bj$ 
defined by 
the Born approximation (\re{SMeeaa}), 
\be\la{SMb} 
{\bj}(t,\x ) 
=\ds\fr e \mu[-i\h\na-\ds\fr ec  \bA(\x)]\psi(t,\x ) 
\cdot\psi(t,\x ) . 
\ee 
The stationary states are given by  (\re{eig}) 
(see Theorem \re{BHS}). Let us write (\re{eig}) in the form 
\be\la{eigg} 
\psi(t,\x )=a_{lmn}(r,\theta)e^{im\vp}e^{-i\om t}, 
\ee 
where $a_{lmn}$ is a real function according to 
Theorem \re{tDl} ii). 
Let us express the gradient operator in 
spherical coordinates (see (\re{nasf})): 
\be\la{gra} 
\na=\e_r\na_r\psi+\e_\theta\fr{\na_\theta}r 
+\e_\vp\fr{\na_\vp}{r\sin\theta} . 
\ee 
Then (\re{SMb}) and (\re{eigg}) give, 
\beqn\la{SMbg} 
{\bj}(t,\x ) 
= 
&&\ds\fr e \mu[-i\h \e_r\na_r a_{lmn}(r,\theta)] 
\cdot a_{lmn}(r,\theta) 
+ 
\ds\fr e \mu[-i\h \e_\theta\fr{\na_\theta}r 
a_{lmn}(r,\theta)] 
\cdot a_{lmn}(r,\theta)\nonumber\\ 
\nonumber\\ 
&+& 
\ds\fr e \mu[m\h \e_\vp\fr{1}{r\sin\theta}] 
\psi(t,\x ) 
\cdot \psi(t,\x ) 
-\ds\fr {e^2} {\mu c} \bA(\x)\psi(t,\x ) 
\cdot\psi(t,\x ) . 
\eeqn 
The first and second term on the RHS are zero since the 
function $ a_{lmn}(r,\theta)$ is real. Hence finally, 
\beqn\la{SMbf} 
~~~~~~~~~{\bj}(t,\x ) 
= 
\ds\fr e \mu[m\h \e_\vp\fr{1}{r\sin\theta}] 
|\psi(t,\x )|^2 
-\ds\fr {e^2} {\mu c} B|\x|\sin\theta\e_\vp|\psi(t,\x )|^2 
={\bj}'(\x )+{\bj}''(\x ) . 
\eeqn 
The magnetic  moment (\re{mg}) becomes 
\be\la{mgb} 
\m=\fr 1{2c} \int \y\times \bj'(\y)d\y+ 
\fr 1{2c} \int \y\times \bj''(\y)d\y=\m'+\m''. 
\ee 
The currents ${\bj}'(\x )$, ${\bj}''(\x )$ are axially symmetric 
with respect to 
the axis $O\x_3$, and 
\be\la{yt} 
(\y\times\e_\vp)_3=|\y|\e_3\sin\theta. 
\ee

\subsection{Langevin Formula for Diamagnetic Susceptibility} 
 By (\re{yt}), 
the direction of $\m''$ is 
opposite to $\bB$, 
hence, $\m''$ describes the 
{\it diamagnetism} of the atom. The corresponding diamagnetic susceptibility 
(per atom) is defined by $\m''=\chi_m\bB$, hence 
\be\la{dis} 
\chi_m:=-\fr{|\m''|}B=-\fr{e^2}{2\mu c^2} 
\int |\y|^2\sin^2\theta|\psi|^2d\y=-\fr{e^2}{2\mu c^2}\Theta_3, 
\ee 
where $\Theta_3$ denotes the moment of inertia of the distribution $|\psi|^2$ 
with respect to the axis $O\x_3$. 
For spherically symmetric stationary states we have 
\be 
\Theta_3=\fr 13\ov\Theta,~~~~~~~~~\ov\Theta:=\int |\y|^2|\psi|^2d\y. 
\ee 
Then (\re{dis}) becomes the {\it Langevin formula} (cf. Exercise 
\ref{ue-Ex-13}) 
\be\la{disb} 
\chi_m=-\fr{e^2\ov\Theta}{6\mu c^2} . 
\ee 
For spherically non-symmetric states the formula also holds 
{\it in the mean} 
due to the {\it random orientation} of the atoms with respect to the direction 
of the magnetic field. 
 
\subsection{Paramagnetism} 
The moment $\m'$ is different from zero even in the absence 
of the magnetic field, 
\be\la{ml} 
~~~~~~~~~\m'=m\h\fr {e}{2c\mu} \int 
\fr{(\y\times \e_\vp)_3\e_3}{|\y|\sin\theta} 
|\psi(t,\y)|^2 
d\y=m\h\e_3\fr {e}{2c\mu} \int 
|\psi(t,\y)|^2 
d\y=\fr {e}{2c\mu}m\h\e_3 
\ee 
by (\re{yt}). 
Hence, $\m'$ describes the 
{\it paramagnetism} of the atom. 
However, its mathematical expectation is zero due to the random 
orientation of the atom since the probabilities of the values $m$ and $-m$ 
are identical. On the other hand, the probabilities are not identical 
if the magnetic field does not vanish. 
The corresponding {\it statistical} 
theory has been  developed by Langevin.

%%%%%%%%%%%%%%%%%%%%%%%%%%%%%%%%%%%%%%%%%%%%%%%%%%%%%%%%%%%%%%%%%%%%%%%%% 
%%%%%%%%%%%%%%%%%%%%%%%%%%%%%%%%%%%%%%%%%%%%%%%%%%%%%%%%%%%% 
%%%%%%%%%%%%%%%%%%%%%%%%%%%%%%%%%%%%%%%%%%%%%%%%%%%%%%% 

\part{Electron Spin and Pauli Equation} 
%%%%%%%%%%%%%%%%%%%%%%%%%%%%%%%%%%%%%%%%%%%%%%%%%%%%%%%%%%%%%%%%%%%%% 
%%%%%%%%%%%%%%%%%%%%%%%%%%%%%%%%%%%%%%%%%%%%%%%%%%%%%%%%%%%%%% 
%%\setcounter{section}{+9} 
\setcounter{subsection}{0} 
\setcounter{theorem}{0} 
\setcounter{equation}{0} 
\section{Electron Spin: Experiments and Interpretation} 
We describe some experiments which have inspired the ideas of a 
{\it spin} and a {\it magnetic moment} of an electron, namely, 
the Einstein-de-Haas experiment, the Stern-Gerlach experiment, 
and the anomalous Zeemann effect. 
 
The concept of spin  and magnetic moment 
of the electron has been introduced by Goudsmith 
and Uhlenbeck in 1925: {\it 
Every electron has an intrinsic angular momentum 
(spin) $\s$ with the magnitude 
 $|\s|=\ds\fr\h2$, and a magnetic moment $\m_\s$ with the magnitude 
$|\m_\s|=\mu_B:=\ds\fr {|e|\h}{2\mu c}$}, 
which is  the {\it Bohr magneton}.
Here the term ``intrinsic'' means that the 
spin angular momentum is 
not related to a rotation of the particle, and 
 spin magnetic moment is 
not related to the corresponding convection current.

Let us consider the experimental data which inspired this conjecture. 
 
\subsection{Einstein-de Haas Experiment} 
Let us consider an iron bar positioned vertically 
in the earth's gravitational field, 
which is attached to a vertical string in such a way that 
it can rotate about its axis. Let us magnetize 
the bar by a vertical external 
weak magnetic field. 
The field orients the elementary Ampere molecular currents, 
so that their magnetic moments increase the external magnetic field. 
Therefore, 
the corresponding elementary angular momenta are also 
oriented in the same direction, 
and the sum of the microscopic angular  momenta increases. 
By axial symmetry, the total 
angular momentum of the bar, macroscopic+microscopic, 
is conserved. Hence, the bar 
as a whole must change its macroscopic angular momentum, too. 
In 1915 Einstein and de Haas measured the changes of magnetic moment 
and macroscopic angular momentum of the bar to check their ratio. 
The classical and quantum theories predict the ratio 
 $\ds\fr {e}{2\mu c}$. 
However, the experimental results contradict this value. 
This suggests the existence of an additional magnetic moment 
of the atoms which is responsible for the anomalous ratio.

\subsubsection{Classical theory} 
Let us assume that in each 
molecule, 
the  currents 
are caused by the rotation of electrons 
with the same angular velocity $\om$. 
Then the angular momentum and 
magnetic moment of the molecule are given by 
\be\la{cra} 
\bL:=\sum \x_k\times \mu(\om\times\x_k),~~~~~~~ 
\m:=\fr1{2c} \sum \x_k\times e(\om\times\x_k), 
\ee 
where $\x_k$ are the electron positions. 
Therefore, the following key relation holds for each molecule, 
\be\la{ker} 
\m=\ds\fr {e}{2\mu c}\bL. 
\ee 
Hence, an increment of the magnetization, $\De\m_3$, of the bar, 
is followed by the corresponding increment, 
$-\De\bL_3=-\ds\fr {e}{2\mu c}\De\m_3$, 
of its macroscopic angular momentum. 
Both quantities, $\De\m_3$ and $\De\bL_3$, can be measured 
experimentally: 
$-\De\bL_3$ by the torsion vibration of the string, 
and $\De\m_3$ by the residual magnetism. The experiment 
was performed by Einstein and de Haas in 1915. However, 
the result was 
\be\la{kerg} 
\m_3=g \ds\fr {e}{2\mu c}\bL_3 
\ee 
with the {\it Land\'e factor} 
$g\approx 2$, which contradicts (\re{ker}). 
This contradiction inspired the ratio $|\s|/|\m_\s|=\ds\fr {e}{\mu c}$ in the 
Stern-Gerlach conjecture. This ratio 
corresponds to 
the Land\'e factor $g=2$ and allows to explain the result 
of the Einstein-de Haas experiment. 
\subsubsection{Quantum theory} 
Let us check that the relation (\re{ker}) also holds for 
the Schr\"odinger equation 
(\re{BSbeb}) for small $B$. 
Indeed, (\re{mg})  and (\re{SMb}) imply for small $B$, 
\beqn\la{mgm} 
~~~~~~~~\m&\approx&\fr 1{2c} \int \y\times \ds\fr e \mu[-i\h\na_\y]\psi(t,\y ) 
\cdot\psi(t,\y )d\y 
\nonumber\\ 
\nonumber\\ 
&=& 
\fr e{2\mu c} \int [-i\h\y\times\na_\y ]\psi(t,\y ) 
\cdot\psi(t,\y )d\y=\fr e{2\mu c}\bL. 
\eeqn 
In particular, 
for the stationary state (\re{eigg}) we have the magnetization 
(cf. (\re{ml})) 
\be\la{m1h} 
~~~~~~~~~\m_3\approx 
\fr {e}{2\mu c}m\h. 
\ee 
On the other hand, the  energy for the stationary state 
is given by 
(\re{elmn}). The external magnetic field causes the transitions 
to stationary states with lower values of the energy 
(\re{elmn}), 
which corresponds to greater values of 
$m$ since $\om_\cL<0$. Therefore, the magnetization 
(\re{ml}) increases 
in the transitions.

\subsection{Double Splitting} 
An additional suggestion for the spin is provided by the 
observation of the splitting of stationary states 
of atoms and molecules in a magnetic field $B$.

\subsubsection{Stern-Gerlach experiment} 
In 1922 Stern and Gerlach sent a beam of silver atoms 
through an inhomogeneous magnetic field. 
Later similar experiments have been performed with 
hydrogen atoms. 
The atoms are in the ground state (\re{ceig}), which 
is non-degenerate 
according to the Schr\"odinger equation (\re{BSbeb}). 
Let us write (\re{BSbeb}) in the form 
\be\la{BSf} 
i\h\pa_t\psi(t,\x)\!\!= 
-\fr 1{2\mu}\h^2\De\psi(t,\x) 
+ 
e\phi(|\x|)\psi(t,\x) 
 - 
\fr {e}{2\mu c} 
 \hat\bL\bB\psi(t,\x) 
, 
~~~~~~\x\in\R^3. 
\ee 
The interaction term 
$\hat\bL \bB\psi(t,\x)$ vanishes for the ground state. 
Therefore, the ground state also satisfies Equation 
(\re{BSf}) with $\bB\ne 0$. 
However, Stern and Gerlach observed 
a splitting of the beam into two components. 
This obviously contradicts the identity of all atoms in the 
ground state and suggests that 
\\ 
i) For $\bB=0$ 
the eigenspace corresponding to the 
ground state has a dimension at least two. 
\\ 
ii) For $\bB\ne 0$ the eigenspace splits in two distinct 
eigenspaces. 
\\ 
This suggests the existence of an additional magnetic moment 
of the atoms which is responsible for the splitting 
in the magnetic field.

\subsubsection{Anomalous Zeemann effect} 
The  Schr\"odinger equation gives a satisfactory 
explanation of  the 'normal' Zeemann effect with 
splitting into three lines. 
However, Zeemann  demonstrated in 1895 that, for $B\ne 0$, 
most of the spectral lines are split into 
a different number of lines: 
two, five, etc. 
This  {\it anomalous} Zeemann effect 
contradicts the Schr\"odinger theory, which only 
predicts the normal Zeemann effect. 
This also suggests the existence of an additional magnetic moment 
responsible for the anomalous splitting.

%%%%%%%%%%%%%%%%%%%%%%%%%%%%%%%%%%%%%%%%%%%%%%%%%%%%%%%%%%%%%%%%%%%%%%%%% 
%%%%%%%%%%%%%%%%%%%%%%%%%%%%%%%%%%%%%%%%%%%%%%%%%%%%%%%%%%%% 
%%%%%%%%%%%%%%%%%%%%%%%%%%%%%%%%%%%%%%%%%%%%%%%%%%%%%%% 

\newpage 
%%%%%%%%%%%%%%%%%%%%%%%%%%%%%%%%%%%%%%%%%%%%%%%%%%%%%%%%%%%%%%%%%%%%% 
%%%%%%%%%%%%%%%%%%%%%%%%%%%%%%%%%%%%%%%%%%%%%%%%%%%%%%%%%%%%%% 
%%\setcounter{section}{+9} 
\setcounter{subsection}{0} 
\setcounter{theorem}{0} 
\setcounter{equation}{0} 
\section{Pauli Equation} 
The Einstein-de Haas, Stern-Gerlach and anomalous 
Zeemann effects 
demonstrate that the Schr\"odinger equation requires a 
modification. 
To the Schr\"odinger equation, 
we introduce an additional spin  magnetic moment 
corresponding to 
spin angular momentum. 
Then the Schr\"odinger equation becomes 
the Pauli equation which explains the Stern-Gerlach effect. 
Let us analyze the details. 
 
\subsection{Additional Magnetic Moment} 
Relation 
(\re{mgm}) implies that, for small $|\bB|$, the 
last term in (\re{BSf}) can be rewritten 
as 
\be\la{lt} 
-\fr {e}{2\mu c} 
\hat\bL \bB\psi(t,\x)\approx-\hat\m \bB\psi(t,\x), 
\ee 
where $\hat\m$ is the quantum observable corresponding to the 
 magnetic moment. 
This corresponds to the fact that $-\m \bB$ 
is 
the energy of the magnetic moment $\m$ in the magnetic field $\bB$. 
The Stern-Gerlach experiment demonstrates that the magnetic moment $\m$ 
does not vanish even for the ground state 
of the hydrogen atom. Therefore, we have 
to postulate the existence of an additional magnetic moment 
$\m_\s\ne 0$. 
The Einstein-de Haas experiment 
suggests the Goudsmith-Uhlenbeck 
conjecture 
 that the additional magnetic moment 
corresponds to an additional 
{\it spin angular  momentum} of the electron, $\s=${\it spin}, with 
the Land\'e factor 2, i.e. 
\be\la{Lf} 
\m_\s=2\fr {e}{2\mu c}\s. 
\ee 
Next, we have to specify 
a mathematical construction for 
this spin angular momentum $\s$.

\subsection{Additional Angular Momentum} 
\subsubsection{Orbital angular momentum} 
First, let us analyze the 
{\it orbital angular momentum} $\bL=(\bL_1,\bL_2,\bL_3)$ 
for the case  $A_k=0$, $k=1,2,3$ and spherically symmetric 
scalar potential $\phi(|\x|)$: 
\be\la{oam} 
i\h\pa_t\psi(t,\x)\!\!=\cH\psi(t,\x):= 
-\fr 1{2\mu}\h^2\De\psi(t,\x) 
+ 
e\phi(|\x|)\psi(t,\x), 
~~~~~~\x\in\R^3. 
\ee 
The angular momentum 
is a conserved quantity which corresponds 
to an invariance of the Lagrangian  (\re{LdNS}) 
with respect to the  {\bf regular representation} 
of the rotation group $SO(3)$. The regular representation  $R_g$ 
acts on the phase space $\E:=L^2(\R^3)$ by the formula 
$R_g\psi(\x):=\psi(g\x),\x\in\R^3$, for $g\in SO(3)$, $\psi\in\E$. 
The Schr\"odinger operator $\cH$ commutes with the representation 
and with the generators  $\bH_k$ 
of the rotations around the axis $O\x_k$ 
(see (\re{amco})). 
The generators automatically satisfy the commutation 
relations (\re{com}) of the Lie algebra of the 
rotation group $SO(3)$. 
The corresponding conserved quantities 
are given by the Noether theorem and 
have the form 
({\re{3mcs}): 
\be\la{adao} 
\bL_k=\langle \psi,\hat\bL_k\psi \rangle,~~~~~~~~~~k=1,2,3, 
\ee 
where $\hat\bL_k=\h\bH_k$. 
An alternative proof of the conservation follows from the commutation 
$[\cH,\bH_k]=0$ (see (\re{com})).

\subsubsection{Spin angular momentum} 
The analysis suggests that the spin angular momentum 
corresponds in a similar way to another action 
of the rotation group, 
which is different from the regular 
representation. 
To construct this  {\bf spinor representation}, 
we denote its generators by $\bh_k$. Then similarly to (\re{com}), we 
necessarily have 
\be\la{comr} 
[\bh_1,\bh_2]=i\bh_3,~~~~~~~[\bh_2,\bh_3]=i\bh_1,~~~~~~~[\bh_3,\bh_1]=i\bh_2. 
\ee 
Further, define $\hat\s_k=-\h\bh_k$ and the spin angular momenta 
\be\la{ada} 
\s_k=\langle \psi,\hat\s_k\psi \rangle,~~~~~~~~~~k=1,2,3. 
\ee 
The momenta 
are conserved quantities if the modified Schr\"odinger  operator 
commutes with $\bh_k$. 
 
The double splitting of the ground state in 
the Stern-Gerlach experiment with the hydrogen atom 
 suggests 
that the dimension of the unperturbed `ground eigenspace' is two. 
Therefore, 
in analogy with the orbital angular momentum, the 
ground state 
{\bf formally} corresponds 
to $2l+1=2$, i.e. $l=1/2$, 
and, for $\bB\ne 0$, it generates two 
stationary states 
which are eigenfunctions of the 
operator $\bh_3$ with the eigenvalues $\pm 1/2$. 
Therefore, the action of the group $SO(3)$ in the 
unperturbed ground eigenspace is an irreducible two-dimensional 
representation $S_g$.

Such a two-dimensional representation is given by 
Proposition \re{pNew}: 
it corresponds to the generators 
$\bh_k:=\ds\fr 12\si_k$, where 
$\si_k$ are the 
 {\bf Pauli matrices} (see (\re{mpa})) 
\be\la{tmm} 
\si_1=\left( 
\ba{rr} 0&1\\1&0\ea 
\right),~~~~~~~ 
\si_2=\left( 
\ba{rr} 0&-i\\i&0\ea 
\right),~~~~~~~ 
\si_3=\left( 
\ba{rr} 1&0\\0&-1\ea 
\right). 
\ee 
\bexe 
Check the commutation relations (\re{comr}) for the generators 
$\bh_k=\ds\fr 12\si_k$. 
\eexe 
 
The simplest way to make the ground eigenspace two-dimensional is to 
define the modified phase space as the tensor product 
$\E^\otimes:=\E\otimes \C^2$ and consider 
the tensor product of the regular representation $T_g$ in $\E$ and 
the spinor representation $S_g$ in $\C^2$. 
\br\la{rcom} 
By the definition of the tensor product of representations, 
all generators $\bH_k$ commute with all generators $\bh_j$. 
\er

\subsection{Pauli Equation. Uniform Magnetic Field} 
Let us summarize our discussion and 
change the Schr\"odinger equation (\re{BSf}) 
to the {\bf Pauli equation} (\re{PE}), 
\beqn\la{BSfm} 
~~~~~~i\h\pa_t\Psi(t,\x)\!\!&=&\cP\Psi(t,\x)\nonumber\\ 
\nonumber\\ 
&:=& 
-\fr 1{2\mu}\h^2\De\Psi(t,\x) 
+ 
e\phi(|\x|)\Psi(t,\x) 
 - 
\fr {e}{2\mu c} 
 \hat\bL \bB\Psi(t,\x) 
- 
\fr {e}{\mu c} 
 \hat\s \bB\Psi(t,\x), 
\eeqn 
where $\Psi(t,\x)=(\psi_1(t,\x),\psi_2(t,\x))\in 
\E^\otimes=\E\otimes 
\C^2$ and the additional factor $2$ in the last term corresponds to 
(\re{Lf}). 
\bd 
i) For the Pauli Equation, 
the total angular momentum is the following mean value: 
\be\la{tam} 
\bJ=\bL+\s. 
\ee 
ii) The total magnetic moment is 
\be\la{tmmi} 
\m=\fr {e}{2\mu c}\bL+\fr {e}{\mu c}\s. 
\ee 
\ed 
 
The unperturbed {\it Pauli operator} $\cP$ with $\bB=0$ commutes 
with all $T_g$ and $S_g$. 
Hence, the orbital and spin momenta, $\bL$ and $\s$, 
are conserved quantities if $\bB=0$. 
Therefore, $\bJ$ is also conserved if $\bB=0$. 
\br\la{rPam} 
For the Pauli equation with $\bB=0$ 
we have three conserved angular momenta 
 $\bL$, $\s$ and $\bJ$. Hence, 
the identification of the total angular momentum with the vector 
$\bJ$, is not well justified. 
\er 
For $\bB=(0,0,B)\ne 0$, the equation becomes 
\beqn\la{BSfmb} 
~~~~~~i\h\pa_t\Psi(t,\x)\!\!&=&\cP\Psi(t,\x)\nonumber\\ 
\nonumber\\ 
&=& 
-\fr 1{2\mu}\h^2\De\Psi(t,\x) 
+ 
e\phi(|\x|)\Psi(t,\x) 
 - 
\fr {e}{2\mu c} 
 \hat\bL_3B\Psi(t,\x) 
- 
\fr {e}{\mu c} 
 \hat\s_3B\Psi(t,\x). 
\eeqn 
\br\la{LsJ} 
The operator $\cP$  commutes with 
$\hat\bL^2:=\hat\bL_1^2+\hat\bL_2^2+\hat\bL_3^2$, 
 $\hat\bL_3$, $\hat\s^2:=\hat\s_1^2+\hat\s_2^2+\hat\s_3^2$, $\hat\s_3$ 
and $\hat\bJ^2:=\hat\bJ_1^2+\hat\bJ_2^2+\hat\bJ_3^2$, 
 $\hat\bJ_3$. 
Therefore, the corresponding mean values 
$\bL^2$, $\bL_3$, $\s^2=3/4$, $\s_3$ and $\bJ^2$, $\bJ_3=\bL_3+\s_3$, 
 are conserved. 
\er

\subsection{Pauli Equation. General Maxwell Field} 
The natural extension of the Pauli equation for a general 
Maxwell field 
reads 
\be\la{PaA} 
[i\h\pa_t-e\phi(t,\x)]\Psi(t,\x)=\fr 1{2\mu} 
[-i\h\na-\ds\fr ec\bA(t,\x)]^2\Psi(t,\x) 
- 
\fr {e}{\mu c} 
 \hat\s  \bB (t,\x)\Psi(t,\x). 
\ee 
It corresponds to the Lagrangian density (cf. (\re{LdNS})) 
\be\la{LdNP} 
\cL_{P}(x,\Psi,\na\Psi) 
= 
[i\h\na_0-e\phi(x)]\Psi\cdot \Psi- 
\fr1{2\mu}|[-i\h\na- \ds\fr ec \bA(x)]\Psi|^2 
- 
\fr {e}{\mu c} 
 [\hat\s  \bB (x)\Psi]\cdot\Psi, 
\ee 
where $\bB (x)=\rot \bA(x)$ and 
'$\cdot$' stands for the scalar product in 
$\C^2\equiv\R^4$. 
This suggests the following extension to the Lagrangian 
density of the coupled Maxwell-Pauli equations (cf. (\re{cLMS})) 
\be\la{cLMP} 
~~~~~~\cL_{MP} 
= 
[i\h\na_0-e\phi(x)]\Psi\cdot \Psi- 
\fr1{2\mu}|[-i\h\na- \ds\fr ec \bA(x)]\Psi|^2- 
\fr {e}{\mu c} 
 [\hat\s\bB (x)\Psi]\cdot\Psi 
-\fr 1{16\pi}\,\F^{\al\beta}\F_{\al\beta} 
\ee 
where $\F^{\mu\nu}:=\pa^\mu\A^\nu-\pa^\nu\A^\mu$ and 
$\F_{\mu\nu}:=\pa_\mu\A_\nu-\pa_\nu\A_\mu$. 
The corresponding Euler-Lagrange equations are 
 the coupled Maxwell-Pauli equations (cf. (\re{SMe})) 
\be\la{PMe} 
~~~~~~~~\left\{\!\! 
\ba{l} 
\ds[i\h\pa_t-e\phi(t,\x)]\Psi(t,\x) 
=\fr 1{2\mu} 
[-i\h\na_\x-\ds\fr ec \bA(t,\x)]^2\Psi(t,\x) 
- 
\fr {e}{\mu c} 
 \hat\s  \bB (t,\x)\Psi(t,\x). 
\\ 
\\ 
\ds\fr 1{4\pi}\,\na_\al\F^{\al\beta}(t,\x) 
\!=\! 
\left(\!\! 
\ba{l} 
\rho:=e|\Psi(t,\x)|^2\\ 
\\ 
\ds\fr{\bj}c 
\!:=\!\ds\fr e{\mu c}[-i\h\na-\ds\fr ec \bA(t,\x)]\Psi(t,\x ) 
\!\cdot\!\Psi(t,\x ) 
\!+\!\fr {e}{\mu c} 
 \rot \Big(\![\hat\s  \Psi(t,\x)]\!\cdot\!\Psi(t,\x)\!\!\Big) 
\ea\!\!\right) 
\ea\right. 
\ee

\subsection{Application to the Stern-Gerlach Experiment} 
Theorem \re{BHS} allows us to construct quantum 
stationary states 
corresponding to the 
Pauli equation (\re{BSfmb}) describing  the hydrogen atom. 
Let us define the vector-functions 
\be\la{Ppm} 
~~~~\Psi_{lmn}^+=\left( 
\ba{l}\psi_{lmn}\\0\ea\right),~~~ 
\Psi_{lmn}^-=\left( 
\ba{l}0\\\psi_{lmn}\ea\right), 
~~~~~n=1,2,3...,~~l\le n-1,~~ m=-l,...,l, 
\ee 
where the functions $\psi_{lmn}$ are given by (\re{eig}). 
They 
are eigenfunctions of the operator $\cP$ corresponding to 
the energies 
\be\la{PpmE} 
E_{mn}^\pm:=-2\pi\h R/n^2+m\h\om_\cL  \pm \h\om_\cL. 
\ee 
In particular, the Schr\"odinger ground state $\psi_{001}$ 
generates two stationary states $\Psi_{001}^\pm$ with distinct energies 
$E_{01}^\pm:=-2\pi\h R \pm \h\om_\cL$ and distinct spin magnetic moments 
$\ds\pm\fr {e\h}{2\mu c}$. 
This explains the double splitting of the  beam in the 
 Stern-Gerlach experiment. 
 
\br\la{rsg} 
{\rm The double splitting of spectral lines is 
observed experimentally {\bf even in the absence of the 
external magnetic field}, when formally 
$\om_\cL:={e B}/({2\mu c})=0$. This means that 
empirically the Larmor frequency $\om_\cL\ne 0$ even for $B=0$. 
It is natural to think that it is produced by the {\bf intrinsic 
magnetic field} 
generated by the electric current 
in the coupled Maxwell-Pauli equations (\re{PMe}). 
Let us recall that 
the formulas (\re{Ppm}), (\re{PpmE}) are obtained 
in the {\it Born approximation} 
neglecting the intrinsic field. 
So the double splitting in the  absence of the 
external magnetic field might be explained by the nonlinear self-interaction 
in the  coupled Maxwell-Pauli equations. 
An alternative mechanism of generation of the magnetic field 
has been proposed by Thomas 
\ci{To} and Frenkel \ci{Fr} (see below). 
} 
\er 
 
\bcom 
{\rm The classical interpretation of 
the splitting term $\pm \h\om_\cL$ in (\re{PpmE}) 
means that the "projection of the spin magnetic moment" 
to the axis $Ox_3$ is equal to $\pm\ds\fr{e}{2\mu c}$. 
The magnitude of the projection is attained "instantly" 
with the magnetic field, 
which contradicts the classical picture (nonzero moment of inertia, etc). 
In the quantum context the instant reaction is not surprising 
in view of the conjecture $\bf A$ (see Preface). 
Namely, 
it should be interpreted 
as the change of the attractor of the system, and 
the formulas (\re{PpmE}) determine the instant bifurcation of 
the (point) attractor. 
} 
 
\ecom

%%%%%%%%%%%%%%%%%%%%%%%%%%%%%%%%%%%%%%%%%%%%%%%%%%%%%%%%%%%%%%%%%%%%%%%%% 
%%%%%%%%%%%%%%%%%%%%%%%%%%%%%%%%%%%%%%%%%%%%%%%%%%%%%%%%%%%% 
%%%%%%%%%%%%%%%%%%%%%%%%%%%%%%%%%%%%%%%%%%%%%%%%%%%%%%% 

\newpage 
%%%%%%%%%%%%%%%%%%%%%%%%%%%%%%%%%%%%%%%%%%%%%%%%%%%%%%%%%%%%%%%%%%%%% 
%%%%%%%%%%%%%%%%%%%%%%%%%%%%%%%%%%%%%%%%%%%%%%%%%%%%%%%%%%%%%% 
%%\setcounter{section}{+9} 
\setcounter{subsection}{0} 
\setcounter{theorem}{0} 
\setcounter{equation}{0} 
\section{Einstein-de Haas and 
Anomalous Zeemann  Effects} 
We consider a 
modification of  the Pauli equation taking 
into account  the interaction 
between the orbital and spin angular momentum. 
The Russell-Saunders method gives 
a satisfactory explanation of the Einstein-de Haas and 
anomalous Zeemann  effects.

The Pauli equation in the form 
(\re{BSfm}) is not sufficient for the explanation of 
the anomalous Zeemann effect. This follows from the 
formulas (\re{PpmE}), (\re{PpmE}). Namely, the selection rules 
$m\mapsto m,m\pm 1$ give the splitting 
$\om_0\mapsto \om_0,\om_0\pm \om_\cL$ 
of the unperturbed spectral line $\om_0$. Hence, the 
splitting gives the 
triplet corresponding to the normal Zeemann effect 
(cf. Remark \re{rZe}). 
 
This situation is related to the fact that the Pauli equation 
(\re{BSfm}) does not take 
into account the interaction between the orbital and spin 
angular momenta (see Remark \re{rPam}), while 
the interaction is suggested by the great 
success of the phenomenological {\it vector model} 
(see Exercise 15). 
That model suggests the interaction through the Maxwell field. 
Namely, the orbital angular momentum is related to the 
rotational motion of the electron, i.e., the circular current 
which generates the magnetic field acting on the spin magntic moment. 
Similarly, the spin angular momentum is related to the `intrinsic rotation' 
of the electron which generates the magnetic field acting 
on the orbital motion of the electron.

\subsection{Spin-Orbital Coupling} 
To explain the anomalous Zeemann effect, we have to modify the 
equation (\re{BSfm}) taking into account 
the interaction between the orbital and spin 
angular momenta. 
In spirit, the interaction might be described 
by the coupled Maxwell-Pauli equations (\re{PMe}) (see Remark \re{rsg}). 
An alternative {\it relativistic} 
modification has been found by Thomas 
\ci{To} and Frenkel \ci{Fr}. 
Namely, the modification takes into account 
the magnetic field 
arising in the ``moving frame of the electron''. 
The magnetic field is expressed in the 
electrostatic radial potential $\phi(|\x|)$ by the Lorentz formulas 
(\re{Maxtr}): 
in the first order approximation w.r.t. 
$\beta=v/c$, we get 
\be\la{wh} 
~~~~\ti\bB =\ds\fr 1c \bE\times\bv=-\ds\fr 1c \na\phi(|\x|)\times\bv= 
-\ds\fr 1{\mu c} \phi'(|\x|)\fr {\x}{|\x|}\times\p= 
-\ds\fr 1{\mu c|\x|} \phi'(|\x|)\hat\bL 
\ee 
where we set {\it formally} $\bv=\p/\mu$. 
This magnetic field produces the corresponding 
correction term $-\ds\fr{e}{\mu c}\hat\s\ti\bB \Psi(t,\x)$ 
in the RHS of (\re{BSfm}). Furthermore, the Thomas and Frenkel 
phenomenological arguments imply an additional factor $1/2$
which is justified by the relativistic Dirac theory \ci{Sak}. 
Hence, finally, the modified equation reads 
\beqn\la{BSfmrs} 
~~~~~~&&i\h\pa_t\Psi(t,\x)\!\!=\cP_m\Psi(t,\x)\nonumber\\ 
&&\\ 
&&:= 
-\fr 1{2\mu}\h^2\De\Psi(t,\x) 
+ 
e\phi(|\x|)\Psi(t,\x)+\ds\fr{e}{2\mu^2 c^2} 
\ds\fr{\phi'(|\x|)}{|\x|} 
  \hat 
\bL 
\hat\s\Psi(t,\x) 
 - 
\fr {e}{2\mu c} 
 \hat\bL \bB\Psi(t,\x) 
- 
\fr {e}{\mu c} 
 \hat\s \bB\Psi(t,\x).\nonumber 
\eeqn 
In the case  $B=0$ the equation becomes 
\beqn\la{BSfmrs0} 
&&i\h\pa_t\Psi(t,\x)\!\!=\ov\cP_m\Psi(t,\x)\nonumber\\ 
&&\\ 
&&:= 
-\fr 1{2\mu}\h^2\De\Psi(t,\x) 
+ 
e\phi(|\x|)\Psi(t,\x)+\ds\fr{e}{2\mu^2 c^2} 
\ds\fr{\phi'(|\x|)}{|\x|} 
  \hat 
\bL 
\hat\s\Psi(t,\x).~~~~~~~~~~~~~~~~~~~~~~~~~~~~~~~~~~~~~~ 
\nonumber 
\eeqn 
\brs 
i) The correction term $\sim \hat\bL\hat\s$ 
 is called the {\bf Russell-Saunders 
spin-orbital coupling}. 
\\ 
ii) 
A "rigorous" justification for 
the correction term follows from the 
Dirac relativistic equation. Namely, just this term 
appears in an asymptotic 
expansion of the Dirac equation  in the  series of $c^{-1}$ 
(see  \ci[Vol. II]{Som} and \ci{BeS,Schiff}). 
\\ 
iii) The term is a genuine relativistic correction 
since it is provided by the Lorentz transformation (\re{Maxtr}) 
of the Maxwell field.

\ers 
We will see that the stationary states 
and the energies for the equation 
(\re{BSfmrs}) depend on the azimuthal quantum number 
(in  contrast to (\re{Ppm}), (\re{PpmE})) 
and describe perfectly the anomalous Zeemann effect as well as 
the Einstein-de Haas experiment. 
First let us 
assume that $\bB=(0,0,B)$ and 
rewrite (\re{BSfmrs}) in the form 
\beqn\la{BSfmf} 
i\h\pa_t\Psi(t,\x)=\ov\cP_m\Psi(t,\x)-\hat\m_3 B\Psi(t,\x), 
\eeqn 
where 
$\ds\hat\m=\fr {e}{2\mu c}\hat\bL+\fr {e}{\mu c}\hat\s$ 
in accordance with the definition  (\re{tmmi}). 
We will apply the perturbation theory to calculate 
the spectrum of the  equation (\re{BSfmf}) for small $|B|$. 
More precisely, we will calculate the corrections to the eigenvalues 
of the unperturbed equation  (\re{BSfmf}) with $B=0$. 
This is sufficient since just these 
corrections describe the splitting of the spectral lines in the 
magnetic field.

We will also calculate the Land\'e factor for quantum stationary states 
of the equation (\re{BSfmf}). 
The factor explains the Einstein-de Haas 
experiment and anomalous Zeemann effect. 
By Definition (\re{tmmi}), the magnetic moment 
for a state $\Psi$ is defined as 
\be\la{tmmh} 
\m=\ds\langle\Psi,\hat\m\Psi\rangle. 
\ee 
The Land\'e factor $g$ for a state $\Psi$ is defined by 
\be\la{dlf} 
\ds\fr{\m_3}{\bJ_3}=g\ds\fr {e}{2\mu c}, 
\ee 
and this factor (or rather its maximal value) 
is measured in the Einstein-de Haas 
experiment.

\subsection{Russell-Saunders Quantum Numbers} 
The main point of the Russell-Saunders method is the analysis of the 
symmetry properties of the problem which 
correspond to the correction term 
$\hat\bL\hat\s$. 
Namely, the operators 
$\cP_m,\hat\bL^2,\hat\bJ^2,\hat\bJ_3:=\hat\bL_3+\hat\s_3$ 
commute with each other. 
On the other hand, 
the momenta $\hat\bL_3$ and $\hat\s_3$ 
do not commute with $\hat\bL\hat\s$ 
and hence with 
the generator $\cP_m$. 
Hence, 
we cannot use the classification of type (\re{Ppm}) 
for the stationary states by the `quantum numbers'  $n,l,m,\pm 1$ 
which correspond to the 
eigenvalues of the operators $\cP,\hat \bL^2,\hat \bL_3,\hat \s_3$. 
Let us  choose 
different quantum numbers.

\bexe 
Check that the operators 
$\cP_m,\hat\bL^2,\hat\bJ^2, \hat\bJ_3$ commute with each other. 
{\bf Hints:}\\ 
i)  (\re{BSfmrs}) implies that 
\be\la{rsqm} 
\cP_m= 
-\fr 1{2\mu}\h^2\De 
+ 
e\phi(|\x|)+ 
\ds\fr{e}{2\mu^2 c^2} 
\ds\fr{\phi'(|\x|)}{|\x|} 
  \hat 
\bL 
\hat\s 
-\fr {e}{2\mu c}[\hat\bL_3+2\hat\s_3]B. 
\ee 
ii) $\hat\bL\hat\s$ commutes with $\hat\bL^2$ since 
each $\hat\bL_n$ and $\hat\s_n$ commutes with $\hat\bL^2$. 
\\ 
iii) $\hat\bJ^2$ commutes with $\hat\bL^2$ since 
$\hat\bJ^2=\hat\bL^2+2\hat\bL\hat\s+\hat\s^2$. 
\\ 
iv) $\hat\bL\hat\s$ commutes with $\hat\bJ^2$ and $\hat\bJ_3$ since 
 $2\hat\bL\hat\s=\hat\bJ^2-\hat\bL^2-\hat\s^2$.

\eexe 
Therefore, we could expect that the operators in  the space 
$\E^\otimes=\E\otimes\C^2$ can be 
simultaneously diagonalized. 
Let us recall that possible eigenvalues of the operators are, respectively, 
$E, \h^2 L(L+1), 
\h^2 J(J+1)$ and $\h M$, where 
$J=0,\fr 12,1,\fr 32,...$ 
and $M=-J,...,J$ 
(see Sections 10.2 and 10.3). 
Let us introduce Dirac's notation. 
 
\bd 
$~~|E,L,J, M\rangle~~$ 
denotes an eigenvector which corresponds to the eigenvalues \\ 
$E, \h^2 L(L+1), 
\h^2 J(J+1),\h M$, 
of the operators $\cP,\hat\bL^2,\hat\bJ^2, \hat\bJ_3$, 
respectively. 
 
\ed

\subsection{Land\'e Formula} 
\bt The Land\'e factor $g=g(\Psi)$ corresponding to 
the stationary state $\Psi=|E,L,J, M\rangle$ is given by 
\be\la{Laf} 
g=\fr 32+ 
\fr{3/4-L(L+1)}{2J(J+1)}. 
\ee 
\et 
\Pr 
The vector $\Psi$ belongs to one of the irreducible 
subspaces invariant with respect 
 to the operators $\hat\bJ_1,\hat\bJ_2,\hat\bJ_3$. 
Indeed, 
$\hat\bJ^2\Psi=\h^2 J(J+1)\Psi$ 
and $\hat\bJ_3\Psi=\h M\Psi$. 
Therefore 
$\Psi$ 
is an element $e_M$ of the canonical basis 
$e_{-J},..., e_{J}$ by Proposition \re{pNew}, 
and all basis vectors $e_{M'}$ are obtained from $e_M$ by an 
application of the operators $\hat\bJ_\pm:=\hat\bJ_1\pm i\hat\bJ_2$.

For a linear  operator $A$ in $\E^\otimes$, let us denote 
the matrix element 
$A^{M,M'}=\langle Ae_M,e_{M'}  \rangle$. Then (\re{tam}) 
and  (\re{tmmi}) imply that 
\be\la{noh} 
\m_3=\hat\m_3^{M,M}=\ds\fr e{2\mu c}(\hat\bJ_3^{M,M}+\hat\s_3^{M,M})= 
\ds\fr e{2\mu c}(\h M+\hat\s_3^{M,M}). 
\ee 
It remains to find the matrix element $\hat\s_3^{M,M}$ and calculate 
$g$ from the definition (\re{dlf}). 
Let us collect the commutators of the operators: 
\be\la{corel} 
\left\{ 
\ba{lll} 
~[\hat\bJ_1,\hat\s_1]=0,&[\hat\bJ_1,\hat\s_2]=i\h\hat\s_3,& 
[\hat\bJ_1,\hat\s_3]=-i\h\hat\s_2, \\ 
~[\hat\bJ_2,\hat\s_2]=0,&[\hat\bJ_2,\hat\s_3]=i\h\hat\s_1,& 
[\hat\bJ_2,\hat\s_1]=-i\h\hat\s_3, \\ 
~[\hat\bJ_3,\hat\s_3]=0,&[\hat\bJ_3,\hat\s_1]=i\h\hat\s_2,& 
[\hat\bJ_3,\hat\s_2]=-i\h\hat\s_1, 
\ea\right. 
\ee 
where the second and the third line follow from the first one by cyclic 
permutations. 
Let us use the commutators  (\re{corel}) to calculate 
the commutators of the operators $\hat\bJ_\pm$ with 
 $\hat\s_\pm:=\hat\s_1\pm i\hat\s_2$. We find 
\be\la{wef} 
[\hat\bJ_-,\hat\s_+]=-2\h\hat\s_3,~~~~~~~~~~~~~~~ 
[\hat\bJ_+,\hat\s_+]=0. 
\ee 
The first formula implies the identity 
\be\la{wefi} 
\hat\bJ_-^{M,M+1}  \hat\s_+^{M+1,M} 
-\hat\s_+^{M,M-1} \hat\bJ_-^{M-1,M} 
=-2\h\hat\s_3^{M,M} , 
\ee 
since all other matrix elements $\hat\bJ_-^{M',M''}=0$ 
for $M''\ne M'+1$ by (\re{actpm}). 
The non-zero matrix elements of $\hat\bJ_\pm$ are known from (\re{actpm}) and 
(\re{actpm'}): 
\be\la{actpm''} 
\hat\bJ_+^{M+1,M}=\h\sqrt{(J-M)(J+M+1)},~~~~~~~ 
\hat\bJ_-^{M-1,M}=\h\sqrt{(J+M)(J-M+1)}. 
\ee 
On the other hand, 
$\hat\s_+$ can be derived from the second 
identity (\re{wef}). Taking the matrix element 
$(\cdot)^{M+1,M-1}$, we derive 
\be\la{wefd} 
\hat\bJ_+^{M+1,M}  \hat\s_+^{M,M-1} 
-\hat\s_+^{M+1,M} \hat\bJ_+^{M,M-1} 
=0. 
\ee 
Then (\re{actpm''}) implies that 
\be\la{wefdi} 
\fr{\hat\s_+^{M+1,M}}{\sqrt{(J-M)(J+M+1)}} 
=\fr{\hat\s_+^{M,M-1}}{\sqrt{(J-M+1)(J+M)}}=:A. 
\ee 
%%%%Then similarly, 
%%%%\be\la{wefì} 
%%%%\fr{\hat\s_+^{j+2,j+1}}{\sqrt{(J-j-1)(J+j+2)}} 
%%%%=A. 
%%%%\ee 
Substituting this into (\re{wefi}) and using 
the matrix elements of 
 $\hat\bJ_-$ from (\re{actpm''}), we get 
\be\la{wefg} 
\hat\s_3^{M,M}=AM. 
\ee 
It remains to calculate the constant $A$. 
We start with the identity 
\be\la{swi} 
\hat\bJ^2=(\hat\bL+\hat\s)^2=\hat\bL^2+2\hat\bL\hat\s+\hat\s^2 
=\hat\bL^2+2\hat\bJ\hat\s-\hat\s^2. 
\ee 
Hence 
\be\la{swm} 
(\hat\bJ\hat\s)^{M,M}=\h^2\ds\fr{J(J+1)-L(L+1)+3/4}2 
\ee 
since $\hat\s^2=3/4$.  On the other hand, 
the same matrix element can be expressed with the help of another identity 
\be\la{swia} 
2\hat\bJ\hat\s=\hat\s_+\hat\bJ_- 
+ 
\hat\s_-\hat\bJ_+ 
+\hat\bJ_3\hat\s_3 
\ee 
as 
\be\la{swmn} 
(\hat\bJ\hat\s)^{M,M}=\ds\fr12 \hat\s_+^{M,M-1}\hat\bJ_-^{M-1,M} 
+ 
\ds\fr12 \hat\s_-^{M,M+1}\hat\bJ_+^{M+1,M} 
+\h M\hat\s_3^{M,M}. 
\ee 
Note that the matrix elements $\hat\s_+^{M+1,M}$ 
and $ \hat\s_-^{M,M+1}$ are complex conjugate. Therefore, we have 
by  (\re{wefdi}) that 
\be\la{wet} 
\hat\s_+^{M+1,M}=A\sqrt{(J-M)(J+M+1)}=\hat\s_-^{M,M+1} , 
\ee 
since the constant $A$ is real by (\re{wefg}). 
Finally, let us substitute (\re{wet}), (\re{swm}), (\re{wefg}) 
and  (\re{actpm''}) 
into (\re{wefdi}). Then we get an equation for $A$  which implies that 
\be\la{Awi} 
A=\h\ds\fr{J(J+1)-L(L+1)+3/4}{2J(J+1)}. 
\ee 
At last, (\re{wefg}) and (\re{noh}) give that 
\be\la{noht} 
\m_3=\ds\fr e{2\mu c}(\h M+AM)= 
\ds\fr e{2\mu c}(1 +A/\h)\h M= 
\ds\fr e{2\mu c}(1 +A/\h)\bJ_3. 
\ee 
Substituting (\re{Awi}) for $A$, we get (\re{dlf}), where $g$ 
coincides with  (\re{Laf}).\bo 
\brs 
i) Our proof follows the calculations in \ci{LVM}. 
\\ 
ii) The formula  (\re{Laf}) has been 
 obtained by Land\'e, \ci{Lan}, 
using the  phenomenological {\it vector model} and the 
Bohr {\it correspondence principle} (see also \ci{Born} 
and \ci[Vol.I]{Som}). 
 
\ers 
\subsection{Application to the Einstein-de Haas and Anomalous 
Zeemann Effects} 
The Land\'e formula  (\re{Laf}) explains satisfactorily 
the Einstein-de Haas experiments 
(see \ci{Born} and 
\ci[Vol.I]{Som}). 
Furthermore, by  {\it perturbation theory} (see \ci{Wig}), 
the Land\'e factor (\re{Laf}) determines the splitting of the 
eigenvalues for    (\re{BSfmf}) 
in the magnetic field $(0,0,B)$ with small $|B|$. 
The resulting correction to the frequency $\om=E/\h$  of the 
eigenstate 
$|E,L,J, M\rangle$ 
is equal to 
\be\la{corL} 
\De \om_{L,J, M}=-\hat\m_3^{M,M} B/\h=-\m_3 B/\h=-g\ds\fr{e} 
{2\mu c}  M B 
=-g\om_\cL M 
\ee 
up to corrections $\cO(B^2)$. 
This formula, together with the 
selection rules  $J\mapsto J'=J\pm 1$, $M\mapsto M'=M,M\pm 1$ 
(suggested by  the Bohr correspondence principle 
\ci{Born} and \ci[Vol.I]{Som}), 
explains the 
multiplet structure in the anomalous Zeemann effect. 
Namely, 
the magnitude of the splitting (\re{corL}), $g\ds\fr{e}{2\mu c} \h B$, 
now depends on the quantum numbers $L,J$ in contrast to 
 (\re{PpmE}) and (\re{nZe}). 
This explains the multiplet structure of the spectrum 
since  (\re{trs}) now becomes 
\be\la{trsb} 
\om_{kk'}=\omega_{kk'}^0 -\om_\cL[g(k)M-g(k')M'] 
=\omega_{kk'}^0 -\om_\cL g(k)[M-M']-\om_\cL[g(k)-g(k')]M', 
\ee 
where $k=(E,L,J,M), k'=(E',L',J',M')$ and 
$\omega_{kk'}^0$ is the 'unperturbed' spectral line corresponding 
to $B=0$. 
Hence, the radiation in the allowed transition 
does not reduce to the normal triplet (\re{trs}). 
 
The calculations are in excellent agreement 
with experimental observations (see \ci{BeS,Born,CS} and 
\ci[Vol.I]{Som}). 
This agreement was one of the greatest successes of quantum theory. 
\br 
The calculation of the splitting  (\re{trsb}) does not depend on 
 the magnitude of the 
unperturbed spectral line $\omega_{kk'}^0$. 
The magnitude can be calculated also with the help 
of the perturbation theory  (see \ci{BeS,CS} and 
\ci[Vol.II]{Som}). 
 
\er

%%%%%%%%%%%%%%%%%%%%%%%%%%%%%%%%%%%%%%%%%%%%%%%%%%%%%%% 
%%%%%%%%%%%%%%%%%%%%%%%%%%%%%%%%%%%%%%%%%%%%%%%%%%%%%%% 
%%%%%%%%%%%%%%%%%%%%%%%%%%%%%%%%%%%%%%%%%%%%%%%%%%%%%%% 
%%%%%%%%%%%%%%%%%%%%%%%%%%%%%%%%%%%%%%%%%%%%%%%%%%%%%%% 
%%%%%%%%%%%%%%%%%%%%%%%%%%%%%%%%%%%%%%%%%%%%%%%%%%%%%%% 
 
\part{Special Relativity}

\setcounter{subsection}{0} 
\setcounter{theorem}{0} 
\setcounter{equation}{0} 
\section{Electromagnetic Nature of Light} 
We deduce the Maxwell equations from 
the Coulomb and Biot-Savart-Laplace 
laws for the interaction of charges and currents. 
We also discuss connections of the Maxwell 
equations to the Einstein ideas on relativity, in particular, 
to the Lorentz transformation. 
 
\subsection{Maxwell Equations. Empirical Synthesis} 
Maxwell stated the equations 
written in the MKS rationalized (or SI) system of units: 
\beqn\la{RmeqMKSr} 
\left\{ 
\ba{ll} 
\dv \bE(t,\x)= \ds\fr{1}{\ve_0}\, \rho(t,\x),& 
\rot \bE(t,\x)= - \dot  \bB(t,\x),\\ 
\dv \bB(t,\x)= 0,&\ds\fr 1{\mu_0}\rot \bB(t,\x)=\bj(t,\x)+ 
\ve_0 \dot \bE(t ,\x). 
\ea 
\right. 
\eeqn 
The equations contain the dielectric permittivity and magnetic 
permeability of the vacuum, 
$\ve_0$ and $\mu_0$, and 
do not contain explicitly the speed of light $c$. 
Maxwell deduced the equations from 
the Coulomb and Biot-Savart-Laplace 
laws for the interaction of charges and currents. 
The Coulomb law states the force of 
an electrostatic interaction of two charges $q_{1,2}$ 
at a distance $r$: 
\be\la{RCoul} 
\bF_2=\ds\fr 1{4\pi\ve_0}\ds\fr{q_1q_2\e_{1,2}}{r^2}, 
\ee 
where $\e_{1,2}$ is the unit vector directed from 
the charge at $q_1$ to the charge at $q_2$, and 
$\bF_2$ is the force acting onto the second elementary charge. 
Similarly, the  Biot-Savart-Laplace 
law states the force of 
a magnetic interaction of two (stationary) 
elementary currents $I_kd\bbl_k$, $k=1,2$, 
at a distance $r$: 
\be\la{RBSL} 
\bF_2=\ds\fr{\mu_0}{4\pi}\ds\fr{I_2d\bbl_2 
\times(I_1d\bbl_1\times \e_{1,2})}{r^2}, 
\ee 
where $\bF_2$ is the force acting onto the 
second elementary current $I_2d\bbl_2$. 
The values of $\ve_0$ and $\mu_0$ 
were measured in a laboratory for electromagnetic experiments 
with a high accuracy. 
In the MKS system 
\be\la{Remu} 
\ve_0\approx \ds\fr 1{4\pi\5\5 9\cdot 10^9}\fr{As}{Vm}, 
~~~~~~~~~~~~~~~~ 
\mu_0\approx 4\pi  10^{-7}\fr{Vs}{Am}. 
\ee 
\subsubsection{The first equation} 
First,  (\re{RCoul}) implies that the electric field 
of the first elementary charge is 
\be\la{RCoulf} 
\bE(\x_2):=\ds\fr{\bF_2}{q_2}= 
\ds\fr 1{4\pi\ve_0}\ds\fr{q_1\e_{1,2}}{|\x_{1,2}|^2}, 
\ee 
where $\x_{1,2}:=\x_2-\x_1$ and $\x_1,\x_2$ are the 
vector-positions of the  charges. Then for a distribution 
of the charges, $\rho(\x_1)d\x_1$, we obtain 
by the principle of superposition, 
\be\la{RCoulfs} 
\bE(\x_2)= 
\ds\fr 1{4\pi\ve_0}\ds\int \fr{\rho(\x_1)\e_{1,2}}{|\x_{1,2}|^2}d\x_1, 
\ee 
Differentiation gives 
\be\la{RCoulfe} 
\dv \bE(\x_2)=\ds\fr 1{\ve_0}\rho(\x_2), 
\ee 
which coincides with the first equation of (\re{RmeqMKSr}). 
\bexe 
Check (\re{RCoulfe}) for $\rho(\x)\in C_0^\infty(\R^3)$. {\bf Hint:} use that 
$\dv\!\!_{\x_2}\fr{\e_{1,2}}{|\x_{1,2}|^2}=\De_{\x_2}\ds\fr 1{|\x_{1,2}|} 
=-4\pi\de(\x_{1,2})$. 
\eexe 
\subsubsection{Last equation for stationary currents} 
Second, (\re{RBSL}) means, by Ampere's law, that the magnetic field 
of the first elementary current  is 
\be\la{RBSLf} 
\bB(\x_2)=\ds\fr{\mu_0}{4\pi}\ds\fr{I_1d\bbl_1\times 
\e_{1,2}}{|\x_{1,2}|^2}. 
\ee 
Integrating, we obtain by the principle of superposition, 
\be\la{RBSLfi} 
\bB(\x_2)=\ds\fr{\mu_0}{4\pi}\int\ds\fr{\bj(\x_1)\times 
\e_{1,2}}{|\x_{1,2}|^2}d\x_1, 
\ee 
where $\bj(\x_1)$ 
is the current density at the point  $\x_1$. 
Differentiating, we get 
\be\la{RBSLfig} 
\rot \bB(\x_2)={\mu_0}\bj(\x_2), 
\ee 
which coincides with the last equation of (\re{RmeqMKSr} 
in the case of the {\bf stationary currents}. 
\bexe 
Check (\re{RBSLfi}) for $\bj(\x)\in C_0^\infty(\R^3)$. 
{\bf Hint:} use the charge continuity 
equation $\dv \bj(\x)\equiv 0$ for the stationary currents. 
\eexe 
\subsubsection{Last equation for nonstationary currents: 
Maxwell's Displacement Current} 
The divergence of the LHS of the last  Maxwell 
equation vanishes since $\dv\rot=0$. On the other hand, 
the divergence of the RHS, $\dv \bj(t ,\x)$ generally 
does not vanish for nonstationary currents. 
To solve this contradiction, 
Maxwell has completed the last 
equation by the {\it displacement current} $\ve_0 \dot \bE(t ,\x)$. 
The current provides a vanishing divergence of the RHS: 
\be\la{R} 
\dv \bj(t ,\x)+\dv\ve_0 \dot \bE(t ,\x)=0 
\ee 
by the charge continuity equation $\dv \bj(t ,\x)+\dot\rho(t ,\x)=0$ 
and the first  Maxwell 
equation. 
\subsubsection{Faraday's equations} 
The second and third  Maxwell 
equations have been divined and checked experimentally by Faraday. 
\subsection{Equations for Potentials} 
Let us repeat the introduction of the 
Maxwell potentials (see Lecture 4) in the MKS system. 
Namely, $\dv \bB(t,\x)= 0$ implies that $\bB(t,\x)= \rot \bA(t,\x)$. 
Then 
$\rot\,\bE(t,\x)= - \dot  \bB\b(t,\x)$ implies 
$\rot\,[\bE(t,\x)+\dot  \bA(t,\x)]=0$ hence 
$\bE(t,\x)+\dot  \bA(t,\x)=-\na_\x\phi(t,\x)$. Finally, 
\be\la{Rpot} 
\bB(t,\x)= \rot \bA(t,\x),\,\,\,\,\,\,\, 
\bE(t,\x)=-\na_\x\phi(t,\x)-\dot  \bA(t,\x),~~~~~~~\,\,\,(t,\x)\in\R^4. 
\ee 
\bexe 
Prove the existence of the potentials (\re{Rpot}). 
{\bf Hint:} 
Use 
the Fourier transform. 
\eexe 
The choice of the potentials is not unique since the {\it gauge transformation} 
\be\la{Rgt} 
\phi(t,\x)\mapsto \phi(t,\x)+\dot\chi(t,\x),\,\,\,\, 
\bA(t,\x)\mapsto \bA(t,\x)-\na_\x\chi(t,\x) 
\ee 
does not change the fields $\bE(t,\x)$,  $\bB(t,\x)$ for any function 
$\chi(t,\x)\in C^1(\R^4)$. 
Therefore, 
it is possible to 
satisfy an additional {\it gauge} condition. Let us choose for example 
the {\it Lorentz gauge} 
\be\la{RLg} 
{\ve_0}\dot\phi(t,\x)+\ds\fr 1{\mu_0}\dv \bA(t,\x)=0,\,\,\,(t,\x)\in\R^4. 
\ee 
Let us express the Maxwell equations (\re{RmeqMKSr}) in terms of 
the potentials. 
Substitution of (\re{Rpot}) into the first Maxwell equation leads to 
$\ds\fr{4\pi}{\ve_0}\rho (t,\x)=\dv \bE(t,\x)= 
-\De\phi(t,\x)-\dv\dot  \bA(t,\x)$. 
Eliminating $\dv\dot  \bA(t,\x)$ 
by 
the differentiation of (\re{RLg}) 
in time, 
${\ve_0}\ddot\phi(t,\x)+\ds\fr 1{\mu_0}\dv \dot \bA(t,\x)=0$, we get 
\be\la{Rdp} 
[{\ve_0}{\mu_0}\pa_t^2-\De]\phi(t,\x)= 
\ds\fr{1}{\ve_0}\rho (t,\x),\,\,\,(t,\x)\in\R^4. 
\ee 
Similarly,  substituting (\re{Rpot}) into 
the last Maxwell equation, we get 
\be\la{RrrA} 
\ds\fr 1{\mu_0}\rot\rot \bA(t,\x)= 
\bj(t,\x)+ 
\ve_0\dot \bE(t ,\x)= 
\bj(t,\x)-\ve_0\na_\x\dot\phi(t,\x) 
-\ve_0\ddot  \bA(t,\x). 
\ee 
\bexe Prove the identity 
\be\la{Rrr} 
\rot\rot =-\De+\na_\x\dv \, . 
\ee 
\eexe 
Substituting  (\re{Rrr}) to (\re{RrrA}) and eliminating 
$\na_\x\dot\phi(t,\x)$ 
by application of $\na_\x$ to (\re{RLg}), we get 
\be\la{RdA} 
[{\ve_0}{\mu_0}\pa_t^2-\De]\bA(t,\x)= 
\mu_0\bj(t,\x),\,\,\,(t,\x)\in\R^4. 
\ee 
\br 
The arguments above show that the 
Maxwell equations (\re{RmeqMKSr}) 
are equivalent to the 
system of two wave equations (\re{Rdp}), (\re{RdA}) for the potentials 
with the Lorentz gauge condition (\re{RLg}). 
\er 
Maxwell deduced the wave equations 
(\re{Rdp}) and (\re{RdA}) with the coefficient $\ve_0\mu_0$ 
and at that time  it  was known that this coefficient 
is equal to $1/v^2$, where $v$ 
is the propagation velocity of the solutions. 
He calculated $v=1/\sqrt{\ve_0\mu_0}$ using the values 
(\re{Remu}). 
The values (\re{Remu}) give $v\approx 3\cdot 10^8 m/s$ which 
almost coincides with the  speed of light $c$. 
Hence the equations (\re{Rdp}) and (\re{RdA}) become 
\be\la{Rdpcm} 
\Box\phi(t,\x)= 
\ds\fr{1}{\ve_0}\rho (t,\x), 
\,\,\,\,\,\,\,\,\,\, 
\Box\bA(t,\x)=\mu_0\bj(t,\x), 
~~~~~~~~~~~~~ 
\,\, 
(t,\x)\in\R^4, 
\ee 
where $\Box:=\ds\fr 1{c^2}\pa_t^2-\De$. 
Furthermore, Maxwell found that the electromagnetic waves 
are transversal like the light waves. 
This is why Maxwell suggested to identify 
 the electromagnetic waves with light. 
Below we will use  {\it Gaussian units} and write (\re{Rdpcm}) 
as 
\be\la{Rdpc} 
\Box\phi(t,\x)= 
4\pi \rho (t,\x), 
\,\,\,\,\,\,\,\,\,\, 
\Box\bA(t,\x)=\fr {4\pi} c\bj(t,\x), 
~~~~~~~~~~~~~ 
\,\, 
(t,\x)\in\R^4. 
\ee 
\medskip\\ 
\subsection{Problem of Luminiferous Ether: 
Michelson and Morley Experiment} 
The discovery of Maxwell led to a new very difficult question. 
The propagation velocity can be equal to $c$ 
only in a {\bf unique preferred reference frame} 
in which the Maxwell equations have the form 
(\re{RmeqMKSr}) or (\re{Rdpc}). 
The preferred reference frame is called the frame of 
the {\it luminiferous ether}. 
In all other frames, 
the propagation velocity 
depends on a direction and  vector of the relative velocity of the frame 
of reference, if space-time is transformed by the  classical 
{\it Galilean transformations} 
\be\la{RGalf} 
\left(\ba{c}t\\x_1\\x_2\\x_3\ea\right) 
\mapsto 
\left(\ba{c}t'\\x_1'\\x_2'\\x_3'\ea\right) 
= 
\left(\ba{c}t\\x_1-vt\\x_2\\x_3\ea\right). 
\ee 
Therefore, the Maxwell equations 
are not invariant with respect to the 
Galilean transformations. 
This is why Michelson and Morley started 
around 1880 the famous experiment to identify the 
 preferred {\it  luminiferous ether frame} with the 
frame in which the Sun is at rest. 
They have tried to check that the Earth moves with respect to the 
luminiferous ether, i.e. the 
velocity of light along and against the velocity 
of the Earth differs by twice the velocity of the Earth. 
Concretely, they 
compared the wavelengths of light along and against the velocity 
of the Earth motion around the Sun. 
However, the result (1887) was negative and very discouraging: 
the wave lengths were 
identical with a high accuracy, hence 
the propagation velocity does not depend on 
the frame of reference! 
Astronomical observations of {\it double stars} 
by de Sitter (1908) 
confirmed the negative result of Michelson and Morley. 
Also the experiment of Trouton and Noble confirmed the 
negative result. 
\medskip\\ 
\subsection{Time in a Moving Frame: Lorentz transformations} 
Various partial explanations of the negative results were proposed 
by Ritz, Fitzgerald, Lorentz and others. The complete explanation 
was provided in 1905 by Einstein, who was able to cumulate 
the Maxwell and  Lorentz ideas into a new complete theory. 
The main novelty was the following postulate of the Einstein theory: 
\medskip 
 
{\bf The time in a moving 
frame is distinct from the time in the rest frame!}\medskip\\ 
We will prove that 
 the transformation of space-time coordinates from the 
rest frame 
to the moving frame of reference is given by the 
{\bf Lorentz formulas} 
\be\la{RLorf} 
\left(\ba{c}ct\medskip\\x_1\medskip\\x_2\medskip\\x_3\ea\right) 
\mapsto 
\left(\ba{c}ct'\medskip\\x_1'\medskip\\x_2'\medskip\\x_3'\ea\right) 
= 
\left(\ba{c}\fr{ct-\beta x_1}{\sqrt{1-\beta^2}}\smallskip\\ 
\fr{x_1-\beta ct}{\sqrt{1-\beta^2}}\smallskip\\x_2\smallskip\\x_3\ea\right)= 
\left( 
\ba{cccc} 
\fr 1{\sqrt{1-\beta^2}}&-\fr \beta{\sqrt{1-\beta^2}}&0&0\\ 
-\fr \beta{\sqrt{1-\beta^2}}&\fr 1{\sqrt{1-\beta^2}}&0&0\\ 
0&0&1&0\\ 
0&0&0&1 
\ea 
\right) 
\left(\ba{c}ct\medskip\\x_1\medskip\\x_2\medskip\\x_3\ea\right) 
\ee 
where $(t,x_1,x_2,x_3)$ stands for the time-space coordinates 
in the rest frame and  $(t',x_1',x_2',x_3')$ corresponds to the 
moving frame if 
the relative velocity is $(v,0,0)$, $|v|<c$ and $\beta:=v/c$ . 
\bexe 
Check that the wave equation 
\be\la{Rweq} 
 [\fr 1{c^2}\,\pa_t^2 -\De]\phi(t,\x)=f(t,\x),\,\,\,\,(t,\x)\in\R^4 
\ee 
 is invariant 
with respect to the transformations (\re{RLorf}). \\ 
{\bf Hints:} 
i) Set $c=1$ and use that $\pa_2=\pa_2'$, $\pa_3=\pa_3'$. 
ii)  Check that the 1D equation 
$ (\pa_t^2-\pa_1^2)\phi(t,x)=f(t,x)$, 
$(t,x)\in\R^2$ is equivalent to 
$ [(\pa_t')^2-(\pa_1')^2)]\phi'(t',x')=f'(t',x')$ if 
$t'=at-bx$ and $x'=ax-bt$ with $a^2-b^2=1$. By definition, 
$\phi'(t',x')=\phi(t,x)$ and $f'(t',x')=f(t,x)$.

\eexe 
 
\bexe\la{R4.5} 
Check that the wave equation (\re{Rweq}) is {\bf not} 
invariant 
with respect to the {\bf Galilean transformation} (\re{RGalf}). 
{\bf Hint:} 
In the new variables, the equation (\re{Rweq}) becomes 
\be\la{Rdpcb} 
[\fr 1{c^2}\,(\pa_{t'}-v\pa_{\x'})^2-\De_{\x'}]\phi'(t',\x')=f'(t',x'), 
\,\,\,(t',\x')\in\R^4. 
\ee 
\eexe 
\br 
For small velocities, $|v|\ll c$, 
the Galilean transformation (\re{RGalf}) is close to the Lorentz one, 
 (\re{RLorf}), and 
the coefficients of the equation (\re{Rdpcb}) are close to 
the ones of (\re{Rweq}). 
\er 
 
%%%%%%%%%%%%%%%%%%%%%%%%%%%%%%%%%%%%%%%%%%%%%%%%%%%%%%%%%%%%%%%%%%%% 
 
%%%%%%%%%%%%%%%%%%%%%%%%%%%%%%%%%%%%%%%%%%%%%%%%%%%%%%%%%%%%%%%%%%%% 
%%%%%%%%%%%%%%%%%%%%%%%%%%%%%%%%%%%%%%%%%%%%%%%%%%%%%%%%%%%%%%%%%%%% 

\newpage 
\setcounter{subsection}{0} 
\setcounter{theorem}{0} 
\setcounter{equation}{0} 
\section{The Einstein Special Relativity  and Lorentz Group} 
The formulas (\re{RLorf}) define 
a particular transformation of the {\it Lorentz group}. 
We define the  Lorentz group as transformations 
of the space-time preserving the form of the wave equation. 
This property follows from the Einstein's postulate 
of Special Relativity Theory extending the {\it Galilei's 
principle of relativity} from mechanics to electrodynamics. 
Namely, the main point of Einstein's {\it special relativity} 
is the following postulate: 
\medskip 
 
{\bf E: Maxwell Equations have an identical form 
in all Inertial Frames} 
\medskip\\ 
The postulate implies  also an invariance of the wave equation 
(\re{Rweq}) in the case $f(x)=0$. 
Let us calculate all possible transformations which 
leave the equation 
with $f(x)\ne 0$ invariant.

First, introduce a new time variable $x_0:=ct$ 
and rewrite the wave equation (\re{Rweq}) in the form 
\be\la{Rweqr} 
 \Box\phi(x):=g^{\al\beta}\pa_\al\pa_\beta 
\phi(x)=f(x),\,\,\,\,x\in\R^4, 
\ee 
where 
$g^{\al,\beta}={\rm diag}\5(1,-1,-1,-1)$, 
$\pa_\al:=\ds\fr{\pa}{\pa x_\al}$ 
and the summation in the repeated indexes is understood, 
$\al,\beta=0,1,2,3$.

We look for maps $\Lam:\R^4\to \R^4$ such that the equation 
keeps its form in new variables $x'=\Lam x$, i.e., (\re{Rweqr}) 
is equivalent to 
\be\la{Rweqre} 
 \Box\phi'(x')=\ 
g_{\al'\beta'}\pa_{\al'}\pa_{\beta'} 
\phi'(x')=f'(x'),\,\,\,\,x'\in\R^4, 
\ee 
where, by definition, $\phi'(x'):=\phi(x)$ and $f'(x'):=f(x)$. 
The map $\Lam$, naturally, i) is continuous and moreover, 
 ii) $\Lam(x+x)=\Lam x+\Lam x$, $x\in\R^4$, 
 which expresses the fact that space-time 
is homogeneous. 
Let us assume, additionally, the normalization condition iii) $\Lam(0)=0$. 
The three conditions imply necessarily that 
$\Lam$ is a {\it linear transformation} of the space $\R^4$ 
defined by a matrix: 
\be\la{Rlinm} 
x_{\al'}'=(\Lam x)_{\al'}=\Lam_{\al'}^\al x_\al, 
~~~~~~~~\al'=0,...,3. 
\ee 
\bexe 
Check that the conditions i), ii), iii),  imply (\re{Rlinm}). 
{\bf Hint:} First, check the corresponding one-dimensional 
 statement with $x\in\R$. 
\eexe 
Finally, rewrite the equation  (\re{Rweqr}) in new variables (\re{Rlinm}). 
By the chain rule 
\be\la{Rchr} 
\pa_\al= 
\fr{\pa x_{\al'}'}{\pa x_\al}\pa_{\al'} = 
 \Lam_{\al'}^\al\pa_{\al'}. 
\ee 
Hence, the equation in the new variables becomes 
\be\la{Rweqrb} 
g_{\al,\beta} 
\Lam_{\al'}^\al 
\Lam_{\beta'}^\beta\pa_{\al'}\pa_{\beta'} 
\phi'(x')=f'(x'),\,\,\,\,x'\in\R^4. 
\ee 
Comparing with  (\re{Rweqre}), 
we get the system of algebraic equations 
\be\la{Raeq} 
g_{\al\beta} 
\Lam_{\al'}^\al 
\Lam_{\beta'}^\beta=g_{\al'\beta'}, 
~~~~~~~~~\al',\beta'=0,1,2,3. 
\ee 
In matrix form, the system is equivalent to the equation 
\be\la{Raeqm} 
\Lam g \Lam^t=g, 
\ee 
where $\Lam^t$ stands for the transposed matrix of $\Lam$. 
\bexe 
Check that 
$|{\rm det}\5\Lam|=1$. {\bf Hint:} take the determinant of 
both sides of  (\re{Raeqm}). 
\eexe 
Hence $\Lam$ is invertible and (\re{Raeqm}) is also equivalent to 
\be\la{Raeqme} 
(\Lam^t)^{-1} g \Lam^{-1}=g, 
\ee 
The matrix equation is equivalent to 
the invariance, with respect to the map $\Lam^{-1}$, 
of the quadratic form 
\be\la{Rqf} 
g(x,x):=( x, gx)= g_{\al\beta}x_\al x_\beta,~~~~~~x\in\R^4 
\ee 
which is called the {\bf Lorentz interval}. 
Then this form is invariant also with respect to the map $\Lam$, hence 
\be\la{Raeqmt} 
\Lam^t g \Lam=g, 
\ee 
or equivalently, 
\be\la{Raeqe} 
g_{\al\beta} 
\Lam_{\al}^{\al'} 
\Lam_{\beta}^{\beta'}=g_{\al'\beta'}, 
~~~~~~~~~\al',\beta'=0,1,2,3. 
\ee

\bd 
 $L$ is the set of all linear maps $\Lam:\R^4\to \R^4$ 
satisfying (\re{Raeqmt}). 
\ed 
 
\bexe 
Check that the set  $L$ is a group. 
\eexe

\bd 
i) 
Minkowski space is the space $\R^4$ endowed with the quadratic form 
(\re{Rqf}). 
\\ 
ii) $L$ is called the {\bf Lorentz group}. 
\ed 
\bex 
The  simplest example of a Lorentz transform is given by the matrices 
\be\la{Rexr} 
\Lam=\hat R:=\left( 
\ba{cc} 
1&0\\ 
0&R 
\ea 
\right), 
\ee 
where $R\in SO(3)$ is a rotation of the 3D space. 
\eex 
\br 
The wave equation (\re{Rweq}) is invariant with respect to 
the transformation $x'=\hat R x$ since the Laplacian $\De$ 
is invariant with respect to the rotations. 
\er 
 
\bexe 
Check (\re{Raeqm}) for the matrix (\re{Rexr}). 
\eexe 
\bexe 
Check that the map $R\mapsto\hat R$ is an isomorphism 
of the rotation group $SO(3)$ onto a subgroup of the Lorentz group $L$. 
\eexe 
 
\bexe 
Construct all Lorentz transformations of the form 
\be\la{Rexrf} 
\Lam=\left( 
\ba{cccc} 
a&b&0&0\\ 
c&d&0&0\\ 
0&0&1&0\\ 
0&0&0&1 
\ea 
\right), 
\ee 
{\bf Solution:} (\re{Raeqm}) is equivalent to the matrix equation 
\be\la{Rmae} 
\left( \ba{cc}a&c\\b&d\ea\right) 
\left( \ba{cc}1&0\\0&-1\ea\right) 
\left( \ba{cc}a&b\\c&d\ea\right) 
= 
\left( \ba{cc}1&0\\0&-1\ea\right) 
\ee 
which is equivalent to a system 
$a^2-c^2=1$, $b^2-d^2=-1$, 
$ab-cd=0$. 
Then $a=\pm\cosh \vp$, $c=\pm\sinh \vp$, 
 $d=\pm\cosh \chi$, $b=\sinh \chi$, and 
$\pm\cosh \vp\sinh \chi\pm\sinh \vp\cosh \chi=0$. 
Therefore, $\tanh\vp = \pm\tanh\chi $ or $\vp = \pm\chi $. 
Finally, we get four one-parametric families of the matrices 
\be\la{Rexrff} 
\Lam^+_\pm:=\left( 
\ba{cccc} 
\cosh \vp&\sinh \vp&0&0\\ 
\pm\sinh \vp&\pm\cosh \vp&0&0\\ 
0&0&1&0\\ 
0&0&0&1 
\ea 
\right),~~~ 
\Lam^-_\pm:=-\left( 
\ba{cccc} 
\cosh \vp&\sinh \vp&0&0\\ 
\pm\sinh \vp&\pm\cosh \vp&0&0\\ 
0&0&1&0\\ 
0&0&0&1 
\ea 
\right) 
\ee 
which are {\bf hyperbolic rotations} 
 (or {\bf boosts}) 
in the angle $\vp$ 
in the plane $x_0,x_1$. 
\eexe 
 
\br 
The Lorentz formulas (\re{RLorf}) 
correspond to the matrix $\Lam_+^+$ from 
(\re{Rexrff}) with $\cosh \vp=1/\sqrt{1-\beta^2}$ 
and $\sinh \vp=-\beta/\sqrt{1-\beta^2}$, so $\tanh \vp=-\beta$. 
\er

%%%%%%%%%%%%%%%%%%%%%%%%%%%%%%%%%%%%%%%%% 
%%%%%%%%%%%%%%%%%%%%%%%%%%%%%%%%%%%%%%%%% 
%%%%%%%%%%%%%%%%%%%%%%%%%%%%%%%%%%%%%%%%% 
 
\newpage 
\setcounter{subsection}{0} 
\setcounter{theorem}{0} 
\setcounter{equation}{0} 
\section{Covariant Electrodynamics} 
Here we complete a justification of 
the Einstein postulate {\bf E}. Namely, 
we still have to determine the transformation of the Maxwell Field 
by the Lorentz group. 
 
Let us introduce the 4D fields and currents 
\be\la{Rr4vf} 
\A(x)=(\phi(t,\x), \bA(x)),~~~~ 
\J(x)=(\rho(x), \ds\fr 1c\,\bj(x)),~~~\,\,\,\,\,\,\,\,\,x\in\R^4. 
\ee 
In this notation, the Maxwell equations 
(\re{Rdpc}) become 
\be\la{R4m} 
\Box \A(x)=4\pi \J(x),\,\,\,\,x\in\R^4. 
\ee 
Hence, the transformations for the potentials 
$\A^\mu(x)$ and the currents $\J^\mu(x)$ must be identical 
since the wave operator $\Box$ is invariant with respect to 
the Lorentz group. We will prove below that the convective 
currents are transformed, 
like the 4-vector $x$, by the matrix $\Lam$. 
Hence, the same holds for the potentials and all currents, so 
\be\la{RrAJ} 
\A'(x')=\Lam\A(x),~~~~~~ 
\J'(x')=\Lam\J(x), ~~~~~~x'=\Lam x. 
\ee 
The Lagrangian approach gives another confirmation for 
(\re{RrAJ}). 
Namely, 
in Lecture 5 we have shown that the Maxwell equations are the canonical 
Euler-Lagrange equations corresponding to the Lagrangian density 
\be\la{RmtL} 
\cL(x,\A,\na\A) 
=-\fr 1{16\pi}\,\F^{\mu\nu}\F_{\mu\nu}-g(\J,\A), 
\,\,\,\,\,\,(x,\A,\na\A)\in\R^4\times\R^4\times\R^{16}. 
\ee 
where $\F^{\mu\nu}:=\pa^\mu\A^\nu-\pa^\nu\A^\mu$, 
$\F_{\mu\nu}:=g_{\mu\mu'}g_{\nu\nu'}\F^{\mu'\nu'} 
$ and 
\be\la{RgJA} 
g(\J,\A):=g_{\mu\nu}\J^\mu\A^\nu. 
\ee 
The Einstein postulate will hold if the density is invariant under 
the corresponding transformations for the fields. 
It is easy to check that the density is invariant if 
the potentials 
$\A^\mu(x)$ and the currents $\J^\mu(x)$ 
are transformed by (\re{RrAJ}). 
\bexe 
Check that  the density (\re{RmtL}) is invariant under the transformations 
 (\re{RrAJ}). 
{\bf Hint:} 
The bilinear form (\re{RgJA}) is symmetric and 
invariant since the corresponding 
quadratic form is invariant. 
\eexe

At last, let us check that the transformation 
law  (\re{RrAJ}) holds for the convective currents. 
It suffices to consider only point charges since a general 
convective current is a combination of moving point charges. 
Let $\x=\x(x_0)$ be a $C^1$ trajectory of a point charge, 
so its  4-current is given by 
\be\la{Rpoc} 
\J^0(x)=e\de(\x-\x(x_0)),~~~~~~~ 
\J^k(x)=e \bv_k(x_0)\de(\x-\x(x_0)), 
\ee 
where $\bv_k(x_0):=\ds\fr{d \x_k(x_0)}{dx_0}$. 
\br 
The expressions (\re{Rpoc}) 
for the convective current 
are equivalent to the postulating 
 of the {\it charge-invariance} 
under the motion 
which is referred  traditionally as to an 
experimental fact (``charge invariance'', \ci{Heitler,Jack}). 
\er
\bp 
The function (\re{Rpoc}) is Lorentz-covariant, i.e. 
for every Lorentz transformation $\Lam\in\bL$, 
\be\la{RLoc} 
\J'(x')=\Lam\J(x),~~~~~~~x'=\Lam x. 
\ee 
\ep 
\Pr 
Since $\J(x)$ is a distribution, (\re{RLoc}) means that 
\be\la{RLocm} 
\langle\J'(x'),\vp'(x')\rangle= 
\langle\Lam\J(x),\vp(x)\rangle 
\ee 
for any test function $\vp(x)\in C_0^\infty(\R^4)$, 
where $\vp'(x'):=\vp(x)$, $x'=\Lam x$. Here we have used that 
$|{\rm det}\5\Lam|=1$ by   (\re{Raeqm}). 
Since  $\J(x)$ is a measure, it suffices to check  (\re{RLocm}) for 
 characteristic functions $\vp(x)$ of any open bounded set 
$\Om\subset\R^4$. Then (\re{RLocm}) becomes, 
\be\la{RLocmb} 
\int_{\Om'}\J'(x')d^4x'=\Lam\int_\Om\J(x) d^4x. 
\ee 
Let us assume that the set $\{x\in\Om:\x=\x(x_0)\}$ 
is an interval of the trajectory $\{(x_0,\x(x_0)): 
a<x_0<b\}$. Then integrating first in $d\x$, 
we get 
\be\la{RLoc1} 
\int_\Om\J(x) d^4x=e\int_a^b (1,\bv(x_0))dx_0=e 
(x_0,\x(x_0))\Bigg|_a^b . 
\ee 
It implies (\re{RLocmb}) since $(x_0,\x(x_0))$ is an invariant 4-vector, i.e. 
$(x_0',\x'(x_0')):=\Lam(x_0,\x(x_0))$.\bo 
\medskip\\ 
\brs i) 
The charge invariance (\re{Rpoc}) 
implies the transformation
(\re{RLoc}) for the convective current. 
For all other currents, the transformation is postulated.
\\ 
ii) The density 
$(1,\bv(x_0))$ is {\bf not} an invariant 4-vector. On the other hand, 
the differential form 
$(1,\bv(x_0))dx_0$ is Lorentz-invariant which follows from 
the integration (\re{RLoc1}). 
\ers 
Let us note that the form $(1,\bv(x_0))dx_0$ 
admits the factorization 
\be\la{Rafa} 
(1,\bv(x_0))dx_0=\fr{(1,\bv(x_0))}{\sqrt{1-\bv^2(x_0)}} 
~dx_0{\sqrt{1-\bv^2(x_0)}}, 
\ee 
where the fraction in RHS is an invariant 4-vector while the 
remaining factor is an invariant differential form. 
Indeed, 
\be\la{Rafai} 
w(x_0):=\fr{(1,\bv(x_0))}{\sqrt{1-\bv^2(x_0)}}= 
\fr{d(x_0,\x(x_0))}{dx_0\sqrt{1-\bv^2(x_0)}}= 
\fr{d(x_0,\x(x_0))}{\sqrt{(dx_0)^2-(d\x(x_0))^2}} 
\ee 
is an invariant 4-vector since 
the Lorentz interval 
$(dx_0)^2-(d\x(x_0))^2$ is Lorentz-invariant. 
Respectively, $dx_0{\sqrt{1-\bv^2(x_0)}}=\sqrt{(dx_0)^2-(d\x(x_0))^2}$ 
is a Lorentz-invariant differential form. 
\bd 
i) The vector $w(x_0)$ from (\re{Rafai}) is the {\bf 4-dimensional velocity} 
 of a particle.\\ 
ii) The function $\tau(b):=\ds\int_a^b\sqrt{(dx_0)^2-(d\x(x_0))^2}= 
\ds\int_a^b{\sqrt{1-\bv^2(x_0)}}dx_0$ 
is the {\bf proper time} of the particle along the trajectory. 
\ed 
Now (\re{Rafa}) can be written as 
\be\la{Rafab} 
(1,\bv(x_0))dx_0=w(x_0)d\tau(x_0). 
\ee 
\bc 
The formulas (\re{RrAJ}) imply the corresponding transformation 
for the Maxwell tensor (\re{tenf}): 
\be\la{RrAJt} 
\F'^{\mu'\nu'}(x)=\Lam_{\mu}^{\mu'}\Lam_{\nu}^{\nu'} 
\F^{\mu\nu}(x), 
 ~~~~~~~~~~~~x'_{\mu'}=\Lam _{\mu'}^\mu x_\mu. 
\ee 
In the matrix form 
\be\la{RrAJtm} 
\F'(x)=\Lam 
\F(x)\Lam^t. 
\ee 
\ec 
\bex 
Applying this formula to the Lorentz bust (\re{RLorf}), 
we get the Lorentz transfromation for the Maxwell fields 
: 
\be\la{Maxtrf} 
\left\{ 
\ba{lll} 
\bE_1'(x')=\bE_1(x),& & \bB_1'(x')=\bB_1(x),\\\\ 
\bE_2'(x')=\ds\fr{\bE_2(x)-\beta \bB_3(x)}{\sqrt{1-\beta^2}}, 
&& \bB_2'(x')=\ds\fr{\bB_2(x)+\beta \bE_3(x)}{\sqrt{1-\beta^2}}, 
\\\\ 
\bE_3'(x')=\ds\fr{\bE_3(x)+\beta \bB_2(x)}{\sqrt{1-\beta^2}}, 
&&\bB_3'(x')=\ds\fr{\bB_3(x)-\beta \bE_2(x)}{\sqrt{1-\beta^2}}. 
\ea 
\right.~~~~~~~~~~~~~~~~~~~~~~~~~ 
\ee

\eex 
Equivalently, 
\be\la{Maxtr} 
\bE'(x')=\ds\fr{\bE(x)+\ds\fr \bv c\times \bB(x)} 
{\sqrt{1-\beta^2}},~~~~~~~~~ 
\bB'(x')=\ds\fr{\bB(x)-\ds\fr \bv c\times \bE(x)} 
{\sqrt{1-\beta^2}}. 
\ee

\bexe 
Check  the formulas (\re{Maxtrf}). {\bf Hints:} 
The formulas follow from  (\re{Maxten}) by 
(\re{RrAJtm}) with 
the Lorentz matrix (\re{RLorf}). 
\eexe

\part{Relativistic Dirac Theory}

%%%%%%%%%%%%%%%%%%%%%%%%%%%%%%%%%%%%%%%%%%%%%%%%%% 
%%%%%%%%%%%%%%%%%%%%%%%%% 
%%%%%%%%%%%%%%%%%%%%%%%%% 
\newpage 
\setcounter{subsection}{0} 
\setcounter{theorem}{0} 
\setcounter{equation}{0} 
\section{Relativistic Equation for Electron Field} 
The Special Relativity of Einstein has to be extended 
to the quantum mechanical theory, as well. 
We introduce the Dirac equation which is the  relativistic 
covariant generalization 
of the Schr\"odinger equation.

The 
Schr\"odinger equation obviously 
is not invariant with respect to the Lorentz group. 
One possible choice is the Klein-Gordon 
equation 
\be\la{RKG} 
\fr {\h^2}{c^2}\pa_t^2\psi(x)=\h^2\De\psi(x) 
-\mu^2c^2\psi(x),\,\,\,~~~~x\in\R^{4}, 
\ee 
which is Lorentz-invariant as well as the wave equation 
(\re{Rweq}). 
However, it leads to negative energies which is not satisfactory 
from a physical point of view. 
Namely, the Klein-Gordon 
equation is obtained from the relativistic energy-momentum 
relation 
\be\la{REp} 
\fr{E^2}{c^2}=\p^2+\mu ^2c^2 
\ee 
by the Schr\"odinger identification 
$E\mapsto i\h\pa_t$, $\p\mapsto -i\h\na_\x$. 
In the Fourier transform, 
$\hat\psi(x^0,\p):=\ds\int e^{i\fr{\p\x}\h}\psi(x_0,\x)d\x$, 
the Klein-Gordon equation becomes 
\be\la{RKGf} 
\h^2\fr {\pa^2}{(\pa x^0)^2}\hat\psi(x_0,\p)=-\p^2\hat\psi(x_0,\p) 
-\mu^2c^2\hat\psi(x_0,\p),\,\,\,~~~~x_0\in\R,~\p\in\R^3. 
\ee 
Then the solutions are linear combinations of 
$\ds e^{ i\fr E\h t}$ where 
$E/c=\pm\sqrt{\p^2+\mu^2c^2}$ coincides with (\re{REp}). 
The solutions with  $E/c=-\sqrt{\p^2+\mu^2c^2}$ seem to 
correspond to negative energies unbounded from below, 
hence the physical interpretation of the Klein-Gordon equation 
requires an additional analysis. 
 
This is why Dirac tried to find a relativistic 
invariant equations of the first order in time, 
like the Schr\"odinger equation, 
to avoid the negative roots. 
Let us follow Dirac's arguments. First, 
the relativistic invariance 
requires then the first order in space. 
Second, we will look for a linear equation 
of the form 
\be\la{Remrd} 
\ga^\al i\h\fr\pa{\pa x^\al}\psi(x)=\mu c\psi(x),~~~~~~x\in\R^4. 
\ee 
Then the corresponding energy-momentum relation can be 
written as 
\be\la{Remr} 
\ga(p)=\mu c,~~~~~~p\in\R^4, 
\ee 
where $p_\al:=(E/c,-\p)$ and $\ga(p)$ is a linear form: 
\be\la{Remrp} 
\ga(p)=\ga^\al p_\al,~~~~~~p\in\R^4. 
\ee 
The third condition is a ``correspondence principle'': 
the equation (\re{Remrd}) must imply 
the Klein-Gordon equation (\re{RKG}). 
Namely, applying the operator $\ds\ga^\al i\h\fr\pa{\pa x^\al}$ 
to both sides, we get 
\be\la{Remrg} 
[\ga^\al i\h\fr\pa{\pa x^\al}]^2\psi(x)=\mu^2 c^2\psi(x),~~~~~~x\in\R^4. 
\ee 
Then the correspondence principle 
is equivalent to the algebraic identity 
\be\la{Remra} 
\ga^2(p)=p_0^2-\p^2,~~~~~~p=(p_0,\p)\in\R^4 
\ee 
since  (\re{REp}) can be written as $p_0^2-\p^2=\mu^2 c^2$. 
Dirac's extra idea was the choice of the 
coefficients $\ga^\al$ in the matrix algebra 
since the scalar coefficients do not exist. 
The existence of the scalar coefficients would mean 
that the polynomial $p_0^2-\p^2$ is reducible which 
is not true. 
\bexe 
Check that (\re{Remra}) is impossible with any scalar coefficients 
$\ga^\al$. 
\eexe 
\bt 
In $2\times 2$ block form, the matrices 
\be\la{Rgp} 
\ga(p)= 
\left(\ba{cc}    p_0&\si\cdot\p\\ 
            -\si\cdot\p&-p_0\ea\right) 
\ee 
satisfy the identity (\re{Remra}), where $\si:=(\si_1,\si_2,\si_3)$ 
are the Pauli spin matrices. 
\et 
\Pr 
By direct multiplication of $2\times 2$ block matrices, we get 
\be\la{Rgpg} 
\ga^2(p)= 
\left(\ba{cc}    p_0^2-(\si\cdot\p)^2& 0 \\ 
                        0            & p_0^2-(\si\cdot\p)^2\ea\right). 
\ee 
It remains to use that $(\si\cdot\p)^2=|\p|^2$.\bo 
\medskip\\ 
Now let us consider the matrices $\ga^\al=\ga(e_\al)$, 
where $e_0=(1,0,0,0)$, etc. From (\re{Rgp}), we get 
\be\la{Rgpgg} 
\ga^0= 
\left(\ba{cc} 1&0\\ 
                0&-1\ea\right), 
~~~~~~~~~~~~ 
\ga^j= 
\left(\ba{cc}    0&\si_j\\ 
            -\si_j&0\ea\right),~~j=1,2,3. 
\ee 
\bp 
The matrices $\ga^\al$ satisfy 
\be\la{Rgco} 
\left\{ 
\ba{l} 
(\ga^0)^2=1,~~~~(\ga^j)^2=-1,~~j=1,2,3,\\ 
\\ 
\ga^\al\ga^\beta+\ga^\beta\ga^\al=0,~~\al\ne\beta. 
\ea 
\right. 
\ee 
\ep 
\Pr 
Let us rewrite (\re{Remra}) in the form 
\be\la{Remrf} 
\ga^2(p)=g(p),~~~~~~p\in\R^4, 
\ee 
where $g(p):=p_0^2-\p^2$. 
It implies 
\be\la{Remri} 
\ga(p)\ga(q)+\ga(q)\ga(p)=2g(p,q), 
\ee 
where $g(p,q)=p_0q_0-\p\q$ is the corresponding 
symmetric bilinear form. 
In particular, for $p=e_\al$ and $q=e_\beta$ 
\be\la{Remrip} 
\ga^\al\ga^\beta+\ga^\beta\ga^\al=2g(e_\al,e_\beta) 
\ee 
which implies (\re{Rgco}).\bo 
\br 
Obviously the matrices (\re{Rgpgg}) are not unique 
solutions to the relations (\re{Rgco}): for example, 
we can replace $\ga_\al$ by  $-\ga_\al$ for certain 
indexes $\al$. We will prove below  the Pauli Theorem: 
the matrices $\ga_\al$ are unique up to 
a change of an orthonormal basis. 
\er 
Let us rewrite  the Dirac equation  (\re{Remrd}) 
in a ``Schr\"odinger form''. First, 
it is equivalent to 
\be\la{Rdirf} 
i\h\ga^0\dot\psi(t,\x)=c(\mu c-i\h\ga^j\fr\pa{\pa x^j})\psi(t,\x). 
\ee 
Second, multiplying by $\ga^0$ and using that $(\ga^0)^2=1$, we get 
\be\la{Rdirs} 
i\h\dot\psi(t,\x)=\cH_D\psi(t,\x):= 
c\ga^0(\mu c-i\h\ga^j\fr\pa{\pa x^j})\psi(t,\x), 
\ee 
where the operator $\cH_D$ is called the {\it free Dirac Hamiltonian}.

%%%%%%%%%%%%%%%%%%%%%%%%%%%%% 
%%%%%%%%%%%%%%%%%%%%%%%%%%%%% 
 
\newpage 
\setcounter{subsection}{0} 
\setcounter{theorem}{0} 
\setcounter{equation}{0} 
\section{Problem of Negative Energies for Dirac Equation} 
We prove the energy conservation for the Dirac equation 
and check that the energy is not bounded from below. 
The Dirac equation (\re{Rdirs}) 
is a Hamiltonian system with the Hamilton 
functional
\be\la{Rede} 
H(\psi):=\langle \psi(\x),\cH_D\psi(\x)\rangle , 
\ee 
in analogy with the Schr\"odinger equation. It is conserved, i.e. 
\be\la{Rencd} 
H(\psi(t,\cdot))=\co,~~~~~t\in\R 
\ee 
for the solutions to 
 (\re{Rdirs}). 
\bexe 
Check (\re{Rencd}). {\bf Hint:} 
Differentiate  (\re{Rede}) and use the Dirac equation 
(\re{Rdirs}) and the symmetry of the Dirac operator $\cH_D$. 
\eexe 
The quadratic form (\re{Rede}) is not positive definite, hence 
energy conservation (\re{Rencd}) does not provide an 
a priori estimate for the solutions. 
To see this, it is useful to split  each Dirac spinor 
into a pair of two-component vectors 
\be\la{Rtcs} 
\psi(\x)=\left(\ba{l}\psi_+(\x)\\\psi_-(\x)\ea\right). 
\ee 
Define the Fourier transform 
\be\la{Rtcsf} 
\hat\psi(\p)=\int e^{i\p\x/\h}\psi(\x)d\x,~~~~~\p\in\R^3. 
\ee 
Then the quadratic form (\re{Rede}) becomes by the Parseval identity 
$$ 
\ba{ll} 
H(\psi)&=(2\pi)^{-3}\langle \hat\psi(\p),c\ga^0 
\left( 
\ba{cc} 
\mu c&-\si\cdot\p\\ 
\si\cdot\p&\mu c 
\ea 
\right) 
\hat\psi(\p)\rangle 
=c(2\pi)^{-3}\langle \hat\psi(\p), 
\left( 
\ba{cc} 
\mu c&-\si\cdot\p\\ 
-\si\cdot\p&-\mu c 
\ea 
\right) 
\hat\psi(\p)\rangle 
\\\\\\ 
~~~~~~~~&=c(2\pi)^{-3}\Big[ 
\mu c 
\langle \hat\psi_+(\p), 
\hat\psi_+(\p)\rangle 
- 
2 
\langle \hat\psi_+(\p), 
\si\cdot\p~ 
\hat\psi_-(\p)\rangle 
- 
\mu c 
\langle \hat\psi_-(\p), 
\hat\psi_-(\p)\rangle 
\Big]. 
\ea 
$$ 
In particular, 
\be\la{Redpm} 
~~~~~~~ 
H\left(\!\!\ba{c}\psi_+(\x)\\0\ea\!\!\right)\!=\!(2\pi)^{-3} 
\mu c^2 
\langle \hat\psi_+(\p), 
\hat\psi_+(\p)\rangle, 
~~ 
H\left(\!\!\ba{c}0\\\psi_-(\x)\ea\!\!\right)\!=\!-(2\pi)^{-3} 
\mu c^2 
\langle \hat\psi_-(\p), 
\hat\psi_-(\p)\rangle 
\ee 
The negative energy 
might lead to an instability of the Dirac dynamics 
due to a possible transition of the solution to the states 
$\left(\!\!\ba{c}0\\\psi_-(\x)\ea\!\!\right)$. 
On the other hand, this instability 
has never been proved. 
 
Dirac suggested that the transition of all particles is forbidden 
by the Pauli 
exclusion principle since almost all states with negative energy 
have been occupied long ago. 
%%CA 
%% I replaced the formulation: 
%% This suggestion is formalized by Quantum Field Theory. 
%% by the one below, because Quantum field Theory works also 
%% for bosons, where there is no Pauli exclusion principle. 
(A more satisfactory solution of the problem of negative energies, 
which is applicable also for bosons, is provided by Quantum Field 
Theory.) 
On the other hand, by the Dirac theory, 
the transitions for certain particles are possible, 
and the 'negative states' can be interpreted as 
the states with positive energy for 
{\it antiparticles} which are {\it positrons}, i.e., 
'electrons with positive charge' $-e$. 
The positrons have been discovered 
in {\it cosmic rays} 
by Anderson in 1932.

%%%%%%%%%%%%%%%%%%%%%%%%%%%%% 
%%%%%%%%%%%%%%%%%%%%%%%%%%%%% 
 
%%%%%%%%%%%%%%%%%%%%%%%%%%%%% 
 
\newpage 
\setcounter{subsection}{0} 
\setcounter{theorem}{0} 
\setcounter{equation}{0} 
\section{Angular Momentum for Dirac Equation} 
We prove the angular momentum conservation 
for the Dirac equation. The conservation justifies the 
Goudsmith-Uhlenbeck conjecture on the electron spin 
as an intrinsic property of the dynamical equations. 
 
The conserved orbital momentum for the Schr\"odinger equation 
is defined by 
(\re{qvf}) and (\re{qvfo}): 
$\bL=\bL(\psi):=\langle\psi,\hat\bL\psi\rangle$, where 
$\hat\bL=-i\h\x\times\na$. For the solutions to the Dirac 
equation the orbital momentum generally is not conserved, 
since the operator $\hat\bL$ does not commute with the Dirac operator 
$H_D$. Hence, for the Dirac equation 
the definition of the angular momentum requires a modification. 
\bd\la{RdamD} 
For the Dirac equation 
the angular momentum is $\bJ=\bJ(\psi)= \langle\psi,\hat\bJ\psi\rangle$ 
where 
\be\la{Rtsi} 
\hat\bJ:=\hat\bL+\ds\fr12 \h\ti\si,~~~~~~~~~~ 
\ti\si:=\left( 
\ba{cc} 
\si&0\\ 
0&\si 
\ea\right). 
\ee 
 
\ed 
 
\bt\la{RtamcD} 
The angular momentum  $\bJ$ is conserved for the 
solutions to the Dirac equation, 
\be\la{RamcD} 
\bJ(\psi(t,\cdot))=\co,~~~~~~~~t\in\R. 
\ee 
\et 
\Pr 
Differentiating, we obtain similarly to 
(\re{Heb}), 
%%CA to what does the above reference refer? 
\beqn\la{Rhre} 
\fr d{dt}\bJ(\psi(t))\!\!&\!\!=\!\!&\!\!\langle\dot\psi(t), 
\hat\bJ\psi(t)\rangle+\langle\psi(t), 
\hat\bJ\dot\psi(t)\rangle=- 
\langle\ds\fr i\h\cH_D\psi(t), 
\hat\bJ\psi(t)\rangle-\langle\psi(t), 
\hat\bJ\fr i\h\cH_D\psi(t)\rangle\nonumber\\ 
\!\!&\!\!=\!\!&\!\! 
-\ds\fr i\h\Big[\langle\cH_D\psi(t), 
\hat\bJ\psi(t)\rangle-\langle\psi(t), 
\hat\bJ\cH_D\psi(t)\rangle\Big] 
=-\ds\fr i\h 
\langle\psi(t),[\cH_D , \hat\bJ] 
\psi(t)\rangle. 
\eeqn 
It suffices to check the commutation $[\cH_D , \hat\bJ]=0$. 
First, we know the commutators $[\hat\bL_k,p_j]=i\h\eps_{kjl}p_l$, 
where $p_j:=i\h\pa_j$ and $\eps_{kjl}$ is the totally antisymmetric tensor. 
Therefore, 
\be\la{Rclh} 
[\hat\bL_k,\cH_D]=-c\h\ga^0\ga^j[\bL_k,p_j]=-ic\h\ga^0\ga^j\epsilon_{kjl}p_l. 
\ee 
This shows that $\cH_D$ does not commute with the orbital angular momentum 
operators $\hat\bL_k$, hence the orbital momentum $\bL$ generally 
is not conserved. It remains to calculate the commutators $[\ti\si,\cH_D]$. 
First we note that 
$$ 
\ga^0\ga^l= 
\left( 
\ba{cc} 
1&0\\ 
0&-1 
\ea\right) 
\left( 
\ba{cc} 
0&\si_l\\ 
-\si_l&0 
\ea\right)= 
\left( 
\ba{cc} 
0&\si_l\\ 
\si_l&0 
\ea\right). 
$$ 
Hence, 
$$ 
[\ti\si_k,\cH_D]=[\ti\si_k,\mu c^2\ga^0-c\ga^0\ga^lp_l]= 
-c 
\left( 
\ba{cc} 
0&    [\si_k,\si_l]\\ 
~[\si_k,\si_l] &0 
\ea\right)p_l. 
$$ 
The commutation relations for the Pauli spin matrices 
allows us to reduce this to 
$$ 
-2ic\eps_{klj}\left( 
\ba{cc} 
0&    \si_j\\ 
\si_j &0 
\ea\right)p_l=-2ic\eps_{klj}\ga^0\ga^jp_l. 
$$ 
Multiplying this by $\h/2$ and adding 
(\re{Rclh}), we get the commutation $[\cH_D , \hat\bJ]=0$ by 
the antisymmetry of $\eps_{klj}$.\bo

\brs 
i) Theorem \re{RtamcD} means that $\ds\fr12\h\ti\si$ represents an intrinsic 
spinor angular momentum of the relativistic electron. 
\\ 
ii) We will demonstrate below that the coupling of the Dirac equation 
to the magnetic field provides 
automatically the correct Land\'e factor 
$g=2$ for the spinor angular momentum. 
\ers 
%%%%%%%%%%%%%%%%%%%%%%%%%%%%% 
%%%%%%%%%%%%%%%%%%%%%%%%%%%%%% 
%%%%%%%%%%%%%%%%%%%%%%%%%%%%%% 
%%%%%%%%%%%%%%%%%%%%%%%%%%% 
 
\newpage 
\setcounter{subsection}{0} 
\setcounter{theorem}{0} 
\setcounter{equation}{0} 
\section{Pauli Theorem} 
We prove the Pauli theorem on the uniqueness of the Dirac matrices. 
The theorem states that the Dirac matrices 
(\re{Rgpgg}) are the unique 
representations of the relations 
(\re{Rgco}), up to equivalence. 
%%%%Let us note  that in (\re{Rgpgg}), 
%%%%the matrix 
%%%%$\ga^0$ is Hermitian, and 
%%%%$\ga^k$, $k=1,2,3$, are anti-Hermitian. 
\bt 
Let $\ga^\al$, $\al=0,...,3$ be operators on a finite-dimensional 
complex vector space $V$, $\dim V\le 4$, 
which satisfy the relations (\re{Rgco}). 
Then $\dim V= 4$ and 
the operators 
$\ga^\al$, for 
a suitable choice of a basis, 
take the matrix form (\re{Rgpgg}). 
%%%%\\ 
%%%%ii) 
%%%%Let additionally, for each $\al$, the operator 
%%%%$\ga^\al$ be either Hermitian or anti-Hermitian. 
%%%%Then $V$ is a direct sum of four-dimensional subspaces 
%%%%in which $\ga^\al$, for 
%%%%a suitable choice of an orthonormal basis, 
%%%%take the matrix form (\re{Rgpgg}) up to a sign. 
\et 
\Pr 
{\it Step i)} A key idea of the proof is the following simple 
characterization of the basis vectors. Namely, 
for the standard Dirac matrices (\re{Rgpgg}), 
the matrix 
$\ga^1\ga^2$ is diagonal and hence commutes with 
$\ga^0$ which is also diagonal: 
$\ga^0$ and $\ga^1\ga^2$ have the diagonal block matrix form 
\be\la{Rbf} 
\ga^0= 
\left(\ba{cc}   1 & 0\\ 
                0 & -1\ea\right),~~~~~~~ 
\ga^1\ga^2= 
\left(\ba{cc}   -i\si_3 & 0 \\ 
                 0      &-i\si_3\ea\right). 
\ee 
Therefore, the basis vectors $e_0,...,e_3$ are common eigenvectors 
of the matrices $\ga^0$ and $\ga^1\ga^2$ with the eigenvalues 
$1$ and $-i$, $1$ and $i$, $-1$ and $-i$, $-1$ and $i$ respectively. 
\\ 
{\it Step ii)} 
Now apply this observation to the general matrices $\ga^\al$ 
from the theorem. 
The anticommutation relations from (\re{Rgco}) imply that 
the matrices $\ga^0$ and $\ga^1\ga^2$ commute with each other: 
\be\la{Rcom} 
\ga^0\ga^1\ga^2=\ga^1\ga^2\ga^0. 
\ee 
\bexe 
Check this commutation. {\bf Solution:} 
$ 
\ga^0\ga^1\ga^2=-\ga^1\ga^0\ga^2=\ga^1\ga^2\ga^0. 
$ 
\eexe 
Hence there exists at least one common eigenvector $\Om_1$ 
for both (since $V$ is a {\it complex vector space}!): 
\be\la{Rcev} 
\ga^0\Om_1=\al\Om_1~~~~~\mbox{and}~~~~~\ga^1\ga^2\Om_1=\beta\Om_1 
\ee 
where $\al$ and $\beta$ are suitable complex numbers. 
\medskip\\ 
{\it Step iii)} 
$\al^2=1$ since $(\ga^0)^2=1$, and similarly 
$\beta^2=-1$ since $(\ga^1\ga^2)^2=-(\ga^1)^2(\ga^2)^2=-1$. 
Hence, $\al=\pm 1$ and $\beta=\pm i$. Let us check that 
all four combinations of the signs are possible for 
suitable eigenvectors $\Om_1$. Namely, 
$$ 
\ga^0\ga^3\Om_1=-\ga^3\ga^0\Om_1=-\al\ga^3\Om_1, 
$$ 
and 
$$ 
\ga^1\ga^2\ga^3\Om_1=\ga^3\ga^1\ga^2\Om_1=\beta\ga^3\Om_1, 
$$ 
hence the vector $\Om_3:=\ga^3\Om_1$ is also a common eigenvector 
with the eigenvalues $-\al$ and $\beta$. Similarly, 
$\Om_2:=-\ga^3\ga^1\Om_1$ resp. $\Om_4:=-\ga^1\Om_1$ are common 
eigenvectors of the operators 
$\ga^0$ and $\ga^1\ga^2$ with the eigenvalues $\al$ and $-\beta$ 
resp. $-\al$ and $-\beta$. Since all four possible 
signs occur, we may permute the four vectors to ensure that 
$\al=1$ and $\beta=-i$. 
\medskip\\ 
{\it Step iv)} 
On the subspace with basis 
$\Om_1$, $\Om_2$, $\Om_3$ and $\Om_4$ the operators 
$\ga^0$ and $\ga^1\ga^2$ have the diagonal block matrix form 
(\re{Rbf}). 
Moreover, in this basis the operators $\ga^1$ and $\ga^3$ have the 
form 
\be\la{Rbfm} 
\ga^1= 
\left(\ba{cc}   0 &  \si_1\\ 
                -\si_1 & 0\ea\right),~~~~~~~ 
\ga^3= 
\left(\ba{cc}   0 & \si_3 \\ 
                -\si_3 & 0\ea\right) 
\ee 
which coincide with (\re{Rgpgg}). From these we may check 
that $\ga^2=-\ga^1(\ga^1\ga^2)$ 
also has the desired form. 
\medskip\\ 
{\it Step v)} 
The vectors $\Om_j$, $j=1,...,4$, 
span the space $V$ since $\dim V\le 4$. 
%%%%If 
%%%%$\dim V=4$, then 
%%%%$V_1= V$ and 
%%%%the item i) is proved. 
%%%%Otherwise, we repeat the process in the 
%%%%orthogonal complement $V_1^\bot$ which is also invariant with 
%%%%respect to all operators $\ga^\al$ by their Hermitian properties. 
\bo 
 
\bexe 
Check the formulae (\re{Rbf}) for $\ga^0$ and  $\ga^1\ga^2$. 
\eexe 
\bexe 
Check the formulae (\re{Rbfm}) 
for $\ga^1$ and $\ga^3$. 
\eexe 
%%%\bexe 
%%%Check that the space $V_1^\bot$ is 
%%% invariant with respect to all operators $\ga^\al$. 
 
%%%\eexe 
 
\bc 
For any Lorentz transformation $\Lam$, 
there exists a nondegenerate  matrix $\Ga(\Lam)\in GL(4,\C)$ 
such that 
\be\la{Rgal} 
\ga(\Lam p)=\Ga(\Lam)\ga(p) \Ga^{-1}(\Lam),~~~~~~~~p\in\R^4. 
\ee 
\ec 
\Pr 
(\re{Remrf}) implies 
\be\la{Remrfi} 
\ga^2(\Lam p)=g(\Lam p)=g(p),~~~~~~~~p\in\R^4, 
\ee 
since $\Lam$ is a Lorentz transformation. 
Hence, the matrices $\ga(\Lam e_\al)$ satisfy the relations 
(\re{Rgco}) as well as $\ga^\al:=\ga(e_\al)$.  Therefore, 
by the Pauli Theorem, we have 
\be\la{Rgalm} 
\ga(\Lam e_\al)=\Ga(\Lam)\ga(e_\al) \Ga^{-1}(\Lam), 
~~~~~~~\al=0,...,3, 
\ee 
where $\Ga(\Lam)$ is an invertible operator in $\R^4$ 
(which 
transforms the vector $e_\al$ into $\Om_\al$, $\al=0,...,3$). 
Then (\re{Rgal}) follows by linearity.\bo 
 
%%%%%%%%%%%%%%%%%%%%%%%%%%%%% 
%%%%%%%%%%%%%%%%%%%%%%%%%%%%%% 
%%%%%%%%%%%%%%%%%%%%%%%%%%%%%% 
%%%%%%%%%%%%%%%%%%%%%%%%%%% 
 
\newpage 
\setcounter{subsection}{0} 
\setcounter{theorem}{0} 
\setcounter{equation}{0} 
\section{Lorentz Covariance} 
We justify the Einstein postulate of Special Relativity Theory 
for the Dirac equation: the equation takes an identical form 
in all Inertial Frames.

Let us consider two frames of reference of two observers 
related by a Lorentz transformation: $x'=\Lam x$. 
By a natural extension of the Einstein postulate ${\bf E}$, 
the Dirac equation (\re{Remrd}) has the same form in both 
frames of reference. This extended postulate 
allows us to determine 
the corresponding transformation of the wave function. 
The next theorem gives a transformation 
of the wave function which leaves the Dirac equation invariant.

\bt 
Let $\psi(x)$ be a solution to the Dirac equation (\re{Remrd}) and 
\be\la{Rcov} 
\psi'(x'):=\Ga(\Lam^\#)\psi(x),~~~~~~x\in\R^4, 
\ee 
where $x'=\Lam x$ and $\Lam^\#:=(\Lam^t)^{-1}$ 
where $\Lam^t$ 
is the transposed matrix to $\Lam$. 
Then the function $\psi'(x')$ is also a solution 
to the Dirac equation. 
\et 
\Pr 
Let us translate the Dirac equation 
(\re{Remrd}) into the Fourier transform 
\be\la{RFt} 
\hat\psi(p):=\int e^{\fr{ip x}\h}\psi(x)dx,~~~~~~p\in\R^4, 
\ee 
where $p x:=p_\al x^\al$. 
Then (\re{Remrd}) becomes 
\be\la{Rdirb} 
\ga(p)\hat\psi(p)=\mu c\hat\psi(p),~~~~~~~~~~p\in\R^4. 
\ee 
The Fourier transform  translates (\re{Rcov}) into 
\be\la{Rcovi} 
\hat\psi'(p')=\Ga(\Lam^\#)\hat\psi(p),~~~~~~p\in\R^4, 
\ee 
where $p=\Lam^t p'$. 
\bexe 
Check (\re{Rcovi}). {\bf Hint:} formally, we have 
\be\la{Rcovs} 
\hat\psi'(p'):= 
\int e^{\fr{ip' x'}\h}\psi'(x')dx' 
= 
\int e^{\fr{ip' \Lam x}\h}\Ga(\Lam^\#)\psi(x)|\det\Lam|dx 
= 
\Ga(\Lam^\#)\int e^{\fr{i\Lam^t p' x}\h}\psi(x)dx=\Ga(\Lam^\#)\hat\psi(p) 
\ee 
since $|\det\Lam|=1$ 
for the Lorentz transformation $\Lam$.

\eexe 
Now express (\re{Rdirb}) in terms of the wave function $\hat\psi'(p')$: 
\be\la{Rdirbe} 
\ga(p) \Ga^{-1}(\Lam^\#)\hat\psi'(p')=\mu c \Ga^{-1}(\Lam^\#)\hat\psi'(p'), 
~~~~~~~~~~p'\in\R^4. 
\ee 
This is equivalent to the Dirac equation (\re{Rdirb}) iff 
\be\la{Riff} 
\Ga(\Lam^\#)\ga(p) \Ga^{-1}(\Lam^\#)=\ga(p'),~~~~~~p'\in\R^4. 
 \ee 
This is equivalent to  (\re{Rgal}) for 
$\Lam^\#$ instead of $\Lam$ 
since $p'=\Lam^\# p$. 
Finally, it is true since $\Lam^\#$ also belongs to the Lorentz group. 
\bo 
 
\bexe 
Check that $\Lam^\#$ is a Lorentz transformation for any $\Lam\in L$. 
 
\eexe

\brs 
{\rm i) The formal calculations (\re{Rcovs}) can be justified 
by the properties of the Fourier transformation of 
tempered distributions. This is necessary since the integrals 
in (\re{Rcovs}) generally never converge for the solutions to the 
Dirac equation by energy and charge conservations (see below). 
\\ 
ii) The theorem implies that the "variance" rule (\re{Rcov}) 
leaves the Dirac equation unchanged. 
This means, by definition, that the Dirac equation is 
{\it covariant} 
with respect to the Lorentz group}. 
\ers

%%%%%%%%%%%%%%%%%%%%%%%%%%%%% 
%%%%%%%%%%%%%%%%%%%%%%%%%%%%%% 
%%%%%%%%%%%%%%%%%%%%%%%%%%%%%% 
%%%%%%%%%%%%%%%%%%%%%%%%%%% 
 
\newpage 
\setcounter{subsection}{0} 
\setcounter{theorem}{0} 
\setcounter{equation}{0} 
\section{Lorentz Transformation of Spinors} 
We give a construction of the operator $\Ga(\Lam)$ 
acting on the spinor wave functions. 
For the rotations the operator will be calculated explicitly. 
 
\subsection{Factorization of 
Lorentz Transformations} 
First let us derive a useful formula for the transformations 
of the Dirac matrices. 
\bl\la{Rl1} 
i) For any two vectors $p,q\in\R^4$ with  $q$ non-null w.r.t. 
$g$ (i.e. $g(q,q)\ne 0$), 
we have 
\be\la{Rgpq} 
\ga(q)\ga(p)\ga^{-1}(q)=\ga(R_q p), 
\ee 
where $R_q$ is a `reflection' $R_q p=2\ds\fr{g(q,p)}{g(q,q)}q-p$. \\ 
ii) 
The transformation 
$R_q$ is in the Lorenz group  but not proper, i.e. $\det R_q=-1$. 
\el 
\Pr 
i) Since $\ga^2(q)=g(q,q)$, we have $\ga^{-1}(q)=\ga(q)/g(q,q)$. 
Therefore, multiplying the relation 
$$ 
\ga(q)\ga(p)+\ga(p)\ga(q)=2g(q,p) 
$$ 
on the right by  $\ga^{-1}(q)$, we obtain 
$$ 
\ga(q)\ga(p)\ga^{-1}(q)+\ga(p)=2\fr{g(q,p)}{g(q,q)}\ga(q). 
$$ 
This implies (\re{Rgpq}) by the linearity of $\ga$. 
\\ 
ii) It is easy to check that $R_q$ is in the Lorentz group, i.e. 
\be\la{Rrql} 
g(R_q p,R_q p)=g(p,p),~~~~~~p\in\R^4. 
\ee 
It remains to check that 
\be\la{Rrqlf} 
\det R_q=-1. 
\ee 
This is obvious since 
\\ 
i) $R_q$ preserves the components  along $q$, i.e. 
$R_q q=q$. 
\\ 
ii) $R_q$ reverses the components  $g$-orthogonal to $q$, i.e. 
$R_q p=-p$ if $g(q,p)=0$.\bo

\bexe 
Check (\re{Rrql}). 
 
\eexe 
 
\bex 
$q=(\pm 1,0)$: then $R_q p=(2p_0,0,0,0)-p=(p_0,-\p)$ 
for $p=(p^0,\p)$. 
\eex 
\bex 
 $q=(0,\q)$ with $\q$ a unit vector: then $g(q,q)=-|\q|^2=-1$ 
and $R_q p=2(\q\cdot \p)q-p=(-p_0,2(\q\cdot \p)\q-\p)$. 
Hence the spatial component along $\q$ is unchanged 
while the orthogonal components are reversed, 
which is precisely the effect of a rotation through $\pi$ about the 
axis $\q$. 
\eex 
 
Now let us factorize the Lorentz  transformations in two reflections. 
\bl\la{Rl2} 
Let $\Lam$ be a proper Lorentz transformation that fixes two linearly 
independent vectors $v_1$ and $v_2$. Then it admits the factorization 
\be\la{Rfac} 
\Lam v=R_wR_u v,~~~~~~v\in\R^4. 
\ee 
Here $u$ is a non-null vector (ie. $g(u,u)\ne 0$), 
$g$-orthogonal to $v_1$ and $v_2$, 
and we set $w=\Lam u+u$ if $\Lam u\ne -u$. Otherwise, we take $w$ 
which is 
$g$-orthogonal to all three vectors $v_1$, $v_2$ and $u$. 
\el 
\Pr 
First let us note that $u$ can be chosen non-null. 
Indeed. 
the plane $\Pi$, $g$-orthogonal to $v_1$ and $v_2$, 
is two-dimensional. Hence, its intersection with the null cone is 
at most one-dimensional cone since the null cone does not contain 
two-dimensional planes. Therefore, all vectors $u\in\Pi$ are 
non-null except for at most two lines. 
 
Second, note that 
\be\la{Rglu} 
g(\Lam u,v_j)=g(\Lam u,\Lam v_j)=g(u,v_j)=0. 
\ee 
since $\Lam$ fixes $v_j$ and preserves $g$. 
Hence, $\Lam u$ and $w=\Lam u+u$ are $g$-orthogonal to  $v_1$ and $v_2$. 
 
Further consider the case  $\Lam u\ne -u$. Then 
$\Lam +1$ is an invertible operator. Namely, $\Lam$ has two eigenvectors 
$v_1$, $v_2$ with the eigenvalue $1$. If it has an eigenvector 
$v_3$ with the eigenvalue $-1$, then also a vector 
$v_4$ $g$-orthogonal to 
$v_1$, $v_2$ and $v_3$ is an eigenvector with the eigenvalue $-1$. 
Then also $\Lam u= -u$ which 
contradicts our assumption. 
 
Therefore, $(\Lam +1)\Pi$ is also a two-dimensional plane. Hence, 
by the same arguments, 
$w=\Lam u+u$ is a non-null vector 
for almost all $u\in\Pi$.

Now let us check the identity (\re{Rfac}). 
First, the composition $R_w R_u $ fixes $v_1$ and $v_2$ since each 
factor reverses both vectors. Hence, (\re{Rfac}) holds for 
$v=v_1$ and $v=v_2$. Further, let us prove  (\re{Rfac}) 
for $v=u$. Namely, 
\be\la{Rgww} 
\ba{ll} 
g(w,w)&=g(\Lam u,\Lam u)+2g(\Lam u,u)+g(u,u)\\ 
&=g(u,u)+2g(\Lam u,u)+g(u,u)\\ 
&=2[g(u,u)+g(\Lam u,u)]=2g(u,w). 
\ea 
\ee 
Therefore, 
\be\la{RRwu} 
R_w u=\fr{2g(u,w)}{g(w,w)}w-u=w-u=\Lam u. 
\ee 
Then 
also $\Lam u=R_w R_u u$ since $R_u u=u$. 
 
So  (\re{Rfac}) holds for 
three linearly independent vectors 
$v_1$, $v_2$ and $u$. Hence their action on the whole space 
$\R^4$ is identical since both $R_w R_u $ and  $\Lam$ 
are proper Lorentz transformations. 
The case  $\Lam u=-u$ can be checked similarly. 
\bo 
\bexe 
Check the lemma for the case $\Lam u=-u$. 
\eexe 
The next lemma demonstrates that each Lorentz transformation admits a 
factorization 
into three ones, each preserving two linearly independent vectors. 
Let us recall that the rotation group $SO(3)$ 
is a natural subgroup of the Lorentz group 
$L$: for each $R\in SO(3)$ the corresponding  Lorentz transformation 
is given by (\re{Rexr}).

\bl\la{Rl3} 
Let $\Lam$ be a proper Lorentz transformation. Then it admits 
the factorization 
\be\la{Raf} 
\Lam=\hat R_1B\hat R_2, 
\ee 
where $R_1,R_2\in SO(3)$ and $B$ is a proper boost of type (\re{Rexrff}). 
\el 
\Pr 
Let us consider the vector $\Lam e_0=(v_0,\bv)$, where $e_0:=(1,0,0,0)$. 
There exists a rotation $Q_1\in SO(3)$ such that $Q_1\bv=(v_1,0,0)$. 
Then the vector 
$e_0':=\hat Q_1\Lam e_0$ has the form $(v_0,v_1,0,0)$ with 
$g(e_0')=v_0^2-v_1^2=1$ 
since $g(e_0)=1$ and $\hat Q_1\Lam\in L$. 
\bexe 
Check that there exists a proper boost $D$ 
of type (\re{Rexrff}) such that $De_0'=e_0$. {\bf Hint:} 
Choose $D$ in the form $\Lam_+^+$ if $v_0>0$ and 
$\Lam_-^-$ if $v_0<0$. 
\eexe 
Now 
$D\hat Q_1\Lam e_0=e_0$, hence 
the Lorentz transformation $B\hat R_2\Lam$ has a matrix of type 
\be\la{RBrl} 
D\hat Q_1\Lam=\left( 
\ba{cc} 
1&\w\\ 
0&Q_2 
\ea 
\right), 
\ee 
\bexe 
Check that $\w=0$. {\bf Hint:} use (\re{Raeq}) and (\re{Raeqe}) for the matrix 
 $D\hat Q_1\Lam$. 
\eexe 
Therefore, $Q_2\in SO(3)$ and (\re{RBrl}) implies that $D\hat Q_1\Lam=\hat Q_2$, 
hence 
\be\la{Rweh} 
~~~~~~~~~~~~~~~~~~~~~~~~~~~~~~~~~~\Lam=\hat Q_2 D^{-1}\hat Q_1^{-1}.~~~~~~ 
~~~~~~~~~~~~~~~~~~~~~~~~~~~~~~~~~~~~~~~~~~~~~~~~~~~~~~~~~\bo 
\ee

Lemmas \re{Rl2} and \re{Rl3} imply that each proper Lorentz transformation 
$\Lam$ 
admits a factorization 
$$ 
\Lam=R_{q_1}...R_{q_6}. 
$$ 
 
\bc\la{Rcgl} 
The matrix $\Ga(\Lam):=\ga(q_1)...\ga(q_6)$ satisfies 
\be\la{Rgalr} 
\ga(\Lam p)=\Ga(\Lam)\ga(p) \Ga^{-1}(\Lam),~~~~~~~~p\in\R^4. 
\ee 
\ec 
\Pr 
Lemma \re{Rl1} implies  by induction that 
\be\la{Rgalp} 
~~~~~~~\Ga(\Lam)\ga(p) \Ga^{-1}(\Lam)= 
\ga(q_1)...\ga(q_6)\ga(p)\ga^{-1}(q_6)...\ga(q_1)^{-1}= 
\ga(R_{q_1}...R_{q_6}p)=\ga(\Lam p).~~~~~\bo 
\ee 
\subsection{Rotations of Dirac Spinors} 
This corollary allows us to construct explicitly the matrix $\Ga(\Lam)$. 
Consider, for example, a rotation $\Lam=\hat R(\theta\n)$ 
through $\theta\in(-\pi,\pi)$ 
about an axis $\n$. 
Then (\re{Rcov}) becomes 
\be\la{Rcovr} 
\psi'(x'):=\Ga(\hat R(\theta\n))\psi(\hat R(-\theta\n)x'),~~~~~~x'\in\R^4 
\ee 
since $\hat R^\#(\theta\n)=\hat R(\theta\n)$ for the orthogonal matrix 
 $R(\theta\n)$. 
The rotation   $\hat R(\theta\n)$ fixes two vectors $\n$ and $e_0$. 
Hence we can apply Lemma \re{Rl2} to factorize  $\hat R(\theta\n)$ 
and then Corollary \re{Rcgl} to construct $\Ga(\hat R(\theta\n))$. 
 
According to Lemma \re{Rl2}, we 
choose any non-null vector $\bu$ which is perpendicular 
to $\n$ and note that 
\be\la{Rrun} 
\hat R(\theta\n)\bu=\cos \theta~\bu-\sin\theta~\n\times\bu 
\ee 
for the rotation about $\n$ in a positive direction. 
According to Lemma \re{Rl2}, we define 
\beqn\la{Rrunt} 
\w=\bu+\hat R(\theta\n)\bu&=&(1+\cos \theta)\bu-\sin\theta~\n\times\bu \nonumber\\ 
&=&2\cos\ds\fr\theta2[\cos\ds\fr\theta2 ~\bu-\sin\ds\fr\theta2 ~\n\times\bu]. 
\eeqn 
It is a nonzero vector since $\theta\in(-\pi,\pi)$, hence 
$\hat R(\theta\n)=\hat R_\w \hat R_\bu$ by Lemma \re{Rl2}. 
Therefore, by Corollary  \re{Rcgl}, we can choose 
\beqn\la{Rrut} 
\Ga(\hat R(\theta\n))=\ga(\w)\ga(\bu)&=& 
\left( 
\ba{cc}0&\si\w\\ 
-\si\w&0\ea 
\right) 
\left( 
\ba{cc}0&\si\bu\\ 
-\si\bu&0\ea 
\right)\nonumber\\ 
\nonumber\medskip\\ 
&=& 
\left( 
\ba{cc}-(\si\w)(\si\bu)&0\\ 
0&-(\si\w)(\si\bu)\ea 
\right). 
\eeqn 
Since the reflection $\hat R_\w$ is unaffected by any normalization of $\w$, 
let us redefine $\w$ as the unit vector 
$-\cos\ds\fr\theta2 ~\bu+\sin\ds\fr\theta2 ~\n\times\bu$. 
Then 
\be\la{Rsvs} 
(\si\w)(\si\bu)=\bu\w+i\si(\w\times\bu) 
=-\cos\fr\theta2-i\sin\fr\theta2\si\n=-\exp(\fr{i\theta}2\si\n) 
\ee 
by the Euler trick. Finally, 
\be\la{Rrutf} 
\Ga(\hat R(\theta\n))= 
\left( 
\ba{cc}\exp(\fr{i\theta}2\si\n)&0\\ 
0&\exp(\fr{i\theta}2\si\n)\ea 
\right). 
\ee 
Finally, (\re{Rcovr}) becomes 
\be\la{Rcovrb} 
\psi'(x'):= 
\left( 
\ba{cc}\exp(\fr{i\theta}2\si\n)&0\\ 
0&\exp(\fr{i\theta}2\si\n)\ea 
\right) 
\psi(\hat R(-\theta\n)x'),~~~~~~x'\in\R^4. 
\ee 
\brs 
i) The matrices (\re{Rrutf}) are unitary, because they represent pure 
rotations. 
\\ 
ii) A map $\hat R(\theta\n)\mapsto \Ga(\hat R(\theta\n))$ is a 
homomorphism of the one-parametric subgroup 
of the Lorentz group $L$, into the 
unitary group $U(4)$. 
\\ 
iii) 
Our calculation prove 
that (\re{Rrutf}) satisfies the identity 
(\re{Rgalr}) 
for $\theta\in(-\pi,\pi)$. 
However it holds for any $\theta\in\R$ by analytic continuation 
and gives, for  $\theta=2\pi$, that 
\be\la{Rrur} 
\Ga(\hat R(2\pi\n))=-1. 
\ee 
Hence, 
(\re{Rcovrb}) implies that 
the Dirac spinor changes sign by a rotation through $2\pi$. 
 
\ers 
\bc 
The infinitesimal generator $G$ of rotations about the axis $\n$ 
satisfies the identity 
\be\la{Rgen} 
-i\h G=\fr \h2 
\left(\ba{cc}\si\n&0\\ 
0&\si\n\ea\right)+\hat\bL\n, 
\ee 
where $\hat\bL$ stands for the standard orbital angular momentum operator. 
\ec 
\Pr 
The infinitesimal generator is defined as the derivative in $\theta$ 
of the rotation around $\n$. Formula (\re{Rcov}) 
implies that the generator is given by 
\be\la{Rcovg} 
G\psi(x):=\left.\fr d{d\theta}\right|_{\theta=0} 
\Big(\Ga(\hat R(\theta\n))\psi(\hat R(-\theta\n)x)\Big),~~~~~~x\in\R^4. 
\ee 
Therefore, $i\h G$ consists of two terms: 
\be\la{Rcog} 
-i\h G\psi(x):=-i\h \Big(\left.\fr d{d\theta}\right|_{\theta=0} 
\Ga(\hat R(\theta\n))\Big) \psi(x) 
-i\h \left.\fr d{d\theta}\right|_{\theta=0} 
\psi(\hat R(-\theta\n)x). 
\ee 
Differentiating (\re{Rrutf}), we obtain 
that the first term coincides with the first summand in (\re{Rgen}). 
\bexe 
Check that the second term in (\re{Rcog}) 
coincides with the second summand in (\re{Rgen}). 
{\bf Hints:} i) Choose the coordinates with $e_3=\n$. Then 
$\hat\bL\n=\hat\bL_3$ and 
$\theta$ is an 
angle of rotation around $e_3$ in a positive direction. ii) 
Use the formula 
\be\la{Rrof} 
\hat\bL_3\psi(x)=i\h\fr\pa{\pa \theta} 
\psi(\hat R(\theta\n)x). 
\ee 
\eexe 
Now (\re{Rgen}) is proved.\bo 
\bexe 
Check the formula (\re{Rrof}). 
\eexe

%%%%%%%%%%%%%%%%%%%%%%%%%%%%% 
%%%%%%%%%%%%%%%%%%%%%%%%%%%%%% 
%%%%%%%%%%%%%%%%%%%%%%%%%%%%%% 
%%%%%%%%%%%%%%%%%%%%%%%%%%% 
 
\newpage 
\setcounter{subsection}{0} 
\setcounter{theorem}{0} 
\setcounter{equation}{0} 
\section{Coupling to Maxwell Field} 
We define the Dirac equation with an external 
Maxwell field and prove gauge invariance.

\subsection{Dirac Equation in the Maxwell Field} 
Let us denote $\ga_0=\ga^0$, $\ga_j=-\ga^j$ and 
\be\la{Rlud} 
P^0=i\h\fr\pa{\pa x^0},~~~~~~~~P^j=-i\h\fr\pa{\pa x^j}. 
\ee 
Write the Dirac equation  (\re{Remrd}) in the form 
\be\la{Rdirff} 
\ga(P)\psi(x)=\mu c\psi(x),~~~~~~x\in\R^4. 
\ee 
where 
the differential operator $\ga(P):=\ga_\al P^\al$ 
is called the {\it Dirac operator}.

Let us recall the Schr\"odinger equation with the Maxwell field: 
\be\la{RSMer} 
\ba{l} 
\ds[i\h\pa_t-e\phi(t,\x)]\psi(t,\x) 
=\fr 1{2\mu} 
[-i\h\na_\x-\ds\fr ec  \bA(t,\x)]^2\psi(t,\x), 
\ea 
\ee 
In the notations (\re{Rlud}) 
\be\la{RSMen} 
\ba{l} 
c\ds[P^0-\ds\fr ec \phi(t,\x)]\psi(t,\x) 
=\fr 1{2\mu} 
[\bP-\ds\fr ec  \bA(t,\x)]^2\psi(t,\x). 
\ea 
\ee 
In other words, in the presence of the  Maxwell field, we change 
$P^0$ to $P^0-\ds\fr ec \phi(t,\x)$ and $\bP$ to $\bP-\ds\fr ec  \bA(t,\x)$. 
This suggests the following generalization of the Dirac equation 
(\re{Rdirff}) 
for the electron field in an external Maxwell field: 
\be\la{Rdirg} 
\ga(P-\ds\fr ec \A(x))\psi(x)=\mu c\psi(x),~~~~~~x\in\R^4, 
\ee 
where $\A(x)=(\phi(t,\x),\bA(t,\x))$. 
\bd 
We accept (\re{Rdirg}) as the definition of the Dirac equation for the 
electron field $\psi(x)$ in the presence of an external Maxwell field 
 $\A(x)$. 
\ed 
\subsection{Gauge Invariance} 
Let us recall that the gauge transformation 
\be\la{Rgtr} 
\phi(t,\x)\mapsto \phi(t,\x)+\fr 1c\,\dot\chi(t,\x),\,\,\,\, 
\bA(t,\x)\mapsto \bA(t,\x)-\na_\x\chi(t,\x) 
\ee 
does not change the Maxwell fields corresponding to the potentials 
$\phi(t,\x)$ and $\bA(t,\x)$. Let us rewrite it as 
\be\la{Rgtrr} 
\phi(x)\mapsto \phi'(x):=\phi(x)+\fr1{i\h}P^0\chi(x),\,\,\,\, 
\bA(x)\mapsto \bA'(x):=\bA(x)+\fr1{i\h}\bP\chi(x) 
\ee 
\bt 
Let $\psi(x)$ be a solution of the Dirac equation (\re{Rdirg}) 
with the potentials $\phi(x),\bA(x)$. Then 
$\psi'(x):=\exp(\ds\fr{e\chi(x)}{i\h c})\psi(x)$ 
satisfies  (\re{Rdirg}) with the potentials  $\phi'(x),\bA'(x)$. 
\et 
\Pr This follows from a direct calculation, since 
$P\exp(\ds\fr{e\chi(x)}{i\h c})=\exp(\ds\fr{e\chi(x)}{i\h c}) 
\ds\fr{e}{i\h c}P\chi(x)$.\bo

%%%%%%%%%%%%%%%%%%%%%%%%%%%%% 
%%%%%%%%%%%%%%%%%%%%%%%%%%%%%% 
%%%%%%%%%%%%%%%%%%%%%%%%%%%%%% 
%%%%%%%%%%%%%%%%%%%%%%%%%%% 
 
\newpage 
\setcounter{subsection}{0} 
\setcounter{theorem}{0} 
\setcounter{equation}{0} 
\section{Pauli Equation as Nonrelativistic Approximation} 
We demonstrate that the Pauli Equation 
corresponds to the nonrelativistic 
approximation of the Dirac equation. 
This justifies the correspondence between relativistic 
and nonrelativistic quantum theories and the Land\'e factor $g=2$, 
predicted by Goudsmith and Uhlenbeck, 
for the spin.

First, rewrite the Dirac equation  (\re{Rdirg}), similarly to 
(\re{RSMen}), as 
\be\la{Rdirr} 
(i\h\pa_t-e\phi)\psi=\ga_0[ 
-c\ga(\bP-\ds\fr ec\bA)\psi+\mu c^2\psi], 
\ee 
where ${\bf g}:=(\ga_1,\ga_2,\ga_3):=(-\ga^1,-\ga^2,-\ga^3)$. 
Let us substitute here 
the block form of the $\ga$ matrices  (\re{Rgpgg}) 
and the splitting of the wave function (\re{Rtcs}). 
Then 
the Dirac equation reduces to the coupled equations 
\be\la{Rdice} 
\left\{ 
\ba{l} 
(i\h\pa_t-e\phi)\psi_+= 
c{\si}(\bP-\ds\fr ec\bA)\psi_-+\mu c^2\psi_+\\ 
\\ 
(i\h\pa_t-e\phi)\psi_-= 
c{\si}(\bP-\ds\fr ec\bA)\psi_+-\mu c^2\psi_- 
\ea\right. 
\ee 
Now let us replace $\psi$ by a gauge transformation 
 $\ti\psi:=\exp(i\mu c^2t/\h)\psi$. 
 This cancels the term $\mu c^2$ in the first equation 
and doubles it in the second one: 
\be\la{Rdiceg} 
\left\{ 
\ba{l} 
(i\h\pa_t-e\phi)\ti\psi_+= 
c{\si}(\bP-\ds\fr ec\bA)\ti\psi_-\\ 
\\ 
(i\h\pa_t-e\phi)\ti\psi_-= 
c{\si}(\bP-\ds\fr ec\bA)\ti\psi_+-2\mu c^2\ti\psi_- 
\ea\right. 
\ee 
Let us assume that the LHS of the last equation is  small compared 
to $\mu c^2$. This is true in the limit $c\to\infty$ 
which corresponds to the 
{\it nonrelativistic approximation}. 
Namely, the Lorentz transformations become then the Galilean ones, 
the retarded potentials becomes the Coulomb ones, etc. 
\br 
The LHS  represents 
a {\bf kinetic energy} and an 
electrostatic potential. Usually both are small compared 
to the {\bf rest energy} $\mu c^2$. 
\er 
So the last equation can be approximated by 
\be\la{Rapp} 
2\mu c^2\ti\psi_-=c{\si}(\bP-\ds\fr ec\bA)\ti\psi_+. 
\ee 
Substituting it into the first equation (\re{Rdiceg}), we obtain 
\be\la{Rdic1} 
(i\h\pa_t-e\phi)\ti\psi_+= 
\fr 1{2\mu}({\si}(\bP-\ds\fr ec\bA))^2\ti\psi_+. 
\ee 
Now we have to evaluate the operator on the RHS. 
\bl 
The identity 
\be\la{Ridh} 
({\si}(\bP-\ds\fr ec\bA))^2=(\bP-\ds\fr ec\bA)^2+\ds\fr ec\h\si\bB, 
\ee 
holds, where $\bB:=\rot\bA$ is the magnetic field. 
\el 
\Pr 
First, we have the standard identity similar to (\re{Rsvs}): 
\be\la{Rsvss} 
(\si(\bP-\ds\fr ec\bA))^2= 
(\bP-\ds\fr ec\bA)(\bP-\ds\fr ec\bA) 
+i\si((\bP-\ds\fr ec\bA)\times(\bP-\ds\fr ec\bA)). 
\ee 
Let us note that its proof does not depend on the commutation of 
the components 
of the vector $\bP-\ds\fr ec\bA$. 
On the other hand, the vector product of the vector $\bP-\ds\fr ec\bA$ 
with itself does not vanish since its components do not commute. 
For example, let us calculate the first component: 
\be\la{Rfco} 
(P_2-\ds\fr ec A_2)(P_3-\ds\fr ec A_3)-(P_3-\ds\fr ec A_3)(P_2-\ds\fr ec A_2) 
=[P_2-\ds\fr ec A_2,P_3-\ds\fr ec A_3]. 
\ee 
The commutator obviously reduces to 
\be\la{Rfcor} 
~~~~~~~~~~~-\ds\fr ec([P_2,A_3]+[A_2,P_3])= 
i\ds\fr e{c\h}(\pa_2A_3-\pa_3A_2)= 
i\ds\fr e{c\h}(\rot A)_1= 
i\ds\fr e{c\h}\bB_1.~~~~~~~~~~~~~~~~~\bo 
\ee 
\medskip\\ 
Now  (\re{Rdic1}) becomes 
\be\la{Rdic2} 
i\h\pa_t\ti\psi_+= 
\fr 1{2\mu}\Big((\bP-\ds\fr ec\bA)^2 -\ds\fr ec\h\si\bB\psi_+ 
+e\phi\Big)\ti\psi_+ 
\ee 
which coincides with the Pauli equation (\re{PaA}). 
\br 
{\rm This agreement with the Pauli spin theory 
was one of the great triumphs of the Dirac theory. It 
means that the Dirac equation automatically explains the 
Land\'e factor $g=2$ for the spinor electron magnetic moment, 
suggested by 
Goudsmith and Uhlenbeck to explain the Einstein-de Haas and Stern-Gerlach 
experiments.} 
\er 
\br 
{\rm For the Dirac equation only the sum  of orbital and spinor 
angular momentum 
is conserved, while for 
the Pauli equation both are conserved separately. 
This means that the Dirac equation 
requires both moments for relativistic invariance, while 
the Pauli equation  is a degenerate version of the 
Dirac one.} 
\er 
 
%%%%%%%%%%%%%%%%%%%%%%%%%%%%% 
%%%%%%%%%%%%%%%%%%%%%%%%%%%%%% 
%%%%%%%%%%%%%%%%%%%%%%%%%%%%%% 
%%%%%%%%%%%%%%%%%%%%%%%%%%% 
 
\newpage 
\setcounter{subsection}{0} 
\setcounter{theorem}{0} 
\setcounter{equation}{0} 
\section{Charge Continuity Equation} 
We define the charge-current densities of 
 the Dirac equation and prove the continuity equation.

\bd 
For a spinor wave function $\psi(x)$ 
\\ 
i) The corresponding charge-current 
 densities are defined by 
\be\la{Rccd} \left\{ \ba{l} ~\rho(x)=e\psi^\da(x)(\ga^0)^2\psi(x) 
:=e\ov\psi_\nu(x)\psi_\nu(x)\\ 
\\ 
\bj^k(x)=e\psi^\da(x)\ga^0\ga^k\psi(x) 
:=e\ov\psi_\nu(x)(\ga^0\ga^k\psi(x))_\nu 
\ea\right|~~~~~~~~x\in\R^4, \ee where $\psi^\da$ stands for the 
conjugate transpose to $\psi$. 
\\ 
ii) The vector $s(x):=(\rho(x),\bj(x))$ with the components 
$s^\al(x)=e\psi^\da(x)\ga^0\ga^\al\psi(x)$ is called {\bf 
four-current density}. 
 
\ed \br The charge density is non-positive and equal to 
\be\la{Rcde} \rho(x)=e\psi^\da(x)\psi(x):=e|\psi(x)|^2. \ee  \er

\bt For any solution $\psi(x)$ to the Dirac equation (\re{Rdirg}), 
the four-current density satisfies the continuity equation 
\be\la{Rcoe} \fr{\pa s^\al}{\pa x^\al}=0,~~~~~~~~~~~~x\in\R^4. \ee 
\et \Pr Substituting the definitions (\re{Rccd}), we obtain 
\be\la{Rsob} i\h\fr{\pa s^\al}{\pa x^\al}=i\h e\Bigg[ \fr\pa{\pa 
x^0}\Big(\psi^\da\psi \Big) +\fr\pa{\pa 
x^k}\Big(\psi^\da\ga^0\ga^k\psi \Big) \Bigg]. \ee Each term 
produces two summands, with derivatives of $\psi$ or $\psi^\da$. 
The contribution  coming from $\psi$ is \be\la{Rctc} i\h 
e\psi^\da\ga^0 \Bigg[ \ga^0\fr\pa{\pa x^0}+ \ga^k\fr\pa{\pa x^k} 
\Bigg] \psi=e\psi^\da\ga^0\ga(P)\psi=e\psi^\da\ga^0 \Big[ \mu 
c+\ds\fr ec\ga(\A) \Big]\psi \ee by the Dirac equation 
(\re{Rccd}). The contribution from $\psi^\da$ can be evaluated 
similarly, with the help of the identities 
$(\ga^0\ga^k)^\da=\ga^0\ga^k$: it is equal to \be\la{Rctct} 
-e\Bigg[i\h\ga^0\Bigg( \ga^0\fr\pa{\pa x^0}+ \ga^k\fr\pa{\pa 
x^k}\Bigg)\psi \Bigg]^\da \psi=-e(\ga^0\ga(P)\psi)^\da\psi = 
-e\Big[\ga^0\Big( \mu c+\ds\fr ec\ga(\A)\Big)\psi \Big]^\da\psi. 
\ee Finally, the sum of the contributions from $\psi$ and 
$\psi^\da$ vanishes by the same identities.\bo 
 
%%%%%%%%%%%%%%%%%%%%%%%%%%%%% 
%%%%%%%%%%%%%%%%%%%%%%%%%%%%%% 
%%%%%%%%%%%%%%%%%%%%%%%%%%%%%% 
%%%%%%%%%%%%%%%%%%%%%%%%%%% 
 
\newpage 
\setcounter{subsection}{0} 
\setcounter{theorem}{0} 
\setcounter{equation}{0} 
\section{Charged Antiparticles} 
We establish a one-to-one correspondence between 
 solutions to the Dirac equations with 
 charge $e$ and $-e$, respectively. 
Namely, consider the Dirac equation (\re{Rdirg}) with $-e$ instead of $e$: 
\be\la{Rdirgg} 
\ga(P+\ds\fr ec\A(x))\psi(x)=\mu c\psi(x),~~~~~~x\in\R^4, 
\ee 
It describes the positron wave field corresponding to  particles 
with positive charge $-e=|e|$. 
We will establish an isomorphism between 
the solutions to (\re{Rdirg}) and (\re{Rdirgg}). 
\bd 
The {\bf charge conjugation operator} $K$ maps a Dirac wave function 
to $\psi_c=K\psi:=\ga^2\ov\psi$, and  $\psi_c$ is called the 
{\bf charge-conjugated wave function}. 
\ed 
Let us note that the operator $K$ interchanges the first two components 
with the last two components of the Dirac spinor. 
Hence it changes the energy sign: 
the factor $\exp(-iEt)$ interchanges with $\exp(iEt)$ 
\bt 
Let $\psi$ satisfy the Dirac equation 
$\ga(P-\ds\fr ec\A(x))\psi(x)=\mu c\psi(x)$ for mass $\mu$ and charge $e$. 
Then  $\psi_c$ satisfies the Dirac equation (\re{Rdirgg}) 
for mass $\mu$ and charge $-e$. 
\et 
\Pr 
{\it Step i)} Let us check that for any vector $p\in\C^4$, 
\be\la{Racv} 
\ga^2\ga(\ov p)\ga^2=\ov{\ga(p)}. 
\ee 
That is, all matrices $\ga^\al$ are real except $\ga^2$ which 
is purely imaginary, hence 
\be\la{Rovj} 
\ov{\ga^\al}= 
\left\{ 
\ba{rl} 
\ga^\al,&\al\ne 2\\ 
-\ga^\al,&\al= 2. 
\ea\right. 
\ee 
On the other hand, the anticommutation relations 
for the Dirac matrices imply 
\be\la{Racr} 
\ga^2\ga^j\ga^2= 
\left\{ 
\ba{rl} 
\ga^\al,&\al\ne 2\\ 
-\ga^\al,&\al= 2. 
\ea\right. 
\ee 
Therefore, (\re{Racv}) follows. 
 
{\it Step ii)} 
Conjugating the Dirac equation for $\psi$, we obtain 
\be\la{Rdirgc} 
\ga^2\ga(\ov{P-\ds\fr ec\A(x)})\ga^2\ov\psi=\mu c\ov\psi. 
\ee 
Here $\ds\fr ec\A(x)$ is real, while $P$ includes the  imaginary factor $i$. 
Hence, multiplying (\re{Rdirgc}) by $\ga^2$, we get 
\be\la{Rdirgcg} 
~~~~~~~~~~~~~~~~~~~~~~~~~~~~~~~~ 
~~~~~~~~~~~~~~~~~~~~\ga(P+\ds\fr ec\A(x))\psi_c=\mu c\psi_c.~~ 
~~~~~~~~~~~~~~~~~~~~~~~~~~~~~~~~~~~~~~\bo 
\ee 
 
%%%%%%%%%%%%%%%%%%%%%%%%%%%%% 
%%%%%%%%%%%%%%%%%%%%%%%%%%%%%% 
%%%%%%%%%%%%%%%%%%%%%%%%%%%%%% 
%%%%%%%%%%%%%%%%%%%%%%%%%%% 
 
\newpage 
\setcounter{subsection}{0} 
\setcounter{theorem}{0} 
\setcounter{equation}{0} 
\section{Hydrogen Atom via Dirac Equation} 
We determine the quantum stationary states 
 of the hydrogen atom from the Dirac equation. 
The main issue is the spherical symmetry 
of the problem. We apply the method of separation of variables 
and take into account the information 
on the irreducible representations of the rotation group 
from Lecture 9.

\subsection{Spectral Problem and Spherical Symmetry} 
Let us consider the Dirac equation for 
the electron field in the hydrogen atom. 
The corresponding four-potential of the nucleus is $\A=(\phi,0,0,0)$ 
with $\phi=-e/|\x|$. Then the corresponding Dirac equation 
becomes 
\be\la{Rdeb} 
i\h\pa_t\psi=\cH_D\psi:=c\ga^0 
(\mu c-\ga(\bP))\psi+e\phi(\x) \psi. 
\ee 
We are going to determine all quantum stationary states which are 
the solutions of the form $\psi_E(\x) e^{-iEt/\h}$ with finite charge 
\be\la{RQE} 
Q(\psi_E)=\ds\int|\psi_E(\x)|^2d\x<\infty 
\ee 
(see  (\re{Rcde})). 
Substituting into the Dirac equation (\re{Rdeb}), we get the 
corresponding stationary eigenvalue problem 
\be\la{Rdebs} 
E\psi_E=\cH_D\psi_E. 
\ee 
It reduces to the coupled equations 
for the components of the spinor 
$\psi_E=\left(\!\!\ba{c}\psi_+\\\psi_-\ea\!\!\right)$: 
\be\la{Rdirco} 
\left\{\ba{l} 
(E-\mu c^2-e\phi)\psi_+=c\si\bP\psi_-\\ 
(E+\mu c^2-e\phi)\psi_-=c\si\bP\psi_+. 
\ea\right. 
\ee 
The main issue for the solution of the problem is its 
spherical symmetry. 
Namely, the nucleus potentials are spherically symmetric and 
it is possible to prove that the angular momentum 
$\bJ=\bJ(\psi)= \langle\psi,\hat\bJ\psi\rangle$ 
is conserved for the solutions to 
(\re{Rdeb}). Here, by Definition \re{RdamD}, 
\be\la{RJLs} 
\hat\bJ:=\hat\bL+\hat\bS 
\ee 
is the total angular momentum operator, 
where $\hat\bL:=-i\h \x\times\na$ and  $\hat\bS:=\ds\fr12 \h\ti\si$ 
are the orbital and spin angular momentum operator, respectively. 
The conservation follows from the commutation 
\be\la{RRamco} 
[\hat \bJ, \cH_D]=0. 
\ee 
\bexe 
Check (\re{RRamco}). 
{\bf Hint:} For the {\it free Dirac equation} 
with $\phi=0$, the commutation is proved in Theorem \re{RtamcD}. 
It remains to check the commutation of $\hat\bJ$ with $\phi(\x)$ 
which follows obviously from 
%%(\re{Ramd}) and 
the spherical symmetry of 
the potential. 
\eexe 
 
\subsection{Spherical Spinors and Separation of Variables} 
Let us recall that we have solved the spectral problem 
for the nonrelativistic Schr\"odinger equation by a general strategy 
of separation of variables (Section 9.1). Now we are going to develop 
it analogously for the relativistic problem (\re{Rdebs}). 
In this case, the role of the orbital angular momentum $\bL$ 
is played by the total angular momentum $\bJ$ since it is conserved. 
Hence, 
the strategy now has to be modified correspondingly: 
\medskip\\ 
{\bf I.} 
First, (\re{RRamco}) implies that the operator 
$\hat\bJ^2:=\hat\bJ_1^2+\hat\bJ_2^2+\hat\bJ_3^2$ commutes 
with $\cH_D$: 
\be\la{RRamco2} 
[\hat \bJ^2, \cH_D]=0. 
\ee 
Second, $\cH_D$ also commutes with each $\hat\bJ_n$. 
Hence, 
each eigenspace of the 
Dirac operator $\cH_D$ is invariant with respect to 
each operator $\hat\bJ_n$ and $\hat\bJ^2$. 
Moreover, 
the operator 
$\hat\bJ^2$ commutes also 
with 
each operator $\hat\bJ_n$, for example, 
\be\la{RRamco3} 
[\hat\bJ^2,\hat\bJ_3]=0. 
\ee 
\bexe 
Check (\re{RRamco3}). {\bf Hint:} 
First, prove the commutation relations 
$[\hat\bJ_k,\hat\bJ_j]=-i\h\epsilon_{kjl}\hat\bJ_l$ 
where $\epsilon_{kjl}$ is a totally antisymmetric tensor. 
The relations follow from similar ones for the orbital 
and spinor angular momenta, and from the commutation 
of the momenta. 
\eexe 
Hence, we could expect that there is a basis 
of common eigenfunctions for the operators $\cH_D$, 
$\hat\bJ_3$ and $\hat\bJ^2$. 
Therefore, it would be helpful to first diagonalize 
simultaneously 
$\hat\bJ^2$ and  $\hat\bJ_3$. 
\medskip\\ 
{\bf II.} The condition (\re{RQE}) means that 
we consider the eigenvalue problem (\re{Rdebs}) 
in the Hilbert space $\E:=L^2(\R^3)\otimes\C^4$. 
On the other hand, 
both operators, $\hat\bJ_3$ and  $\hat\bJ^2$, 
act only on the spinor variables and angular variables 
 in spherical coordinates. 
Hence, the operators act also in the Hilbert space 
$\E_1:=L^2(S,dS)\otimes\C^4$, 
where $S$ stands for the two-dimensional sphere 
$|\x|=1$. 
Moreover, 
both operators $\hat\bJ_3$ and $\hat\bJ^2$ have the block form 
\be\la{Rbbf} 
\hat\bJ_3=\left(\ba{cc} 
\hat\bL_3+\hat\s_3&0\\ 
0&\hat\bL_3+\hat\s_3 
\ea\right),~~~~~~~~~~~~ 
\hat\bJ^2=\left(\ba{cc} 
(\hat\bL+\hat\s)^2&0\\ 
0&(\hat\bL+\hat\s)^2 
\ea\right), 
\ee 
where $\hat\s:=\ds\fr12\h\si$. 
Therefore, we can split the space 
$\E_1$ into the sum $\E_1=E_1^+\oplus E_1^-$, where 
$ E_1^\pm=E_1:=L^2(S,dS)\otimes\C^2$, and 
the action of the operators is identical in each summand. 
In the following section we will prove
\bl\la{Rlssh} 
i) In the space $E_1$ 
there exists an orthonormal basis 
of {\it Spinor Spherical Harmonics} $\cS_{jk}(\theta,\vp)$ 
which are 
common eigenfunctions 
of the operators $\hat\bJ_3$ and $\hat\bJ^2$: 
\be\la{RRsev} 
~~~~~~~~~~\hat\bJ_3\cS_{jk}(\theta,\vp)\!=\!\h k \cS_{jk}(\theta,\vp),~~ 
\hat\bJ^2\cS_{jk}(\theta,\vp)\!=\!\h^2 j(j+1) \cS_{jk}(\theta,\vp),~~ 
k\!=\!-j,\!-\!j\!+\!1,...,j, 
\ee 
where $j=\ds\fr12,\ds\fr32,...$. 
\\ 
ii) 
The space of the solutions to (\re{RRsev}) 
is two-dimensional for each fixed $j,k$ except for a one-dimensional space 
for $j=1/2,k=-1/2$. 
\el 
{\bf III.} 
The lemma suggests that we could construct the eigenfunctions of the 
Dirac operator $\cH_D$, by separation of 
variables, in the form 
\be\la{Rsevf} 
\psi_E=
\left(\ba{c} 
\psi_+ 
\\ 
\psi_- 
\ea\right)= 
\left(\ba{c} 
R^+_+(r)\cS^+_{jk}(\theta,\vp) 
+ 
R^-_+(r)\cS^-_{jk}(\theta,\vp) 
\\ 
\\ 
R^+_-(r)\cS^+_{jk}(\theta,\vp) 
+ 
R^-_-(r)\cS^-_{jk}(\theta,\vp) 
\ea\right) 
\ee 
where $\cS^+_{jk},\cS^-_{jk}$ is the basis of the  solutions to the equations
(\re{RRsev})
with the fixed $j,k$. 
The following theorem justifies this particular choice 
for the eigenfunctions. 
\bl\la{Rlsv} (On Separation of Variables) 
Each solution to the spectral problem (\re{Rdebs}) is a sum 
(or a series) of solutions of the particular form (\re{Rsevf}). 
\el 
{\bf Formal Proof} 
Let $\pi_{jk}$ denote the orthogonal projection in $E_1$ 
onto the linear span $E_{jk}$ of $\cS_{jk}$. 
Let us define its action also in 
$\E=E_1\otimes L^2(\R^+)\oplus E_1\otimes L^2(\R^+)$ as 
$\pi_{jk}\otimes 1\oplus\pi_{jk}\otimes 1$, or equivalently, 
\be\la{Rape} 
[\Pi_{jk} 
\left(\ba{c} 
\psi_+\\ 
\psi_-\ea\right)](r,\cdot,\cdot):= 
\left(\ba{c} 
\pi_{jk}[\psi_+(r,\cdot,\cdot)]\\ 
\pi_{jk}[\psi_-(r,\cdot,\cdot)]\ea\right), 
~~~~~~~~r>0. 
\ee 
in the spherical coordinates 
$r,\theta,\vp$. 
Then $\Pi_{jk}$ 
commutes with 
the Schr\"odinger operator $\cH$ since the latter commutes with 
 $\bH_3$ and $\bH^2$: 
\be\la{Rprc} 
[\cH,\Pi_{jk}]=0. 
\ee 
Hence, applying $\Pi_{jk}$ to (\re{Rdebs}), we get {\it formally}
\be\la{RSMssbg} 
E_\om\Pi_{jk}\psi_\om=\cH \Pi_{jk}\psi_\om. 
\ee 
It remains to note that 
\\ 
i) The function $\Pi_{jk}\psi_\om$ 
has the form (\re{Rsevf}) 
since 
$\pi_{jk}[\psi_\pm(r,\cdot,\cdot)] 
\in E_{jk}$, and 
$\dim\E_{jk}\le 2$; 
\\ 
ii) 
$\psi_\om=\sum_{j,k}\Pi_{jk}\psi_\om$. 
\bo

\brs 
i) A complete solution to the spectral problem (\re{Rsevf}) relies on 
an investigation of 
all commutation relations of the operators 
$\hat\bJ_k$, $\hat\bL_k$ and $\hat\bS_k$, 
$k=1,2,3$, i.e.  the Lie algebra generated by them. 
\\ 
ii) It still remains to determine the radial functions 
in (\re{Rsevf}). We will substitute (\re{Rsevf}) into the equation 
(\re{Rdirco}). This  gives a radial eigenvalue problem 
which will be solved explicitly. 
\ers

%%%%%%%%%%%%%%%%%%%%%%%%%%%%%%%%%%%%%%%%55 

\subsubsection{Tensor product and Clebsch-Gordan Theorem} 
We prove Lemma \re{Rlssh}.
In (\re{RRsev}), we consider the action of the operator 
$\hat\bJ=\hat\bL+\hat\s$ in the space 
$E_1:=L^2(S,dS)\otimes\C^2$. Let us note  that the 
 operator $\hat\bL$ acts on $L^2(S,dS)$ while $\hat\s$ acts on 
the second factor $\C^2$, and 
$\hat\bL$ obviously commutes with  $\hat\s$. 
Therefore, the operator $\hat\bJ$ is a generator 
of the tensor product of the regular and spinor representations 
of the rotation group $SO(3)$ since $\hat\bL$, $\hat\s$ are 
the generators of the representations. 
Then the eigenfunctions and eigenvalues of $\hat\bJ^2$ can be found 
by the Clebsch-Gordan theorem \ci{GMS}. 
 
Namely, we know the spectral decomposition of the operator 
$\hat\bL^2$ in the space $L^2(S,dS)$: \be\la{RsdL} 
L^2(S,dS)=\oplus_{l=0}^\infty L(l). \ee Here the $L(l)$ are 
finite-dimensional orthogonal eigenspaces of the operator 
$\hat\bL^2$ corresponding to the eigenvalues $\h^2 l(l+1)$, where 
$l=0,1,...$. In $L(l)$ there is an orthonormal basis 
$e_{-l},...,e_{l}$ where $e_m=\hat\bH_+^{m+l}e_{-l}$ (here 
 $\hat\bH_+:=\hat\bH_1+ i\hat\bH_2$, where $\bH_k:=\h^{-1}\bL_k$) 
and (cf. (\re{RRsev})) 
\be\la{ReL} 
\hat\bL_3 e_{lm}=\h m e_{lm},~~~~ 
 \hat\bL^2 e_{lm}=\h^2 l(l+1)m e_{lm},~~~~~~~~~m=-l,...,l. 
\ee 
Namely, $e_{lm}=\h Y_l^m$, where $Y_l^m$ are 
Spherical Harmonics (\re{sev}).
Similarly,  in $\C^2$ 
there is an orthonormal basis $f_{-1/2},f_{1/2}$ 
where $f_{1/2}=\hat\s_+f_{-1/2}$ and 
\be\la{Res} 
\hat\s_3 f_{s}=\h s f_{s},~~~~ 
 \hat\s^2 f_{s}=\h^2 s(s+1) f_{s},~~~~~~~~~s={-1/2},{1/2}. 
\ee 
Namely, $f_{-1/2}=\left(\ba{c} 0\\1\ea\right)$ and 
$f_{1/2}=\left(\ba{c} 1\\0\ea\right)$.
Therefore, we have 
\be\la{RsdJ} 
L^2(S,dS)\otimes\C^2=\oplus_{l=0}^\infty L(l)\otimes\C^2, 
\ee 
and the tensor products $e_{lm}\otimes f_{s}$ with 
$m=-l,...,l$ and $s=-1/2,1/2$ 
form an orthonormal basis 
in the space $E(l):=L(l)\otimes\C^2$. The relations (\re{ReL}),  (\re{Res}) 
imply that 
\be\la{ReLs} 
\hat\bJ_3 e_{lm}\otimes f_{s}=\h (l+s) e_{lm}\otimes f_{s}. 
\ee

Now let us state the Clebsch-Gordan theorem for our particular case. 
It is known as the "addition of angular momenta''. 
 
\bl 
For each $l=0,1,...$ 
\\ 
i) 
The space $E(l)$ 
is an orthogonal sum of two eigenspaces $E_\pm(l)$ of the operator 
$\hat\bJ^2$: 
\be\la{Raam} 
E(l)=E_+(l)\oplus E_-(l),~~~~~~~~~~~~ 
\hat\bJ^2|_{E_\pm(l)}=\h^2l_\pm(l_\pm+1), 
\ee 
where 
$l_\pm=l\pm 1/2$ and $\dim E_\pm(l)=2\l_\pm+1$. 
\\ 
ii) 
For $l\ge 1$, 
in the space $E_\pm(l)$ there exists a basis 
$\cS_{k}^\pm$, 
$k\!=\!-l_\pm,-l_\pm+1,...,l_\pm$, 
satisfying the eigenvalue problem (\re{RRsev}) with $j=l_\pm$. 
For $l=0$, 
the space $E_-(0)=0$ and in $E_+(0)$ 
there exists a basis $\cS_{k}^+$, 
$k=-1/2,1/2$, satisfying (\re{RRsev}) with $j=1/2$. 
 
\el 
\bexe 
Prove the lemma. {\bf Hints:} 
{\rm {\it i)} Let us denote $\hat\bJ_+:=\hat\bJ_1+i\hat\bJ_2$. 
Then 
$\cS_{k}^\pm:=\hat\bJ_+^{k+l_+}\cS_{-l_+}^\pm$, 
$k\!=\!-l_\pm,-l_\pm+1,...,l_\pm$. 
\\ 
{\it ii)} 
Hence, the space $E_\pm(l)$ is uniquely determined by 
$\cS_{-l_\pm}^\pm\in E(l)$ which is an eigenvector 
of the operator $\hat\bJ_3$ with the eigenvalue $-\h l_\pm$. 
\\ 
{\it iii)} 
Obviously, $\cS_{-l_+}^+=e_{-l}\otimes f_{-1/2}$ and 
it remains to construct $\cS_{-l_-}^-$. 
It is orthogonal to $E_+(l)$ and 
belongs to the subspace $F\subset E(l)$ 
which consists of all eigenvectors of the operator 
$\hat\bJ_3$, with the eigenvalue $-\h l_-$, in the space $E(l)$. 
We have to choose a nonzero  vector 
$\cS_{-l_-}^-\in F$ which is orthogonal to $E_+(l)$. 
Further, consider two cases, $l\ge 1$ and $l=1$, separately: 
 
$l\ge 1$. In this case 
the eigenspace $F$ 
is the two-dimensional linear span of the vectors 
$e_{-l}\otimes f_{1/2}$ and $e_{-l+1}\otimes f_{-1/2}$ 
if $l\ge 1$. 
The intersection $E_+(l)\cap F$ is the one-dimensional 
linear span of the vector $\hat\bJ_+\cS_{-l_+}^+$. 
Hence, $\cS_{-l_-}^-\in F$ 
is determined uniquely (up to a factor) 
as a vector orthogonal 
to 
$\hat\bJ_+\cS_{-l_+}^+$.

$l=0$. Now the space $F$ is the one-dimensional span of 
$e_{0}\otimes f_{1/2}$ since the  vector $e_{1}$ 
does not exist in this case. Hence, $E_-(0)=0$.

} 
\eexe 
{\bf Proof of Lemma \re{Rlssh} } 
{\it i)} 
The functions $\cS_{l_\pm,k}^\pm:=\bJ_3^{k+l_\pm}\cS_{-l_-}^\pm$, 
$k=-l_\pm,...,l_\pm$, are solutions to the problem 
(\re{RRsev}) with $j=l_\pm$, and 
form an orthogonal basis in  the space $E(l)$. 
\\ 
{\it ii)} The solutions are linear combinations of 
the orthogonal functions 
$\cS_{j,k}^+$ and $\cS_{j,k}^-$ if $(j,k)\ne (1/2,-1/2)$. 
Otherwise, all solutions are proportional to 
 $\cS_{1/2,-1/2}^+$ since the function 
$\cS_{1/2,-1/2}^-$ does not exist.\bo 
\medskip 
 
It turns out that the spaces $E_ \pm(l)$ are 
eigenspaces also for the operator $\si\bL$: 
\bl\la{Rlsl} 
$\si\hat\bL$ takes the value $\h l$ resp.  $-\h (l+1)$ on the space 
$E_+(l)$ resp. $E_-(l)$. 
\el 
\Pr 
This follows immediately from the identity 
\be\la{Rifi} 
~~~~~~~~~~~~~~~ 
~~~~~~~~~~~~~~~~~~~~~~~~~~~ 
\si\hat\bL=[\hat\bL+\ds\fr12 \h\si]^2-\hat\bL^2-\ds\fr14\h^2\si^2 
~~~~~~~~~~~~~~~~~~~~~~~~~~~~~~~~~~~~~~~~~ 
\ee 
since $l_\pm(l_\pm+1)-l(l+1)$ equals either $\h l$ or $-\h (l+1)$.\bo

\subsection{Radial Equations} 
We are going to substitute the expansion 
(\re{Rsevf}) into the coupled equations (\re{Rdirco}) to derive 
ordinary differential equations for the radial functions 
$R^\pm_\pm(r)$. 
For this purpose we need an expression of the operator $\si\bP$ 
in terms of the orbital angular momentum and related operators. 
The following lemma gives the necessary relations. 
 
\bl 
The following relations hold: 
\beqn 
\si\bP=|\x|^{-2} \si\x(\x\bP+i\si\bL),~~~~\la{Rrsp}\\ 
~~~\si\x(\si\bL+\h)+(\si\bL+\h)\si\x=0.\la{Rrsx} 
\eeqn 
\el 
\Pr 
The formula for products of spin matrices gives 
\be\la{Rsxp} 
(\si\x)(\si\bP)=\x\bP+i\si(\x\times\bP)=\x\bP+i\si\bL. 
\ee 
Now the equation (\re{Rrsp}) follows on multiplying 
this equation on the left by $\si\x$. 
The  equation (\re{Rrsx}) follows on multiplying the commutation relations 
$[\bL_j,\x_k]=i\h\eps_{jkl}\x_l$ by $\si_j\si_k=2\de_{jk}-\si_k\si_j$ 
and simplifying.\bo 
 
Substituting  the expression (\re{Rrsp}) into 
 the equations (\re{Rdirco}) and using  (\re{Rrsx}), we get 
\be\la{Rdircog} 
\left\{\ba{ll} 
(E-\mu c^2-e\phi)\psi_+&=c 
|\x|^{-2} \si\x(\x\bP+i\si\bL) 
\psi_-\\ 
&=c 
|\x|^{-2}(\x\bP-i(\si\bL+\h))\si\x 
\psi_-\medskip\\ 
(E+\mu c^2-e\phi)\psi_-&=c|\x|^{-2} \si\x(\x\bP+i\si\bL) 
\psi_+. 
\ea\right. 
\ee 
The last equation can be rewritten as 
\be\la{Rcbr} 
(E+\mu c^2-e\phi)\si\x\psi_-=c (\x\bP+i\si\bL) 
\psi_+. 
\ee 
Together with the second equation in (\re{Rdircog}), 
this suggests the substitution $\Psi_-:=\ds\fr{\si\x}{|\x|}\psi_-$ 
and $\Psi_+:=\psi_+$. 
Then we get, rewriting the equations (\re{Rdircog}) in 
spherical coordinates, 
\be\la{Rdircogs} 
\left\{\ba{ll} 
(E-\mu c^2-e\phi)\Psi_+&=c 
(-i\h\ds\fr d{dr}-ir^{-1}(\si\bL+2\h)) 
\Psi_-\\ 
\\ 
(E+\mu c^2-e\phi)\Psi_-&=c 
(-i\h\ds\fr d{dr}+ir^{-1}\si\bL) 
\Psi_+. 
\ea\right. 
\ee 
By Lemma \re{Rlsv}, it suffices to construct 
all nonzero 
solutions to (\re{Rdirco}) in the form (\re{Rsevf}). 
For example, we can assume that 
\be\la{Rvca} 
R^+_+(r)\not\equiv 0. 
\ee 
Let us denote by $\pi_{jk}^+$ the orthogonal projection in 
the space $E_1$ onto the linear span of the function $\cS_{jk}^+$. 
Denote by  $\Pi_{jk}^+:=\pi_{jk}^+\otimes 1$ the corresponding 
projector 
in the space 
$E_1\otimes\L^2(\R)=\L^2(\R^3)\otimes\C^2$. 
Then $\Pi_{jk}^+$ commutes with the operator $\si\hat\bL$ 
by Lemma \re{Rlsl}. Hence, applying $\Pi_{jk}^+$ to the equations 
(\re{Rdircogs}), we get 
\be\la{Rdircop} 
\left\{\ba{l} 
(E-\mu c^2-e\phi)\Pi_{jk}^+\Psi_+=c 
(-i\h\ds\fr d{dr}-ir^{-1}(\h l+2\h)) 
\Pi_{jk}^+\Psi_-\\ 
\\ 
(E+\mu c^2-e\phi)\Pi_{jk}^+\Psi_-=c 
(-i\h\ds\fr d{dr}+ir^{-1}\h l) 
\Pi_{jk}^+\Psi_+, 
\ea\right. 
\ee 
where $l=j-\ds\fr12$ by the lemma. Let us note that 
\be\la{Rpjk} 
\Pi_{jk}^+\Psi_+=R_1(r)\cS_{jk}^+(\theta,\vp),~~~~~~~ 
\Pi_{jk}^+\Psi_-=R_2(r)\cS_{jk}^+(\theta,\vp), 
\ee 
where 
\be\la{Rnzs} 
R_1(r)\equiv R^+_+(r)\not\equiv 0 
\ee 
by (\re{Rvca}). Let us denote by $R(r):= 
\left(\ba{c} R_1(r)\\R_2(r)\ea\right)$ and 
substitute the representations 
(\re{Rpjk}) into the equations (\re{Rdircop}). Then (\re{Rdircop}) 
is equivalent to 
the following {\it radial equation} for the vector-function 
$R(r)$: 
\be\la{Rmre} 
(E-e\phi-\mu c^2\si_3)R(r)= 
-ic\h 
\Bigg[ 
\Big(\ds\fr d{dr}+\fr 1r\Big)+\fr{(l+1)}r \si_3 
\Bigg]\si_1 R(r),~~~~~~~~~~r>0, 
\ee 
where $\si_k$ are the Pauli matrices. 
 
\bc 
The eigenvalue problem for the Dirac equation 
with a spherically symmetric electrostatic 
potential can be written in the form 
\be\la{Rmreb} 
(E-e\phi-\mu c^2\si_3)R(r)= 
-ic\h 
\Bigg[ 
\Big(\ds\fr d{dr}+\fr 1r\Big)\si_1+i\fr{(l+1)}r \si_2 
\Bigg]R(r),~~~~~~~~~~r>0 
\ee 
which follows from  (\re{Rmre}) since $\si_3\si_1=i\si_2$. 
\ec

\subsection{Hydrogen Spectrum} 
Here we calculate the eigenvalues $E$ of the problem (\re{Rdebs}). 
As in the nonrelativistic case, substitute 
$R(r)=e^{-\ka r}P(r)$. 
Then the equation (\re{Rmre}) reduces to 
\be\la{Rmrebr} 
~~~~~~~~~~~~~~(E-e\phi-\mu c^2\si_3)P(r)= 
-ic\h 
\Bigg[ 
\Big(\ds\fr d{dr}+\fr 1r-\ka\Big)\si_1+i\fr{(l+1)}r \si_2 
\Bigg]P(r),~~~~~~r>0, 
\ee 
or equivalently, to 
\be\la{Rmrebre} 
~~~~~~~~~~~(E-\mu c^2\si_3  -ic\h\ka\si_1)P(r)= 
-ic\h 
\Bigg[ 
\Big(\ds\fr d{dr}+\fr 1r\Big)\si_1+i\fr{(l+1)}r \si_2 
+\fr{ie\phi}{c\h} 
\Bigg]P(r),~~~r>0. 
\ee 
For the above matrix we introduce the notation 
\be\la{Rdma} 
M\equiv \fr i{c\h}(E-\mu c^2\si_3  -ic\h\ka\si_1) , 
\ee 
because it frequently appears in the calculations. Let us also rewrite 
the Coulombic potential as $e\phi=-c\h\al/r$, where 
$$ 
\al:=\fr{e^2}{c\h}\approx\fr 1{137} 
$$ 
is the dimensionless 
Sommerfeld {\it fine structure constant}. 
 
We find the parameter $\ka$ from the asymptotic condition at infinity. 
Namely, we suggest that $R(r)$ is a ``polynomial'' 
\be\la{RRr} 
R(r)=r^\de\sum_{0}^n R_kr^k 
\ee 
with an $R_n\ne 0$. 
Then 
the equation  (\re{Rmrebre}) implies that 
$MR_n=0$, hence $\det M=0$: 
\be\la{Rdema} 
\det(E-\mu c^2\si_3  -ic\h\ka\si_1)=0. 
\ee 
This is equivalent to 
\be\la{Rdemae} 
c^2\h^2\ka^2=\mu^2c^4-E^2, 
\ee 
so, in particular, $E<\mu c^2$. 
We have to choose the positive root for $\ka$ to have 
a solution satisfying (\re{RQE}). 
To justify (\re{RRr}), we seek a solution in the general form 
\be\la{RRrg} 
R(r)=r^\de\sum_{0}^\infty R_k r^k 
\ee 
where we can assume that $R_0\ne 0$ for a nontrivial 
solution. 
Substituting into (\re{Rmrebre}), we get the equation 
\be\la{Rsge} 
\sum_{0}^\infty r^{k+\de}MR_k=\sum_{0}^\infty 
\Big[ 
(k+\de+1)\si_1+i(l+1)\si_2+i\al 
\Big]r^{k+\de-1}R_k. 
\ee 
This gives the recurrence equation 
\be\la{Rrece} 
MR_{k-1}=\Big[ 
(k+\de+1)\si_1+i(l+1)\si_2+i\al 
\Big]r^{k+\de-1}R_k, ~~~~~k=0,1,... 
\ee 
This equation with $k=0$ implies 
that 
\be\la{Rine} 
\Big[ 
(\de+1)\si_1+i(l+1)\si_2+i\al 
\Big]R_0=0. 
\ee 
This implies the "indicial equation'' 
\be\la{Rineq} 
\det\Big[ 
(\de+1)\si_1+i(l+1)\si_2+i\al 
\Big]=0 
\ee 
since $R_0\ne 0$. 
It is equivalent to 
\be\la{Rineqe} 
(\de+1)^2=(l+1)^2-\al^2. 
\ee 
Then $|\de+1|\approx l+1$ 
since $\al$ is small. 
Therefore, 
we have to choose the positive root for $\de+1$ 
since for the negative root we get $-\de\approx l+2\ge 2$ 
while 
$\de>-3/2$ 
by the condition (\re{RQE}). 
 
Finally, an investigation of the recurrence 
equation (\re{Rrece}) shows that the series (\re{RRrg}) 
should terminate by the condition (\re{RQE}) 
as in the case of  the Schr\"odinger equation. 
Hence, we arrive at  (\re{RRr}) 
with an $R_n\ne 0$. This implies again 
 (\re{Rdema})  and  (\re{Rdemae}), however it is not sufficient 
to determine the eigenvalues $E$ since we have the additional 
unknown parameter $\ka$. Therefore, we need an additional equation 
which we will derive from the  recurrence 
equation (\re{Rrece}) with $k=n$: 
\be\la{Rrecen} 
MR_{n-1}=\Big[ 
(n+\de+1)\si_1+i(l+1)\si_2+i\al 
\Big]r^{n+\de-1}R_n. 
\ee 
Namely, the characteristic equation for the matrix $M$ reads, 
\be\la{RceM} 
(M-2iE/c\h)M=0 
\ee 
since its determinant is zero and the trace is $2iE/c\h$. 
Therefore, multiplying both sides of (\re{Rrece}) by 
$M-2iE/c\h$, we get 
\be\la{Rreceg} 
0=(M-2iE/c\h)\Big[ 
(n+\de+1)\si_1+i(l+1)\si_2+i\al 
\Big]R_n. 
\ee 
Multiplying here the Pauli matrices, we arrive at 
\be\la{Rsee} 
0=\Big[2\ka(n+\de+1)-2\al E/c\h\Big]R_n. 
\ee 
This gives us the new quantization condition 
\be\la{Rsqc} 
\al E=c \h(n+\de+1), 
\ee 
which together with the equation (\re{Rdemae}) 
determines the eigenvalues $E$: 
solving the system of equations, 
we get 
\be\la{REde} 
E=E_{ln}=\fr{\mu c^2}{\sqrt{1+\Big[\al^2/(n+\de+1)^2\Big]}}, 
\ee 
where $\de=\de(l)$ is given by (\re{Rineqe}). 
 
Since $\al$ is small, we can approximate the eigenvalues by the 
binomial expansion: 
\be\la{REdea} 
E_{ln}\approx \mu c^2 
-\fr {\mu c^2\al^2}{2(n+\de+1)^2}. 
\ee 
\brs 
i) {\rm The approximation (\re{REdea}) with $\de=0$ coincides 
with the nonrelativistic spectrum of the hydrogen atom up to the 
unessential additive constant $\mu c^2$. 
} 
\\ 
ii) {\rm The relativistic formula depends on the angular momentum 
$j$ through $\de=\de(l)$, while the nonrelativistic formula does 
not depend on the angular momentum. 
This was another  triumph of the Dirac theory 
since it corresponds to the experimental observation of the 
{\it fine structure}. 
} 
\ers 

\br
The above analysis gives also the corresponding eigenfunctions
\be\la{Rsevfg} 
\psi_E=
\left(\ba{c} 
\psi_+ 
\\~\\ 
\psi_- 
\ea\right)=
\ds\fr{\si\x}{|\x|}
\left(\ba{c} 
\Psi_+ 
\\~\\ 
\Psi_- 
\ea\right)=
\ds\fr{\si\x}{|\x|}
\left(\ba{c} 
R_1(r)\cS^+_{jk}(\theta,\vp) 
\\~\\
R_2(r)\cS^+_{jk}(\theta,\vp) 
\ea\right)
\ee

\er

\part{Mathematical Appendices}

\setcounter{subsection}{0} 
\setcounter{equation}{0} 
\section{Newton Mechanics} 
 
We recall the Newton mechanics of one and many particles, 
for potential force fields. In the case of a certain symmetry 
of the potential we derive the corresponding conservation laws.

\subsection{One Particle} 
\subsubsection{Newton equation} 
The motion of one particle of mass $m>0$ is governed by 
the Newton differential equation 
\be\la{N1} 
m\ddot \x(t)=F(\x(t),t),\,\,\,\,t\in\R. 
\ee 
Here $\x(t)\in\R^3$ is the particle position at time $t$ and 
$F(\cdot)$ is the force field. Let us assume that 
$F\in C^1 (\R^3\times\R,\R^3)$. Then 
the solution $\x(t)$ is defined uniquely by the initial 
conditions $\x(0)=\x_0\in\R^3$, $\dot \x(0)=\bv_0\in\R^3$ 
by the {\em main theorem 
of ordinary differential equations}. 
The solution exists for $|t|\le \ve$, where $\ve>0$ 
depends on the initial data $\x_0,v_0$. 
\subsubsection{Energy conservation} 
Let us assume that 
the force field $F$ has a {\it potential function} 
(or simply {\it potential}) $V(\cdot)\in C^2(\R^3\times\R)$, 
\be\la{p1} 
F(\x,t)=-\na V(\x,t),\,\,\,\,\x\in\R^3,\,\,\,t\in\R. 
\ee 
\bd 
i) $\E:=\R^3\times\R^3$ is the {\bf phase space} of the Newton equation, 
$\E^+:=\R^3\times\R^3\times\R$ is the {\bf extended phase space} 
of the Newton equation. 
\\ 
ii) The energy $E(\x,\bv,t)$ is the function on the extended phase space 
defined by 
\be\la{E1} 
E(\x, \bv,t)=\fr{m\bv^2}2+ V(\x,t),\,\,\,\,(\x,\bv,t)\in \E^+. 
\ee 
\ed 
\bt 
Let the condition (\re{p1}) hold. Further, we assume that 
the potential does not depend on $t$, 
\be\la{Vt} 
V(\x,t)\equiv V(\x),\,\,\,\,\,\,(\x,t)\in\R^3\times\R. 
\ee 
 Then for every solution 
$\x(t)\in C^2([t_0, t_1],\R^3)$ to the Newton equation, 
the  energy is conserved, 
\be\la{Ec} 
E(t):=E(\x(t), \dot \x(t))=\co ,\,\,\,t\in [t_0, t_1]. 
\ee 
\et 
\Pr By the chain rule of differentiation, 
the Newton equation (\re{N1}), and (\re{p1}), 
\be\la{Ecp} 
\dot E(t)=m\dot \x(t)\cdot \ddot \x(t)+\na V(\x(t))\cdot \dot \x(t) 
= [m\ddot \x(t)+\na V(\x(t))]\cdot \dot \x(t)=0,\,\,\,t\in [t_0, t_1]. 
\ee 
 
\subsubsection{Well-posedness condition} 
 
\bt 
Let the condition (\re{Vt}) hold, 
and let the potential be bounded from below 
by some constant $C\in\R$, 
\be\la{wp} 
V(\x)\ge C,\,\,\,\,\x\in\R^3. 
\ee 
Then 
every solution $\x(t)$ to the Newton equation (\re{N1}) 
exists globally in time, i.e., for all $t\in \R$. 
\et 
\Pr The energy conservation (\re{Ec}) implies that the 
velocity is bounded, $|\dot \x(t)|\le \ov v$. Then also $|\x(t)|$ 
is bounded by $\ov vt+\co$. 
This provides the existence of the global solution 
for all $t\in \R$.\bo 
 
\subsection{Many Particles} 
\subsubsection{Newton equations} 
The motion of $n$ particles with masses $m_i>0$ is governed by 
the Newton differential equation 
\be\la{NN} 
m_i\ddot \x_i(t)=F_i(x(t),t),\,\,t\in\R,\,\,\,\,\,i=1,...,n. 
\ee 
Here 
\\ 
i) $\x_i(t)\in\R^3$ is the position of the $i$-th particle 
at time $t$, $x(t)=(\x_1(t),...,\x_n(t))\in\R^{3n}$ and 
\\ 
ii) $F_i(x(t),t)\in\R^3$ is the force acting on the  $i$-th particle. 
 
Let us assume that the force field 
$F(x,t):=(F_1(x,t),...,F_n(x,t)) 
\in C^1 (\R^{3n}\times\R,\R^{3n})$. 
Then 
the solution $x(t)=(\x_1(t),...,\x_n(t))$ to the system  (\re{NN}) 
is defined uniquely by the initial 
conditions $x(0)=x_0\in\R^{3n}$, 
$\dot x(0)=v_0\in\R^{3n}$ by the {\em main theorem 
of ordinary differential equations}. 
The solution exists for $|t|\le \ve$, where $\ve>0$ 
depends on the initial data $x_0,v_0$.

\subsubsection{Energy conservation} 
Let us assume that 
the force field $F(x,t)$ has a {\it potential function} 
(or simply {\it potential}) $V(\cdot)\in C^2(\R^{3n})$, 
\be\la{pN} 
F_i(x,t)=-\na_{\x_i} V(x,t),\,\,\,\,x\in\R^{3n},\,\,\,i=1,...,n. 
\ee 
\bd 
i) $\E=\R^{3n}\times\R^{3n}$ is the 
{\bf phase space} of the Newton system (\re{NN}), 
$\E^+:=\R^{3n}\times\R^{3n}\times\R$ is the {\bf extended phase space} 
of the Newton system (\re{NN}). 
\\ 
ii) The energy $E(x,v ,t)$ is the function on the extended 
phase space defined by 
\be\la{EN} 
E(x, v,t)=\sum_i\fr{m_i\bv_i^2}2+ V(x,t),\,\,\,\,(x,v)\in \E, 
\ee 
where $v=(\bv_1,...,\bv_n)$. 
\ed 
 
Let us call {\it a trajectory} any solution 
$x(t)$ to the Newton system 
(\re{NN}). 
\bt 
Let the condition (\re{pN}) hold. Further, we assume that 
the potential does not depend on $t$, 
\be\la{Vtn} 
V(x,t)\equiv V(x),\,\,\,\,\,\,(x,t)\in\R^{3n}\times\R. 
\ee 
Then for any trajectory 
$x(t)\in C^2([t_0, t_1],\R^{3n})$, 
the  energy is conserved, 
\be\la{EcN} 
E(t):=E(x(t), \dot x(t))=\co ,\,\,\,t\in [t_0, t_1]. 
\ee 
\et 
\Pr By the chain rule of differentiation, 
the Newton system (\re{NN}), and (\re{pN}), 
\beqn\la{EcNp} 
\dot E(t)&=&\sum_i m_i\dot \x_i(t)\cdot \ddot \x_i(t)+ 
\sum_i \na_{\x_i} V(x(t))\cdot \dot \x_i(t)\nonumber\\ 
&= &\sum_i [m_i\ddot \x_i(t)+\na_{\x_i} V(x(t))]\cdot \dot \x_i(t)=0, 
\,\,\,\,\,\,\,\,\,\,\,\,t\in [t_0, t_1]. 
\eeqn 
 
\subsubsection{Well-posedness condition} 
 
\bt 
Let the condition (\re{Vtn}) hold, and 
let the potential be bounded from below 
by some constant $C\in\R$: 
\be\la{wpN} 
V(x)\ge C,\,\,\,\,x\in\R^{3n}. 
\ee 
Then 
every solution $x(t)$ to the Newton equation (\re{NN}) 
exists globally in time, i.e. for all $t\in \R$. 
\et 
\Pr The energy conservation (\re{EcN}) implies that the 
velocity is bounded, $|\dot x(t)|\le$const. Then also $|x(t)|$ 
is bounded by $\ov vt+\co$. This provides the existence of the global solution 
for all $t\in \R$.\bo 
 
\subsection{Symmetry Theory} 
The invariance of the potential $V$ with respect to translations in time, 
(\re{Vtn}), provides the energy conservation (\re{EcN}). 
Let us show that the invariance of the potential 
$V(x)$ with respect to some 
transformations of the {\it configuration space} 
$Q:=\R^{3n}$ leads to new conservation laws. 
\subsubsection{Translation group} 
Let us fix a vector $\bh\ne 0$ in $\R^3$ and 
consider the translations $x\mapsto x+\bh s$ of $\R^3$ 
 and the corresponding 
action in $\R^{3n}$: 
\be\la{tr} 
T_s(\x_1,,,.,\x_n)=(\x_1+\bh s,,,.,\x_n+\bh s), 
\,\,\, (\x_1,,,.,\x_n)\in\R^{3n}. 
\ee 
\bd\la{tin} 
The system (\re{NN}) is invariant with respect to 
the translations (\re{tr}) if 
\be\la{trin} 
V(T_s(x),t)= V(x,t), 
\,\,\, (x,t)\in\R^{3n+1}, 
\,\,\,\,\,\forall s\in\R. 
\ee 
\ed 
\bex The Newton system (\re{NN}) is invariant with respect to 
the translations (\re{tr}) with every $\bh \in\R^3$, if 
the potential energy has the structure 
\be\la{trine} 
V(\x_1,...,\x_n,t)= W (\x_1-\x_n,..., \x_{n-1}-\x_n,t),\,\,\,\,\, 
(\x_1,...,\x_n,t)\in \R^{3n+1} 
\ee 
with a function $W$ of $3n-2$ variables. 
\eex 
\bd\la{dmN} 
i) The momentum $\p_i$ of the $i$-th particle 
is the vector function 
on the phase space $\E$ defined by 
\be\la{miN} 
\p _i:=m_i \bv_i\in\R^3,\,\,\,\,\,(x,v)\in \E. 
\ee 
ii) the (total) momentum $\p $ of the system 
(\re{NN}) is the vector function 
on the phase space $\E$ defined by 
\be\la{mN} 
\p :=\sum_i \p _i=\sum_i m_i \bv_i\in\R^3,\,\,\,\,\,(x,v)\in \E. 
\ee 
iii) 
The center of mass of the system of $n$ particles is 
\be\la{cemd} 
\X:=\fr 1M \sum_i m_i \x_i, \,\,\,\,\,x=(\x_1,...,\x_n)\in\R^{3n}, 
\ee 
where 
$M:=\sum_i m_i$ is the total mass of the system. 
\ed 
\bt\la{tmc} 
Let  (\re{pN}) hold and 
the system (\re{NN}) be invariant with respect to 
the translations (\re{tr}) along 
 a fixed vector $\bh \in\R^3$. Then for any trajectory 
$x(t)\in C^2([t_0, t_1],\R^{3n})$, 
the projection of the momentum 
$\p (t)$ onto $\bh$ is conserved, 
\be\la{mNc} 
\p _\bh (t):=\p (t)\cdot \bh =\co , 
\,\,\,\,\,t\in [t_0, t_1]. 
\ee 
\et 
\Pr 
By (\re{pN}), (\re{NN}),  and 
the chain rule of differentiation, 
\be\la{mcNp} 
~~~~~~~~~\dot \p _\bh (t)=\sum_i m_i\ddot \x_i(t)\cdot \bh 
=-\sum_i \na_{\x_i} V(x(t),t)\cdot \bh 
=-\fr {d}{ds}\Bigg|_{s=0} V(T_s x(t),t)=0,\,\,t\in [t_0, t_1] 
\ee 
by (\re{trin}).\bo 
 
\bc\la{ccem} 
Let the  Newton system (\re{NN}) be invariant with respect to 
the translations (\re{tr}) along 
 all vectors $\bh \in\R^3$. Then for any trajectory, 
the momentum 
$\p (t)$ is conserved, 
$\p (t)=\co $, 
 and 
the center of mass  $\X(t)= \sum_i m_i \x_i(t)/M$ moves 
uniformly: $\X(t)=\bv t+\X(0)$. 
\ec 
\Pr Since (\re{mcNp}) holds for every $\bh$, we have 
$\p (t)=\co $ and 
$$ 
~~~~~~~~~~~~~~~~~~~~~~~~~~~~~~~~~~~\ddot \X(t)=\fr 1M \sum_i m_i \ddot \x_i=0. 
~~~~~~~~~~~~~~~~~~~~~~~~~~~~~~~~~~~~~~~~~~~~~~~\loota 
$$ 
 
\subsubsection{Rotation group} 
Let us fix a unit vector $\e\in\R^3$ and 
consider the rotation around $\e$ in $\R^3$ 
with an angle of $s$ {\it radian}. Let us denote by 
$R_\e(s)\in SO(3)$ the corresponding orthogonal matrix 
and define the corresponding  transformation in $\R^{3n}$ by 
\be\la{rot} 
R_s(\x_1,...,\x_n)= (R_\e(s)\x_1,...,R_\e(s)\x_n),\,\,\, (\x_1,...,\x_n)\in\R^{3n}. 
\ee 
\bd\la{rin} 
The system (\re{NN}) is invariant with respect to 
the rotations (\re{rot}) if 
\be\la{atrin} 
V(R_s(x),t)= V(x,t),\,\,\,(x,t)\in\R^{3n+1}, 
\,\,\,\,\,\forall s\in\R. 
\ee 
\ed 
\bex The system (\re{NN}) is invariant with respect to 
the rotations (\re{tr}) with every $\e\in\R^3$, if 
the potential energy has the structure 
\be\la{rotine} 
V(\x_1,...,\x_n,t)= W(\{|\x_i-\x_j|:1\le i < j\le n\},t),\,\,\,\,\, 
(\x_1,...,\x_n,t)\in \R^{3n+1}. 
\ee 
\eex 
\bd\la{damN} 
i) The angular momentum $\bL_i$ of the $i$-th particle 
is the vector function 
on the phase space $\E$ defined by 
\be\la{amiN} 
\bL_i(x,v):=\x_i\times \p_i\in\R^3,\,\,\,\,\,(x,v)\in \E. 
\ee 
ii) the angular momentum $\bL$ of the Newton system (\re{NN}) 
is the vector function 
on the phase space $\E$ defined by 
\be\la{amN} 
\bL(x,v):=\sum_i \bL_i=\sum_i \x_i\times \p_i\in\R^3,\,\,\,\,\,(x,v)\in \E. 
\ee 
\ed 
\bt\la{tamc} 
Let the  Newton system (\re{NN}) be invariant with respect to 
the rotations (\re{rot}) around 
 a fixed vector $\e\in\R^3$. Then for any trajectory 
$x(t)\in C^2([t_0, t_1],\R^{3n})$, 
the projection of the angular momentum 
$\bL(x(t),\dot x(t))$ onto $\e$ is conserved, 
\be\la{amNcr} 
\bL_r(t):=\bL(x(t),\dot x(t))\cdot \e=\co , 
\,\,\,\,\,t\in [t_0, t_1]. 
\ee 
\et 
\Pr 
The differentiation gives, 
\beqn\la{amcNp} 
\dot \bL_\e(t)&=& 
[\sum_i \dot \x_i(t)\times \p_i(t)+\sum_i \x_i(t)\times \dot \p_i(t)] 
\cdot \e\nonumber\\ 
&=& 
[\sum_i \dot \x_i(t)\times m_i\dot \x_i(t) + 
\sum_i \x_i(t)\times m_i\ddot \x_i(t)] 
\cdot \e\nonumber\\ 
&=&\sum_i [\x_i(t)\times m_i\ddot \x_i(t)]\cdot \e 
=\sum_i m_i\ddot \x_i(t)\cdot[\e\times \x_i(t)]. 
\eeqn 
Therefore, the Newton system (\re{NN}) and (\re{amN}) 
imply by 
the chain rule of differentiation, 
\be 
\dot \bL_\e(t)=-\sum_i \na_{\x_i} V(x(t),t)\cdot[\e\times \x_i(t)] 
=-\fr {d}{ds}\Bigg|_{s=0} V(R_s x(t),t)=0,\,\,\,\,t\in [t_0, t_1]. 
\ee 
by (\re{atrin}), since 
\be\la{rotv} 
\e\times \x_i(t)=\fr{d}{ds}\Bigg|_{s=0} R_\e(s) \x_i(t). 
\ee 
This identity follows from the fact that the vectors on 
both sides are orthogonal 
to the plane containing $e$ and $\x_i(t)$, and have the same length. 
Indeed, the length of the LHS is $|\x_i(t)|\sin\al$, 
where $\al$ is the angle between $e$ and $\x_i(t)$, 
and the length of the RHS is the radius of the circle 
$\{R_\e(s) \x_i(t):s\in [0,2\pi]\}$ which is equal to 
$|\x_i(t)|\sin\al$.\bo 
 
\newpage 
%%%%%%%%%%%%%%%%%%%%%%%%%%%%%%%%%%%%%%%%%%%%%%%%%%%%%%%% 
 
%%%%%%%%%%%%%%%%%%%%%%%%%%%%%%%%%%%%%%%%%%%%%%%%%%%%%%%% 
 
%%%%%%%%%%%%%%%%%%%%%%%%%%%%%%%%%%%%%%%%%%%%%%%%%%%%%%%% 

%%\setcounter{section}{+3} 
\setcounter{subsection}{0} 
\setcounter{theorem}{0} 
\setcounter{equation}{0} 
\section{Lagrangian Mechanics} 
 
We introduce Lagrangian systems corresponding to one particle 
and to many particles, formulate the Hamilton least action 
principle, derive the Euler-Lagrange equations and check that 
the Newton equations are of the Euler-Lagrange form. 
In the case of a certain symmetry of the Lagrangian function 
we derive the corresponding conservation laws. 
 
\subsection{One Particle} \label{one-p} 
\subsubsection{Lagrangian function} 
We expose the Lagrangian form of the Newton equation 
(\re{N1}) with the potential (\re{p1}), 
\be\la{N1L} 
m\ddot \x(t)=-\na V(\x(t),t),\,\,\,\,t\in\R. 
\ee 
\bd 
The Lagrangian $L(\x,\bv,t)$ of the system 
is the following function on the extended phase space 
$\E^+=\R^3\times\R^3\times\R$ 
(cf. (\re{E1})), 
\be\la{E1L} 
L(\x, \bv,t)=\fr{m\bv^2}2- V(\x,t),\,\,\,\,(\x,\bv,t)\in \E^+. 
\ee 
\ed 
\bexe 
Check that the Newton equation 
(\re{N1L}) can be represented in the {\bf Euler-Lagrange} form, 
\be\la{EL1} 
\fr d{dt}L_\bv(\x(t),\dot \x(t),t)=L_\x(\x(t),\dot \x(t),t) 
,\,\,\,\,t\in\R. 
\ee 
\eexe 
Let us consider more general Lagrangian systems with 
an arbitrary function $L(\x,\bv,t)$. 
\bd \la{LmE1} 
i) The Lagrangian system for one particle 
is the dynamical system described by the 
Lagrangian equation (\re{EL1}) with a 
function $L(\x,\bv,t)\in C^2(\E^+)$. 
\\ 
ii) The momentum of the Lagrangian system 
is the vector-function on the extended phase space $\E^+$ defined by 
\be\la{mLd} 
\p=L_\bv(\x,\bv,t),\,\,\,\,(\x,\bv,t)\in \E^+. 
\ee 
iii) The energy of the Lagrangian system 
is the function on the extended phase space $\E^+$ defined by 
\be\la{eLd} 
E(\x,\bv,t)=\p\bv-L(\x,\bv,t),\,\,\,\,(\x,\bv,t)\in \E^+. 
\ee 
\ed 
\bex 
The Newton equation (\re{N1L}) 
results from the Lagrangian system with the 
Lagrangian functional (\re{E1L}), momentum $\p=m\bv$, and 
 energy $E=\ds\fr{m\bv^2}2+V(\x,t)$. 
\eex 
\bt\la{ECL} Let the Lagrangian not depend on time, 
\be\la{Lt1} 
L(\x, \bv,t)=L(\x, \bv),\,\,\,\,(\x,\bv,t)\in \E^+. 
\ee 
Then for any trajectory 
$\x(t)\in C^2([t_0, t_1],\R^3)$, 
the energy is conserved, (\re{Ec}). 
\et 
\Pr The differentiation of (\re{eLd}) with $\x=\x(t)$ and $\bv=\dot \x(t)$ 
gives, 
\be\la{pec} 
\dot E(t)=\dot \p \bv+\p\dot \bv-L_\x\dot \x-L_\bv\dot \bv= 0 
\ee 
by Equations (\re{EL1}) and Definition (\re{mLd}). 
\bo 
\bexe 
Calculate the momentum and the energy 
for the Lagrangian $L(\x,\bv)=-m\sqrt{1-\bv^2}$. 
\eexe 
\subsubsection{Action functional} 
\bd\la{C1} 
$C^1=C^1([0,\infty), \R^3)$ is the space of all paths in 
three-dimensional space. 
\ed 
We will consider the real-valued functionals ${\cal F}$ on $C^1$. 
By definition, ${\cal F}$ is a map $C^1\to\R$. 
\bex ${\cal F}(\x)=\ds\int_0^T |\dot \x(t)|dt$ is the 
length of the path $\x(\cdot)\in C^1$, $t\in[0,T]$. 
\eex 
\bd \la{dGd} 
The Gateau differential $D{\cal F}(\x)$ is the {\bf linear} functional 
$C^1\mapsto\R$ defined by 
\be\la{Gd} 
\langle D{\cal F}(\x), \bh  \rangle=\fr{d}{d\ve}\Bigg|_{\ve=0} 
{\cal F}(\x+\ve \bh ), 
\,\,\,\,\,\bh (\cdot)\in C^1 
\ee 
if the derivative on the RHS exists. 
\ed 
Let us fix a $T>0$. 
\bd The action is the functional on $C^1$ defined by 
\be\la{S} 
S_T(\x(\cdot))=\int_0^T L(\x(t),\dot \x(t),t)dt,\,\,\,\,\x(\cdot)\in C^1. 
\ee 
\ed 
Note that the functional is defined on the whole of $C^1$ 
if $L(\x,\bv,t)\in C(\E^+)$. Moreover, the functional is 
differentiable if $L(\x,\bv,t)\in C^1(\E^+)$: 
\bl 
The Gateau differential $DS_T(\x)$ exists for $\x\in C^1$. 
\el 
\Pr From Definition \re{dGd} we get by the theorem 
of the differentiation of integrals, 
\beqn\la{DS} 
\langle DS_T(\x), \bh  \rangle:&=&\fr{d}{d\ve}\Bigg|_{\ve=0} 
\int_0^T L(\x(t)+\ve \bh (t), \dot \x(t)+\ve \dot \bh (t),t)dt\nonumber\\ 
&=&\int_0^T [L_\x(\x(t), \dot \x(t)) \bh (t)+ 
L_\bv(\x(t), \dot \x(t),t)\dot \bh (t)]dt , 
\eeqn 
since $L(\x,\bv,t)\in C^2(\E^+)$ by our basic assumptions.\bo 
\subsubsection{Hamilton least action principle} 
Let us introduce the space of {\it variations}. 
\bd\la{C10} 
$C^1(T)=\{\bh \in C^1:\bh (0)=\bh (T)=0 \}$. 
\ed 
\bd The function $\x\in C^1$ satisfies the 
Hamilton least action principle (LAP) if, for any $T>0$, 
\be\la{LAP} 
\langle DS_T(\x), \bh  \rangle=0,\,\,\,\,\forall \bh (\cdot)\in C^1(T) 
\ee 
\ed 
\bt\la{tLAP} 
For $\x\in C^2([0,\infty),\R^3)$ the Hamilton LAP is equivalent to the 
Euler-Lagrange equations (\re{EL1}) with $t\in [0,T]$. 
\et 
\Pr The partial integration in (\re{DS}) 
gives 
\be\la{paLAP} 
\langle DS_T(\x), \bh  \rangle= 
\int_0^T [L_\x(\x(t), \dot \x(t),t)-\fr d{dt} 
L_\bv(\x(t), \dot \x(t),t)] \bh (t)dt,\,\,\,\, \bh \in C^1_0(T). 
\ee 
Therefore, (\re{LAP}) is equivalent to (\re{EL1}) 
by the following lemma: 
\bl 
{\bf Main lemma of the calculus of variations} 
(du Bois-Reymond). 
\\ 
Let a function $f(t)\in C[0,T]$ and $\int_0^T f(t)\bh (t)dt=0$ 
for any function $\bh (t)\in C[0,T]$ with the boundary values 
$\bh (0)=\bh (T)=0$. Then $f(t)=0$, $t\in [0,T]$. 
\el 
 
\bexe 
Prove the lemma. 
\eexe

\subsection{Many Particles} \label{many-p} 
\subsubsection{Lagrangian function} 
We extend the Lagrangian formalism  to the Newton equations 
(\re{NN}) with the potential (\re{pN}), 
\be\la{NNL} 
m_i\ddot \x_i(t)=-\na_{\x_i} V(x(t),t),\,\,\,\,t\in\R. 
\ee 
We introduce the 
Lagrangian $L(x,v,t)$ of the system (\re{NNL}) 
as the following function on the extended phase space 
$\E^+=\R^{3n}\times \R^{3n}\times \R$ 
(cf. (\re{E1L})), 
\be\la{ENL} 
L(x, v,t)=\sum_i\fr{m_i\bv_i^2}2- V(x,t),\,\,\,\,(x,v,t)\in \E^+, 
\ee 
where $v=(\bv_1,...\bv_n)$. 
\bexe 
Check that the Newton equation 
(\re{NNL}) can be represented in the 
 {\bf Euler-Lagrange} form, 
\be\la{ELN} 
\fr d{dt}L_{v}(x(t),\dot x(t),t)=L_{x}(x(t),\dot x(t),t) 
,\,\,\,\,t\in\R. 
\ee 
\eexe 
 
Let us consider more general Lagrangian systems with the 
 extended phase space 
$\E^+:=\R^N\times\R^N\times\R$, where $N=1,2...$, 
and an arbitrary function $L(x,v,t)$. 
\bd \la{LmEN} 
i) The Lagrangian system in the extended phase space 
$\E^+:=\R^N\times\R^N\times\R$ 
is the dynamical system described by the 
equations (\re{ELN}) with a 
function $L(x,v,t)\in C^2(\E)$. 
\\ 
ii) The momentum of the Lagrangian system 
is the vector function on the extended phase space $\E^+$ defined by 
\be\la{mLdN} 
p=L_{v}(x,v,t),\,\,\,\,(x,v,t)\in \E^+. 
\ee 
iii) The energy of the Lagrangian system 
is the function on the phase space $\E^+$ defined by 
\be\la{eLdN} 
E(x,v,t)= p v-L(x,v,t),\,\,\,\,(x,v,t)\in \E^+. 
\ee 
\ed 
\bexe Check that 
the Newton equations (\re{NNL}) 
result from the Lagrangian system with the 
Lagrangian functional (\re{ENL}), momentum 
$p=(\p_1,...,\p_n)$ where 
$\p_i=m_i\bv_i$, and 
the energy 
\be\la{En} 
E=\sum_i\ds\fr{m_i\bv_i^2}2+V(x,t). 
\ee 
\eexe 
\bt\la{ECLN} 
Let the Lagrangian not depend on time, 
\be\la{LtN} 
L(x, v,t)=L(x, v),\,\,\,\,(x,v,t)\in \E^+. 
\ee 
Then for any trajectory 
$x(t)\in C^2([t_0, t_1],\R^N)$, 
the energy is conserved, (\re{EcN}). 
\et 
\Pr The differentiation of (\re{eLdN}) with $x=x(t)$ and $v=\dot x(t)$ 
gives, 
\be\la{pecN} 
\dot E(t)=\dot p v+p\dot v 
-L_{x}\dot x-L_{v}\dot v= 0 
\ee 
by Equations (\re{ELN}) and Definition (\re{mLdN}). 
\bo

\subsubsection{Action functional} 
 
\bd\la{CN} 
$C^1=C^1([0,\infty), \R^{N})$ is the space of all paths in 
$N$-dimensional space. 
\ed 
We will consider the real-valued functionals ${\cal F}$ on $C^1$. 
By definition, ${\cal F}$ is a map $C^1\to\R$. 
\bex ${\cal F}(x)=\ds\int_0^T |\dot x(t)|dt$ is the 
length of the path $x(\cdot)\in C^1$, $t\in[0,T]$. 
\eex 
\bd \la{dGdN} 
The Gateau differential $D{\cal F}(x)$ is the {\bf linear} functional 
$C^1\mapsto\R$ defined by 
\be\la{GdN} 
\langle D{\cal F}(x), h \rangle=\fr{d}{d\ve}\Bigg|_{\ve=0} {\cal F}(x+\ve h) 
\ee 
for $h(\cdot)\in C^1$ 
if the derivative on the RHS exists. 
\ed 
Let us fix a $T>0$. 
\bd The action is the functional on $C^1(T)$ defined by 
\be\la{SN} 
S_T(x)=\int_0^T L(x(t),\dot x(t))dt,\,\,\,\,x(\cdot)\in C^1. 
\ee 
\ed 
Note that the functional is defined on the whole of $C^1(T)$ 
if $L(x,v)\in C(\E)$. Moreover, the functional is 
differentiable if $L(x,v)\in C^1(\E)$: 
\bl 
The Gateau differential $DS_T(x)$ exists for $x(\cdot)\in C^1$. 
\el 
\Pr From Definition \re{dGd} we get by the theorem 
of the differentiation of integrals, 
\beqn\la{DSN} 
\langle DS_T(x), h \rangle:&=&\fr{d}{d\ve}\Bigg|_{\ve=0} 
\int_0^T L(x(t)+\ve h(t), \dot x(t)+\ve \dot h(t))dt\nonumber\\ 
&=&\int_0^T [L_{x}(x(t), \dot x(t)) h(t)+ 
L_{v}(x(t), \dot x(t))\dot h(t)]dt , 
\eeqn 
since $L(x,v)\in C^2(\E)$ by our basic assumptions.\bo 
\subsubsection{Hamilton least action principle} 
Let us introduce the space of {\it variations}. 
\bd\la{CN0} 
$C^1(T)=\{h(\cdot)\in C^1:h(0)=h(T)=0 \}$. 
\ed 
\bd The function $x\in C^1$ satisfies the 
Hamilton least action principle (LAP) if for any $T>0$ 
\be\la{LAPN} 
\langle DS_T(x), h \rangle=0,\,\,\,\,\forall h(\cdot)\in C^1(T) . 
\ee 
\ed 
\bt\la{tLAPN} 
For $x\in C^2([0,\infty),\R^N)$ the Hamilton LAP is equivalent to the 
Euler-Lagrange equations (\re{ELN}). 
\et 
\Pr The partial integration in (\re{DSN}) 
gives 
\be\la{paLAPN} 
\langle DS(x), h \rangle= 
\int_0^T [L_{x}(x(t), \dot x(t))-\fr d{dt} 
L_{v}(x(t), \dot x(t))] h(t)dt. 
\ee 
Therefore, (\re{LAPN}) is equivalent to (\re{ELN}) 
by the main lemma of the calculus of variations. 
\bo 
\br 
The expositions in Sections \ref{one-p}  and \ref{many-p} 
formally are almost identical. 
\er

\newpage 

\setcounter{subsection}{0} 
\setcounter{theorem}{0} 
\setcounter{equation}{0} 
\section{Noether Theory of Invariants} 
We construct a Noether invariant for a system whose 
Lagrangian is invariant with respect to a one-parametric 
transformation group. This construction is applied to one 
particle and many particles mechanical systems. 
 
The invariance of 
the Lagrangian $L$ with respect to translation in time, 
(\re{LtN}), 
provides the conservation of energy (\re{Ec}). 
Let us show that the 
invariance of the Lagrangian with respect to some 
transformations of the {\it configuration space} 
$X:=\R^N$ leads to new conservation laws. 
 
\subsection{Symmetry and Noether Theorem on Invariants} 
 
Consider a group $G=\{g\}$ of differentiable 
transformations $g\in C^2(X,X)$ of the configuration 
space $X$ of a Lagrangian system.

\bd\la{dsym} 
$G$ is a {\bf symmetry group} of the Lagrangian system if 
the identity 
\be\la{sym} 
L(g(x),dg(x)v,t)=L(x,v,t), \,\,(x,v,t)\in \E^+, 
\,\,\,\,\,\forall g\in G 
\ee 
holds, where $dg:\R^N\to\R^N$ is the differential of $g$. 
\ed 
Let us recall the definition of the differential: 
\be\la{dif} 
dg(x)v:=\fr {dg(X(\tau))}{d\tau}\Bigg|_{\tau=0} 
\ee 
if $X(0)=x$ and $\dot X(0)=v$. 
 
Consider a {\bf one-parametric subgroup} $\{g_s\in G: s\in\R\}$ 
of the symmetry group $G$, 
\be\la{sym1} 
L(g_s(x),dg_s(v),t)=L(x,v,t), \,\,(x,v,t)\in \E^+,\,\,\,\,\, 
s\in \R. 
\ee 
\br 
Since $g_s$ is a one-parametric subgroup, we have 
$g_0=Id$, hence 
\be\la{s0q} 
(g_0(x),dg_0(v))=(x,v),\,\,\,\,\,\,\,~~~~~(x,v)\in \E. 
\ee 
\er 
 
\bd The Noether invariant is a function on the extended 
phase space $\E^+$, which is defined by 
\be\la{NI} 
I(x,v,t)=L_v(x,v,t)\fr{dg_s(x)}{ds}\Bigg|_{s=0}, \,\,\,\, 
(x,v,t)\in \E^+. 
\ee 
\ed 
\bt (E.Noether \ci{EN}) 
Let $x(t)\in C^2(\R,\R^N)$ be a solution to the Euler-Lagrange equations 
(\re{ELN}) and $\{g_s: s\in\R\}$ a one-parametric symmetry group 
of the Lagrangian system. Then 
$I(t):=I(x(t),\dot x(t),t)=$\co , $t\in\R$. 
\et 
\Pr Differentiation gives 
\be\la{NId} 
\dot I(t)=\fr{d}{dt}L_v(x(t),\dot x(t),t)\,\,\, 
\fr{d}{ds}\Bigg|_{s=0}g_s(x(t)) 
+L_v(x(t),\dot x(t),t)\,\,\,\fr{d}{dt}\fr{d}{ds} 
\Bigg|_{s=0}g_s(x(t)). 
\ee 
For the first summand on the RHS, Eqns. (\re{ELN}) give 
\be\la{NId1} 
\fr{d}{dt}L_v(x(t),\dot x(t),t)\,\,\,\fr{d}{ds}\Bigg|_{s=0}g_s(x(t)) 
=L_x(x(t),\dot x(t),t)\,\,\,\fr{d}{ds}\Bigg|_{s=0}g_s(x(t)). 
\ee 
For the second summand we have 
\be\la{NId2} 
L_v(x(t),\dot x(t),t)\,\,\,\fr{d}{dt}\fr{d}{ds}\Bigg|_{s=0}g_s(x(t)) 
=L_v(x(t),\dot x(t),t)\,\,\,\fr{d}{ds}\Bigg|_{s=0}\fr{d}{dt}g_s(x(t)) , 
\ee 
since 
$\ds\fr {\pa^2 }{\pa t\pa s}=\fr {\pa^2 }{\pa s\pa t}$. 
At last, 
\be\la{NId2l} 
\fr{d}{dt}g_s(x(t))=dg_s(x(t))\dot x(t) 
\ee 
by Definition 
(\re{dif}) 
of the differential of the map $g_s$. 
Now (\re{NId})-(\re{NId2l}) gives, 
by the chain rule and (\re{s0q}), 
\beqn\la{NIdl} 
\dot I(t)&=&L_x(x(t),\dot x(t),t)\fr{d}{ds}\Bigg|_{s=0}g_s(x(t)) 
+L_v(x(t),\dot x(t),t)\fr{d}{ds}\Bigg|_{s=0}dg_s(x(t))\dot x(t)\nonumber\\ 
&=&\fr{d}{ds}\Bigg|_{s=0}L(g_s(x(t)),dg_s(x(t))\dot x(t),t)=0 
\eeqn 
according to (\re{sym}). 
\bo

\subsection{Application to Many-Particle Systems} 
Let us apply the Noether theorem to a Lagrangian system 
of $n$ particles, i.e., with $N=3n$ and 
$x=(\x_1,...,\x_n)$, where $\x_i\in\R^3$. 
\\ 
\subsubsection{Translation group} 
Let us fix a vector $h\ne 0$ in $\R^3$ and 
consider the transformation $T_s$ 
from (\re{tr}). 
By Definition \re{dsym} 
the Lagrangian system is invariant with respect to 
the translations (\re{tr}) if 
\be\la{trinL} 
L(T_s(x),dT_s(x)v,t)= L(x, v,t), 
\,\,\, (x,v,t)\in\E^+, 
\,\,\,\,\,\forall s\in\R. 
\ee 
\bexe 
Check that for the translations $T_s$, $s\in\R$, the differential 
is given by 
\be\la{dTs} 
dT_s(x)v=v,\,\,\,\,\,x,v\in\R^{3n}. 
\ee 
\eexe 
 
\bexe Check that  the Lagrangian system 
is invariant with respect to 
the translations (\re{tr}) for every $h\in\R^3$, if 
the Lagrangian has the structure (cf. (\re{trine})), 
\be\la{trineL} 
L(\x_1,...,\x_n, v,t)= \Lam (\x_1-\x_n,..., \x_{n-1}-\x_n, v,t),\,\,\,\,\, 
(\x_1,...,\x_n,v,t)\in \E^+. 
\ee 
\eexe 
\bd\la{dmNL} 
i) The momentum $p_i$ of the $i$-th particle of the Lagrangian system 
is the vector function 
on the space $\E^+$ defined by 
\be\la{miNL} 
\p_i(x,v,t):=L_{\bv_i}(x,v,t)\in\R^3,\,\,\,\,\,(x,v,t)\in \E^+. 
\ee 
ii) the momentum $p$ of the Lagrangian system is the vector function 
on the space $\E^+$ defined by 
\be\la{mNL} 
p(x,v,t):=\sum_i \p_i=\sum_i L_{\bv_i}\in\R^3,\,\,\,\,\,(x,v,t)\in \E^+. 
\ee 
\ed 
\bt\la{tmcL} 
Let the Lagrangian system (\re{NN}) be invariant with respect to 
the translations (\re{tr}) along 
 a fixed vector $h\in\R^3$. Then for any trajectory 
$x(t)\in C^2([t_0, t_1],\R^{3n})$, 
the projection of the momentum 
$p(x(t),\dot x(t))$ onto $h$ is conserved, 
\be\la{mNcL} 
p_h(t):=p(x(t),\dot x(t),t) h=\co , 
\,\,\,\,\,t\in [t_0, t_1]. 
\ee 
\et 
{\bf Proof 1} The conservation follows from the Noether theorem 
for the one-parametric symmetry group $g_s=T_s$ since $ph$ 
coincides with the corresponding Noether invariant. Indeed, 
the invariant reads 
\be\la{NIT} 
I:=L_v\fr{d}{ds}\Bigg|_{s=0}T_s(x) 
=\sum_i L_{\bv_i}\fr{d}{ds}\Bigg|_{s=0}(\x_i+hs)= 
\sum_i \p_i h=ph. ~~~~~~~~~\loota 
\ee 
{\bf Proof 2} 
By Definition \re{dmNL}, 
the Euler-Lagrange equations  (\re{NNL}), and the 
chain rule of differentiation, 
\beqn\la{mcNpL} 
\dot p_h(t)=\sum_i \dot \p_i(t)h 
&=&-\sum_i \na_{\x_i} L(x(t), \dot x(t),t)h\nonumber\\ 
&=&-\fr {d}{ds}\Bigg|_{s=0} L(T_s (x(t)), dT_s(x(t),t) \dot x(t))=0, 
\,\,\,\,t\in [t_0, t_1] 
\eeqn 
by (\re{dTs}) and (\re{trinL}).\bo 
\subsubsection{Rotation group} 
Let us fix a unit vector $e\in\R^3$ and 
consider the transformation $R_s$ 
from (\re{rot}). 
By Definition \re{dsym} 
the Lagrangian system is invariant with respect to 
the rotations (\re{rot}) if 
\be\la{rinL} 
L(R_s(x),dR_s(x)v,t)= L(x, v,t), 
\,\,\, (x,v,t)\in\E^+, 
\,\,\,\,\,\forall s\in\R. 
\ee 
\bexe 
Check that for the rotations $R_s$, $s\in\R$,  the differential 
is given by 
\be\la{dRs} 
dR_s(x)v=R_s v,\,\,\,\,\,v\in\R^{3n}. 
\ee 
\eexe

\bexe Check that the Lagrangian system (\re{NN}) 
is invariant with respect to 
the rotations (\re{rot}) with every $r\in\R^3$, if 
the Lagrangian has the structure (cf. (\re{rotine})) 
\beqn\la{rotineL} 
L(\x_1,...,\x_n, \bv_1,...,\bv_n,t)&=& \Lam_1(\{|\x_i-\x_j|:1\le i < j\le n; 
\,\,|\bv_i|:1\le i \le n\},t),\nonumber\\ 
&&(\x_1,...,\x_n,\bv_1,...,\bv_n,t)\in \E^+. 
\eeqn 
\eexe

\bd\la{damNL} 
i) The angular momentum $\bL_i$ of the $i$-th particle 
is the vector function 
on the space $\E^+$ defined by 
\be\la{amiNL} 
\bL_i(x,v,t):=\x_i\times \p_i,\,\,\,\,\,(x,v,t)\in \E^+. 
\ee 
ii) the angular momentum $\bL$ of the Newton system (\re{NN}) 
is the vector function 
on the space $\E^+$ defined by 
\be\la{amNL} 
\bL(x,v,t):=\sum_i \bL_i=\sum_i \x_i\times \p_i,\,\,\,\,\,(x,v,t) 
\in \E^+. 
\ee 
\ed 
\bt\la{tamcL} 
Let the  Lagrangian system be invariant with respect to 
the rotations (\re{rot}) around 
 a fixed vector $\e\in\R^3$. Then, for any trajectory 
$x(t)\in C^2([t_0, t_1],\R^{3n})$, 
the projection of the angular momentum 
$\bL(x(t),\dot x(t))$ onto $\e$ is conserved, 
\be\la{amNcL} 
\bL_\e(t):=\bL(x(t),\dot x(t)) \e=\co , 
\,\,\,\,\,t\in [t_0, t_1]. 
\ee 
\et 
{\bf Proof 1} 
The conservation follows from the Noether theorem 
for the one-parametric symmetry group $g_s=R_s$, since $\bL_e$ 
coincides with the corresponding Noether invariant. Indeed, 
the invariant reads 
\beqn\la{NIL} 
I:&=&L_v\fr{d}{ds}\Bigg|_{s=0}R_s(x) 
=\sum_i L_{v_i}\fr{d}{ds}\Bigg|_{s=0}(R_\e(s) \x_i)\nonumber\\ 
~&&\nonumber\\ 
&=& 
\sum_i \p_i (\e\times \x_i)=\sum_i \e (\x_i\times \p_i)= 
\e\bL 
\eeqn 
according to (\re{rotv}).\bo 
\\ 
{\bf Proof 2} 
By Definition \re{damNL}, 
the Euler-Lagrange equations  (\re{NNL}),  and the 
chain rule of differentiation, 
\beqn\la{amcNpL} 
\dot \bL_\e(t)&=& 
[\sum_i \dot \x_i(t)\times \p_i(t)+\sum_i \x_i(t)\times \dot \p_i(t)] 
 \e\nonumber\\ 
&=& 
[\sum_i \dot \x_i(t)\times  L_{\bv_i}(x(t),\dot x(t))+ 
 \sum_i  \x_i(t)    \times  L_{\x_i}(x(t),\dot x(t))] 
\e\nonumber\\ 
&=& 
\sum_i (\e\times \dot \x_i(t))  L_{\bv_i}(x(t),\dot x(t))+ 
\sum_i (\e\times \x_i(t))       L_{\x_i}(x(t),\dot x(t)). 
\eeqn 
However,  (\re{rotv}) and (\re{dRs}) 
imply that 
\be\la{rotvL} 
\e\times \x_i(t)=\fr{d}{ds}\Bigg|_{s=0} R_\e(s) \x_i(t),~~~~~~~ 
\e\times \dot \x_i(t)=\fr{d}{ds}\Bigg|_{s=0} dR_\e(s) \dot \x_i(t). 
\ee 
Therefore, (\re{amcNpL}) implies by the chain rule and (\re{rinL}), 
\be\la{amcNpLi} 
~~~~~~~~~~~~~~\dot \bL_\e(t)=\fr d{dt}L(R_s(x(t)),dR_s(x(t))\dot x(t))=0. 
~~ 
~~~~~~~~~~~~~~~~~~~~~~\loota 
\ee

%%%%%%%%%%%%%%%%%%%%%%%%%%%%%%%%%%%%%%%%%%%% 
 
%%%%%%%%%%%%%%%%%%%%%%%%%%%%%%%%%%%%%%%%%%%% 
%%%%%%%%%%%%%%%%%%%%%%%%%%%%%%%%%%%%%%%%%%%% 
 
%%%%%%%%%%%%%%%%%%%%%%%%%%%%%%%%%%%%%%%%%%%% 

\newpage 

\setcounter{subsection}{0} 
\setcounter{theorem}{0} 
\setcounter{equation}{0} 
\section{Hamilton Mechanics} 
 
%%\section{Hamilton Mechanics} 
\subsection{Legendre Transform } 
 
We introduce the Legendre transform which translates the 
Euler-Lagrange equations to the Hamiltonian form and study 
the Hamilton-Jacobi equation. 
Let us consider a Lagrangian system with  the extended phase space 
$\E^+:=\R^N\times\R^N\times\R$ 
and the Lagrangian functional 
$L\in C^2(\E^+)$. 
We will identify $\R^N$ with its dual space. 
 
\bd\la{dLt} 
i) The {\it Legendre} transform corresponding 
to the Lagrangian $L$ 
is the map of the extended phase space 
$\E^+$ into 
itself which is defined by 
$\lam:(x,v,t)\mapsto(x,p,t)$ with $p:=L_v(x,v,t)$. 
\\ 
ii) The {\it Legendre} transform of the function $L(x,v,t)$ 
on $\E^+$ is the function 
on $\lam\E^+$ defined by 
$(\Lam L)(x,p,t)\equiv pv-L(x,v,t)$ 
with 
$(x,v,t)=\lam^{-1}(x,p,t)$, 
if the 
Legendre map $\lam$ is a $C^1$-diffeomorphism $\E^+\to \lam\E^+$. 
\ed 
\bexe 
Prove that $\lam: \E^+\to \lam\E^+$ 
 is a $C^1$-diffeomorphism 
iff the following inequality holds: 
$\ds|L_{vv}(x,v,t)|\ne 0$, $(x,v,t)\in \E^+$. 
\eexe 
\bex 
The inequality holds for the Lagrangian (\re{ENL}) because then 
the Jacobian matrix 
$J:=L_{vv}(x,v,t)$ is diagonal and 
$|J|=m_1...m_n\ne 0$ since all $m_i>0$. 
\eex

\bex 
$\Lam v^2=p^2/4$, $\Lam v^4=3v^4=3(p/4)^{4/3}$,... 
\eex 
 
\bexe 
Prove  that $\Lam( \Lam L)=L$ if $\lam: \E^+\to \lam\E^+$ 
 is a $C^1$-diffeomorphism. 
\eexe

\bt\la{tH} 
Let the Legendre transform $\Lam$ 
be a $C^1$-diffeomorphism $\E^+\to \Lam\E^+$. Then 
$\Lam$ transforms 
 the Euler-Lagrange equations (\re{ELN}) into the 
{\it Hamiltonian} form, 
\be\la{HE} 
\dot x(t)=H_p(x(t),p(t),t),\,\,\,\,\,\,\,\,\, 
\dot p(t)=-H_x(x(t),p(t),t), 
\ee 
where $H(x,p,t)$ is the  Legendre transform of the Lagrangian, 
\be\la{HL} 
H(x,p,t)=pv-L(x,v,t),\,\,\,\,\,\,\,\,\,\,p=L_v(x,v,t). 
\ee 
\et 
\Pr The first equation of (\re{HE}) follows by 
differentiation of the identity $H(x,p,t)\equiv pv-L(x,v,t)$: 
\be\la{1H} 
H_p=v+pv_p-L_x x_p-L_v v_p=v=\dot x 
\ee 
since $p=L_v$ by definition, and $x_p=0$. 
The second equation  of (\re{HE}) follows from the Euler-Lagrange 
equation (\re{ELN}): 
\be\la{2H} 
H_x=p_x v+pv_x-L_x-L_v v_x=-L_x=-\dot p. 
\ee 
since $p=L_v$ by definition, and $p_x=0$. 
\bo 
\br 
In (\re{1H}) and (\re{2H}) the derivatives $L_x$ 
mean the derivatives with fixed $v$ and $t$, 
but in all other terms the derivatives in $x$ mean the 
derivatives with fixed $p$ and $t$. 
\er 
\bex 
For the Lagrangian (\re{ENL}) the energy has the form  (\re{En}), 
hence $H(x,p,t)=\sum_i\ds\fr {\p_i^2}{2m_i}+V(x,t)$. 
\eex 
 
\bexe\la{exrp} 
Calculate the momentum, energy and the Hamilton function 
for the Lagrangian of the relativistic particle 
\be\la{relp} 
L(\x,\bv)=-\mu c^2\sqrt{1-\left(\fr\bv c\right)^2},\,\,\,\,(\x,\bv)\in\R^3\times\R^3. 
\ee 
\eexe 
{\bf Solution:} 
\be\la{relps} 
~~~~~\p:=\ds\fr{\mu\bv}{\ds\sqrt{1-\left(\fr\bv c\right)^2}},~~~ 
\bv=\fr\p{\sqrt{\ds\mu^2+\left(\fr\p c\right)^2}},~~~ 
E=\ds\fr{\mu c^2}{\ds\sqrt{1-\left(\fr\bv c\right)^2}},~~~ 
H=c^2\sqrt{\mu^2+\left(\fr\p c\right)^2}. 
\ee

\subsection{Hamilton-Jacobi Equation} 
Let us consider the 
Lagrangian  function $L(x,v,t)\in C^2(\E^+)$ 
with the corresponding 
Hamilton function $H(x,p,t)\in C^2(\E^+)$ 
on the extended phase space $\E^+:=\R^N\times\R^N\times\R$, and the 
Cauchy problem of the type 
\beqn\la{Cp} 
\left. 
\ba{rcl} 
-\dot S(t,x)&=&H(x,\na S(t,x),t),\,\,\,\,(t,x)\in \R^{N+1}\\ 
~&&\\ 
S|_{t=0}   &=&S_0(x),\,\,\,\, x\in\R^N, 
\ea 
\right| 
\eeqn 
where $S_0(x)\in C^1(\R^N)$ is a given function. 
Let us describe the Hamilton-Jacobi method of the construction 
of the solution to the problem (\re{Cp}). 
 
First, consider the corresponding  Cauchy problem 
for the Hamilton system, 
\be\la{CHE} 
\left. 
\ba{rclrcl} 
\dot x(t)&=&H_p(x(t),p(t),t),&\,\,\,\,\,\,\,\,\, 
\dot p(t)&=&-H_x(x(t),p(t),t)\\ 
~&&&&&\\ 
x|_{t=0}\!\!\!&=&x_0,& 
p|_{t=0}\!\!\!&=&\na S_0(x_0) 
\ea\right| 
\ee 
with $x_0\in\R^N$. 
Let us denote the solution 
by $(x(t,x_0),p(t,x_0))$. The solution 
exists and is  $C^1$-smooth  for small $|t|$ depending on $x_0$. 
Let us define the function 
$\cS$ by the action integral 
\be\la{acS} 
\cS(t,x_0)= 
S_0(x_0)+\int_0^t L(x(s,x_0), \dot x(s,x_0),s)ds 
\ee 
for $x_0\in\R^N$ and small $|t|$. At last, let us express 
$x_0$ in $x(t,x_0)$ for small $|t|$: this is possible 
since the Jacobian 
$x_{x_0}(t,x_0)=E$ for $t=0$. 
Thus, $x_0=x_0(t,x)$ where $x_0(t,x)\in C^1(\R\times\R^N)$, 
hence we can define 
\be\la{aS} 
S(t,x)=\cS(t,x_0(t,x)),\,\,\,\,x\in\R^N 
\ee 
for small $|t|$. 
\bt\la{tHJ} 
Let $T>0$ and the map $x_0\mapsto x(t,x_0)$ 
be a $C^1$-diffeomorphism 
of $\R^N$ for $t\in[0,T]$. Then 
the function $S(t,x)$ from  (\re{aS}) 
is the unique solution to the Cauchy problem 
(\re{Cp}) for $t\in[0,T]$. 
\et 
\Pr 
The theorem follows from the 
properties of the 
differential $1$-form 
$\om^1=pdx-Hdt$ in $\R^{N+1}$ 
called the {\it Poincar\'e-Cartan integral invariant} 
\ci{A}. 
 
{\it Step i)} 
Let us consider $x_0,x_0+\De x_0\in\R^N$ and 
$\tau,\tau+\De\tau\in[0,T]$. 
Let 
$\M_\tau$ denote the following two-dimensional submanifold 
in the extended phase space 
$\E^+$, 
\be\la{dM} 
\M_\tau=\{x(t,x_0+s\De x_0),x(t,x_0+s\De x_0),t): 
s\in[0,1], t\in[0,\tau+s\De\tau] 
\}. 
\ee 
The boundary $\pa\M_\tau$ is the union 
$\al\cup\ga_1\cup\beta\cup\ga_0$, where 
\be\la{pM} 
\left.\ba{l} 
\al~:=\{(x_0+s\De x_0,\na S_0(x_0+s\De x_0),0): 
s\in[0,1]\},\\ 
\beta~:=\{(x(t+s\De\tau,(x_0+s\De x_0)), 
p(t+s\De\tau,(x_0+s\De x_0)),t+s\De\tau): 
s\in[0,1]\},\\ 
\ga_0:=\{(x(t,x_0)), 
p(t,x_0),t): 
t\in[0,\tau]\},\\ 
\ga_1:=\{(x(t,x_0+\De x_0)), 
p(t,x_0+\De x_0),t): 
t\in[0,\tau+\De\tau]\} 
\ea\right| 
\ee 
are oriented according to the increment of the parameters 
$s,t$. 
Therefore, by the Stokes theorem, 
\be\la{ST} 
\int\limits_{\M_\tau} d\om^1=\int\limits_{\al}\om^1 
+\int\limits_{\ga_1}\om^1 
-\int\limits_{\beta}\om^1 
-\int\limits_{\ga_0}\om^1. 
\ee 
 
{\it Step ii)} 
The central point of the proof is the observation that 
the restriction of the form $d\om^1$ onto the submanifold 
${\M_\tau}$ vanishes, 
\be\la{omv} 
d\om^1|_{\M_\tau}=0. 
\ee 
This follows from two facts: 
i) 
the Hamilton vector field 
$\cH:=(H_p,-H_x,1)$ 
in the extended phase space $\E^+$ is tangent to 
$\M_\tau$ at every point, and ii) $d\om^1(\cH,\V)\equiv 0$ 
for every vector field $\V$ in $\E^+$. The last identity 
follows from the following calculations: first, 
$d\om^1=dp\we dx-dH\we dt= 
dp\we dx-(H_pdp+H_xdx)\we dt$ is the antisymmetric bilinear form 
with the matrix 
\be\la{ssm} 
A= 
\left(\ba{rrr} 
0&E&H_x\\ 
-E&0&H_p\\ 
-H_x&-H_p&0 
\ea\right), 
\ee 
where $E$ is the $N\times N$ identity matrix. 
Second, $A\cH\equiv 0$, hence 
\be\la{dHV} 
d\om^1(\cH,\V)= 
\langle 
A\cH 
,\V 
\rangle\equiv 0. 
\ee 
 
{\it Step iii)} 
Now  (\re{ST}) reads 
\be\la{STr} 
\int\limits_{\beta}(pdx-Hdt)=\int\limits_{\al}\na S_0(x)dx 
+\int\limits_{\ga_1}Ldt 
-\int\limits_{\ga_0}Ldt , 
\ee 
since $dt|_{\al}=0$, $p|_{\al}=\na S_0(x)$ and 
$\om^1|_{\ga_i}=Ldt$. The first term on the 
RHS of  (\re{STr}) is equal to 
$S_0(x_0+\De x_0)-S_0(x_0)$. Therefore, (\re{STr}) 
becomes by (\re{acS}), (\re{aS}), 
\beqn\la{STre} 
\int\limits_{\beta}(pdx-Hdt)&=& 
S_0(x_0+\De x_0)+\int\limits_{\ga_1}Ldt 
-(S_0(x_0)+\int\limits_{\ga_0}Ldt) 
\nonumber\\ 
~\nonumber\\ 
&=&S(x+\De x,\tau+\De\tau)- 
S(x,\tau) , 
\eeqn 
where $x+\De x=x(\tau+\De\tau, x_0+\De x_0)$ and 
 $x=x(\tau, x_0)$. Finally, (\re{STre}) 
implies $\dot S(t,x)=-H(x,p,t)$ and $\na S(t,x)=p$, hence 
Equation (\re{acS}) follows.\bo

%%%%%%%%%%%%%%%%%%%%%%%%%%%%%%%%%    %%%%%%%%%%%%%%%%%%%%%%%%%%% 
%%%%%%%%%%%%%%%%%%%%%%%%%%%%%%%%%    %%%%%%%%%%%%%%%%%%%%%%%%%%% 
%%%%%%%%%%%%%%%%%%%%%%%%%%%%%%%%%    %%%%%%%%%%%%%%%%%%%%%%%%%%% 
%%%%%%%%%%%%%%%%%%%%%%%%%%%%%%%%%    %%%%%%%%%%%%%%%%%%%%%%%%%%% 
 
\newpage 

\setcounter{subsection}{0} 
\setcounter{theorem}{0} 
\setcounter{equation}{0} 
\section 
{Theory of Noether Currents} 
We consider a generalization of the symmetry theory 
to Lagrangian fields. 
Let the corresponding  Lagrangian density 
be invariant with respect to a symmetry group. 
Then the Noether free divergent currents and the corresponding 
invariants can be constructed. 
We prove the Noether theorem providing the construction. 
 
\subsection{Field Symmetry} 
Consider the one-parametric group of 
transformations $g_s:\R^{4}\times\R^N\to\R^{4}\times\R^N$ 
of the form 
\be\la{gs} 
g_s:\left(\ba{c}x\\ \psi\ea\right)\mapsto 
\left(\ba{c}y\\ \psi_s\ea\right)= 
\left.\left(\ba{c}a_s (x)\\ b_s(\psi)\ea\right)\right|\,\,\,s\in\R 
\ee 
where $a_s$ and $b_s$ are some differentiable 
transformations $a_s:\R^{4}\to\R^{4}$ and 
$b_s:\R^N\to\R^N$, respectively. 
Let us define the 
corresponding transformations of the fields 
\be\la{gsf} 
\psi(x)\mapsto\psi_s(y):=b_s(\psi(x)). 
\ee 
This definition implies a corresponding transformation for the 
derivatives: by the chain rule, 
\be\la{npm} 
\na\psi(x)\mapsto\na_y\psi_s(y):=\na_\psi b_s(\psi(x))\na\psi(x) 
\fr{\pa x(y)}{\pa y}. 
%%%=(\na b_s\psi)(a_s^{-1}y)\fr{\pa x(y)}{\pa y}. 
%%%\Bigg(\fr{\pa y(x)}{\pa x}\Bigg)^{-1}. 
\ee 
\br 
At $s=0$ all transformations are identities since $g_0=Id$ for 
the {\bf group} $g_s$ (cf. (\re{s0q})): 
\be\la{s0} 
(a_0x, b_0\psi)=(x,\psi),\,\,\,\, 
(x,\psi)\in\R^{4}\times\R^N. 
\ee 
\er 
 
\bd 
The transformation  $g_s$, $s\in \R$, is a symmetry of a 
Lagrangian field with the Lagrangian density 
$\cL(x,\psi,\na\psi)$ if the following identity holds, 
\be\la{symf} 
\cL(x,\psi,\na\psi)=\cL(y,\psi_s,\na_y\psi_s) 
\Big|\fr{\pa y(x)}{\pa x}\Big|,\,\,\,\,\,\,\,\,\,\,\,\,\,\,\, 
(x,\psi,\na\psi)\in \R^{d+1}\times \R^N\times \R^{4N}, 
\ee 
where $y:=a_s(x)$,  $\psi_s:=b_s(\psi)$ and 
$\na_y\psi_s:=\na_\psi b_s(\psi)\na\psi 
\ds\fr{\pa x(y)}{\pa y}$. 
\ed

\bex\la{e1} {\bf Time-translations} 
Consider the 
time translation along $e_0=(1,0,0,0)$, 
\be\la{gst0} 
g_s:\left(\ba{c}x\\ \psi\ea\right)\mapsto 
\left(\ba{c}y\\ \psi_s\ea\right)= 
\left.\left(\ba{c}x-se_0\\ \psi\ea\right)\right|\,\,\,s\in\R. 
\ee 
Then 
\be\la{gstp0} 
\psi_s(y)=\psi(y+se_0),\,\,\,\,\na_y\psi_s(y)=(\na\psi)(y+se_0), 
\,\,\,\,\,\,\,\,\Big|\fr{\pa y(x)}{\pa x}\Big|\equiv 1. 
\ee 
Hence, (\re{symf}) for this transformation 
is equivalent to (\re{Lt}). 
\eex 
 
\bex \la{e2}{\bf Space-translations} 
Consider the 
space-translation along $e_1=(0,1,0,0)$, 
\be\la{gst} 
g_s:\left(\ba{c}x\\ \psi\ea\right)\mapsto 
\left(\ba{c}y\\ \psi_s\ea\right)= 
\left.\left(\ba{c}x+se_1\\ \psi\ea\right)\right|\,\,\,s\in\R. 
\ee 
Then 
\be\la{gstp} 
\psi_s(y)=\psi(y-se_1),\,\,\,\,\na_y\psi_s(y)=(\na\psi)(y-se_1), 
\,\,\,\,\,\,\,\,\Big|\fr{\pa y(x)}{\pa x}\Big|\equiv 1. 
\ee 
Hence, (\re{symf}) for this transformation 
means that the Lagrangian density $\cL$ does not depend on $\x_1$. 
\eex 
 
\bex \la{e3}{\bf Space-rotations} 
Consider the 
group 
\be\la{gsr} 
g_s:\left(\ba{c}(x_0,\x )\\ \psi\ea\right)\mapsto 
\left(\ba{c}y\\ \psi_s\ea\right)= 
\left.\left(\ba{c} (x_0,R_n(-s)\x )\\ \psi\ea\right)\right|\,\,\,s\in\R, 
\ee 
where $R_n(s)$ is the rotation of $\R^3$ around $\e_n$ with an angle 
of $s$ {\it radian}. 
Then 
\be\la{gsrp} 
\!\!\!\!\psi_s(y)\!=\!\psi(y_0,R_n(s){\bf y}),\,\,\,\,\,\, 
\na_y\psi_s(y)\!=\!(\na\psi) 
\!\left(\! 
\ba{cc} 
1&0\\0& R_n(s) \ea\!\! 
\right)\!\! 
, 
\,\,\,\,\,\Big|\fr{\pa y(x)}{\pa x}\Big|\equiv 1. 
\ee 
Hence, (\re{symf}) for this transformation is equivalent to (\re{LrN}). 
\eex 
 
\bex \la{e4}{\bf Phase rotations} 
For $\psi\in\C^M$ define 
\be\la{gsir} 
g_s:\left(\ba{c}x\\ \psi\ea\right)\mapsto 
\left(\ba{c}y\\ \psi_s\ea\right)= 
\left.\left(\ba{c}x\\ e^{is}\psi\ea\right)\right|\,\,\,s\in\R. 
\ee 
Then 
\be\la{gstir} 
\psi_s(y)=e^{is}\psi(y),\,\,\,\,\na_y\psi_s(y)=e^{is}\na\psi(y), 
\,\,\,\,\,\,\,\,\Big|\fr{\pa y(x)}{\pa x}\Big|\equiv 1. 
\ee 
Hence, (\re{symf}) for this transformation is equivalent  to (\re{Lri}). 
\eex 
\subsection{Noether Current and Continuity Equation} 
\bd 
For a given one-parametric group 
$g_s$ 
of transformations 
(\re{gs}) and a given trajectory $\psi(x)$, 
let us define the vector fields 
\be 
\la{dvw} 
 v(x)= \fr{\pa a_sx}{\pa s}\Bigg|_{s=0}, 
\,\,\,\,\,\,\,\,\,\,\,\,\,\,\,\, 
 w(x)=\fr{\pa  \psi_s(x)}{\pa s}\Bigg|_{s=0}, 
\,\,\,\, 
\,\,\,\,\,\,\,\,\,\,x\in\R^{d+1}. 
\ee 
\ed 
 
\bd The Noether current corresponding to a given one-parametric group 
$g_s$ 
of transformations 
(\re{gs}) is the following  vector field 
\be\la{Nc} 
S_\al(x)\!=\!\pi_\al(x) 
w(x)\!+\!\cL(x,\psi(x),\na\psi(x))v_\al(x), 
\,\,\,\,\,x\in\R^{d+1},~~~~\al=0,...,3. 
\ee 
\ed 
 
\bt \la{tN2} 
{\rm (E.Noether \ci{EN})} 
Let $g_s$ be a one-parametric symmetry group, 
i.e. 
(\re{symf}) holds for $s\in\R$. 
Let 
 $\psi(x)\in C^2(\R^{4},\R^N)$ be a solution to the 
equations (\re{ELfp}), and $w(x)\in  C^1(\R^{4},\R^N)$, 
$v(x)\in C^1(\R^{4},\R^{4})$ 
are defined by 
(\re{dvw}). Then the continuity equation holds, 
\be\la{ceq} 
\pa_\al S_\al(x)=0,\,\,\,\,x\in\R^{4}. 
\ee 
\et 
\bc\la{ctN2} 
Let i) all conditions of Theorem \re{tN2} hold, 
ii) the bounds (\re{bLd}) hold, 
iii) $\psi(x)\in C^2_\si$, $w(x)\in  C^1_\si$ 
with a $\si>3/2$, 
and $v(x)\in C^1_0$. Then 
the conservation law holds, 
\be\la{clS} 
\cS_0(t):=\int_{\R^3} S_0(t,\x)d\x=\co,\,\,\,\,t:=x_0\in\R. 
\ee 
\ec 
\Pr We have $\cS_0(t)=\lim_{R\to \infty}\cS_0^R(t)$ where 
\be\la{clSR} 
\cS_0^R(t):=\int_{|\x|\le R} S_0(t,\x)d\x,\,\,\,\,t\in\R. 
\ee 
Differentiating, we get by (\re{ceq}) and the Stokes theorem, 
\be\la{clSRd} 
\dot\cS_0^R(t):=-\int_{|\x|\le R} \na_\al S_\al(t,\x)d\x 
=-\int_{|\x|= R} n_\al(\x)S_\al(t,\x)d\5\Si 
,\,\,\,\,t\in\R, 
\ee 
where $n_\al(\x):=x_\al/|\x|$ and $d\5\Si$ is the Lebesgue measure 
on the sphere $|\x|= R$. Therefore, $\dot\cS_0^R(t)\to 0$ 
as $R\to\infty$ since $S_\al(x)\in C^1_\si$ 
with $\si>3$. 
Hence (\re{clS}) follows.\bo 
\br 
The integral identity (\re{clSRd}) means that the vector field 
$S_k(t,\x)$ is the current density of the field $S_0(t,\x)$. 
\er 
{\bf Proof of Theorem \re{tN2}} (cf. \ci{GF,Z}) 
Consider an arbitrary open region  $\Om\subset\R^3$ with a smooth boundary. 
Integrating the symmetry 
condition (\re{symf}) over 
$\Om$, we get 
\be\la{omi} 
\int_\Om \cL(x,\psi(x),\na\psi(x))dx= 
\int_{\Om_s} \cL(y,\psi_s(y),\na_y\psi_s(y))dy, 
~~~~~~~~s\in\R, 
\ee 
where  $\Om_s:=a_s(\Om)$. 
Let us make the change of variables $y=a_s(x)$ on the RHS. 
Then we get the identity 
\be\la{omic} 
\int_{\Om} \cL(a_s(x),b_s(\psi(x)),D_s(x)) 
I_s(x)dx=\co, 
~~~~~~~~s\in\R, 
\ee 
where $\ds I_s(x):=\Big|\fr{\pa a_s(x)}{\pa x}\Big|$ and 
\be\la{defp} 
[D_s(x)]_\al:= 
\fr{\pa \psi_s(y)}{\pa y_\al}\Bigg|_{y=a_s(x)},\,\,\,\,\,\al=0,...,d. 
\ee 
Differentiating (\re{omic}) in $s$, we get by (\re{s0}), 
\beqn\la{omid} 
%%\hspace{-21mm} 
&&\int_{\Om} \Big[\cL_x (x,\psi(x),\na\psi(x))\cdot 
\fr d{ds}\Bigg|_{s=0}\!a_s(x) 
\!+\! 
\cL_\psi (x,\psi(x),\na\psi(x))\cdot 
\fr d{ds}\Bigg|_{s=0}\!b_s(\psi(x)) 
\nonumber\\ 
\hspace{-21mm}&&\!\!+ \cL_{\na_\al\psi}(x,\psi(x),\na\psi(x))\cdot 
\fr d{ds}\Bigg|_{s=0}\! 
[D_s(x)]_\al\!+\! 
\cL (x,\psi(x),\na\psi(x))\fr d{ds}\Bigg|_{s=0}\!I_s(\psi(x))\Big] 
dx\!=\!0. 
\eeqn 
Let us calculate the four  derivatives in $s$.\\ 
i) By Definition (\re{dvw}), 
\be\la{1d} 
\ds\fr d{ds}\Bigg|_{s=0}a_s(x)=v(x). 
\ee 
ii) By Definition (\re{gsf}),  the chain rule and (\re{dvw}), (\re{s0}), 
\be\la{icr} 
~~~~~~~~~\fr d{ds}\Bigg|_{s=0}b_s(\psi(x))= 
\fr d{ds}\Bigg|_{s=0} \psi_s(a_sx)= 
\fr d{ds}\Bigg|_{s=0}[ \psi_s(a_0x)+ \psi_0(a_sx)]= 
w(x)+\na\psi(x)\cdot v(x). 
\ee 
iii) By definition, $[D_s(x)]_\al:=\ds\fr{\pa b_s(\psi(x))}{\pa[a_s(x)]_\al} 
=\fr{\pa \psi_s(a_s(x))}{\pa[a_s(x)]_\al}$. 
Hence the same arguments imply  
\beqn\la{icD} 
\fr d{ds}\Bigg|_{s=0}[D_s(x)]_\al 
&=& 
\fr d{ds}\Bigg|_{s=0} 
\fr{\pa \psi_s(a_s (x))}{\pa [a_s(x)]_\al} 
\nonumber\\ 
&=& 
\fr d{ds}\Bigg|_{s=0} 
\Bigg[ 
\fr{\pa \psi_s(a_0(x))}{\pa[a_0(x)]_\al} 
+ 
\fr{\pa \psi_0(a_s(x))}{\pa[a_0(x)]_\al} 
+ 
\fr{\pa \psi_0(a_0(x))}{\pa[a_s(x)]_\al} 
\Bigg]\nonumber\\ 
&=&\na_\al w(x)+\na_\al(\na\psi(x)v(x))+ 
\fr d{ds}\Bigg|_{s=0} 
\fr{\pa\psi(x)}{\pa[a_s(x)]_\al}. 
\eeqn 
To calculate the last derivative, let us use the matrix identity 
\be\la{ide} 
\fr{\pa\psi(x)}{\pa[a_s(x)]_\beta} 
\fr{\pa[a_s(x)]_\beta}{\pa x_\al}= 
\fr{\pa\psi(x)}{\pa x_\al} . 
\ee 
Differentiating in $s$, we get by 
(\re{s0}), 
\be\la{ided} 
\fr d{ds}\Bigg|_{s=0}\fr{\pa\psi(x)}{\pa[a_s(x)]_\al} 
+ 
\fr{\pa\psi(x)}{\pa x_\beta} 
\fr{\pa v_\beta}{\pa x_\al} 
=0. 
\ee 
Therefore, the last derivative in (\re{icD}) equals $-\na\psi\na_\al v$. 
Hence (\re{icD}) becomes, 
\be\la{icDb} 
\fr d{ds}\Bigg|_{s=0}[D_s(x)]_\al 
=\na_\al w(x)+\na(\na_\al\psi(x))\cdot v(x). 
\ee 
iv) Finally, the derivative of the determinant $I_s$ 
of the Jacobian matrix is the trace 
of the derivative, 
\be\la{dd} 
\fr d{ds}\Bigg|_{s=0}I_s(\psi(x))= 
\fr d{ds}\Bigg|_{s=0}\Big|\fr{\pa a_s(x)}{\pa x}\Big| 
= 
\tr \fr d{ds}\Bigg|_{s=0}\fr{\pa a_s(x)}{\pa x}=\tr\fr{\pa v(x)}{\pa x}= 
\na\cdot v(x) , 
\ee 
since the Jacobian matrix  is diagonal: 
$\ds \fr{\pa a_0(x)}{\pa x}=E$. 
\medskip\\ 
Collecting all calculations i) -- iv) in (\re{omid}), we get 
\beqn\la{omidc} 
&&\int_{\Om} \Big[\cL_x (x,\psi(x),\na\psi(x))\cdot v(x) 
\,+\, 
\cL_\psi (x,\psi(x),\na\psi(x))\cdot (w(x)+\na\psi(x)\cdot v(x)) 
\nonumber\\ 
&&+\pi_\al(x) 
\cdot (\na_\al w(x)+\na(\na_\al\psi(x))\cdot  v(x)) 
\nonumber\\ 
&&+ 
\cL (x,\psi(x),\na\psi(x))\na\cdot v(x) 
\Big] 
dx=0. 
\eeqn 
Since the region $\Om$ is arbitrary, the integrand is zero 
by the main lemma of the calculus of variations. 
We rewrite it as follows, 
\be\la{oint} 
\cL_\psi \cdot w(x) 
+\pi_\al(x)\cdot\na_\al w(x)+ 
\na\cdot\Big[\cL (x,\psi(x),\na\psi(x))v(x)\Big] 
=0. 
\ee 
Finally, let us substitute $\cL_\psi=\na_\al\pi_\al(x)$ 
from the Euler-Lagrange equations (\re{ELfp}). Then (\re{oint}) becomes 
\be\la{oel} 
\na_\al\Big[\pi_\al(x)\cdot w(x)\Big]+ 
\na\cdot\Big[\cL (x,\psi(x),\na\psi(x))v(x)\Big] 
=0 , 
\ee 
which coincides with (\re{ceq}) by 
(\re{Nc}).\bo 
\br 
The justification of the formal  proof (\re{icD}) -- (\re{ided}) 
of 
(\re{icDb}) follows from the identity of type (\re{ide}), 
\be\la{idet} 
[D_s(x)]_\beta 
\fr{\pa[a_s(x)]_\beta}{\pa x_\al}= 
\fr{\pa \psi(a_s x)}{\pa x_\al} , 
\ee 
by differentiation similar to (\re{ided}) and (\re{icD}). 
\er 
 
%%%%%%%%%%%%%%%%%%%%%%%%%%%%%%%%%%%%%%%%%%%%%%%%%%%%%%%%%%%% 
%%%%%%%%%%%%%%%%%%%%%%%%%%%%%%%%%%%%%%%%%%%%%%%%%%%%%%%%%%%% 
%%%%%%%%%%%%%%%%%%%%%%%%%%%%%%%%%%%%%%%%%%%%%%%%%%%%%%%%%%%% 
%%%%%%%%%%%%%%%%%%%%%%%%%%%%%%%%%%%%%%%%%%%%%%%%%%%%%%%%%%%% 
 
\newpage 

\setcounter{subsection}{0} 
\setcounter{theorem}{0} 
\setcounter{equation}{0}

\section{Application to Four Conservation Laws} 
We obtain the formulas for Noether currents corresponding to concrete 
symmetry groups: time- and space translations, space- and phase 
 rotations. The formulas imply, by the Noether 
theorem, 
the corresponding conservation laws for 
energy, momentum, angular momentum and charge. 
We consider first general Lagrangian fields and then 
specify 
the formulas for the Klein-Gordon and Schr\"odinger equations.

\subsection{General Lagrangian Fields} 
Let us apply the Noether theorem \re{tN2} to 
 the four groups of the examples \re{e1} -- \re{e4}. 
\medskip\\ 
 {\bf I. Proof of Theorem 
\re{ECLf}} For the group (\re{gst0}), 
Definition (\re{dvw}) implies by 
(\re{gstp0}), 
\be 
\la{dvw1} 
 v(x)= -e_0, 
\,\,\,\,\,\,\,\,\,\,\,\,\,\,\,\, 
 w(x)=\na  \psi(x)e_0=\na_0\psi(x), 
\,\,\,\, 
\,\,\,\,\,\,\,\,\,\,x\in\R^4. 
\ee 
Hence, the Noether current (\re{Nc}) becomes, 
\be\la{Nc1} 
\left\{\ba{rcl} 
S_0(x)\!&=&\!\pi_0(x)\na_0  \psi(x) 
\!-\!\cL(x,\psi(x),\na\psi(x)),\\ 
S_k(x)\!&=&\!\pi_k(x) 
\na_0\psi(x),\,\,\,\,k=1,2,3. 
\ea\right. 
\ee 
The identity (\re{Lt}) implies that the 
Lagrangian density satisfies the invariance condition 
(\re{symf}) with the group  (\re{gst0}). 
Therefore, 
Theorem \re{tN2} implies 
the continuity equation (\re{ceq})  for the current (\re{Nc1}), and 
Corollary \re{ctN2} implies (\re{clS}), 
which means the conservation of energy by 
Definition (\re{eLdf}). 
\medskip\\ 
{\bf II. Proof of Theorem 
\re{mCLf}} 
Let us consider the case $n=1$ for concreteness. 
For the group (\re{gst}), 
Definition (\re{dvw}) implies by 
(\re{gstp}), 
\be 
\la{dvw2} 
 v(x)= e_1, 
\,\,\,\,\,\,\,\,\,\,\,\,\,\,\,\, 
 w(x)=-\na  \psi(x)e_1=-\na_1\psi(x), 
\,\,\,\, 
\,\,\,\,\,\,\,\,\,\,x\in\R^4. 
\ee 
Hence, the Noether current (\re{Nc}) becomes, 
\beqn\la{Nc2} 
\left\{\ba{rcl} 
S_0(x)\!&=&\!-\pi_0(x)\na_1  \psi(x),\\ 
S_1(x)\!&=& 
\!-\pi_1(x)\na_1  \psi(x) 
\!+\!\cL(x,\psi(x),\na\psi(x)),\\ 
S_k(x)\!&=&\!-\pi_k(x) 
\na_1\psi(x),\,\,\,\,k=2,3. 
\ea\right. 
\eeqn 
The 
Lagrangian density satisfies the invariance condition 
(\re{symf}) with the group  (\re{gst}). 
Therefore, 
Theorem \re{tN2} implies 
the continuity equation (\re{ceq})  for the current (\re{Nc2}), and 
Corollary \re{ctN2} implies (\re{clS}), 
which means the conservation of the first component of momentum 
by 
Definition (\re{mLdf}). 
\medskip\\ 
{\bf III. Proof of Theorem 
\re{MCLf}} 
For the group (\re{gsr}), 
Definition (\re{dvw}) implies by 
(\re{gsrp}) and (\re{rotv}), 
\be 
\la{dvw3} 
 v(x)= (0,\e_n\times\x), 
\,\,\, 
 w(x)=\na\psi(x)(0,\e_n\times\x) 
=(\x\times\na_\x)_n\psi(x), 
\,\,\, 
\,\,\,\,x\in\R^4. 
\ee 
Hence, the Noether current (\re{Nc}) becomes, 
\beqn\la{Nc3} 
\left\{\ba{rcl} 
S_0(x)\!&=&\!\pi_0(x)(\x\times\na_\x)_n\psi(x),\\ 
S_k(x)\!&=&\!\pi_k(x) 
(\x\times\na_\x)_n\psi(x)+\cL \e_n\times\x, 
\,\,\,\,k=1,2,3. 
\ea\right. 
\eeqn 
The identity (\re{LrN}) implies that the 
Lagrangian density satisfies the invariance condition 
(\re{symf}) with the group  (\re{gsr}). 
Therefore, 
Theorem \re{tN2} implies 
the continuity equation (\re{ceq})  for the current (\re{Nc3}), and 
Corollary \re{ctN2} implies (\re{clS}), 
which means the conservation of the $n$-th component of angular 
momentum 
by 
Definition (\re{MLdf}). 
\medskip\\ 
{\bf IV. Proof of Theorem 
\re{Qc}} 
 For the group (\re{gsir}), 
Definition (\re{dvw}) implies by 
(\re{gstir}), 
\be 
\la{dvw4} 
 v(x)=0, 
\,\,\, 
 w(x)=i\psi(x), 
\,\,\, 
\,\,\,\,x\in\R^4. 
\ee 
Hence, the components of the Noether current (\re{Nc}) equal 
the charge-current densities, 
\be\la{Nc4} 
S_\al(x)\!=\pi_\al(x)\cdot i\psi(x), 
\,\,\,\,\,\,\,\,\,\,\,\,\al=0,...,3. 
\ee 
The identity (\re{Lri}) implies that the 
Lagrangian density satisfies the invariance condition 
(\re{symf}) with the group  (\re{gsir}). 
Therefore, 
Theorem \re{tN2} implies 
the continuity equation (\re{ceq})  for the current (\re{Nc4}), and 
Corollary \re{ctN2} implies (\re{clS}), 
which means the conservation of charge 
by Definition (\re{LQ}). 
 
%%%%%%%%%%%%%%%%%%%%%%%% 
 
%%\pi_0(x)&\!\!\!\!=-i\h(i\h\na_0\!-\!\ds\fr ec\phi(x))\psi(x), 
%%\,\,\,\, 
%%\pi_k(x)\!=\!-i\h(\!-\!i\h\na_k\!-\!\ds\fr ec A_k(x))\psi(x), 

%%%%%%%%%%%%%%%%%%%%%%%%%% 
\subsection{Klein-Gordon Equation} 
~\medskip\\ 
Let us substitute Expressions (\re{piKG}) and (\re{LdNKG}) 
into (\re{Nc1}), (\re{Nc2}), (\re{Nc3}) and (\re{Nc4}). Then 
Theorem \re{tN2} implies 
for solutions to Equation (\re{NKG}): 
\\ 
 {\bf I. Energy flux} The continuity equation (\re{ceq}) holds 
for  the energy- 
and energy current densities 
\beqn\la{Nc1ec} 
\left\{ 
\ba{rcl} 
S_0(x)\!&=&\! 
-i\h(i\h\na_0\!-\!\ds\fr ec\phi(x))\psi(x)\cdot\na_0\psi(x) 
\!-\!\cL(x,\psi(x),\na\psi(x))\\ 
&&\\ 
&=&\!\ds\fr{(i\h\na_0\!-\!\ds\fr ec\phi(x))\psi(x)\cdot 
(i\h\na_0\!+\!\ds\fr ec\phi(x))\psi(x)}2,\\ 
&&\\ 
&&\!+\!\sum\limits_{k=1}^3 
\ds\fr{|(\!-\!i\h\na_k\!-\!\ds\fr ec \bA_k(x))\psi(x)|^2}2, 
\\ 
&&\\ 
S_k(x)\!&=&\!-i\h(\!-\!i\h\na_k\!-\!\ds\fr ec \bA_k(x))\psi(x)\cdot 
\na_0\psi(x),\,\,\,\,k=1,2,3, 
\ea\right. 
\eeqn 
if the potentials $\phi(x), \bA(x)$ do not depend on time $x_0=ct$. 
\\ 
{\bf For the free equation (\re{NKGf}):} 
\beqn\la{Nc1ecf} 
\left\{ 
\ba{rcl} 
S_0(x)\!&=&\!\ds\fr{|\na_0\psi(x)|^2}2+ 
\ds\sum\limits_{k=1}^3\fr{|\na_k\psi(x)|^2}2+\mu^2\fr{|\psi(x )|^2}2,\\ 
&&\\ 
S_k(x)\!&=&\!-\na_k\psi(x)\cdot 
\na_0\psi(x),\,\,\,\,k=1,2,3. 
\ea\right. 
\eeqn 
 {\bf II. Momentum flux} 
 The continuity equation (\re{ceq}) holds 
for  the first components of the 
momentum- 
and momentum current densities 
\be\la{Nc2ec} 
~~~~~~~\left\{ 
\ba{rcl} 
S_0(x)\!&\!=\!&\!\!\! 
i\h(i\h\na_0\!-\!\ds\fr ec\phi(x))\psi(x)\cdot 
\na_1  \psi(x),\\ 
&&\\ 
S_1(x)\!&\!=\!&\!\!\!\ds 
i\h(-i\h\na_1\!-\!\ds\fr ec \bA_1(x))\psi(x)\cdot 
\na_1\psi(x)+\cL(x,\psi(x),\na\psi(x))\\ 
&&\\ 
&=&\!\!\! 
\ds\fr{|(\!-\!i\h\na_0\!-\!\ds\fr ec \phi(x))\psi(x)|^2}2 
\!+\!\ds\fr{(\!-i\h\na_1\!-\!\ds\fr ec \bA_1(x))\psi(x)\cdot 
(\!-\!i\h\na_1\!+\!\ds\fr ec \bA_1(x))\psi(x)}2\\ 
&&\\ 
&&\!\!\!+\sum\limits_{k=2}^3 
\ds\fr{|(\!-\!i\h\na_k\!-\!\ds\fr ec \bA_k(x))\psi(x)|^2}2, 
\\ 
&&\\ 
S_k(x)\!&=&\!\!\! 
i\h(-i\h\na_k\!-\!\ds\fr ec \bA_k(x))\psi(x) 
\cdot 
\na_1\psi(x),\,\,\,\,k=2,3, 
\ea\right. 
\ee 
if the potentials $\phi(x), \bA(x)$ 
do not depend on $x_1$. 
\\ 
{\bf For the free equation (\re{NKGf}):} 
\beqn\la{Nc2ecf} 
\left\{ 
\ba{rcl} 
S_0(x)\!&\!=\!&\!-\na_0\psi(x)\cdot \na_1 \psi(x),\\ 
&&\\ 
S_1(x)\!&\!=\!&\!\ds\frac{|\na_0\psi(x)|^2}2+\fr{|\na_1\psi(x)|^2}2 
- 
\ds\sum\limits_{k=2,3}\fr{|\na_k\psi(x)|^2}2-\mu^2\fr{|\psi(x )|^2}2,\\ 
&&\\ 
S_k(x)\!&=&\!\na_k\psi(x)\cdot 
\na_1\psi(x),\,\,\,\,k=2,3. 
\ea\right. 
\eeqn 
\\ 
{\bf III. Space rotations} 
 The continuity equation (\re{ceq}) holds 
for  the $n$-th component of the 
angular momentum- 
and angular momentum current densities 
\be\la{Nc3ec} 
~~~\left\{ 
\ba{rcl} 
S_0(x)\!&=&\!-i\h(i\h\na_0\!-\!\ds\fr ec\phi(x))\psi(x) 
\cdot(\x\times\na_\x)_n\psi(x),\\ 
&&\\ 
S_k(x)\!&=&\! 
-i\h(\!-\!i\h\na_k\!-\!\ds\fr ec \bA_k(x))\psi(x) 
\cdot 
(\x\times\na_\x)_n\psi(x)+\cL (\e_n\times\x)_k, 
\,\,\,\,k=1,2,3, 
\ea\right. 
\ee 
if Eq. (\re{Lr}) holds for the density (\re{LdNKG}). 
\\ 
{\bf For the free equation (\re{NKGf}):} 
\beqn\la{Nc3ecf} 
\left\{ 
\ba{rcl} 
S_0(x)\!&=&\!\na_0\psi(x)(\x\times\na_\x)_n\psi(x),\\ 
&&\\ 
S_k(x)\!&=&\!-\na_k\psi(x)\cdot 
(\x\times\na_\x)_n\psi(x)+\cL (\e_n\times\x)_k, 
\,\,\,\,k=1,2,3. 
\ea\right. 
\eeqn 
\\ 
{\bf IV. Phase rotations} 
 The continuity equation (\re{ceq}) holds 
for  the charge- 
and charge current densities 
\be\la{Nc4ec} 
\left\{ 
\ba{rcl} 
S_0(x)&=&-i\h(i\h\na_0-\ds\fr ec\phi(x))\psi(x)\cdot i\psi(x), 
\\ 
&&\\ 
S_k(x)&=&-i\h(-i\h\na_k-\ds\fr ec \bA_k(x))\psi(x)\cdot i\psi(x), 
\,\,\,\,\,\,\,k=1,2,3. 
\ea\right. 
\ee 
{\bf For the free equation (\re{NKGf}):} 
\be\la{Nc4ecf} 
S_0(x)\!=\ds\na_0\psi(x)\cdot i\psi(x), 
\,\,\,\,\, 
S_k(x)\!=\ds-\na_k\psi(x)\cdot i\psi(x), 
\,\,\,\,\,\,\,k=1,2,3. 
\ee

%%%\pi_0(x)= 
%%%\ds\fr{i\h\psi(x)}2\cdot 
%%%\pi_k(x)= 
%%%-\fr 1{2\mu}i\h(-i\h\na_k-\ds\fr ec \bA_k(x))\psi(x)\cdot 

\subsection{Schr\"odinger Equation} 
~\medskip\\ 
Let us substitute Expressions (\re{piS}) and (\re{LdNS}) 
into (\re{Nc1}), (\re{Nc2}), (\re{Nc3}) and (\re{Nc4}). Then 
Theorem \re{tN2} implies 
for solutions to Equation (\re{NS}):\\ 
 {\bf I. Energy flux} The continuity equation (\re{ceq}) holds 
for  the energy- 
and energy current densities 
\beqn\la{Nc1ecS} 
\left\{ 
\ba{rcl} 
S_0(x)\!&=&\!-i\h\psi(x)\cdot \na_0\psi(x) 
\!-\!\cL(x,\psi(x),\na\psi(x))\\ 
&&\\ 
&=&\!e\phi(x)\psi(x)\cdot\psi(x) 
+\ds\fr 1{2\mu}\sum\limits_{k=1}^3 
|(-i\h\na_k-\ds\fr ec \bA_k(x))\psi(x)|^2 
\\ 
&&\\ 
S_k(x)\!&=&\ds\!-\fr 1{m} i\h(-i\h\na_k-\ds\fr ec \bA_k(x))\psi(x)\cdot 
\na_0\psi(x),\,\,\,\,k=1,2,3, 
\ea\right. 
\eeqn 
if the potentials 
 $\phi(x), \bA(x)$ 
do not depend on time $x_0=t$. 
\\ 
{\bf For the free equation (\re{NSf}):} 
\beqn\la{Nc1ecSf} 
\left\{ 
\ba{rcl} 
S_0(x)\!&=&\! 
\ds\fr 1{2\mu}\,\ds\sum\limits_{k=1}^3|\na_k\psi(x)|^2,\\ 
&&\\ 
S_k(x)\!&=&\ds\!-\fr 1{m}\,\na_k\psi(x)\cdot 
\na_0\psi(x),\,\,\,\,k=1,2,3. 
\ea\right. 
\eeqn 
\\ 
 {\bf II. Momentum flux} 
 The continuity equation (\re{ceq}) holds 
for  the first component of the 
momentum- 
and momentum current densities 
\be\la{Nc2ecS} 
~~~~~~\left\{ 
\ba{rcl} 
\!\!\!S_0(x)\!\!\!&\!=\!&\!\!\! 
i\h\psi(x)\cdot \na_1\psi(x),\\ 
&&\\ 
\!\!\!S_1(x)\!\!\!&\!=\!&\!\!\! 
\ds\fr 1{m} i\h(-i\h\na_1-\ds\fr ec \bA_1(x))\psi(x) 
\cdot\na_1\psi(x)+ 
\cL(x,\psi(x),\na\psi(x)) 
\\ 
\!\!\!&&\\ 
\!\!\!\!&\!=\!&\!\!\!e\phi(x)\psi(x)\cdot\psi(x) 
\!+\!\ds\fr 1{2\mu} 
(\!-\!i\h\na_1\!-\!\ds\fr ec \bA_1(x))\psi(x)\cdot 
(\!-\!i\h\na_1\!+\!\ds\fr ec \bA_1(x))\psi(x) 
\\ 
&&\\ 
\!\!\!&&\!\!\!-\ds\fr 1{2\mu}\sum\limits_{k=2}^3 
|(-i\h\na_k-\ds\fr ec \bA_k(x))\psi(x)|^2, 
\\ 
&&\\ 
\!\!\!S_k(x)\!\!\!&=&\!\!\! 
\ds\fr 1{m} i\h(-i\h\na_k-\ds\fr ec \bA_k(x))\psi(x) 
\cdot 
\na_1\psi(x),\,\,\,\,k=2,3, 
\ea\right. 
\ee 
if the potentials $\phi(x), \bA(x)$  do not depend on $x_1$. 
\\ 
{\bf For the free equation (\re{NSf}):} 
\beqn\la{Nc2ecSf} 
\left\{ 
\ba{rcl} 
S_0(x)\!&\!=\!&\! 
i\psi(x)\cdot \na_1(x),\\ 
&&\\ 
S_1(x)\!&\!=\!&-i \na_0(x)\cdot \psi(x)+\ds\fr 1{2\mu}\, 
|\na_1\psi(x)|^2- 
\ds\fr 1{2\mu}\,\ds\sum\limits_{k=2,3}|\na_k\psi(x)|^2,\\ 
&&\\ 
S_k(x)\!&=&\!\ds\fr 1{m}\,\na_k\psi(x)\cdot 
\na_1\psi(x),\,\,\,\,k=2,3. 
\ea\right. 
\eeqn 
\\ 
{\bf III. Space rotations} 
 The continuity equation (\re{ceq}) holds 
for  the $n$-th component of the 
angular momentum- 
and angular momentum current densities 
\be\la{Nc3ecS} 
~~~~~~\left\{ 
\ba{rcl} 
S_0(x)\!\!&\!=\!&\!\! 
-i\h\psi(x) 
\cdot 
(\x\times\na_\x)_n\psi(x),\\ 
&&\\ 
S_k(x)\!\!&\!=\!&\!\! 
-\ds\fr 1{m} i\h(-i\h\na_k\!-\!\ds\fr ec \bA_k(x))\psi(x) 
\cdot 
(\x\times\na_\x)_n\psi(x)\!+\!\cL (\e_n\times\x)_k, 
\,\,\,\,k=1,2,3 
\ea\right. 
\ee 
if Eq. (\re{Lr}) holds for the density (\re{LdNS}). 
\\ 
{\bf For the free equation (\re{NSf}):} 
\beqn\la{Nc3ecSf} 
\left\{ 
\ba{rcl} 
S_0(x)\!&=&\! 
-i\psi(x)\cdot 
(\x\times\na_\x)_n\psi(x),\\ 
&&\\ 
S_k(x)\!&=&\!-\ds\fr 1{m}\,\na_k\psi(x)\cdot 
(\x\times\na_\x)_n\psi(x)+\cL (\e_n\times\x)_k, 
\,\,\,\,k=1,2,3. 
\ea\right. 
\eeqn 
\\ 
{\bf IV. Phase rotations} 
 The continuity equation (\re{ceq}) holds 
for  the charge- 
and charge current densities 
\be\la{Nc4ecS} 
S_0(x)\!= 
-\h\psi(x) 
\cdot \psi(x), 
\,\,\,\,\, 
S_k(x)\!= 
\fr 1{m} \h(i\h\na_k+\ds\fr ec \bA_k(x))\psi(x) 
\cdot \psi(x), 
\,\,\,\,\,\,\,k=1,2,3. 
\ee 
{\bf For the free equation (\re{NSf}):} 
\be\la{Nc4ecSf} 
S_0(x)\!=-\psi(x) 
\cdot \psi(x), 
\,\,\,\,\, 
S_k(x)\!=\fr 1{m}\,i\na_k\psi(x)\cdot \psi(x), 
\,\,\,\,\,\,\,k=1,2,3. 
\ee 
 
%%%%%%%%%%%%%%%%%%%%%%%%%%%%%%%%%%%%%%%%%%%%%%%%%%%%%%%%%%% 
%%%%%%%%%%%%%%%%%%%%%%%%%%%%%%%%%%%%%%%%%%%%%%%%%%%%%%%%%%%%% 
 
\newpage 

\setcounter{subsection}{0} 
\setcounter{theorem}{0} 
\setcounter{equation}{0} 
\section 
{Cauchy Problem for  Maxwell Equations} 
%%\setcounter{equation}{0} 
%%%\subsection{Dynamics of  Maxwell Field} 
We prove the existence of dynamics for inhomogeneous Maxwell 
equations and construct an integral representation 
for the solutions.

Let us consider the Cauchy problem for the Maxwell 
equations (\re{meq}) with the 
initial conditions 
\be\la{A1} 
\bE|_{t=0}=\bE_0(\x),   ~~\bB|_{t=0}=\bB_0(\x),~~~~\x\in\R^3. 
\ee 
We assume $(\bE_0(\x),\bB_0(\x)) \in L^2\oplus L^2$, 
where $L^2=L^2(\R^3)\otimes\R^3$, 
$\rho(t,\x)\in C(\R, L^2(\R^3))$, 
 $\bj(t,\x) \in C(\R, L^2)$ and also 
$(\bE(t,\x),\bB(t,\x))\in C(\R, L^2\oplus L^2).$ 
 Then the system (\re{meq}) leads to the identities 
\be 
\dv \bE_0(\x)= 4\pi\rho (0,\x)\,\,~~~~~\,\, 
\dv \bB_0(\x)=0, ~~~\x\in\R^3,\la{A2} 
\ee 
which are necessary constraints for the existence of solutions 
to the overdetermined system (\re{meq}). 
%%%Let $T$ be an arbitrary positive number. 
%%%\bd 
%%%Let $C_T=C(0,T;L^2(\R^3))$, $\ov C_T=C(0,T;L^2)$ 
%%%and 
%%%$D_T=\{(E_0,B_0,4\pi\rho, \fr 1c\bj)$ $\in L^2\oplus L^2\oplus C_T\oplus \ov C_T:$ 
%%% (\re{cce}) and (\re{A2}) hold for $0\leq t\leq T\}$. 
%%%\ed 
 
%%%Note that $D_T$ is a linear Banach space. 
 
\bt\la{LA1} 
 Let $\bE_0(x),\bB_0(x)$ 
and $\rho(t,\x),\bj(t,\x)$ satisfy 
all conditions mentioned above and 
the constraints (\re{cce}) and (\re{A2}). Then\\ 
i) The Cauchy problem (\re{meq}), (\re{A1}) has a unique solution 
$(\bE(t,\x),\bB(t,\x))\in C(\R, L^2\oplus L^2)$.\\ 
%%%ii) For every $T>0$ the map 
%%%$(E_0,B_0,\rho,\fr 1c\bj)\mapsto (E(t,\x),B(t,\x))|_{0\leq t\leq T}$ 
%%%is a linear continuous operator 
%%%$D_T\to \ov C_T\oplus \ov C_T$ 
%%%with  norm 
%%%${\cal O}(T)$.  \\ 
ii) Let $\bj(t,\x)\equiv 0$. Then the 
energy is conserved: 
\be\la{ecM} 
\int_{\R^3}[\bE^2(t,\x)+\bB^2(t,\x)]d\x=\co,~~~~~t\in\R^3. 
\ee 
iii) The convolution representation holds 
\beqn\la{A3} 
\left(\ba {c}\bE(t)\\\bB(t)\ea\right) 
=\bM(t) *\left(\ba{c} \bE_0\\\bB_0\ea\right)+ 4\pi 
\int_0^t \bG(t-s)*\left(\ba{c} \rho(s)\\\ds\fr 1c\, 
\bj(s)\ea\right)ds,\, \,\,\,\,t\in\R, 
\eeqn 
where $\bE(t):=\bE(t,\cdot)$ etc, and 
$\bM(t)$ resp. $\bG(t)$ is 
%%% 
$6\times 6$  resp. $6\times 4$ 
%%%% 
matrix-valued 
distribution concentrated on the sphere $|\x|=|t|$, for every fixed $t\in\R$:
\be\la{A4} 
\bM(t)(\x)=0,\,\,\,\,\,\,\bG(t,\x)=0,\,\,~~~~~~{\rm for}\,\,\,|\x|\not= |t|. 
\ee 
\et 
{\bf Proof} {\it ad i)} 
We introduce the complex field $\bC(t,\x)=\bE(t,\x)+i\bB(t,\x)$ 
and rewrite (\re{meq}) as 
\beqn 
%%% 
&&\fr 1c\,\dot \bC(t,\x)=-i\, 
%%%% 
\rot\, \bC(t,\x)-   \ds\fr {4\pi}c\,\bj(t,\x),\,\,\bC|_{t=0}=\bC_0(\x),\la{A6}\\ 
&&\dv \bC(t,\x)= 4\pi\rho(t,\x),\la{A6'} 
\eeqn 
where $\bC_0(\x)=\bE_0(\x)+i\bB_0(\x)$. Fourier transformation 
$\ds\hat \bC(\bk,t)=\int\exp(i\bk\cdot x) \bC(t,\x)d\x$ leads to 
the equations 
\beqn 
&&\dot{\hat \bC}(t,\bk)=c\hat m(\bk) \hat \bC(t,\bk)- 4\pi\hat \bj(t,\bk), 
\,\,\hat \bC|_{t=0}=\hat \bC_0(\bk),\la{A7}\\ 
&&-i \bk\cdot \hat \bC(\bk,t)=  4\pi\hat \rho(\bk,t),\la{A7'} 
\eeqn 
where $\hat m(\bk)$ denotes the $3\times 3$ skew-adjoint matrix of the 
 operator $-\bk\times $ in $\C^3$. 
The solution ${\hat C}(t,\bk)$ is defined uniquely from the first 
equation (\re{A7}) of the overdetermined system  (\re{A7}), (\re{A7'}), 
\be\la{A8} 
\hat \bC(t,\bk)=\exp(c\hat m(\bk)t)\hat \bC_0(k)-  4\pi 
\int_0^t\exp(c\hat m(\bk)(t-s)) 
\hat\bj(s,\bk)ds, 
\,\,~~~~~~\bk\in\R^3. 
\ee 
We still have to show that (\re{A8}) satisfies the constraint (\re{A7'}). 
Indeed, the Fourier transformed equations (\re{A2}), (\re{cce}) are 
\beqn 
&&-i\bk\cdot\hat \bC_0(\bk)= 4\pi\hat\rho (0,\bk),~~~~~~~\bk\in\R^3,\la{A8'}\\ 
&& 
\dot {\hat\rho}(t,\bk)- 
ik\cdot\hat \bj(t,\bk)=0,~~~~~\bk\in\R^3,\,t\in\R. 
\la{A8''} 
\eeqn 
With $S(t,\bk)=4\pi{\hat\rho}(t,\bk)+i\bk\cdot \hat \bC(t,\bk)$ they 
imply by 
(\re{A7}) 
\be\la{A9} 
~~~~~~~~S(\bk,0)\!=\!4\pi\hat\rho(\bk,0)\!+\!i\bk\cdot \hat C_0(\bk)\!=0, 
~~~~ 
\dot S(t,\bk)\!= \!4\pi\dot{\hat\rho}(t,\bk)\!-\! 
4\pi i \bk\cdot\hat \bj(t,\bk)\!=\!0, ~~~\bk\in\R^3 , 
\ee 
since $k\cdot \hat m(\bk) \hat \bC(t,\bk)=0$. 
Therefore,  $S(t,\bk)=0$ which means  (\re{A7'}). 
Since $\hat m(\bk)$ is a skew-adjoint 
matrix, its exponent 
$\exp(c\hat m(\bk)t)$ is unitary. 
Now  i) follows from (\re{A8}). 
 
{\it ad ii)} The Parseval 
identity implies 
\be\la{ecMC} 
\int_{\R^3}[\bE^2(t,\x)+\bB^2(t,\x)]d\x= 
\int_{\R^3}|\bC(t,\x)|^2d\x=(2\pi)^{-3}\int_{\R^3}|\hat \bC(t,\bk)|^2d\bk. 
\ee 
Therefore, 
(\re{ecM}) follows from (\re{A8}) 
since $\bj(t,\bk)\equiv 0$ and $\exp(c\hat m(\bk)t)$ 
is a unitary matrix. 
 
{\it ad iii)} 
We have to transform (\re{A8}) back to position space in order to 
check (\re{A4}). We have $\hat m=\hat m(\bk)=-\bk\times $, 
$\hat m^2=-\bk^2+|\bk><\bk|$, $\hat m^3=-|\bk|^2\hat m,\dots$. Hence, 
\begin{eqnarray*} 
&&\hat m^{2j+1}=(-1)^j |\bk|^{2j} \hat m=(-1)^j \fr {\hat m} {|\bk|}  |\bk|^{2j+1} 
{\rm \,\,for\,\,}j \geq 0,\\ 
&&\hat m^{2j}=\hat m^{2j-1} \hat m= 
(-1)^{j-1} |\bk|^{2j-2} \hat m^2= 
-(-1)^j  \left( \fr {\hat m} {|\bk|}\right)^2  |\bk|^{2j}, 
~~~~~~~j\geq 1, 
\end{eqnarray*} 
which yields by Euler's trick for the exponential 
\beqn 
&&~~~~~~~~~~\exp(\hat m(\bk) t)=\sum_0^\infty (\hat m  t)^n/n!= 
\sum_0^\infty (\hat m  t)^{2j}/(2j)!+ 
\sum_0^\infty (\hat m  t)^{2j+1}/(2j+1)!\\ 
\!\!\!\!\!\!\!\!\!\!\!\!&& 
=1+\left( \fr {\hat m} {|\bk|}\right)^2(1-\cos |\bk| t) 
+\fr {\hat m} {|\bk|}\sin |\bk| t= 
\cos |\bk| t+\hat m\fr {\sin |\bk| t}{|\bk|}+(1-\cos |\bk| t) 
\fr{|\bk><\bk|}{|\bk|^2}.\nonumber 
\eeqn 
%%% 
%%%% 
Let us denote by 
$\hat K(t,\bk)=\sin |\bk| t /|\bk|$, 
$\hat \m(t,\bk)=\pa_t{\hat K}(t,\bk)+ \hat m\hat K(t,\bk)$,
and $\hat D(t,\bk)=1-\cos |\bk| t$.
 Then  we finally obtain, 
\be\la{A12} 
\exp(c\hat m(\bk) t)=\hat \m(ct,\bk)+|\bk>\fr{\hat D(ct,\bk)} 
{|\bk|^2}<\bk|. 
\ee 
Inserting this into (\re{A8}) and using the constraints 
(\re{A8'}) and (\re{A8''}), we get 
\beqn\la{A13} 
\hat \bC(t,\bk)&=&\hat \m(ct,\bk)\hat \bC_0(\bk)+4\pi i|\bk> 
\fr{\hat D(ct,\bk)}{|\bk|^2}\hat\rho (\bk,0)\\ 
&&- 
  4\pi\int_0^t  [\hat \m(c(t-s),\bk)\hat \bj(s,\bk)- i|\bk> 
\fr{\hat D(c(t-s),\bk)}{|\bk|^2} 
  \dot{\hat\rho}(s,\bk)]ds, 
\eeqn 
which through integration by parts becomes 
\be\la{A14} 
\hat \bC(t,\bk)=\hat \m(ct,\bk)\hat \bC_0(\bk)- 
  4\pi c\int_0^t  [\hat \m(c(t-s),\bk)\hat \bj(s,\bk)- 
i|\bk>\fr{\pa_t\hat D(c(t-s),\bk)}{|\bk|^2} 
\hat\rho(s,\bk)]ds. 
\ee 
Using $\pa_t\hat D(t,\bk)=|\bk|\sin |\bk| t=|\bk|^2 \hat K(t,\bk)$, we get 
the inverse Fourier transforms, 
\beqn\la{A15} 
&& \m(t,\x):= 
F^{-1}_{\bk\mapsto\x}\hat \m(t,\bk)= \pa_tK(t,\x)-i\5\rot\!\!\circ K(t,\x),\\ 
&& 
 \g(t,\x):=  F^{-1}_{\bk\mapsto\x} 
\Big(i|\bk>\hat K(t,\bk),\,\,-\hat \m(t,\bk)\Big) 
= (-\nabla K(t,\x),\,\,-  \m(t,\x)), 
\eeqn 
where $K(t,x)$ denotes the Kirchhoff kernel 
\be\la{A21} 
K(t,\x):=F^{-1} \hat K(t,\bk) = \fr 1 {4\pi  t}\de(|\x|-| t|). 
\ee 
With these notations, 
 (\re{A14}) implies (\re{A3}) in the ``complex'' form 
\beqn\la{A16} 
\bC(t)= \m(ct)*\bC_0+ 
4\pi\int_0^t \g(c(t-s))* 
\left(\ba{c} c\rho(s)\\ \bj(s)\ea\right)ds\,  \,\,\,\,\,t\in\R. 
\eeqn 
Separating into real and imaginary parts, we obtain 
\be\la{A18} 
\,\bE(t,\x)=\bE_{(r)}(t,\x)+\bE_{(0)}(t,\x),\,\,\,\,\,\,B\b(t,\x)=\bB_{(r)}(t,\x)+\bB_{(0)}(t,\x), 
\ee 
where we denote 
\beqn\la{A19} 
\left(\ba {c}\bE_{(0)}(t)\\\bB_{(0)}(t)\ea\right) 
=\left(\ba{cc} \pa_tK(ct)&\rot\!\!\circ K(ct)\\ 
-\rot\!\!\circ K(ct)&\pa_tK(ct)\ea\right) 
*\left(\ba{c} \bE_0\\\bB_0\ea\right)\,, 
\eeqn 
and
\beqn\la{A20} 
\left(\ba {c}\bE_{(r)}(t)\\\bB_{(r)}(t)\ea\right) 
=4\pi\int_0^t 
\left(\ba{cc} -\nabla K(c(t-s))&-\pa_tK(c(t-s))\\0&\rot\!\! 
\circ K(c(t-s))\ea\right) 
*\left(\ba{c} c\rho(s)\\ \bj(s)\ea\right)ds. 
\eeqn 
 Here $K(t,x)$ coincides with the Kirchhoff kernel 
\be\la{A212} 
K(t,\x)=F^{-1}_{\bk\to\x} \hat K(t,\bk) = \fr 1 {4\pi t}\de(|\x|-|t|). 
\ee 
Now (\re{A3}) and (\re{A4}) follow immediately. \hfill$\Box$ 
\medskip\\ 
{\bf Remark} 
The formula (\re{A20}) 
coincides with standard 
Lienard-Wiechert
representation 
of the ``retarded'' fields $\bE_{(r)}(t,\x )$ and $\bB_{(r)}(t,\x )$ 
through the 
Kirchhoff retarded potentials $\phi(t,\x ),\bA(t,\x )$, \ci{Scharf}: 
%%% 
\beqn\la{A222} 
\ba{ll} 
\bE_{(r)}(t,\x )=-\nabla\phi(t,\x )-\dot \bA(t,\x ), 
&\bB_{(r)}(t,\x )=\rot\, \bA(t,\x ),\\ 
\ds\phi(t,\x )=\int d^3y\fr {\Theta(\tau)} {4\pi |\x-\y|}\rho(\y,\tau), 
&\ds \bA(t,\x )=\int d^3y\fr {\Theta(\tau)} {4\pi |\x-\y|} \bj(\y,\tau), 
\ea 
\eeqn 
where $\tau=t-|\x-\y|$ is the {\it retarted time}, and $\Theta(\tau)$
is the Heaviside step function. 
%%%% 
We emphasize, that 
$(\bE_{(r)}(t,\x ),\bB_{(r)}(t,\x ))$ is not a solution to Maxwell equations 
(\re{A1}) with prescribed $\rho(t,\x )$ and $\bj(t,\x )$, 
since $\bE_{(r)}|_{t=0}=0$, 
and hence $\dv \bE(t,\x )=\rho(t,\x )$ is not satisfied  
at $t=0$. For the same reason, 
$(\bE_{(0)}(t,\x ),\bB_{(0)}(t,\x ))$ is not a solution to 
the Maxwell equations 
(\re{A1}) with $\rho = 0, \bj = 0$. 
 \newpage 
%%%%%%%%%%%%%%%%%%%%%%%%%%%%%%%%%%%%%%%%%%%%%%%%%%%%% 
%%%%%%%%%%%%%%%%%%%%%%%%%%%%%%%%%%%%%%%%%%%%%%%%%%%%%% 
%%%%%%%%%%%%%%%%%%%%%%%%%%%%%%%%%%%%%%%%%%%%%%%%%%%% 
 
%%\setcounter{section}{14} 
\setcounter{subsection}{0} 
\setcounter{theorem}{0} 
\setcounter{equation}{0} 
\section{Lorentz Molecular Theory of Polarization and Magnetization} 
We analyze the {\it macroscopic} 
Maxwell field generated by charge and current 
distributions of a molecule at rest.

\subsection{Constitutive 
Equations} 
 
The Maxwell equations (\re{meq}) 
 define the electromagnetic field 
generated by a given charge and current distribution. 
On the other hand, the Lorentz equations (\re{Le}), 
(\re{Ler}) define 
the motion of charged particles in a given 
Maxwell field. 
However, the classical theory cannot explain the structure 
of matter, i.e., the stability of particles, 
the constitution of atoms and molecules, solid states etc. 
This is related to the fact that the coupled system 
(\re{meq}), (\re{Le}) (or  (\re{meq}), (\re{Le})) 
is not well defined for point particles. 
Hence, we miss the correct dynamical equation for matter. 
In particular, we need an additional hypothesis to 
get a satisfactory theory of matter in a Maxwell field. 
 
Such a theory has been constructed by Lorentz \ci{Lor} 
to justify the transition from a {\it microscopic} "electron theory" 
to the {\it macroscopic} Maxwell equations. 
First, the theory postulates 
that matter is a collection of 
{\it identical small cells}  called  {\it molecules}. 
Second,  it is necessary to introduce an additional hypothesis 
concerning the molecular response to an external Maxwell field. 
The state of a neutral molecule is characterized by its 
{\it dipole moment and magnetic moment}. The parameters 
allow to describe the field generated by the molecule, 
at large distances from 
the molecule, with high precision. Hence, the parameters give a 
complete description of the molecular field for 
any macroscopic observation. Indeed, 
 observations can be made only at 
distances which are much larger than the size of a molecule. 
Hence, it is sufficient to specify the influence of the external 
fields onto the parameters by the corresponding {\it constitutive 
equations}. 
 
We start with an analysis of the distant 
Maxwell field generated by the 
charge and current distributions of a molecule at rest.

\subsection{Stationary Molecular Fields in Dipole Approximation} 
Let us denote by $a>0$ the size of the molecule 
and choose the origin 'in its center', i.e. assume 
that 
\be\la{r0} 
\rho(t,\y)=0,\,\,\,\bj(t,\y)=0,\,\,\,\,\,\,|x|>a,\,\,\,t\in\R. 
\ee 
Let us assume that 
\be\la{r1} 
\rho(t,\cdot)\in L^1(\R^3),\,\,\,\bj(t,\cdot)\in L^1(\R^3) 
\otimes\R^3,\,\,\,\,\,\,t\in\R. 
\ee 
{\bf Static fields} 
First, 
let us consider the static case when 
the densities do not depend on time. Then Equations 
(\re{dp}), (\re{dA}) 
become the stationary Poisson equations and 
their solutions are the Coulomb potentials 
\be\la{cpo} 
\phi(\x)=\int \fr{\rho(\y)d\y}{|\x-\y|}, 
~~~~~~\,\,\,\, 
\bA(\x)= \ds\fr 1c\,\int \fr{\bj(\y)d\y}{|\x-\y|}, 
~~~~~~~~~~\,\,\,\,\x\in\R^3. 
\ee 
Let us expand $1/|\x-\y|$ in a Taylor series 
for small $|\y|\le a$: 
\be\la{Ts} 
\fr1{|\x-\y|}=\fr 1{\sqrt{\x^2+\y^2-2\y\x}} 
=\fr 1{|\x|}+\fr{\y\x}{|\x|^3}+{\cal O} 
\left(\fr 1{|\x|^3}\right),~~~~~\,\,\,\,|\x|\to\infty. 
\ee 
Then (\re{cpo}) becomes, 
\be\la{cpob} 
\left\{ 
\ba{l} 
\phi(\x) 
=\ds\fr Q{|\x|}+\fr{\p\x}{|\x|^3}+{\cal O} 
\left(\fr 1{|\x|^3}\right) 
~\\ 
\\ 
\bA(\x) 
=\ds\fr \bJ{c|\x|}+\fr{\M\x}{|\x|^3}+{\cal O} 
\left(\fr 1{|\x|^3}\right) 
\ea 
\right| 
~~~~~\,\,\,\,|\x|\gg a, 
\ee 
where we denote 
\be\la{QpJM} 
\ba{ll} 
Q=\ds\int \rho(\y)d\y,& \p=\ds\int \y\rho(\y)d\y,\\ 
~\\ 
\bJ=\ds\int \bj(\y)d\y,&  \M_{kl}=\ds\fr 1c \ds\int 
\bj_k(\y)\y_l d\y.\\ 
\ea 
\ee 
We will identify the molecular fields with the 
first two terms in the expansions (\re{cpob}), 
since $|\x|/a\gg 1$ in 
all {\it macroscopic observations}. 
 
Let us note that the remainder in (\re{Ts}) 
is 
 ${\cal O}(\y^2)$. Therefore, the 
first two terms in the expansions (\re{cpob}) 
correspond to the following {\it dipole approximations} for 
$\rho(\y)$ and  $\bj(\y)$: 
\be\la{dap} 
\rho_d(\y)=Q\de(\y)-\p\cdot\na_\y\de(\y), 
~~~~~~~~~~~ 
\bj_d(\y)=\bJ\de(\y)-c \M\na_\y \de(\y). 
\ee

\subsection{Non-Stationary Fields in Multipole Approximations} 
It is easy to see that an asymptotic behavior of the type (\re{cpob}) 
holds for the 
{\it retarded} Kirchhoff potentials (\re{A22b}) 
generated by non-stationary 
localized densities satisfying (\re{r0}): 
%%In this case the particular solutions 
%%to the equations (\re{dp}), (\re{dA}) are given by the 
%%{\it retarded} Kirchhoff potentials (\re{A22b})). 
\be\la{rp} 
~~~~~~~~~~~\ds\phi(t,\x)=\int \fr  {\rho(t-|\x-\y|/c,\y)}{ |\x-\y|}d\y,~~~ 
\ds A(t,\x)=\ds\fr 1c\int \fr {\bj(t-|\x-\y|/c,\y)}{ |\x-\y|}d\y, 
\,\,\,~~~~(t,\x)\in\R^4. 
\ee 
For this purpose, 
let us continue the  Taylor expansion 
(\re{Ts}) and obtain a complete expansion of the type 
(\re{cpob}), 
including all negative powers of 
$|\x|$: 
\be\la{Tsn} 
\fr1{|\x-\y|} 
=\fr 1{|\x|}+\fr{\y\x}{|\x|^3}+ 
\sum\limits_{|\al|\ge 2}\fr {\y^\al P_\al(\n)}{|\x|^{|\al|+1}}, 
\ee 
where we denote $\n:=\x/|\x|$ and $P_\al$ is a polynomial. 
The expansion is a  convergent series for $|\x|>a$. 
Hence, the retarded potentials 
may be expressed by 
the converging series 
\be\la{cpor} 
\left\{ 
\ba{l} 
\phi(t,\x) 
=~~\ds\fr {Q(t)}{|\x|}+~\fr{\p(t)\x}{|\x|^3}+ 
\sum_{|\al|\ge 2} \fr{\phi_\al(t)P_\al(\n)}{|\x|^{|\al|+1}}~\\ 
\\ 
\bA(t,\x) 
= \ds\fr {\bJ(t)}{c|\x|}+\fr{\M(t)\x}{|\x|^3}+ 
\sum_{|\al|\ge 2} \fr{A_\al(t)P_\al(\n)}{|\x|^{|\al|+1}}~ 
\ea 
\right| 
~~~~~\,\,\,\,|\x|>a. 
\ee 
These expansions correspond to 
the following 
{\it multipole approximations} for $\rho(t,\y)$ and  $\bj(t,\y)$: 
\beqn\la{mdapn} 
\left\{ 
\ba{l} 
\rho_m(t,\y)= Q(t)\de(\y)-\p(t)\cdot\na_\y\de(\y)+ 
\sum_{|\beta|\ge 2} \rho_\beta(t)\na_\y^\beta\de(\y), 
\\ 
~\\ 
\bj_m(t,\y)= \bJ(t)\de(\y)-c \M(t)\na_\y\de(\y) 
+\sum_{|\beta|\ge 2}\5\bj_\beta(t)\na_\y^\beta\de(\y). 
\ea\right. 
\eeqn 
The coefficients are defined by 
\beqn 
&&~~~~\left\{ 
\ba{ll} 
Q(t)=\ds\int \rho(t-|\x-\y|/c,\y)d\y,& ~~~~ 
\p(t)=\ds\int \y\rho(t-|\x-\y|/c,\y)d\y,\\ 
~\\ 
\bJ(t)=\ds\int \bj(t-|\x-\y|/c,\y)d\y,&  \M_{kl}(t)=\ds\fr 1c \ds\int 
\bj_k(t-|\x-\y|/c,\y)\y_l d\y,\la{mdac}\\ 
\ea\right. 
\\ 
&&~~~~ 
\rho_\beta(t)=\ds\int_{\R^3}\fr{(-\y)^\beta}{\beta!} 
\rho(t-|\x-\y|/c,\y)d\y, 
~~ 
 \bj_\beta(t)=\ds\int_{\R^3}\fr{(-\y)^\beta}{\beta!} 
\bj (t-|\x-\y|/c,\y)d\y.\la{mdaca} 
\eeqn 
Let us justify the convergence of the series (\re{mdapn}) in the sense 
of distributions. 
 
\bd 
The space $\cH_a(\R^3)$ consists of test functions $\psi(\y)$ 
which are real analytic in the ball $B_a=\{\y\in\R^3:|\y|< a\}$, 
and the Taylor series 
$\psi(\y)=\ds{\sum}_\beta \fr{\y^\beta}{\beta!}\psi^{(\beta)}(0)$ 
converges uniformly in $B_a$. 
\ed 
\bexe 
Check that 
the function $\psi_\x(\y):=1/|\x-\y|$ belongs to the space 
 $\cH_a(\R^3)$  for $|\x| >a$. 
\eexe 
 
\bp\la{pc} 
Let (\re{r0}) hold and $\rho(t,\cdot),\bj(t,\cdot)\in L^1(\R^3)$. 
Then the series (\re{mdapn}) converges and coincides with $\rho(t,\y)$ 
or $\bj(t,\y)$ 
in the following sense: 
\beqn\la{mdas} 
~~~~~\left\{ 
\ba{l} 
\langle\rho(t,\y),\psi(\y)\rangle= 
\langle Q(t)\de(\y)-\p(t)\cdot\na_\y\de(\y)+ 
\sum_{|\beta|\ge 2} \rho_\beta(t)\na_\y^\beta\de(\y),\psi(y)\rangle 
\\ 
~\\ 
\langle \bj(t,\y),\Psi(\y)\rangle=\5 
\langle \bJ(t)\de(\y)-c \M(t)\na_\y\de(\y) 
+\sum_{|\beta|\ge 2}\5\bj_\beta(t)\na_\y^\beta\de(\y),\Psi(y)\rangle 
\ea\right. 
\eeqn 
for every test function from the space 
$\psi\in \cH_a(\R^3)$ or $\Psi\in \cH_a(\R^3)\otimes\R^3$, respectively. 
\ep 
\bexe 
Prove Proposition \re{pc}. 
{\bf Hint:} 
Substitute the Taylor expansion for $\psi$  and $\Psi$ into the LHS of 
(\re{mdas}) and use (\re{r0}). 
\eexe

\bc 
The multipole approximations (\re{mdapn}) satisfy the charge 
continuity equation 
\be\la{cema} 
\dot\rho_m(t,\y)+\na_\y 
\cdot\bj_m(t,\y)=0,~~~~\,\,\,\,(t,\y)\in\R^4. 
\ee 
\ec 
\bexe 
Prove the corollary. 
{\bf Hint:} Use the identity 
$\langle\dot\rho_m(t,\y)+\na_\y 
\cdot\bj_m(t,\y),\psi(t,\y)\rangle=\langle\dot\rho(t,\y)+\na_\y 
\cdot\bj(t,\y),\psi(t,\y)\rangle$ for 
$\psi\in C_0^\infty(\R^4)$ such that 
$\psi(t,\cdot),\dot\psi(t,\cdot) \in\cH_a(\R^3)$, $t\in\R$. 
\eexe 
Substituting the series (\re{mdapn}) into (\re{cema}), we get 
\be\la{ces} 
\dot Q(t)\de(\y)-\dot \p(t)\cdot\na_\y\de(\y) 
+\bJ(t)\cdot\na_\y\de(\y)+ 
\sum_{|\beta|\ge 2} C_\beta(t)\na_\y^\beta\de(\y) 
=0. 
\ee 
Therefore, we have 
\be\la{QJ} 
\dot Q(t)\equiv 0,~~~~~~~~~~~\bJ(t)\equiv \dot \p(t). 
\ee

\subsection{Magnetic Moment of a Molecule} 
Let us consider a molecule in a stationary state, i.e. 
$\rho(t,\y)\equiv\rho(\y)$ and  $\bj(t,\y)\equiv\bj(\y)$. 
Then the multipole expansions 
(\re{mdapn}) do not depend on time, i.e., 
\beqn\la{mdapns} 
\left\{ 
\ba{l} 
\rho_m(t,\y)\equiv\rho_m(\y)= Q\de(\y)-\p\cdot\na_\y\de(\y)+ 
\sum_{|\beta|\ge 2} \rho_\beta\na_\y^\beta\de(\y) 
\\ 
~\\ 
\bj_m(t,\y)\equiv\bj_m(\y)= \bJ\de(\y)-c \M\na_\y\de(\y) 
+\sum_{|\beta|\ge 2}\5\bj_\beta\na_\y^\beta\de(\y). 
\ea\right. 
\eeqn 
 
\bp \la{pmagmom} 
Let the multipole charge-current densities (\re{mdapns}) 
correspond to a stationary state 
of the molecule. Then the matrix $\M$ is skewsymmetric, and 
\be\la{mag} 
 \M\na_\y =\m\times\na_\y, 
\ee 
 where $\m\in\R^3$ is the 
{\bf magnetic moment} of the molecule, i.e., of the current density 
$\bj(\y)$ in the stationary state. 
\ep 
\Pr Substituting (\re{mdapns}) into 
(\re{cema}), we get 
\be\la{cemas} 
\na_\y\cdot\bj_m(\y)=0,~~~~\,\,\,\,\y\in\R^3. 
\ee 
Therefore, 
in particular  $\bJ=0$ and 
$\na_\y\cdot[\M\na_\y\de(\y)]=0$. Then 
$\M_{kl}+\M_{lk}=0$ and the vector $\m\in\R^3$ 
is defined by the following matrix identity: 
\be\la{mom} 
~~~~~~~~~~~~~~~~~~~~~~~~~\M=\left( 
\ba{rrr} 
0~~~~& -\m_3& \m_2\\~\\ 
\m_3& 0~~~~& -\m_1\\~\\ 
-\m_2& \m_1& 0~~~~ 
\ea\right) . ~~~~~~~~~~~~~~~ 
~~~~~~~~~~~~~~~~~~~\loota 
\ee 
Formulas  (\re{mom}) and (\re{QpJM}) imply that 
\be\la{mome} 
\m=\fr 1{2c} \int \y\times \bj(\y)d\y. 
\ee 
\br 
The integral does not depend on the choice of the origin 
since $\bJ\!\!:=\!\!\ds\int \bj(\y)d\y\!=\!0$ by (\re{cemas}). 
\er 
~\smallskip\\ 
{\bf Adiabatic Condition}\,\, 
Let us assume that the molecular 
dynamics can be described as 
an {\it adiabatic evolution} of stationary states with 
corresponding dipole electric moment 
$\p(t)$ and 
magnetic moment $\m(t)$. 
~\bigskip\\ 
Finally, let us assume that the molecule is neutral, i.e., $Q=0$. 
We will identify $\rho(t,\x)$ and $\bj(t,\x)$ with the first two terms 
on the  RHS of (\re{mdapns}). 
Then, by (\re{QJ}) and (\re{mag}), we get 
the dipole approximation 
\be\la{dapnb} 
~~~~~~~~~~\rho(t,\x)\!\approx\!\rho_d(t,\x)\!:=\!\!-\p(t)\cdot\na_\x\de(\x), 
~~~~ 
\bj(t,\x)\!\approx\!\bj_d(t,\x)\!:=\! 
\dot\p(t)\5\de(\x)+ c\na_\x\times \m(t)\de(\x). 
\ee 
Let us stress that the approximations are sufficient for any macroscopic 
observation of the molecular fields at distances much larger than the size 
of the molecules. 
 
%%%%%%%%%%%%%%%%%%%%%%%%%%%%%%%%%%%%%%%%%%%%%%%%%%%%%%%%%%%%%%%%5 
\subsection{Macroscopic Limit: Maxwell Equations in Matter} 
 
{\bf Macroscopic Limit} 
The total 
charge-current densities 
$\rho(t,\x), \rho(t,\x)$ 
in matter are sums of contributions of 
 all molecules concentrated 
at the points $\x_n\in\R^3$. In the dipole approximation, 
the densities are 
\be\la{tom} 
\left\{ 
\ba{ll} 
\rho_{\rm mol}(t,\x)&=-\sum_n 
\p_n(t)\cdot\na_\x\de(\x-\x_n) 
\\~\\ 
\bj_{\rm mol}(t,\x)&=\sum_n\Big[ 
\dot \p_n(t)\5\de(\x-\x_n)+c\na_\x\times 
\m_n(t)\de(\x-\x_n) 
\Big] . 
\ea\right. 
\ee 
Now let us consider the {\it macroscopic limit}, when the diameter $a$ 
of a molecule converges to zero, and each singular density 
converges to a limit distribution. 
More precisely, 
let us  assume that for every fixed $t$ 
we have the following asymptotics 
in the sense of distributions of $\x$: 
\be\la{ass} 
\left\{ 
\ba{l} 
\sum_n\p_n(t)\de(\x-\x_n)\,\to \,\,\bP(t,\x)\\ 
~\\ 
\sum_n\dot\p_n(t)\de(\x-\x_n)\,\to \,\,\dot \bP(t,\x),~~~~~~ 
\sum_n\m_n(t)\de(\x-\x_n)\to \bM(t,\x) 
\ea\right|\,\,\,\,a\to 0. 
\ee 
Then by the continuity of the differentiation of distributions, 
we have in the limit $a\to 0$, 
\be\la{toml} 
\left\{ 
\ba{lll} 
\rho_{\rm mol}(t,\x)&\approx 
-\na_\x\cdot 
~\bP(t,\x) 
\\~\\ 
\bj_{\rm mol}(t,\x)&\approx 
\dot \bP(t,\x)+c\na_\x\times 
\bM(t,\x) 
\ea\right. . 
\ee 
\bd\la{dDH} 
i) The vector functions $\bP(t,\x)$ and $\bM(t,\x)$ are called the 
{\bf electric polarization} and 
 {\bf magnetization} of the molecules at point $\x$ and time $t$, respectively. 
\medskip\\ 
ii) 
$\bD(t,\x)=\bE(t,\x)+4\pi \bP(t,\x)$ is called the   {\bf dielectric 
displacement} 
and $\bH(t,\x)=\bB(t,\x)-4\pi \bM(t,\x)$ is called the  {\bf magnetic 
field intensity}. 
\ed 
Let us separate the macroscopic and molecular charge and current densities: 
\be\la{mami} 
\rho(t,\x)=\rho_{\rm mac}(t,\x)+\rho_{\rm mol}(t,\x), 
\,\,\,\,~~~~~~~~~~~ 
\bj(t,\x)=\bj_{\rm mac}(t,\x)+\bj_{\rm mol}(t,\x) , 
\ee 
where the molecular densities are identified with the 
macroscopic 
limits (\re{toml}). 
Let us  substitute the expressions (\re{mami}), (\re{toml}) 
into the Maxwell equations 
(\re{meq}) and express the fields $E,B$ in terms of 
$D,P,H,M$. Then we obtain the 
{\it Maxwell equations in matter}: 
\be\la{meqm} 
~~~~~~~~~~~~\left\{ 
\ba{ll} 
\dv \bD(t,\x)\!=\! 4\pi\rho_{\rm mac} (t,\x),&~~ 
\rot \bE(t,\x)\!=\! - \ds\fr 1c\dot  \bB(t,\x),\\ 
~&\\ 
\dv \bB(t,\x)\!=\! 0,&~~\rot \bH(t,\x)\!= \!\ds\fr 1c \, 
\dot \bD(t ,\x)\!+\!\ds\fr{4\pi}c 
\,\bj_{\rm mac}(t,\x) 
\ea 
\right|(t,\x)\in\R^4. 
\ee 
{\bf Constitutive Equations} 
Equations (\re{meqm}) contain two additional unknown vector fields 
$\bD,\bH$. 
Therefore, we need two  additional vector equations. 
For {\bf isotropic} materials they are the {\bf constitutive equations} 
\be\la{coe} 
\bD(t,\x)=\ve \bE(t,\x),~~~~~~~~~~~~\bB(t,\x)=\mu \bH(t,\x), 
\ee 
where $\ve$ is called the {\bf permittivity} and 
$\mu$ is called the {\bf permeability} of matter. 
The {\bf constitutive equations} are equivalent to 
\be\la{coee} 
\bP(t,\x)=\chi_e\bE(t,\x),~~~~~~~~~~~~\bM(t,\x)=\chi_m \bH(t,\x), 
\ee 
where $\chi_e$ is called the {\bf electric susceptibility} and 
$\chi_m$ is called the {\bf magnetic susceptibility} of matter. 
The following relations follow from Definition \re{dDH} ii): 
\be\la{coeeo} 
\ve=1+4\pi\chi_e ,~~~~~~~~~~~~\mu=1+4\chi_m. 
\ee

%%%%%%%%%%%%%%%%%%%%%%%%%%%%%%%%%%%%%%%%%%%%%%%%%%%%%%%%%%%%%%%%%%%%%%%% 
 
\newpage 
 
%%%%%%%%%%%%%%%%%%%%%%%%%%%%%%%%%%%%%%%%%%%%%%%%%%%% 
%%%%%%%%%%%%%%%%%%%%%%%%%%%%%%%%%%%%%%%%%%%%%%%%%%%%% 
%%%%%%%%%%%%%%%%%%%%%%%%%%%%%%%%%%%%%%%%%%%%%%%%%%%%% 

%%\setcounter{section}{12} 
\setcounter{subsection}{0} 
\setcounter{theorem}{0} 
\setcounter{equation}{0} 
\section 
{Long-Time Asymptotics and Scattering} 
%%\setcounter{equation}{0} 
%%\subsection{Retarded Potentials and Long-Time Asymptotics} 
The {\em retarded potentials} are 
particular  solutions of the wave and Maxwell equations. 
Here we want to explain the outstanding role of these potentials. 
 
\subsection{Retarded Potentials } 
Formula (\re{A20}) may be rewritten as 
%%% 
\be\la{A22EB} 
 \bE_{(r)}(t,\x)=-\nabla\phi_{(r)}(t,\x)-\fr 1c\,\dot \bA_{(r)}(t,\x), 
~~~~~~~~~~~~~~\bB_{(r)}(t,\x)=\rot\, \bA_{(r)}(t,\x), 
\ee 
where the potentials are given by 
\be\la{A22} 
\ds\phi_{(r)}(t,\x)=\int \fr {\Theta(t_{ret})} { |\x-\y|}\rho(t_{ret},\y)d\y, 
~~~~~~~~~~~~~~ 
\ds \bA_{(r)}(t,\x)=\ds\fr 1c\,\int \fr {\Theta(t_{ret})} { |\x-\y|} 
\bj(t_{ret},\y)d\y, 
\ee 
and $t_{ret}=t-|\x-\y|/c$. 
%%%% 
Let us assume that the charge and current densities are continuous and 
localized in space, 
\be\la{R0} 
\rho(t,\x)=0,\,\,\,~~~~~\bj(t,\x)=0,\,\,\,\,\,\, 
~~~~~~~~|\x|>R,\,\,\,t\in\R. 
\ee 
%%%%%%%%%%%%%%%%%%%%% 
Then (\re{A22}) for large $t>0$ 
become the standard {\it retarded potentials} 
\ci{Jack}, 
\be\la{A22b} 
\left\{\ba{lll} 
\ds\phi_{(r)}(t,\x)&=\ds\phi_{ret}(t,\x)&:= 
~~\ds\int \fr{\rho(t-|\x-\y|/c,\y)} { |\x-\y|}d\y\\~ \\ 
\ds \bA_{(r)}(t,\x)&=\ds \bA_{ret}(t,\x)&:=\ds\fr 1c\,\int 
\fr {\bj(t-|\x-\y|/c,\y)} { |\x-\y|} 
d\y 
\ea\right|~~~~~~~~~~t>R+|\x|. 
\ee 
Further, the fields  (\re{A22EB}) 
become the {\it retarded fields} 
\be\la{A22EBr} 
 \left\{\ba{ll} 
\bE_{(r)}(t,\x)&=-\nabla\phi_{ret}(t,\x)-\dot A_{ret}(t,\x), \\ 
\bB_{(r)}(t,\x)&=\rot\, \bA_{ret}(t,\x), 
\ea\right|~~~~~~~~~~t>R+|\x|. 
\ee 
The distinguished  role of the particular retarded solutions (\re{A22b}) 
to the wave equations (\re{dp}), (\re{dA}) 
is justified in {\it scattering theory}. 
In general, the solutions to the wave equations  (\re{dp}), (\re{dA}) 
are defined uniquely by the initial conditions at time zero: 
\beqn 
~\phi|_{t=0}=\phi_0(\x),&&~~~~~\dot\phi|_{t=0}=\pi_0(\x), 
~~~~~~~~~~~~\x\in\R^3.~~~~~~~~\la{incp}\\ 
~\nonumber\\ 
\bA|_{t=0}=\bA_0(\x),&&~~~~~\dot\bA|_{t=0}=\Pi_0(\x), 
~~~~~~~~~~~~\x\in\R^3.~~~~~~~~\la{incA} 
\eeqn 
However, the {\bf asymptotic behavior} of the solutions for $t\to+\infty$ 
and any {\bf fixed} point $\x$ 
does not 
depend on the initial data $\phi_0,\pi_0,\A_0,\Pi_0$ and coincides 
with the retarded 
potentials (\re{A22b}). 
\medskip\\ 
{\bf Space-localized initial data} 
For example let us consider 
the initial functions with compact supports. 
\bp 
Let (\re{R0}) hold, and let 
the initial functions $\phi_0(x),\pi_0(x),\bA_0(x),\Pi_0(x)$ be 
continuous and localized in space, 
\be\la{R00} 
\phi_0(\x)=\pi_0(\x)=0,\,\,\,~~~\bA_0(\x)=\Pi_0(\x)=0, 
~~~~\,\,\,\,\,\,|\x|>R. 
\ee 
Then for large time the solutions to the Cauchy problems 
(\re{dp}), (\re{incp}) and (\re{dA}), (\re{incA}) 
coincide with the retarded potentials 
(\re{A22b}): 
\be\la{A22c} 
\phi(t,\x)=\phi_{ret}(t,\x),~~~~~~ 
 \bA(t,\x)= \bA_{ret}(t,\x), 
~~~~~~~~~~t>R+|\x|. 
\ee 
\ep 
\Pr 
Let us prove the proposition for the scalar potential $\phi(t,\x)$. 
The Kirchhoff formula for the solution reads 
\be\la{A22d} 
~~~~~~~\phi(t,\x)=\ds\fr 1{4\pi t} 
\int_{S_t(\x)}\pi_0(\y)dS(\y)+\pa_t\left(\ds\fr 1{4\pi t} 
\int_{S_t(\x)}\phi_0(\y)dS(\y)\right)+ 
\phi_{ret}(t,\x),~~~~~~ 
t>0,~~\x\in\R^3, 
\ee 
where $S_t(\x)$ denotes the sphere $\{\y\in\R^3:|\x-\y|=t\}$ and 
$dS(\y)$ is the Lebesgue measure on the sphere. Now 
(\re{A22c}) follows from (\re{R00}).\bo 
\medskip\\ 
A similar theorem 
holds for the Maxwell equations. 
\bt 
Let (\re{R0}) and 
the conditions of Theorem \re{LA1} hold, and 
\be\la{R000} 
\bE_0(\x)=\bB_0(\x)=0,\,\,\, 
~~~~\,\,\,\,\,\,|\x|>R. 
\ee 
Then 
\be\la{A22e} 
\bE(t,\x)=\bE_{ret}(t,\x),~~~~~~~ 
\bB(t,\x)=\bB_{ret}(t,\x) 
~~~~~~~t>R+|\x|. 
\ee 
\et 
\Pr 
(\re{A22e}) follows from (\re{A18}) and (\re{A22EBr}) 
since $\bE_{(0)},\bB_{(0)}$ vanish 
for $t>R+|\x|$ 
by (\re{R000}), (\re{A19}) and 
(\re{A4}). 
\bo 
\medskip\\ 
{\bf Finite energy initial data} 
For the Maxwell equations with $\bj=0$ the energy is conserved, 
(\re{ecM}). 
For the wave equations (\re{dp}) and (\re{dA}) 
with $\rho=0$ and $\bj=0$ 
the energy conservations 
read (see (\re{Nc1ec})), 
\be\la{ecw} 
~~~~~\int_{\R^3}[\ds\fr 1{c^2}|\dot\phi(t,\x)|^2\!+ \!|\na\phi(t,\x)|^2]d\x\! 
=\!\co, 
~~~ 
\int_{\R^3}[\ds\fr 1{c^2}|\dot \bA(t,\x)|^2\!+\!  |\na \bA(t,\x)|^2]d\x\!=\!\co. 
\ee 
For solutions with finite initial energy, (\re{A22c}) and (\re{A22e}) 
hold 
in the local energy semi-norms 
in the limit $t\to+\infty$. 
\bt\la{trp} 
Let (\re{R0}) hold and, further, 
$\phi(t,\x)$, $\bA(t,\x)$, and $\bE(t,\x),\bB(t,\x)$ are finite energy 
solutions to the wave equations (\re{dp}) and (\re{dA}), 
or to the Maxwell Eqns (\re{meq}), respectively. Then for any $R>0$, 
\be\la{reta} 
\left\{\ba{l} 
\ds\int_{|\x|<R}[|~\dot\phi(t,\x)-\dot\phi_{ret}(t,\x)|^2+ 
|\,\na\phi(t,\x)-\na\phi_{ret}(t,\x)|^2]d\x\to 0\\~\\ 
\ds\int_{|\x|<R}[|\dot \bA(t,\x)-\dot \bA_{ret}(t,\x)|^2+ 
|\na \bA(t,\x)-\na \bA_{ret}(t,\x)|^2]d\x\to 0\\~\\ 
\ds\int_{|\x|<R}[| \bE(t,\x)- \bE_{ret}(t,\x)|^2+ 
| \bB(t,\x)- \bB_{ret}(t,\x)|^2]d\x\to 0 
\ea\right|~~~~t\to +\infty. 
\ee 
\et 
\Pr 
Let us split the initial functions into two components: 
a first, 
space-localized component similar to 
(\re{R00}), (\re{R000}), and the rest. 
For the first component we construct a 
solution to the non-homogeneous equations, and 
for the rest we construct a solution 
to the homogeneous equations. 
Then, for the 
first component 
the convergence (\re{reta}) follow from 
(\re{A22c}) and (\re{A22e}). 
The contribution of 
 the second component is uniformly small in time 
by energy conservation. 
\bo 
%%%%%%%%%%%%%%%%%%%%%%%%%%%%%%%%%%%%%%%%%%%%%%%%%%% 
 
\subsection{Limiting Amplitude Principle in Scattering Problems} 
{\bf Time-periodic source and spectral problem} 
Let us consider a wave problem with an external 
periodic source 
\be\la{wla} 
(\pa_t^2-\De)\phi(t,\x)=b(\x)e^{-i\nu t}, 
\ee 
where $\nu\in\R$ and the amplitude $b(\x)$  decays rapidly as 
$|\x|\to\infty$ 
(see for example (\re{SMeasb})). 
We assume $c=1$ for simplicity. 
Similar problems arise in scattering problems for the Schr\"odinger 
equation 
\be\la{Sla} 
i\h\dot\psi(t,\x)= H\psi(t,\x)+b(\x)e^{-i\nu t} 
\ee 
(see for example (\re{SMes1}),  (\re{SMes2})). 
Here $H$ is the Schr\"odinger operator corresponding to a 
{\it static} Maxwell field with the potentials 
$\phi(t,\x)\equiv\phi(\x)$ and $\bA(t,\x)\equiv \bA(\x)$ 
(see (\re{NS})), 
\be\la{SMsla} 
H:=\fr 1{2 m} (-i\h\na_\x-\ds\fr ec \bA(\x))^2 
+e\phi(\x). 
\ee 
Let us write Equations (\re{wla}), (\re{Sla}) in a unified 
form, 
\be\la{Sl} 
i\dot\Psi(t,\x)= A\Psi(t,\x)+B(\x)e^{-i\nu t}. 
\ee 
For the wave equation (\re{wla}) we set 
$\Psi(t,\x):=(\phi(t,\x),\dot\phi(t,\x))$ and 
\be\la{Am} 
A=i\left(\ba{cc} 
0& 1\\ 
\De& 0  \ea \right),~~~~~~~~~ 
B(\x)=i\left(\ba{c} 
0\\ 
b(\x)  \ea \right). 
\ee 
The corresponding  {\it free equation} reads 
\be\la{Slf} 
i\dot\Psi(t,\x)= A\Psi(t,\x). 
\ee 
Let us denote by $U(t)$ the corresponding dynamical group if it exists: 
\be\la{dg} 
\Psi(t)= U(t)\Psi(0), 
\ee 
where $\Psi(t):=\Psi(t,\cdot)$.

We consider {\it finite energy solutions} to Equations 
(\re{Sl}), (\re{Slf}). This means that 
$\Psi(t,\cdot)\in C(\R, E)$, where $E$ is the corresponding Hilbert 
{\it phase space}. That is, $E:=L^2(\R^3)$ for the Schr\"odinger 
equation (\re{Sla}), and $E:=\{(\psi(\x),\pi(\x)): 
\na\psi(\x),\pi(\x)\in L^2(\R^3)\}$ is the Hilbert space with the 
norm 
\be\la{dgn} 
\Vert (\psi(\x),\pi(\x)) \Vert_E^2:=\Vert \na\psi(\x) \Vert^2 
+\Vert \phi(\x) \Vert^2, 
\ee 
where $\Vert \cdot \Vert$ stands for the norm in $L^2(\R^3)$.

\br\la{Un} 
 $U(t)$, $t\in\R$, is a unitary operator in the corresponding  Hilbert space 
by energy conservation. 
\er

Let us look for a solution of the type $\Psi_\nu(\x)e^{-i\nu t}$ to 
Eq. (\re{Sl}). Substituting, we get the {\it Helmholtz} stationary 
equation, 
\be\la{Sls} 
(A-\nu)\Psi_\nu(\x)=-B(\x). 
\ee

\bd 
i) The {\bf spectrum} of Equation (\re{Sl}), $\spec A$, is the set 
of all $\om\in\C$ such that the operator $A+i\om$ is not invertible 
in the corresponding Hilbert space $E$. 
\\
ii) The  {\bf resolvent} of the operator $A$ is defined by
\be\la{resol}
R(\om):=(A+i\om)^{-1},~~~~~~\om\in\C\setminus \spec A.
\ee

\ed 
Equation (\re{Sls}) admits a unique solution $\Psi_\nu(\x)\in E$ 
for every $B\in E$ if $\nu\not\in \spec A$:
\be\la{Slsr} 
\Psi_\nu=-R(\nu)B,~~~~~~~~\nu\in\C\setminus \spec A.
\ee

\bexe Calculate $\spec A$ for the free 
 Schr\"odinger 
equation corresponding to (\re{Sla}) with $\h=1$ and $m=1$: 
\be\la{Slae} 
i\dot\psi(t,\x)= \De \psi(t,\x). 
\ee 
{\bf Hint:} Use the Fourier transform $\hat\psi(\bk):= 
\ds\int e^{i\bk\x}\psi(\x)d\x$. \\ 
{\bf Solution:} 
The operator $A=\De$ 
becomes the multiplication operator $\hat A(\bk)=-\bk^2$, 
and the phase space $\hat E=L^2(\R^3)\approx E$ by the 
Parseval theorem. Therefore, 
$\hat R(\om,\bk)$ is multiplication by $-(\bk^2+\om)^{-1}$. 
It is bounded in $\hat E$ iff the function $(\bk^2+\om)^{-1}$ 
is bounded in $\R^3$. Hence, 
 $\spec A=\ov\R_-=\{\om\in\R:\om \le 0\}$. 
\eexe 
\bexe Calculate the resolvent $R(\om)$ corresponding to (\re{Slae}). 
\\ 
{\bf Hint:} $R(\om)$ is a convolution with a fundamental solution 
$\E_\om(\x)$ of the operator $\De-\om$. 
Choose the fundamental solution from the space of tempered 
distributions. 
\\ 
{\bf Solution:} The tempered fundamental solution is unique, 
\be\la{fus} 
\E_\om(\x)=\E_\om^-(\x):= 
\fr{e^{-\sqrt{\om}|\x|}}{4\pi|\x|},~~~~~~~\om\in\C\setminus \ov\R_-, 
\ee 
where we choose $\rRe\sqrt{\om}>0$ for all $\om\in\C\setminus\ov\R_-$. 
 
\eexe 
\br 
i) The distribution 
$\E_\om(\x)$ is tempered since its Fourier transform is a bounded function. 
\\ 
ii) The distribution 
$\E_\om^+(\x):= 
\ds\fr{e^{\sqrt{\om}|\x|}}{4\pi|\x|}$ 
also is a fundamental solution, however, 
it is not tempered for all $\om\in\C\setminus\ov\R_-$. 
\\ 
iii) For $\om<0$, both fundamental solutions, $\E_\om^\pm(\x)$, 
are tempered. 
\er 
{\bf Limiting amplitude principle} 
The principle states 
(see \ci{Eid}) 
the 
following long-time asymptotics for a general class of solutions to 
 (\re{Sl}) with $\psi(0,\x)\in E$: 
\be\la{lapr} 
\Psi(t,\x)= \ov\Psi_\nu(\x)e^{-i\nu t} 
+\sum_l C_l\Psi_l(\x)e^{-i\om_l t}+r(t,\x), 
\ee 
where $\ov\Psi_\nu(\x)$ is a {\it limiting amplitude}, 
$\Psi_l(\x)$ stand for the eigenfunctions of the discrete 
spectrum of the operator $A$, and 
$r(t,\x)\to 0$, $t\to\infty$, in an appropriate norm. 
\medskip\\ 
For the wave equation  (\re{wla}), the asymptotics  (\re{lapr}) follows 
immediately from the Kirchhoff formula (\re{A22d}), if the initial functions 
decay rapidly. Let us assume, for example, that 
\be\la{exd} 
|\na\phi(0,\x)|+|\phi(0,\x)|+|\pi(0,\x)|=\cO(e^{-\ve|x|}),~~~~~~|\x|\to\infty, 
\ee 
with an $\ve>0$. Then the 
first and second terms on the RHS of (\re{A22d}) decay exponentially, 
like $e^{-\ve t}$, while 
the last is just proportional to $e^{-i\nu t}$. 
Indeed, substituting $\rho(t,\x)=\ds\fr 1{4\pi}b(\x)e^{-i\nu t}$ 
and $c=1$ 
into (\re{A22b}), we get 
\be\la{jex} 
\ds\phi_{ret}(t,\x):= 
~~\ds\int \fr{b(\y)e^{-i\nu (t-|\x-\y|)}} { 4\pi |\x-\y|}d\y 
=e^{-i\nu (t)}\ov\phi(\x),~~~~~\ov\phi(\x)= 
\ds\int \fr{b(\y)e^{i|\x-\y|}} { 4\pi |\x-\y|}d\y. 
\ee 
This gives (\re{lapr}) without the sum. This 
corresponds to the absence of the discrete spectrum for the wave 
equation with constant coefficients. 
 
For the Schr\"odinger equation (\re{Sla}) the asymptotics  (\re{lapr}) 
is proved now for a restricted class of potentials. For example, 
it can be deduced from the Kato and Jensen results \ci{KJ} for 
\be\la{KJ} 
\phi(\x)=\cO(|\x|^{-4-\ve}),~~~~\bA(\x)\equiv 0, 
\ee 
where $\ve >0$. 
 The asymptotics holds 
for the initial function $\psi(0,\x)$ from the {\it Agmon spaces}, 
$H_\si=\{\psi(\x): (1+|\x|)^\si(1-\De)\in L^2(\R^3) \}$ 
with $\si>1/2$. Then the  remainder $r(t,\x)\to 0$ in a dual space 
$H_{\si}^*$ 
to $H_{\si}$. 
\bp \la{pLAPS} 
i) For the Schr\"odinger equations (\re{Sla}), (\re{SMsla}), 
with  potentials satisfying (\re{KJ}), 
$\psi(0,\x)\in  H_\si$, $\si>1/2$, 
and 
$\nu\ne \om_l$, $\nu\ne 0$, 
the asymptotics (\re{lapr}) holds and 
$\Vert r(t,\cdot)\Vert_{H_{\si}^*} \to 0$, $t\to\infty$. \\ 
ii) 
The limiting 
amplitude is given by  (cf. (\re{Slsr})) 
\be\la{LAbP} 
\ov\Psi_\nu=-\lim_{\ve\to 0+}R(\nu+i\ve)B,~~~~~~~\nu\in\R, 
~~~~~~B\in H_\si, ~~\si>1/2, 
\ee 
where the limit holds in the space $H_\si^*$.

\ep 
\brs\la{rlia} 
i) The limiting amplitude $\ov\Psi_\nu\in E:=L^2(\R^3)$ 
if $\nu\not\in \spec A$. 
The results \ci{KJ} imply that $\ov\Psi_\nu\in  H_\si^*$
for $\nu\ge 0$, though $\nu\in \spec\!_c\5 A$
then, 
where $\spec\!_c\5 A$ stands for the {\bf continuous spectrum} 
of the operator $A$.
However
generally  $\ov\Psi_\nu\not\in E$: this is related to a slow decay 
of  $\ov\Psi_\nu(\x)$ at infinity, $|\x|\to\infty$, 
which physically means a radiation of electrons 
to infinity. 
\\
ii) The Coulomb potential $\phi(\x)=C|\x|^{-1}$ 
does not fit the bound (\re{KJ}). 
\\ 
iii) 
Formula (\re{LAbP}) 
is called the ``{\it limiting absorption principle}'' 
since 
$R(\nu+i\ve)=(A-\ve+i\nu)^{-1}$ is the resolvent 
of the ``damped'' equation 
\be\la{Slab} 
i\dot\Psi(t,\x)= A\Psi(t,\x)-\ve i\Psi(t,\x)+B(\x)e^{-i\nu t}, 
\ee 
where the term $-\ve i\Psi(t,\x)$ describes an absorption of energy. 

\ers 
Let us sketch a {\it formal} proof of the proposition 
using the 
results \ci{KJ}. Namely, 
the Duhamel representation 
for the solutions to the inhomogeneous equation   (\re{Sl}) 
reads, 
\be\la{dgr} 
\Psi(t)= U(t)\Psi(0)-i\int_0^tU(t-s)Be^{-i\nu s}ds, 
\ee 
where $U(t):=e^{-iAt}$ is the dynamical group of the corresponding
free equation. 

The results \ci{KJ} imply that 
the 
first term on the RHS admits an asymptotics (\re{lapr}) with 
$\ov\Psi_\nu=0$. 
This is a central achievement  of the Kato-Jensen theory 
\ci{KJ}: the contribution of the continuous spectrum to the 
solution  $U(t)\Psi(0)$ of the homogeneous equation 
vanishes as $t\to\infty$. 
 
 It remains to analyze the integral term. 
It can be rewritten 
as 
\be\la{dgri} 
I(t)=-ie^{-i\nu t}\int_0^tU(s)Be^{i\nu s}ds. 
\ee 
Therefore, we have asymptotically, 
\be\la{dgrb} 
I(t)=-ie^{-i\nu t}\int_0^t e^{-iAs}Be^{i\nu s}ds\sim 
e^{-i\nu t}\ov\Psi_\nu,~~~~~~~t\to\infty, 
\ee 
since
\be\la{dgrbo} 
-i\int_0^\infty e^{-iAs}Be^{i\nu s}ds 
=-i\lim_{\ve\to 0+}\int_0^\infty e^{-iAs}Be^{i(\nu+i\ve) s}ds 
=-\lim_{\ve\to 0+}(A-\nu-i\ve)^{-1}B=:\ov\Psi_\nu, 
\ee 
where the last limit exists 
in the space $H_\si^*$ by the results of \ci{KJ}
since $\nu\ne \om_l$ and  $\nu\ne 0$. 
\bo 
\br \la{rNH} 
Our arguments above demonstrate that\\ 
i) The term 
 $\ov\Psi(\x)e^{-i\nu t}$  on the RHS of (\re{lapr}) 
characterizes the {\bf response} of a physical system 
 to the external periodic source $B(\x)e^{-i\nu t}$, 
 and  it 
does not depend on the  initial data. 
\\ 
ii) On the other hand, the sum over the discrete 
spectrum on the RHS depends only on the initial data 
and does not depend on the external source. 
\medskip\\ 
Hence, in an experimental observation of a scattering problem, 
the response, in principle, can be singled out 
if the experiment is concentrated 
on the field which is built by the 
external source. 
 
\er

%%%%%%%%%%%%%%%%%%%%%%%%%%%%%%%%%%%%%%55 
 
\part{Exercises}

%% %%CA 
%% 
%% Exercises 
%% 
%% 
 
\newpage 
%%%%%%%%%%%%%%%%%%%%%%%%%%%%%%%%%%%%%%%%%%%%%%%%%%%%%%%%%%%%%%%%%%%%% 
%%%%%%%%%%%%%%%%%%%%%%%%%%%%%%%%%%%%%%%%%%%%%%%%%%%%%%%%%%%%%% 
%%\setcounter{section}{+9} 
\setcounter{subsection}{0} 
\setcounter{theorem}{0} 
\setcounter{equation}{0} 
%%\section{Exercises} 
\section{Exercises for Part I} 
 
\subsection{Exercise 1: 
Main Lemma of the Calculus of Variations} 
\begin{itemize} 
\item[] 
\bd 
$C[a,b]$ is a space of continuous functions 
on the interval $[a,b]$. 
$$ 
C_0[a,b]:=\{h(x)\in C[a,b]:\,h(a)=h(b)=0\}. 
$$ 
\ed 
\begin{lemma} 
Let $f(x)\in C[a,b]$ and $\ds\int\limits_0^T f(x)h(x)\,dx=0$ 
for every $h(x)\in  C[a,b]$. Then $f(x)\equiv0$. 
\end{lemma} 
 
\item[a)] Prove this lemma. 
\item[b)] Prove the analogous lemma for $h(x)\in C_0 [a,b]$. 
\item[] {\bf Proof of a)} 
Let there exist a point $x_0$ such that 
$f(x_0)\not=0$. 
Without loss of generality, we assume $f(x_0)=a>0$. 
Since $f(x)\in C[a,b]$, then for any $\ve >0$ 
there exists a $\delta>0$ such that 
$|f(x)-a|<\ve$ for all $x\in {\cal O}_\de(x_0)\equiv 
(x_0-\de,x_0+\de)$. 
We choose $\ve=a/2$. 
Then $f(x)>a-\ve=a/2$,  $\forall x\in {\cal O}_\de(x_0)$. 
Now we choose the function $h(x)\in C[a,b]$ such 
that supp $h(x)\in{\cal O}_\de(x_0)$ and $h(x)\ge 0$. 
For example,  $h(x)=0$ for $|x-x_0|\ge \de$ 
and $h(x)=\de-|x-x_0|$ for $|x-x_0|\le \de$. 
Then 
$$ 
\int\limits_0^T f(x)h(x)\,dx 
\int\limits_{x_0-\de}^{x_0+\de} f(x)h(x)\,dx> 
\frac{a}{2}\int\limits_{x_0-\de}^{x_0+\de} h(x)\,dx>0.\,\,\,\,\Box 
$$ 
\item[] The Proof of a) covers the case b), because the function $h(x)$ 
used in the proof  $h(x) \in C_0 [a,b]$. 
\end {itemize} 
 
%%%%%%%%%%%%%%%%%%%%%%%%%%%%%%%%%%%%%%%%%% 
 
\subsection{Exercise 2: Euler-Lagrange Equations} 
\begin{itemize} 
\item[a)] Given  a functional $F: C^1[a,b]\to \R$, 
$$ 
F[y]:=\int\limits_a^b f(x,y(x),y'(x))\,dx, 
$$ 
introduce a space 
$$ 
L[a,b]:=\{y\in C^1[a,b],\quad y(a)=y_a,\quad y(b)=y_b\}. 
$$ 
Show that for a $y$ which minimizes $F$, the Euler-Lagrange equation 
$$ 
f_y(x,y,y')-\frac{d}{dx}f_{y'}(x,y,y')=0 
$$ 
holds. Here $f_y:=\pa f/\pa y$.\\ 
\item[] 
{\bf Proof:} 
Let $F(y_*):= \min\limits_{y\in L[a,b]} F(y)$. 
Let $h(x)\in C_0^1[a,b]:=\{h(x)\in C^1[a,b]:\,h(a)=h(b)=0\}.$ 
Fix such a function $h(x)$. 
Introduce the function 
$$ 
S(\ve):=F[y_*+\ve h]=\int\limits_a^b 
f(x,y_*(x)+\ve h(x),y'_*(x)+\ve h'(x))\,dx,\,\,\,\,\ve\in\R . 
$$ 
Since 
$y_*(x)+\ve h(x)\in L[a,b]$ for any $\ve$, 
$S(\ve)\ge S(0)=F(y_*)$. 
Hence, $\min S(\ve)=S(0)$ and $S'(0)=0$. 
$$ 
S'(\ve)\Big|_{\ve=0}= 
\int\limits_a^b[f_y(x,y_*(x),y'_*(x)) h(x)+ 
f_{y'}(x,y_*(x),y'_*(x)) h'(x)]\,dx=0. 
$$ 
Integrating by parts we obtain 
$$ 
\int\limits_a^b[f_y(x,y_*(x),y'_*(x)) - 
\frac{d}{dx}f_{y'}(x,y_*(x),y'_*(x))] h(x)\,dx=0. 
$$ 
Exercise 1 implies  $f_y(x,y_*(x),y'_*(x)) - 
\ds\frac{d}{dx}f_{y'}(x,y_*(x),y'_*(x))=0$.\\ 
%%%%%%%%%%%%%%%%%%%%%%%%%%%%%%%%%%%%%%%%% 
\item[b)] 
{\bf Geodesic line:} 
Use the Euler-Lagrange equations 
to calculate the minimum 
of the length functional in the plane 
$$ 
L[y]=\int\limits_a^b 
\sqrt{1+|y'(x)|^2}\,dx 
$$ 
with fixed boundary conditions $y(a)=y_a$, 
$y(b)=y_b$.\\ 
\item[] 
{\bf Solution:} 
$f(x,y,y'):= \sqrt{1+|y'(x)|^2}$. 
Then $f_y=0$. 
Substituting into the  Euler-Lagrange equations 
we get 
$$ 
\frac{d}{dx}f_{y'}(x,y(x),y'(x))=0. 
$$ 
Hence, $f_{y'}(x,y(x),y'(x))=\frac{y'}{\sqrt{1+|y'(x)|^2}}=$const. 
Hence, 
$y'=$const, and $y(x)=a+bx$. 
\bigskip\\ 
%%%%%%%%%%%%%%%%%%%%%%%%%%%%%%%%%%%% 
\end{itemize}

\subsection{Exercise 3: Light Propagation in a Stratified Medium} 
\begin{itemize} 
\item[] 
Consider a medium where the index of refraction 
only depends on one coordinate $y$, 
$n(y)=1/v(y)$ (here $v(y)$ is the speed of light.) 
The resulting system is effectively two-dimensional, 
i.e., the $z$ coordinate may be suppressed. 
 
\item[a)] Use the {\bf Fermat principle} (minimization of the flight time) 
$$ 
T[y]:=T_{y(a),y(b)}:=\int\limits_a^b\frac{dl}{v(y)} 
$$ 
with fixed starting point $(x=a, y=y_a)$ and end point $(x=b, y=y_b)$) 
to prove the {\bf Snellius law} of refraction 
$$ 
n(y)\sin\al(y)=\const. 
$$ 
where $\al(y)$ is the angle between the light ray and the $y$-axis. 
\item[b)] 
Use the Snellius law to sketch the light rays of a wave guide with 
$n(y)=2+\cos y$. 
\item[c)] 
Use the Snellius law to sketch the light rays for ordinary 
refraction 
$n(y)=n_1$ for $y<0$, and $n(y)=n_2$ for $y>0$. 
\item[] 
{\bf Proof of a)} Since $dl=\sqrt{1+(y'(x))^2}$ and $n(y)=1/v(y)$, 
we rewrite $T[y]$ in the form 
 $$ 
T[y]=\int\limits_a^b  n(y)\sqrt{1+(y'(x))^2}\,dy. 
$$ 
Applying the Euler-Lagrange equations to 
$f(x,y,y'):= n(y)\sqrt{1+|y'(x)|^2}$, we get 
$$ 
n'(y)\sqrt{1+|y'(x)|^2}- 
\frac{d}{dx}\frac{ n(y)y'}{\sqrt{1+|y'(x)|^2}}=0. 
$$ 
Differentiating in $x$ we get 
$$ 
\frac{ n'(y)}{\sqrt{1+|y'(x)|^2}} 
- n(y)\frac{ y''}{(1+|y'(x)|^2)^{3/2}}=0. 
$$ 
This is equivalent to 
$$ 
\frac{d}{dx}\left[ 
 n(y)\cdot\frac{ 1}{\sqrt{1+|y'(x)|^2}}\right]=0. 
$$ 
Hence, $n(y)\cdot\ds\frac{ 1}{\sqrt{1+|y'(x)|^2}}=\const$. 
Since $\sin \al(y)=\ds\frac{ 1}{\sqrt{1+\ctg^2\al}}= 
\frac{ 1}{\sqrt{1+|y'(x)|^2}}$, 
we get $n(y)\sin\al(y)=\const$. 
\hfill$\Box$ 
\end{itemize} 
 
%%%%%%%%%%%%%%%%%%%%%%%%%%%%%%%%%%%%%%%%%%%%% 
 
%%%%%%%%%%%%%%%%%%%%%%%%%%%%%%% 
%%%%%%%%%%%%%% 
\section{Exercises for Part III} 
 
%%%%%%%%%%%%%%%%%%%%%%%%%%%%%%%%%%%%%%%%%%%%% 
 
\subsection{Exercise 4: The Kepler Problem in 5 Easy Steps} \label{ue-Ex-4} 
\begin{itemize} 
\item[] 
The Kepler problem is defined by the two-particle Lagrangian 
 
$$L=\frac{m_1}{2}\dot x_1^2 + \frac{m_2}{2} 
\dot x_2^2 + \frac{\gamma m_1 m_2}{|x_1 
-x_2|} . 
$$ 
 
Solve it by going through the following steps: 
 
\item[a)] 
Reduction to a one-particle problem (coordinate $x$) 
via separation of the center-of-mass coordinate $X$. 
 
\item[b)] 
Reduction to a two-dimensional problem by the observation that the plane 
spanned by $(x(0),\dot x (0))$ is invariant. 
 
\item[c)] 
Reduction to a one-dimensional problem (coordinate $r$) by the 
introduction of polar coordinates $(r,\varphi)$. 
Show that Kepler's third law holds, find the effective 
potential of the one-dimensional problem, and check for energy conservation. 
 
\item[d)] 
Investigate the orbits of the one-dimensional problem. 
 
\item[e)] 
Integrate the orbit equation to find the possible orbits. Show that, 
depending on the values of the integration constants (angular momentum, 
energy), the orbits are indeed ellipses, parabolae or hyperbolae. 
 
\item[] 
{\bf Solution:} 
\item[a)] 
Denote by $X(t)$ the center of mass, 
$X(t):=\ds\frac{ m_1 x_1(t)+ m_1 x_1(t)}{M}$, 
where the total mass is $M:=m_1+m_2$, and $x:=x_1-x_2$. 
Then 
$$ 
\sum\limits_{i=1}^2 m_i \dot x^2_i= 
\frac{ m_1 m_2 \dot x^2(t)}{M}+ 
 M\dot X^2. 
$$ 
Finally, we can rewrite the Lagrangian in the form 
$$ 
L=\frac{ m_1 m_2\dot x^2}{2M}+ 
 \frac{M\dot X^2}{2}+ \frac{\ga m_1 m_2}{|x_1-x_2|} 
= L_1(x,\dot x,\dot X). 
$$ 
Applying the Euler-Lagrange equation to the coordinate $X$ 
$$ 
\frac{d}{dt}\frac{\pa L}{\pa \dot X}-\frac{\pa L}{\pa X}=0, 
$$ 
we get 
$\ds\frac{d}{dt} M\dot X=0.$ 
Hence, 
\be\la{CA-consQ} 
\dot X=\const, 
\ee 
 then 
$X(t)= at+b$, where $a,b$ are some constants. 
 
\begin{remark} 
{\rm We can obtain Eqn (\ref{CA-consQ}) from the following 
arguments. 
Note that the potential 
$$ 
V(x_1,x_2):=-\frac{\ga m_1 m_2}{|x_1-x_2|} 
$$ 
depends only on the difference $x_1-x_2$. 
Hence, $V$ is invariant with respect to translations 
$$ 
T_s:(x_1,x_2)\to (x_1+hs,x_2+hs),\,\,\,s\in\R,\,\,\,\, 
h\in\R^3. 
$$ 
Hence, the total momentum (see Section 3.2) is conserved, 
$p:=\sum\limits_{i=1}^2 m_i \dot x_i=\const$, 
and the center of mass moves uniformly, 
$X(t)= At+B$.} 
\end{remark} 
%%%%%%%%%%%%%%%%%%%%%%%%%%%% 
Finally, we get that it suffices to consider 
the Lagrangian 
$$ 
L_1\equiv L_1(x,\dot x):= 
\frac{ m_1 m_2\dot x^2}{2M}+ 
  \frac{\ga m_1 m_2}{|x|}. 
$$ 
%%%%%%%%%%%%%%%%%%%%%%%%%%%%%%%%%%%%%% 
\item[b)] 
We can omit the 
constants $m_1 m_2$ and consider the equivalent Lagrangian 
$$ 
L\equiv L(x,\dot x):= 
\frac{ \dot x^2}{2M}+ 
  \frac{\ga }{|x|}, 
$$ 
which gives the same trajectories. 
Denote by $M(t)$ the angular momentum 
$ 
M(t):= x(t)\times p(t), 
$ 
where $p=\ds\frac{\pa L}{\pa \dot x}=\frac{\dot x}{M}$. 
Then, 
\be\la{CA-consM} 
M(t)= x(t)\times \frac{\dot x(t)}{M}=\const,\quad t\in \R. 
\ee 
Indeed, 
$$ 
\dot M(t)=\dot x(t)\times p(t)+ 
 x(t)\times \dot p(t)= \dot x(t)\times \frac{\dot x(t)}{M}+ 
 x(t)\times \frac{\ddot x(t)}{M} 
=x(t)\times \frac{\ddot x(t)}{M}. 
$$ 
On the other hand, the Euler-Lagrange equation implies 
$ 
\ds\frac{\ddot x(t)}{M}+ \frac{\ga x}{|x|^3}=0. 
$ 
Hence, $\ddot x\Vert x$ and 
$\dot M(t)=x(t)\times \ds\frac{\ddot x(t)}{M}=0$. 
\hfill$\Box$ 
%%%%%%%%%%%%%%%%%%%%%%%%%%%%%%%%%%%%%%%%%%%%% 
\begin{remark} 
{\rm We can obtain Eqn (\ref{CA-consM}) from the following 
arguments. 
Note that the potential 
$ 
V(x):=-\ds\frac{\ga }{|x|} 
$ 
depends only on $|x|$. 
Hence, $V(x)$ is invariant with respect to rotations 
(see section 2.3.2), 
and the angular momentum $M$ is conserved 
(see Theorem 2.15).} 
\end{remark} 
%%%%%%%%%%%%%%%%%%%%%%%%%%%% 
%%%%%%%%%%%%%%%%%%%%%%%%%%%% 
Finally, Eq. (\ref{CA-consM}) implies 
$$ 
x(t)\times \dot x(t)=x(0)\times \dot x(0)=: \bar n. 
$$ 
Hence, $x(t)$ belongs to the plane perpendicular to the vector 
$\bar n$. 
We can choose the coordinates such that $x_3\Vert \bar n$. 
Then, $x(t)=(u(t),v(t),0)$ and the Lagrangian for 
the variables $u(t), v(t)$ is 
\be\la{CA-L3} 
L(u,v,\dot u,\dot v):= 
\frac{\dot u^2+\dot v^2}{2M}+\frac{\ga}{\sqrt{u^2+v^2}}. 
\ee 
%%%%%%%%%%%%%%%%%%%%%%%%%%%%%%%%%%%%%% 
\item[c)] 
We choose polar coordinates $(r,\varphi)$: 
\be\la{CA-pc} 
\left\{ 
\ba{l} 
u=r\cos\varphi,\\ 
v=r\sin\varphi. 
\ea 
\right. 
\ee 
Then 
$u^2+v^2=r^2$, $\dot u^2+\dot v^2=\dot r^2+r^2\dot\varphi^2$. 
We substitute into the Lagrangian and get 
\be\la{CA-Lpc} 
L(r,\varphi,\dot r,\dot \varphi):= 
\frac{\dot r^2+r^2\dot \varphi^2}{2M}+\frac{\ga}{r}. 
\ee 
Applying the Euler-Lagrange equation to the coordinate $\varphi$, 
$ 
\ds\frac{d}{dt}\frac{\pa L}{\pa \dot\varphi}- 
\frac{\pa L}{\pa \varphi}=0, 
$ 
we get 
$ 
\ds\frac{d}{dt}\Big[\frac{r^2\dot\varphi}{M}\Big]=0. 
$ 
Hence, we proved Kepler's third law 
\be\la{CA-Kl} 
r^2\dot\varphi=\const=:I. 
\ee 
\begin{remark} 
{\rm 
(\re{CA-Kl}) is equivalent to (\re{CA-consM}). Namely, 
if we calculate the angular momentum $M(t)$ 
in polar coordinates, we get} 
$$ 
M(t)=\frac{u\dot v-\dot u v}{M}= 
\frac{r^2\dot\varphi}{M}=M(0). 
$$ 
\end{remark}

Now applying the Euler-Lagrange 
equation to the coordinate $r$: 
$\ds 
\frac{d}{dt}\frac{\pa L}{\pa \dot r}- 
\frac{\pa L}{\pa r}=0, 
$ 
we get the reduced radial equation 
$$ 
{\ddot r} 
={r\dot \varphi^2}-\frac{\ga M}{r^2}. 
$$ 
By (\ref{CA-Kl}) we rewrite it as follows: 
$$ 
\ds\ddot r=\frac{I^2}{r^{3}}-\frac{\ga M}{r^2}=-V'(r),~~~~~~~~ 
V(r):=\ds\frac{I^2}{2r^{2}}-\frac{2\ga M}{r}. 
$$ 
Multiplying this equation by $\dot r$ 
and 
integrating, we obtain the reduced energy conservation 
\be\la{CA-r2} 
\dot r^2+V(r)=E_0. 
\ee 
\begin{remark} 
{\rm Since the Lagrangian (\ref{CA-L3}) does not depend on 
$t$, the energy is conserved, 
$$ 
E(t):= 
\frac{\dot u^2+\dot v^2}{2M}-\frac{\ga}{\sqrt{u^2+v^2}}=\co. 
$$ 
In polar coordinates, 
taking into account (\ref{CA-Kl}), we get 
$$ 
E(t)= 
\frac{\dot r^2+r^2\dot \varphi^2}{2M}-\frac{\ga}{r}= 
\frac{\dot r^2}{2M}+\frac{I^2}{2Mr^2}-\frac{\ga}{r}=\co, 
$$ 
which coincides with (\ref{CA-r2}).} 
\end{remark} 
\item[d)] 
Now we find $r(\varphi)$. 
Since $\ds\frac{dr}{d\varphi}=\frac{\dot r}{\dot \varphi}$, 
we get by (\ref{CA-r2}) and (\ref{CA-consM}) 
$$ 
\frac{dr}{d\varphi}= 
\frac{\pm\sqrt{-\ds\frac{I^2}{r^{2}}+ 
\frac{2\ga M}{r}+2E_0}}{\ds\frac{I}{r^2}}. 
$$ 
Hence 
$$ 
\int \frac{I dr}{r^2\sqrt{-\ds\frac{I^2}{r^{2}}+ 
\frac{2\ga M}{r}+2E_0}}=\pm\int d\varphi . 
$$ 
We introduce the {\bf Clerot substitution} 
$\rho:=1/r$. Then $dr=-\rho^{-2}d\rho$. 
Hence, 
$$\int \frac{I dr}{r^2\sqrt{-\ds\frac{I^2}{r^{2}}+\frac{2\ga M}{r}+2E_0}}= 
-\int \frac{I d\rho}{\sqrt{-I^2\rho^{2}+2\ga M\rho+2E_0}} 
=\int \frac{I d\rho}{\sqrt{D-I^2(\rho-B)^{2}}}. 
$$ 
Here $D:=2E_0+(\ga M)^2 I^{-2}$, 
$B:=\ga M I^{-2}$. 
Note that the constant $D$ must be positive for any non-empty trajectory. 
Further, 
$$ 
\int \frac{I d\rho}{\sqrt{D-I^2(\rho-B)^{2}}} 
=\arcsin\{\frac{I(\rho-B)}{\sqrt D}\}=\pm\varphi+\vp_0. 
$$ 
Finally, we get 
$ 
\ds\frac{I(\rho-B)}{\sqrt D}=\sin(\varphi+\vp_0). 
$ 
Since $\rho:=1/r$, we get 
\be\la{CA-r} 
\frac{A_1}{r}=A_2+\sin(\varphi+\vp_0), 
\ee 
where 
\beqn 
A_1&:=&\ds\frac{I}{\sqrt D},\la{CA-A1}\\ 
A_2&:=&\ds \frac{I B}{\sqrt D}=\frac{\ga M}{\sqrt{2E_0I^2+(\ga M)^2}}. 
\la{CA-A2} 
\eeqn 
Note that $r$ is bounded below by (\ref{CA-r}). 
\item[e)] 
Let, for simplicity, $\vp_0=0$. 
Then (\ref{CA-r}) becomes 
\be\la{CA-r1} 
A_2 r=A_1-r\cos\vp. 
\ee 
We remember that $r\sin\varphi=v$ and $r=\sqrt{u^2+v^2}$, 
and now we re-write (\ref{CA-r1}) in the form 
$ 
(u^2+v^2)A_2^2=(A_1-v)^2. 
$ 
Hence, 
$$ 
u^2+v^2\Big(1-\frac{1}{A_2^2}\Big)+2\frac{A_1}{A_2^2}v=\frac{A_1^2}{A_2^2}. 
$$ 
Obviously, $A_1 \ge 0$. 
Let us assume that $A_1 >0$. 
Then, if $A_2>1$ we have an ellipse, 
if  $A_2<1$ we have a hyperbola, and 
if  $A_2=1$ we have a parabola ($A_2$ is a positive constant). 
By (\ref{CA-A2}) we  get: 
if $E_0<0$ we have an ellipse, 
if  $E_0>0$ we have a hyperbola, and 
if  $E_0=0$ 
we have a parabola. 
 
When $A_1 =0$ (i.e. $I=0$) then we have either $r=0$, $\varphi$ arbitrary 
(the point mass is in the origin) or $\sin \varphi =-A_2 ={\rm const}$ 
(the point mass moves along a line through the origin). 
\end{itemize} 
 
%%%%%%%%%%%%%%%%%%%%%%%%%%%%%%%%%%%%%%%%%% 
 
\subsection{Exercise 5: Bohr-Sommerfeld Quantization of the $H$ 
Atom} 
\begin{itemize} 
\item[] 
The Hamiltonian for an electron with charge $-e$ 
and mass $m$ in the Coulomb field of an infinitely 
heavy nucleus with charge $e$ is 
\be\la{CA-3.1} 
H=\frac{1}{2\mu}\left( 
p_r^2+\frac{p_\varphi^2}{r^2}\right)- 
\frac{e^2}{r}, 
\ee 
where the reduction to the effective two-dimensional problem 
(in polar coordinates $(r,\varphi)$, see (\ref{CA-pc})) 
has already been performed (cf (\ref{CA-Lpc})). 
Find the first two integrals of motion 
(conservation of angular momentum and energy). 
Assume that the energy $E$ is less that zero to have 
periodic orbits (ellipses). Then 
use the Bohr--Sommerfeld quantization rules 
\be\la{CA-3.2} 
\int  p_r \, dr=kh, \quad \quad \int 
 p_\varphi \,d\varphi =lh 
\ee 
(here both integrals are over one period, $k$ and $l$ are integers, and $h$ 
is Planck's constant) to derive the quantization of angular momentum and 
the Balmer formula for the energy levels of the electron. Use the 
principal quantum number $n=k+l$ in the latter case. 
 
\item[] 
{\bf Solution:} 
By (\ref{CA-3.1}) it follows that 
\beqn 
\dot\varphi&=&H_{p_{\varphi}}= 
\frac{p_\varphi}{\mu r^2},\la{CA-3.3}\\ 
\dot p_\varphi&=&-H_\varphi=0,\la{CA-3.4}\\ 
\dot r&=&H_{p_{r}}=\frac{p_r}{\mu }, 
\la{CA-3.5}\\ 
\dot p_r&=&-H_r= 
\frac{p_\varphi^2}{\mu r^3}-\frac{e^2}{r^2},\la{CA-3.6} 
\eeqn 
By (\ref{CA-3.4}) we have 
\be\la{CA-3.7} 
p_\varphi=M=\const. 
\ee 
Then by (\ref{CA-3.5}) we get 
\be\la{CA-3.8} 
\dot p_r=\frac{M^2}{\mu r^3}-\frac{e^2}{r^2}. 
\ee 
Since $\ds\frac{dp_r}{dr}=\frac{\dot p_r}{\dot r}$, 
(\ref{CA-3.5}), (\ref{CA-3.8}) imply 
\be\la{CA-3.9} 
\ds\frac{dp_r}{dr}=\frac{\ds\frac{M^2}{\mu r^3}-\frac{e^2}{r^2}}{\ds\frac{p_r}{\mu }}, \quad {\rm or} \quad\ds\frac{p_r\,dp_r}{\mu }= 
\Big(\frac{M^2}{\mu r^3}-\frac{e^2}{r^2}\Big)\,dr. 
\ee 
Hence, 
\be\la{CA-3.10} 
\frac{p^2_r}{2\mu }+\frac{M^2}{2\mu r^2} 
-\frac{e^2}{r}={\cal E}=\const. 
\ee 
Let ${\cal E}<0$. Then (see Exercise 4) 
the orbits are ellipses. 
Hence, 
\be\la{CA-3.11} 
p_r=\sqrt{2\mu (\frac{e^2}{r}-E)-\frac{M^2}{r^2}}. 
\ee 
where $E\equiv |{\cal E}| >0$ was introduced for convenience. 
 
Now we find the extrema of $r$. 
(\ref{CA-3.5}) implies that $\dot r=0$ 
iff $p_r=0$. 
By (\ref{CA-3.11}) we have 
$$ 
2\mu (\frac{e^2}{r}-E)-\frac{M^2}{r^2}=0. 
$$ 
$$ 
r^2-\frac{e^2r}{E}+\frac{M^2}{2\mu E}=0. 
$$ 
$$ 
r=\frac{e^2}{2E}\pm 
\sqrt{\frac{e^4}{4E^2}-\frac{M^2}{2\mu E}}=\frac{e^2}{2E}\pm 
\frac{1}{2E}\sqrt{e^4-\frac{2M^2E}{\mu }}=:r_\pm. 
$$ 
Hence, $r_{max}=r_+$ and $r_{min}=r_-$. 
Now we use (\ref{CA-3.2}). 
At first, (\ref{CA-3.2}) and 
(\ref{CA-3.7}) imply 
\be\la{CA-3.12} 
\int  p_\varphi \, d\varphi= 
\int M\,d\varphi= 
2\pi M=hl. 
\ee 
Hence, $M=hl$, $l\in\Z$. 
Then (\ref{CA-3.11}) implies 
\beqn\la{CA-3.13} 
\int  p_r \, dr&=& 
2\int\limits_{r_{min}}^{r_{max}} 
p_r\,dr= 
2\int\limits_{r_{min}}^{r_{max}} 
\sqrt{2\mu (\frac{e^2}{r}-E)-\frac{M^2}{r^2}}\,dr\nonumber\\ 
&=& 
2\int\limits_{r_{min}}^{r_{max}} 
\sqrt{-2\mu Er^2+2\mu e^2r-M^2}\,\frac{dr}{r}. 
\eeqn 
We use the formula 
\beqn 
\int\sqrt{ax^2+bx+c}\frac{dx}{x}&=& 
\sqrt{ax^2+bx+c}- 
\frac{b}{2\sqrt-a}\arcsin\frac{2ax+b}{b^2-4ac} 
\nonumber\\ 
&&-\sqrt{-c}\arcsin\frac{bx+2c}{x\sqrt{b^2-4ac}}, 
\,\,\,a<0,\,\,\,\,c<0.\nonumber 
\eeqn 
We apply this formula with 
$a=-2\mu E<0$, $b=2\mu e^2$, $c=-M^2<0$ 
to 
\beqn\la{CA-3.14} 
\int  p_r \, dr&=& 
2\Big[ 
\sqrt{2\mu (\frac{e^2}{r}-E)-\frac{M^2}{r^2}}\Big|_{r_{min}}^{r_{max}}- 
\frac{2\mu e^2}{2\sqrt{2\mu E}} 
\arcsin\frac{-4\mu Er+2\mu e^2}{\sqrt{4\mu ^2e^4-8M^2\mu E}}\Big|_{r_{min}}^{r_{max}}\nonumber\\ 
&&-|M|\arcsin\frac{2\mu e^2r-2M^2}{r\sqrt{4\mu ^2e^4-8M^2\mu E}}\Big|_{r_{min}}^{r_{max}} 
\Big]=2\Big[0+\pi\frac{\mu e^2}{\sqrt{2\mu E}}-|M|\pi\Big]\nonumber\\ 
&=& 
2\pi\Big[\sqrt{\frac{\mu e^4}{2E}}-|M|\Big]. 
\eeqn 
Now (\ref{CA-3.2}) implies 
$$ 
2\pi\Big[\sqrt{\frac{\mu e^4}{2E}}-|M|\Big]=hk,\,\,\,\,k\in\Z. 
$$ 
Applying (\ref{CA-3.12}), we get 
$$ 
\sqrt{\frac{\mu e^4}{2E}}=\h(k+|l|). 
$$ 
Denoting $k+|l|=n$, we get 
$$ 
\frac{\mu e^4}{2E}=\h^2n^2. 
$$ 
Hence, (cf. Theorem \re{HS}) 
$$ 
E=\frac{\mu e^4}{2\h^2n^2}=\frac{R}{n^2},\quad R:=\frac{\mu e^4}{2\h^2}. 
$$ 
\end{itemize} 
%%%%%%%%%%%%%%%%%%%%%%%%%%%%%%%%%%%%%%%%%%%%%%%%%%%%%%%%%% 

%%%%%%\subsection{Exercise 6: Bohr Correspondence Principle 
%%%%and Selection Rules} 

\subsection{Exercise 6: Rutherford Scattering Formula} 
\begin{itemize} 
\item[] 
The orbit equation for a particle in a central potential $U(r)$ is 
\be\la{CA-3.15} 
\varphi =\int_{r_0}^r \frac{M /r'{}^2 dr' }{\sqrt{2\mu  (E-U(r')) - 
M^2 /r'^2}} , 
\ee 
where $E$ is the energy and $M$ is the angular momentum. For the Coulomb 
potential $U(r)=\alpha /r$ (with $\alpha \in {\rm I \hspace{-0.1cm}R}$), and 
for $E>0$ this is a scattering orbit (a hyperbola). 
 
\item[a)] 
Calculate the angle $\varphi_0$, which is obtained from Eq. (1) by integrating 
from $r_{\rm min}$ to $\infty$. Express the total angle of deflection of the 
orbit for a particle starting at $r=\infty$ and going back to $r=\infty$ 
(i.e., the scattering angle $\chi$) with the help of $\varphi_0$. 
 
\item[b)] 
Re-express the constants of motion $(E,M)$ by $(E,b)$, where b is the 
scattering parameter, i.e., the normal distance between the asymptote 
of the incident particle and the scattering center at $r=0$. 
 
Then, assume that a constant flux of particles (i.e., a constant number 
$n$ of particles per area and time) with fixed energy and direction 
is approaching the scattering center. The number of particles per area 
and per time in a ring between $b$ and $b+db$ is then $dN = 2\pi n \, b 
\, db $. If there is a one-to-one functional relation between $b$ and the 
scattering angle $\chi$, then $dN$ is at the same time the number of 
particles per time that is scattered in an angle between $\chi$ and 
$\chi +d\chi$. Use this to calculate the differential cross section 
for Coulomb scattering (i.e., the Rutherford scattering formula) 
$$ 
d\sigma \equiv \frac{dN}{n}= 2\pi b(\chi) \left| \frac{db(\chi)}{d\chi} 
\right| d\chi. 
$$ 
 
\item[] 
{\bf Solution:} 
\item[a)] 
For the Lagrangian 
$L=\frac{\mu \bar x^2}{2}+\frac{\al}{|x|}$ 
in polar coordinates, we get (\ref{CA-3.15}) (see Exercise 4). 
Let $E>0$, then the orbits are hyperbolae. 
The scattering angle is $\chi=|\pi-2\varphi_0|$, where 
\be\la{CA-3.16} 
\varphi_0: =\int_{r_{min}}^\infty 
 \frac{M dr'}{r'\sqrt{2\mu  Er'^2 -2\mu \al r'-M^2 }}. 
\ee 
We use the formula 
\beqn 
\int\frac{dx}{x\sqrt{ax^2+bx+c}}&=& 
\frac{1}{\sqrt-c}\arcsin\frac{bx+2c}{x\sqrt{b^2-4ac}},\,\,\,c:=-M^2<0.\nonumber 
\eeqn 
Hence, 
\be\la{CA-3.17} 
\varphi_0: =M\frac{1}{M} 
 \arcsin\frac{-2\mu \al r' -2M^2}{r'\sqrt{4\mu ^2 \al^2 +8M^2\mu E }} 
|_{r_{min}}^{\infty}. 
\ee 
Now we find $r=r_{min}$: 
$$ 
2\mu  Er^2 -2\mu \al r-M^2=0 
$$ 
Since $r>0$, we find 
$$ 
r=\frac{\al}{E}+\sqrt{\frac{\al^2}{4E^2}+\frac{M^2}{2\mu E}} 
=\frac{\al}{E}+\frac{1}{2E} 
\sqrt{\al^2+2M^2E/\mu } 
$$ 
\beqn\la{CA-3.18} 
\varphi_0&:=& 
\arcsin\frac{-\al -M^2/(\mu r)}{\sqrt{\al^2 +2M^2E/\mu }} 
|_{r_{min}}^{\infty}= 
 \arcsin\frac{-\al}{\sqrt{\al^2 +2M^2E/\mu }}-\arcsin(-1)\nonumber\\ 
&=&-\arcsin\frac{\al}{\sqrt{\al^2 +2M^2E/\mu }}+\pi/2. 
\eeqn 
Hence, 
\be\la{CA-3.19} 
\chi=|\pi-2\varphi_0|=2\arcsin\frac{|\al|}{\sqrt{\al^2 +2M^2E/\mu }}. 
\ee 
%%%%%%%%%%%%%%%%%%%%%%%%%%%%%% 
\item[b)] 
Since $E=\mu v_\infty^2/2$, $v_\infty=\sqrt{2E/\mu }$. 
Hence, 
$M=\mu |\bar r\times \bar v|=\mu v_\infty b=\sqrt{2\mu E}b$. 
(\ref{CA-3.19}) implies 
\be\la{CA-3.20} 
\chi=2\arcsin\frac{|\al|}{\sqrt{\al^2 +4E^2b^2}}. 
\ee 
Hence, 
$$ 
\frac{\al^2}{\sqrt{\al^2 +4E^2b^2}}=\sin^2\frac{\chi}{2}. 
$$ 
Hence, 
$$ 
b^2=\frac{\al^2}{4E^2}\left[ 
\frac{1}{\sin^2(\chi/2)}-1\right]= 
\frac{\al^2}{4E^2}\ctg^2\frac{\chi}{2}, 
$$ 
\be\la{CA-3.21} 
b=\frac{\al}{2E}\ctg\frac{\chi}{2}. 
\ee 
At first, note that 
 $dN = 2\pi n b\, db= 
2\pi nb(\chi) \ds\left| \frac{db(\chi)}{d\chi} 
\right| d\chi$ and 
$ 
d\sigma \equiv \ds\frac{dN}{n}= 
2\pi b(\chi) \ds\left| \frac{db(\chi)}{d\chi} 
\right| d\chi$. 
On the other side, 
then (\ref{CA-3.21}) implies 
$$ 
\frac{db}{d\chi}=\frac{\al}{2E}\frac{1}{2} 
\left(-\frac{1}{\sin^2(\chi/2)}\right). 
$$ 
Hence, 
$$ 
d\sigma= 
2\pi  \frac{\al}{2E}\ctg\frac{\chi}{2} 
\frac{\al}{2E}\frac{1}{2}\frac{1}{\sin^2(\chi/2)} d\chi 
=\pi\frac{\al^2}{4E^2}\frac{\cos \chi/2}{\sin^3\chi/2}\,d\chi . 
$$ 
Hence, 
since 
$ 
d\Omega=2\pi\sin\chi\,d\chi= 
4\pi\sin\chi/2\cos\chi/2\,d\chi, 
$ 
we get for the differential scattering cross section 
$$ 
d\sigma= \frac{\al^2}{16E^2}\frac{1}{\sin^4\chi/2}\,d\Omega. 
$$ 
\end{itemize}

%%%%%%%%%%%%%%%%%%%%%%%%%%%%%%%%%%%%%%%%%% 

\subsection{Exercise 7: 
Energy Flow in Maxwell Field} 
 
\begin{itemize} 
\item[] 
Use the Maxwell equations in matter and the fact that 
$\int_\Sigma d^3 xE\cdot \bj $ is the change of kinetic energy per time 
of the charged matter contained in the volume $\Sigma$ to show that 
$u= \frac{1}{8\pi}({\bf E}^2 +{\bf B}^2)$ is the energy density of the 
electromagnetic field, and 
${\bf S}\equiv \frac{c}{4\pi}{\bf E}\times {\bf B}$ is the density of 
energy flow. 
 
\item[] {\bf Solution:} 
Let us recall the Maxwell equations 
\beqn\label{CA-meq2} 
\left\{ 
\ba{ll} 
\dv {\bf E}(t,\x)= 4\pi\rho (t,\x),&\rot {\bf E}(t,\x)= - \ds\fr 1c\dot 
{\bf B}(t,\x),\\ ~\\ 
\dv {\bf B}(t,\x)= 0,&\rot {\bf B}(t,\x)= \ds\fr 1c \,\dot {\bf E}(t ,\x) 
+\ds\fr{4\pi}c 
\,\bj(t,\x), 
\ea 
\right|~~~~~~(t,\x)\in\R^4 , \nonumber 
\eeqn 
where $\rho (t,\x)$ and $\bj(t,\x)$ stand for the charge 
and current density, respectively. 
Subtracting ${\bf B}\cdot \nabla\times {\bf E}$ from ${\bf E}\cdot 
\nabla\times {\bf B}$ we obtain 
\be 
{\bf E}\cdot \rot {\bf B} -{\bf B}\cdot \rot {\bf E} = \frac{1}{c} 
\left( {\bf E}\cdot \dot {\bf E} +{\bf B}\cdot \dot {\bf B} 
\right) +\frac{4\pi}{c}{\bf E}\cdot \bj . 
\ee 
The l.h.s. may be rewritten like 
\be 
{\bf E}\cdot \rot {\bf B} -{\bf B}\cdot \rot {\bf E} = (\nabla \cdot 
{\bf B})\times {\bf E} -(\nabla \cdot {\bf E}) 
\times {\bf B} = -\nabla \cdot ({\bf E}\times {\bf B}), 
\ee 
where the brackets indicate the action of the derivative. 
We arrive at 
\be 
\partial_t \frac{1}{8\pi} ({\bf E}^2 + {\bf B}^2)= - 
{\bf E}\cdot \bj -\nabla \cdot {\bf S}, 
\ee 
where 
\be 
{\bf S}\equiv \frac{c}{4\pi}{\bf E}\times {\bf B} 
\ee 
is called Poynting vector. Integration over a region $\Sigma$ of space 
leads to 
\be \label{CA-int-e-dens} 
\partial_t \frac{1}{8\pi} \int_\Sigma d^3 x({\bf E}^2 + {\bf B}^2)= - 
\int_\Sigma d^3 x{\bf E}\cdot \bj -\int_{\partial \Sigma} da \, {\bf n}\cdot 
 {\bf S}. 
\ee 
Next, we need the fact that $\int_\Sigma d^3 x{\bf E}\cdot \bj$ is the change 
of kinetic energy per time unit of the matter described by $\bj$ in 
the electric field ${\bf E}$, 
\be 
\int_\Sigma d^3 x{\bf E}\cdot \bj =\partial_t {\cal E}_{\rm kin}. 
\ee 
This we will prove below. Now we integrate over the whole space in 
(\ref{CA-int-e-dens}) and use that the fields go to zero for large distances 
to find 
\be 
\partial_t \left( \int d^3 x \frac{1}{8\pi}({\bf E}^2 +{\bf B}^2) 
+{\cal E}_{\rm kin} 
\right) =0, 
\ee 
which is the conservation of the total energy. Therefore, 
\be 
u= \frac{1}{8\pi}({\bf E}^2 +{\bf B}^2) 
\ee 
is the energy density of the electromagnetic field, and the integral 
${\cal E}_{\rm em}=\int_\Sigma d^3 x u$ is its energy in the volume 
$\Sigma$. For finite volume $\Sigma$ we finally find 
\be 
\partial_t ({\cal E}_{\rm em} + {\cal E}_{\rm kin} )= - 
\int_{\partial \Sigma} da \, {\bf n}\cdot 
 {\bf S}, 
\ee 
therefore $\int_{\partial \Sigma} da \, {\bf n}\cdot {\bf S}$ is the 
flow of electromagnetic energy through the surface $\partial \Sigma$, and 
${\bf S}$ is the density of the energy flow. 
 
We still have to prove that $\int_\Sigma d^3 x{\bf E}\cdot \bj$ is the power 
(change of kinetic energy per time unit) of the charged matter in 
the volume $\Sigma$ in the electric field ${\bf E}$. Here we assume that the 
matter is composed of point particles and prove the assumption for a 
single particle, where $\bj = e{\bf v}\delta(x-x(t))$. We find 
\be 
\int_\Sigma d^3 x{\bf E}\cdot \bj = e{\bf v}\cdot {\bf E}(x), 
\ee 
if the particle is contained in the volume $\Sigma$. Further, we have in 
the non-relativistic case 
\be 
{\cal E}_{\rm kin} =\frac{m}{2}{\bf v}^2 \quad \Rightarrow \quad 
\partial_t {\cal E}_{\rm kin} = m{\bf v}\cdot \dot {\bf v} 
\ee 
and the expression for the Lorentz force 
\be 
m\dot {\bf v} =e{\bf E} + \frac{e}{c}{\bf v}\times {\bf B} . 
\ee 
Inserted into the expression for $\partial_t {\cal E}_{\rm kin}$ this 
gives 
\be 
\partial_t {\cal E}_{\rm kin} = e{\bf v}\cdot {\bf E} , 
\ee 
because the second (magnetic) term in the Lorentz force does not contribute. 
This is what we wanted to prove. The proof for relativistic matter particles 
is similar. \bo 
 
\end{itemize}

\subsection{Exercise 8: Electromagnetic Plane Waves} 
\begin{itemize} 
\item[] 
Find the electric and magnetic fields of a plane electromagnetic 
wave by inserting 
the plane wave ansatz $\bE = \bE_0 \exp (i\bk \cdot \x 
-i\omega t) $ into the free Maxwell equations. Specialize to $\bk = k 
\hat e_3$, find the two linearly independent solutions and construct 
a circularly polarized wave as an example. Calculate the corresponding 
energy flow per time unit. 
 
\item[] {\bf Solution:} 
The Maxwell equations in vacuum are 
\begin{displaymath} 
\nabla \times \bB=\frac{1}{c}\dot{\bE} \quad ,\quad \nabla 
\times \bE =-\frac{1}{c}\dot{\bB} 
\end{displaymath} 
\be 
\nabla \cdot \bB=0 \quad ,\quad \nabla \cdot \bE =0 \, . 
\ee 
>From here the wave equation may be derived easily, e.g., 
\begin{displaymath} 
\nabla \times (\nabla \times \bE )= -\frac{1}{c}\nabla \times 
\cdot {\bB} , 
\end{displaymath} 
\begin{displaymath} 
\nabla (\nabla \cdot \bE) -\Delta \bE =-\frac{1}{c^2}\ddot{\bE} , 
\end{displaymath} 
\be 
\Delta \bE =\frac{1}{c^2}\ddot{\bE} , 
\ee 
and, in a similar fashion 
\be 
\Delta \bB =\frac{1}{c^2}\ddot{\bB} \, . 
\ee 
Inserting the plane-wave ansatz 
\be \label{CA-plane-wave} 
\bE =\bE_0 e^{i\bk\cdot \x -i\omega t} 
\ee 
into the wave equation leads to the dispersion relation 
\be 
\bk^2 =\frac{\omega^2}{c^2} \, . 
\ee 
Further, the plane wave ansatz may be verified for the magnetic field 
with the help of the Maxwell equations. From 
\be 
\nabla \times \bE =i\bk\times \bE_0 
e^{i\bk\cdot \x -i\omega t} =-\frac{1}{c}\dot{\bB} 
\ee 
we find upon integration 
\be 
\bB=\frac{c}{\omega}\bk\times \bE +\bB_0 (\x) 
\ee 
where $\bB_0$ is a constant of integration which may depend on $\x$ 
only. Further, we have 
\be 
\nabla \times \bB=\frac{1}{c}\bE , 
\ee 
\be 
-i\frac{c}{\omega}\bk^2 \bE +\nabla \times\bB_0 =-i\frac{\omega}{c} 
\bE , 
\ee 
and, therefore, $\nabla \times \bB_0 =0$, which, together with 
$\nabla \cdot \bB_0=0$, implies $\bB_0 ={\rm const}$. If we assume 
that such a constant magnetic field in all space is absent, then 
\be 
\bB=\frac{\bk}{|\bk|}\times \bE = 
\frac{\bk}{|\bk|}\times \bE_0 e^{i\bk\cdot \x -i\omega t} 
\ee 
follows. Obviously, $|\bE| =|\bB|$ and $\bE \perp \bB$ hold. 
 
Inserting the plane wave ansatz into the Maxwell equations, we get 
\begin{displaymath} 
\bk\times\bB=-\frac{\omega}{c}\bE \quad ,\quad 
\bk\times\bE=\frac{\omega}{c}\bB \, , 
\end{displaymath} 
\be \label{CA-k-max} 
\bk\cdot \bB=0 \quad ,\quad \bk\cdot \bE=0 \, . 
\ee 
Here we allow for a complex notation for the plane wave field for convenience, 
but one should keep in mind that only the real part of the field is physical. 
For a plane wave field (\ref{CA-plane-wave}) with $\bE_0 =\n E_0 
=\n |E_0| e^{i\phi}$ the physical field is 
\be 
\bE_{\rm ph}=\rRe \left( \bE_0 e^{i\bk\cdot\x -i\omega t} \right) 
=\n |E_0| \cos (\bk\cdot \x -\omega t +\phi) 
\ee 
where $\n$ is a real constant unit vector. 
 
Now let us assume without loss of generality 
that $\bk =k\hat e_z$. Then Equation (\ref{CA-k-max}) 
has two linearly independent solutions pointing into the $x$ and $y$ 
directions, and, consequently, the general solution is 
\be 
\bE=E_{0,x}\hat e_x e^{ikz -i\omega t} + 
E_{0,y}\hat e_y e^{ikz -i\omega t} 
\ee 
\be 
\bE_{\rm ph}=|E_{0,x}|\hat e_x \cos (kz -\omega t +\phi_1) + 
|E_{0,y}|\hat e_y \cos (kz -\omega t +\phi_2) \, . 
\ee 
If $\phi_1 =\phi_2$, then the electromagnetic wave is linearly polarized. 
If we assume that $\phi_1 =\phi_2 =0$ and focus on the $x,y$ plane 
$z=0$, then the electric field points into the $|E_{0,x}|\hat e_x + 
|E_{0,y}|\hat e_y$ direction at a time $t=0$, is zero at a time 
$t=(\pi /2\omega)$, points into the minus $|E_{0,x}|\hat e_x + 
|E_{0,y}|\hat e_y$ direction at $t=(\pi /\omega)$, etc. Therefore, the electric 
field vector oscillates along a line parallel to the 
$|E_{0,x}|\hat e_x + |E_{0,y}|\hat e_y$ direction. 
 
Circular polarization we have for $|E_{0,x}| =|E_{0,y}|\equiv E_0$ 
and $\phi_2 -\phi_1 
=(\pi /2)$. For $\phi_1=0$, e.g., we have in the $z=0$ plane an electric 
field vector with length $E_0$ pointing into the $\hat e_x $ direction 
at time $t=0$, a vector with the same length pointing into the $\hat e_y$ 
direction at time $t=(\pi /2\omega )$, etc. Therefore, the field vector 
rotates in a counter-clockwise direction with angular velocity $\omega$ 
without changgng its length. We find for the physical fields in this case 
\be 
\bE_{\rm ph}=E_0 \left( \hat e_x \cos (kz -\omega t) -\hat e_y 
\sin (kz -\omega t)\right) 
\ee 
\be 
\bB=\frac{c}{\omega}\bk\times \bE =\hat e_z \times \bE = 
 E_0 \left( \hat e_y \cos (kz -\omega t) +\hat e_x 
\sin (kz -\omega t)\right) 
\ee 
and, therefore, for the energy flow density 
\be \label{CA-S-circ} 
\bS=\frac{c}{4\pi} \bE\times \bB =\frac{c}{4\pi} 
E_0^2 \hat e_z =\frac{c}{4\pi} 
E_0^2 \frac{\bk}{|\bk|} . 
\ee 
The energy flow density points into the direction of the wave vector $\bk$. 
 
\begin{remark} 
{\rm For general polarizations the energy flow is still time 
dependent. E.g., for linear polarization $\bE =E_0 \n \cos 
(kz -\omega t)$ ($E_0$ real, $\n$ in the $x,y$ plane), the energy flow 
density is 
\be 
\bS =\frac{c}{4\pi} E_0^2 \frac{\bk}{|\bk|} \cos^2 (kz -\omega t). 
\ee 
Here, one is more interested in the average energy flow, where a time 
average for times $t>>\omega^{-1}$ is performed, 
\be 
\lim_{T\to \infty} 
\frac{1}{T}\int_0^T \cos^2 (kz -\omega t)dt =\frac{1}{2}, 
\ee 
which leads to 
\be \label{CA-S-lin} 
\overline{\bS}=\frac{c}{8\pi} E_0^2 \frac{\bk}{|\bk|} 
\ee 
(the factor of $(1/2)$ compared to (\ref{CA-S-circ}) is due to the fact that 
in (\ref{CA-S-lin}) we took into account only one of the two degrees of 
freedom).} 
\end{remark} 
 
\end{itemize} 
 
%%%%%%%%%%%%%%%%%%%%%%%%%%%%%%%%%%%%%%%%%% 
 
\subsection{Exercise 9: Hertzian Dipole Radiation} \label{ue-Ex-9} 
\begin{itemize} 
\item[] 
For a time-dependent dipole, $\rho (t,\x)=-\p(t)\cdot \nabla 
\delta^3 (\x)$, calculate the radiation field, i.e., the electric 
and magnetic fields in the radiation (far) zone. Calculate the 
corresponding energy flux and discuss its direction dependence. Hint: use 
the retarded potentials and the large distance expansion. 
 
\item[] {\bf Solution:} 
With the help of the current conservation equation $\dot \rho+\nabla \cdot \bj 
 =0$, the corresponding dipole current density may be calculated, 
\be 
\bj = \dot{\p} (t) \delta^3 (\x). 
\ee 
The retarded vector potential generated by this dipole is 
\begin{displaymath} 
\bA (\x,t)= \frac{1}{c}\int d^3 x' \int dt' \frac{\bj (x',t')} 
{|\x -\x'|} \delta (t' +\frac{|\x -\x'|}{c} -t) = 
\end{displaymath} 
\be \label{ue-di-A} 
\frac{1}{c} \int dt' \dot{\p} (t') \frac{1}{|\x|} 
\delta (t' +\frac{|\x|}{c} -t) = \frac{1}{c} \frac{\dot{\p} 
(t-(r/c))}{r}, 
\ee 
where $r=|\x|$. The scalar potential may be calculated in an easy 
way by employing the Lorentz gauge condition $\nabla \cdot \bA + (1/c) 
\dot \phi =0$, 
\begin{displaymath} 
\dot \phi =-c\nabla \cdot \bA = -\partial_t \nabla \cdot 
\frac{\p (t-(r/c))}{r}, 
\end{displaymath} 
\begin{displaymath} 
\phi =  -\partial_t \nabla 
\frac{\p (t-(r/c))}{r} = -\frac{1}{r}\nabla \cdot \p (t-(r/c)) 
-e\p \cdot\nabla \frac{1}{r} = 
\end{displaymath} 
\be 
\frac{\x\cdot \dot{\p} (t-(r/c))}{cr^2} + \frac{\x\cdot 
\p}{r^3} \simeq \frac{\x\cdot \dot{\p} (t-(r/c))}{cr^2}, 
\ee 
where the large distance approximation was used in the last step. 
 
Now the magnetic field may be calculated, 
\begin{displaymath} 
\bB = \nabla \times \bA = \frac{1}{c}\nabla \times \frac{\dot{\p} 
(t-(r/c))}{r} = 
\end{displaymath} 
\be 
\frac{1}{c}\dot{\p}(t-(r/c))\times \frac{\x}{r^3} + 
\frac{1}{cr}\nabla (t-\frac{r}{c}) \times \ddot{\p}(t-\frac{r}{c}) 
\simeq \frac{1}{c^2}\frac{\ddot{\p} (t-\frac{r}{c}) \times \x}{r^2}, 
\ee 
and, in a similar fashion, the electric field 
\begin{displaymath} 
\bE = -\frac{1}{c}\dot{\bA} -\nabla \phi \simeq 
-\frac{1}{c^2}\frac{\ddot{\p}}{r} -\nabla \frac{\x\cdot 
\dot{\p}(t-\frac{r}{c})}{cr^2} = 
\end{displaymath} 
\begin{displaymath} 
-\frac{1}{c^2}\frac{\ddot{\p}}{r} -\frac{1}{cr^2}(\dot{\p} + 
x_j \nabla \dot p_j (t-\frac{r}{c})) -\frac{1}{c}\x\cdot \dot{\p} 
\nabla r^{-2} \simeq 
\end{displaymath} 
\be 
-\frac{1}{c^2}\frac{\ddot{\p}}{r} +\frac{\x (\x\cdot \ddot{ 
\p})}{c^2r^3} = \frac{1}{c^2r^3}(\ddot{\p} (t-\frac{r}{c})\times 
\x)\times \x , 
\ee 
where for $\bE$ only the first and third terms in the second line 
contribute in the large distance limit. Observe that $\x\perp\bE\perp 
\bB$, i.e., this is a radiation field. 
 
For the energy flow (Poynting vector) we find 
\begin{displaymath} 
\bS = \frac{c}{4\pi}\bE\times \bB = \frac{1}{4\pi c^3 r^5} 
\left( (\ddot{\p} \cdot \x)\x -\ddot{\p} r^2 \right) 
\times (\ddot{\p}\times \x) = 
\end{displaymath} 
\be 
\frac{1}{4\pi c^3 r^5} \left( r^2 \ddot{\p}^2 -(\ddot{\p}\cdot 
\x)^2\right) \x 
\ee 
If we assume that the dipole is oriented toward the $z$ direction, 
$\p =(0,0,p)$ then we find 
\be \label{ue-hertz} 
\bS=\frac{\sin^2 \theta}{4\pi c^3 r^2}\ddot p^2 \n , 
\ee 
where 
\be 
\n =\frac{\x}{r} \, ,\quad \cos\theta =\frac{\n\cdot\p} 
{|\p|}. 
\ee 
Therefore, $\bS$ behaves like $r^{-2}$ for large distances (which it 
must due to energy conservation), and there is no radiation (energy flow) 
in the direction of the dipole (for $\theta =0,\pi$). 
 
\end{itemize}

\subsection{Exercise 10: Polarizability in an External Electromagnetic 
Wave} \label{ue-Ex-10} 
\begin{itemize} 
\item[] 
In many instances, a classical, point-like 
electron in matter may be described approximately as 
a damped harmonic oscillator which oscillates around a fixed, 
positively charged center. In an external electromagnetic field the 
corresponding equation of motion for the position of the electron is 
$$ 
m(\ddot {\bf x} (t) + \gamma \dot {\bf x} (t) + \omega_0^2  {\bf x}(t))= 
e({\bf E} +\frac{1}{c}\dot {\bf x} \times {\bf B})\, . 
$$ 
Here $m$ is the mass and $e$ the charge of the electron. 
 $\gamma$ is a damping constant and ${\bf x}$ is the distance 
vector from 
the center. For $|{\bf v}|<<c$ the second, magnetic 
part of the Lorentz force may be neglected. Under this 
assumption, c 
alculate 
the electric dipole moment ${\bf p}=e{\bf x}$ of the 
electron in the 
direction of the external field for an 
oscillating field ${\bf E}={\bf E_0}\exp (-i\omega t)$. 
Assume that the 
dipole oscillates with the same frequency, ${\bf p}= 
{\bf p_0}\exp (-i 
\omega t)$. Discuss the frequency dependence of the resulting 
polarizability. 
 
\item[] {\bf Solution:} 
Inserting the dipole moment into the above equation and 
assuming that the 
velocity is sufficiently small (neglecting the magnetic 
part of the Lorentz 
force), we get 
\be 
\ddot{\p} +\gamma \dot{\p} +\omega_0^2 \p =\frac{e^2}{m} 
\bE_0 e^{-i\omega t}. 
\ee 
Further assuming that the distance vector (and the dipole 
moment) oscillates 
with the same frequency like the external electric field 
(driven oscillator) 
leads to 
\be 
(-\omega^2 -i\omega \gamma +\omega_0^2 )\p_0 =\frac{e^2}{m} 
\bE_0 
\ee 
or 
\be 
\p_0 =\frac{e^2}{m}\frac{1}{\omega_0^2 -\omega^2 -i\omega \gamma} 
\bE_0 \equiv \alpha (\omega) \bE_0 
\ee 
with the atomic polarizability $\alpha$. The complex nature of the 
polarizability means that there exists a phase difference between 
$\bE$ and $\p$. This phase difference is 0 for low frequency 
and $\pi$ in the limit of very high frequency. The real and imaginary parts of 
$\alpha$ are 
\be 
{\rm Re} \alpha = \frac{e^2}{m}\frac{\omega_0^2 -\omega^2} 
{(\omega_0^2 -\omega^2)^2 +\omega^2 \gamma^2 } 
\ee 
\be 
{\rm Im} \alpha = \frac{e^2}{m}\frac{\gamma\omega} 
{(\omega_0^2 -\omega^2)^2 +\omega^2 \gamma^2 }  . 
\ee 
This shows that there is anomalous dispersion (weaker refraction for light 
with higher frequency) for values of $\omega$ near the eigenfrequency 
$\omega_0$. 
 
If there are $n$ atoms per volume and $f_k$ denotes the fraction of 
electrons per atom that oscillate with eigenfrequency $\omega_k$, then 
the total polarization per volume is 
\be 
\p =\frac{ne^2}{m} \sum_k 
\frac{f_k}{\omega_k^2 -\omega^2 -i\omega \gamma} 
\bE , 
\ee 
and, with $\p=\chi_e \bE$, $\epsilon =1+4\pi \chi_e$, we find for 
the electric permittivity $\epsilon$ 
\be \label{ue-perm} 
\epsilon =1+ \frac{4\pi ne^2}{m} \sum_k 
\frac{f_k}{\omega_k^2 -\omega^2 -i\omega \gamma}, 
\ee 
which is the formula of Drude for the permittivity. 
For many materials $\mu \sim 1$ and $n\sim \sqrt{\epsilon}$. For frequencies 
near the eigenfrequencies anomalous dispersion occurs. 
 
\end{itemize} 
 
%%%%%%%%%%%%%%%%%%%%%%%%%%%%%%%%%%%%%%%%%% 
 
\subsection{Exercise 11: 
Fresnel's Formulae for Reflection and Refraction} 
\begin{itemize} 
\item[] 
Two homogeneous,  sotropic transparent media (with constant, real 
permittivities $\epsilon_1 ,\epsilon_2$ and permeabilities $\mu_1, \mu_2$) 
are separated by a plane. 
Use the Maxwell equations in macroscopic matter, the resulting boundary 
conditions, and the plane wave ansatz 
to derive the laws of reflection and refraction. 
 
\item[] {\bf Solution:} 
The Maxwell equations in macroscopic matter, but without further microscopic 
charge and current distributions, are 
\be \label{CA-meq-d} 
\nabla \cdot \bD =0 
\ee 
\be \label{CA-meq-b} 
\nabla \cdot \bB=0 
\ee 
\be \label{CA-meq-e} 
 \nabla \times \bE =-\frac{1}{c}\dot{\bB} 
\ee 
\be \label{CA-meq-h} 
\nabla \times \bH =\frac{1}{c}\dot{\bD} 
\ee 
where 
\be \label{CA-lin-rel} 
\bB =\mu \bH \quad ,\quad \bD=\epsilon \bE . 
\ee 
Now we assume that we have a medium 1 with 
permittivity $\epsilon_1$ and permeability $\mu_1$ in the region $z<0$, 
and a medium 2 with 
permittivity $\epsilon_2$ and permeability $\mu_2$ in the region $z>0$, 
so the separating plane is the $(x,y)$ plane $z=0$. Then the following 
continuity conditions for the fields infinitesimally below and above the 
separating plane follow from the Maxwell equations: 
\be \label{CA-cont-d} 
\bD^{(2)}_\perp =\bD^{(1)}_\perp 
\ee 
\be \label{CA-cont-b} 
\bB^{(2)}_\perp =\bB^{(1)}_\perp 
\ee 
\be \label{CA-cont-e} 
\bE^{(2)}_{||} =\bE^{(1)}_{||} 
\ee 
\be \label{CA-cont-h} 
\bH^{(2)}_{||} =\bH^{(1)}_{||}. 
\ee 
Here, e.g., $\bD^{(2)}_\perp$ is the component perpendicular to the 
separating plane of the $\bD$ field infinitesimally above the plane 
(i.e., in region 2), etc. 
 
To prove Eq. (\ref{CA-cont-d}), we integrate the Maxwell equation 
(\ref{CA-meq-d}) over the volume 
of a cylinder, such that the bottom and top of the cylinder are 
parallel discs 
of the same radius infinitesimally below and above the plane $z=0$, and the 
cylinder mantle is perpendicular to the plane and has infinitesimal height. 
We get 
\begin{displaymath} 
0=\int d^3 x \nabla \cdot \bD =\left( \int_{\rm bottom} + 
\int_{\rm mantle} +\int_{\rm top}\right) d\f\cdot \bD 
\end{displaymath} 
\be 
=\int_{\rm disc}df (\n_2 + \n_1 )\cdot \bD , 
\ee 
where we used the facts that the mantle area is infinitesimal and that 
the two discs have the same area. The two discs have opposite orientation, 
therefore $\n_2 =-\n_1$, 
\be 
0=\int_{\rm disc}d\f \cdot (\bD_2 - \bD_1 )   , 
\ee 
and Eq. (\ref{CA-cont-d}) follows 
($n_i$ are perpendicular to the plane $z=0$). 
The proof of Eq. (\ref{CA-cont-b}) is analogous. 
 
To prove Eq. (\ref{CA-cont-e}), we integrate Eq. 
(\ref{CA-meq-e}) over a rectangle with corners at 
$(x_1, y_1, -\delta)$, $(x_1,y_1,\delta)$, $(x_2,y_2,-\delta)$, 
and $(x_2,y_2,\delta)$. 
Here $\delta$ is infinitesimal, whereas the distance $l=[(x_2 -x_1)^2 + 
(y_2 -y_1)^2]^{1/2}$ is finite. We find 
\be 
\int_{\rm rect} df \n \cdot \nabla \times \bE = 
-\frac{1}{c} \int_{\rm rect} df \n \cdot \nabla \times \dot{\bB} 
\ee 
where $\n$ is the unit vector perpendicular to the rectangle. 
Here the r.h.s. may be neglected, because the area of the rectangle 
is infinitesimal, 
\be 
\lim_{\delta \to 0} 
\int_{\rm rect} df \n \cdot \nabla \times \dot{\bB} =0. 
\ee 
The l.h.s. is then 
\be 
0= \int_{\rm rect} df \n \cdot \nabla \times \bE = 
\int d{\bf l} \cdot (\bE^{(2)} -\bE^{(1)}), 
\ee 
where $d{\bf l}$ is the line element along the upper finite side of the 
rectangle, and we neglected the line integrals along the infinitesimal 
sides. As the orientation of the finite sides of the rectangle in 
the $(x,y)$ plane is arbitrary, Eq. (\ref{CA-cont-e}) follows. 
The proof of eq. (\ref{CA-cont-b}) is 
analogous. 
 
Within a given medium (i.e., with fixed $\epsilon$ and $\mu$), the 
Maxwell equations may be re-written like 
\be 
\nabla \cdot \bE =0 
\ee 
\be 
\nabla \cdot \bB =0 
\ee 
\be 
 \nabla \times \bE =-\frac{1}{c}\dot{\bB} 
\ee 
\be 
\nabla \times \bB =\frac{n}{c}\dot{\bE} 
\ee 
where $n$ stands for the index of refraction:
\be\la{nemu}
n\equiv \sqrt{\epsilon \mu}.
\ee

 As for the 
Maxwell equations in vacuum it follows that $\bE$ (as well as 
 $\bB$, $\bD $ and $\bH$) 
obeys the wave equation 
\be 
\Delta \bE -\frac{n^2}{c^2}\ddot{\bE} =0 
\ee 
with propagation velocity $v=c/n \le c$. Therefore, we may use the plane 
wave ansatz for the electromagnetic field in a region with fixed index of 
refraction, 
\be 
\bE = \bE_0 e^{i\bk\x -i\omega t} 
\ee 
where the dispersion relation is $|\bk| =(n/c)\omega$. 
 
Now we want to study reflection and refraction. We assume that an incident 
plane wave in region 1 ($z<0$) propagates toward the separating plane 
$z=0$ (i.e., $k_z >0$), and we want to determine a 
further plane wave in region 1 with 
$k_z^{''}<0$ (``reflected wave'') and a plane wave in region 2 (``refracted 
wave'') from the boundary conditions (\ref{CA-cont-d})--(\ref{CA-cont-b}). 
Therefore we make the ansatz 
\beqn 
\bE = \bE_0 e^{i\bk\x -i\omega t}\, , 
& \bk\cdot \bE_0 =0\,  , & z<0 \\ 
\bE^{'} = \bE_0^{'} e^{i\bk^{'}\x -i\omega^{'} t}\, , 
& \bk^{'} \cdot \bE_0^{'} =0\, , &z>0 \\ 
\bE^{''} = \bE_0^{''} e^{i\bk^{''}\x -i\omega^{''} t}\, , 
& \bk^{''} \cdot \bE_0^{''} =0\, , & z<0 
\eeqn 
where 
$|\bk|=(n_1 /c)\omega$, $|\bk^{'}|=(n_2 /c)\omega^{'}$ and 
$|\bk^{''}|=(n_1 /c)\omega^{''}$. The other fields may be computed 
easily from the Maxwell equation (\ref{CA-meq-b}) 
\be 
\bB=\frac{c}{\omega} \bk\times \bE , 
\ee 
\be 
\bB^{'}=\frac{c}{\omega^{'}} \bk^{'}\times \bE^{'} , 
\ee 
\be 
\bB^{''}=\frac{c}{\omega^ {''}} \bk^{''}\times \bE^{''} , 
\ee 
and from relations (\ref{CA-lin-rel}). 
 
The total solution in region 1 is now $\bE^{(1)}=\bE +\bE^{''}$, 
and the solution in region 2 is $\bE^{(2))}= \bE^{'}$, and analogously 
for the other fields.  Now we may use the continuity conditions 
(\ref{CA-cont-d})--(\ref{CA-cont-b}). 
As they must hold everywhere along the $z=0$ plane and for all times, the 
phase functions multiplying the constant vectors must be the same, i.e., 
\be 
\omega =\omega^{'}=  \omega^{''} , 
\ee 
\be 
k_x =k_x^{'}=k_x^{''} \quad ,\quad k_y =k_y^{'}=k_y^{''} . 
\ee 
Therefore, all three wave vectors must lie in the same plane (``plane of 
incidence''), and we may choose without loss of generality 
$k_y =k_y^{'}=k_y^{''}=0$ (incidence plane $=(x,z)$ plane). 
The dispersion relations are now 
\be 
|\bk| =|\bk^{''}| =(n_1 /c)\omega \quad ,\quad 
|\bk^{'}| =(n_2 /c)\omega 
\ee 
and imply 
\be 
k^{''}_z =-k_z . 
\ee 
Defining the angles between the wave vectors and the z axis, 
\be 
\cos\alpha = \frac{k_z}{|\bk|}\, ,\quad 
\cos\alpha^{'} = \frac{k_z^{'}}{|\bk^{'}|}\, ,\quad 
\cos\alpha^{''} = -\frac{k_z^{''}}{|\bk|} \, , 
\ee 
we have therefore $\alpha = \alpha^{''}$, i.e., the incidence angle 
equals the reflection angle. We further have 
\be 
\frac{\sin\alpha}{\sin\alpha^{'}}=\frac{k_x /|\bk|}{k_x^{'}/|\bk^{'}|} 
=\frac{n_2}{n_1} , 
\ee 
which is the law of Snellius. 
 
With the help of the unit vector $\hat e_z$ (which is perpendicular to the 
plane $z=0$) the continuity conditions (\ref{CA-cont-d})--(\ref{CA-cont-b}) 
may be expressed like 
follows: for the tangential component of $\bE$ 
\be \label{CA-ft-cont-e} 
 (\bE_0 +\bE_0^{''} -\bE_0^{'})\times \hat e_z =0 , 
\ee 
normal component of $\bB$ 
\be \label{CA-ft-cont-b} 
(\bk \times \bE_0 +\bk^{''}\times \bE_0^{''} 
-\bk^{'}\times \bE_0^{'})\cdot \hat e_z =0 , 
\ee 
tangential component of $\bH=(1/\mu)\bB$ 
\be \label{CA-ft-cont-h} 
\left( \frac{1}{\mu_1}(\bk \times \bE_0 +\bk^{''}\times \bE_0^{''} 
) - \frac{1}{\mu_2} \bk^{'}\times \bE_0^{'} \right) \times \hat e_z=0 , 
\ee 
normal component of $\bD=\epsilon \bE$ 
\be \label{CA-ft-cont-d} 
\left( \epsilon_1 (\bE_0 +\bE_0^{''} )- \epsilon_2 \bE_0^{'} 
\right) \cdot \hat e_z =0 . 
\ee 
For a further investigation we specialize to two linearly independent, 
linearly polarized incident waves. This is sufficient, because a general 
incident electromagnetic plane wave may always be expressed as a linear 
combination of these two. The first case is a vector $\bE_0$ 
perpendicular to the incidence plane, i.e., $\bE_0 =E_0 \hat e_y$. 
First, we find that $\bE_0^{'}$ and $\bE_0^{''}$ also have only 
components perpendicular to the incidence plane, $\bE_0^{'}=E_0^{'} 
\hat e_y$, $\bE_0^{''}=E_0^{''} \hat e_y$. For assume that 
$\bE_i^{'}=E_i^{'}\hat k^{'}\times \hat e_y$ (and analogous for 
$\bE^{''}$) are nonzero. Then we find from (\ref{CA-ft-cont-e}) 
\be 
\left( E_i^{''} \hat k^{''}\times \hat e_y - 
E_i^{'} \hat k^{'}\times \hat e_y \right) \times \hat e_z =0 , 
\ee 
\be 
-E_i^{''} \cos \alpha - E_i^{'}\cos \alpha^{'} =0 \quad \Rightarrow \quad 
E_i^{''}=-\frac{\cos \alpha^{'}}{\cos\alpha}E_i^{'} . 
\ee 
$\cos\alpha$ and $\cos\alpha^{'}$ are greater than zero by definition, 
therefore the two amplitudes have different sign. On the other hand, we find 
from (\ref{CA-ft-cont-b}) 
\be 
\left( \frac{E_i^{''}}{\mu_1} \bk^{''}\times (\hat k^{''}\times \hat e_y) 
-  \frac{E_i^{'}}{\mu_2} \bk^{'}\times (\hat k^{'}\times \hat e_y) 
\right) \times \hat e_z =0 , 
\ee 
\be 
\left( -\frac{E_i^{''}}{\mu_1} |\bk^{''}| 
+  \frac{E_i^{'}}{\mu_2} |\bk^{'}| 
\right) \hat e_y\times \hat e_z =0 , 
\ee 
\be 
-\frac{E_i^{''}n_1}{\mu_1} 
+  \frac{E_i^{'}n_2}{\mu_2}=0 \quad \Rightarrow \quad E_i^{''}= 
\frac{n_2\mu_1}{n_1\mu_2}E_i^{'} , 
\ee 
which implies that the two amplitudes must have the same sign. This requires 
that both amplitudes are zero. 
 
For the perpendicular components we find from (\ref{CA-ft-cont-e}) 
\be \label{CA-perp-sol1} 
 ( E_0 + E_0^{''} - E_0^{'})\hat e_y \times \hat e_z =0 \quad \Rightarrow \quad 
E_0 + E_0^{''} - E_0^{'} =0 , 
\ee 
and from (\ref{CA-ft-cont-h}) 
\begin{displaymath} 
\left( \frac{1}{\mu_1}(E_0 \bk \times \hat e_y + 
E_0^{''} \bk^{''} \times \hat e_y )- \frac{1}{\mu_2} 
E_0^{'} \bk^{'} \times \hat e_y \right) \times \hat e_z =0 , 
\end{displaymath} 
\begin{displaymath} 
\left( \frac{1}{\mu_1}(E_0 \bk \cdot \hat e_z + 
E_0^{''} \bk^{''} \cdot \hat e_z )- \frac{1}{\mu_2} 
E_0^{'} \bk^{'} \cdot \hat e_z \right)  \hat e_y =0 , 
\end{displaymath} 
\be \label{CA-perp-sol2} 
\frac{n_1}{\mu_1}\cos \alpha (E_0 -E_0^{''})- \frac{n_2}{\mu_2}\cos \alpha^{'} 
E_0^{'} =0 . 
\ee 
Eq. (\ref{CA-ft-cont-b}) gives the same result like Eq. (\ref{CA-ft-cont-e}), 
and Eq. (\ref{CA-ft-cont-d}) is trivially zero. 
Inserting (\ref{CA-perp-sol1}) into (\ref{CA-perp-sol2}) leads to 
\begin{displaymath} 
\sqrt{\frac{\epsilon_1}{\mu_1}}\cos \alpha (E_0 -E_0^{''}) = 
\sqrt{\frac{\epsilon_2}{\mu_2}}\cos \alpha^{'} (E_0 +E_0^{''}) , 
\end{displaymath} 
\begin{displaymath} 
\frac{E_0^{''}}{E_0} = \frac{\sqrt{\epsilon_1 /\mu_1}\cos\alpha - 
\sqrt{\epsilon_2 /\mu_2}\cos\alpha^{'}}{ 
\sqrt{\epsilon_1 /\mu_1}\cos\alpha + 
\sqrt{\epsilon_2 /\mu_2}\cos\alpha^{'}} = 
\end{displaymath} 
\be 
\frac{1 - \sqrt{\frac{\epsilon_2\mu_1}{\epsilon_1\mu_2}}\frac{\cos\alpha^{'} 
}{\cos\alpha}} 
{1 + \sqrt{\frac{\epsilon_2\mu_1}{\epsilon_1\mu_2}}\frac{\cos\alpha^{'} 
}{\cos\alpha}} = 
\frac{1 - \frac{\mu_1}{\mu_2}\frac{\tan\alpha}{\tan\alpha^{'}}} 
{1 + \frac{\mu_1}{\mu_2}\frac{\tan\alpha}{\tan\alpha^{'}}} . 
\ee 
Further, 
\be 
\frac{E_0^{'}}{E_0}=1+\frac{E_0^{''}}{E_0}=\frac{2} 
{1 + \sqrt{\frac{\epsilon_2\mu_1}{\epsilon_1\mu_2}}\frac{\cos\alpha^{'} 
}{\cos\alpha}} = \frac{2} 
{1 + \frac{\mu_1}{\mu_2}\frac{\tan\alpha}{\tan\alpha^{'}}} . 
\ee 
For $\mu_1 =\mu_2$ (which is true for many substances) we find 
\be 
\frac{E_0^{''}}{E_0} = \frac{1 -\frac{\sin\alpha \cos\alpha'}{\cos\alpha 
\sin\alpha'}} 
{1 +\frac{\sin\alpha \cos\alpha'}{\cos\alpha 
\sin\alpha'}} = 
\frac{\sin (\alpha -\alpha')}{\sin (\alpha +\alpha')} , 
\ee 
\be 
\frac{E_0^{'}}{E_0} = \frac{2} 
{1 +\frac{\sin\alpha \cos\alpha'}{\cos\alpha 
\sin\alpha'}} = 
\frac{2\cos\alpha \sin \alpha'}{\sin (\alpha +\alpha')} , 
\ee 
which are the Fresnel formulae. 
 
The second case to study is for $\bE_0$ parallel to the incidence plane, 
$\bE_0 =E_0 \hat k \times \hat e_y$. Again, it may be proven easily that 
both $\bE_0^{'}$ and  $\bE_0^{''}$ must be parallel to the incidence 
plane, as well, in this case,  $\bE_0^{'}=  E_0^{'}\hat k^{'}\times \hat 
e_y$ and $\bE_0^{''}=  E_0^{'}\hat k^{''}\times \hat e_y$. 
We find from (\ref{CA-ft-cont-e}) 
\begin{displaymath} 
\left( E_0 \hat k\times \hat e_y + E_0^{''} \hat k^{''} \times \hat e_y 
- E_0^{'} \hat k^{'} \times \hat e_y \right) \times \hat e_z =0 , 
\end{displaymath} 
\begin{displaymath} 
\left( E_0 \hat k\cdot \hat e_z + E_0^{''} \hat k^{''} \cdot \hat e_z 
- E_0^{'} \hat k^{'} \cdot \hat e_z \right)  \hat e_y =0 , 
\end{displaymath} 
\be 
(E_0 - E_0^{''}) \cos\alpha - E_0^{'}\cos\alpha' =0 , 
\ee 
and from (\ref{CA-ft-cont-h}) 
\begin{displaymath} 
\left( \frac{1}{\mu_1}[ E_0 \bk\times (\hat k\times \hat e_y) + 
E_0^{''} \bk^{''}\times (\hat k^{''} \times \hat e_y) ] - 
\frac{1}{\mu_2}E_0^{'} \bk^{'}\times (\hat k^{'} \times \hat e_y) 
\right) \times \hat e_z =0 , 
\end{displaymath} 
\begin{displaymath} 
\left( \frac{1}{\mu_1}[ E_0 |\bk| + 
E_0^{''} |\bk^{''}| ] - 
\frac{1}{\mu_2}E_0^{'} |\bk^{'}| 
\right) \hat e_y \times \hat e_z =0 , 
\end{displaymath} 
\be 
\frac{n_1}{\mu_1}(E_0 + E_0^{''}) -  \frac{n_2}{\mu_2} E_0^{'} =0 
\quad \rightarrow \quad 
E_0^{'}=\sqrt{\frac{\epsilon_1\mu_2}{\epsilon_2\mu_1}}(E_0 +E_0^{''}) 
\ee 
(again (\ref{CA-ft-cont-b}) and (\ref{CA-ft-cont-d}) 
do not contain additional information). 
Inserting this into the above equation leads to 
\begin{displaymath} 
(E_0 -E_0^{''})\cos\alpha = 
\sqrt{\frac{\epsilon_1\mu_2}{\epsilon_2\mu_1}}(E_0 +E_0^{''}) \cos\alpha' , 
\end{displaymath} 
\begin{displaymath} 
\frac{E_0^{''}}{E_0} = \frac{\cos\alpha - 
\sqrt{\epsilon_1\mu_2 /\epsilon_2\mu_1}\cos\alpha^{'}}{ 
\cos\alpha + 
\sqrt{\epsilon_1\mu_2 /\epsilon_2\mu_1}\cos\alpha^{'}} = 
\end{displaymath} 
\be 
\frac{1 - \sqrt{\frac{\epsilon_1\mu_2}{\epsilon_2\mu_1}}\frac{\cos\alpha^{'} 
}{\cos\alpha}} 
{1 + \sqrt{\frac{\epsilon_1\mu_2}{\epsilon_2\mu_1}}\frac{\cos\alpha^{'} 
}{\cos\alpha}} = 
\frac{1 - \frac{\epsilon_1}{\epsilon_2}\frac{\tan\alpha}{\tan\alpha^{'}}} 
{1 + \frac{\epsilon_1}{\epsilon_2}\frac{\tan\alpha}{\tan\alpha^{'}}} . 
\ee 
Further, 
\be 
\frac{E_0^{'}}{E_0}= 
\sqrt{\frac{\epsilon_1\mu_2}{\epsilon_2\mu_1}}(1 +\frac{E_0^{''}}{E_0}) 
= \frac{2 \sqrt{\frac{\epsilon_1\mu_2}{\epsilon_2\mu_1}}}{ 
1 + \frac{\epsilon_1}{\epsilon_2}\frac{\tan\alpha}{\tan\alpha^{'}}} . 
\ee 
For $\mu_1 =\mu_2$ these formulae simplify to 
\be \label{CA-fres-3} 
\frac{E_0^{''}}{E_0} = 
\frac{1 - \frac{n_1^2}{n_2^2}\frac{\tan\alpha}{\tan\alpha^{'}}} 
{1 + \frac{n_1^2}{n_2^2}\frac{\tan\alpha}{\tan\alpha^{'}}} 
= \frac{1-\frac{\sin\alpha' \cos\alpha'}{\sin\alpha \cos \alpha}} 
{1-\frac{\sin\alpha' \cos\alpha'}{\sin\alpha \cos \alpha}} 
=\frac{\sin\alpha \cos\alpha -\sin\alpha' \cos\alpha'} 
{\sin\alpha \cos\alpha +\sin\alpha' \cos\alpha'} , 
\ee 
\be 
\frac{E_0^{'}}{E_0}= 
\frac{2 \frac{n_1}{n_2}}{ 
1 + \frac{n_1^2}{n_2^2}\frac{\tan\alpha}{\tan\alpha^{'}}} = 
\frac{2\frac{\sin\alpha'}{\sin\alpha}}{1+ \frac{\sin\alpha' 
\cos\alpha'}{\sin\alpha \cos\alpha}} = 
\frac{2\sin\alpha' \cos\alpha}{\sin\alpha \cos\alpha + \sin\alpha' 
\cos\alpha'} , 
\ee 
which are the Fresnel formulae for the case of electric field parallel to 
the incidence plane. 
 
From these results some physical applications immediately follow. 
For $n_2<n_1$ total reflection for $\alpha >\alpha_{\rm max}$ 
follows from the Snellius law, where 
$\sin \alpha_{\rm max}=n_2 /n_1$. 
 
Another interesting consequence is the possibility to generate 
linearly polarized light from unpolarized light by choosing the 
so-called Brewster angle for the incident light beam. Indeed, 
it follows from (\ref{CA-fres-3}) that the component parallel to the 
incidence plane 
is not reflected at all if the condition 
\be 
\sin\alpha \cos\alpha =\sin\alpha' \cos \alpha' 
\ee 
holds. The ``solution'' $\alpha =\alpha'$ is of course forbidden if 
$n_2 \ne n_1$. But there exists the possible solution $\alpha' =\pi/2 - 
\alpha$ which leads to $\sin \alpha' =\cos\alpha$, $\cos\alpha' =\sin\alpha$ 
and 
\be 
\frac{n_2}{n_1}=\frac{\sin \alpha}{\sin\alpha'} = \frac{\sin\alpha}{\cos\alpha} 
=\tan\alpha . 
\ee 
Therefore, if an unpolarized light beam is incident with an angle 
$\alpha_{\rm B}$ such that $\tan \alpha_{\rm B} =n_2 /n_1$, then the 
reflected beam will contain only light linearly polarized into the 
direction perpendicular to the incidence plane, $\bE_0^{''} 
\sim \hat e_y$. 
 
\end{itemize} 
 
%%%%%%%%%%%%%%%%%%%%%%%%%%%%%%%%%%%%%%%%%% 
 
\subsection{Exercise 12: Classical Zeemann Effect } \label{ue-Ex-12} 
\begin{itemize} 
\item[] 
Again electrons in matter are described by spherical harmonic 
oscillators which oscillate around a fixed center (nucleus), 
as in Exercise \re{ue-Ex-10}. 
Here we assume that the electrons are excited with their 
eigenfrequency by some unspecified mechanism (e.g. heat or radiation), 
therefore we ignore a possible damping. These harmonic oscillators 
are now exposed to a constant, external magnetic field, therefore the 
equation of motion for the position of one electron is 
\be 
m(\ddot{\x}(t) +\omega_0 \x (t))=\frac{e}{c}\dot{\x}\times \bB . 
\ee 
Find the three eigenfrequencies and eigenmodes of this equation of motion. 
(Hint: use the symmetry of the problem and the known behavior of a 
charged particle in a constant magnetic field to guess the right ansatz.) 
Use the resulting dipole moments $\p =e\x$ to calculate the 
dipole radiation of all three modes and their polarization patterns 
both parallel and perpendicular to the constant, external magnetic field.

\item[] {\bf Solution:} 
We assume without loss of generality that the constant, external magnetic 
field points into the positive $z$ direction, $\bB = B\hat e_3$, $B>0$. 
Then we have the following equations of motion, 
\beqn 
\ddot x_1 +\omega_0^2 x_1 &=& \frac{eB}{mc}\dot x_2 \label{CA-x-eq} , \\ 
\ddot x_2 +\omega_0^2 x_2 &=& -\frac{eB}{mc}\dot x_1 \label{CA-y-eq} , \\ 
\ddot x_3 +\omega_0^2 x_3 &=& 0. 
\eeqn 
Obviously, the third, decoupled equation is solved by $x_3 = x_0\sin 
(\omega_0 t -\phi_0)$ where $\phi_0$ is an irrelevant integration 
constant and is set to zero in the sequel. Therefore the first 
eigenmode is 
\be 
\x_{||} =x_0 \sin (\omega_0 t) \hat e_3 
\ee 
with eigenfrequency $\omega_0$ (the eigenfrequency is not altered by the 
magnetic field). Now we use the following facts about a charge moving in 
a constant magnetic field: a charge moving parallel to the magnetic 
field feels no force at all, whereas a charge moving perpendicular to 
the magnetic field is deflected such that it moves along a circle 
with constant angular velocity. 
Further, the free two-dimensional harmonic oscillator composed of Eqs. 
(\ref{CA-x-eq}) and (\ref{CA-y-eq}) has circles with constant angular 
velocity as possible eigenmodes, 
therefore we try a circle with constant angular velocity as an ansatz: 
\be 
x_1 = x_0 \cos\omega t \quad ,\quad \dot x_1 =-x_0 \omega \sin \omega t 
\quad ,\quad \ddot x_1 =-x_0 \omega^2 \cos\omega t , 
\ee 
\be 
x_2 = x_0 \sin\omega t \quad ,\quad \dot x_2 =x_0 \omega \cos \omega t 
\quad ,\quad \ddot x_2 =-x_0 \omega^2 \sin\omega t . 
\ee 
We find from (\ref{CA-x-eq}) 
\be 
-x_0 \omega^2 \cos\omega t +\omega_0^2 x_0 \cos \omega t =\frac{eB}{mc} 
x_0 \omega \cos\omega t 
\ee 
and an analogous equation with $\cos \omega t\rightarrow \sin \omega t$ 
from (\ref{CA-y-eq}). Therefore the ansatz is a solution provided that 
\be 
-\omega^2 +\omega_0^2 =\frac{eB}{mc}\omega 
\ee 
\be 
\Rightarrow \quad \omega_\pm =-\frac{eB}{2mc}\pm \sqrt{\omega_0^2 + 
\frac{e^2B^2}{4m^2c^2}} \simeq -\frac{eB}{2mc}\pm \omega_0 , 
\ee 
where in the last step we assumed that the eigenfrequency $\omega_0$ is 
much larger in absolute value than the Larmor frequency 
$\om_\cL =-\frac{eB}{2mc}$. The minus 
sign for $\omega_-$ means that this eigenmode rotates in a clockwise 
direction, and the minus sign of $\om_\cL$ is the law of induction. 
Therefore, we find the following two perpendicular eigenmodes 
\be 
\x_1 =x_0 \cos (\omega_1 t)\hat e_1 +x_0 \sin (\omega_1 t) \hat e_2 
\ee 
\be 
\x_2 =x_0 \cos (\omega_2 t)\hat e_1 -x_0 \sin (\omega_2 t) \hat e_2 
\ee 
where $\omega_1 =\omega_+$ and $\omega_2 =-\omega_-$. 
 
Next, we want to calculate the dipole radiation emitted by these 
three eigenmodes. Here we use the formulae for Hertzian dipole radiation 
(see Exercise \re{ue-Ex-9}), and $\p =e\x$, 
\be 
\bB (t,\x)=\frac{e}{c^2r^2}\ddot{\p} (t-\frac{r}{c})\times \x , 
\ee 
\be 
\bE (t,\x)=\frac{e}{c^2r^3}[-r^2 \ddot{\p}(t-\frac{r}{c})+ 
(\x\cdot \ddot{\p}(t-\frac{r}{c}))\x ] , 
\ee 
\be 
\bS =\frac{e^2}{4\pi c^3 r^5}[r^2 \ddot{\p}^2 -(\x\cdot \ddot{ 
\p})^2 ]\x . 
\ee 
For the parallel eigenmode, $\p_{||}=ex_0 \sin (\omega_0 t)\hat e_3$, 
we find linearly polarized radiation with polarization direction parallel 
to the external $\bB$ field (the radiation $\bB$ field is always 
perpendicular to the $\hat e_3$ direction). The linear polarization is 
extracted most easily from the Poynting vector 
\be 
\bS=\frac{e^4\omega_0^4x_0^2}{4\pi c^3 r^5}(r^2 -z^2)\sin^2 (\omega_0 t') 
\x = \frac{e^4\omega_0^4x_0^2}{4\pi c^3 r^5} \sin^2 \theta \sin^2 
(\omega_0 t')\n , 
\ee 
where we used spherical polar coordinates, $\n \equiv  (\x/r)$, and 
$t'\equiv t-(r/c)$. For linear polarization at each point in space in 
the radiation zone the 
electric (and magnetic) field vector oscillates and vanishes for certain 
times (instead of rotating along an ellipse or a circle), therefore 
also the energy flow vanishes at certain times. 
Further , there is no radiation in  the $\hat e_3$ direction. 
 
For the eigenmodes perpendicular to the constant, external magnetic field 
the Poynting vector is (where $\omega =\omega_\pm$) 
\begin{displaymath} 
\bS= \frac{e^4\omega^4x_0^2}{4\pi c^3 r^5} [r^2 -(x\cos (\omega t') + 
y \sin (\omega t'))^2]\x = 
\end{displaymath} 
\be 
\frac{e^4\omega^4x_0^2}{4\pi c^3 r^2}[1-\sin^2 \theta \cos^2 
(\omega t' -\varphi) ] \n . 
\ee 
In the $(x,y)$ plane perpendicular to $\hat e_3$ we get (with $\n_\perp 
=\cos\varphi \hat e_1 + \sin\varphi \hat e_2$) 
\be 
\bS =\frac{e^4\omega^4x_0^2}{4\pi c^3 r^2}\sin^2 (\omega t' -\varphi) 
\n_\perp 
\ee 
and, therefore, linearly polarized light. Further, the polarization 
direction is perpendicular to the $\hat e_3$ direction (because the 
magnetic radiation field is parallel to $\hat e_3$). 
For the radiation in $\hat e_3$ direction we get 
\be 
\bS=\frac{e^4\omega^4x_0^2}{4\pi c^3 r^2}\hat e_3 
\ee 
and the energy flow does not depend on time. Therefore, the radiation is 
circularly polarized (the radiation field vectors rotate without changing 
their length). Further, the $+$ mode has left circular polarization 
(the $\bE$ field rotates counter-clockwise) and the $-$ mode 
has right circular polarization.

\end{itemize} 
 
\subsection{Exercise 13: Diamagnetism and Paramagnetism } \label{ue-Ex-13} 
\begin{itemize} 
\item[] 
The relation between the magnetic induction $\bB$, the magnetic field 
intensity $\bH$ and the magnetization 
${\bf M}$ of matter is (for isotropic matter) 
$$ 
\bB=\bH +4\pi {\bf M} \, ,\quad \bB =\mu \bH  \, ,\quad 
{\bf M}=\chi_m \bH \, ,\quad \mu =1+4\pi \chi_m . 
$$ 
Here a substance is called diamagnetic if $\mu <1$. Use again the 
harmonic oscillator model in a constant  external magnetic field (like 
in Exercise \re{ue-Ex-12}) to derive the diamagnetic behavior. 
Here the electrons 
moving on circular orbits have to be interpreted as currents which 
induce a magnetic field. Assume that the net induced magnetic field 
without external magnetic field is zero. 
 
\item[] {\bf Solution:} 
We again assume that the external magnetic field is along the $\hat e_3$ 
direction, $\bB =B\hat e_3$. Further, we are interested only in the 
d.o.f. perpendicular to the external magnetic field. We know from Exercise 
\re{ue-Ex-12} 
that there are two degrees of freedom, $+,-$, with frequencies 
\be 
\omega_\pm =-\frac{eB}{2mc}\pm \omega_0 . 
\ee 
Without external field these are just two modes rotating with the same 
angular velocity $\omega_0$, one (the + mode) in the counter-clockwise 
direction, the other in the clockwise direction. 
%%We now assume e.g., that 
%%50\% of the relevant electrons move counter-clockwise and the other 
%%50\% clockwise, then the net contribution to the induction is zero. 
The two modes induce magnetic moments of equal strengths but 
opposite orientations, therefore, macroscopically, their net 
contribution to the magnetization is zero. 
With external field, however, the additional contribution is in the 
clockwise (negative) direction in both cases, and a net contribution 
remains. Further, the electron current and induced magnetic field depend 
linearly on $\omega$, therefore we may just omit the $\omega_0$ piece 
and use just $\om_\cL =-\frac{eB}{2mc}$. 
The magnetic moment of a current distribution is 
\be 
{\bf m} =\frac{1}{2c}\int d^3 x \x\times \bj , 
\ee 
where $\bj$ is the current density. In our case the electron is in a 
circular orbit in the $(x,y)$ plane, therefore 
\be 
\bj =e \delta (\x -\x(t)) \dot{\x} 
\ee 
(with $x_3 =0$).  With 
\be 
\x \times \dot{\x} = (x_1 \dot x_2 +x_2 \dot x_1 )\hat e_3 
=x_0^2 \om_\cL \hat e_3 
\ee 
we get 
\be 
{\bf m} = \frac{1}{2c} e\om_\cL x_0^2 \hat e_3 =-\frac{e^2 Bx_0^2}{4mc^2} 
\hat e_3 , 
\ee 
and, therefore, the magnetic moment is opposite to the external 
magnetic field. For general orbits of the electrons, $x_0^2$ must be 
replaced by the orbit average $\overline{x_1^2 +x_2^2}$; if the 
force law for the electron is spherically symmetric, as in our case, 
then $\overline{x_1^2 +x_2^2}=(2/3)\overline{\x^2}$. If there 
are several electrons per atom with different average orbit radii 
and if there are $n$ atoms per volume, then the magnetization 
is 
\be 
{\bf M} = -n\frac{e^2 B}{6mc^2}\sum_j \overline{\x_j^2} 
\hat e_3 = -n\frac{e^2}{6mc^2}\sum_j \overline{\x_j^2} 
\bB 
\ee 
and the susceptibility is approximately (if it is small) 
\be 
\chi_m =  -n\frac{e^2 }{6mc^2}\sum_j \overline{\x_j^2} \, . 
\ee 
The assumptions here were rather general (electrons orbiting around 
nuclei) and, 
in fact, all materials are diamagnetic, but the diamagnetism 
may be over-compensated by 
other effects with positive magnetic susceptibility (like 
paramagnetism). 
 
Paramagnetism is present when the following conditions hold: 
1) the atoms or molecules of the material 
already have a fixed magnetic moment 
${\bf m}_0$ (even without external magnetic field), 
and 2) the orientations of these magnetic moments are randomly 
distributed 
in the absence of an external magnetic field. Without external 
magnetic 
field there is, therefore, no macroscopic magnetization of the 
paramagnetic material. However, in an external magnetic field the 
magnetic moments are partially aligned along the direction of 
the external 
field, because this aligned position is energetically favorable. This 
alignment is thwarted by thermal fluctuations, and the average 
magnetic moment per molecule is 
\be \label{CA-para} 
\overline{{\bf m}}=\eta \frac{|{\bf m}_0|^2}{kT}\bB_{\rm ext} , 
\ee 
where $\eta$ is some constant depending on the molecule type, 
$T$ is the temperature and $k$ the Boltzmann constant. Both 
quantum mechanical 
and thermodynamic considerations are needed to derive equation 
(\ref{CA-para}), which is beyond the scope of the present discussion. 
 
\end{itemize} 
 
%%%%%%%%%%%%%%%%%%%%%%%%%%%%%%%%%%%%%%%%%% 
 
\subsection{Exercise 14: Stark Effect} 
\begin{itemize} 
\item[] 
A hydrogen atom is exposed to a constant electric field. Use 
perturbation 
theory to calculate the energy shifts for the ground state 
$\psi_{100}$ 
of the hydrogen atom 
(with no energy degeneracy of the unperturbed system) and for the 
states $\psi_{2lm}$ (with a four-fold energy degeneracy). 
 
\item[] {\bf Solution:} 
The Hamiltonian for the electron wave function in a Coulomb field and 
in a constant electric field is 
\be 
H=-\frac{\hbar^2}{2\mu}\Delta -\frac{e^2}{r}-e\bE\cdot \x , 
\ee 
and we assume $\bE =f\hat e_3$. The eigenfunctions $\psi_{nlm} 
=R_{nl}(r)F_l^m(\theta) e^{im\varphi}$ 
of the hydrogen atom are no longer eigenfunctions of the 
full Hamiltonian, therefore perturbation theory in the perturbing term 
$V=-efz=-efr\cos\theta$ is needed. 
 
For the ground state, no degeneracy of the unperturbed system occurs, 
because there is only one ground state $\psi_{100}$ with energy 
$E_{100}=-R$. In this case, the perturbation series for the energy is 
\be 
E=E_k^0 +V_{kk} -\sum_{i\ne k}\frac{|V_{ki}|^2}{E_i^0 -E_k^0}+\cdots 
\ee 
where $E_i^0$ are the eigenenergies of the unperturbed Hamiltonian 
$H_0 =H-V$, $H_0 \psi_i = E_i^0 \psi_i$, and the perturbation about the 
level $k$ is calculated. Further, 
\be 
V_{ki} =\int d^3 x \overline{\psi_k}(\x) V \psi_i (\x) 
\ee 
is the matrix element of the perturbation $V$ w.r.t. the eigenstates 
of the unperturbed system. 
In our case $\psi_k =\psi_{100}$, 
\be 
V_{ki} =-ef\int_0^\infty dr r^3 \overline{R_{10}}(r)R_{nl}(r)\int 
d\Omega \cos\theta \overline{F_0^0}(\theta) F_l^m(\theta ) 
e^{im\varphi} , 
\ee 
which is non-zero only for $m=0$, $l=1$. Therefore, $V_{kk}$ is zero, 
and there is no first order (linear in the external field) contribution. 
Up to second order in the perturbation, we get 
\be 
E= -R -e^2 f^2 \co , 
\ee 
which is quadratic in the external field (quadratic Stark effect). 
For the electric dipole moment we find 
\be 
\p =e\langle \x\rangle =-\frac{\partial}{\partial \bE} 
\langle V\rangle = -\frac{\partial}{\partial \bE}E = 
\co \, e^2 \bE . 
\ee 
The interpretation of this result is as follows: the spherically symmetric 
ground state $\psi_{100}$ has no permanent electric dipole moment. 
But the external electric field induces a dipole moment proportional to 
its strength by polarizing the atom (displacing the charge center of 
the electron wave function relative to the nucleus). 
 
For higher states (e.g., $n=2$) degeneracy occurs because 
all states with 
quantum number $n$ have energy $-(R/n^2)$. E.g. for $n=2$ 
there are four states 
\be 
\psi_1 \equiv \psi_{200}\, ,\quad \psi_2 \equiv \psi_{210} 
\, ,\quad 
\psi_3 \equiv \psi_{211}\, ,\quad \psi_4 \equiv \psi_{21-1} 
\, ,\quad 
\ee 
with unperturbed energy $E_2^0 =-(R/4)$. When degeneracy 
occurs, the 
energies within the degenerate subspace have to be 
determined exactly 
from the degeneracy condition 
\be 
{\rm det} \left( (E^0_2 -E)\delta_{\alpha\beta}+ 
V_{\beta\alpha} \right) =0, 
\ee 
where $\alpha$ labels the degeneracy subspace. 
Here 
\be 
V_{\beta\alpha}= -ef\int_0^\infty dr r^3 
\overline{R_{2l_\beta}}(r) 
R_{2l_\alpha}(r)\int 
d\Omega \cos\theta \overline{F_{l_\beta}^{ m_\beta}} 
(\theta) 
F_{l_\alpha}^{m_\alpha}(\theta ) 
e^{i(m_\alpha -m_\beta )\varphi} . 
\ee 
It follows that 
\be 
m_\alpha =m_\beta =0 \, ,\quad l_\beta =l_\alpha \pm 1 
\ee 
and, therefore, only $V_{12}$ and $V_{21}$ are non-zero. 
They may be calculated 
and one finds 
\be 
V_{12}=V_{21} =\ldots =3efa_0 , 
\ee 
where $a_0$ is the Bohr radius. The determinant condition is 
\be 
{\rm det} \left( \ba{llll} E_2^0 -E & V_{12} & 0&0 \\ 
V_{12} & E_2^0 -E &0 &0 \\ 0&0&E_2^0 -E &0 \\ 
0&0&0&E_2^0 -E \ea \right) = 0 , 
\ee 
\be 
(E_2^0 -E)^2 [(E_2^0 -E)^2 -V_{12}^2 ]=0 , 
\ee 
with the solutions 
\be 
E=E_2^0 
\ee 
(two-fold degenerate) and 
\be 
E\equiv E_\pm =E_2^0 \pm V_{12} . 
\ee 
This expression is linear in the external field 
(linear Stark effect). 
The electric dipole moment for the two non-degenerate 
modes $E_\pm$ 
is 
\be 
\p =\frac{\partial}{\partial \bE} E_\pm =\pm 3ea_0 \hat e_3 . 
\ee 
The interpretation is as follows: The spherically 
non-symmetric wave functions 
give rise to a permanent electric dipole moment.

\end{itemize}

\subsection{Exercise 15: Vector Model for Spin-Orbital Interaction} 
A phenomenological 
{\it vector model} has been used in the {\it old quantum theory} 
for {\it an addition of the magnetic moments} in many-electron 
atoms and molecules. It is not rigorous and contradicts the 
Pauli equation, however, sometimes it gives 
exact results. 
 
\subsubsection{Precession of angular momentum} 
Let us consider the orbital angular momentum 
  $\bL$. 
\bt 
In the uniform magnetic field $\bB=(0,0,B)$ the components 
$\bL_3(t)$ and $\s_3(t)$ 
are conserved, while the vectors $(\bL_1(t), \bL_2(t))\in\R^2$ 
and $(\s_1(t), \s_2(t))\in\R^2$ 
rotate with angular velocity 
$\om_\cL$ 
and $2\om_\cL$, respectively, 
where $\om_\cL=-\ds\fr {eB}{2\mu c}$ is the Larmor frequency. 
\et 
\Pr 
For concreteness we consider the orbital momentum. 
The conservation of $\bL_3$ is already proved. 
Since $\hat\bL_k=-\h\bH_k$, we have by 
(\re{com}), 
\be\la{comL} 
[\hat\bL_1,\hat\bL_2]=i\h\hat\bL_3,~~~~~~~[\hat\bL_2,\hat\bL_3] 
=i\h\hat\bL_1,~~~~~~~[\hat\bL_3,\hat\bL_1]=i\h\hat\bL_2. 
\ee 
Therefore, 
\be\la{comLt} 
[\cP,\hat\bL_1]=-i\om_\cL\h\hat\bL_2,~~~~~~~ 
[\cP,\hat\bL_2]=i\om_\cL\h\hat\bL_1. 
\ee 
since all $\hat\bL_k$ commute with all $\hat\s_j$ by 
Remark \re{rcom}. 
Hence, analogously to the Heisenberg equation (\re{He}), 
we get 
\beqn\la{M1bd} 
\dot \bL_1(t)&=&\langle\Psi(t),i\h^{-1}[\cP\hat\bL_1-\hat\bL_1 
\cP]\Psi(t)\rangle 
\nonumber\\ 
&=&\langle\Psi(t),\om_\cL\hat\bL_2\Psi(t)\rangle=\om_\cL\bL_2(t). 
\eeqn 
Similarly, we have $\dot \bL_2(t)=-\om_\cL\bL_1(t)$, hence 
\beqn\la{M2bd} 
~~~~~~~~~~~~~~~~~~~~~~~~\fr d{dt} (\bL_1(t)+i\bL_2(t)) 
=-i\om_\cL(\bL_1(t)+i\bL_2(t)).~~~~~~ 
~~~~~~~~~~~~~~~~~~~~~~~~~~~~~~~~~~~~~\bo 
\eeqn 
The theorem implies the {\it precession} of the vectors $\bL(t)$ and 
$\s(t)$ 
with the angular velocities $\om_\cL$ and $2\om_\cL$, respectively, 
around the magnetic field $\bB$. 
\br 
{\rm A similar precession can be proved for a classical 
system of electrons rotating as rigid bodies in the 
uniform magnetic field \ci{Jack}.} 
\er 
 
 \subsubsection{Vector model} 
Let us apply the idea of the precession to the addition 
of the orbital and 
spin magnetic moments. This addition explains 
the Einstein-de Haas and anomalous Zeemann experiments. 
 
The model explains the magnetization  in the 
Einstein-de Haas experiment by the classical mechanism of the 
reorientation in a magnetic field of a total magnetic moment 
$\m$ which exists 
even in the absence of the magnetic field. 
 $\m$  is the sum 
of the  orbital and spin magnetic moments 
$\m_o:=\ds\fr e{2\mu c}\bL$ and $\m_s:=\ds\fr e{\mu c}\s$: 
\be\la{mmm} 
\m=\m_o+\m_s=\ds\fr e{2\mu c}\bL+\ds\fr e{\mu c}\s . 
\ee 
The moments 
are defined by the 
orbital and spin 
 angular momenta 
$\bL$ and $\s$, respectively,  with the 
corresponding distinct gyromagnetic ratios 
$\ds\fr e{2\mu c}$ and $\ds\fr e{\mu c}$. 
 
In the absence of an external magnetic field, 
the total angular momentum $\bJ=\bL+\s$ is conserved, while 
$\bL$ and $\s$ are generally not conserved. For example, 
the spin angular momentum $\s$ 
precesses in the magnetic field 
generated by the orbital angular momentum. 
Similarly, the orbital  angular momentum $\bL$ 
precesses in the magnetic field 
generated by the spin angular momentum. 
Therefore, the lengths of the vectors $\bL$ and $\s$ 
are conserved. Hence,  the conservation of the sum $\bJ=\bL+\s$ 
implies that 
the vectors $\bL$ and $\s$ rotate around 
 $\bJ$. 
Then the  total magnetic moment 
$\m=\ds\fr e{2\mu c}\bL+\ds\fr e{\mu c}\s$ also rotates around 
$\bJ$. 
Since the angular velocity of the rotation is very high, we 
have to take into account only the effective value of the 
magnetic moment, $\m_{\rm eff}$,  which is the projection of 
the  total magnetic moment 
onto $\bJ$. Let us calculate this projection. 
 
The angle $\al$ between the vectors  $\bJ$ and $\bL$ 
is conserved as well as 
the angle $\beta$ between the vectors  $\bJ$ and $\s$, and 
\be\la{alb} 
\cos\al=\fr{|\bJ|^2+|\bL|^2-|\s|^2}{2|\bJ||\bL|}, 
~~~~~~~~~~~~~~~~~~\cos\beta=\fr{|\bJ|^2+|\s|^2-|\bL|^2}{2|\bJ||\s|}. 
\ee 
Then the projection equals 
\be\la{pron} 
\m_{\rm eff}=\ds\fr e{2\mu c}|\bL|\cos\al+\ds\fr e{\mu c}|\s|\cos\beta. 
\ee 
A final, highly nontrivial approximation is as follows: we 
redefine the lengths of the vectors $\bJ$,  $\bL$ and $\s$ as 
\be\la{red} 
~~~~~~~|\bJ|^2\!:=\!\langle\Psi,(\hat\bJ_1^2\! 
+\!\hat\bJ_2^2\!+\!\hat\bJ_3^2)\Psi\rangle,~~ 
|\bL|^2\!:=\!\langle\Psi,(\hat\bL_1^2\!+\!\hat\bL_2^2\! 
+\!\hat\bL_3^2)\Psi\rangle,~~~ 
|\s|^2\!:=\!\langle\Psi,(\hat\s_1^2\!+\!\hat\s_2^2\!+\!\hat\s_3^2)\Psi\rangle. 
\ee 
The operators $\hat\bJ_1^2\!+\!\hat\bJ_2^2\!+\!\hat\bJ_3^2$, 
$\hat\bL_1^2\!+\!\hat\bL_2^2\!+\!\hat\bL_3^2$, and 
$\hat\s_1^2\!+\!\hat\s_2^2\!+\!\hat\s_3^2=3/4$ 
commute. Hence, the quantum stationary states can be classified by the 
eigenvalues of the operators which are equal 
to $J(J+1)$, $L(L+1)$, 
and $3/4$, 
where $J,L=0,1,2,...$ . For the states we have $|\bJ|^2=J(J+1)$, 
$|\bL|^2=L(L+1)$, and $|\s|^2=3/4$. Substituting this into (\re{pron}), we 
obtain {\it the Land\'e formula} (\re{Laf}) 
for the effective gyromagnetic ratio 
\beqn\la{pgr} 
g_{\rm eff}:=\ds\fr{\m_{\rm eff}}{|\bJ|\ds\fr e{2\mu c}}&=& 
\fr{|\bJ|^2+|\bL|^2-|\s|^2}{2|\bJ|^2} 
+2\fr{|\bJ|^2+|\s|^2-|\bL|^2}{2|\bJ|^2} 
\nonumber\\ 
\nonumber\\ 
&=&\fr 32+ 
\fr{|\s|^2-|\bL|^2}{2|\bJ|^2}=\fr 32+ 
\fr{3/4-L(L+1)}{2J(J+1)}. 
\eeqn 
The formula is confirmed experimentally 
by the Einstein-de Haas and anomalous Zeemann 
effects.

%%%%%%%%%%%%%%%%%%%%%%%%%%%%% 
%%%%%%%%%%%%%%%%%%%%%%%%%%%%% 


\newpage 
 
\begin{thebibliography}{99} 
 \bibitem{Ab} 
M.Abraham, Theorie der Elektrizit\"at, 
Bd.2: Elektromagnetische Theorie der Strahlung, Teubner, 
Leipzig, 1905. 
 
\bibitem{A} 
 V.Arnold, Mathematical Methods of Classical Mechanics, 
 Springer, Berlin, 1978. 
 

\bibitem{BV} 
 A.V.Babin, M.I.Vishik, 
Attrac\-tors of Evo\-lu\-tio\-na\-ry  Equa\-tions, North-Holland, 
 Amsterdam, 1992. 
 



\bibitem{Bek} 
R.Becker, Elektronentheorie, Teubner, Leipzig, 1933. 
 
 
 
\bibitem{K10}   A.Bensoussan, C.Iliine, A.Komech, 
Breathers for a relativistic  nonlinear wave 
     equation, {\em Arch.  Rat. Mech. Anal.} {\bf 165} (2002), 
317-345. 
 
 
\bibitem{BL} H.Berestycki, P.L.Lions, 
{\em  Arch. Rat. Mech. and Anal. }  {\bf 82} (1983), 
no.4, 313-375. 
 
 
\bibitem{BSh}  F.A.Berezin, M.A.Shubin, 
The Schr\"odinger equation, 
 Kluwer Academic Publishers, Dordrecht, 
1991. 
 
\bibitem{Be} 
H.Bethe, Intermediate Quantum Mechanics, Benjamin, NY, 1964. 
 
\bibitem{BeS} 
H.Bethe, E.Salpeter, 
Quantum Mechanics of One- and Two-Electron Atoms, 
Berlin, Springer, 1957. 
 
 
\bibitem{BD} 
J.D.Bjorken, S.D.Drell, 
Relativistic Quantum Mechanics, McGraw-Hill, NY, 1964;
Relativistic Quantum Fields, McGraw-Hill, NY, 1965.



\bibitem{Bo} 
N.Bohr, 
{\em Phil. Mag.}  {\bf 26} (1913) 1; 476; 857. 
 
\bibitem{Born}  M.Born, 
Atomic Physics, 
Blackie \& Son,  London-Glasgow, 1951. 
 
\bibitem{Bour} 
N.Bournaveas, 
Local existence for the Maxwell-Dirac equations in three space dimensions, 
{\em Commun. Partial Differ. Equations} {\bf 21 } (1996), 
no.5-6, 693-720. 
 
 
 
 \bibitem{Br} 
L. de Broglie, 
The Current Interpretation of Wave Mechanics, 
Elsevier, NY, 1964. 
 
 
 \bibitem{BP} 
V.S.Buslaev, G.S.Perelman, 
On the stability of solitary waves for nonlinear Schr\"odinger 
equations, {\em Amer. Math. Soc. Trans.} (2) {\bf 164} (1995), 
75-98. 
 
 
 
\bibitem{BS} 
V.S.Buslaev, C.Sulem, 
 On asymptotic stability of solitary waves for nonlinear 
Schr\"odinger equations, 
{\em  Ann. Inst. Henri Poincaré, Anal. Non Linéaire} 
{\bf 20} (2003), no.3, 
419-475. 
 
 
 
\bibitem{CS} 
E.U.Condon, E.U.,  G.H.Shortley, 
The Theory of Atomic Spectra, 
Cambridge University Press, Cambridge, 1963. 
 
\bibitem{Cu} 
S.Cuccagna, 
Stabilization of solutions to nonlinear Schr\"odinger equations, 
{\em  Comm. Pure Appl. Math.}  {\bf 54} (2001), 
 no.9, 
1110-1145. 
 
 \bibitem{Cukin}
S.Cuccagna,
On asymptotic stability in 3D of kinks for the $\phi^4$
model, preprint 2004.

\bibitem{DG} 
 C.Davisson, L.Germer, 
{\em Nature} {\bf 119} (1927), 558. 
 
\bibitem{DB} 
M.Defranceschi, C.Le Bris, 
Mathematical Models and Methods for ab Initio Quantum Chemistry, 
Lecture Notes in Chemistry. 74, Springer, Berlin, 2000. 
 
 
 
\bibitem{DKKS} 
T.Dudnikova, A.Komech, E.A.Kopylova, Yu.M.Suhov, 
On convergence to equilibrium distribution, I. 
Klein-Gordon equation with mixing, 
{\em Comm.  Math. Phys.} {\bf 225} (2002), no.1, 1-32. 
 
\bibitem{DKMdir} 
T.Dudnikova, A.Komech, N.Mauser, 
On the convergence to a statistical equilibrium 
for the Dirac equation, 
{\em Russian Journal of Math. Phys.} 
{\bf 10} (2003), 
no.4, 399-410. 
 
\bibitem{DKM2hc} 
T.Dudnikova, A.Komech, N.Mauser, 
On two-temperature problem for harmonic crystals, 
{\em Journal of Statistical Physics} 
{\bf 114} (2004), no.3/4, 1035-1083. 
 
 
 
 
\bibitem{DKRS} 
T.Dudnikova, A.Komech, N.E.Ratanov, Yu.M.Suhov, 
On convergence to equilibrium distribution, II.  Wave 
equation with mixing, 
{\em Journal of Statistical Physics} {\bf 108} (2002), no.4, 
1219-1253. 
 
\bibitem{DKSj} 
T.Dudnikova, A.Komech, H.Spohn, 
 On a two-temperature problem for wave  equation with mixing, 
{\em  Markov Processes and Related Fields} {\bf 8} (2002), 
no.1, 43-80. 
 
\bibitem{DKSc} 
T.Dudnikova, A.Komech, H.Spohn, 
On convergence  to statistical equilibrium 
for harmonic crystals, 
  {\em Journal of Mathematical Physics} 
 {\bf 44} (2003), no.6, 2596-2620. 
 
 
 
 
 
 
\bibitem{edks} 
T.Dudnikova, A.Komech, H.Spohn, 
Energy-momentum relation for solitary waves 
of relativistic wave equation, 
{\em Russian Journal Math. Phys.} {\bf 9} (2002), no.2, 153-160. 
 
 
 
 
 
 
\bibitem{Eid} 
D.M.Eidus, 
The principle of limit amplitude, 
{\it Russ. Math. Surv.} {\bf 24} No.3, 97-167 (1969). 
 
 
\bibitem{Ei} A.Einstein, 
\"Uber einen die Erzeugung und Verwandlung des Lichtes betreffenden 
heuristischen Gesichtspunkt 
(Concerning the generation and transformation of light as seen 
 from a heuristic point of view), Annalen der Physik, March 18, 1905. 
 
 
 
\bibitem{EGS} 
M.Esteban, V.Georgiev, E.Sere, 
 Stationary solutions of the Maxwell-Dirac 
and the Klein-Gordon-Dirac equations, 
{\em Calc. Var. Partial Differ. Equ.}  {\bf  4} (1996), no.3, 
265-281. 
 
 
 
 
\bibitem{F} M. Fedoriuk, 
Asymptotic Methods for Partial Differential Equations, 
Encyclopaedia of Mathematical Sciences Vol. 34, 
Springer, Berlin, 1999. 
\bibitem{FeQED} R.Feynman, 
Quantum Electrodynamics, Addison-Wesley, Reading, Massachusetts, 1998. 
 
\bibitem{Fr} 
Ya.Frenkel, {\em Zs. Phys.} {\bf 37} (1926), 43. 
 
 
\bibitem{FGJS} 
J.Fr\"ohlich, S.Gustafson, B.L.G.Jonsson, I.M.Sigal, 
Solitary wave dynamics in an external potential, 
arXiv:math-ph/0309053v1. 
 
\bibitem{FTY} 
J.Fr\"ohlich, T.P.Tsai, H.T.Yau, 
On the point-particle (Newtonian) limit of the nonlinear Hartree 
equation, {\em Comm. Math. Physics} {\bf 225} (2002), no.2, 223-274. 
 
 
 
 
\bibitem{GF}  I.Gelfand, S.Fomin, 
Calculus of Variations, Dover Publications, 
Mineola, NY,  2000. 
 
\bibitem{GMS} I.M.Gelfand, R.A.Minlos, Z.Ya.Shapiro, 
Representations of the Rotation and Lorentz Groups 
and their Applications,  Pergamon Press, Oxford, 1963. 
 
 
\bibitem{Gell} 
M.Gell-Mann, 
{\em Phys. Rev.}  {\bf 125} (1962), 1067. 
 
 
 
\bibitem{GSS} M.Grillakis, J.Shatah, W.A.Strauss, Stability theory of 
solitary 
waves in the presence of symmetry, I; II. 
{\em J. Func. Anal.} {\bf 74} (1987), 
no.1, 160-197; {\bf 94} (1990), no.2, 308-348. 
 
\bibitem{GS} 
S.J.Gustafson, I.M.Sigal, 
Mathematical Concepts of Quantum Mechanics, 
Springer, Berlin, 2003. 
 
\bibitem{GNS} 
Y.Guo, K.Nakamitsu, W.Strauss, 
Global finite-energy solutions of the Maxwell-Schr\"odinger system, 
{\em  Comm. Math. Phys.}  {\bf 170}  (1995),  no. 1, 181--196. 
 
 
\bibitem{Hale} 
J.Hale, 
Asymptotic Behavior of Dissipative Systems, 
 AMS, Providence, 1988. 
 

 

  
 
 
\bibitem{Han} 
K.Hannabus, 
 An Introduction to Quantum Theory, 
Clarendon Press, Oxford, 1997. 
 
\bibitem{Heis}
W.Heisenberg,
Der derzeitige Stand der nichtlinearen Spinortheorie der
Elementarteilchen, {\em  Acta Phys. Austriaca} {\bf 14} (1961),
328-339.




\bibitem{Heitler} 
W.Heitler, 
The Quantum Theory of Radiation, 
Oxford University Press, Oxford, 1954. 
 
\bibitem{Hen} 
D.Henry, 
Geometric Theory of Semilinear Parabolic 
Equations, Lecture Notes in Mathematics, 840, 
Springer, Berlin, 1981. 

\bibitem{HS} 
P.D.Hislop,  I.M.Sigal, 
Introduction to Spectral Theory. 
With applications to Schr\"odinger operators, 
Springer, NY, 1996. 
 
\bibitem{K11} V.Imaikin, A.Komech, P.Markowich, 
Scattering of solitons of the Klein-Gordon equation 
coupled to a classical particle, 
{\em Journal of Mathematical Physics} 
{\bf 44} (2003), no.3, 1202-1217. 
 
\bibitem{K9} V.Imaikin, A.Komech, H.Spohn, 
Soliton-like asymptotics and scattering for a particle coupled to 
 Maxwell field, 
 {\em Russian Journal of Mathematical Physics} {\bf 9} (2002), 
no.4, 428-436. 
 
 
\bibitem{K12} V.Imaikin, A.Komech, H.Spohn, 
Scattering theory for a particle coupled to a scalar field, 
{\em Journal of Discrete and 
Continuous Dynamical Systems} {\bf 10} (2003), 
no.1$\&$2, 387-396. 
 
\bibitem{IZ} 
 C.Itzykson, J.B.Zuber, Quantum Field Theory,
McGraw-Hill, NY, 1980.



\bibitem{Jack} 
Jackson, 
Classical Electrodynamics, 
 3rd ed., 
John Wiley \& Sons, 
New York, 1999. 
 
 
\bibitem{JP} V.Jaksic, C.-A.Pillet, 
Ergodic properties of classical dissipative systems. I. 
{\em  Acta Math.} {\bf 181} (1998), no.2, 245-282. 
 
 
\bibitem{Jam} 
M.Jammer, 
The Conceptual 
Development of Quantum Mechanics, McGraw-Hill, NY, 1966. 
 
\bibitem{JKV}   P.Joly,  A.Komech, O.Vacus, 
On transitions to stationary states in a 
Maxwell-Landau-Lifschitz-Gilbert system, 
{\em SIAM J. Math. Anal.} {\bf 31} (1999), no.2, 346-374. 
 

 
\bibitem{KJ} 
T.Kato, A.Jensen, 
Spectral properties of Schr\"odinger operators 
 and time-decay of the wave functions, {\em Duke Math.~J.} {\bf 46} 
(1979), 
 583-611. 
 
 
\bibitem{Ka} 
W.Kauffmann, 
{\em Ann. Physik} {\bf 61} (1897), 544. 
 
 
 
 
 
 
 
 
\bibitem{K1}  A.Komech, 
On stabilization of string-nonlinear oscillator interaction, 
 {\em J. Math. Anal. Appl.} {\bf 196} (1995), 384-409. 
 
\bibitem{Kpla}    A.Komech, 
On transitions to stationary states in Hamiltonian 
 nonlinear  wave equations, 
{\em Phys. Letters A} {\bf 241} (1998), 311-322. 
 
\bibitem{K3}   A.Komech, 
On transitions to stationary states in 
one-dimensional nonlinear wave equations, 
{\em Arch. Rat. Mech. Anal.} {\bf 149} (1999), 
no.3, 213-228. 
 
\bibitem{K4}   A.Komech, 
Attractors of nonlinear Hamiltonian 
one-dimensional  wave equations, 
{\em Russ. Math. Surv.} {\bf 55} (2000), 
no.1, 43-92. 
 
\bibitem{Kis3} A.Komech, 
On attractor of a singular nonlinear 
$U(1)$-invariant Klein-Gordon equation, 
p. 599-611 in: 
Proceedings of the $3^rd$ ISAAC Congress, 
Freie Universit\"at Berlin, Berlin, 2003. 
 
 \bibitem{Kanwe} A.Komech, 
 On Global Attractors of Hamilton Nonlinear Wave Equations,
preprint 2005.
 
 
 
\bibitem{K5}    A.Komech, M.Kunze, H.Spohn, 
Effective Dynamics for a mechanical particle 
coupled to a wave field, 
{\em Comm. Math. Phys.} 
{\bf 203} (1999), 1-19. 
 
 
 
\bibitem{K7}   A.Komech, H.Spohn, 
Soliton-like asymptotics for a classical 
 particle interacting with a scalar wave field, 
{\em Nonlinear Analysis} 
 {\bf 33} (1998), no.1, 13-24. 
 
\bibitem{sp4}   A.Komech, H.Spohn, 
Long-time asymptotics for the coupled Maxwell-Lorentz equations, 
{\em Comm. Partial Diff. Equs.} {\bf 25} (2000), 
no.3/4, 558-585. 
 
\bibitem{K6}   A.Komech, H.Spohn, M.Kunze, 
Long-time asymptotics for a classical 
 particle interacting with a scalar wave field, 
 {\em Comm. Partial Diff. Equs.} {\bf 22} (1997), no.1/2, 
307-335. 
 
\bibitem{KVin}  
A.Komech, N.Mauser, A.Vinnichenko, 
On attraction to solitons in relativistic  
nonlinear wave equations, 
to appear in {\em Russian J. Math. Phys.}



 
\bibitem{Lan} A.Land\'e, 
{\em Zs. f. Phys.} {\bf 5} (1921), 5. 
 
 
\bibitem{LVM} 
V.G.Levich, Yu.A.Vdovin, V.A.Mamlin, Course of Theoretical 
Physics, v. II, Nauka, Moscow, 1971. [Russian] 
 
 
 
 
\bibitem{Lieb} 
E.H.Lieb, 
The Stability of Matter: from Atoms to Stars, 
Springer, Berlin, 2001. 
 
 
\bibitem{Lor} 
H.A.Lorentz, 
The Theory of Electrons and its Applications to the Phenomena of Light 
and Radiant Heat. Lectures from a course held at Columbia 
University, New York, NY, USA, March and April 1906, 
Repr. of the 2nd ed., 
Sceaux: Éditions Jacques Gabay, 1992. 
 
 

 
 
 
\bibitem{JvN} 
J. von Neumann, 
Mathematical Foundations of Quantum Mechanics, 
Princeton University Press 
Princeton, 1955. 
 
 
 
\bibitem{New} 
R.Newton, 
Quantum Physics, 
Springer, NY, 2002. 
 
 
\bibitem{EN} E.Noether, 
Gesammelte Abhandlungen. Collected papers, 
Springer, Berlin,  1983. 
 
\bibitem{NMPZ} 
 S.P.Novikov, S.V.Manakov, L.P.Pitaevskii, V.E.Zakharov, 
     Theory of Solitons: The Inverse Scattering Method, 
Consultants Bureau [Plenum], New York, 1984. 
 
 
 
 
 
\bibitem{Per} 
J.Perrin, New experiments on the cathode rays, 
{\em  Nature}, {\bf 53} (1896), 298-299. 
 
 
\bibitem{PW} C.-A.Pillet, C.E.Wayne, 
Invariant manifolds for a class of dispersive, 
Hamiltonian, partial differential equations, 
{\em J. Differ. Equations} {\bf 141} (1997), No.2, 310-326. 
 
\bibitem{PSW}
R.Pyke, A.Soffer, M.I.Weinstein,
Asymptotic stability of the kink soliton of the $\vp^4$
nonlinear equation, preprint, 2003.


\bibitem{RS} 
M.Reed, B.Simon, 
Methods of Modern Mathematical Physics, 
Academic Press, NY, I (1980), II (1975), III (1979), 
IV (1978). 



\bibitem{Sak}
J.J.Sakurai, Advanced Quantum Mechanics,
Addison-Wesley, Reading, Massachusets, 1967.

 
\bibitem{Scharf} 
G.Scharf, Finite Quantum Electrodynamics. The Causal Approach, 
Springer, Berlin, 1995. 
 
 
\bibitem{Schiff} 
L.Schiff, 
Quantum Mechanics, McGraw-Hill, NY, 1955. 
 
\bibitem{Sch} 
E.Schr\"odinger, 
Quantisierung als Eigenwertproblem, 
{\em Ann. d. Phys.} 
I, II {\bf 79} (1926) 361, 489; 
III {\bf 80} (1926) 437; 
IV. {\bf 81} (1926) 109. 
 
 
 
 
 
 
 
 
\bibitem{Sh} M.A.Shubin, 
Pseudodifferential Operators and Spectral Theory, 
Springer-Verlag, Berlin, 2001. 
 
\bibitem{SW1} 
A.Soffer, M.I.Weinstein, Multichannel nonlinear scattering for 
nonintegrable equations, {\em Comm. Math. Phys.} {\bf 133} (1990), 119-146. 
\bibitem{SW2} 
A.Soffer, M.I.Weinstein, 
Multichannel nonlinear scattering for nonintegrable equations. II. The case 
of anisotropic potentials and data. 
{\em J. Differential Equations}  {\bf 98} 
(1992) no. 2 376. 
 
 
 
 
\bibitem{Som} 
A.Sommerfeld, 
Atombau und Spektrallinien, Vol. I and II, Friedr. Vieweg $\&$ 
Sohn, Brounschweig, 1951. 
 
\bibitem{SS} 
A.Sommerfeld, G.Schur, {\em Ann. d. Phys.} {\bf 4} (1930), 409. 

 \bibitem{Sp} H.Spohn,
Dynamics of Charged Particles and Their Radiation Field,
Cambridge University Press, Cambridge, 2004.

\bibitem{St} 
 W.A.Strauss, Nonlinear Invariant Wave Equations, Lecture 
Notes in Phys.73 (1978), Springer, Berlin, 197-249. 
 
\bibitem{SuSu} 
C.Sulem, P.L.Sulem, 
The Nonlinear Schr\"odinger Equation. Self-Focusing and Wave Collapse, 
Springer, NY, 1999. 
 
\bibitem{SSA} 
A.Szabo, 
N.S.Ostlund, 
Modern Quantum Chemistry: 
Introduction to Advanced Electronic Structure Theory, 
Dover, 1996. 

\bibitem{Te} 
R.Temam, 
Infinite-Dimensional Dynamical Systems in Mechanics and Physics, 
 Applied Mathematical Sciences, Springer, Berlin, 1988. 
 
 \bibitem{Thal} 
B.Thaller, 
The Dirac Equation, 
Springer, Berlin, 1991. 
 
\bibitem{To} 
L.H.Thomas, {\em Nature} {\bf 117} (1926), 514. 
 
\bibitem{T} 
J.J.Thomson, 
{\em Phil. Mag.} {\bf 44} (1897), 298. 
 
 
 
 
\bibitem{Va} 
B.R.Vainberg, 
Asymptotic Methods in Equations of Mathematical Physics, 
Gordon and Breach, New York, 1989. 
 
 
 
\bibitem{We} 
G.Wentzel, {\em ZS. f. Phys.} {\bf 43} (1927), 1, 779. 
 
\bibitem{Weyl} 
H.Weyl, 
The Theory of Groups and Quantum Mechanics, 
Dover, NY, 1949. 
 
\bibitem{Wig} 
E.P.Wigner, Theory of Group, Academic Press, NY, 1959. 
 
\bibitem{Z} 
 E.Zeidler,
Applied Functional Analysis. 
Main Principles and Their Applications,
Applied Mathematical Sciences Vol.109, 
Springer, Berlin, 1995.



\end{thebibliography}
\end{document}